%% file: thesis.tex
\pdfoutput=1

\documentclass{phdthesis}

\definecolor{orange}{rgb}{1,0.647,0}
\definecolor{magenta}{rgb}{1,0,1}
\definecolor{cyan}{rgb}{0,1,1}
\definecolor{grey}{rgb}{0.25,0.25,0.25}

\newcommand{\eln}{El Ni\~{n}o\xspace}
\newcommand{\lan}{La Ni\~{n}a\xspace}
\newcommand{\degree}{\ensuremath{^{\circ}}}

\newcommand{\ztw}{\ensuremath{Z_{20}}\xspace}
\newcommand{\zgr}{\ensuremath{Z_{\mathrm{grad}}}\xspace}

\newcommand{\legenda}[1]{\color{#1}\begin{picture}(12,4)
    \thicklines\drawline(0,2)(12,2)
\end{picture}}
\newcommand{\legendb}[1]{\color{#1}\begin{picture}(12,4)
    \thicklines\dashline[+40]{3}(0,2)(12,2)
\end{picture}}
\newcommand{\legendc}[1]{\color{#1}\begin{picture}(12,4)
    \thicklines\dashline[+120]{1}(0,2)(12,2)
\end{picture}}

\DeclareMathOperator{\diag}{diag}
\DeclareMathOperator{\mathspan}{span}
\DeclareMathOperator{\Tr}{Tr}
\DeclareMathOperator{\Corr}{Corr}
\DeclareMathOperator*{\argmin}{arg\,min}
\DeclareMathOperator*{\argmax}{arg\,max}
\DeclareMathOperator{\sgn}{sgn}
\DeclareMathOperator{\var}{Var}
\DeclareMathOperator{\vol}{vol}
\DeclareMathOperator{\im}{im}

\DeclareRobustCommand*{\vec}[1]{\ensuremath{%
    \mathchoice{\mbox{\boldmath$\displaystyle#1$}}
               {\mbox{\boldmath$\textstyle#1$}}
               {\mbox{\boldmath$\scriptstyle#1$}}
               {\mbox{\boldmath$\scriptscriptstyle#1$}}}}

\newindex{not}{ndx}{nnd}{Table of Notation}

\begin{document}

\DeclareGraphicsExtensions{.pdf,.jpg,.png}

\thesistitle{Nonlinear Dimensionality Reduction Methods in Climate
  Data Analysis}
\thesisyear{2008}
\thesismonth{September}
\thesisauthor{Ian~Ross}
\thesisdegree{Doctor of Philosophy}
\thesisfaculty{Science}
\thesisdepartment{School of Geographical Sciences}
\thesiswordcount{ca. 80,000 words}
\thesisstartdate{October~2005}
\thesislastdate{September~2008}

%
%

\maketitlepages

\newpage
\begin{new-abstract}
  Linear dimensionality reduction techniques, notably principal
  component analysis, are widely used in climate data analysis as a
  means to aid in the interpretation of datasets of high
  dimensionality.  These linear methods may not be appropriate for the
  analysis of data arising from nonlinear processes occurring in the
  climate system.  Numerous techniques for nonlinear dimensionality
  reduction have been developed recently that may provide a
  potentially useful tool for the identification of low-dimensional
  manifolds in climate data sets arising from nonlinear dynamics.  In
  this thesis I apply three such techniques to the study of El
  Ni\~{n}o/Southern Oscillation variability in tropical Pacific sea
  surface temperatures and thermocline depth, comparing observational
  data with simulations from coupled atmosphere-ocean general
  circulation models from the CMIP3 multi-model ensemble.

  The three methods used here are a nonlinear principal component
  analysis (NLPCA) approach based on neural networks, the Isomap
  isometric mapping algorithm, and Hessian locally linear embedding.
  I use these three methods to examine El Ni\~{n}o variability in the
  different data sets and assess the suitability of these nonlinear
  dimensionality reduction approaches for climate data analysis.

  I conclude that although, for the application presented here,
  analysis using NLPCA, Isomap and Hessian locally linear embedding
  does not provide additional information beyond that already provided
  by principal component analysis, these methods are effective tools
  for exploratory data analysis.
\end{new-abstract}

\cleardoublepage
\begin{author-declaration}
  \setlength\parindent{0pt}

  I declare that the work in this dissertation was carried out in
  accordance with the Regulations of the University of Bristol.  The
  material presented here is the result of my own independent research
  performed at the University of Bristol, School of Geographical
  Sciences, between \thesisstartdate \hspace*{1pt} and
  \thesislastdate, and no part of the dissertation has been submitted
  for any other academic award.  Sections of
  Chapters~\ref{ch:data-and-models},~\ref{ch:enso}~and~\ref{ch:obs-cmip-enso}
  and all of Chapter~\ref{ch:isomap} have previously appeared as:

  \begin{center}
    \begin{minipage}{0.85\textwidth}
      I. Ross, P. J. Valdes and S. Wiggins.  ENSO dynamics in current
      climate models: an investigation using nonlinear dimensionality
      reduction.  \textit{Nonlin. Processes Geophys.}, 15(2):339--363,
      April 2008.
    \end{minipage}
  \end{center}

  Any opinions expressed in this thesis are those of the author. \\

  \vspace{2cm}
  \thesisauthor \\
  \thesismonth~\thesisyear
\end{author-declaration}

\cleardoublepage
\begin{acknowledgments}
  Many thanks to my supervisors, Paul Valdes and Steve Wiggins.  Paul,
  first of all, gave me a job and provided a nurturing and congenial
  environment, in the form of the BRIDGE group.  Paul's easy-going
  leadership really set the tone for BRIDGE (``field trips'' that
  consist of a weekend camping and surfing in Devon, anyone?), which
  ended up being a very productive arrangement for everyone concerned.
  I've certainly enjoyed being part of that and the group will be
  something I'll miss a lot when I leave Bristol.

  As for Steve, during the three years of my Ph.D., we missed just a
  handful of our weekly meetings due to his absence or other
  engagements.  For someone covering head of department
  responsibilities while maintaining an active research programme,
  that's an extraordinary level of commitment, one for which I am very
  grateful.  The only thing that (slightly) tempers this gratitude is
  Steve's habit of sending emails in the dead of night with wads of
  papers attached to them, all with the comment ``You really should
  know about this stuff...''.  As a result of Steve's
  ``encouragement'', I've probably read about five times as much as I
  otherwise would have done.  I even enjoyed some of it.

  Of the other BRIDGE-ites, special mention has to go to Rupes (for
  always refusing to understand things in the most enlightening
  fashion possible), Dan (``I trust David Blunkett!''), Gethin (a boy
  from Wales more interested in computers than sheep) and Rachel (her
  door is always open, she's always ready for a chat, and she lives at
  the bottom of the steepest hill in Somerset).  Also, apologies to
  anyone who's had to give a group seminar with me in the front row
  heckling (that's nearly everyone!).

  This is a thesis about climate data analysis, so we need some
  climate data.  I've used data from the NCEP atmospheric and ocean
  reanalyses, both truly excellent resources, I've used the NOAA ERSST
  v2 data set, and I've used GCM simulations archived for the IPCC
  Fourth Assessment Report.  There's a blurb that goes with the IPCC
  data: ``I acknowledge the modelling groups, the Program for Climate
  Model Diagnosis and Intercomparison (PCMDI) and the WCRP's Working
  Group on Coupled Modelling (WGCM) for their roles in making
  available the WCRP CMIP3 multi-model data set.  Support for this
  data set is provided by the Office of Science, U.S. Department of
  Energy''.  Those official words don't capture just how useful these
  multi-model ensemble databases are and what a job it is to organise
  them.  All kudos to the people involved!  On another official note,
  I should mention that my Ph.D. work was funded by an e-Science
  studentship from NERC, number NER/S/G/2005/13913.

  Finally, of course, an enormous thank you to Rita.  She lives with
  me, shares her life with me, sometimes works with me, even puts up
  with my ``jokes'', and yet through all of this, she maintains the
  sunniest of dispositions, the happiest of smiles.  As anyone who
  knows me will attest, this must mean that she is a very angel.
  We've had a lot of fun over the last four and a half years,
  including some things that were more ``fun'' than fun (the
  completion of two Ph.D.s, broken collarbones, invisible fishbones,
  immigration anxieties), but some that were absolute unalloyed
  \textbf{FUN} (holidays in Ireland, Greece, even Austria, and every
  everyday day).  I am absolutely sure that we will have many years
  more.  Life is good, and the reason is Rita.
\end{acknowledgments}

\cleardoublepage
\tableofcontents

\cleardoublepage
\lotTOC

\cleardoublepage
\lofTOC

%
%

\pagenumbering{arabic}
\renewcommand\chapterheadstartvskip{\vspace*{-5\baselineskip}}

\cleardoublepage
\include{01-intro}

\cleardoublepage
\include{02-nldr-overview}

\cleardoublepage
\include{03-data-and-models}

\cleardoublepage
\include{04-enso}

\cleardoublepage
\include{05-obs-cmip-enso}

\cleardoublepage
\include{06-nlpca}

\cleardoublepage
\include{07-isomap}

\cleardoublepage
\include{08-hessian-lle}

\cleardoublepage
\include{09-conclusions}

%
%

\renewcommand\chapterheadstartvskip{\vspace*{2.3\baselineskip}}

\cleardoublepage
\addcontentsline{toc}{chapter}{Table of Notation}
\label{pg:notation}
\printindex[not][This table of notation gives the page of definition
  of all special notation used in this thesis.]

\cleardoublepage
\chaptertoc{Glossary}
\markboth{GLOSSARY}{GLOSSARY}
\input{glossary.tex}

\cleardoublepage
\addcontentsline{toc}{chapter}{Bibliography}
\bibliographystyle{plainnat}
\bibliography{thesis}

\end{document}

%% file: 01-intro.tex
\chapter{Introduction}
\label{ch:intro}

Recent advances in observational and modelling technology have led to
a situation in climate science that would have been unthinkable even a
few years ago.  Our problem?  We have too much data! Satellite
instruments and improved {\it in situ} monitoring networks observe
variations in the climate system in unprecedented detail, while modern
general circulation models (GCMs) simulate atmospheric, ocean and land
surface processes at high spatial and temporal resolution.  New
methods and novel tools are needed to analyse the resulting glut of
data.

The problem is not simply the quantity of data, but that the data is
represented as points in high-dimensional space, recording many
simultaneous measurements.  As an example, consider an atmospheric GCM
that outputs a time series of geopotential height on several pressure
levels.  Each entry in this time series can be viewed as a single
vector in $\mathbb{R}^m$ (\index[not]{R@$\mathbb{R}$, the real
  numbers}$\mathbb{R}$ is the real numbers\footnote{References to
  definitions of all non-standard notation can be found in the table
  of notation on page~\pageref{pg:notation}.}, and
\index[not]{Rm@$\mathbb{R}^m$, $m$-dimensional Euclidean
  space}$\mathbb{R}^m$ is $m$-dimensional Euclidean space), where $m$
is the number of spatial points in the model grid.  For the UK Met
Office HadCM3 model \citep{gordon-ukmo-hadcm3}, $m = 133,\!152$,
representing a $96 \times 73$ horizontal grid with horizontal spatial
resolution of $3.75^\circ \times 2.5^\circ$ (longitude $\times$
latitude), with 19 vertical levels through the atmosphere.  Collecting
data on only three pressure levels
($\mathrm{850\,hPa}$\footnote{$\mathrm{1\,hPa = 100\,Pa}$.  This is a
  convenient unit for measurement of atmospheric pressure --- sea
  level pressure is around $\mathrm{1000\,hPa}$, while
  $\mathrm{500\,hPa}$ represents a vertical level approximately
  half-way through the atmosphere, in terms of mass.},
$\mathrm{500\,hPa}$ and $\mathrm{250\,hPa}$, say, for a view of the
lower, middle and upper troposphere), one still has $m = 21,\!024$.

Thinking of such a model as a dynamical system, high-dimensional data
of this type is difficult to interpret.  Although two-dimensional
geographical maps of geopotential height can be plotted on a single
pressure level at a single time, this view is not appropriate for
considering the dynamics of the system.  To do this, the whole state
of the model at a given timestep should be considered as a single
point in the phase space $\mathbb{R}^m$.  The modern dynamical systems
approach then considers evolution of the system as controlled by
geometrical structures in phase space, such as periodic orbits, saddle
points, and so on \citep{wiggins-applied-nonlin}.

However, our situation is far from hopeless.  It is a commonplace of
observational meteorology and climatology that the evolution of the
atmosphere and ocean is characterised by recognisable and recurrent
coherent structures, such as synoptic weather systems in the
atmosphere or mesoscale eddies in the ocean.  The existence of these
coherent structures represents a coupling between many individual
degrees of freedom, couplings that persist for extended periods of
time.  This coherent behaviour leads to the hope that it may be
possible to derive a simplified representation of the evolution of the
atmosphere or ocean, eliminating degrees of freedom that are in some
sense ``uninteresting'', to concentrate on degrees of freedom that
capture the large-scale coherent structures.  This simplification
represents a reduction of the dimensionality of the system: we go from
our original high-dimensional representation including all of the
degrees of freedom of the system to a lower-dimensional representation
capturing the essential features of interest.

Some encouragement for the project of constructing lower-dimensional
representations of phenomena of interest in the climate system can be
drawn from results in the rigorous functional analysis of partial
differential equations (PDEs).  Here, the long term behaviour of these
infinite dimensional systems is found to be confined to a finite
dimensional global attractor \citep{robinson-finite-pdes}.  In some
cases, it can be proven that this attractor is embedded in a finite
dimensional manifold, called the inertial manifold of the system.  In
this case, the long term dynamics of the infinite dimensional PDE
system is rigorously equivalent to a finite dimensional system of
ordinary differential equations describing a flow on the inertial
manifold.  Bounds on the dimensionality of the inertial manifold can
sometimes be derived in terms of system parameters.

While these results are both theoretically appealing and consonant
with our intuitive notions of long term coherent behaviour in fluid
systems, they are of relatively limited practical applicability.  The
bounds on attractor and inertial manifold dimension are typically very
high, and the existence of inertial manifolds has only been proven for
a restricted set of problems, a set excluding most of the equations of
interest in applications to geophysical fluids and the climate system
\citep{foias-ns-turb}.

Another source of encouragement in the project of dimensionality
reduction for climate dynamics lies in empirical observations of
coherent structures in other fluid flows and, more generally, the
existence of coherent dissipative structures for a wide range of
nonlinear partial differential equation systems
\citep{cross-hohenberg}.  The presence of these coherent patterns is a
strong indication that aspects of the behaviour of these systems may
be represented by an effective low-dimensional model.

In this thesis, I report on the application of a number of methods of
dimensionality reduction to a problem in climate data analysis, namely
the study of interannual tropical Pacific climate variability and the
\eln/Southern Oscillation (ENSO).  This problem is approached in the
context of an inter-model comparison using the World Climate Research
Programme's (WCRP's) Coupled Model Intercomparison Project phase 3
(CMIP3) multi-model data set.  The primary goal here is to explore the
applicability of some nonlinear dimensionality reduction methods to a
relatively well understood problem in climate data analysis.  It is
unlikely that such a study will discover anything new about ENSO
itself, but it is likely to help elucidate differences in behaviour
between the models examined.

Our question here is, given high-dimensional data from observations or
model simulations, what is the best way to characterise
low-dimensional behaviour?  We are interested in attempting to infer
low-dimensional dynamics from relatively limited amounts of data.
Observational time series from the Pacific provide around 100 years of
monthly sea surface temperatures, and less than 30 years of
comprehensive coverage of sub-surface ocean temperature and current
fields.  Time series of several hundred years are available from
coupled GCM simulations.  Throughout this thesis, in order to
facilitate inter-model comparison, we will proceed in a ``black box''
fashion, adopting a purely data-driven approach without using
information about the internal features of the models we are studying.

The goal of all dimensionality reduction techniques is to construct a
lower-dimensional representation of a data set or dynamical system
that, in some sense, captures the important characteristics of the
variability of the original system.  This rather vague formulation
clearly encompasses a vast range of problems and techniques in
different fields.  The literature on dimensionality reduction reflects
this range, both in methods and in applications.  To provide some
context for the selection of the nonlinear dimensionality reduction
methods used here, Chapter~\ref{ch:nldr-overview} provides a
reasonably extensive survey of the literature on nonlinear
dimensionality reduction, with some emphasis, from the point of view
of applications, on earlier work in climate data analysis.
Chapter~\ref{ch:data-and-models} describes the observational data
sets, the models and some test data sets used here.
Chapter~\ref{ch:enso} describes the basic phenomenology of ENSO and
reviews theoretical ideas about the mechanisms underlying ENSO
variability, as well as describing approaches to the modelling of ENSO
and previous applications of nonlinear dimensionality reduction to
this problem.  Chapter~\ref{ch:obs-cmip-enso} describes some basic
results concerning interannual tropical climate variability in
observations and the CMIP3 models, in order to provide a background
for the interpretation of the later nonlinear dimensionality reduction
results.

Chapters~\ref{ch:nlpca}--\ref{ch:hessian-lle} present results of the
application of three ``geometrical/statistical'' nonlinear
dimensionality reduction methods to the climate data sets considered
here, namely nonlinear principal component analysis (NLPCA), Isomap
and Hessian locally linear embedding (Hessian LLE, also known as
Hessian eigenmaps).  Finally, Chapter~\ref{ch:summary} provides a
summary of results and some suggestions for further work.

The three dimensionality reduction methods examined here were chosen
from the large number of methods described in
Chapter~\ref{ch:nldr-overview} for a number of reasons.  NLPCA
(Chapter~\ref{ch:nlpca}) has been applied to many different climate
data analysis applications
\citep[e.g.,][]{monahan-enso,hamilton-nlpca-qbo,wu-enso-interdecadal,hsieh-review-2004,casty-nh-regimes},
but does not previously appear to have been used for an inter-model
comparison of the type performed here.  Most previous studies using
NLPCA have applied the method to only a single observational data set.
It is of interest to determine how well NLPCA (and the other methods
explored here) can capture the differences in behaviour seen in
different models, and to see whether these nonlinear methods can
represent those differences in an intuitively accessible way.  Isomap
is selected as the second method applied here
(Chapter~\ref{ch:isomap}) because it is one of the two most commonly
applied nonlinear dimensionality reduction methods, the other being
locally linear embedding (LLE).  Isomap and LLE are the most
frequently used representatives of, respectively, \emph{global} and
\emph{local} geometrical/statistical dimensionality reduction methods.
Isomap has seen only one previous application in climate data analysis
\citep{gamez-enso,gamez-enso-2}, also in the context of analysis of
ENSO behaviour, and it is again of interest to see how it performs in
a model comparison setting.  The final method examined here, Hessian
LLE or Hessian eigenmaps (Chapter~\ref{ch:hessian-lle}), is a
relatively new method that has not, as far as I know, been applied to
any serious applications since its initial description by
\citet{donoho-hessian}.  However, it shares some computational
features with the important LLE method, while being significantly more
amenable to analysis than the original LLE algorithm.  Despite its
clear theoretical appeal, Hessian LLE has some characteristics that
lead one to expect that it might be rather numerically unstable, and
these issues are explored in Chapter~\ref{ch:hessian-lle}, as well as
describing the application of the method to the processing of ENSO
data.


%% file: 02-nldr-overview.tex
\chapter{Overview of Nonlinear Dimensionality Reduction}
\label{ch:nldr-overview}

The literature on nonlinear dimensionality reduction methods is vast.
A huge range of methods for analysing high-dimensional dynamical
systems and data sets have been developed in a number of different
fields.  In this chapter, I attempt to review some of this literature,
to provide an overview of previous work and to draw some links between
the rather disparate communities that have developed these methods.
No claim is made that the treatment here is comprehensive ---
applications, in particular, are referenced relatively sparsely, with
just a few indicative studies being mentioned for each method treated.
Also, here I treat in detail only methods developed for the analysis
of data sets (\emph{geometrical/statistical methods}), neglecting
methods developed for the analysis of dynamical systems represented
explicitly as equations (\emph{dynamical methods}).  This focus
reflects the methods most likely to be useful for climate data
analysis.  Although the use of simplified models of the atmosphere is
widespread in studies of low-dimensional behaviour in the climate
system, it is inevitable that analysis of both observations and
results from more complex models will require the adoption of a
data-centred viewpoint.  Among the dynamical dimensionality reduction
methods that have been used in climate science applications are a
range of Galerkin projection approaches
\citep{hasselmann-pips-pops,achatz-opsteegh-i,kwasniok-pips-2,kwasniok-pips-3,crommelin-bases},
methods based on stochastic averaging
\citep{majda-stoch-climate-models-2,franzke-barotropic,franzke-gcm}
and methods based on hidden Markov models and other Markov methods
\citep{pasmanter-cyclic-markov,crommelin-mc,horenko-hmm-pca-sde}.
Theoretical ideas concerning the existence of global attractors and
inertial manifolds for dissipative partial differential equations are
also important for understanding the relationship between variability
on different timescales in the atmosphere and ocean and the existence
of a well-defined notion of ``climate''
\citep{temam-turb-ims,foias-ns-turb,dymnikov-attractors}.

As well as dimensionality reduction itself, there are a number of
related problems often treated by comparable methods, such as
clustering and classification.  A good example of the cross-over
between clustering and dimensionality reduction methods is the work of
\citet{kushnir-multiscale-clustering}, who developed a method for
simultaneous dimensionality reduction and cluster identification in
high-dimensional data.  Other application areas that can be viewed
from a dimensionality reduction viewpoint include synchronisation,
where the relationship between different parts of a coupled system can
often be represented by a so-called synchronisation manifold, an
invariant manifold of the coupled system \citep[e.g.,][]{josic-sync},
and control theory, where the control of high-dimensional systems is
often simplified by dimensionality reduction
(\citet{montgomery-adaptive-control} and
\citet{kreuzer-dimred-control} provide simple examples) and where
``equation free'' methods seem to offer the possibility of applying
linear feedback control theory to systems defined by very
high-dimensional microscopic models \citep{siettos-coarse-bifurc}.
Space limitations prevent further exploration of these areas here.

The diversity of dimensionality reduction methods renders a direct
intercomparison between different methods very difficult.  There are
very few studies comparing the performance of different methods on the
same realistic problem, and the simple test problems used for
demonstrating the performance of new dimensionality reduction methods
vary widely between the different fields for which new methods are
developed.  This leads to a frustrating situation, for both the reader
and the author of a review such as this, since it appears to be
impossible to provide a clear answer to the question ``Well, which
method is better for application X?'' without going quite far beyond a
simple literature review.  The information to answer this question in
most cases simply does not exist.  Some ideas for future work to help
alleviate this problem are presented in Chapter~\ref{ch:summary}, but
this handicap should be borne in mind in what follows.  In most cases
where no comparison between methods is offered, this is because no
such comparison has ever been conducted in a realistic setting.

There are a number of reviews of geometrical/statistical methods of
dimensionality reduction available, most of which are slightly more
narrowly focused than the coverage here.  Among the most useful of
these are \citep{burges-review}, \citep{fodor-review} and
\citep{cayton-review}, each of which describes most of the more common
dimensionality reduction methods.

We will begin by reviewing the notion of \emph{dimension} as it
appears in different fields of mathematics before settling on a simple
operational definition to be used in the following discussion.

\section{Definitions of dimensionality}

The notion of \emph{dimension} is fundamental to many areas of
mathematics, and there are consequently a number of different
definitions in common use.

The most basic and intuitive ideas of dimensionality arise in geometry
and the study of vector spaces.  In this context, the dimension counts
the number of independent ``directions'' in a space.  This informal
idea is made precise, in the context of a vector space, by defining
the \emph{Hamel dimension} to be the cardinality of a basis for the
vector space \citep[e.g.,][]{strang-linear-algebra}.  This definition
extends naturally to manifolds\footnote{Definitions of common terms
  and concepts from differential geometry required to treat manifolds
  are given in Section~\ref{sec:nonlinear-diff-geom} below.}: one can
either consider the dimensionality of the tangent spaces at each point
in the manifold (which are vector spaces), or one can observe that
local coordinate charts for a manifold are homeomorphisms between open
neighbourhoods of the manifold and open subsets of Euclidean space, so
that the dimensionality of the manifold is simply the dimensionality
of the appropriate Euclidean space \citep[Chapter III]{choquet-bruhat-amp}.

In more abstract settings, the dimensionality of other mathematical
structures (topological spaces, for example) may be defined in a
variety of ways.  These definitions are not particularly relevant to
our main interest here, which is in the dimensionality of the phase
spaces of various dynamical systems, represented by either vector
spaces or manifolds.  For some further discussion and speculation on
ideas of dimensionality for structures in algebraic geometry, see
\citep{manin-dimension}.

Beyond simple definitions based on the cardinality of bases, several
dimensionality measures have been developed for characterising the
``size'' of point sets embedded in Euclidean space.  These methods are
of great relevance to dynamical systems theory because of the tendency
of trajectories of dissipative dynamical systems to accumulate on
attractors, sets of measure zero in the state space of the dynamical
system \citep[Section~8.2]{wiggins-applied-nonlin}.  A natural way to
distinguish between different types of attractor is by determining
their dimension.  In the case of attractors that are fixed points,
periodic orbits or invariant tori, this characterisation by dimension
is straightforward and corresponds to the simple definition of the
dimensionality of a manifold presented above.  In cases where chaotic
dynamics are encountered, attractors may be \emph{strange}, and can,
in some sense, be considered to have non-integral dimension.  Several
definitions of dimension have been developed in this context, with the
intention of providing a finer distinction between point sets of
different ``size'' than traditional notions of dimensionality which
always yield an integral dimension, and which may not apply at all in
the case of more complicated sets.  These methods for measuring the
dimensionality of point sets generally rely on scaling behaviour of
some function of a cover of the set in the limit as the cover becomes
infinitely fine.

A typical and useful example is the Hausdorff-Besicovich dimension,
defined for an arbitrary subset $S$ of some metric space $M$.  Here, I
follow the presentation of \citet{manin-dimension}.  A $d$-dimensional
ball in Euclidean space, $B_\rho$, of radius $\rho$, with $d$ a
natural number, has volume\footnote{Here,
  \index[not]{Gamma@$\Gamma(z)$, gamma function}$\Gamma(z)$ is the
  standard gamma function, defined by the integral $\Gamma(z) =
  \int_0^\infty e^{-s} s^{z-1} \, ds$ for any complex $z$ (except for
  the negative integers, where $\Gamma(z)$ has poles and where the
  integral does not converge).  The gamma function has the property
  that $z \Gamma(z) = \Gamma(z + 1)$.}
\begin{equation}
  \vol_d(B_\rho) = \frac{\pi^{d/2}}{\Gamma(1 + d/2)} \rho^d.
\end{equation}
We now \emph{define} the volume of a $d$-dimensional ball for any real
$d$ via this formula.  We cover our set $S$ with a finite number of
balls of radii $\rho_m$ and try to count the $d$-dimensional volume of
$S$ as if it were truly a $d$-dimensional object for some real $d$:
\begin{equation}
  v_d(S) = \lim_{\rho \to 0} \inf_{\rho_m < \rho} \sum_m
  \vol_d(B_{\rho_m}).
\end{equation}
Here, for a given $\rho$, we find the minimum total volume of balls
covering $S$ with $\rho_m < \rho$, and consider the limit of this
quantity as the maximum radius of the balls in the cover goes to zero,
giving a finer and finer cover of $S$.  The surprising outcome of this
procedure is that, for any compact closed $S \subset M$, there exists
a value $D_{\mathrm{HB}}$, called the Hausdorff-Besicovich dimension,
such that $v_d(S) = 0$ for $d < D_{\mathrm{HB}}$ and $v_d(S) = \infty$
for $d > D_{\mathrm{HB}}$.  Unlike definitions of dimension based on
counting elements in a set (e.g., the number of elements in a basis,
or the number of overlapping open sets to which a point can belong in
a minimal finite cover, as for the topological dimension), this
definition can give non-integral dimension values.  The classic
example is the Cantor middle-thirds set
\citep[Section~11.2]{strogatz-book}, for which $D_{\mathrm{HB}} = \log
2 / \log 3$, but the attractors of many dynamical systems are also
known to have non-integral $D_{\mathrm{HB}}$.

Although it has a certain intuitive appeal, in practice the
Hausdorff-Besicovich dimension is rather difficult to calculate.  In
particular, it is desirable to have a definition of dimensionality
that is applicable not only to systems defined analytically, but also
to time series, derived either from observations or from numerical
computation of the evolution of some system.  The correlation
dimension, introduced by \citet{grassberger-dimension}, is such a
definition.  Calculate the correlation sum $C(\varepsilon)$ for a time
series of points $\vec{x}_i \in \mathbb{R}^m$ (throughout, we use a
bold italic font to indicate \index[not]{vector@$\vec{x}$, vector
  quantities}vector quantities), with $i = 1, \dots, N$, as
\begin{equation}
  C(\varepsilon) = \frac{2}{N(N - 1)} \sum_{i=1}^N \sum_{j=1}^N
  \Theta(\varepsilon - || \vec{x}_i - \vec{x}_j ||),
\end{equation}
where \index[not]{Theta@$\Theta(x)$, Heaviside step function}$\Theta$
is the Heaviside step function\footnote{$\Theta(x) = 0$ for $x \leq 0$
  and $\Theta(x) = 1$ for $x > 0$.}, and $|| \bullet ||$ denotes the
usual \index[not]{Euclidean norm@$|| \vec{x} ||$, Euclidean norm of a
  vector}Euclidean norm of a vector, $||\vec{x}|| = ( \sum_i x_i^2
)^{1/2}$.  The correlation sum counts the number of pairs of points
$(\vec{x}_i, \vec{x}_j)$ whose point-to-point distance is less than
$\varepsilon$.  In the limit where $N \to \infty$ (infinite amount of
data) and $\varepsilon \to 0$, we expect to see scaling behaviour, so
that $C(\varepsilon) \propto \varepsilon^{D_{\mathrm{corr}}}$ for some
definite value $D_{\mathrm{corr}}$.  We thus define the correlation
dimension $D_{\mathrm{corr}}$ as
\begin{equation}
  D_{\mathrm{corr}} = \lim_{N \to \infty} \lim_{\varepsilon \to 0}
  \frac{d \log C(\varepsilon)}{d \log \varepsilon}
\end{equation}
This basic definition is easily applied to common geometrical objects
and yields the expected geometrical dimensions.  For time series data,
some care is required in the computation of $D_{\mathrm{corr}}$, and a
number of techniques have been developed to avoid problems due to time
correlation in the input data, sampling issues and noise.
\citet{kantz-ts-book} provide fairly exhaustive coverage of the
relevant methods, including an extensive bibliography, and also
provide software to apply these and other nonlinear time series
analysis methods \citep{hegger-tisean}.

There are several other approaches to assigning a dimension to a point
set or time series in the same spirit as the correlation dimension,
such as capacity dimension, box counting dimension, R\'{e}nyi
dimensions and information dimension, each of which provides a more or
less fine distinction between sets of different ``size''.  A number of
relationships are known between the different definitions, and these,
as well as practical computational methods are again described by
\citet{kantz-ts-book}.

One final dimension definition deserving of mention is the
Kaplan-Yorke or Lyapunov dimension, which is based on the Lyapunov
spectrum of a dynamical system
\citep{farmer-dimensions,frederickson-ky-dim}.  The idea here is to
use the Lyapunov exponents of a system, $\lambda_i$, sorted in
descending order of magnitude, to determine the dimension of the
system's attractor by considering the balance between stretching and
compression of phase space volumes as the system evolves between
states on the attractor.  Since the attractor of the system is an
invariant set, when considered as a $D$-dimensional sub-volume of the
system state space, it neither shrinks nor expands in volume as the
system evolves.  Knowing the Lyapunov exponents of the system, we can
then seek a value of $D$, $D_{\mathrm{KY}}$, such that this
volume-preserving property is true.  A finite one-dimensional subset
of phase space in the neighbourhood of the attractor will be stretched
exponentially by the evolution of the system at a rate $e^{\lambda_1
  t}$ determined by the first Lyapunov exponent.  Assuming that
$\lambda_1 > 0$, such a one-dimensional subset thus does not have
constant volume.  If the second Lyapunov exponent is negative, with
$\lambda_2 < -\lambda_1$, then a typical two-dimensional area is
stretched in one direction at rate $e^{\lambda_1 t}$ and shrinks in
the orthogonal direction at rate $e^{\lambda_2 t}$, giving a total
rate of shrinkage in area of $e^{-(|\lambda_2|-\lambda_1)t}$.  (Here,
\index[not]{absolute value@$| x |$, absolute value}$| \bullet |$
denotes the absolute value of a real number.)  If the attractor is
fractal in nature, then its projection onto the contracting direction
in state space may be a Cantor-like set with dimension
$D_{\mathrm{KY}} - 1 < 1$.  This fractal object will have a volume
invariant under the flow of the system if $\lambda_1 +
(D_{\mathrm{KY}} - 1)\lambda_2 = 0$, i.e. if $D_{\mathrm{KY}} = 1 +
\lambda_1/|\lambda_2|$.  The natural generalisation of this idea to
higher-dimensional cases is based on the suggestion that the integer
part of the dimension of the attractor should be identified with the
maximal number of Lyapunov exponents, in descending order of
magnitude, that can be added to give a positive sum: this identifies
the highest dimensionality subsets of the state space of the system
that are stretched in volume by the evolution of the system.  The
fractional part of the dimension is found by a simple linear
interpolation, as above.  The Kaplan-Yorke dimension is thus defined,
in a fairly intuitive way, as
\begin{equation}
  D_{\mathrm{KY}} = k + \frac{\sum_{i=1}^k \lambda_i}{| \lambda_{k+1}
    |},
\end{equation}
where $\sum_{i=1}^k \lambda_i \geq 0$ and $\sum_{i=1}^{k+1} \lambda_i
< 0$.  This definition and conjectures relating it to dimension
definitions based on scaling computations, particularly the
information dimension, provide a close link between the dynamics of a
system and the dimensionality of its attractor.

Despite the nice theoretical links to be made between dynamics and
attractor dimension represented by some of the definitions presented
above, we will take a simpler view of the dimension of a dynamical
system throughout the rest of this thesis.  We will be considering
methods for reducing the dimensionality of either dynamical systems or
data sets nominally resulting from integration of dynamical systems.
It therefore seems beneficial to adopt a simple and widely applicable
operational definition of dimensionality that measures the number of
independent parameters required to uniquely identify states of our
dynamical system, i.e., to uniquely identify points in the state space
of our system.

For a typical continuous-time dynamical system defined as
\begin{equation}
\frac{d\vec{x}}{dt} = f(\vec{x}),
\end{equation}
with $\vec{x} \in \mathbb{R}^m$, the state space of the system is
simply $m$-dimensional Euclidean space.  The initial temptation is to
identify $m$ as the dimensionality of the system.  Similarly, in cases
where evolution of the system occurs on some manifold $M$, we might
identify the dimensionality of the system with the dimensionality of
the manifold.  However, this is only a starting point, since the
behaviour of the system may effectively lie in a lower-dimensional
subset of phase space.  There are several cases where this situation
arises.  In dissipative dynamical systems, the long-term dynamics of
the system occur on an attractor, a lower-dimensional subset of the
original phase space that can in many cases be embedded in a
submanifold of the phase space.  In a slightly more general sense, in
systems with a timescale separation between fast and slow degrees of
freedom, dynamics on a \emph{slow manifold} may be seen.  A slow
manifold is an invariant manifold of the motion of the system, not
necessarily an attractor, to which motions of the system are attracted
quickly (on the fast timescale of the system).  Such a slow manifold
provides a dimensionality reduction of the system in the sense that
the dynamics of the system quickly decays to a motion on the slow
manifold, and, once the dynamics lie effectively in the slow manifold,
the fast degrees of freedom can be expressed as a function of the slow
degrees of freedom.

Low-dimensional behaviour can also be observed in non-dissipative
systems: it is not the case that low-dimensional behaviour implies the
existence of a low-dimensional attracting subset.  Consider a
conservative dynamical system whose phase space is a smooth Riemannian
manifold $M$ with inner product \index[not]{Riemannian metric@$\langle
  u, v \rangle_M$, inner product}$\langle \bullet, \bullet \rangle_M$,
i.e. a smooth bilinear form $\langle \bullet, \bullet \rangle_M :
T_p{}M \times T_p{}M \to \mathbb{R}$, with \index[not]{tangent
  space@$T_p{}M$, tangent space}$T_p{}M$ being the tangent space to
$M$ at a point $p \in M$.  Let $W$ be a smooth potential function $W:
M \to \mathbb{R}$.  Then the dynamics of this system are governed by
the Lagrangian
\begin{equation}
  \mathcal{L}(x, \dot{x}) = \frac{1}{2} \langle \dot{x}, \dot{x}
  \rangle_M - W(x),
\end{equation}
where $\dot{x} \in T_x{}M$ is the system velocity at point $x$, lying
in the tangent space $T_x{}M$.  Consider a family of singularly
perturbed potentials of the form
\begin{equation}
  W_\varepsilon(x) = V(x) - \varepsilon^{-2} U(x),
\end{equation}
where $\varepsilon \ll 1$ parameterises the family and where the
``strong'' potential $U$ acts to constrain the motion of the system to
a submanifold $N \subset M$, i.e. $U(x) = 0$ for $x \in N$.  For
initial conditions with uniformly bounded energy, the solutions to
this system, $x_\varepsilon$, oscillate within a distance
$O(\varepsilon)$ of $N$ on a timescale of
$O(\varepsilon)$\footnote{Formally, the notation \index[not]{big-O
    notation@$O(\varepsilon)$, big-O notation}$O(\varepsilon)$ is a
  case of the more general usage that $f(x) \sim O(g(x))$ as $x \to
  \infty$ if, for some $x_0$, there exists a value $A$ such that
  $|f(x)| < A |g(x)|$ for all $x > x_0$, i.e. $f(x)$ is bounded above
  asymptotically by $g(x)$, up to a constant factor.  In the simple
  case here, this simply means that variations about the submanifold
  $N$ are bounded by $\varepsilon$.}.  In the limit $\varepsilon \to
0$, the sequence of solutions $x_\varepsilon$ converges uniformly to
some function of time, $x_0$, taking values in $N$.  One can then seek
a dynamical description of this limit, asking the question: is there a
dynamical system with phase space $N$ such that $x_0$ is the solution
of the corresponding equations of motion?  This type of problem is
generally referred to as a \emph{homogenisation problem}
\citep{bornemann-homogenization}.  Note that the submanifold $N$ is
\emph{not} an invariant manifold of the original problem, so is not a
slow manifold of the system, but that the solution to the
homogenisation problem clearly provides a reduced dimensionality
representation of the original system.  This example demonstrates the
importance of distinguishing between the attracting set/attractor
behaviour seen in dissipative systems and other ways of looking at
low-dimensional behaviour in other kinds of systems.

\begin{figure}
  \begin{center}
    \includegraphics[width=\textwidth]{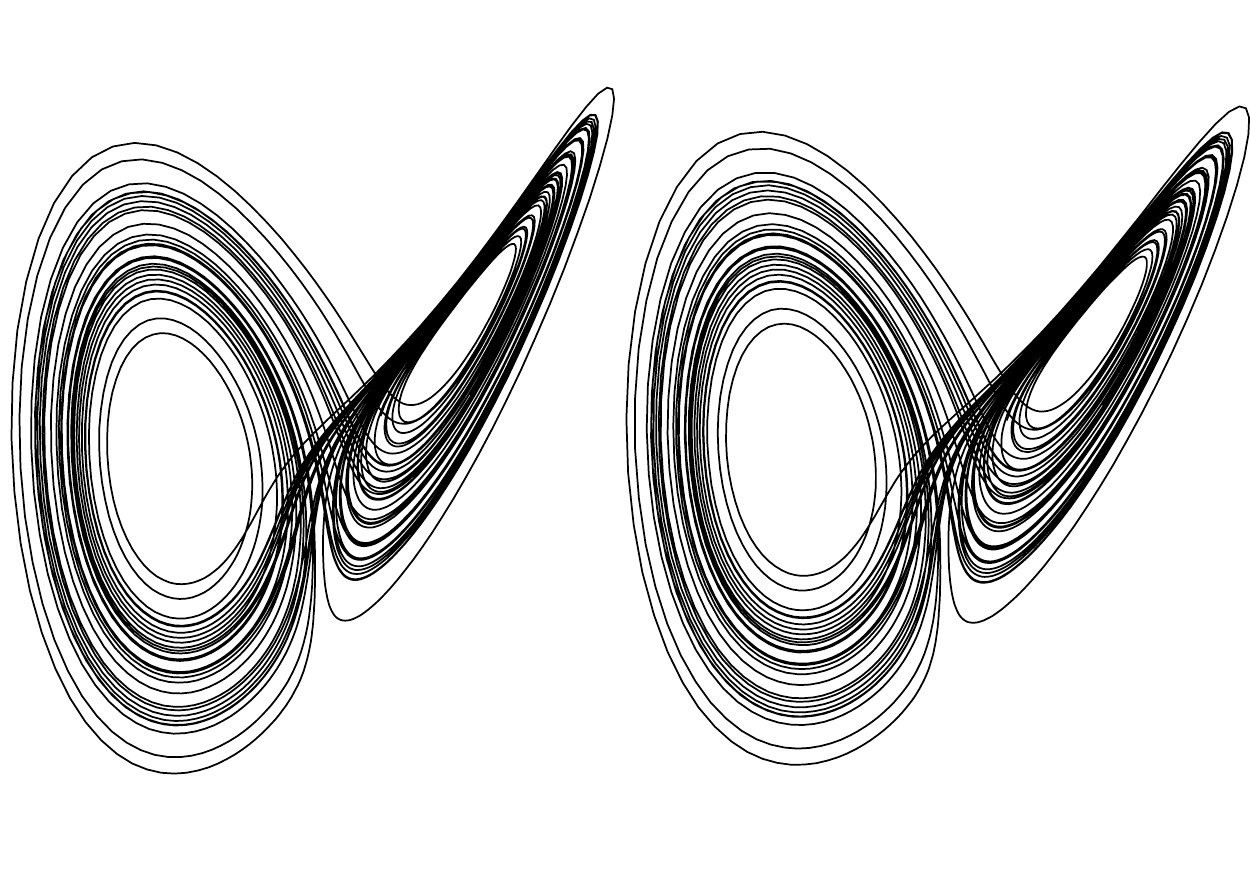}
  \end{center}
  \caption[Lorenz attractor]{Stereo pair of a segment of a trajectory
    lying in the attractor of the Lorenz system.}
  \label{fig:lorenz-attractor}
\end{figure}

One justification for adopting the seemingly unsophisticated
definition of dimension used here, disregarding all the other
possibilities presented above, is the following.  For a
finite-dimensional dissipative dynamical system, the long term
evolution of the system lies on an attractor embedded in the system
state space.  To some extent, it would be useful to identify the
dimensionality of the system with the dimensionality of that
attractor, particularly in situations where the dimensionality of the
attractor is very much lower than that of the state space.  However,
the structure of the attractors of dissipative dynamical systems is
not ``nice''.  (This is the primary reason for the proliferation of
methods for measuring their dimensionality.)  Generically, if a system
exhibits chaotic dynamics, the attractor, while smooth in some
directions (associated with stretching due to positive Lyapunov
exponents) will have complicated self-similar fractal structure in
other directions \citep[Chapter~30]{wiggins-applied-nonlin}.  This
means that it is difficult to envisage a natural way to parameterise
points on the attractor that does not make use of a coordinate system
for a linear subspace of the system state space in which the attractor
can be embedded.  As an example, consider the attractor for the
classic Lorenz system \citep{lorenz-1963}.  A segment of a trajectory
lying in the attractor of this system is shown in
Figure~\ref{fig:lorenz-attractor}.  This is a three-dimensional
ordinary differential equation system, which for the standard choice
of parameters has a Hausdorff-Besicovich dimension of around 2.06.  In
this case, the attractor is, in some sense, almost two-dimensional,
composed of two two-dimensional sheets that surround unstable steady
states, with the two sheets appearing to merge in the lower part of
Figure~\ref{fig:lorenz-attractor}.  Because of this merging ``two
sheet'' structure, there is no straightforward scheme for assigning
coordinates to points on the attractor that uses less than three
dimensions.  A scheme could be constructed to assign two-dimensional
coordinates on each of the ``sheets'', but additional information is
then needed to record on which of the two sheets a particular
trajectory of the system lies at any given point in time.  In this
simple example, the ``linear subspace of the system state space in
which the attractor can be embedded'' is the whole of the
three-dimensional state space, but for higher-dimensional examples,
this need not be the case, and the linear subspace of the state space
containing the attractor may be of strictly lower dimension than the
original state space.  In these cases, the number of coordinates
needed to uniquely identify a point on the attractor, while smaller
than the dimensionality of the original high-dimensional phase space,
is still greater than the dimensionality of the attractor as reported
by any of the scaling-based dimensions defined above.

%
%
%

When we come to consider dimensionality reduction in the context of
dynamical systems with state space $\mathbb{R}^m$, we will generally
think of some sort of projection $\phi : \mathbb{R}^m \to M$, where $M
\subset \mathbb{R}^m$ is an $n$-dimensional manifold.  The dimension
of our reduced system will then be $n$, the number of coordinates
required to parameterise points in the reduced state space, $M$.  One
observation to be made here is that, being a projection, $\phi$ is
non-injective, so that multiple states of our original system are
identified with a single state of the reduced system.  This
multiplicity of ``microstates'' of the original system corresponding
to ``macrostates'' of the reduced system mirrors the situation in
statistical mechanics, where one averages over microscopic degrees of
freedom to derive a representation in terms of macroscopic order
parameters.

\section{Dimensionality reduction}

So, what does \emph{dimensionality reduction} mean?  And why would we
want to do it?  The answer to the first question is simple.  We wish
to take a high-dimensional dynamical system, either in the form of a
set of equations, or in the form of a data set produced by the
evolution of our system, and produce a lower-dimensional
representation of the system, again either as a set of equations or as
some form of data set, that captures the essential characteristics of
the evolution of our dynamical system.  Here, what is meant by
``captures the essential characteristics'' is very much dependent on
the dimensionality reduction method used, and differs greatly between
equation-based dynamical methods and data-based
geometrical/statistical methods.

As an example, consider a continuous time dynamical system with only
quadratic nonlinearities, as is often encountered in climate modelling
applications \citep[e.g.,][]{majda-stoch-climate-models-2}.  We write
the system state as $\vec{x} \in \mathbb{R}^m$ and the system as
\begin{equation}
  \label{eq:dyn-dim-red-ex}
  \frac{d \vec{x}}{dt} = \mathbf{L} \vec{x} + B(\vec{x}, \vec{x}) +
  \vec{f}(t).
\end{equation}
where $\mathbf{L}$ is a linear operator acting on the system state
(here and throughout, \index[not]{matrix values@$\mathbf{A}$, matrix
  values}matrices are indicated by a bold roman font, and individual
entries of a matrix $\mathbf{A}$ are written as $A_{ij}$), $B(\bullet,
\bullet)$ is a quadratic term, and $\vec{f}(t)$ a forcing term.
Suppose that we can partition the state vector $\vec{x}$ as $\vec{x} =
(\vec{y}, \vec{z})$ with $\vec{y} \in \mathbb{R}^n$, $\vec{z} \in
\mathbb{R}^p$ with $m = n + p$.  We can then write
\eqref{eq:dyn-dim-red-ex} as
\begin{equation}
  \label{eq:dyn-sys-red-1}
  \begin{gathered}
    \frac{d\vec{y}}{dt} = \mathbf{L}_{11} \vec{y} + \mathbf{L}_{12}
    \vec{z} + B^1_{11}(\vec{y}, \vec{y}) + B^1_{12}(\vec{y}, \vec{z})
    + B^1_{22}(\vec{z}, \vec{z}) + f_1(t), \\
    \frac{d\vec{z}}{dt} = \mathbf{L}_{12} \vec{y} + \mathbf{L}_{22}
    \vec{z} + B^2_{11}(\vec{y}, \vec{y}) + B^2_{12}(\vec{y}, \vec{z})
    + B^2_{22}(\vec{z}, \vec{z}) + f_2(t),
  \end{gathered}
\end{equation}
where we have partitioned the operators $\mathbf{L}$ and $B$ in the
obvious way.  Now, assume that there is a separation of timescales
between the $\vec{y}$ and the $\vec{z}$ degrees of freedom, so that
the evolution of $\vec{y}$ is slow and that of $\vec{z}$ is,
comparatively, fast.  In climate applications, $\vec{y}$ might
represent slower ``climate'' variability, while the $\vec{z}$ degrees
of freedom represent faster ``weather''.  In this setting, the goal is
to find an effective evolution equation for the slow degrees of
freedom, in some sense eliminating the fast degrees of freedom.  One
approach is to average over the fast degrees of freedom
\citep{majda-stoch-climate-models-2,kifer-coupled-averaging}, treating
the averaged effect of the fast degrees of freedom on the slow degrees
of freedom as a stochastic forcing, resulting in an effective
stochastic differential equation for $\vec{y}$,
\begin{equation}
  d\vec{y} = \hat{\mathbf{L}} \vec{y} dt + \hat{B}(\vec{y}, \vec{y})
  dt + G(\vec{y}) d\vec{\xi}(t),
\end{equation}
where $d\vec{\xi}(t)$ is a noise process representing part of the
averaged effect of the fast degrees of freedom on the slow degrees of
freedom, and $\hat{\mathbf{L}}$ and $\hat{B}$ are the linear and
quadratic components of a modified slow vector field, reflecting the
fact that the influence of the averaged fast degrees of freedom may
shift the mean state of the slow degrees of freedom from that
represented by $\mathbf{L}_{1j}$ and $B^1_{ij}$ in
\eqref{eq:dyn-sys-red-1}.  In this case, we have gone from an original
deterministic system to a stochastic reduced system.  This example
shows just one of a number of possible routes for reducing the
dimensionality of dynamical systems.

Alternatively, consider a typical climate data analysis task.  We may
have space-time output from a general circulation model (GCM) of, say,
500\,hPa geopotential height.  Here, the data points lie on the
model's computational grid and we may have daily or twice-daily time
resolution.  For a typical modern GCM, running at a global horizontal
spatial resolution of $128 \times 64$, this equates, for hemispheric
data, to $128 \times 32 = 4096$ spatial grid points per time step.
Naively, if we want to examine the variation in time of the state of
the Northern Hemisphere mid-troposphere, our state space is thus
$\mathbb{R}^{4096}$.  For daily data, this equates to about 1.5
million data values per year of simulated time.  However, we know that
coherent spatial structures are seen in this type of data, on
timescales ranging from a few days (synoptic scale) to years.  We
might thus hope to extract these coherent modes of variability from
our data set and use them to provide a reduced representation of at
least some portion of the total variability in the data.

As to why we might choose to attempt to develop reduced dimensionality
representations of dynamical systems or data sets, the simplest answer
is to help to understand the systems we study.  In this context, to
``understand'' means to be able to develop simpler analytical or
semi-analytical models of aspects of the system of interest, either to
prove rigorous results, or to facilitate experimentation that will
develop insight that we can then apply to the original system.  For
example, \citet{majda-paradigm} studied a 104-dimensional
deterministic system composed of four main degrees of freedom
nonlinearly coupled to a ``heat bath'' constructed from 100 modes of a
truncated Galerkin projection of a chaotic partial differential
equation.  The chaotic dynamics of the bath modes in this problem make
the dynamics of the overall system very difficult to understand.
\citeauthor{majda-paradigm} used a systematic averaging procedure to
produce a four-dimensional system with stochastic forcing that
reproduced important features of the dynamics of the original chaotic
system.  In a very real sense, the dynamics of this reduced
dimensionality stochastic system is easier to understand than the
dynamics of the original 104-dimensional system.  Even when reduction
of a system to a lower-dimensional form leads to a less accurate
representation of the physical processes of interest, the improved
ability to visualise trajectories of the system and to understand the
dynamics in the reduced phase space may compensate for this loss.
Lower-dimensional representations are also useful for feature
identification and clustering applications.

Another reason for attempting to reduce the dimensionality of systems
that we study is to aid computational analysis.  The phrase ``the
curse of dimensionality'' was first used by
\citet{bellman-dynamical-prog} to refer to the exponential increase in
volumes of spaces with increasing dimension, and the resulting
difficulties of sampling such spaces.  For example, to sample the unit
interval $[0, 1]$ so that no two sample points are separated by a
distance greater than $\tfrac{1}{10}$ requires only 11 points.  (The
notation \index[not]{closed interval@$[a, b]$, closed interval in
  $\mathbb{R}$}$[a, b]$ indicates a closed interval in $\mathbb{R}$:
$\{ x \,|\, x \in \mathbb{R}, a \leq x \leq b \}$.)  To achieve the
same sampling condition in a unit hypercube in $\mathbb{R}^{10}$, $[0,
  1]^{10}$, requires $11^{10}$ points.  This effect makes search and
optimisation problems in high-dimensional spaces essentially
intractable in many cases.  In a more general sense, this ``curse of
dimensionality'' also encompasses some of the non-intuitive features
of the geometry of high-dimensional spaces \citep{verleysen-curse}.
Phenomena like concentration of norms, where the distances of points
in a distribution from the mean become more and more tightly
distributed as the dimensionality increases, invalidate many
intuitions developed from low-dimensional geometry.

\section{Classifying dimensionality reduction methods}
\label{sec:nldr-classification}

In this section, I outline the main characteristics used to classify
the dimensionality reduction methods examined in the rest of the
chapter.  The sheer diversity of methods makes it difficult to imagine
a coherent framework that would allow all methods to be assessed
together.  Some of the following categorisations are thus only
relevant for a subset of dimensionality reduction methods.  The
methods reviewed in this chapter are (approximately) classified
according to some aspects of this scheme in Table~\ref{tab:methods}
--- this table is intended to act as a rough guide and preview to what
follows below.

%
%
\begin{table}
  \begin{center}
    \begin{tabular}{lccc}
      \toprule
      \textbf{Method} & \textbf{D/G} & \textbf{Linear?} & \textbf{Section} \\
      \midrule
      \multicolumn{4}{l}{\textit{Classical projection methods}} \\
      \textbf{Principal component analysis (PCA)} & D/G & Yes &
      \ref{sec:pca-definition} \\
      Canonical correlation analysis (CCA) & G & Yes &
      \ref{sec:pca-definition} \\
      Singular spectrum analysis (SSA) & G & No &
      \ref{sec:pca-definition} \\
      Spectral methods & D/G & Yes & \ref{sec:spectral-methods} \\
      Multidimensional scaling (MDS) & G & Yes &
      \ref{sec:mds-description}, \ref{sec:isomap-mds} \\
      & & & \\
      \multicolumn{4}{l}{\textit{Other projection methods}} \\
      Principal interaction patterns & D & No &
      \ref{sec:nldr-classification} \\
      Random projections & G & Yes & \ref{sec:random-proj} \\
      Kernel PCA & G & No & \ref{sec:kernel-methods} \\
      & & & \\
      \multicolumn{4}{l}{\textit{Differential geometry methods}} \\
      \textbf{Isomap} & G & No & \ref{sec:nonlinear-diff-geom},
      Chapter~\ref{ch:isomap} \\
      Locally linear embedding (LLE) & G & No &
      \ref{sec:nonlinear-diff-geom} \\
      \textbf{Hessian LLE} & G & No & \ref{sec:nonlinear-diff-geom},
      Chapter~\ref{ch:hessian-lle} \\
      Riemannian normal coordinates & G & No &
      \ref{sec:nonlinear-diff-geom} \\
      Riemannian manifold learning (RML) & G & No &
      \ref{sec:nonlinear-diff-geom} \\
      & & & \\
      \multicolumn{4}{l}{\textit{Neural network methods}} \\
      \textbf{Nonlinear PCA (NLPCA)} & G & No & \ref{sec:ann-methods},
      Chapter~\ref{ch:nlpca} \\
      Self-organising maps (SOMs) & G & No & \ref{sec:ann-methods} \\
      & & & \\
      \multicolumn{4}{l}{\textit{Spectral graph theory methods}} \\
      Laplacian eigenmaps & G & No & \ref{sec:spectral-graph-theory}
      \\
      Diffusion maps & G & No & \ref{sec:spectral-graph-theory} \\
      & & & \\
      \multicolumn{4}{l}{\textit{Miscellaneous other methods}} \\
      Independent component analysis (ICA) & G & Yes & \ref{sec:ica}
      \\
      Principal curves and surfaces & G & No &
      \ref{sec:principal-curves} \\
      Computational homology & G & No &
      \ref{sec:computational-topology} \\
      \bottomrule
    \end{tabular}
  \end{center}
  \caption[Table of dimensionality reduction methods]{List of
    dimensionality reduction methods considerd in this review.
    Methods used for analysis within this thesis are highlighted in
    bold.  Methods are classified according to whether they are
    dynamical (D) or geometrical/statistical methods (G), and whether
    they are linear or nonlinear.  References to the sections of the
    thesis where individual methods are described are provided.}
  \label{tab:methods}
\end{table}
%

\subsection*{Dynamical versus geometrical/statistical}

The principal distinction we will draw here is between
\emph{dynamical} and non-dynamical or \emph{geometrical/statistical}
dimensionality reduction methods.  As the name implies, dynamical
methods provide a reduced representation of the dynamics of a
high-dimensional system, usually in the form of a low-dimensional
dynamical system whose trajectories in some sense approximate the
trajectories of the original system.  This reduced dimensionality
dynamical system can then be analysed using all the methods of
dynamical systems theory.  Geometrical/statistical methods provide
only a lower-dimensional parameterisation of a data set without any
consideration of the dynamics that may have produced the data set.
The main difference in the use of these methods arises from the
general requirement for dynamical methods that the original system be
available as a set of equations.  This is obviously not possible for
experimental and observational data, but it may also be impractical
when analysing complex environmental models such as climate models.
Although one can theoretically write down the evolution equations for
such models as a discrete time dynamical system (perhaps with some
elements of stochastic forcing), in practice, for any model with any
semblance of realism, this is all but impossible, and one must treat
the system by the same methods as used for observational data (with
the proviso that models offer perfect observability of a sort that is
difficult to achieve in even the cleanest experimental arrangements).

There is a certain degree of overlap between these categories, but not
much.  The most important example is the classic proper orthogonal
decomposition (POD) or principal component analysis (PCA) method
(Section~\ref{sec:linear-methods}).  This is used as a
geometrical/statistical method, identifying linear subspaces
containing the greatest fractions of the variance of a data set, but
also as a Galerkin projection method for the reduction of dynamical
systems, where the equations of the system are projected onto the
linear subspaces found by the POD/PCA procedure.  The latter approach
was popularised by the work of \citet{holmes-turb-book} who applied
POD to the modelling of shear layer turbulence.  One can argue that
this approach is slightly different to other dynamical dimensionality
reduction methods, in that, in some sense, it is not a ``predictive''
method.  The eigenfunctions spanning the subspace into which the model
equations are projected are determined from a statistical analysis of
trajectories of the original system, rather than from a direct
analysis of the model equations themselves, as is done, for example,
in centre manifold reduction methods.  This does not detract from the
usefulness of the approach, but it does mean that it is more
empirically based than some other dynamical methods.

\subsection*{Model reduction versus data reduction}

A comparable but slightly different distinction can be made between
methods of \emph{model reduction} and methods of \emph{data
  reduction}, the former category mostly corresponding to dynamical
methods and the latter to geometrical/statistical methods.  The basic
distinction here is between starting from a model expressed as a set
of equations and getting as output a reduced model, also expressed as
a set of equations (perhaps with some numerical parameters determined
from integrations of the original model) and starting from a data set
from some source (perhaps a model, perhaps observations) and getting
as output a reduced dimensionality representation of the data.

We can distinguish four possible cases: model-to-model, model-to-data,
data-to-data and data-to-model, where by ``model'' we mean an explicit
system of equations that can be manipulated analytically, and by
``data'' we mean a data set representing either trajectories of a
system or an approximation to some geometrical object in the system's
state space.  ``Model-to-model'' corresponds to the basic case of what
was referred to as a dynamical method above, where, given a set of
equations, we derive another, lower-dimensional, set of equations by
averaging, asymptotic analysis or some other means.  Examples include
slow manifold methods \citep[e.g.][]{rhodes-maas-pope}, stochastic
averaging \citep[e.g.,][]{majda-stoch-climate-models-2} and
homogenisation methods
\citep[e.g.,][]{pavliotis-multiscale,bornemann-homogenization}, the
latter two approaches often being subsumed under the label of
``multi-scale methods''.  ``Model-to-data'' and ``data-to-data'' both
refer to the application of geometrical/statistical methods.  These
methods can be applied to an observational or simulated data set but,
given a set of equations for a system, we can also integrate to
produce a set of trajectories and then apply our
geometrical/statistical method to this data set.
Geometrical/statistical methods are thus of very general
applicability.  The final ``data-to-model'' category refers to methods
that attempt to identify a dynamical system that is in some sense the
best fit to a given data set.  These model fitting methods can be more
or less sophisticated and the results more or less convincing
depending on the application and the exact approach followed.  A good
example is the principal interaction patterns (PIPs) method
\citep{hasselmann-pips-pops,kwasniok-pips-3}, where a low-dimensional
dynamical system describing time evolution and a set of patterns
describing spatial variability are simultaneously fitted to a
space-time data set using a variational method.  The overall state of
the system is represented as a linear combination of the spatial
patterns, and the only nonlinearity in the reduced model appears in
the evolution equation for the coefficients of the spatial patterns in
the representation of the system state.  The basic structure of the
dynamical system governing the evolution of the coefficients is fixed
in advance, in the sense that the overall evolution of the expansion
coefficients of the system state is represented as a sum over simple
monomial modal interaction terms.  The coefficients of the interaction
terms can be used to frame the optimisation to be performed to find
the reduced model as a parametric optimisation problem.  This method
has found some success in the representation of mid-latitude
atmospheric variability \citep{kwasniok-pips-2,kwasniok-pips-3}, where
there are dynamical reasons to expect the interactions between modes
in the atmospheric flow to be representable mostly in terms of triad
interactions, i.e. interactions that involve only quadratic
nonlinearities in the modal expansion coefficients.  Selection rules
restricting the possible mode-to-mode interactions further constrain
the nonlinear terms that may appear in the evolution equations.

\subsection*{Linear versus nonlinear}

The distinction between a linear dimensionality reduction method and a
nonlinear one is fairly simple.  A linear reduction method projects
points in the state space of a system to a linear subspace of the
state space, while a nonlinear method projects state space points to a
more general lower-dimensional manifold.  Examples of linear reduction
methods include geometrical/statistical methods such as PCA, and also
dynamical methods, since all conventional Galerkin methods are
essentially linear.  In practice, the distinction between linear and
nonlinear methods is not always helpful, since it is relatively common
to use an initial linear dimensionality reduction step before applying
a nonlinear method.  This is the case for the nonlinear PCA method
described in detail in Chapter~\ref{ch:nlpca}, and, in the form it is
applied here, also for the Hessian LLE method used in
Chapter~\ref{ch:hessian-lle}.

\subsection*{Deterministic versus stochastic}

This distinction is mostly only meaningful for model reduction
methods.  For data-to-data methods, the nature of the system that was
the source of the input data is often irrelevant, although information
about noise in the input data can be propagated through the reduction
method to give some idea of error bounds on the reduced dimensionality
representation of the inputs.  For data-to-model methods, it is
possible to attempt to fit either a deterministic or a stochastic
model to the input data: instances of both approaches exist.

For model reduction methods, there are four possibilities, based on
whether the original high-dimensional model is deterministic or
stochastic and whether the reduced model is deterministic or
stochastic.  Deterministic-to-deterministic reduction methods are
common and include centre manifold, slow manifold and singular
perturbation theory methods.  Deterministic-to-stochastic methods have
started to receive much more attention in recent years, based on
theoretical advances in stochastic averaging and homogenisation theory
for partial differential equations.  Many of these methods,
particularly those based on averaging over fast degrees of freedom,
are as applicable to high-dimensional stochastic systems as they are
to deterministic systems, providing stochastic-to-stochastic reduction
methods.

The stochastic-to-deterministic case is slightly unusual.  There is at
least one dimensionality reduction method that can be considered to be
of this form, known as diffusion maps \citep{coifman-diffusion-maps}.
This method makes links between the theory of random walks on graphs
and the theory of diffusion processes on manifolds and, as such, has
links to the Laplacian eigenmaps and Hessian LLE methods described in
Section~\ref{sec:spectral-graph-theory} and
Chapter~\ref{ch:hessian-lle}.  However, in a much more general sense,
a link can be made between stochastic processes, represented as
stochastic differential equations, and deterministic diffusions, via
the relationship between a stochastic differential equation and its
associated Fokker-Planck equation \citep{oksendal-sdes}.  Although the
Fokker-Planck equation describes the deterministic evolution only of
the probability distribution of the states of a system, in a sense it
provides a maximal deterministic description of the dynamics of the
system: the evolution of the probability distribution of the system is
the only thing that can be predicted deterministically.  Further, from
the point of view of dimensionality reduction, since the Fokker-Planck
equation is linear, it may be possible to consider modal
decompositions of the evolution of the distribution of the system
states.  The work of \citeauthor{coifman-diffusion-maps} and others is
closely related to this more general viewpoint.

\subsection*{Theoretical underpinnings}

Much of the development of dynamical dimensionality reduction methods
has been based on rigorous results in dynamical systems theory,
perturbation methods or the theory of stochastic processes.  Examples
include methods based on centre manifolds and normal forms, both
well-known and well-studied areas with strong results, ideas relating
renormalisation group theory, normal forms and singular perturbation
theory
\citep{chen-renorm-group,omalley-amplitude-eqs,deville-renorm-group},
and stochastic averaging methods based on rigorous convergence results
for averaging in systems with timescale separation
\citep[e.g.,][]{kifer-coupled-averaging}.  These theoretical
underpinnings provide a good basis for the development of practical
dimensionality reduction methods and help to provide confidence that
the methods really do work as advertised and are well understood.

The situation for geometrical/statistical methods is somewhat
different.  There do exist methods with strong theoretical backing,
particularly the classical linear methods, but a more common situation
is for a method to be developed on the basis of intuitions about the
properties of data sets in some field of application.  Any theoretical
justification is supplied post hoc, if the method works.  If this
sounds like a grossly unfair characterisation of an immense body of
work, consider this: it is surprisingly hard, in many applications, to
do much better than linear reduction of a data set using PCA.  Almost
any method that does a better job will rely on idiosyncratic features
in the input data set and so will be at least partially
application-dependent.

This strongly applications-oriented view has meant that rigorous
theoretical work, which would have to be conducted on simplified
models of problems of interest, has been less common for
geometrical/statistical methods.  If a method can extract relevant and
interesting information from a large data set in some application
area, this may be enough for the method to become more widely adopted.
Questions of rigour or the development of a solid theoretical basis
for the method may be perceived to be of secondary importance.
Unfortunately, this is particularly the case in the fields where
dimensionality reduction methods are most needed.  If the problem at
hand is sufficiently complex, any help in unravelling the complexity
may be welcome, and rigorous analysis can seem unnecessary, especially
if such analysis can only be performed on simple ``toy'' problems
related to the central problem the method addresses.  This can be the
case even if analysis of such toy problems can help to build intuition
about more complicated systems.

In the sections that follow, I have tried to assess the theoretical
background for each of the methods considered, not only to provide an
indication of how well understood each method is, but also to help
draw parallels between different methods and to point to instances
where theoretical results for one method or set of methods might be
applied or adapted for another.

\subsection*{Applications}

The range of applications to which dimensionality reduction techniques
have been applied is huge, and there has been a tendency towards
ghettoisation, with different terminology and slightly different
approaches to the same basic problems used in different fields.
Witness the number of terms used to refer to the most common
dimensionality reduction method, principal component analysis (PCA),
proper orthogonal decomposition (POD), empirical orthogonal function
(EOF) analysis, the Karhunen-Lo\`{e}ve decomposition, and so on, all
of which denote essentially the same procedure, although developed in
slightly different mathematical settings for different applications.

This fragmentation of the subject of dimensionality reduction into
many disparate fields is one reason for the dichotomy between the
level of rigour applied to dynamical and geometrical/statistical
methods mentioned above.  Dynamical methods have traditionally been
developed by applied mathematicians with relatively little concern for
the ultimate application of the methods, which has led to the
existence of a number of methods that are conceptually useful but
difficult to apply to larger problems.  (There are, of course,
honourable exceptions.)  On the geometrical/statistical side of
things, many algorithms have been developed in the machine learning
community, where new methods are presented based on a few simple
geometrical test problems and a relatively small number of well-known
larger data sets.  There is little emphasis on rigorous analysis of
these new methods although again, exceptions exist.  One distinct
problem from the point of view of trying to develop a unified view of
dimensionality reduction techniques is that many of the
geometrical/statistical methods that have been developed, originally
for applications in machine learning such as clustering, indexing,
feature recognition or manifold learning, have not been applied to
dynamical systems.  Particularly in the case of the clustering and
manifold learning methods, this is rather surprising, since it would
seem that these methods may have something to offer for attempts to
identify coherent structures in the state spaces of dynamical systems.

In the treatment of individual methods presented below, I have
attempted to present a few indicative examples of applications of each
method, paying particular attention to cases where links can be drawn
between different fields, and to the honourable exceptions mentioned
above.  I have also attempted an assessment of how successful or
otherwise methods were in the different problems to which they were
applied.

\subsection*{Other points}

A few other issues are relevant for some methods.

\paragraph{Local versus global}

In literature on geometrical/statistical dimensionality reduction
methods, a distinction is made between local and global methods ---
local methods patch together a reduced dimensionality data manifold
from a set of solutions to local optimisation problems (often local
approximations to the tangent spaces of the data manifold found via
local PCA or singular value decomposition) while global methods solve
a single large problem incorporating global information from the whole
of the input data set (the eigendecomposition of the data covariance
matrix in PCA is the archetypal example).  This distinction has
implications for the scaling of computational requirements of
different methods, and there is also some disagreement about how
faithful nonlinear reductions based on local techniques can be
\citep{wu-lle-mapping}.

\paragraph{Numerical issues}

For some of the geometrical/statistical methods, scaling of
computational requirements with data set size can be of concern for
larger problems.  For some methods, adaptations have been developed to
deal with these scaling issues (e.g., for the Isomap method,
\citet{bachmann-remote-sensing-2} developed a number of refinements to
allow them to use the method for large hyperspectral imagery data
sets).  Also for some geometrical/statistical methods, there may be
particular pathologies associated with data sampling density or
uniformity.  This is a particular problem for methods that attempt to
construct approximations to the Laplacian or Hessian of a manifold by
finite differencing \citep{belkin-niyogi,donoho-hessian}.

\paragraph{Tolerance of noise}

A related issue for data-based reduction methods is the question of
how well the method behaves with noisy data.  In the absence of strong
theoretical results for many of these methods, the only approach seems
to be to try them on realistic data sets to see what happens.  This is
clearly less than satisfactory, but some general statements can be
made about some of the methods simply on the basis of reasoning about
obvious characteristics, such as the finite differencing issue
mentioned above.

\section{Linear methods}
\label{sec:linear-methods}

We begin our survey of geometrical/statistical methods of
dimensionality reduction by examining some classical linear methods.
The primary disadvantage of linear methods for analysing data sets
from integration of dynamical systems is that they are able to project
only into linear subspaces of the original high-dimensional data
space.  If our data points, instead of lying in a linear subspace, lie
in a curved low-dimensional submanifold of the data space, a linear
method will not detect the full structure of the data manifold,
instead approximating it by the linear subspace that is in some sense
nearest.  Despite this flaw, linear methods find extensive use in many
applications.  These methods are easier to understand and analyse than
some nonlinear methods that have been developed more recently, and
they serve as an excellent test case against which newer and more
complex nonlinear methods can be compared.  In terms of the
classification of methods described in
Section~\ref{sec:nldr-classification}, all linear methods are clearly
global in nature, since they project all data points to the same
linear subspace, determined using a combination of information from
all data points.

\subsection{Principal component analysis}
\label{sec:pca-definition}

Principal component analysis (PCA) is perhaps the most commonly used
of all dimensionality reduction methods, applied in every field where
multivariate time series need to be analysed.  The method is common
enough that it has several names, including proper orthogonal
decomposition (POD), empirical orthogonal function (EOF) analysis and
the Karhunen-Lo\`{e}ve decomposition.  All of these names refer to
essentially the same computation, although the exact mathematical
setting differs between the different uses.  This profusion of terms
seems to have arisen due to independent discoveries of the underlying
idea of principal component analysis in different fields of
mathematics: PCA and EOF analysis in statistics, POD in fluid dynamics
and functional analysis and the Karhunen-Lo\`{e}ve decomposition in
the theory of random functions.  Brief notes on the relationship
between these different settings are given below, following an
explanation of the basic concepts behind PCA.

As well as the different terms used for the overall method,
terminology also varies between fields for the different elements of
the data decomposition provided by the method.  This can be
exceedingly confusing when trying to read literature across different
fields.  Here, we will use the terminology most common in the climate
data analysis community, described by \citet{vonstorch-zwiers}.  We
will generally use the term PCA in the data reduction context, and POD
in the dynamical context, since this seems to be a sensible
distinction that is maintained through most of the literature.  (PCA
is usually referred to as a Karhunen-Lo\`{e}ve decomposition when
dealing with continuous functions rather than discrete data sets.)  An
impression of the importance of PCA can be gained from a search on ISI
Web of Science for the terms listed above, which turns up more than
30,000 references.

PCA treats an $m$-dimensional multivariate time series, represented as
a set of vectors $\vec{x}_i \in \mathbb{R}^m$, with $i = 1, \dots, N$,
and identifies a sequence of mutually orthogonal directions in
$\mathbb{R}^m$ that correspond to the directions of greatest variance
in the input data.  (We will use \index[not]{input data
  vectors@$\vec{x}_i$, input data vectors}$\vec{x}_i$ throughout to
refer to input data vectors in the original high-dimensional data
space.)  Consider the first such direction, which we denote by
$\vec{q}_1$, a unit vector in $\mathbb{R}^m$.  Let
\index[not]{ensemble mean@$\langle \vec{x} \rangle$, ensemble
  mean}$\langle \vec{x} \rangle$ denote the ensemble mean of the set
of input data vectors, i.e. $\langle \vec{x} \rangle = N^{-1}
\sum_{i=1}^N \vec{x}_i$, and let \index[not]{variance@$\var(x)$,
  variance}$\var(x)$ denote the sample variance over a set of values,
i.e. for $u_i \in \mathbb{R}$ with $i = 1, \dots, N$, $\var(u) =
(N-1)^{-1} \sum_{i=1}^N (u_i - \langle u \rangle)^2$.  Then, writing
\index[not]{vector dot product@$\vec{x} \cdot \vec{y}$, vector dot
  product}$\vec{x} \cdot \vec{y}$ for the usual vector dot product,
\begin{equation}
  \label{eq:pca-q1-max-prob}
  \vec{q}_1 = \argmax_{||\vec{q}||=1} \, \var(\vec{x} \cdot \vec{q}),
\end{equation}
where \index[not]{argmax@$\argmax_x f(x)$, $x$ maximising
  $f(x)$}$\argmax_x f(x)$ denotes the value of $x$ that maximises a
function $f(x)$.  We can project each of the input vectors onto this
direction of maximum variance to give a scalar sequence $\alpha_{i1} =
\vec{x}_i \cdot \vec{q}_1$.  Here, $\vec{q}_1$ tells us the direction
of maximum variance in the data, while $\alpha_{i1}$ measures how much
of each data vector $\vec{x}_i$ lies in this direction.  We can
calculate residual vectors $\tilde{\vec{x}}_i = \vec{x}_i -
\alpha_{i1} \vec{q}_1$ and then find the next orthogonal direction
explaining the most variance in the data, $\vec{q}_2$, as
\begin{equation}
  \vec{q}_2 = \argmax_{||\vec{q}||=1, \vec{q} \cdot \vec{q}_1 = 0}
  \var(\tilde{\vec{x}} \cdot \vec{q}).
\end{equation}
A sequence $\alpha_{i2} = \vec{x}_i \cdot \vec{q}_2$ measuring how
much of each data vector lies in the direction of $\vec{q}_2$ can then
be calculated, in the same way as for the first direction.  Again,
residuals can be found, and the process repeated to eventually find an
orthonormal basis $\{ \vec{q}_j \}$ for $\mathbb{R}^m$.  The basis
vectors lie in the directions of the greatest variance in the input
data and are ordered by the proportion of the total data variance
lying along those directions.  The sequences $\alpha_{ij}$ give the
coordinates of the data vectors in this new basis: $\alpha_{ij} =
\vec{x}_i \cdot \vec{q}_j$.

PCA has a variance partitioning property, in the sense that, if we
denote the variance of a set of vectors $\vec{y}_i$ as
\index[not]{variance (vector)@$\var(\vec{y})$, variance
  (vector)}$\var(\vec{y})$ (this is simply the sum of the variance of
the individual vector components), then thinking just of the first PCA
direction found above, $\vec{q}_1$, and the residuals,
$\tilde{\vec{x}}_i$,
\begin{equation}
  \label{eq:pca-variance-partition-1}
  \var(\vec{x}) = \var(\tilde{\vec{x}}) + \var(\alpha_{\bullet1}
  \vec{q}_1),
\end{equation}
or, in terms of the full basis,
\begin{equation}
  \label{eq:pca-variance-partition-2}
  \var(\vec{x}) = \sum_{j=1}^m \var(\alpha_{\bullet{}j} \vec{q}_j),
\end{equation}
where \index[not]{variance (vector
  component)@$\var(\vec{y}_{\bullet{}j})$, variance
  (component)}$\var(\vec{y}_{\bullet{}j})$ denotes the sample variance
of the $j$th component of a set of vectors $\vec{y}_i$.  These
relations imply that the data variance in each of the directions found
by PCA is uncorrelated, meaning that one can think of each of these
directions independently ``explaining'' a portion of the total data
variance.

\citet{burges-review} gives a clear explanation of how this view is
related to the most common method of computation used for PCA, which
is based on an eigendecomposition of the input data covariance matrix.
This computation breaks the input data set, with variability in both
space and time, into a set of mutually orthogonal spatial patterns
(which correspond to the mutually orthogonal axes of greatest variance
in the input data), the $\vec{q}_j$, and a set of scalar time series,
the $\alpha_{ij}$, finding all of the $\vec{q}_j$ at once, instead of
in the step-wise fashion described above.  Subtracting the data mean
from each of the the $\vec{x}_i$ input data vectors, we can construct
the covariance matrix as
\begin{equation}
  \label{eq:pca-covariance}
  \mathbf{C} = \langle (\vec{x} - \langle \vec{x} \rangle) (\vec{x} -
  \langle \vec{x} \rangle)^T \rangle,
\end{equation}
where \index[not]{transpose@$\vec{y}^T, \mathbf{A}^T$, vector/matrix
  transpose}$\vec{y}^T$ denotes the transpose of a vector $\vec{y}$
(and similarly for the transpose of a matrix).  The covariance matrix
$\mathbf{C}$ is a symmetric matrix, meaning that we can write the
eigenvector decomposition of $\mathbf{C}$ as $\mathbf{C} = \mathbf{Q}
\bm{\Lambda} \mathbf{Q}^T$, with $\bm{\Lambda} = \diag(\lambda_1,
\dots, \lambda_m)$ a \index[not]{diagonal matrix@$\diag(\dotsc)$,
  diagonal matrix}diagonal matrix of the eigenvalues $\lambda_i$ in
descending order of magnitude, and $\mathbf{Q}$ a matrix whose columns
are the corresponding orthogonal eigenvectors $\vec{q}_i$ (the
orthogonality of the eigenvectors is also a consequence of the
symmetry of $\mathbf{C}$).  The eigenvectors $\vec{q}_i \in
\mathbb{R}^m$ are patterns of variation in the data (the directions of
greatest variance found above), called empirical orthogonal functions
(EOFs) in the climate community.  The first of these, $\vec{q}_1$,
represents the direction in data space with the greatest variance,
$\vec{q}_2$ the direction orthogonal to $\vec{q}_1$ with the next
greatest variance in the data, and so on.  As noted above, all linear
methods, including PCA, are global methods: the directions of greatest
variance in the data found from the eigendecomposition of the data
covariance matrix are shared by the whole data set.

The input data time series, $\vec{x}_i$, can then be expanded in terms
of the orthogonal basis provided by the EOFs as
\begin{equation}
  \label{eq:pca-expansion}
  \vec{x}_i = \sum_j \alpha_{ij} \vec{q}_j.
\end{equation}
The coefficients $\alpha_{ij}$ are called the principal component (PC)
time series and give the temporal variation in the data in each of the
orthogonal directions in data space spanned by the EOFs\footnote{Note
  that this is one of the points where there is greatest variability
  in terminology.  Sometimes the EOFs are referred to as ``loadings''
  and the PCs as ``scores'', and sometimes other names are used for
  both elements of the decomposition.  We will stick with ``EOFs'' and
  ``PCs'' throughout.}.  The eigenvalue associated with each EOF
measures the proportion of the total variance of the input data
explained by that EOF.  With the EOFs in descending eigenvalue order,
we may extract an EOF subset explaining some pre-selected proportion
of the total variance, $V_p = \{ \vec{q}_i \, | \, 1 \leq i \leq p \}$
say, where $p$ is the number of EOFs required to explain the required
proportion of the total variance.  By projecting the input data into
the subspace $\mathcal{V}_p = \mathspan(V_p)$, the linear subspace
\index[not]{span@$\mathspan(S)$, span of set of vectors}spanned by the
eigenvectors in $V_p$, we arrive at a reduced dimensionality
representation of variability in the input data.  Compared to the
original data, this reduced representation has the minimum squared
error totalled over all data points of any choice of projection basis
of dimension $p$.

The earliest presentations of the ideas behind PCA are found in
\citep{pearson-pca}, where the basic idea is presented, and
\citep{hotelling-pca-1,hotelling-pca-2}, which give the first
systematic treatment of the method and also appear to represent the
first use of the phrase ``principal components''.  Particularly in the
latter references, the form of the analysis presented above can be
perceived quite clearly, although not in exactly the modern form used
here.  A recent review of both conventional PCA and some related
methods is given by \citet{hannachi-eofs}, who concentrate
particularly on modifications of PCA used in meteorological and
climate data analysis applications.  These include the widely used
method of \emph{rotated EOFs}, which attempts to produce spatial
patterns that are, in some sense, more ``physical'' than basic EOFs by
selecting a small number of leading spatial patterns and linearly
transforming the PCs in the space spanned by these leading EOFs so as
to minimise some functional that represents ``simplicity'' of
structure of the spatial patterns.  \citet{jolliffe-simplified-eofs}
present some alternatives to the conventional rotation procedure that
may be somewhat less subjective.

The setting for proper orthogonal decomposition (POD) is slightly
different to that for PCA, being based on analysis of continuous
functions, rather than finite-dimensional data vectors.  The monograph
of \citet{holmes-turb-book} did much to popularise the use of these
ideas in fluid dynamical applications.  \citet{smith-pod-tutorial}
provide a recent tutorial introduction to the method.  Consider square
integrable functions defined over some domain of interest $\Omega$,
i.e. functions in the space \index[not]{L2omega@$L^2(\Omega)$,
  square-integrable functions}$L^2(\Omega)$, where $f \in L^2(\Omega)$
implies that $\int_\Omega | f |^2 \, dx < \infty$.  The key point of
POD is to find a sequence of orthogonal functions $\phi_j(x) \in
L^2(\Omega)$ that solve a maximisation problem paralleling
\eqref{eq:pca-q1-max-prob} for the PCA case,
\begin{equation}
  \label{eq:pod-max-prob}
  \max \frac{\langle | (u, \phi_1)_{L^2(\Omega)} |^2 \rangle}{||
    \phi_1 ||_{L^2(\Omega)}^2},
\end{equation}
where the $u_i(x) \in L^2(\Omega)$ are the input data functions, and
\index[not]{L2 inner product@$(f, g)_{L^2(\Omega)}$, inner product on
  $L^2(\Omega)$}$(\bullet, \bullet)_{L^2(\Omega)}$ and \index[not]{L2
  norm@$||\bullet||_{L^2(\Omega)}$, norm on
  $L^2(\Omega)$}$||\bullet||_{L^2(\Omega)}$ respectively represent an
inner product and norm on $L^2(\Omega)$, i.e.
\begin{equation}
  (f, g)_{L^2(\Omega)} = \int_\Omega f \cdot g^* dx, \qquad
  ||f||_{L^2(\Omega)}^2 = (f, f) = \int_\Omega |f|^2 dx,
\end{equation}
where \index[not]{complex conjugation@$z^*$, complex
  conjugation}${}^*$ represents complex conjugation.  In practical
applications appeals to ergodicity are usually used to replace the
ensemble average in \eqref{eq:pod-max-prob} by a time average over
trajectories of the system under study.  This setting is conceptually
identical to the maximum variance condition defining the first
principal component in PCA, except that here the input data are
functions rather than finite-dimensional data vectors.  By using
calculus of variations, the minimisation problem
\eqref{eq:pod-max-prob} can be transformed into an eigenvalue problem
for the functions $\phi_j(x)$, although here the eigenvalue problem is
expressed as a Fredholm integral equation:
\begin{equation}
  \int_\Omega \langle u(x) \otimes u^*(x') \rangle \phi_j(x') dx' =
  \lambda_j \phi_j(x),
\end{equation}
with \index[not]{tensor product@$u \otimes v$, tensor product}$u
\otimes v$ denoting the tensor product in $L^2(\Omega)$.  Here, the
kernel of the integral equation is the covariance tensor of the input
data, averaged over the data ensemble, closely comparable to the use
of the eigendecomposition of the covariance matrix to determine the
principal components in PCA.  As should be clear from this outline,
much of the analysis of POD parallels the development of PCA quite
closely, although in a functional analysis setting.

The third term in common use for the method that we are calling PCA is
the \emph{Karhunen-Lo\`{e}ve decomposition}, which arose in the theory
of stochastic processes \citep{loeve-book}.  It is possible to
represent any centred stochastic process $X_t$ defined on an interval
$t \in [a, b]$, i.e. a process for which, writing
\index[not]{expectation operator@$\mathbb{E}$, expectation
  operator}$\mathbb{E}$ for the expectation operator, $\mathbb{E}(X_t)
= 0$ for all $t \in [a, b]$, via a decomposition in terms of a set of
functions $e_k(t)$ and a set of random variables $Z_k$ as
\begin{equation}
  \label{eq:kl-expansion}
  X_t = \sum_{k=0}^\infty Z_k e_k(t),
\end{equation}
where the $Z_k$ are pairwise uncorrelated, and the $e_k(t)$ are
continuous functions over the interval $[a, b]$ that are pairwise
orthogonal in $L^2([a, b])$.  As for POD, the parallels with the
description of PCA are clear, with the Karhunen-Lo\`{e}ve expansion
\eqref{eq:kl-expansion} being directly comparable to the expansion in
terms of the PCA eigenvectors \eqref{eq:pca-expansion}.

The relative simplicity of PCA makes it more susceptible to detailed
analysis than some of the complex nonlinear manifold learning methods
described below.  Chapter~13 of \citep{vonstorch-zwiers} provides
information about confidence interval estimates for the explained
variance fractions, along with a detailed analysis of the statistical
properties of the EOFs and PCs. An interesting example of rigorous
analysis of PCA is \citep{north-eofs}, a study providing some physical
insight into what PCA means for linear stochastic models.  For a
particular class of model, it is shown that the EOFs at individual
Fourier frequencies correspond to the orthogonal normal modes of the
system.  There are some parallels between this approach and that of
\citet{donoho-isometry}, who analysed the Isomap algorithm, described
in detail in Chapter~\ref{ch:isomap}, to determine the class of
manifolds that could be faithfully represented in Isomap reductions.
These approaches are potentially a rather fruitful way to think about
dimensionality reduction methods: is it possible to determine the
class of data structures or models that can be faithfully represented
by any particular projection scheme?  Even if this is not possible for
a particular method, it seems that thinking about the dimensionality
reduction process in these terms may be of some use.  Another example
of a detailed analytical result for PCA, this time in the context of a
dynamical reduction, is the work of \citet{homescu-errors}, who
provide detailed error estimates for POD reduction of a dynamical
system, based on the use of an adjoint model --- the linearity of the
method is crucial in the construction of the adjoint model here.

Applications of PCA are commonplace in fields as disparate as fluid
dynamics and psychology.  Here, I mention a few scattered examples to
give a flavour of the diversity of uses.

In climate science, PCA is used extensively for defining modes of
variability in the atmosphere and ocean.  I use it here
(Chapter~\ref{ch:obs-cmip-enso}) to examine patterns of spatial
variability in tropical Pacific sea surface temperature and
thermocline depth associated with the \eln/Southern Oscillation.
Other applications are common, both in the analysis of observational
data and in the processing of output from climate models
\citep[e.g.,][]{mo-persistent,gladstone-pmip-nao,vonstorch-zwiers}.

PCA is often used as a preprocessing step before applying other
methods of dimensionality reduction.  This is done here in
Chapters~\ref{ch:nlpca}~and~\ref{ch:hessian-lle}, where tropical
Pacific sea surface temperature and thermocline data are preprocessed
using PCA before applying nonlinear methods to analyse behaviour in a
reduced space spanned by the leading EOFs of the field of interest.
Another example, in an entirely different field, is the work of
\citet{hegger-protein-folding}, who analysed results from molecular
dynamics simulations of short amino acid chains in water.  As input
data, they use the dihedral angles of the bonds in the molecules they
study, and use an initial PCA step to reduce this high-dimensionality
data to a more manageable size.  Degrees of freedom in the data that
do not correspond to conformational changes in the molecules are
eliminated by examining the distribution of values in their principal
component time series --- degrees of freedom with Gaussian PC
distributions are assumed to be ``uninteresting'' oscillations within
a single conformation.  The time series of the ``interesting''
principal components are then passed on to a more sophisticated
nonlinear time series analysis, but the initial PCA dimensionality
reduction step is important to make the later analysis tractable.

In computer science, PCA has found applications in clustering,
indexing and classification problems
\citep[e.g.,][]{shen-indexing,villalba-classification} as well as an
extremely interesting application in computer graphics, where PCA was
used to provide a basis for the efficient representation of fluid flow
\citep{treuille-fluids}.  The latter study used a novel method where
an empirical basis for the representation of fluid flows was derived
from a training set of simulations, using PCA to project the different
terms of the Navier-Stokes equations in a computationally efficient
way.  PCA was also used to construct bases to assist in handling
moving boundaries in the flow, by representing the pressure changes
responsible for maintaining free-slip conditions at the boundaries.
The overall result was a system able to simulate realistic
three-dimensional imagery of fluid flows, albeit in relatively
constrained settings.

Many elaborations of the basic PCA approach have been developed for
different purposes.  Two of these are of particular note, one
developed for the purposes of analysing correlations between coupled
input data sets and the other for extracting coherent patterns of
spatiotemporal variance, so incorporating a temporal dimension that is
not present in standard PCA analysis.

The first of these approaches is canonical correlation analysis (CCA),
which finds linear transformations of two coupled input fields that
maximise the correlation between the principal component time series
associated with the two variables
\citep[e.g.,][Chapter~14]{vonstorch-zwiers}.  If we have two input
data time series, $\vec{x}_i \in \mathbb{R}^{m_x}$, $\vec{y}_i \in
\mathbb{R}^{m_y}$, with $i = 1, \dots, N$, then the first CCA mode is
defined as the pair $\vec{a}^{(1)} \in \mathbb{R}^{m_x}$,
$\vec{b}^{(1)} \in \mathbb{R}^{m_y}$ such that the correlation between
the scalar time series $u^{(1)}_i = \vec{a}^{(1)} \cdot \vec{x}_i$ and
$v^{(1)}_i = \vec{b}^{(1)} \cdot \vec{y}_i$ is maximised.  Subsequent
CCA modes $(\vec{a}^{(j)}, \vec{b}^{(j)})$ are defined by requiring
them to be the patterns giving the best correlation between the time
series $u^{(j)}_i = \vec{a}^{(j)} \cdot \vec{x}_i$ and $v^{(j)}_i =
\vec{b}^{(j)} \cdot \vec{y}_i$ subject to the condition that
$u^{(j)}_i$ and $v^{(j)}_i$ are uncorrelated with $u^{(k)}_i$ and
$v^{(k)}_i$ for $k < j$.  \citet{bretherton-coupled} present a clear
account of CCA and related methods for the analysis of coupled
spatiotemporal data sequences, fitting CCA, a form of coupled field
PCA and singular value decomposition \citep{stewart-svd} into a common
theoretical framework to aid intercomparison.  In \citet{wallace-svd},
the same authors present a comparison of the results of applying these
methods to some climate data sets.

The second method of importance derived from PCA is called singular
spectrum analysis (SSA).  This is a method for extracting coherent
modes of temporal or spatiotemporal variability from time series
(univariate or multivariate) by applying PCA to a lag-covariance
matrix of the input data.  SSA shares characteristics with nonlinear
time series methods based on embedding theorems of Whitney and Takens
\citep{kantz-ts-book} and also with spectral methods.  The method was
originally proposed by \citet{broomhead-ssa} and described in more
detail in an early review that presented SSA as a ``toolkit'' for
signal extraction applications \citep{vautard-ssa-toolkit}.
\citet{ghil-ssa} provide a more recent systematic review of SSA and
associated spectral methods in the context of the analysis of climate
time series.  Applications of SSA have included studies of
paleoclimate time series \citep{vautard-ssa-paleo}, adaptive filtering
and prediction of the time series of the Southern Oscillation index
\citep{keppenne-enso-ssa} and Northern Hemisphere weather regimes
\citep{plaut-ssa-regimes}.  In the original approach of
\citet{broomhead-ssa}, \citet{vautard-ssa-paleo} and
\citet{vautard-ssa-toolkit}, SSA modes are selected for use in
dimensionality reduction on the basis of their eigenvalues only.  More
recent modifications of the original SSA algorithm have been developed
that incorporate significance tests against arbitrary coloured noise
models \citep{allen-ssa-signif,allen-mcssa,allen-ssa-opt-filt},
permitting the use of a more objective hypothesis testing approach for
mode selection.

\subsection{Multidimensional scaling}
\label{sec:mds-description}

Multidimensional scaling (MDS) \citep{borg-mds} is a statistical
dimensionality reduction method taking as its input distance or
dissimilarity measures for a set of data points; in the simplest case,
this means a matrix of point-to-point distances between each of the
data points, measured using some metric.  It then attempts to find
points in a lower-dimensional Euclidean space such that the Euclidean
distances between the output points correspond to the distance or
dissimilarity values between the input points.  MDS is an essential
component of the Isomap algorithm used in Chapter~\ref{ch:isomap}, and
is explained in detail there.

The initial reduction in data dimensionality involved in going from
$N$ data vectors $\vec{x}_i \in \mathbb{R}^m$ to a matrix $D_{ij} = ||
\vec{x}_i - \vec{x}_j ||$ of inter-point distances can be
considerable.  This step effectively eliminates details in the
original data vectors that depend only on the embedding of the data
manifold in the original high-dimensional space, since the distance
matrix captures all of the intrinsic geometrical structure in the
input data set.  The use of such a distance matrix forms the basis of
a number of dimensionality reduction and related methods based on
elucidating such intrinsic geometrical or topological structures in
data.  As well as MDS and Isomap, treated in Chapter~\ref{ch:isomap},
other examples are some of the differential geometry based nonlinear
methods described in Section~\ref{sec:nonlinear-diff-geom} and methods
based on computational topology described in
Section~\ref{sec:computational-topology}.

\subsection{Spectral methods}
\label{sec:spectral-methods}

Although not customarily considered as dimensionality reduction
methods, spectral decompositions of time series data can be used for
this purpose, although great care is required in assessing the
statistical and physical significance of any ``signals'' seen in noisy
data sets.  Along with conventional Fourier methods
\citep[e.g.,][Chapter~13]{press-nr}, wavelet methods
\citep{torrence-wavelets}, SSA \citep{ghil-ssa} and other spectral or
spectral-like signal decomposition methods
\citep[e.g.,][]{huang-hilbert-spectrum} can be used for the purposes
of dimensionality reduction.

\subsection{Random projections}
\label{sec:random-proj}

This is a rather counter-intuitive method of dimensionality reduction
relying on a result known as the Johnson-Lindenstrauss lemma
\citep{jl-lemma,dasgupta-jl}.  This states that a set of $N$ points in
high-dimensional Euclidean space can be projected into a
$d_{\mathrm{rand}}$-dimensional Euclidean space with
\begin{equation}
  \label{eq:rand-proj}
  d_{\mathrm{rand}} \sim O(\log N/\varepsilon^2)
\end{equation}
so that the distance between any two points changes only by a factor
of $1\pm\varepsilon$.  A corollary of this result is that almost any
random projection from a high-dimensional Euclidean space to a
lower-dimensional Euclidean space approximately preserves inter-point
distances.  This provides the basis for the random projections method,
where the original high-dimensional data points are simply replaced
with their projections to a random lower-dimensional linear subspace.
(Note that the logarithmic dependence of $d_{\mathrm{rand}}$ on the
number of points and the $\varepsilon^{-2}$ factor in
\eqref{eq:rand-proj} places rather strong restrictions on practical
applications of this method: the original data dimension must be very
high to gain much benefit from such a projection.)  In fact, it is not
even strictly necessary to ensure that the transformation from the
high-dimensional space to the low-dimensional space is a projection.
For reduction from $\mathbb{R}^m$ to $\mathbb{R}^d$ with $d \ll m$,
any random matrix $\mathbf{R} \in \mathbb{R}^{d \times m}$ whose
columns, considered as vectors in $\mathbb{R}^d$, have unit lengths is
suitable, and reduced coordinates can be calculated as $\vec{y}_i =
\mathbf{R} \vec{x}_i$ with $\vec{x}_i \in \mathbb{R}^m$ the original
data vectors and $\vec{y}_i \in \mathbb{R}^d$ the reduced vectors.
This method was used by \citet{bingham-rand-proj} who compared random
projections to PCA, SVD and a discrete cosine transform for
dimensionality reduction of image data, and compared random
projections and SVD for dimensionality reduction of text data.  The
random projections method appears to work rather well for these
applications.

\section{Nonlinear methods}

Many of the nonlinear methods presented here, including most of the
differential geometry based methods and some of the neural network
based methods, were originally developed in the machine learning and
machine vision communities, for the purposes of extracting
low-dimensional information from data sets or image streams for such
applications as object identification and feature tracking.  A typical
example might be the discovery of an orientation manifold for an
object represented as a number of bitmapped images, each image showing
a different orientation of the object, perhaps with variations in
lighting conditions or other extraneous characteristics.  The term
\emph{manifold learning}, often seen in the literature in these
fields, evokes the process of learning the structure of this
low-dimensional manifold from the higher-dimensional input data.
Although some dimensionality reduction methods construct reduced
representations of data sets without learning the manifold on which
the data lie, the two problems of dimensionality reduction and
manifold learning are very closely linked, and most methods that do
one also do the other.

If we represent a parameterised manifold as a map $f : \Omega \to
\mathbb{R}^m$ from a subset $\Omega$ of $\mathbb{R}^d$ into
$\mathbb{R}^m$, with $d < m$, so that $\vec{x}_i =
\vec{f}(\vec{y}_i)$, where the $\vec{x}_i$ are our data points in the
high-dimensional space, and the $\vec{y}_i$ are the corresponding
points in the low-dimensional ``feature'' space, ``learning the
manifold'' means developing a reconstruction of the map $\vec{f}$ that
can be used to associate points in the reduced coordinate space
$\mathbb{R}^d$ with points in the original data space.  This
definition needs to be qualified by the observation that none of the
dimensionality reduction methods discussed here are able to develop a
global parameterisation for data manifolds with non-trivial topology,
and so are not capable of discovering the global structure of the data
manifold in either a topological sense or in the sense of global
Riemannian geometry.  The only methods that attempt to reconstruct
global structures of data manifolds from point cloud data are the
computational topology methods discussed in
Section~\ref{sec:computational-topology}, and these methods seek to
determine \emph{only} global topological invariants of the data
manifold, without producing any sort of geometrical parameterisation
of the manifold.

\subsection{Differential geometry methods}
\label{sec:nonlinear-diff-geom}

Many data-oriented nonlinear dimensionality reduction techniques can
be placed into a common framework, along with PCA, by considering them
as seeking a transformation that preserves ``interesting'' information
in the input data, where these ``interesting'' features are derived
from some sort of discretised differential geometric analysis of the
input data.  In the case of PCA, this ``interesting'' information is
simply the Euclidean distances between data points; the required
transformation is thus a linear orthogonal transformation.  A more
complex example is Isomap \citep{tenenbaum-isomap}, which finds a
nonlinear transformation that preserves not Euclidean distances
between data points, but an approximation to distances between data
points as measured along geodesics in the data manifold\footnote{Much
  of the following material describing differential geometry based
  methods assumes some familiarity with the concepts behind the Isomap
  algorithm.  Isomap is described in detail in
  Chapter~\ref{ch:isomap}.  In particular,
  Section~\ref{sec:isomap-geodesics} describes the calculation of
  approximate geodesics in a data manifold from a nearest neighbour
  graph, an approach used by several other methods treated here.}.
Although distances along individual geodesics may change under
reparameterisation of coordinates on the manifold, the totality of all
geodesic distances between points on the manifold encodes the global
Riemannian structure of the manifold, a structure that is an intrinsic
feature of the dynamics of the system under study, and is independent
of the details of the embedding in the observation space.  The hope is
thus that a method like Isomap might be better able to identify
intrinsic geometrical structures in the input data than methods based
on calculations dependent on the details of the embedding of the data
manifold in the high-dimensional input data space.  Further
elaborations of the idea of using approximate geodesic distances in
dimensionality reduction derive transformations that preserve other
geometrical structure in the data as well as geodesic distances, for
example local curvature \citep{lin-rml}.

Of the large number of data-driven dimensionality reduction methods
that have been developed fitting the pattern described above, by far
the most widely used are Isomap and locally linear embedding (LLE)
which respectively serve as canonical examples of global and local
nonlinear dimensionality reduction methods, and provide a basis for
assessment of more recently developed methods based on ideas from
differential geometry.  Isomap, which finds an approximate global
isometry of the input data, i.e. a transformation that preserves
distances within the data manifold, was originally described by
\citet{tenenbaum-isomap}; several extensions have since been developed
to help with treating larger data sets
\citep{desilva-landmarks,bachmann-remote-sensing-2} and to provide a
slight generalisation of the type of transformations representable by
the method \citep{desilva-global-local}.  LLE was first presented in
\citep{roweis-lle} and explained in more detail by \citet{saul-lle}.
Several generalisations and adaptations of the original idea have
since been developed, notably the Hessian LLE approach of
\citet{donoho-hessian}.

Isomap and LLE have had a great influence on work in the field of
geometrical/statistical dimensionality reduction, although these ideas
have had little impact in communities more concerned with dynamical
methods.  To give some impression of the extent of this influence, a
search on ISI Web of Science shows that, as of July 2008,
\citep{tenenbaum-isomap} has 441 citations and \citep{roweis-lle} has
451.  For both methods, around half of the citations are from the
fields of computer science and artificial intelligence (mostly machine
learning), while another 20\% or so are devoted to related
applications in image analysis and machine vision.  The remaining
20--30\% of citations are spread across applications primarily in
neuroscience and neuroimaging, biology and mathematics.  (The strong
bias towards computer science, artificial intelligence and machine
learning citations is indicative more of the development of further
dimensionality reduction methods on the basis of ideas taken from
Isomap and LLE than of applications per se.)

Isomap is explained in some detail in Chapter~\ref{ch:isomap}, where
it is applied to the analysis of interannual tropical Pacific climate
variability.  Very briefly, Isomap uses an approximation to geodesic
distances in the data manifold to construct a global isometry
transforming the original input data to a lower-dimensional Euclidean
space.  The coordinates in this lower-dimensional space then serve as
coordinates for the original manifold.  (Note that the isometry, as an
invertible transformation, is a transformation between the original
data manifold and a lower-dimensional Euclidean space of the same
dimension, not between the original high-dimensional input data space
and a low-dimensional Euclidean space.)  This arrangement, using a
global isometry, is rather restrictive in terms of what kinds of
manifolds Isomap can represent in a faithful fashion, as illustrated
very nicely by \citet{donoho-isometry}.  Isomap is a global method, in
the sense that it constructs a Gramian matrix, i.e. a matrix of inner
products of a set of vectors, and uses an eigendecomposition of this
matrix to find the embedding transformation from the lower-dimensional
feature space to the higher-dimensional data space.  In this sense, it
has much in common with PCA, and in fact, both Isomap and PCA can be
considered in a common framework using the ideas of multidimensional
scaling (again, see Chapter~\ref{ch:isomap} for the details).  LLE, as
its name suggests, takes a rather different approach.

The idea of LLE is to approximate each data point by a linear
combination of its nearest neighbour points, and then to assume that
these linear reconstructions also apply in the reduced dimensionality
representation of the data.  Starting with points in the
high-dimensional data space, $\vec{x}_i \in \mathbb{R}^m$, with $i =
1, \dots, N$, we define a weight matrix $\mathbf{W} \in \mathbb{R}^{N
  \times N}$.  This matrix is constrained by the condition that
$W_{ij} = 0$ unless $\vec{x}_j \in \mathcal{N}(\vec{x}_i)$, where
\index[not]{neighbourhood@$\mathcal{N}(p)$, neighbourhood of a point
  $p$}$\mathcal{N}(\vec{x}_i)$ is the set of nearest neighbours to
point $i$, usually defined simply as the $k$ nearest neighbours as
measured by Euclidean distances in $\mathbb{R}^m$.  This condition
reflects the fact that each point will be reconstructed as a linear
combination of its nearest neighbours only.  Further, we require that
$\sum_j W_{ij} = 1$, a simple normalisation condition.  Under these
constraints, we define a cost function as
\begin{equation}
  \label{eq:lle-weight-cost}
  J_1(\mathbf{W}) = \sum_i || \vec{x}_i - \sum_j W_{ij} \vec{x}_j
  ||^2.
\end{equation}
This measures the mismatch between the original data points
$\vec{x}_i$ and the linear reconstructions built from each points'
nearest neighbours.  We can find an optimum value for the weight
matrix $\mathbf{W}$ by solving a least-squares optimisation problem as
\begin{equation}
  \mathbf{W}^{(\text{opt})} = \argmin_{\mathbf{W}} J_1(\mathbf{W}),
\end{equation}
where \index[not]{argmin@$\argmin_x f(x)$, $x$ minimising
  $f(x)$}$\argmin_x f(x)$ denotes the value of $x$ that minimises a
function $f(x)$.  The local weight values $W^{(\text{opt})}_{ij}$
satisfy local symmetries: they are invariant under rotations,
rescalings and translations of each point $\vec{x}_i$ and its nearest
neighbours.  Invariance under rotations and rescalings follows
directly from the form of \eqref{eq:lle-weight-cost}, while invariance
under translations follows from the row-sum condition on the weight
matrix, i.e. $\sum_j W_{ij} = 1$.  These symmetries reflect the fact
that the weight matrix encodes local geometric properties that are not
dependent on a particular choice of coordinate frame.  So far, there
is little to distinguish LLE from many other methods to reconstruct
manifolds using locally linear approximations.  The crucial step comes
when we assume that points in the $d$-dimensional reduced space, with
$d < m$, which we write as $\vec{y}_i \in \mathbb{R}^d$, are related
by the same linear relations as the points in the original data space.
The justification for this assumption is that, if the data points lie
on or near to a $d$-dimensional manifold embedded in the data space,
there should be an affine map (actually, a composition of rotation,
rescaling and translation only) that transforms high-dimensional
coordinates in each point neighbourhood to global intrinsic
coordinates on the manifold.  The weights $\mathbf{W}$ are constructed
so as to be invariant under such linear transformations, so one would
expect the same relationships to hold between the reduced coordinates
on the manifold as between the original data space coordinates.  Note
that this condition of preservation of the local weights is
\emph{assumed}: no strong results are offered in either
\citep{roweis-lle} or \citep{saul-lle} concerning the conditions under
which this condition is satisfied --- the justification is purely
heuristic (although reasonable).  Some stronger results exist for the
Hessian LLE method, derived from LLE (Chapter~\ref{ch:hessian-lle}).
The condition that the same weights relate points in the
low-dimensional representation of the data as do the high-dimensional
data points can be expressed through a second cost function, which is
a function of the reduced coordinate vectors $\vec{y}_i$,
parameterised by the optimum weight matrix found above:
\begin{equation}
  J_2(\{ \vec{y}_i \}; \mathbf{W}^{(\text{opt})}) = \sum_i ||
  \vec{y}_i - \sum_j W^{(\text{opt})}_{ij} \vec{y}_j ||^2.
\end{equation}
We can minimise this cost function with respect to the reduced
coordinates $\vec{y}_i$ by solving an $N \times N$ eigenvalue
decomposition problem (which is sparse because the weight matrix
$\mathbf{W}$ is sparse), with the bottom $d$ eigenvectors associated
with nonzero eigenvalues forming an orthogonal basis for the reduced
representation.

This concentration on local geometric properties is the main
distinction between Isomap and LLE.  Isomap solves a single large
eigenvalue problem that aims to preserve global geometric information
about the data manifold (geodesic distances between each pair of data
points) in the reduction to a lower-dimensional representation, while
LLE solves a number of local optimisation problems to develop the
weight matrix $\mathbf{W}$, which is then used in a global eigenvalue
problem to construct a representation of the global structure of the
manifold that preserves local geometric information (the relative
locations of the nearest neighbours to each point) from overlapping
subsets of the data.  The rather different theoretical bases of the
two approaches make it difficult to compare them directly.  It might
be of some use to attempt to characterise the set of problems for
which Isomap and LLE give equivalent results, perhaps adopting ideas
from the analysis of image manifolds performed by
\citet{donoho-isometry} for Isomap.

As for Isomap, applications of LLE have been numerous.  LLE was
originally developed for the purposes of image classification in
machine learning contexts, and has been used in many studies in this
field, particularly for face recognition in static and video images
\citep[e.g.,][]{fan-lle-face-recog,jiang-lle-face-tracking,kadoury-lle-face-det}.
Other image analysis applications include the processing of
hyperspectral imagery \citep{mohan-lle-hyperspectral} and
classification of geophysical data: \citet{boschetti-dimred} shows
examples of the analysis of both gravity anomaly data and hand-drawn
images of geological sections.  Other applications are more closely
tied to the geometrical basis of the LLE algorithm, such as the study
of \citet{sun-lle-meshing}, who use LLE to develop a new algorithm for
producing two-dimensional meshes of surfaces embedded in
three-dimensional Euclidean space, for applications in computer
graphics and animation.  Another animation application of LLE is the
work of \citet{jin-lle-animation}, who describe a scheme for
calculating ``in between'' frames in animations of deformable shapes
from a small set of key frames, applying constraints that maintain
volume or other relevant geometrical invariants of the shape.  LLE has
also seen some application in classification and modelling problems in
chemistry and molecular biology
\citep[e.g.,][]{lheureux-lle-qsar,wang-lle-mem-proteins}.

While Isomap and LLE are by far the most commonly applied of this
class of dimensionality reduction methods, many other methods have
been proposed.  Assessment of these methods is difficult, as new
methods are frequently presented with test cases consisting only of
rather simple geometrical data sets, or one of a few standard pattern
recognition examples widely used in the machine learning community.
Absent a coherent theoretical framework for the analysis of these
methods, the only way to assess their performance in more complex
problems is to try them.  To some extent, for even the more commonly
used methods such as Isomap and LLE, theoretical insights are rather
lacking.  For Isomap, there are some asymptotic convergence results
\citep{bernstein-isomap-proofs} and a very interesting examination of
exactly what manifolds can be faithfully represented by Isomap in a
constrained image manifold context \citep{donoho-isometry}, but little
else.  For LLE, the formal results that do exist are for a
modification of the algorithm called Hessian LLE
(Chapter~\ref{ch:hessian-lle}), which appears somewhat easier to
analyse than the original LLE algorithm \citep{donoho-hessian}.  In
\citet{saul-lle}, the original inventors of the LLE approach observe
that ``[n]otwithstanding these recent results, our theoretical
understanding of algorithms for nonlinear dimensionality reduction is
far from complete.''

Several other nonlinear dimensionality reduction methods have been
proposed that have similarities to Isomap and/or LLE.  Two methods,
proposed by \citet{lin-rml} and \citet{brun-rnc}, rely on the idea of
Riemannian normal coordinates in a manifold.  We recall some notions
from differential geometry to allow us to define Riemannian normal
coordinates.  A manifold $M$ of dimension $m$ is a topological space
for which every point has a neighbourhood $U$ homeomorphic to an open
set $V$ of $\mathbb{R}^m$ with $\phi: U \to V \subset \mathbb{R}^m$.
Then $(U, \phi)$ is called a local coordinate chart.  This simply
means that a manifold everywhere looks like a subset of $\mathbb{R}^m$
locally.  An atlas for a manifold $M$ is a collection of charts $\{
  (U_\alpha, \phi_\alpha) \, | \, \alpha \in J \}$ with $J$ some index
set, such that $\{ U_\alpha \, | \, \alpha \in J \}$ is an open cover
of $M$.  A manifold $M$ is called a differential manifold of class
$C^r$ if there is an atlas of $M$, $\{ (U_\alpha, \phi_\alpha) \, | \,
  \alpha \in J \}$, such that for any $\alpha, \beta \in J$, denoting
function composition by \index[not]{function composition@$f \circ g$,
  function composition}$\circ$, i.e. $(f \circ g)(x) = f(g(x))$, the
composite $\phi_\alpha \circ \phi_\beta^{-1} : \phi_\beta(U_\alpha
\cap U_\beta) \to \mathbb{R}^m$ is differentiable of class
\index[not]{Cr@$C^r$, differentiability class}$C^r$, i.e. it has at
least $r$ continuous derivatives.  A smooth ($C^\infty$) differential
manifold $M$ endowed with a smooth inner product $g(u, v) = \langle u,
v \rangle_M$ (called the Riemannian metric) on each tangent space
$T_p{}M$ is called a Riemannian manifold $(M, g)$.

The exponential map $\exp_p(v)$ transforms a tangent vector $v \in
T_p{}M$ to a point $q \in \gamma \subset M$, with $\gamma$ being the
unique geodesic through $p$ whose tangent vector at $p$ is $v$, such
that $\mathrm{dist}(p, q) = ||v|| = g(v, v)^{1/2}$, $\mathrm{dist}(p,
q)$ being the distance between $p$ and $q$ measured along $\gamma$.
The exponential map at a point $p$ thus takes elements of the tangent
space at $p$ into points on the manifold, mapping along geodesics
through $p$, with the displacement of the resulting point given by the
magnitude of the element of the tangent space, as measured by the
metric.

Riemannian normal coordinates (RNCs) with centre $p$ are defined to be
the local coordinates defined by the chart $(U, \exp_p^{-1})$. Here,
the chart mapping $\exp_p^{-1}: U \to T_p{}M$ assigns local
coordinates to points in $M$ via an isomorphism $E: \mathbb{R}^m \to
T_p{}M$ that establishes a basis for $T_p{}M$.  The full chart
function is thus $\phi = E^{-1} \circ \exp_p^{-1} : U \to
\mathbb{R}^m$.  Riemannian normal coordinates are unique up to the
choice of the orthonormal basis used in the definition of the
isomorphism $E$.  A very loose way of thinking about the definition of
RNCs is that they represent a ``closest reasonable approximation'' for
local coordinates in $M$: if you imagine taking part of $M$ and trying
to smooth it out as nicely as possible to make it look like part of
$\mathbb{R}^m$, Riemannian normal coordinates are the most natural
coordinate system that you would end up with.  This can be made
precise by looking at a number of conditions on the components of the
metric at $p$ as represented in RNCs \citep{choquet-bruhat-amp}.  For
the purposes of explaining the two dimensionality reduction methods
considered here, only two properties of Riemannian normal coordinates
are important.  For a neighbourhood $U$ of the point $p \in M$ (there
are some technical conditions on the neighbourhood $U$ that are not
important here), these conditions state that:
\begin{enumerate}
  \item{The coordinates of $p$ are $(0, ..., 0)$.}
  \item{Given a vector $v \in T_p{}M$ with components $v^i$ in local
    coordinates, define $\gamma_v$ to be the geodesic with starting
    point $p$ and velocity vector $v$. Then $\gamma_v$ is represented
    in local coordinates as $\gamma_v(t) = (tv^1, ..., tv^m)$ so long
    as it is confined to $U$. In words, geodesics through $p$ are
    locally linear functions of $t$, the arclength along the
    geodesic.}
\end{enumerate}

%
%
\begin{figure}
  \begin{center}
    \includegraphics[width=\textwidth]{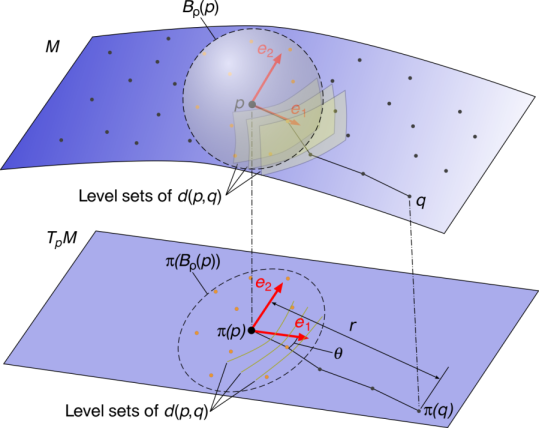}
  \end{center}
  \caption[RNC method]{Schematic view of Riemannian normal coordinates
    method of \citet{brun-rnc}.  Refer to text for explanation.}
  \label{fig:rnc-diagram}
\end{figure}
%

The first method we consider, presented by \citet{brun-rnc}, is based
on a direct calculation of approximate Riemannian normal coordinates
based at a single point for all points in the data manifold.  The
approximate Riemannian normal coordinates are calculated as polar
coordinates in two steps, one to determine the geodesic distance from
the base point to each point in the data set, and another to determine
an angular coordinate.  The following explanation should be read in
conjunction with Figure~\ref{fig:rnc-diagram}, where projection from
the original data manifold $M$ to the Riemannian normal coordinate
space, which is essentially $T_p{}M$ for some base point $p \in M$, is
denoted by $\pi: M \to T_p{}M$.  Consider a set of data points $X
\subset \mathbb{R}^m$, which we assume to be sampled from a manifold
$M$ (shown in the top part of Figure~\ref{fig:rnc-diagram}).  A base
point $p \in X$ is selected and local PCA is performed in a ball
$B_p(\rho) \subset X$ of radius $\rho$ surrounding $p$, with all
distances measured using the Euclidean distance in the data space.
The eigenvalue spectrum from the local PCA calculation is used to
determine the dimensionality of the manifold $M$ by taking only the
leading larger magnitude eigenvalues.  The eigenvectors from the PCA
analysis are then used to form an approximate basis $\{ \vec{e}_i \}$
for the tangent space at $p$.  In Figure~\ref{fig:rnc-diagram}, $M$ is
a two-dimensional manifold and the vectors $\vec{e}_1$ and $\vec{e}_2$
provide a basis for the tangent space $T_p{}M$.  All points $y \in
B_p(\rho)$ (orange points in Figure~\ref{fig:rnc-diagram}) are then
mapped to an approximation to $T_p{}M$ by projection onto the $\{
  \vec{e}_i \}$ as $y \mapsto \pi(y)$.  This approximation to $T_p{}M$
is displayed in the bottom part of Figure~\ref{fig:rnc-diagram}.
Next, the geodesic distances from each point $q \in X$ to the base
point, denoted by $d(p, q)$ are approximated using the same type of
nearest neighbour graph plus shortest paths approach as in Isomap ---
a nearest neighbour graph is constructed to serve as a skeleton for
the manifold $M$, based on Euclidean distances in the data space and a
nearest neighbour count or neighbourhood radius.  Geodesic distances
from the base point $p$ to all points that are not in $B_\rho(p)$ are
estimated by calculating shortest paths through the graph using
Dijkstra's algorithm \citep{ahu-algorithms}.  In
Figure~\ref{fig:rnc-diagram}, a single shortest path is shown from the
base point $p$ to a typical point $q \in M$: the distance along this
shortest path is used to determine the projection radius $r$ for the
projected point $\pi(q) \in T_p{}M$.  Finally, the direction of each
point $q \in X$ with respect to $p$ in the approximate Riemannian
normal coordinates is found by numerically estimating the gradient
$\vec{g} = \nabla_y d^2(y, q)|_{y=p}$.  The idea here is simply that
the gradient of the distance function $d(y, q)$, at point $y = p$,
will point in the direction of the geodesic that runs from $p$ to $q$.
In the top part of Figure~\ref{fig:rnc-diagram}, level sets of the
distance function $d(y, q)$ are shown as transparent surfaces, and the
projections of these level sets to the approximate tangent space are
shown as yellow curves in the lower part of
Figure~\ref{fig:rnc-diagram}.  In the calculations for the
\citet{brun-rnc} method, $d^2(y, q)$ is used rather than $d(y, q)$ for
numerical stability, and $d^2(y, q)$ is interpolated using a second
order polynomial, since we only have values for $d(y, q)$ at a finite
sample of points close to $p$ (the number of points in $B_p(\rho)$).
Determination of the direction of the gradient vector $\vec{g}$ then
provides the angular component $\theta$ of the coordinates of the
projected point.  The approximate Riemannian normal coordinates of the
point $q$ are then given by $\pi(q) = r \vec{g} / |\vec{g}|$.

This method is slightly unusual compared to some of the other
differential geometry based methods in that it really is a
\emph{local} method.  Although geodesic distances (used as the radial
component $r$ of the Riemannian normal coordinates) are approximated
using calculations involving all of the data points, the direction of
each data point from the base point (used as the angular component
$\theta$ of the Riemannian normal coordinates: see
Figure~\ref{fig:rnc-diagram}) is estimated using only points in
$B_p(\rho)$, i.e. points within a distance $\rho$ of the base point.
Intuitively, it would appear that this would make the method extremely
sensitive to noise in the data.  It seems as though this method may
only be of real use in cases where the data points are known to lie on
a low-dimensional manifold with relatively little sampling noise.  In
\citep{brun-rnc}, only examples of this type are presented, meaning
that it is difficult to assess the usefulness of this method for more
difficult problems.  In its favour, because it does not require the
solution of large eigenproblems and relies primarily on calculations
for points in the ball $B_p(\rho)$ only, this method is fast, and so
may find application in the simple cases described above.  It also has
the capability to clearly identify manifolds of non-trivial topology
--- \citet{brun-rnc} illustrate this with an example of a set of
images sampled from a manifold homeomorphic to the Klein bottle.

The second method based on Riemannian normal coordinates, described by
\citet{lin-rml} and called Riemannian manifold learning (RML), bears
the same resemblance to Isomap as does the \citet{brun-rnc} method, in
that it uses a nearest neighbour graph and shortest paths through the
graph to define approximate geodesics in the data manifold.
Transformations from the data space to a lower-dimensional space are
sought that preserve these geodesic distances.  Where RML differs from
Isomap is that it also attempts to preserve local curvature
information, by seeking local transformations that preserve the angles
between line segments joining adjacent points in the nearest neighbour
graph.  RML goes beyond the simple nearest neighbour graph used in
Isomap by building a simplicial complex
\citep[Section~2.1]{hatcher-alg-top} that approximates the data
manifold, using the dimensionality of this complex as a guide to
estimating the data dimension.  The construction of the approximating
simplicial complex is based on the ideas of
\citet{freedman-manifolds}, with some optimisations, and is organised
so as to generate a complex constructed from well-shaped simplices.

%
%
\begin{figure}
  \begin{center}
    \includegraphics[width=\textwidth]{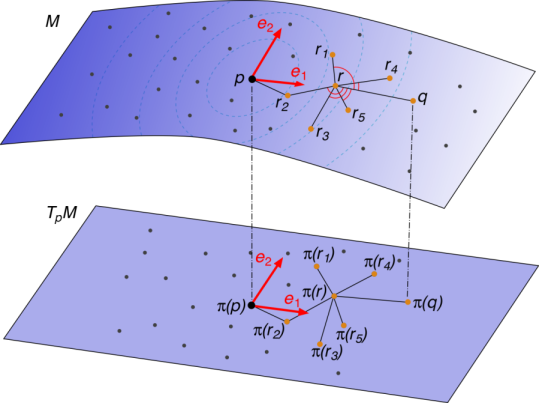}
  \end{center}
  \caption[RML method]{Schematic view of Riemannian manifold learning
    method of \citet{lin-rml}.  Refer to text for explanation.}
  \label{fig:rml-diagram}
\end{figure}
%

The computation of Riemannian normal coordinates for points in the
data manifold then proceeds in a slightly different fashion to the
method of \citet{brun-rnc}.  The following explanation refers to
Figure~\ref{fig:rml-diagram}, and as before, projection from the
original data manifold $M$ to the Riemannian normal coordinate space
is denoted by $\pi: M \to T_p{}M$.  First, a base point $p$ is
selected based on the geodesic radius of each point --- if the
approximate distance function defined by shortest paths through the
nearest neighbour graph is $d(x, y)$ for any two data points $x, y \in
X$, then the geodesic radius of a point $\rho_g(x) = \max_{y \in X}
d(x, y)$ is the maximum distance from the point $x$ to any other point
in the data set.  The base point $p$ is selected as the point with the
minimal geodesic radius, making it, in some sense, the point closest
to the ``centre'' of the data set.  Once the base point $p$ has been
selected, a basis for $T_p{}M$, the tangent space at $p$, is
constructed.  Since we know the dimensionality of the manifold from
the dimensionality of the approximating simplicial complex, we can
simply pick a suitable number of point-to-point vectors connecting $p$
to its nearest neighbours and orthogonalise this set to provide a
basis for $T_p{}M$.  In Figure~\ref{fig:rml-diagram}, $M$ is
two-dimensional and the two vectors $\vec{e}_1$ and $\vec{e}_2$ form a
basis for $T_p{}M$.  (In the \citet{brun-rnc} method, local PCA is
used at this point to simultaneously determine a dimensionality
estimate for $M$ and to produce a basis for $T_p{}M$.)  Once a basis
has been constructed for the tangent space, Riemannian normal
coordinates are assigned to data points one by one, in order of
increasing geodesic distance from the base point $p$: the dashed
curves in the upper part of Figure~\ref{fig:rml-diagram} represent
contours of constant geodesic distance from the base point $p$, and
indicate the order in which points are dealt with.

Points directly connected to $p$ in the simplicial complex are
assigned coordinates by a simple projection onto the basis for
$T_p{}M$, while points further from $p$ are assigned coordinates by
solving constrained local optimisation problems, one for each point.

One of these local optimisation problems is illustrated in
Figure~\ref{fig:rml-diagram}, where the orange points in the upper
part of the diagram, $r$ and $r_1$ to $r_5$, are used to constrain the
projection of the next point to be handled, $q$.  The projection onto
the $\{ \vec{e}_j \}$ basis is constrained so that angular
relationships between the connections in the nearest neighbour graph
are maintained as closely as possible in the reduced representation of
the manifold: the angular constraints are indicated by the angles
marked in red in the upper part of Figure~\ref{fig:rml-diagram}.  The
optimisation problem is thus set up so that the angles between the
vectors $q-r$ and each of the $q-r_i$ are maintained as closely as
possible in the projected coordinates.  Each local optimisation uses
Riemannian normal coordinates computed for points closer to $p$ to
compute the equivalent angles in the projected coordinates, these
closer points being the $r$ and $r_i$ in Figure~\ref{fig:rml-diagram}.
\citet{lin-rml} explain the details of this optimisation scheme, but
the main point is that reduced coordinates are assigned to each data
point by solving a small local optimisation problem.  At no point is
it necessary to solve a large global optimisation or
eigendecomposition problem.  This means that the RML method is
potentially very fast, even for large data sets, with the most
expensive step probably being the single-source shortest paths
calculation using Dijkstra's algorithm.  However, the potential
efficiency of the method is rather balanced out by its significantly
greater complexity compared to methods such as Isomap or LLE.
Experiments also seem to indicate that it may be rather sensitive to
data sampling issues.  I implemented the RML method with a view to
applying it to some of the tropical Pacific climate data analysed in
Chapters~\ref{ch:obs-cmip-enso}--\ref{ch:hessian-lle}, but was only
able to produce reasonable results for the simplest of test cases.
The method is rather complex, and the breadth-first search of the
nearest neighbour graph implied by assigning coordinates in order of
geodesic distance from the base point can, in some cases, introduce
circular dependencies between the computations for different data
points.  It is not quite clear how to reliably lift this dependency in
practice.

Neither of the methods based on Riemannian normal coordinates has, as
far as I know, been applied to any significant problems, which would
have made a comparison of the \citet{lin-rml} method with Isomap and
nonlinear PCA of some interest.  Both \citet{brun-rnc} and
\cite{lin-rml} display the results of applying their methods to simple
geometrical test cases, and \citet{lin-rml} also show results from
some of the commonly used face recognition data sets.  In theory, at
least computationally, these methods should display good scaling
behaviour when applied to larger data sets.  However, both methods use
Riemannian normal coordinates based at a single point to represent the
whole of the data manifold, and it is not at all clear that this
approach can work well for data manifolds with complex structure.  The
theoretical basis of both methods is well defined, at least for data
points sampled from smooth manifolds, although the RNC method of
\citet{brun-rnc} may have problems in practical applications because
of the use of a finite-difference approximation to the gradient
$\nabla_y d^2(x, y)$, and the computational problems with
\citeauthor{lin-rml}'s RML method mentioned above require careful
treatment to make the method work in the general case.

\subsection{Kernel methods}
\label{sec:kernel-methods}

One very useful way to think about several of the
geometrical/statistical dimensionality reduction methods is as
variants of kernel PCA, a nonlinear extension of standard principal
component analysis that relies on a transformation known as ``the
kernel trick''.  For centred data, the covariance matrix $\mathbf{C}$
used in the calculation of the PCA eigendecomposition
\eqref{eq:pca-covariance} is a Gramian matrix, i.e. a matrix of inner
products of the input vectors, $\vec{x}_i \in \mathbb{R}^m$, as
$C_{ij} = \vec{x}_i \cdot \vec{x}_j$.  In kernel PCA, one assumes the
existence of a mapping $\Phi : \mathbb{R}^m \to \mathcal{F}$, where
$\mathcal{F}$ is some vector space called the \emph{feature space}
(which may be an infinite-dimensional function space), and a
\emph{kernel}, a nonlinear function $k(\vec{x}_i, \vec{x}_j) =
\Phi(\vec{x}_i) \cdot \Phi(\vec{x}_j)$.  The kernel $k$ and mapping
$\Phi$ allow us to define a matrix $\mathbf{C}'$ as $C'_{ij} =
k(\vec{x}_i, \vec{x}_j)$.  The matrix $\mathbf{C}'$ is a Gramian
matrix, since it is formed of inner products taken in the vector space
$\mathcal{F}$ and we can thus use it as the basis of a PCA-like
eigendecomposition.  The eigendecomposition in the vector space
$\mathcal{F}$ has a \emph{nonlinear} relationship to our original
input data $\vec{x}_i$ through the nonlinear mapping $\Phi$.  By
suitable definition of $k$ and $\Phi$, it is possible to represent
Isomap (Chapter~\ref{ch:isomap}), LLE
(Section~\ref{sec:nonlinear-diff-geom}), Laplacian eigenmaps
(Section~\ref{sec:spectral-graph-theory}) and other similar algorithms
as instances of kernel PCA.  \citet{scholkopf-kernel-pca} give a good
review of kernel PCA and related approaches, while
\citet{ham-kernel-view} construct explicit representations of Isomap,
LLE and Laplacian eigenmaps as kernel methods.

As originally envisaged, kernel PCA used simple Gaussian or polynomial
kernels, but experiments have shown that these kernels do not provide
a suitable basis for dimensionality reduction in practical problems
(\citet{scholkopf-kernel-pca} show some examples).  The idea of
constructing a kernel to match the characteristics of other
dimensionality reduction methods gives much better results, and is
presented very clearly in \citep{weinberger-kernel}, where another
dimensionality reduction method, based on semi-definite programming,
is developed as a kernel method, with the constraints for the
semi-definite programming problem being derived from simple geometric
notions of what relationships between points in the original space
should be preserved in the reduced space.

\subsection{Neural network methods}
\label{sec:ann-methods}

There are two different classes of methods that use artificial neural
networks \citep{haykin-nns} as a means of nonlinear dimensionality
reduction.  The first class covers methods that use multilayer
perceptron neural networks as function approximators, to approximate a
nonlinear function mapping input data values in a high-dimensional
space to reduced values in a low-dimensional space, and then to
reconstruct an approximation to the original data from the reduced
representation.  The networks are trained by comparing the
reconstructed data to the original data.  Methods in this class
include the nonlinear PCA method of \citet{kramer-nlpca} which is
treated in detail in Chapter~\ref{ch:nlpca}, as well as other methods
based on such autoassociative networks \citep{hinton-ann}.

%
%
\begin{figure}
  \begin{center}
    \includegraphics[width=\textwidth]{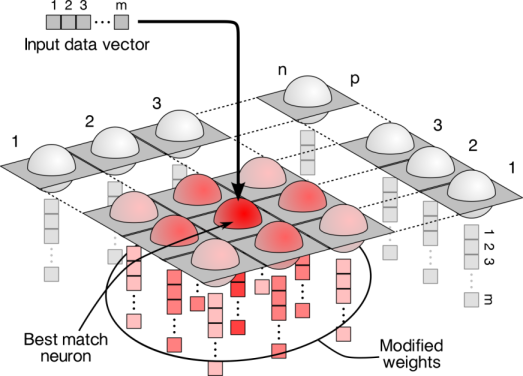}
  \end{center}
  \caption[Self-organising map]{Schematic view of self-organising map
    neural network.  Here, the input data is $m$-dimensional and the
    neurons in the network are laid out in an $n \times p$ rectangular
    grid.  Each neuron has an associated $m$-dimensional weight vector
    and training proceeds by identifying the neuron whose weights best
    match each input vector (the best matching neuron is highlighted
    in red here) and updating the weights of that neuron and its near
    neighbours in the grid to more closely match the input vector.}
  \label{fig:som-diagram}
\end{figure}
%

The second class of neural network methods is based on self-organising
maps (SOMs) \citep{kohonen-soms}, a neural network architecture
designed to project high-dimensional data to a lower-dimensional
(usually two-dimensional) discrete representation, preserving the
locality between data vectors in the original high-dimensional space.
SOMs differ from many of the other dimensionality reduction methods
here in that they produce a discrete characterisation of the input
data.  A typical SOM network architecture is illustrated in
Figure~\ref{fig:som-diagram}.  The network is typically arranged as a
rectangular or hexagonal grid of neurons, each neuron having an
associated weight vector, of the same dimensionality as the input data
vectors $\vec{x}_i \in \mathbb{R}^m$.  The SOM training algorithm
compares each input data vector in turn to all the weight vectors in
the network, selecting the neuron with the closest matching weight
vector according to some comparison criterion, then adjusts the
weights of the best matching neuron and its near neighbours in the
network to be closer to the input data item.  The basic update rule
for each weight vector $\vec{w}_i^t \in \mathbb{R}^m$ at training step
$t$ is
\begin{equation}
  \label{eq:som-update}
  \vec{w}_i^{t+1} = \vec{w}_i^t + \Theta(i, t) \alpha(t) \, (\vec{x}_t
  - \vec{w}_i^t).
\end{equation}
Here $\vec{x}_t$ is the input data vector presented at training step
$t$, $\alpha(t)$ is a learning coefficient that decreases
monotonically with $t$, and $\Theta(i, t)$ is a function that
describes the influence on other neurons in the network of the neuron
whose weight vector best matches the input data vector.  $\Theta(i,
t)$ might be set to one for the best-matching neuron and all of its
neighbours within a particular radius in the network and zero
elsewhere, or might use a Gaussian function centred on the
best-matching neuron: the latter is the choice shown in schematic form
in Figure~\ref{fig:som-diagram}.  Whichever form is used for
$\Theta(i, t)$, its effective radius is set to decrease with $t$.  The
update process \eqref{eq:som-update} is repeated for a number of
training cycles, during each of which each of the input data vectors
is used to update the network weights.  The training parameters
$\alpha(t)$ and $\Theta(i, t)$ are varied between training cycles to
reduce the degree to which an individual data vector is able to affect
the network weights as the training process progresses.  The final
result is a configuration of network weights that associates patterns
in the input data with groups of neurons in the network with similar
weights.  The use of the neighbourhood function $\Theta(i, t)$ in the
weight update rule \eqref{eq:som-update} constrains the network
weights to cluster similar input data vectors together in the map.
Further data vectors, not included in the training set, may then be
classified by comparing them to each of the neurons in the network.
There appear to be relatively few theoretical results justifying the
SOM approach.  \citet{fort-som-maths} collects some of the known
results, but the stronger results apply primarily to the
one-dimensional case, which is of little practical interest.
Confidence in the method is based more on its empirical success than
on any strong theoretical underpinnings.

Self-organising maps have seen application in geophysics --- for
instance, \citet{klose-soms} used SOMs in the interpretation of
seismic survey data --- and also in climate science.
\citet{leloup-som} used SOMs to study decadal variability in
\eln/Southern Oscillation (ENSO) indexes based on tropical Pacific sea
surface temperature data.  A self-organising map was trained using
these indexes from observational data between 1950 and the present.  A
hierarchical clustering algorithm was then used to partition the cells
of the SOM to capture coherent regions of different types of behaviour
in the map.  Composite spatial patterns of variability for each
cluster were then calculated by forming the mean of the weight vectors
associated with each of the neurons in each cluster: since the weight
vectors lie in the same space as the input data, this results in a
composite pattern that is directly comparable to the input data.  The
classification of patterns of variability of ENSO produced by the SOM
was then compared to corresponding variability seen in thermocline
depth and sea level pressure patterns, and time trajectories of
evolution between different SOM clusters were presented as a method of
following the evolution of ENSO variability.  A clear split in
behaviour is seen around the mid-1970s climate shift.
\citet{leloup-som-2} present a similar sort of analysis, examining
ENSO variability in the same set of IPCC models as used in this
thesis.  Finally, \citet{reusch-soms} present a SOM analysis of
monthly mean and monthly standard deviation sea level pressure data in
the North Atlantic.  Here, the self-organising map was trained
directly on spatial maps of sea level pressure data, rather than on a
small set of indexes as in \citep{leloup-som,leloup-som-2}.  The
results are compared to a PCA analysis of the same data.  The SOM
approach is able to pick out the North Atlantic Oscillation as the
main mode of variability in the data more clearly than PCA, due to
some coupling of distinct spatial patterns of variability in the PCA
analysis.

\subsection{Spectral graph theory methods}
\label{sec:spectral-graph-theory}

Many of the differential geometry based methods described in
Section~\ref{sec:nonlinear-diff-geom} rely on the construction of a
weighted graph based on nearest neighbour relationships between data
points as measured in the high-dimensional input space.  One might
then ask whether methods from graph theory might have something to
offer in terms of determining low-dimensional structures in data.  One
method based on this idea is called Laplacian eigenmaps
\citep{belkin-niyogi}.  This method works by constructing a nearest
neighbour graph from the input data with edge weights chosen so as to
make the graph Laplacian an approximation to the Laplace-Beltrami
operator on the data manifold.  The nearest neighbour graph is
constructed in the usual manner using either a neighbourhood radius
$\varepsilon$ or neighbour count $k$.  For input data vectors
$\vec{x}_i \in \mathbb{R}^m$, assumed to lie in a Riemannian manifold
$M$, edge weights are given by a matrix $\mathbf{W}$, where
\begin{equation}
  W_{ij} = \exp ( - || \vec{x}_i - \vec{x}_j ||^2 / \tau )
\end{equation}
if points $i$ and $j$ are connected, and $W_{ij} = 0$ otherwise.
Here, $\tau \in \mathbb{R}$ is a parameter of the method.  The graph
Laplacian is then calculated as $\mathbf{L} = \mathbf{D} - \mathbf{W}$
where $\mathbf{D}$ is a diagonal matrix with $D_{ii} = \sum_j W_{ij}$.
The Laplacian $\mathbf{L}$ can be thought of as an operator on
functions defined over the vertices of the graph, and with the edge
weights selected here, is closely analogous to the Laplace-Beltrami
operator over the data manifold $M$ in which the data points lie.  The
final step of the algorithm is to solve the generalised eigenvalue
problem
\begin{equation}
  \label{eq:le-eigenproblem}
  \mathbf{L} \vec{q}_j = \lambda_j \mathbf{D} \vec{q}_j,
\end{equation}
ordering the eigenvalues $\lambda_j$ and eigenvectors $\vec{q}_j$ in
ascending order of eigenvalue.  There is always one zero eigenvalue,
denoted by $\lambda_0 = 0$, which we ignore, since it relates to a
constant function over the graph.  The next $d$ eigenvectors are then
used to define an embedding in $d$-dimensional Euclidean space as
\begin{equation}
  \vec{x}_i \to ( \, \vec{q}_1(i), \dots, \vec{q}_d(i) \, ).
\end{equation}

The analogy between the graph Laplacian and the Laplace-Beltrami
operator on the manifold approximated by the graph is the key point to
this method \citep{belkin-niyogi}.  Consider the manifold $M$ as being
embedded in $\mathbb{R}^m$, which means that a Riemannian structure
(i.e. an inner product and corresponding metric tensor) on $M$ is
induced by the standard Riemannian structure on $\mathbb{R}^m$.  We
seek a map from $M$ to $\mathbb{R}$, denoted $f: M \to \mathbb{R}$,
that maps points close together on $M$ to points close together on the
real line.  In the following, we assume that $f \in C^2(M)$.  Consider
two points $\vec{x}, \vec{z} \in M$, close together as measured by the
metric on $M$ (note that we use vector notation for $\vec{x}$ and
$\vec{z}$ to emphasise that the points have coordinates in
$\mathbb{R}^m$ by virtue of the embedding of $M$ in $\mathbb{R}^m$).
It can be shown \citep{belkin-niyogi}, that
\begin{equation}
  \label{eq:le-gradient-mapping}
  | f(\vec{z}) - f(\vec{x}) | \leq || \nabla f(\vec{x}) || \,
  ||\vec{z} - \vec{x} || + o(|| \vec{z} - \vec{x} ||),
\end{equation}
where the notation \index[not]{little-O notation@$o(\varepsilon)$,
  little-O notation}$o(\dotsc)$ is a case of the more general usage
that $f(x) \sim o(g(x))$ as $x \to \infty$ if $\lim f(x)/g(x) = 0$,
i.e. $f(x)$ is asymptotically dominated by $g(x)$.  The relation
\eqref{eq:le-gradient-mapping} is essentially a Taylor expansion of
the function $f$ about the point $\vec{x}$.  Here, the gradient
$\nabla f(\vec{x})$ is a vector in the tangent space $T_x{}M$ such
that, for any vector $\vec{v} \in T_x{}M$, $df(\vec{v}) = \langle
\nabla f(\vec{x}), \vec{v} \rangle_M$, with \index[not]{exterior
  derivative@$df$, exterior derivative}$df$ denoting the exterior
derivative on $M$, i.e. $\nabla f(\vec{x})$ is dual to $df(\vec{x})$
for any $f$.  The norms on the right hand side of
\eqref{eq:le-gradient-mapping} are Euclidean norms in the space in
which the data manifold is embedded, $\mathbb{R}^m$, i.e. from the
point of view of dimensionality reduction, the original
high-dimensional data space.  The relationship
\eqref{eq:le-gradient-mapping} thus shows that the norm of the
gradient $||\nabla f||$ gives an estimate of how far apart $f$ maps
neighbouring points in $M$.  Heuristically, an embedding that
preserves locality in some average sense can then be found by solving
the minimisation problem
\begin{equation}
  \label{eq:le-minimisation}
  \min_{||f||_{L^2(M)}=1} \int_M ||\nabla f||^2
\end{equation}
where the constraint on the $L^2(M)$ norm of $f$ here is simply to
prevent the degenerate solution $f = 0$.  Stokes' theorem can then be
used to relate the integrand in \eqref{eq:le-minimisation} to
\index[not]{Laplace-Beltrami operator@$\Delta f$, Laplace-Beltrami
  operator}$\Delta f$, the result of applying the Laplace-Beltrami
operator on $M$ to $f$, and the solution of the minimisation problem
turns out to be to let $f$ be the eigenfunction of $\Delta$ with the
smallest non-zero eigenvalue.  This situation directly parallels the
solution of the eigenproblem \eqref{eq:le-eigenproblem} for the graph
Laplacian in the Laplacian eigenmaps case.

The Laplacian eigenmaps computation is similar to LLE in that it is
primarily based on local computations, which makes it relatively
insensitive to outliers and noise.  As with LLE, this means that the
Laplacian eigenmaps method has rather different behaviour to global
algorithms such as Isomap.  Laplacian eigenmaps is a relatively
lightweight computation, involving only computations in local
neighbourhoods and a single sparse eigenproblem.  As for most of the
other differential geometry based methods, the determination of the
nearest neighbour graph is also required.  The analogy with LLE is in
fact, rather deeper, as LLE can to some extent be interpreted in the
same theoretical framework as Laplacian eigenmaps, with the operator
in the final global LLE eigenproblem being closely related to the
graph Laplacian, as shown by \citet{belkin-niyogi}.  The relationships
between spectral graph theory and diffusion processes on manifolds are
developed further in \citep{coifman-diffusion-maps} and references
therein.

In \citep{belkin-niyogi}, the Laplacian eigenmaps method is applied
only to relatively simple test data sets, but there are some
interesting studies applying the method to much more complex problems.
One of these studies, in the field of biological signalling networks,
by \citet{barbano-bio-networks}, examined a particular neurochemical
signalling cascade, based around a protein involved in a complex
network of interactions.  The signalling network was modelled as a
system of ODEs connecting the concentrations of 102 different chemical
species.  This 102-dimensional system was integrated to quasi-steady
state from random initial conditions when forced by pulses of some of
the main external controlling factors in varying ratios, also
selecting rate kinetic constants randomly.  The final concentrations
of species were taken to define points in 102-dimensional space
characterising the state of the system.
\citeauthor{barbano-bio-networks} were then able to use Laplacian
eigenmaps to reduce their 102-dimensional data to three dimensions and
to pick out coherent characteristics of the signalling cascade with
biological significance.  In particular, they were able to test the
robustness of their results to a number of variations in the topology
of the signalling network --- the real network appears to hold
together as a coherent whole, with all of its parts being necessary to
ensure robustness of the overall response of the system to variations
in external conditions.

A second interesting application of Laplacian eigenmaps is the study
of \citet{shen-fmri-eigenmaps}, who used a variant of the technique to
post-process functional magnetic resonance imaging (fMRI) data showing
brain activity in the presence of different external stimuli.  Here,
the input data is in the form of a time series of three-dimensional
fMRI images of the brain of an experimental subject under controlled
stimulus conditions.  A variant of Laplacian eigenmaps was used to
reduce the dimensionality of this input data, while preserving the
functional connectivity between different voxels (three-dimensional
volume elements).  The reduced dimensionality data was then processed
further using a clustering algorithm to identify functional regions of
the brain activated by different external stimuli.

Another dimensionality reduction method closely related to both
Laplacian eigenmaps and LLE is Hessian LLE \citep{donoho-hessian},
also called Hessian eigenmaps, described in detail in
Chapter~\ref{ch:hessian-lle}.  This method has some computational
similarities to LLE, but its theoretical basis has more in common with
the spectral graph theory methods exemplified by Laplacian eigenmaps.

\subsection{Computational topology}
\label{sec:computational-topology}

Another set of ideas that are of interest here, although not strictly
dimensionality reduction methods, are approaches based on
computational algebraic topology, in particular the computation of
homology groups for point cloud data \citep{kaczynski-homology}.  Most
of the dimensionality reduction methods described so far rely on the
approximation of \emph{geometrical} information from point cloud data,
such as geodesic distances, local tangent spaces, curvature and so on.
The computation of \emph{topological} invariants offers some prospect
of providing a robust characterisation of dynamics and structure in a
more highly summarised form than some of the geometrical methods.  A
full treatment of the ideas of algebraic topology and homology theory
can be found in \citep{hatcher-alg-top}.  Here I confine myself to
brief definitions and a few motivating comments.

The fundamental idea of homology theory is that, given a topological
space $X$, one can associate with the space certain Abelian groups,
the homology groups of $X$, denoted $H_k(X)$ with $k = 0, 1, 2,
\dots$, that encode important topological properties of the space $X$
in an algebraic form.  Let us give an abstract definition of the
homology groups before providing a more concrete example.  For a
topological space $X$, we compute a \emph{chain complex}, $A = C(X)$,
a sequence of Abelian groups, $A_0, A_1, A_2, \dots$, connected by
group homomorphisms \index[not]{chain map@$\partial_k$, chain
  map}$\partial_k : A_k \to A_{k-1}$, with the property that
$\partial_k \circ \partial_{k+1} = 0$ for all $k$.  This means that
the image of the mapping $\partial_{k+1}$ (denoted by
\index[not]{image@$\im f$, image of a map}$\im \partial_{k+1}$) is
contained within the kernel of the mapping $\partial_k$ (denoted by
\index[not]{kernel@$\ker f$, kernel of a map}$\ker \partial_k$).  If
we now define $Z_k(X) = \ker \partial_k$, the $k$-cycles of $X$, and
$B_k(X) = \im \partial_{k+1}$, the $k$-boundaries of $X$, we can
finally define the $k$th homology group of $X$, $H_k(X)$, as the
quotient group
\begin{equation}
  H_k(X) = Z_k(X) / B_k(X).
\end{equation}

\begin{figure}
  \begin{center}
    \includegraphics[width=\textwidth]{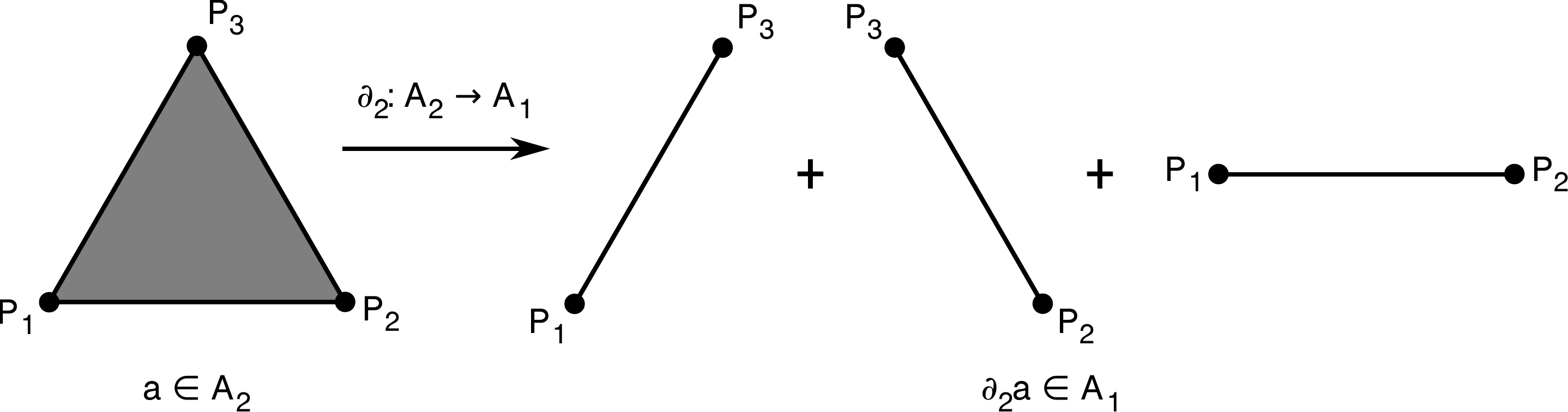}
  \end{center}
  \caption[Simplicial homology example]{Example of simplicial homology
    calculation.  The 2-chain, $a \in A_2$, consisting of a single
    2-simplex (i.e. a triangle) is mapped to its boundary $\partial_2
    a \in A_1$ by the boundary operator $\partial_2: A_2 \to A_1$.
    The boundary $\partial_2 a$ consists of three line segments,
    i.e. three 1-simplices.}
  \label{fig:homology-diagram}
\end{figure}

This setting is extremely general, and a large number of homology
theories have been developed for different types of topological
spaces.  Perhaps the simplest example, and an example that allows for
the use of some geometrical intuition to help understand the
homological approach, is that of simplicial homology, where the
topological space $X$ is a simplicial complex.  Here, the $k$-chains
of $X$, $A_k$, are free Abelian groups whose generators are the
$k$-dimensional simplices of $X$.  An element of $A_k$ is thus, in a
formal sense, a combinatorial sum of $k$-simplices from $X$.  In this
case, the mapping $\partial_k : A_k \to A_{k-1}$ is a boundary map:
for $a \in A_k$, $\partial_k a \in A_{k-1}$ represents the boundary of
the simplices in the $k$-chain $a$.  As a simple example, suppose that
$a \in A_2$ is a single $2$-simplex, i.e. a triangle (see
Figure~\ref{fig:homology-diagram}).  The boundary of a triangle is
made up of three lines, i.e. three $1$-simplices.  This combination of
$1$-simplices is a $1$-chain, $\partial_2 a \in A_1$.  (There is a
technical point that should be mentioned here: in the full theory, the
simplices must be oriented, to allow for the correct behaviour when
combining $k$-chains.)  In the simplicial homology setting, the
condition that $\partial_k \circ \partial_{k+1} = 0$ is thus an
instance of the fact that the boundary of a boundary is empty for both
manifolds and simplicial complexes.

%
%
%

Consider now the $k$-boundaries of $X$, $B_k(X) = \im \partial_{k+1}$.
We see that these are precisely those $k$-chains of $X$ that are the
boundaries of some $(k\!+\!1)$-chain.  For instance, the $1$-chain
found as the boundary of the triangle above is a member of $B_1(X)$.
Similarly, the $k$-cycles of $X$, $X_k(X) = \ker \partial_k$ are those
$k$-chains of $X$ that have an empty boundary.  Again, the $1$-simplex
found as the boundary of the triangle above has an empty boundary
(since it is a boundary itself), so is a member of $Z_1(X)$.  We now
see that the homology groups $H_k(X) = Z_k(X) / B_k(X)$ represent
precisely those $k$-cycles of $X$ that are \emph{not} $k$-boundaries.
We know that all $k$-chains that are $k$-boundaries are also
$k$-cycles, but the $k$-cycles that are not $k$-boundaries encode
interesting and useful information about the simplicial complex $X$.

\begin{figure}
  \begin{center}
    \begin{tabular}{ccc}
      \includegraphics[width=0.3\textwidth]{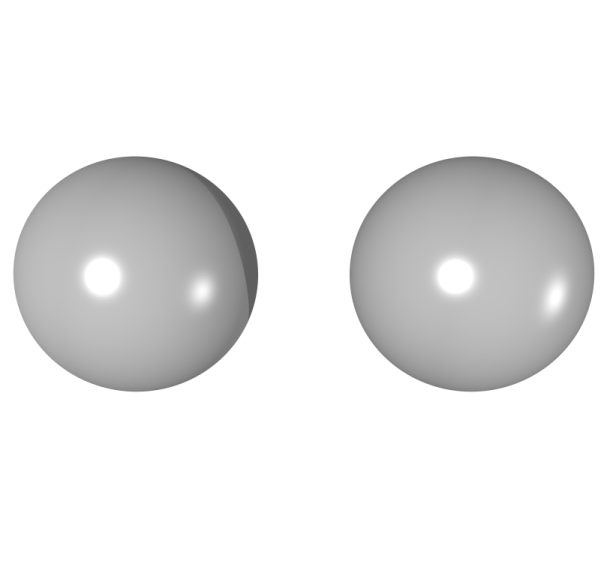} &
      \includegraphics[width=0.25\textwidth]{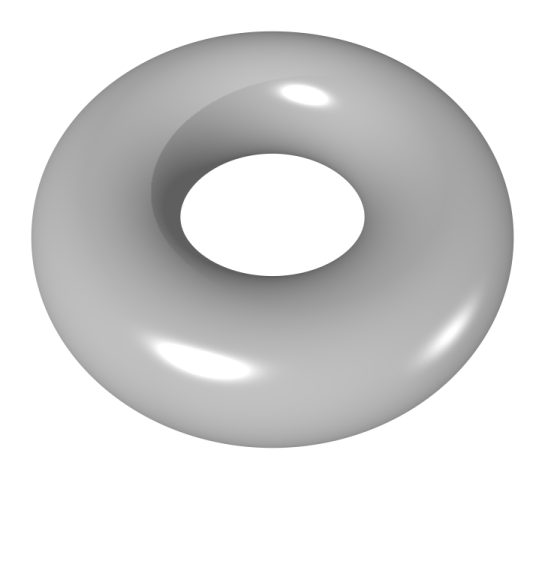} &
      \includegraphics[width=0.3\textwidth]{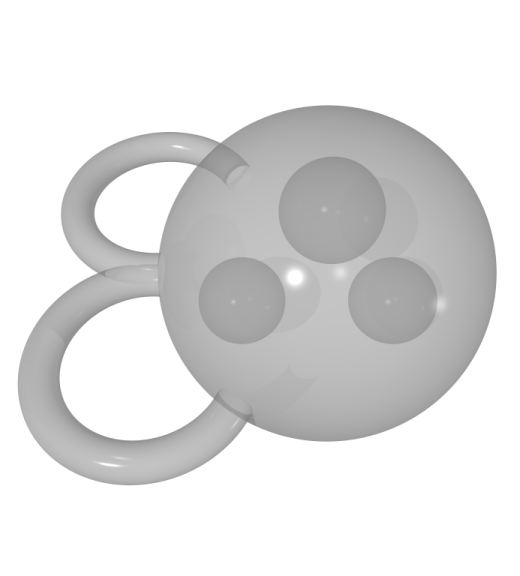}
      \\
      $H_0 \cong \mathbb{Z}^2, \beta_0 = 2$ &
      $H_0 \cong \mathbb{Z}, \beta_0 = 1$ &
      $H_0 \cong \mathbb{Z}, \beta_0 = 1$ \\
      $H_1 = 0$ &
      $H_1 \cong \mathbb{Z}, \beta_1 = 1$ &
      $H_1 \cong \mathbb{Z}^2, \beta_1 = 2$ \\
      $H_2 = 0$ &
      $H_2 = 0$ &
      $H_2 \cong \mathbb{Z}^3, \beta_2 = 3$ \\
      & & \\
      (a) Two spheres & (b) Torus & (c) Modified sphere
    \end{tabular}
  \end{center}
  \caption[Homology examples]{Homology groups for some simple example
    point sets in $\mathbb{R}^3$: a pair of spheres (a), a torus (b),
    and a sphere with two handles and three interior cavities (c).}
  \label{fig:homology-pictures}
\end{figure}

Although for general topological spaces $X$, the homology groups
$H_k(X)$ may be arbitrary Abelian groups, in most of the applications
of interest here, the situation is simpler.  For subsets of Euclidean
space $X \subset \mathbb{R}^m$, it can be shown that
\begin{equation}
  H_k(X) = 0 \text{ for $k \geq m$}
\end{equation}
and, writing \index[not]{Z@$\mathbb{Z}$, the integers}$\mathbb{Z}$ to
denote the integers, and \index[not]{group isomorphism@$A \cong B$,
  group isomorphism}$A \cong B$ to denote a group isomorphism between
two groups $A$ and $B$,
\begin{equation}
  H_k(X) \cong \mathbb{Z}^{\beta_k} \text{ for $0 \leq k < m$},
\end{equation}
where the $\beta_k$ are non-negative integers called the \emph{Betti
  numbers} of $X$.  For this type of situation, the totality of the
homological structure of $X$ is encoded in $m$ integers.  Things are
even better than this though, since simple geometrical interpretations
can be applied to the homology groups in this setting.  First,
$H_0(X)$ counts the number of connected components in $X$: if $\beta_0
= n$, then $X$ has exactly $n$ connected components.  The higher
homology groups count different types of ``holes'' in $X$.  For the
sake of simplicity, consider the case $m = 3$, and consider $X$ to be
some three-dimensional solid object in $\mathbb{R}^3$.  In this case,
we know that $H_k(X) = 0$ for $k \geq 3$, so we only need to consider
$H_1(X)$ and $H_2(X)$.  $H_1(X)$ counts the number of ``tunnels''
through $X$, i.e. the number of distinct ways that a curve could be
threaded through void spaces in $X$ from one outer surface to another,
while $H_2(X)$ counts the number of internal cavities in $X$.  Some
examples are shown in Figure~\ref{fig:homology-pictures} to help
clarify this description.

%
%
%

In practice, it is extremely difficult to compute homology groups for
realistic examples.  Until recently, this rendered the use of
computational topology methods for data analysis rather infeasible.
Recently though, a new method has been developed, cubical homology
\citep{kaczynski-homology}, which employs a discretisation of point
sets in terms of cubical elements, rather than the triangulation
required to construct the simplicial complex used in simplicial
homology.  This approach is computationally much more tractable than
simplicial and other homology theories, and software has been
developed to compute homology groups for point sets represented as
pixellated images or discretely sampled volume elements.  It has been
demonstrated that cubical homology is strictly equivalent to
simplicial homology, meaning that the Betti numbers calculated from
the cubical homology procedure are the same as those that would be
found via simplicial homology.

Two recent studies indicate the potential of these topological methods
for the characterisation of complex data sets that arise from the
integration of dynamical systems.  \citet{gameiro-betti} used the
Betti numbers of spatiotemporal patterns from the integration of
partial differential equation systems to compute Lyapunov exponents
measuring the development of spatiotemporal chaos.  The second study
examined the evolution of the complexity of patterns in solutions to
the Cahn-Hilliard equation, a nonlinear PDE model for phase separation
in the formation of alloys and other compound materials
\citep{gameiro-cahn-hilliard}.  In both of these studies, a
thresholding procedure was used to produce point sets for analysis by
cubical homology.  Denoting the domain of integration of the PDE
system of interest by $\Omega$ and solutions to the PDE by $u(t, x)$
with $x \in \Omega$, $t \in \mathbb{R}$, then the point sets
$X^\pm(t)$ are defined by
\begin{equation}
  \begin{gathered}
    X^+(t) = \{ x \in \Omega \, | \, u(t, x) > m \}, \\
    X^-(t) = \{ x \in \Omega \, | \, u(t, x) < m \},
  \end{gathered}
\end{equation}
where $m$ is a threshold value.  For the Cahn-Hilliard equation, the
sets $X^\pm(t)$ represent the regions of $\Omega$ where one phase or
other of the multi-phase material being modelled predominates.
Computation of homology groups $H_k(X^\pm(t)) \cong
\mathbb{Z}^{\beta_k^\pm(t)}$ can then characterise the topological
structure of the distribution of the two phases as a function of time,
with the time series of the Betti numbers $\beta^\pm_k(t)$ providing a
concise summary of the structure of the solutions.

Many further extensions of these ideas exist, but one idea of
particular interest is \emph{persistent homology}
\citep{zomorodian-persistent,ghrist-barcodes}, an adaptation of
simplicial homology.  Generally, when constructing a simplicial
complex from a discrete point cloud data set, some spatial scale has
to be selected, usually in the form of a neighbourhood radius used to
decide which points should be connected by $1$-simplices, which then
form the skeleton of the simplicial complex.  If one computes homology
groups for simplicial complexes based on different neighbourhood radii
for the same point cloud data set, in general they will not be the
same, since topological features can appear and disappear as the scale
at which the data set is sampled varies (the topological features
recorded by the homology groups are discrete).  \emph{Persistent
  homology} is a rigorous approach to this problem of
scale-dependence, based on the construction of a filtered simplicial
complex, an increasing sequence of simplicial complexes constructed
from the point cloud data set by varying the neighbourhood radius
used.  The simplicial complexes composing the sequence are connected
by \emph{chain maps}, homomorphisms between the chain complexes
defined on the individual simplicial complexes in the sequence.  The
filtered simplicial complex thus represents structure from the point
cloud data set at a range of different spatial scales in an organised
way.  It is then possible to construct an algebraic structure called a
\emph{persistence complex}, a family of chain complexes over the
simplicial complexes, connected both via boundary maps on each
simplicial complex and via the chain maps relating the different
simplicial complexes in the sequence.  Suitably constructed homology
groups calculated from this persistence complex capture all of the
variation in homological structure of the filtered simplicial complex
in a coherent fashion, in a sense including both the structure that
arises from the boundary maps in each individual chain complex, and
also the chain maps defined between the different simplicial complex
elements of the filtered simplicial complex.  The information encoded
in these homology groups can be summarised in an easily visualised
form as \emph{barcodes}, sets of intervals of the neighbourhood radius
over which different topological features of the data are present.
One can then immediately identify which features of the input data set
exhibit \emph{topological persistence}, i.e. features that are most
insensitive to data sampling resolution and the scale at which the
data is examined.  The barcode for a data set can be thought of as the
persistent analogue of the Betti numbers for simplicial homology
groups.

Approaches to data analysis for dynamical systems applications based
on computational topology are in their infancy, but they hold out a
great deal of promise, capturing as they do discrete topological
invariants of data sets in a highly summarised way.

\subsection{Miscellaneous geometrical/statistical methods}

A number of other methods deserve mention here, but cannot easily be
allocated to any of the categories above.

\subsubsection{Independent component analysis}
\label{sec:ica}

This is a method based on a modification of the ideas of PCA and
related methods, originally developed for time series analysis in
signal processing applications.  While PCA produces principal
component time series that are \emph{uncorrelated}, independent
component analysis (ICA) seeks to find components of a time series
that are \emph{statistically independent}, a much stronger condition
that relies on relationships between the higher moments of the data.
A number of different methods of finding such independent components
have been developed, some of which are described by
\citet{aires-climate-ica}, who applied ICA to the analysis of tropical
sea surface temperature variability.  In the approach implemented by
\citeauthor{aires-climate-ica}, the component independence criterion
is expressed using an information-theoretic measure, and the
relationship between the input data (which is reduced using an initial
PCA step) and the output components is represented by a feedforward
neural network.  Other recent applications of ICA have included the
work of \citet{ikeda-ica}, who used ICA to process
magnetoencephalography data, and \citet{li-process-trends-ica}, who
applied ICA to the analysis of chemical process trend data.  There
does not appear to have been a comprehensive comparison of ICA to
other dimensionality reduction methods, and the different approaches
to defining a data-based criterion for statistical independence
between data sets makes comparison between different studies
difficult.

\subsubsection{Principal curves and surfaces}
\label{sec:principal-curves}

Principal curves and surfaces (PCS) is a non-parametric nonlinear
generalisation of PCA based on a simple intuitively appealing notion
of what it means for an approximation to lie through the ``middle'' of
a data set.  The method was originally presented by
\citet{hastie-pcs}, who concentrated on one-dimensional reduced
representations of data sets, i.e. principal curves.

The idea of a principal curve is initially most easily explained by
considering data drawn from a known probability distribution.  Let
$\vec{X}$ denote a random $m$-dimensional vector with density $\phi$,
with finite second moments and $\mathbb{E}(\vec{X}) = 0$.  Let
$\vec{f}(s)$ be a smooth non-self intersecting curve in
$\mathbb{R}^m$, parameterised by arclength $s$ (meaning that
$||\vec{f}'(s)|| = 1$), with $s \in S \subset \mathbb{R}$.  Define the
projection index $s_{\vec{f}}: \mathbb{R}^m \to \mathbb{R}$ as
\begin{equation}
  \label{eq:pcs-projection}
  s_{\vec{f}}(\vec{x}) = \sup_s \{ s : || \vec{x} - \vec{f}(s) || =
    \inf_t || \vec{x} - \vec{f}(t) || \}.
\end{equation}
The projection index $s_{\vec{f}}(\vec{x})$ of a point $\vec{x} \in
\mathbb{R}^m$ is simply the value of $s$ for which $\vec{f}(s)$ is
closest to $\vec{x}$, taking the largest value of $s$ when there are
several candidate values.

The essential idea behind PCS is then to define a consistency
condition on the curve $\vec{f}$.  A curve $\vec{f}$ is said to be
self-consistent, or a principal curve of the density $\phi$, if
\begin{equation}
  \label{eq:pcs-consistency}
  \mathbb{E}_\phi(\vec{X} | s_{\vec{f}}(\vec{X}) = s) = \vec{f}(s)
\end{equation}
for almost all $s$, where \index[not]{expectation with respect to
  density@$\mathbb{E}_\phi$, expectation with density
  $\phi$}$\mathbb{E}_\phi$ denotes the expectation with respect to the
density $\phi$.  This condition means that each point on the curve,
$\vec{f}(s)$, is the mean (weighted by the density $\phi$) of all
points that project to $\vec{f}(s)$ under the projection operator
$s_{\vec{f}}$.  Intuitively, a principal curve is simply a
parameterised curve that best lies through the ``middle'' of the
centres of the distribution $\phi$.

\citet{hastie-pcs} present a number of theoretical results concerning
principal curves, showing the relationship between principal curves
and principal components (e.g., if a straight line is a
self-consistent curve, then it is a principal component) and showing
that principal curves are critical points of the distance between the
curve and the density $\phi$, in the following sense.  Let $d(\vec{x},
\vec{f}) = || \vec{x} - \vec{f}(s_{\vec{f}}(\vec{x})) ||$ denote the
distance from a data point $\vec{x}$ to its projection onto the curve
$\vec{f}$, and let $D^2(\phi, \vec{f}) = \mathbb{E}_\phi[ d^2(\vec{X},
  \vec{f})]$ denote the expectation of the squared distance with
respect to the density $\phi$.  Now let $\mathcal{G}$ be a class of
curves parameterised over $S$, and for $\vec{g} \in \mathcal{G}$,
consider perturbations of $\vec{f}$ of the form $\vec{f}_t = \vec{f} +
t\vec{g}$.  Then the curve $\vec{f}$ is a critical point of the
distance function $D^2(\phi, \vec{f})$ if
\begin{equation}
  \left. \frac{d D^2(\phi, \vec{f}_t)}{dt} \right|_{t = 0} = 0, \text{
    for all } \vec{g} \in \mathcal{G}.
\end{equation}
Now, consider a class $\mathcal{G}_B$ of smooth ($C^\infty$) curves
parameterised over $S$, such that $||\vec{g}|| \leq 1$ and
$||\vec{g}'|| \leq 1$ for all $\vec{g} \in \mathcal{G}_B$, i.e. the
perturbations are all bounded with bounded derivative.
\citeauthor{hastie-pcs} prove that for perturbations in
$\mathcal{G}_B$, a principal curve is a critical point of the distance
function.

Principal curves may be found using an iterative algorithm presented
by \citet{hastie-pcs}, which they extend from the problem of finding a
principal curve of a distribution to that of finding a principal curve
of a discretely sampled data set by using a number of different
smoothing methods in the computation of the conditional expectation in
\eqref{eq:pcs-consistency}.

The extension of the ideas of principal curves to higher-dimensional
reduced representations of data sets, touched on only very briefly by
\citet{hastie-pcs}, is explored in more detail by \citet{leblanc-pcs}.
Formally, all of the definitions presented above for principal curves
extend over to the case of a fully parameterised surface $\vec{f}(s,
t)$, but practical difficulties arise because of the lack of a
canonical parameterisation for surfaces equivalent to the arclength
parameterisation for curves.  Although there is no conceptual problem,
this redundancy does present organisational and numerical problems for
the representation of parameterised surfaces.  \citet{leblanc-pcs} use
a spline-based multivariate regression modelling framework to
represent surfaces in their adaptation of the PCS algorithm.  This
framework allows them to construct parameterised surface
representations in a systematic way, and appears to be quite
successful.

There does not appear to have been any direct comparison between the
performance of PCS and other nonlinear dimensionality reduction
methods, although \citet{malthouse-nlpca-problems} demonstrated that
there are strong similarities between PCS and the nonlinear PCA method
described in Chapter~\ref{ch:nlpca}.  \citet{hastie-pcs} describe two
applications of PCS, one being the computation of optimal magnet
positions for the alignment of particle accelerator beamlines and the
other the comparison of different chemical assay methods for
electronic waste recycling.  \citet{dong-pcs} present an application
of PCS to chemical process monitoring, where they use a combination of
PCS and neural network methods to build a nonlinear control model for
the so-called Tennessee Eastman problem \citep{downs-tennessee}.
\citet{banfield-pcs} used principal curves to aid in the
identification of ice floes in satellite imagery, while
\citet{jacob-pcs} used them to characterise phase plots of human
respiratory data under different physiological conditions.
\citet{hicks-pcs} use principal curves to model the shape, size and
texture of different diatom species as seen in photographs and
drawings, developing a system for automated identification of species
from photographs.  All of these applications are based only on
principal \emph{curves} --- the higher dimensional extensions of the
method seem to have received much less attention.

\subsection{Out-of-sample extensions}

One question that arises for several nonlinear dimensionality
reduction methods is the treatment of \emph{out-of-sample} points.
Suppose that we are given an initial data set, to which we apply a
nonlinear dimensionality reduction procedure to produce a reduced
dimensionality representation of the input data.  If we are then
confronted with further input data items, naively we would need to
incorporate these new data points into the original data set and rerun
our nonlinear dimensionality reduction analysis to produce a new
reduced dimensionality representation incorporating the new data
points.  Because the computational requirements of the methods
described here can be rather onerous, there is an incentive to develop
a cheaper approach to handling extra data points than adding the new
points to the original data set and completely redoing the
dimensionality reduction analysis.

This question of out-of-sample data is most natural in the setting of
neural networks, where a ``training'' data set is used to determine
network weights (and sometimes architecture) by adjusting the weights
to optimise a cost function based on the training data (e.g., for the
NLPCA method of Chapter~\ref{ch:nlpca}, \eqref{eq:nlpca-cost-function}
shows the cost function optimised over the training data set to
determine the network weights).  After the training phase is over,
further data items can be applied to the inputs of the neural network:
in the dimensionality reduction case, the network outputs then provide
a reduced dimensionality reduction of the input data.

Although neural networks provide the most obvious setting for the idea
of an out-of-sample extension to a dimensionality reduction method,
the concept is applicable to most of the methods described here.  The
exact form of computations required to treat out-of-sample points
varies from method to method, and is generally rather more complicated
than neural network methods (where new data items are simply applied
to the inputs of the network, relying on the generalisation capability
of the network to produce a sensible output for items outside the
training set), but there has been some work to develop such extensions
for a number of methods.

For LLE, some consideration of out-of-sample extensions is included in
\citep{saul-lle}, by the original developers of LLE, but a more
general approach is offered by \citet{bengio-out-of-sample}, who
developed out-of-sample extensions of several nonlinear dimensionality
reduction methods in the common framework of Nystr\"{o}m's method.
Nystr\"{o}m's method was originally developed as a method of smoothly
interpolating solutions to certain classes of integral equations
\citep[Section~18.1]{press-nr}, but the basic idea has since been
adapted to linear algebra problems, where for an $n \times n$ matrix
$\mathbf{K}$ with rank $r \ll n$, Nystr\"{o}m's method provides a way
to estimate the eigenvalues and eigenvectors of $\mathbf{K}$ from a
small matrix $\mathbf{A}$ --- if the rank of $\mathbf{A}$ is $r$, then
the decomposition is exact.  The algebraic details of this are spelled
out in \citet[Section~3.1]{burges-review}.
\citet{bengio-out-of-sample} rely on a kernel representation of the
nonlinear dimensionality reduction methods
(Section~\ref{sec:kernel-methods}) to make the connection to
Nystr\"{o}m's method, and provide explicit formulae for out-of-sample
extensions of a number of common dimensionality reduction methods,
including Isomap and LLE.  The ideas presented in
\citet{bengio-out-of-sample} seem to be applicable quite generally to
dimensionality reduction methods expressible in terms of kernel PCA.

\section{Discussion}

Perhaps the strongest impression that one gains from a survey of the
literature on geometrical/statistical nonlinear dimensionality
reduction methods is just how ad hoc are many of the methods that have
been developed.  This arises, I believe, from the largely
applications-oriented approach taken to developing such methods.  Most
of the methods described in this chapter have been developed for
dealing with extremely difficult problems of classification and
clustering in the fields of machine learning, machine vision,
molecular biology and climate science.  In many of these applications,
any method that is empirically shown to perform better than linear
methods is welcome, and the development of a strong theoretical
framework to understand what is happening in these methods is a
secondary concern.  This is particularly clear in the machine learning
literature, where a plethora of dimensionality reduction methods have
appeared, with each new method generally being applied only to simple
geometrical test data sets and one or two of a standard set of
handwriting recognition or face recognition test cases.  Some of the
better-known methods have been applied in more complex settings in
this field \citep[e.g.,][]{jiang-lle-face-tracking}, but there is
little guidance available for which is the ``best'' method to use in
different circumstances.

This situation contrasts quite strongly with the case for the
development of dynamical dimensionality reduction methods, where the
emphasis has been much more on developing well-founded mathematical
notions of what it means to seek ``simplified'' or reduced dynamics
for a system.  These approaches are typified by slow manifold theory
\citep[e.g.,][]{rhodes-maas-pope} and averaging methods of various
kinds
\citep[e.g.,][]{kifer-coupled-averaging,majda-stoch-climate-models-2}
where much rigorous work has been devoted to understanding clearly how
these dimensionality reduction methods work and under what
circumstances the approximations involved in them are valid.

These rather different approaches to the development of dimensionality
reduction methods in the dynamical and geometrical/statistical
situations appear difficult to reconcile.  It would clearly be useful
to develop stronger theoretical foundations for understanding the
geometrical/statistical methods, but where results of this type have
been found, for instance for Isomap
\citep{bernstein-isomap-proofs,donoho-isometry}, they appear to be of
limited use in the practical situations where one might want to apply
these methods.  Coming from the opposite direction, there is still
quite a gap between the complexity of systems to which dynamical
dimensionality reduction methods can be applied and the models and
observational data sets that are of practical interest.

There do appear to be some ideas that might form the basis for
developing a more integrated theoretical view of at least some
geometrical/statistical dimensionality reduction methods.  Among these
are the use of kernel PCA as a framework for understanding different
methods, as demonstrated by \citet{scholkopf-kernel-pca},
\citet{ham-kernel-view} and \citet{weinberger-kernel}.  It is not
clear whether this approach can offer any insights into the
susceptibility of different methods to problems with data sampling and
noise properties of the input data, but it at least places otherwise
rather disparate dimensionality reduction methods into a shared
mathematical setting in which their commonalities and differences can
be explored.

A second setting in which a number of different dimensionality
reduction methods can be explored involves the relationships between
spectral graph theory and differential geometry
\citep{belkin-niyogi,donoho-hessian} and between random walks on
graphs and diffusion processes on manifolds
\citep{coifman-diffusion-maps}.  The important point in these analyses
is the relationship (sometimes, as in \citet{belkin-niyogi}, not much
more than an analogy) between discrete and continuous mathematical
structures and the dynamics on them: graphs versus manifolds, random
walks versus diffusions.  These relationships would appear to be the
best hope for developing stronger mathematical machinery for studying
geometrical/statistical dimensionality reduction methods, particularly
in cases where the data set of interest is produced by measurements on
some dynamical system, perhaps contaminated with noise.

A situation where some very promising work has been done in developing
methods for extracting coherent mathematical structures from point
cloud data is in the field of computational topology, mentioned in
Section~\ref{sec:computational-topology}.  Here the notion of
topological persistence has turned out to be key: the homology groups
of discretisations of a manifold at different spatial scales can be
related in a single mathematical structure that captures the important
topological properties of the data set as seen at the different
scales.  It appears that these methods are applicable to the analysis
of dynamical systems in a variety of settings and probably deserve
more attention.


%% file: 03-data-and-models.tex
\chapter{Data and Models}
\label{ch:data-and-models}

In this study, I examine ENSO variability in a number of observational
and model data sets, concentrating primarily on tropical Pacific sea
surface temperatures (SSTs).  Since ENSO is a coupled ocean-atmosphere
phenomenon, involving interactions between the ocean surface, ocean
equatorial wave dynamics and the wind fields across the Pacific basin,
a full examination of ENSO variability would require consideration of
other variables as well as SST, in particular thermocline depth and
surface wind stress.  However, in
Chapters~\ref{ch:nlpca}--\ref{ch:hessian-lle}, my intent is to study
the suitability of some of the geometrical/statistical dimensionality
reduction methods described in Chapter~\ref{ch:nldr-overview} for the
analysis of climate data.  For this purpose, a simple inter-model
comparison exercise seems most appropriate, so I restrict the bulk of
the analysis to sea surface temperature data only, with some
consideration of thermocline depth variability.  Additionally,
restriction of the analysis to SST data only sidesteps the issue of
the relative scalings to be applied to different fields, a factor that
could have an impact on the results for all of the analysis methods
considered here, but which is not particularly germane to the issues I
am trying to explore.

In this chapter, I describe the sources of data that I use, both
observational and model simulations.  In addition, I also describe a
number of simple geometrical test data sets used to help characterise
the behaviour of the dimensionality reduction methods explored in
Chapters~\ref{ch:nlpca}--\ref{ch:hessian-lle}.

\section{Observational data}

\subsection{Sea surface temperature}

\begin{figure}
  \begin{center}
    \includegraphics[width=\textwidth]{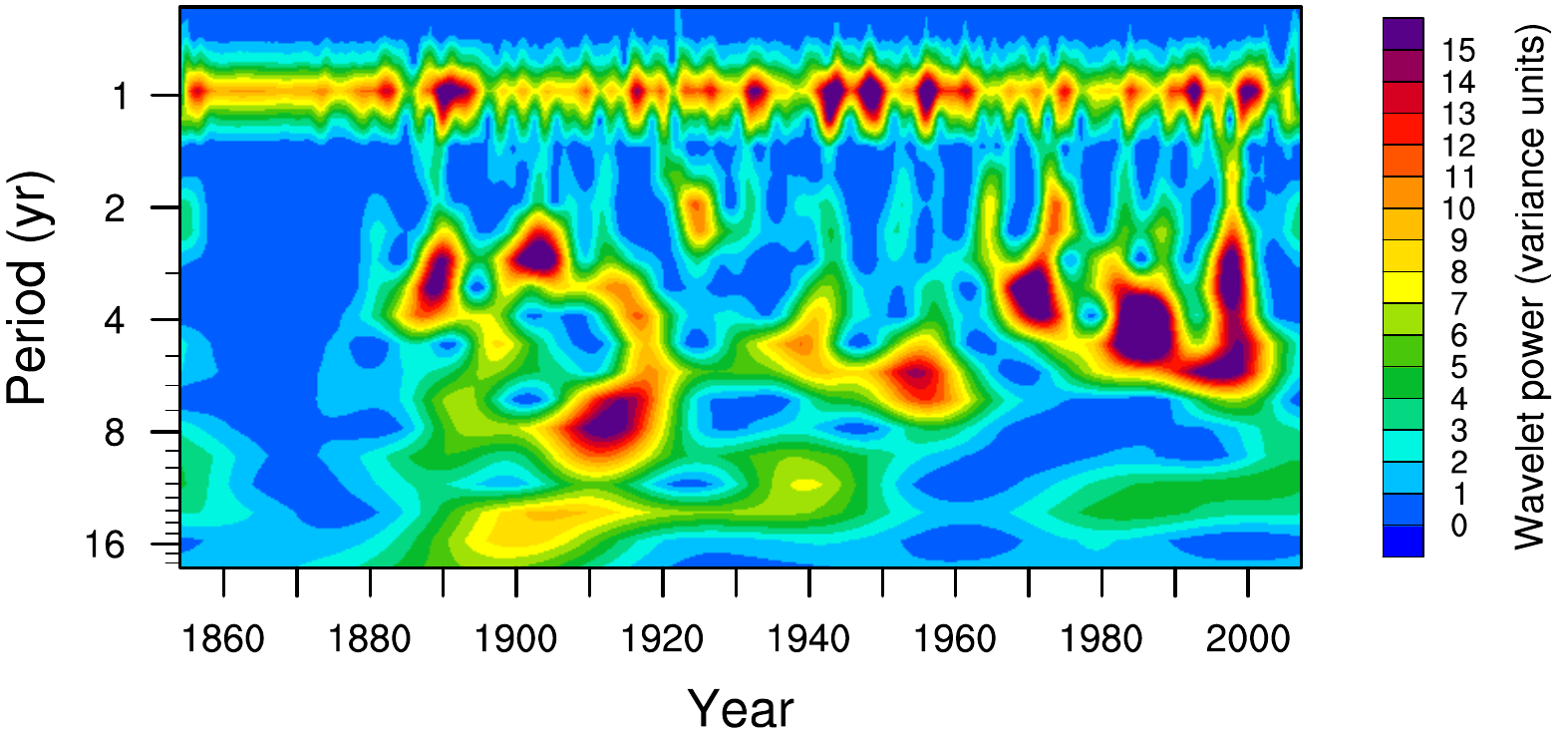}
  \end{center}
  \caption[ERSST v2 SST wavelet power spectrum]{Morlet wavelet power
    spectrum of ERSST v2 sea surface temperature data set, averaged
    over NINO3 SST index region (150\degree{}W--90\degree{}W,
    5\degree{}S--5\degree{}N), plotted in units of total SST
    variance.}
  \label{fig:ersst-wavelets}
\end{figure}

As observational sea surface temperature data, I use the NOAA ERSST v2
data set \citep{ersst-v2}, obtained from the NOAA/OAR/ESRL Physical
Sciences Division at \url{http://www.cdc.noaa.gov/}.  This is a global
data set running from 1854 to the present day at $2^\circ \times
2^\circ$ resolution, constructed from SST observations, using
statistical reconstructions for data-poor regions and time periods.
Because of a paucity of observations in the equatorial Pacific before
about 1900, most variability in this region in the early part of the
time series is due solely to the climatological annual cycle.  This is
demonstrated by Figure~\ref{fig:ersst-wavelets}, which plots the
wavelet power spectrum for the mean ERSST sea surface temperature
across the NINO3 SST index region (150\degree{}W--90\degree{}W,
5\degree{}S--5\degree{}N).  Before about 1890, significant power
appears only at annual frequencies.  No interannual variability that
might be associated with ENSO processes appears in the data before
this time.  For the purposes of this study, I extract a 100-year
subset of the full ERSST v2 time series, running from 1900--1999, in
order to reduce problems due to this non-stationarity.  There is still
residual non-stationarity in the SST observations associated with
changes in ENSO behaviour over time, but this is much smaller.  For
comparison with the models shown below in Table~\ref{tab:models}, the
number of ocean grid points for the ERSST v2 data set in the region
125\degree{}W--65\degree{}W, 20\degree{}S--20\degree{}N, i.e. the
dimensionality $m$ of the SST data considered as a set of vectors in
$\mathbb{R}^m$, is 1626.

%
%
%

\subsection{Thermocline depth and warm water volume}
\label{sec:thermocline-calc-desc}

Thermocline depth in the ocean is usually defined as the depth of the
20\degree C isotherm of ocean temperature, dividing warmer, mixed,
surface waters from cooler, stratified, underlying waters.  This
definition makes sense from the point of view of capturing changes in
the near-surface heat content of the ocean, but, for model studies,
may not be the best measure to use, since it is not closely related to
the dynamics of the model under study.  An alternative definition
identifies the point in the water column of maximum vertical
temperature gradient as the location of the dividing line between
surface and deep waters.  I will refer to these two measures of
thermocline depth as \ztw and \zgr, respectively.  There are
significant differences between \ztw and \zgr in some data sets, for
some regions and some times of the year, and this raises questions
over which definition should be used.
Section~\ref{sec:thermocline-var-results} presents analysis of which
differences are important for investigating the link between
equatorial thermocline variability and ENSO, and which can be
neglected.

For both modelled and observed ocean temperature data sets, \ztw and
\zgr are computed from cubic spline fits to the vertical temperature
profile at each horizontal point.  Both cases treat boundary cases
identically (surface temperature less than 20\degree C, bottom
temperature greater than 20\degree C), then the 20\degree C isotherm
is located using Brent's method \citep[Section~9.3]{press-nr}, and the
location of the point of maximum temperature gradient is found using
an explicit expression for the gradient of the spline interpolation,
again using Brent's method.  This latter calculation, involving what
is essentially a numerical differencing step, is sensitive to
numerical problems in the input data, and results for some of the
model data sets display numerical artefacts (see
Section~\ref{sec:thermocline-var-results} for details).

Following \citet{meinen-wwv}, I define the Pacific warm water volume
(WWV) as the volume of water lying above the thermocline in the region
120\degree{}E--80\degree{}W, 5\degree{}S--5\degree{}N.  This region is
extensive enough to capture variations in thermocline structure due to
equatorial wave dynamics important for ENSO, without extending so far
meridionally as to be excessively affected by seasonal variations in
the temperature structure of the ocean mixed layer
(Section~\ref{sec:thermocline-var-results}).  I calculate WWV from
both modelled and observational ocean temperature time series by a
Simpson's rule integration of the water volume above the thermocline.
                                   
To provide a basis for comparing model results to observations, I use
thermocline depth and warm water volume calculated from the NCEP GODAS
(Global Ocean Data Assimilation System) data set
\citep{behringer-godas-2004,behringer-godas-2007}.  This provides a
reanalysis of a number of ocean variables, including potential
temperature, as time series on 40 depth levels (with 10 metre vertical
resolution in the upper 200 metres of the ocean), running from January
1980 to March 2008.  This data is derived from simulations with the
GFDL MOMv3 ocean model \citep{pacanowski-mom3}, forced by momentum,
heat and freshwater fluxes from the NCEP atmospheric reanalysis
\citep{kalnay-ncep-reanalysis}, and further constrained by the
assimilation of temperature and salinity profiles from ship
observations and fixed buoy moorings in the Pacific
\citep{mcphaden-toga-obs} and Atlantic \citep{servain-pirata}.  This
time series is rather short, but is the best observational ocean
volume temperature data available in the Pacific.  I derive
thermocline depth and warm water volume time series from this data
using the same procedures as for the model data, substantially
following \citet{meinen-wwv}.  The NCEP reanalysis time series is
slightly longer than the data presented by \citeauthor{meinen-wwv};
the WWV time series derived from the NCEP \ztw thermocline depth gives
a good match to their results (not shown) during the period of overlap
between the time series.

\section{The CMIP3 models}
\label{sec:cmip3-models}

Model simulations from a range of coupled ocean-atmosphere GCMs were
used for this study, exploiting results from the World Climate
Research Programme's (WCRP) Coupled Model Intercomparison Project
phase 3 (CMIP3) multi-model data set (Table~\ref{tab:models}).  In
this study, I use data from pre-industrial control simulations in the
CMIP3 database.  I do not use all of the CMIP3 models, excluding from
consideration simulations that show little or no interannual
variability in the tropical Pacific, either because of model structure
or due to other unidentified problems (e.g., the GISS-AOM and GISS-ER
models).  For all model simulations, monthly time series of SST and
ocean body temperature are used, the length of the time series
available for each model being shown in Table~\ref{tab:models}.  Warm
water volume time series were calculated for all models where ocean
body temperature data was available in the same manner as for the
observational data (Section~\ref{sec:thermocline-calc-desc}).

%
%
\begin{table}
  \caption[Models used in this study]{Models used in this study,
    listing atmosphere and equatorial ocean spatial resolutions,
    lengths of simulation available ($L$), NINO3 SST index standard
    deviations ($\sigma_{\mathrm{NINO3}}$), number of ocean grid
    points in the region 125\degree{}W--65\degree{}W,
    20\degree{}S--20\degree{}N ($m$) and line style used in later
    plots.  Model horizontal resolution is expressed as degrees
    longitude $\times$ degrees latitude or a spectral grid designation
    and vertical resolution as $\mathrm{L}n$, where $n$ is the number
    of model levels.}

  \label{tab:models}

  \vskip4mm

  \begin{center}
    {\setlength\tabcolsep{4.5pt}
      \begin{tabular*}{\textwidth}{@{}lcccccl@{}}
        \toprule

        \multirow{2}{1.5cm}{Model} & Atmosphere & Ocean & $L$ &
        $\sigma_{\mathrm{NINO3}}$ & \multirow{2}{0.4cm}{\centering $m$} &
        \multirow{2}{0.8cm}{Legend} \\

        & resolution & resolution & (yr) & (\degree{}C) & & \\

        \midrule

        BCCR-BCM2.0\footnotemark[1] &
        T63 L31 &                                   
        $1.5^\circ\!\times\!0.5^\circ\,$L35 &        
        250 &                                       
        1.44 &                                      
        6133 &                                      
        \legenda{red} \\

        CCSM3\footnotemark[2] &
        T85 L26 &                                   
        $1.125^\circ\!\times\!{}0.27^\circ\,$L40 &   
        500 &                                       
        1.06 &                                      
        19550 &                                     
        \legenda{green} \\
	
        CGCM3.1(T47)\footnotemark[3] &
        T47 L31 &                                   
        $1.85^\circ\!\times\!{}1.85^\circ\,$L29 &    
        500 &                                       
        0.59 &                                      
        1742 &                                      
        \legenda{blue} \\
	
        CGCM3.1(T63)\footnotemark[3] &
        T63 L31 &                                   
        $1.4^\circ\!\times\!{}0.94^\circ\,$L29 &     
        400 &                                       
        0.64 &                                      
        4473 &                                      
        \legenda{magenta} \\
	
        CNRM-CM3\footnotemark[4] &
        T63 L45 &                                   
        $2^\circ\!\times\!{}0.5^\circ\,$L31 &        
        430 &                                       
        1.90 &                                      
        3049 &                                      
        \legenda{orange} \\

        CSIRO-Mk3.0\footnotemark[5] &
        T63 L18 &                                   
        $1.875^\circ\!\times\!{}0.84^\circ\,$L31 &   
        380 &                                       
        1.26 &                                      
        3395 &                                      
        \legenda{cyan} \\

        ECHO-G\footnotemark[6] &
        T30 L19 &                                   
        $2.75^\circ\!\times\!{}0.5^\circ\,$L20 &     
        341 &                                       
        1.51 &                                      
        3418 &                                      
        \legendb{red} \\

        FGOALS-g1.0\footnotemark[7] &
        T42 L26 &                                   
        $1^\circ\!\times\!{}1^\circ\,$L33 &          
        350 &                                       
        1.98 &                                      
        6281 &                                      
        \legendb{green} \\

        GFDL-CM2.0\footnotemark[8] &
        $2.5^\circ\!\times\!{}2^\circ\,$L24 &        
        $1^\circ\!\times\!{}1/3^\circ\,$L50 &        
        500 &                                       
        1.37 &                                      
        10073 &                                     
        \legendb{blue} \\

        GFDL-CM2.1\footnotemark[8] &
        $2.5^\circ\!\times\!{}2^\circ\,$L24 &        
        $1^\circ\!\times\!{}1/3^\circ\,$L50 &        
        500 &                                       
        1.52 &                                      
        10073 &                                     
        \legendb{magenta} \\

        GISS-EH\footnotemark[9] &
        $5^\circ\!\times\!{}4^\circ\,$L20 &          
        $2^\circ\!\times\!{}2^\circ\,$L16 &          
        400 &                                       
        1.03 &                                      
        6172 &                                      
        \legendb{orange} \\

        INM-CM3.0\footnotemark[10] &
        $5^\circ\!\times\!{}4^\circ\,$L21 &          
        $2^\circ\!\times\!{}2.5^\circ\,$L33 &        
        330 &                                       
        1.29 &                                      
        1276 &                                      
        \legendb{cyan} \\

        IPSL-CM4\footnotemark[11] &
        $2.5^\circ \!\times\!3.75^\circ\,$L19 &      
        $2^\circ\!\times\!{}1^\circ\,$L31 &          
        500 &                                       
        1.19 &                                      
        3078 &                                      
        \legendc{red} \\

        MIROC3.2(hires)\footnotemark[12] &
        T106 L56 &                                  
        $0.28^\circ\!\times\!{}0.187^\circ\,$L47 &   
        100 &                                       
        1.20 &                                      
        9944 &                                      
        \legendc{green} \\

        MIROC3.2(medres)\footnotemark[12] &
        T42 L20 &                                   
        $1.4^\circ\!\times\!{}0.5^\circ\,$L43 &      
        500 &                                       
        1.14 &                                      
        6527 &                                      
        \legendc{blue} \\

        MRI-CGCM2.3.2\footnotemark[13] &
        T42 L30 &                                   
        $2.5^\circ\!\times\!{}0.5^\circ\,$L23 &      
        350 &                                       
        1.06 &                                      
        2583 &                                      
        \legendc{magenta} \\

        UKMO-HadCM3\footnotemark[14] &
        $3.75^\circ\!\times\!{}2.5^\circ\,$L19 &     
        $1.25^\circ\,$L20 &                          
        341 &                                       
        1.13 &                                      
        3926 &                                      
        \legendc{orange} \\

        UKMO-HadGEM1\footnotemark[15] &
        $1.875^\circ\!\times\!{}1.25^\circ\,$L38 &   
        $1^\circ\!\times\!{}1/3^\circ\,$L40 &        
        240 &                                       
        0.97 &                                      
        11337 &                                     
        \legendc{cyan} \\

        \bottomrule
      \end{tabular*}}

    \vspace{0.25cm}
    \begin{tabular}{lll}
      \footnotemark[1] \citep{furevik-bcm} &
      \footnotemark[6] \citep{min-echo-g} &
      \footnotemark[11] \citep{marti-ipsl-cm4} \\

      \footnotemark[2] \citep{collins-ccsm3} &
      \footnotemark[7] \citep{yu-lasg-gcm} &
      \footnotemark[12] \citep{k-1-miroc} \\

      \footnotemark[3] \citep{kim-cgcm3-1} &
      \footnotemark[8] \citep{delworth-gfdl-cm2} &
      \footnotemark[13] \citep{yukimoto-mri-cgcm} \\

      \footnotemark[4] \citep{salas-melia-cnrm-cm3} &
      \footnotemark[9] \citep{schmidt-giss-modele} &
      \footnotemark[14] \citep{gordon-ukmo-hadcm3} \\

      \footnotemark[5] \citep{gordon-csiro-mk3} &
      \footnotemark[10] \citep{volodin-inmcm3} &
      \footnotemark[15] \citep{johns-hadgem1}
    \end{tabular}
    \end{center}
\end{table}
%

\section{Geometrical test data sets}
\label{sec:test-data-sets}

\begin{figure}
  \begin{center}
    \begin{tabular}{cc}
      \includegraphics[width=0.48\textwidth]{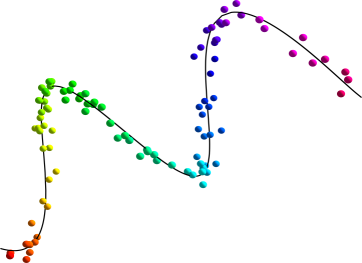} &
      \includegraphics[width=0.48\textwidth]{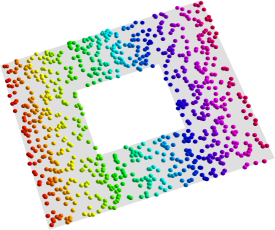} \\
      (a). Helix & (b). Plane with hole \\
       & \\
      \includegraphics[width=0.48\textwidth]{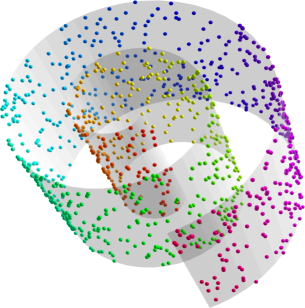} &
      \includegraphics[width=0.48\textwidth]{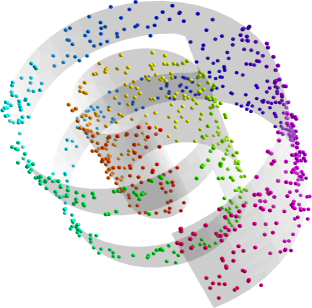} \\
      (c). Swiss roll & (d). Swiss roll with hole
    \end{tabular}
  \end{center}
  \caption[Three-dimensional views of geometrical test data
    sets]{Examples of simple geometrical data sets used for testing
    dimensionality reduction methods.  Data points are shown as
    coloured spheres, with the hue of the points varying linearly
    along one of the intrinsic directions in the data manifolds.  The
    data manifolds themselves are indicated in these views in grey.
    Of the four data sets shown here, the helix (a) and the Swiss roll
    with hole (d) are shown with added noise, while the plane with
    hole (b) and Swiss roll (c) are shown without added noise.}
  \label{fig:test-data-3d}
\end{figure}

In order to characterise the basic behaviour of the dimensionality
reduction methods considered in
Chapters~\ref{ch:nlpca}--\ref{ch:hessian-lle}, simple geometrical data
sets are used.  These test data sets consist of subsets of simple one-
and two-dimensional manifolds embedded in three-dimensional Euclidean
space (Figure~\ref{fig:test-data-3d}).  All examples are used both
with and without small random perturbations to test the noise
sensitivity of the dimensionality reduction methods.  In all figures
showing these data sets, both the three-dimensional views of
Figure~\ref{fig:test-data-3d} and subsequent representations of
results from nonlinear dimensionality reduction methods, points in the
data set are labelled by hue to provide landmarks related to the
intrinsic geometry of the data manifolds.  This provides immediate
visual feedback about the quality of dimension reduction results
(e.g. Figure~\ref{fig:nlpca-test-data-sets-reduce} on
page~\pageref{fig:nlpca-test-data-sets-reduce}).

%
%
%

The test examples are generated by uniform random sampling of points
(100 for the helix, 1000 for the two-dimensional examples) from the
following sets.

\paragraph{A segment of a helix}

$\vec{r} = t \vec{h}_1 + r [ \cos (2\pi t / p) \vec{h}_2 + \sin (2\pi
  t / p) \vec{h}_3 ]$, where $t \in [ 0, 1 ]$ is a variable
parameterising points along the helix, $r$ is the radius of the helix
and $p$ its pitch.  The orthonormal vectors ${ \vec{h}_1, \vec{h}_2,
  \vec{h}_3 }$ give the directions of the axis of the helix and a
basis in a plane normal to the direction of the axis respectively.
Here, $\vec{h}_1$ is in the direction $(1, 1, \frac{1}{2})$ and
$\vec{h}_2$ and $\vec{h}_3$ are found by orthonormalising the set $\{
  (1, 1, \frac{1}{2}), (1, 0, 0), (0, 1, 0) \}$
(Figure~\ref{fig:test-data-3d}a).

\paragraph{A rectangular segment of a plane}

$\vec{r} = u \vec{e}_1 + v \vec{e}_2$, where $u \in [0, 1]$ and $v \in
[0, 1]$ parameterise points in the rectangle and the vectors
$\vec{e}_1$ and $\vec{e}_2$ are a basis in the plane.  Here,
$\vec{e}_1$ and $\vec{e}_2$ are found by orthonormalising the set $\{
  (-1, 1, \frac{1}{2}), (\frac{1}{2}, 1, \frac{1}{2}), (0, 0, 1) \}$
to give $\{ \vec{e}_1, \vec{e}_2, \vec{e}_3 \}$.

\paragraph{A rectangular segment of the plane with a hole}

This is formulated in the same way as the plane, except that points
for which $0.3 < u < 0.7$ and $0.3 < v < 0.7$ are excluded
(Figure~\ref{fig:test-data-3d}b).

\paragraph{A ``Swiss roll''}

This is the product of a segment of an Archimedean spiral (in polar
coordinates, $r = s \theta$, for some constant $s$) and a line
segment.  It is parameterised in a similar way to the plane, as $(u,
v)$ with $u \in [0, 1]$ and $v \in [0, 1]$, where $u$ parameterises
distance along the line segment and $v$ arclength along the spiral.
An arclength parameterisation along the spiral facilitates uniform
distribution of random points across the surface.  Writing $(x, y) =
(s \theta \cos \theta, s \theta \sin \theta)$ for the Cartesian
coordinates of points along the spiral, the arclength from the centre
of the spiral through an angle $\phi$ is
\begin{equation}
  a = l(\phi) = \int_0^\phi \sqrt{\left( \frac{\partial x}{\partial
      \theta} \right)^2 + \left( \frac{\partial y}{\partial \theta}
    \right)^2} \, d\theta = s \int_0^\phi \sqrt{1 + \theta^2} \,
  d\theta = \frac{s}{2} \left( \phi \sqrt{1 + \phi^2} + \sinh^{-1}
  \phi \right).
\end{equation}
The function $l(\phi)$ is monotonic, and can be inverted to give the
angle $\phi$ as a function of arclength, $\phi = l^{-1}(a)$.  I
perform this inversion numerically, as there does not appear to be a
closed form for the inverse.  Using this, it is then straightforward
to distribute points uniformly in arclength between two angles
$\phi_0$ and $\phi_1$, by calculating $a_0 = l(\phi_0)$, $a_1 =
l(\phi_1)$, $\Delta a = a_1 - a_0$, and then calculating angular
coordinates as $\theta(v) = l^{-1}(a_0 + v \Delta a)$ for (uniformly
distributed) arclength coordinates $v \in [0, 1]$.  Using this
approach, points in the Swiss roll data set used here are given by
$\vec{r} = u \vec{e}_3 + s \theta(v) (\cos \theta(v) \vec{e}_1 + \sin
\theta(v) \vec{e}_2)$, where the orthonormal basis $\{ \vec{e}_1,
  \vec{e}_2, \vec{e}_3 \}$ is the same as that used for generating the
plane data set (Figure~\ref{fig:test-data-3d}c).

\paragraph{A ``Swiss roll'' with a hole}

This is formulated in the same way as the Swiss roll, except that
points for which $0.3 < u < 0.7$ and $0.3 < v < 0.7$ are excluded
(Figure~\ref{fig:test-data-3d}d).

\paragraph{A ``fishbowl''}

This is a sphere with a cap removed: $\{ (x, y, z) \, | \, x^2 + y^2 +
  z^2 = R^2, z \leq z_{\mathrm{max}} \}$, with $R=0.5$ and
$z_{\mathrm{max}} = 0.4$.  Points distributed uniformly on this
surface are generated by sampling from a symmetrical 3-D Gaussian
distribution, projecting the sampled points down to the surface of the
sphere, and clipping points that have $z > z_{\mathrm{max}}$.  (This
gives a uniform distribution of points on the sphere because of the
spherical symmetry of the 3-D Gaussian distribution.)


%% file: 04-enso.tex
\chapter{The \eln/Southern Oscillation}
\label{ch:enso}

The \eln/Southern Oscillation (ENSO) is the most important mode of
interannual variability in the Earth's climate, driven by
atmosphere-ocean interactions in the equatorial Pacific, but with
effects reaching as far as north-eastern North America and Europe
\citep{philander-book,mcphaden-enso,liu-teleconnections}.  ENSO events
(\eln and \lan) occur irregularly at intervals of 2--7 years, and
individual events are variable in their evolution and effects.  The
clearest manifestation of ENSO variability occurs in the eastern
tropical Pacific, where South American fishermen have long noticed an
irregular warming of coastal waters, with a consequent impact on
upwelling of nutrients and fish stocks, occurring in boreal winter
around Christmas time, leading to the name \eln, ``Christ child''.
These variations in sea surface temperature observed near the American
coast are part of a coherent pattern of changes in SST, sea surface
height, thermocline depth and surface winds across the equatorial
Pacific, related to large-scale coupled dynamics of the ocean and
atmosphere in the tropical Pacific.  The variations in sea surface
height and winds are associated with changes in the large-scale sea
level pressure distribution across the Pacific basin and beyond.
Indeed, the ``Southern Oscillation'' part of the designation ENSO
refers to a dipolar variation in pressure between Tahiti in the
central Pacific and Darwin in the western Pacific, a variation first
noted by \citet{walker-1924} and others in the context of studies of
the predictability of the Indian monsoon.

In this chapter, I describe the climatology of the tropical Pacific
and the observed phenomenology and underlying mechanisms of ENSO.
This background is helpful for the interpretation of the nonlinear
dimensionality reduction results presented in
Chapters~\ref{ch:nlpca}--\ref{ch:hessian-lle}.  I also describe the
different approaches taken to the modelling of ENSO, to put the
results from the CMIP3 GCM ensemble into context.

\section{Tropical Pacific climatology}

\begin{figure}
  \begin{center}
    \includegraphics[width=\textwidth]{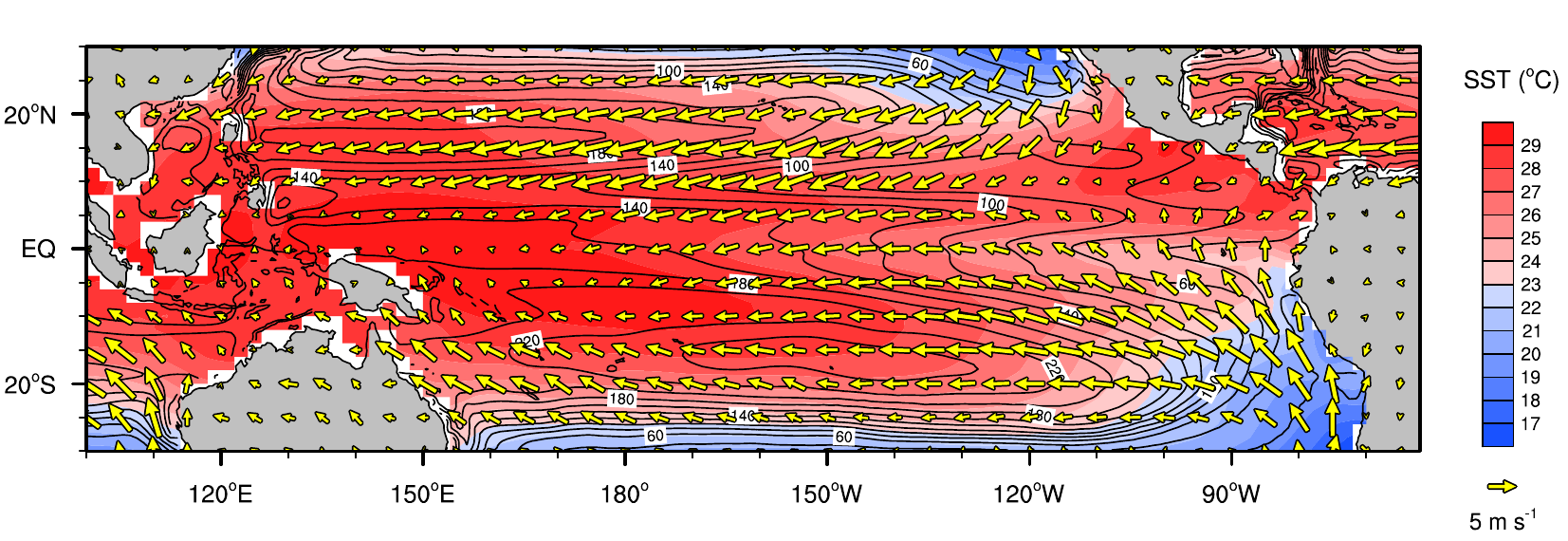}
  \end{center}
  \caption[Mean climatological SST, thermocline and wind state of
    equatorial Pacific]{Mean climatological sea surface temperature
    (colours, from ERSST v2), thermocline depth (contours, in metres,
    from NCEP GODAS) and surface wind field (arrows, from the NCEP
    reanalysis).  All data is averaged over the period January 1980 to
    December 2006, the longest period of overlap between the different
    data sets.}
  \label{fig:eq-pacific-clim}
\end{figure}

The climatological mean state in the equatorial Pacific
(Figure~\ref{fig:eq-pacific-clim}) has high sea surface temperatures
in the western Pacific, often called the Western Warm Pool, and lower
temperatures in the eastern Pacific, particularly in a tongue of cold
upwelling water lying along the equator and stretching south along the
western coast of South America.  The trade winds blow easterly along
the equator over the eastern and central Pacific, and this wind stress
is balanced by a zonal thermocline and sea level gradient, with a deep
thermocline in the west and a shallower thermocline, in some cases
shoaling to the surface, in the eastern part of the basin.

%
%
%

The strong asymmetry of this mean state arises primarily from
constraints on the atmospheric circulation imposed by conservation of
angular momentum.  Air upwelled by thermally direct convection over
the Intertropical Convergence Zone in the rising branch of the Hadley
cell near the equator moves away from the equator at height, falls
back to the surface in the sub-tropics and causes a surface level
return flow that leads to trade winds with a strong easterly
component.  The nature of this north-easterly (in the northern
hemisphere) and south-easterly (in the southern hemisphere) trade wind
return flow was essentially explained by \citet{hadley-trade-winds}.
\citeauthor{hadley-trade-winds}'s reasoning was based on the simple
observation that, although a point on the surface of the earth at the
equator is moving at a speed of around $\mathrm{1668\,km\,hr^{-1}}$
($\mathrm{463\,m\,s^{-1}}$) with respect to the centre of the earth,
the winds that we observe never attain such great speeds.  The bulk of
the atmosphere must, therefore, be rotating along with the solid
earth.  Consider then, a mass of air at 20\degree{}N, in the
downwelling branch of the Hadley circulation.  Due to the
comparatively shorter line of latitude at this point compared to the
equator, this mass of air is moving at a lesser speed with respect to
the centre of the earth, namely $\mathrm{1567\,km\,hr^{-1}}$
($\mathrm{435\,m\,s^{-1}}$).  As this mass of air moves southwards
towards the equator, it thus has an excess velocity of
$\mathrm{101\,km\,hr^{-1}}$ ($\mathrm{27.9\,m\,s^{-1}}$) to the west
relative to a point on the equator.  This difference in velocities is
the source of the easterly component of the trade winds, and leads to
an easterly component both north and south of the equator.  (This
argument is essentially a paraphrase of the principle of conservation
of angular momentum for our air mass.)  Of course, observed zonal
winds are slower than this total velocity difference, because the
difference is dissipated gradually through friction as the air mass
moves south (in the NCEP reanalysis data set
\citep{kalnay-ncep-reanalysis}, within the central Pacific between
30\degree{}S and 30\degree{}N, the region of sub-tropical downwelling,
the absolute maximum wind speed is $\mathrm{13.2\,m\,s^{-1}}$, while
at the equator, the maximum wind speed is $\mathrm{10.8\,m\,s^{-1}}$).

%
%
\begin{figure}
  \begin{center}
    \includegraphics[width=\textwidth]{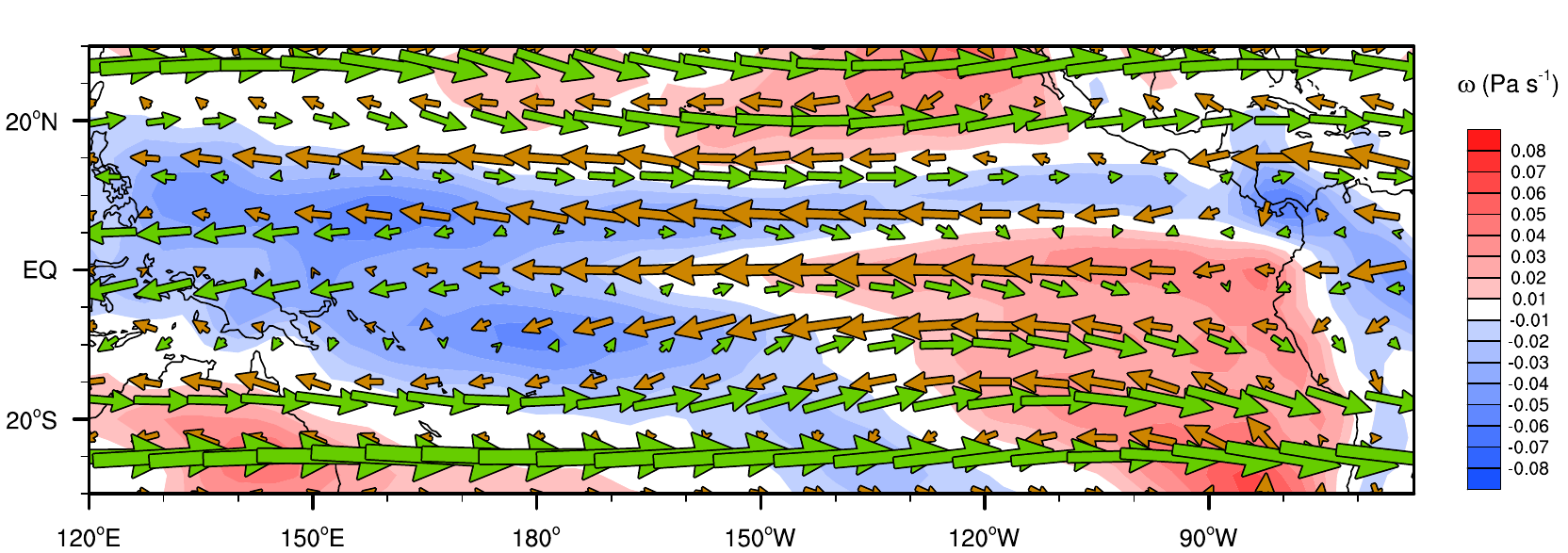}
  \end{center}
  \caption[Walker circulation]{Walker circulation in the Pacific:
    climatological mean 500\,hPa pressure velocity $\omega$ (colours,
    with negative [blue] pressure velocity denoting rising air, and
    positive [red] values descending air), low level (850\,hPa) winds
    (brown arrows) and upper level (200\,hPa) winds (green arrows),
    all from the NCEP reanalysis.}
  \label{fig:walker-circulation}
\end{figure}
%

Another important feature of the climatological mean state in the
equatorial Pacific is the so-called Walker circulation
(Figure~\ref{fig:walker-circulation}).  This is a zonal circulation
along the equatorial band of the atmosphere, representing a
perturbation to the zonally symmetric Hadley cell.  In the Pacific,
the Walker circulation has its rising branch in the west over the warm
waters of the Western Warm Pool, and its descending branch in the
eastern Pacific.  The low level winds are easterly (as already noted
above), and the upper level winds are westerly, closing the
circulation cell.  The Walker circulation is driven by the zonal
contrast in SST across the Pacific, and provides a connection between
the tropical Indian and Pacific Oceans, with the climatological mean
state in the Indian Ocean having westerly low level winds feeding into
the rising branch of the Walker circulation in the west Pacific.  This
thermally driven circulation is susceptible to SST perturbations in
the Pacific and shows a strong connection to ENSO variability.  During
\eln, when warm waters cover most of the central and eastern
equatorial Pacific, the centre of convective activity in the
equatorial Pacific shifts eastwards, following the warmer waters, and
the rising branch of the Walker circulation thus moves east.  This has
effects on atmospheric circulation over both the Indian and Atlantic
Oceans, and, in particular, provides a mechanism for ENSO variability
to have an influence on the Indian monsoon system.

\section{ENSO phenomenology}
\label{sec:enso-phenomenology}

\begin{figure}
  \begin{center}
    \includegraphics[width=\textwidth]{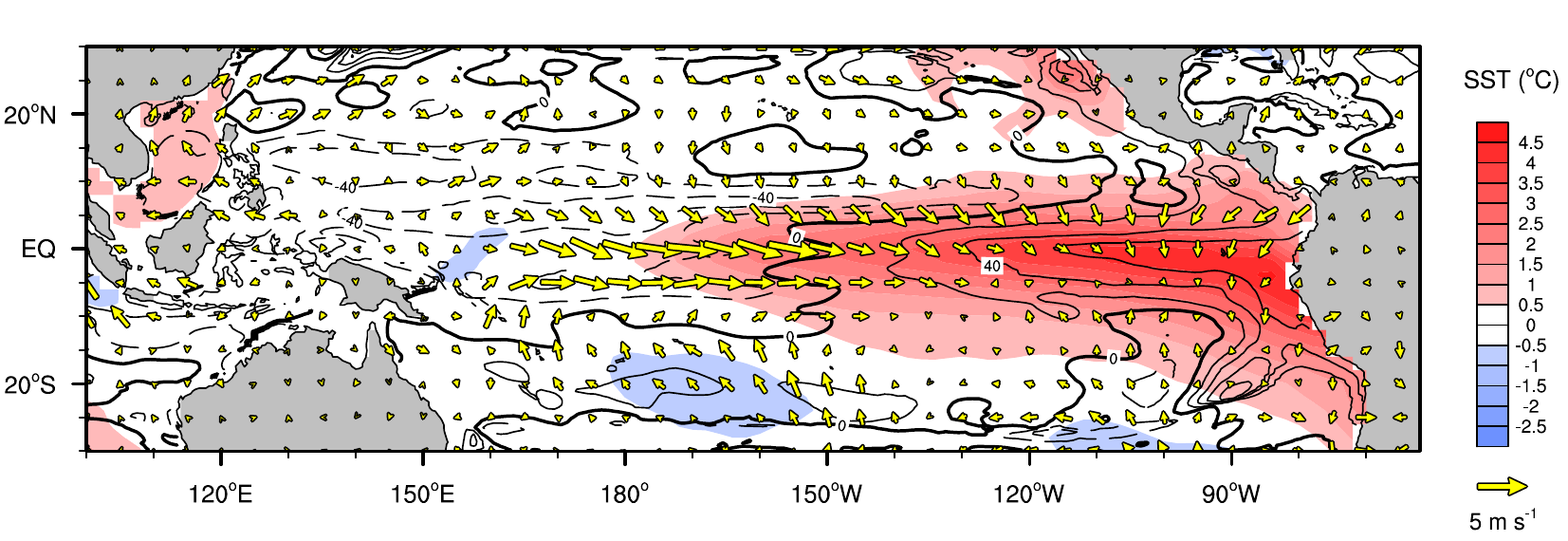}
  \end{center}
  \caption[\eln conditions in equatorial Pacific]{Sea surface
    temperature anomalies\footnote{Throughout this thesis, use of the
      word ``anomaly'' refers to anomalies with respect to a mean
      climatological annual cycle, calculated independently at each
      spatial grid point for a data set.}  (colours, from ERSST v2),
    thermocline depth anomalies (contours, in metres, from NCEP GODAS)
    and surface wind anomalies (arrows, from the NCEP reanalysis) for
    a fully developed \eln event.  The data shown is a composite for
    the period September 1997--January 1998, with anomalies taken with
    respect to the annual cycle over the period used for the
    climatology in Figure~\ref{fig:eq-pacific-clim}.}
  \label{fig:eq-pacific-el-nino}
\end{figure}

ENSO variability in the equatorial and tropical Pacific is associated
with large fluctuations in SST, surface winds and thermocline depth
and structure.  The fully developed \eln state
(Figure~\ref{fig:eq-pacific-el-nino}) has anomalously warm waters in
the eastern Pacific, replacing the climatological cold SST tongue, and
anomalous westerly winds along the equator to the west of the SST
anomaly (in some cases, in the western and central Pacific, these
anomalous westerlies lead to overall westerly winds rather than the
normal easterlies).  The modified wind stress leads to a change in the
zonal thermocline gradient, with a deepened thermocline in the east
and shoaling in the western Pacific.

%
%
%

\begin{figure}
  \begin{center}
    \includegraphics[width=\textwidth]{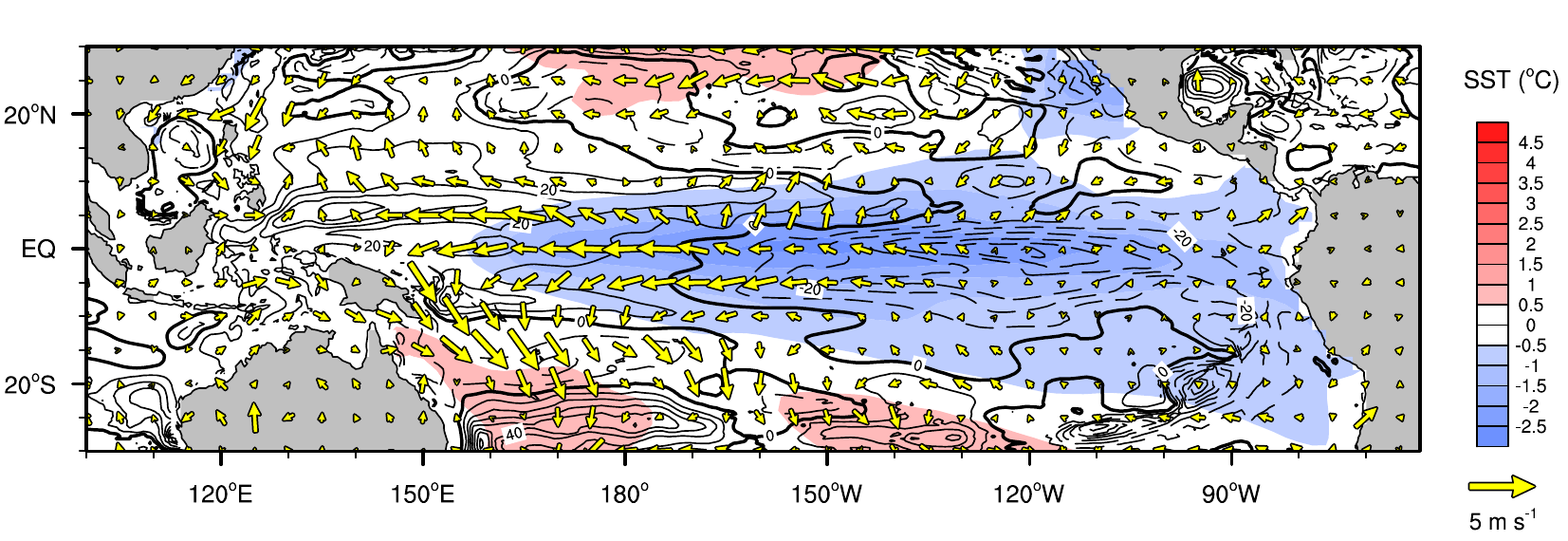}
  \end{center}
  \caption[\lan conditions in equatorial Pacific]{Sea surface
    temperature anomalies (colours, from ERSST v2), thermocline depth
    anomalies (contours, in metres, from NCEP GODAS) and surface wind
    anomalies (arrows, from the NCEP reanalysis) for a fully developed
    \lan event.  The data shown is a composite for the period October
    1988--January 1989, with anomalies taken with respect to the annual
    cycle over the period used for the climatology in
    Figure~\ref{fig:eq-pacific-clim}.}
  \label{fig:eq-pacific-la-nina}
\end{figure}

In some respects, the situation during a \lan event is the opposite of
that during an \eln (Figure~\ref{fig:eq-pacific-la-nina}).  There is a
negative SST anomaly in the eastern Pacific, with easterly zonal wind
anomalies to the west of the SST anomaly and an associated enhancement
of the zonal thermocline gradient, with a deeper thermocline in the
west and shoaling in the east (during strong \lan events, the
thermocline may shoal to the surface in the eastern Pacific).
However, more careful examination reveals that there is a notable
asymmetry between \eln and \lan.  First, the size of the SST anomalies
associated with \eln is larger.  For the 1900--1999 ERSST v2 data, the
NINO3 SST index (mean anomalous SST in the region
150\degree{}W--90\degree{}W, 5\degree{}S--5\degree{}N, which is
positive during \eln and negative during \lan) has quite different
ranges of variability for positive and negative excursions: the
standard deviation of positive values is 0.64\degree{}C, and that of
negative values 0.46\degree{}C.  The second major difference between
\eln and \lan is that the centre of the positive SST anomaly during an
\eln is rather further to the east than the centre of the negative SST
anomaly during a \lan.  This spatial asymmetry is one of the main
reasons for expecting that a nonlinear approach to dimensionality
reduction for ENSO might be of some worth.  In conventional PCA,
individual modes are able to represent only symmetric standing
oscillations, so cannot capture this asymmetry.  Although a
combination of PCA modes can represent any form of variability, the
expectation is that a nonlinear dimensionality reduction method may be
able to represent more complex patterns of variability in a single
mode.

%
%
%

There are other asymmetries between \eln and \lan, including important
differences in the off-equatorial wind field in the western Pacific
associated with equatorial wave dynamics involved in recharging the
Western Warm Pool, but these are rather variable between different
events and are thus difficult to pick out from composites such as
Figures~\ref{fig:eq-pacific-el-nino}~and~\ref{fig:eq-pacific-la-nina}.
The asymmetry between \eln and \lan events is not completely
understood, but the asymmetry of the underlying climatological mean
state is almost certainly a contributing factor.  The observed
asymmetry in ENSO variability has previously been investigated using a
variety of measures based on SST variance and skewness, nonlinear
dynamical heating and the explicit characterisation of symmetric and
asymmetric structures in SSTs during ENSO events
\citep{an-enso-asym,an-enso-nonlin,an-enso-cmip,monahan-dai}.

The picture presented here is a simplification of reality.  In fact,
there is strong variability between the temporal evolution of
different \eln events.  For different events, different interactions
between the Pacific and other ocean basins and between the equatorial
Pacific and the mid-latitudes are important.  \citet{mcphaden-97enso}
and \citet{mcphaden-02enso} give detailed descriptions of the
evolution of two rather different \eln events, in 1997-98 and 2002-03.

\section{ENSO mechanisms}
\label{sec:enso-mechanisms}

The basic mechanisms of the evolution of \eln and \lan events are
relatively well understood, as a result of comprehensive observing,
modelling and theoretical programmes over the last 20 years
\citep{mcphaden-02enso,latif-ensip,neelin-enso-theory,dijkstra-book},
although interactions between ENSO and the annual cycle, and the
source of the irregularity of ENSO variability are less well
understood.  There is an interesting contrast between the situation in
the tropics, where atmosphere-ocean interactions are relatively
straightforward to model and understand, and the situation in the
mid-latitudes, where things are much more difficult.  The primary
reason for this, as noted in \citet{neelin-enso-theory}, is that,
because the tropical ocean spatially integrates wind stress forcing
over quite extensive regions, the coupled response of the
atmosphere-ocean system is quite forgiving to variations in the
atmospheric response to SST anomalies.  Little more is required of the
atmospheric component of a model than that it reproduce the westerly
wind anomalies that are observed to the west of a positive SST
anomaly.  Even very simple analytically solvable models are able to
meet this condition \citep{gill-atmos-response}.  In the
mid-latitudes, the atmospheric response to ocean surface heat
anomalies is more complicated, operates on a faster timescale, and the
coupled response of the atmosphere-ocean system is much more sensitive
to deficiencies in the modelling of atmospheric processes and
ocean-atmosphere exchanges of heat and moisture.

The oscillation that is ENSO requires two main physical components, a
positive feedback to destabilise perturbations to the climatological
state, and a mechanism causing the perturbation to saturate and to
carry the oscillation over to its next phase.  The mechanism providing
a positive feedback to perturbations was first described by
\citet{bjerknes-feedback}.  Suppose that there exists a localised
positive anomalous SST perturbation in the eastern equatorial Pacific.
This produces anomalous westerly winds to the west of the SST
perturbation, acting to weaken the climatological easterly winds
there.  Changes in the wind field along the equator excite wave modes
in the equatorial subsurface ocean (equatorially trapped Kelvin waves
propagating eastwards and off-equatorial Rossby waves propagating
westwards) and propagation of these waves leads to a relatively quick
adjustment of the zonal thermocline gradient across the Pacific basin,
with a deepening of the thermocline in the eastern Pacific and a
concordant shoaling in the west.  (In this context, ``relatively
quick'' means within the space of a few months: an equatorially
trapped Kelvin wave propagates at a speed of around
$\mathrm{2\,m\,s^{-1}}$, so takes approximately 3 months to cross the
Pacific from 140\degree{}E to 80\degree{}W).  The deepening
thermocline in the eastern Pacific inhibits upwelling of cooler
subsurface waters there, so reinforcing the initial positive SST
perturbation.  The same mechanism works to reinforce negative SST
perturbations, and so provides a positive feedback amplifying any
anomalous SST perturbation in the eastern Pacific.

As noted in Bjerknes's original 1969 paper, this feedback mechanism is
enough to allow a transition between an \eln state and a \lan state or
vice versa, once things get started.  A second component to the ENSO
oscillation is then required to allow the perturbations amplified by
the Bjerknes feedback to dissipate, permitting the oscillation to turn
over into its next phase.  This mechanism is connected to changes in
the zonal mean thermocline structure in the Pacific, which can be
measured in terms either of ocean heat content (usually defined in
terms of an integral or average of ocean temperature in the top 300\,m
of the ocean) or thermocline depth (which I use here).  This mechanism
was originally proposed by \citet{wyrtki-1985} following earlier work
examining sea level changes across the Pacific during \eln events
\citep{wyrtki-1975}.  These variations in zonal mean thermocline depth
connect changes in zonal wind stress over the central and eastern
Pacific to changes in Ekman flow of surface waters between the
equatorial and off-equatorial Pacific.

\begin{figure}
  \begin{center}
    \includegraphics[width=\textwidth]{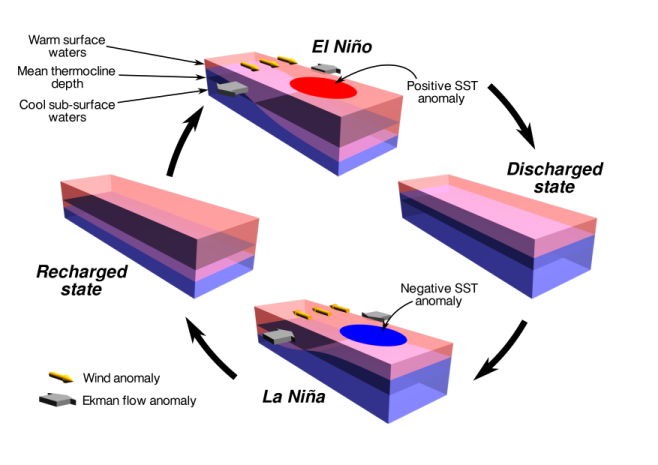}
  \end{center}
  \caption[Cartoon view of ENSO cycle]{Schematic view of cycling
    between \eln, discharged, \lan and recharged states in the
    Bjerknes-Wyrtki picture of ENSO variability.  Each image shows a
    cartoon of the equatorial region of the Pacific, with thermocline
    anomalies shown as warm red water overlying cooler blue water, and
    the mean thermocline depth (i.e. the zero anomaly surface) picked
    out for comparison.  SST anomalies in the eastern Pacific are
    indicated by coloured patches on the surface, while wind anomalies
    and anomalies in meridional transport of warm surface waters to
    the west of the SST anomalies are indicated by arrows.  (Based on
    Figure~1 of \citep{jin-1997a}.)}
  \label{fig:enso-cartoons}
\end{figure}

The basic idea is that the timescale of ENSO variability is set by the
amount of time it takes for a pool of warm water to accumulate in the
western Pacific, which then moves across the Pacific basin to give
\eln conditions.  (See Figure~\ref{fig:enso-cartoons} for a cartoon
representation of the combined effect of these zonal mean thermocline
variations and the Bjerknes feedback.)  The accumulation of warm water
in the equatorial region serves to make positive SST anomalies in the
eastern Pacific more likely, and to precondition the ocean-atmosphere
system for a transition to \eln conditions.  During an \eln event,
warm equatorial waters are transported away from the equator by Ekman
transport in the surface layer, decreasing the mean thermocline depth,
i.e. reducing the overall reservoir of warm water in the equatorial
Pacific.  This reduced mean thermocline depth makes the eastern
Pacific susceptible to the production of negative SST anomalies,
leading, via the Bjerknes feedback, to \lan conditions.  During this
time, the Western Warm Pool recharges with warm water, transported
there by westward-moving off-equatorial Rossby waves and the
equatorial current systems, leading to a greater zonal mean
thermocline depth and a larger reservoir of equatorial warm water,
ready to start the next phase of the oscillation.

%
%
%

This cycle of SST, thermocline depth, wind and meridional transport
anomalies, affecting the enormous area of the Pacific basin, has
knock-on effects on climate across the whole globe.  The existence of
teleconnections related to variability in the tropical Pacific has
been known at least since the work of \citet{walker-1924}.  Some of
the earliest references describing ENSO teleconnections from a global
point of view include \citep{ropelewski-enso-precip} and
\citep{halpert-enso-temp}.  The review of \citet{liu-teleconnections}
provides a recent overview of teleconnection processes throughout the
global atmosphere and ocean, with a particular emphasis on
ENSO-related effects.  A few of the more obvious processes that occur
deserve some comment here.  First, the amount of warm water in the
equatorial surface ocean controls the strength of the upwelling branch
of the atmospheric Hadley cell, and this in turn controls the angular
momentum flux in the atmosphere exported from the tropics into the
mid-latitudes.  This affects the location and strength of the Pacific
storm tracks and allows ENSO variability to have impacts on the
climate across the North American continent
\citep{bjerknes-feedback,philander-book,cole-us-tele,hu-us-tele}.  The
Walker circulation is another aspect of the atmospheric circulation
strongly influenced by ENSO variability.  Under climatological
conditions, the rising branch of the Pacific part of the Walker
circulation lies over the Western Warm Pool in the far west of the
Pacific basin.  The extensive warming of equatorial surface waters
across the Pacific during \eln events causes the rising branch of the
Walker circulation to be displaced eastwards, leading to adjustments
in the pattern of the Walker circulation in other ocean basins.  This
mechanism provides a linkage between ENSO variability and the Asian
monsoon system and aspects of the climate further west over the Indian
Ocean, and also potentially influences the course of the African
monsoon
\citep{soman-enso-monsoon,lau-enso-monsoon,meehl-enso-monsoon}.  There
is also some evidence that the linkage between the Pacific and Indian
Oceans via the Walker circulation can be responsible for triggering
weaker ENSO events confined to the western and central Pacific, due to
the effects of an internal mode of variability in the Indian Ocean
\citep{clarke-indian-ocean}.

The relatively simple picture of the Bjerknes and Wyrtki feedbacks
causing oscillations of ocean and atmospheric conditions across the
Pacific captures the basic mechanisms leading to vacillation between
warm \eln and cold \lan states in the equatorial Pacific, but there
are several aspects of observed ENSO variability not explained by this
mechanism.  The irregularity of ENSO variability, the partial phase
locking of \eln and \lan events to the boreal winter
\citep{rasmusson-enso,galanti-phase}, the potential role of westerly
wind bursts in the western Pacific and possible connections to the
Madden-Julian Oscillation \citep{mcphaden-97enso,eisenman-enso-wwb},
and aspects of the interdecadal variability of ENSO, particularly the
apparent shift in behaviour observed in the mid-1970s
\citep{fedorov-change}, are all still very much open questions.  Much
modelling and theoretical effort has gone into addressing these
questions, and substantial progress has been made.

For example, as regards the interaction between ENSO and the annual
cycle in the Pacific, work with conceptual ENSO models with seasonally
varying forcing seems to indicate the presence of a ``Devil's
staircase'' pattern of frequency and phase locking characteristic of
the quasi-periodic route to chaos observed in driven nonlinear
oscillators \citep{tziperman-enso-chaos}.  This phenomenon has also
been observed in intermediate complexity models of ENSO variability
\citep{tziperman-quasi-per,jin-enso,jin-subharmonic}. If this
behaviour also occurs in the real ocean-atmosphere system, it may
offer a clear explanation for the irregularity of ENSO.

This work also touches on a deeper question concerning the exact
source of ENSO variability, a question that has led to a long-running
disagreement in the literature on ENSO theory and modelling.  One
point of view is that ENSO arises from unstable modes of variability
in the tropical ocean-atmosphere system, with limits to predictability
determined by growth in errors in initial conditions associated with
chaotic dynamics
\citep[e.g.,][]{zebiak-cane,jin-enso,tziperman-enso-chaos,chen-enso-predict}.
The other possibility is that ENSO is a damped linear oscillation
excited by stochastic forcing, the limits to predictability being
inherent in the stochastic nature of the forcing
\citep[e.g.,][]{burgers-stoch-enso,moore-stoch-enso,thompson-stoch-enso}.
As of the time of writing, this question is far from being resolved,
despite the extensive work that has been done to develop conceptual
and intermediate complexity models of ENSO.  The likelihood is that
there is at least some truth to both viewpoints, with ENSO being a
manifestation of a weakly damped mode, whose evolution (\eln) can be
initiated by external noise forcing and whose exact timing is
determined both by deterministic processes (the recharge of warm water
in the western Pacific, for example) and by interactions with the
annual cycle in the Pacific \citep{philander-sporadic-enso}.

It is clear from the account here that ENSO is a complex phenomenon,
involving, in an essential way, both oceanic and atmospheric
variability.  The description of the physical processes involved in
ENSO is necessarily rather abbreviated here.  Some further aspects of
the phenomena involved, in the context of ENSO models, are described
in the following section.  Given the complex and coupled nature of
ENSO behaviour in the Pacific, it is unsurprising that the different
mechanisms described above receive different emphasis in different
studies of overall ENSO behaviour.  This is particularly true in the
case of the development of conceptual models of ENSO, where some
models focus on equatorial wave dynamics, some on ocean heat content
variations and some on nonlinear advection in equatorial regions
(Section~\ref{sec:conceptual-models}).

\section{Modelling ENSO}
\label{sec:modelling-enso}

There are at least four different types of model commonly used to
represent ENSO variability: global general circulation models (GCMs),
statistical models, intermediate complexity models (ICMs) and
conceptual models.  Chapter 9 of \citep{dijkstra-book} has a
comprehensive discussion of ENSO variability in these different types
of model and includes derivations of many of the most important model
equations.  The different approaches to ENSO modelling are employed
for rather different purposes, the clearest distinction being between
models constructed expressly for the purposes of prediction of ENSO
variability (statistical models and ICMs) and models intended to
improve our understanding of ENSO processes (conceptual models and
some simpler ICMs).  The subject of the simulation of ENSO by general
circulation models, which we will examine first, is a slightly
different case from the specially constructed ENSO models.

\subsection{General circulation models}
\label{sec:modelling-enso-gcms}

General circulation models are a special case here, since they are not
constructed for the express purpose of simulating ENSO variability.
ENSO in GCMs arises as an emergent consequence of physically based
modelling of fundamental processes in the atmosphere and ocean, and
the interaction between these processes in and around the Pacific.
Coupled ocean-atmosphere GCMs consist of dynamical cores for the
atmosphere and ocean, along with ``physics'' parameterisations
\citep{henderson-sellers-primer,mote-numerical-modelling}.  The
dynamical cores are schemes to solve the evolution equations for
velocity, pressure, temperature, salinity (in the ocean), humidity (in
the atmosphere) and so on for both domains, using appropriate
numerical methods.  The physics parameterisations in each domain are
physically based or empirical representations of processes that occur
on spatial resolutions unresolved by the dynamical core.  For example,
in the atmosphere, parameterisations are required for radiative
transfer of solar and thermal radiation, cloud microphysics,
small-scale convection, sub-grid scale dissipative mechanisms, and so
on.  In the ocean, dissipation, convection and mixing processes need
to be parameterised.  In addition, in coupled ocean-atmosphere GCMs,
boundary-layer mechanisms coupling atmospheric and ocean dynamics must
be represented, such as oceanic responses to surface wind stress,
fluxes of water and heat between the atmospheric boundary layer and
the ocean surface, and so on \citep{garratt-boundary-layer}.  In a
comprehensive coupled atmosphere-ocean GCM, there are some other
processes that might also be considered as parameterisations, such as
river routing and representations of vegetation and other land surface
processes.  The end result of all of these different factors is a more
or less faithful representation of some aspects of the behaviour of
the climate system.  ENSO variability, if exhibited at all by a GCM,
arises, as it does in the real climate system, as an emergent property
of the mechanisms at work in the equatorial Pacific ocean and
atmosphere.  At no point is ENSO variability ``put in'' to the model,
although interannual variability in the tropical Pacific can obviously
be affected by choices made in the different model parameterisations.
This generalised modelling approach makes coupled processes such as
ENSO a stringent test of model performance, but it does theoretically
allow GCMs to represent any process that occurs in the real atmosphere
or ocean, and to represent connections between different processes in
different parts of the world.  This is essential if details of the
processes leading to atmospheric and oceanic teleconnections are to be
understood.  GCMs are also the best vehicle for more complex
experiments examining the effects of variations in climatic conditions
on ENSO behaviour, for either future climate change
\citep[e.g.,][]{collins-enso-change,merryfield-co2-enso} or
paleoclimate applications
\citep[e.g.,][]{liu-midhol-enso,ob-glacial-enso}.

Only recently have coupled GCMs begun to be able to simulate ENSO
variability in a realistic fashion: the results of
\citet{vanoldenborgh-enso}, \citet{achutarao-enso-better} and
\citet{capotondi-cmip3-enso} examining the CMIP3 ensemble simulations
compare favourably to those of, for instance, \citet{neelin-comp} and
\citet{latif-ensip}, who studied earlier generations of GCMs, or even
\citet{achutarao-enso-cmip}, who used models from the previous round
of CMIP experiments.  However, even with the progress that has been
made, many GCMs still have great difficulty in producing a realistic
looking ENSO, and as the results in Chapter~\ref{ch:obs-cmip-enso}
will show, out of the models in the CMIP3 ensemble, relatively few do
a truly convincing job in the equatorial Pacific.  Current models have
improved in terms of the overall frequency of simulated \eln events
and the enhanced temperature variability over the eastern Pacific, but
they still display significant deficiencies in the representation of
ocean-atmosphere coupling mechanisms important for ENSO variability
--- see particularly \citep{vanoldenborgh-enso} on this point, where
the individual feedback mechanisms relating wind stress, thermocline
depth and sea surface temperature are examined in detail in current
models.

One particularly difficult aspect of modelling ENSO in GCMs, compared
to the simpler models described below, is that in GCMs, not only the
interannual variability that is of interest for ENSO, but also the
mean background state and seasonal cycle are simulated.  In simplified
models, both the mean background state and seasonal cycle are
frequently fixed to observed values.  The effect of variations in the
background state of the ocean and atmosphere on the coupling between
the two systems makes GCM modelling of ENSO processes particularly
challenging \citep{philander-sporadic-enso,guilyardi-cmip3}.
Furthermore, the simple fact that ICMs can use parameterisations of
physical processes specialised for the equatorial Pacific imposes
another comparative disadvantage on GCMs.  In this context,
consideration of the parameterisation of vertical mixing in the ocean
is quite illuminating.  It has been shown \citep{battisti-hirst-do}
that the most important nonlinear process in the equatorial Pacific,
in terms of setting the amplitude of ENSO-related vacillations, is the
dependence of vertical upwelling of water across the thermocline on
the vertical temperature structure of the ocean, in particular the
thermocline depth.  The thermocline depth in the eastern equatorial
Pacific, where these upwelling processes are of most importance, is
highly variable, and it is difficult for a GCM to capture the strong
dependence of vertical diffusion processes on the thermocline depth.
Even though modern ocean GCMs generally use a more sophisticated
parameterisation of vertical diffusion than simple constant diffusion
coefficients between ocean model layers, such as the widely used
scheme of \citet{gent-mcwilliams}, these parameterisations are applied
globally, and cannot easily be modified to take account of the unique
features of the highly variable thermocline structure in the eastern
equatorial Pacific.  In most ICMs, the rate of upwelling of cooler
water into the surface layer across the thermocline is carefully
parameterised to match the observed behaviour in the eastern
equatorial Pacific.  This specialisation of intermediate complexity
models to one particular region of the globe, and to the processes of
most importance for that region, has a strong influence on the success
of ICMs in simulating ENSO variability compared to GCMs.

\subsection{Statistical models}
\label{sec:statistical-models}

The main use for statistical models in the study of ENSO is for
prediction of the future evolution of ENSO events.  This application
is somewhat orthogonal to the goals of this study, so I mention only a
few details here.  \citet{latif-enso-predict} provide a comprehensive
review of studies in the predictability and prediction of ENSO up to
1998, while \citet{chen-enso-predict} report on more recent work in
this field.  Statistical models are competitive in terms of predictive
skill with ICMs, and do a better job of ENSO prediction than GCMs and
unsurprisingly, conceptual models.  Statistical models are generally
simpler to construct than dynamical models \citep{xue-markov-enso}, so
are likely to remain a subject of interest even if further progress is
made in dynamical modelling of ENSO.

Most of the statistical models of ENSO that have been developed to
date are linear models based on some form of regression of predictand
variables (e.g. eastern equatorial Pacific SST one season into the
future) onto a set of predictor variables (e.g. a history of wind
stress fields in the equatorial Pacific).  A variety of different
schemes have been used for this regression, including simple linear
regression \citep{tang-enso-nns,vonstorch-zwiers}, canonical
correlation analysis \citep{barnston-enso-cca,bretherton-coupled},
principal oscillation patterns \citep{xu-enso-pops}, linear inverse
modelling techniques
\citep{penland-optimal-growth,penland-stochastic-sst} and other Markov
methods of varying levels of sophistication
\citep{johnson-markov-sst-1,johnson-markov-sst-2,pasmanter-cyclic-markov,xue-markov-enso},
and singular spectrum analysis to improve the fitting of simple
auto-regressive processes to ENSO variability indexes
\citep{keppenne-enso-ssa,ghil-ssa}.

Beyond these linear statistical methods, there has been some
application of nonlinear methods to the statistical modelling of ENSO,
notably neural networks \citep{tang-enso-nns,tangang-enso-nns} and
some sophisticated nonlinear regression techniques
\citep{kondrashov-data-based,timmermann-ace-model}.  There has also
been some application of hierarchical Bayesian methods to the
development of statistical models able to provide a characterisation
of forecast uncertainty \citep{berliner-bayesian-enso}, but this seems
to be a relatively unexplored area, and there is potential for further
development here.  \citet{mason-prob-enso} review some other
statistical methods of producing probabilistic forecasts.

In terms of prediction skill, it is interesting to observe that some
studies show that there is currently little to be gained from the
application of the more complex statistical (or dynamical) models
\citep{halide-simple-enso-model,chen-enso-predict}.  In
\citep{halide-simple-enso-model}, a very simple ENSO prediction model
based on a linear neural network using lagged NINO3 SST index values
as predictor variables showed prediction skill comparable to several
more complex models.  The results depend on the forecast lead time,
but the simple model is at least in the same range of forecast skill
as the more complex models.

\subsection{Intermediate complexity models}

A different approach to modelling ENSO, oriented more towards gaining
mechanistic understanding of the processes underlying ENSO
variability, is embodied in intermediate complexity models (ICMs).
These are mechanistic models based on simplifications of the equations
of motion of the atmosphere and ocean that emphasise the processes
most important for ENSO.  (There is also a slightly different class of
hybrid models, that use a mechanistic ocean and statistical
atmosphere, but these will not be covered here; some are described in
the reviews mentioned below.)

The development of ICMs for modelling ENSO is reviewed in
\citet{neelin-enso-theory}, while Chapter~7 of \citet{dijkstra-book}
provides a more recent pedagogical presentation.  The best known ENSO
ICM is the Zebiak-Cane model \citep{zebiak-cane}, which was the first
such model to be used successfully for ENSO prediction.  This model
simulates anomalies with respect to an observed annual cycle of
low-level winds, sea surface temperature and thermocline depth.  The
atmosphere is represented using a Gill-type shallow water
approximation \citep{gill-atmos-response}, forced by heating anomalies
that depend on SST and low-level moisture convergence.  The ocean
model covers the Pacific basin only and is based on a linearised
shallow water approximation, incorporating a well-mixed surface layer
overlying the deeper shallow water layer, in order to represent the
effect of surface wind stress on near-surface currents.  The ocean
model is forced by surface wind stress anomalies using a standard bulk
aerodynamic relation between the modelled winds and wind stress.  Both
the atmosphere and ocean use an equatorial beta-plane approximation to
the Coriolis force \citep{majda-lecture-notes}, a factor critical for
realistically modelling the dynamics of equatorial wave modes in both
the atmosphere and ocean.  The dynamics of these wave modes were
explored in detail in a series of papers by
\citet{cane-sarachik-1,cane-sarachik-2,cane-sarachik-3}, which formed
the basis of our understanding of the behaviour of the ocean component
of the Zebiak-Cane and other similar models.  The model produces
realistic-looking ENSO behaviour, with ocean warming events occurring
with irregular amplitude and irregular spacing in time.  Later
versions of the model have used data assimilation techniques to help
in model initialisation and thus improve predictive skill
\citep{chen-zc-improved-1,chen-zc-improved-2}.

Most other ICMs follow a similar approach to the Zebiak-Cane model,
using equatorial beta-plane shallow water equation or similar
approximations to the primitive equations.  These approximations
capture aspects of the atmospheric response to SST anomalies most
important to ENSO variability and represent the propagation of
internal wave modes in the ocean that are believed to provide the
memory for the ENSO oscillator
\citep[e.g.,][]{battisti-icm,kleeman-icm}.

As well as being used for operational seasonal prediction, ICMs have
been applied to a number of questions surrounding ENSO.  A few example
applications include investigation of the so-called ``predictability
barrier'' related to the growth phase of \eln conditions
\citep{samelson-zc-predict}, the examination of mechanisms governing
interactions between ENSO and the annual cycle in the Pacific
\citep{tziperman-zc-seasonal}, studies looking at the possibility that
the irregularity and seasonal phase locking of ENSO events are both
related to a quasi-periodicity route to chaos in the coupled
ocean-atmosphere system \citep{tziperman-quasi-per,jin-enso}, and
investigations of the influence of intraseasonal synoptic-scale
variability on ENSO \citep{moore-stoch-enso}.  ICMs have even been
used for paleoclimate simulations of ENSO variability
\citep[e.g.,][]{an-lgm-enso}.

\subsection{Conceptual models}
\label{sec:conceptual-models}

Conceptual models of ENSO are highly simplified models, usually in the
form of ordinary differential equation or delay differential equation
systems.  These simple models represent basic mechanisms of ENSO
variability in a highly summarised form, suitable for analytical
investigation, and are generally derived from intermediate complexity
models via heuristic reasoning.  In this sense, conceptual models of
ENSO have something in common with simplified models of other
phenomena in the climate system, used to build physical intuition and
to explore the influence of different processes on the evolution of
these phenomena.  Examples include simplified models of the
Madden-Julian Oscillation in the Indian and western equatorial Pacific
Oceans \citep{majda-mjo-icm} and box models of the ocean circulation
used in the study of the thermohaline circulation, an approach that
began with the work of \citet{stommel-box-model}.  Although these
highly simplified models can represent only a few aspects of a complex
fluid system like the climate system, they do provide a framework for
experimentation and can help to develop physical understanding of the
most important processes for the phenomenon of interest.  This is
particularly the case when results from these simple models can be
compared to results from more complex models or to observations.  A
good example is the extended study of bifurcations of states of the
Atlantic thermohaline circulation conducted by Henk Dijkstra and
coworkers, summarised by \citet{dijkstra-equilibria} and described in
more detail in the monograph of \citet{dijkstra-book}.

As for highly simplified models of the other phenomena mentioned,
conceptual models of ENSO cannot, in general, be rigorously derived
from the full equations for fluid flow in the atmosphere and ocean, or
from intermediate complexity models relying on shallow water or other
approximations to the tropical Pacific ocean and atmosphere.  Instead,
they are derived from such models by heuristic physical arguments,
often involving spatial averaging of the equations for more complex
models and the application of physically reasonable (although not
rigorously justifiable) parameterisations for physical processes
represented explicitly in the more complex models.

Four main conceptual models of ENSO variability have been developed,
each based on a different view of which phenomena in the tropical
Pacific are most important for controlling interannual variability
associated with ENSO.  These are the delay oscillator
\citep{suarez-schopf-do,battisti-hirst-do}, the recharge oscillator
\citep{jin-1997a}, the advective-reflective oscillator
\citep{picaut-adv-refl} and the western Pacific oscillator
\citep{weisberg-wpac-osc}.  It is not necessarily the case that ENSO
variability in the real ocean-atmosphere system is confined to a
single one of the mechanisms invoked in these different models.  It is
conceivable that all of the processes represented in these models (and
perhaps others) play a part in producing observed ENSO variability.
These models should be thought of more as a means to explore the
possible coupled ocean-atmosphere mechanisms of variability at work in
the Pacific that may contribute to ENSO.

The first conceptual ENSO model to be developed and the model that has
had the most influence was the so-called \emph{delayed oscillator}
model of \citet{suarez-schopf-do} and \citet{battisti-hirst-do}.  The
ideas behind this model arose from results of a study
\citep{schopf-suarez-icm} examining coupled ocean-atmosphere
variability in an ICM similar to the classic Zebiak-Cane model
\citep{zebiak-cane}.  This ICM displays self-sustained oscillations
with a period of 3--5 years, with many of the characteristics of
observed ENSO variability.  The behaviour of the ICM highlights the
importance of ocean-atmosphere coupling for the existence of ENSO
variability.  From their experiments, \citeauthor{schopf-suarez-icm}
conclude that some positive feedback mechanism is required to simulate
the observed variability, and that this feedback arises from
ocean-atmosphere coupling.  Further, \citeauthor{schopf-suarez-icm}
demonstrated the importance of the propagation of signals across the
Pacific Ocean basin by equatorially trapped waves.  Lagged
cross-correlation plots between winds, SST and sea surface elevation
in the Pacific clearly show propagation of signals that seem to be
closely related to the cycle of ENSO variability.

Based on these results, \citet{suarez-schopf-do} developed a
conceptual model of ENSO as a single delay differential equation for
the average SST over a region of the eastern/central Pacific, relying
on the existence of a strong local positive feedback in the
ocean-atmosphere system, an unspecified nonlinear mechanism limiting
the growth of SST perturbations, and a treatment of equatorial wave
propagation via a time delayed negative feedback:
\begin{equation}
  \label{eq:enso-delayed-osc}
  \frac{dT}{dt} = - b T(t - \tau) + c T - e T^3.
\end{equation}
Here, $T(t)$ is the modelled SST value as a function of time, $b$, $c$
and $e$ measure the strengths of the delayed linear, local linear and
local nonlinear feedback terms respectively, and $\tau$ is the time
delay for the delayed linear feedback.  This model is formulated on
the basis that strong ocean-atmosphere coupling in the eastern Pacific
(essentially the \citet{bjerknes-feedback} feedback mechanism
described in Section~\ref{sec:enso-mechanisms}) causes emission of
westward propagating signals in the form of equatorial Rossby waves,
which, following reflection at the western boundary of the basin,
propagate back into the eastern Pacific (as equatorial Kelvin waves)
to influence the ocean-atmosphere coupling there.

Study of the delayed oscillator model was extended significantly by
the work of \citet{battisti-hirst-do}, who used an ICM similar to the
Zebiak-Cane model, again showing ENSO-like self-sustained
oscillations.  Their first interesting result came from a comparison
of results of a linearised version of their ICM with results of the
original (nonlinear) ICM.  This showed that the main effect of the
nonlinearities in the ICM was to restrict the growth of SST
disturbances, leading to finite amplitude oscillations.  The spatial
and temporal structure of solutions to both the original ICM and its
linearisation were very similar.  This seems to indicate that the
primary cause of ENSO variability is well represented by a linear
instability, and confirms the more heuristic conclusions of
\citet{suarez-schopf-do}.

\citeauthor{battisti-hirst-do} constructed what they called
\emph{analogue} models based on their ICM, designed to capture the
essential features of the model.  The construction of these analogues
relies on careful spatial averaging of the fields of the original ICM,
along with some heuristic reasoning to develop parameterisations for
processes that cannot be represented in the highly summarised form of
the analogue models.  This approach allowed
\citeauthor{battisti-hirst-do} to examine the influence of individual
factors arising from the ICM equations in the linear and nonlinear
instability terms in their analogue models.  They could thus identify
the most important nonlinear process in their full model by
selectively removing each term and examining the effect on the
modelled oscillations.  The most important nonlinearity, and the one
that sets the amplitude of oscillations, is the term associated with
vertical upwelling across the thermocline.  Removal of other nonlinear
processes from the model (e.g. zonal SST advection, surface heating)
had little effect on the model oscillations, while removing the
vertical upwelling nonlinearity completely changed the nature of the
model behaviour.

The simplest of the analogue models developed by
\citet{battisti-hirst-do} has the same model equation as the model of
\citet{suarez-schopf-do}, i.e. \eqref{eq:enso-delayed-osc}, although
the physical balance of terms is different.  In contrast to
\citep{suarez-schopf-do}, where the fundamental balance is between
local linear instability and local nonlinear damping, in
\citep{battisti-hirst-do} the fundamental balance is between local
linear instability and the delayed negative linear feedback.  The
nonlinearity here plays only a secondary role.

Adaptations of this simple delayed oscillator model have been used for
a number of studies of other aspects of ENSO variability, including
the existence of chaotic dynamics in a delay oscillator with seasonal
forcing \citep{tziperman-enso-chaos,ghil-enso-dde}, phase locking of
ENSO events to the annual cycle
\citep{tziperman-do-locking,galanti-phase} and the effect of realistic
stochastic forcing (through the wind field) on a delayed oscillator
model \citep{saynisch-do-noise}.

The second of the major conceptual models of ENSO is the
\emph{recharge oscillator} of \citet{jin-1997a}.  This model is
constructed explicitly to be consistent with the feedback mechanisms
proposed by \citet{bjerknes-feedback} and \citet{wyrtki-1985},
described in Section~\ref{sec:enso-mechanisms}, and makes use of the
idea of the ``recharge'' of the equatorial warm water volume as a
required precondition for the occurrence of \eln, an idea originally
proposed by \citet{cane-zebiak-1985}.  The processes modelled in the
recharge oscillator are essentially those illustrated in
Figure~\ref{fig:enso-cartoons}, which is adapted from Figure~1 of
\citep{jin-1997a}.

The recharge oscillator is based on modelling two processes, the fast
adjustment of the zonal thermocline slope in the equatorial Pacific to
variations in the zonal wind stress, and the slower adjustment of the
mean equatorial thermocline depth across the Pacific as a result of
transfers of warm water into and out of the equatorial latitude band.
Unlike the delayed oscillator and the western Pacific oscillator to be
treated below, which are expressed as delay differential equations,
the recharge oscillator that is the final result of the analysis of
\citet{jin-1997a} is a simple second order ordinary differential
equation system, making analysis rather simpler than for the models
involving delays.  Indeed, \citet{jin-1997a} not only produces
explicit analytic expressions for oscillation frequency as a function
of the system parameters, but goes on to examine the effects of
stochastic excitation on the model, a situation of interest in the
real ocean-atmosphere system, where it may be the case that ENSO
manifests itself as a damped stable mode of oscillation, excited by
stochastic forcing due to winds in the western Pacific or influences
from outside the equatorial Pacific region.  As for the other
conceptual models, suitable selection of model parameters produces
oscillations that replicate some of the features of observed ENSO
variability.

One interesting application of the recharge oscillator was the study
of \citet{mechoso-simple-models}, who examined ENSO variability in the
output of a coupled ocean-atmosphere GCM simulation by fitting
parameters of the recharge oscillator equations to the first SSA mode
of the GCM sea surface temperature and thermocline depth anomalies.
(This approach could be considered an example of data-to-model
dimensionality reduction, in the terminology of
Chapter~\ref{ch:nldr-overview}.)  \citeauthor{mechoso-simple-models}
considered how well the recharge oscillator was able to represent the
leading mode of SST and thermocline variability in their GCM, based
either on simply fitting the parameters in the recharge oscillator
equations, or fitting the parameters related to individual physical
processes of importance in the recharge oscillator.

The third conceptual model of ENSO is the \emph{advective-reflective
  oscillator} of \citet{picaut-adv-refl}.  Although the delayed
oscillator model succeeds in representing some of the features of
observed ENSO variability, one aspect of the model is less than
satisfactory.  Wave reflection at the western boundary of the Pacific
basin is essential to the dynamics of the delayed oscillator, but some
studies performed after the initial development of the delayed
oscillator model \citep[e.g.,][]{delcroix-waves,boulanger-waves}
indicate that the reflection coefficient for reflection of Rossby
waves at the western boundary of the Pacific may be rather small.
This, along with observations of coherent variations of the Southern
Oscillation Index (the normalised pressure difference between Tahiti
and Darwin, a common measure for the atmospheric aspect of ENSO
variability) and the zonal location of the eastern edge of the Western
Warm Pool in the Pacific \citep{picaut-wwp}, led
\citeauthor{picaut-adv-refl} to propose a rather different mechanism
for ENSO variability, with less emphasis on processes in the eastern
Pacific and the resulting dynamics of equatorial waves emitted there,
and more emphasis on zonal advection in the central Pacific.

The essential idea behind the model of \citeauthor{picaut-adv-refl} is
that variations in the position of the eastern edge of the Western
Warm Pool due to zonal advection modify the region where SSTs are
above the threshold for the maintenance of organised atmospheric
convection.  The position of the eastern edge of the warm pool is
determined by a complex system of surface currents in the central
equatorial Pacific generated by local wind forcing (primarily as a
result of the Bjerknes feedback), free equatorial Kelvin and Rossby
waves, and equatorial waves reflected from the western and eastern
boundaries.  The interaction of these effects can lead to
self-sustaining oscillations in the position of the eastern edge of
the Western Warm Pool, and consequent fluctuations in the eastern
Pacific SST, thermocline structure and wind field.
\citeauthor{picaut-adv-refl} do not present simple equations for their
model, as has been done for the other conceptual model described here
(although the ``unified oscillator'' model of \citet{wang-uo} can, for
suitable parameter choices, simulate the relevant behaviour).
Instead, they report on experiments with a restricted linear ICM that
helps to elucidate the mechanisms of oscillation.

Because it emphasises interactions between equatorial current systems
and the location of the eastern edge of the Western Warm Pool, the
\citet{picaut-adv-refl} model is less sensitive to the strength of
western boundary reflections than the delayed oscillator model.
\citeauthor{picaut-adv-refl} find that their model produces
oscillations with no or very little reflection at the western
boundary, although reflection at the eastern boundary is required (the
eastern boundary of the Pacific appears to be a much more efficient
wave reflector than the western boundary, so this is a reasonable
condition).  As with all of the conceptual models reported here, the
\citep{picaut-adv-refl} model produces regular oscillations.  The
irregularity and annual phase locking of ENSO arise from some other
mechanism not considered by these models.

The \emph{Western Pacific oscillator} of \citet{weisberg-wpac-osc} is,
like the model of \citet{picaut-adv-refl}, based on a hypothesis
concerning the mechanism of operation of ENSO derived from
observational data: in this case, correlations between SST and sea
level pressure variation in the off-equatorial western Pacific.  These
observations suggest a model where off-equatorial variations of
thermocline depth and wind field in the far western Pacific provide
the necessary negative feedback that, along with the usual Bjerknes
positive feedback in the eastern/central Pacific, leads to ENSO
oscillations.

One part of the western Pacific oscillator scenario is the usual
\citet{gill-atmos-response} atmospheric response to a heating anomaly
in the eastern Pacific, where a pair of off-equator cyclones is formed
to the west of the heating anomaly.  The westerly wind anomalies
associated with this cyclone pair cause increases in thermocline depth
and SST in the eastern Pacific, producing \eln conditions there.  The
main innovation of the \citet{weisberg-wpac-osc} model is to consider
the effect of this cyclone pair on conditions in the western Pacific.
The primary direct effect of the cyclones in the western Pacific is to
reduce the thermocline depth (and hence the SST) in off-equatorial
regions via Ekman pumping.  These changes enhance off-equatorial high
sea level pressure anomalies in the western Pacific.  Through
interaction with variations in the region of peak atmospheric
convection (which shifts eastwards during \eln events) the sea level
pressure anomalies initiate equatorially convergent easterly winds in
the far western Pacific.  Finally, these easterly winds trigger an
upwelling Kelvin wave propagating eastwards that raises the
thermocline and reduces SST anomalies in the eastern Pacific, so
permitting the coupled ocean-atmosphere system to oscillate.

This rather complex chain of interactions is captured in four
equations relating variations in the equatorial thermocline depth in
the NINO3 region, $h_1$, the off-equatorial thermocline depth in the
western Pacific, $h_2$, the equatorial westerly wind stress in the
west-central Pacific, $\tau_1$ and the equatorial easterly wind stress
in the western Pacific, $\tau_2$:
\begin{subequations}
  \begin{gather}
    \frac{dh_1}{dt} = a \tau_1 + b_2 \tau_2(t - \delta) -
    \varepsilon_1 h_1^3, \label{eq:wpo-1} \\
    \frac{dh_2}{dt} = -c \tau_1(t - \lambda) - \varepsilon_2
    h_2^3, \label{eq:wpo-2} \\
    \frac{d \tau_2}{dt} = d h_2 - \varepsilon_3
    \tau_2^3, \label{eq:wpo-3} \\
    \frac{d \tau_1}{dt} = e h_1 - \varepsilon_4
    \tau_1^3. \label{eq:wpo-4}
  \end{gather}
\end{subequations}
In each of these equations, cubic nonlinearities act to limit the
amplitude of oscillations.  The first term on the right hand side of
\eqref{eq:wpo-1} represents the local forcing of thermocline anomalies
by westerly wind stress, similar to the second term in the delayed
oscillator, \eqref{eq:enso-delayed-osc}, while the second term
represents the negative feedback induced by easterly winds over the
far western Pacific, with a delay ($\delta$) since the effect of these
winds takes some time to be felt in the central Pacific.  The first
term on the right hand side of \eqref{eq:wpo-2} represents the forcing
of off-equatorial thermocline anomalies in the western Pacific by the
zonal wind stress due to the pair of cyclones associated with the Gill
atmospheric response to heating anomalies in the central Pacific,
again with a delay ($\lambda$) since it takes time for this influence
to propagate into the western Pacific.  Finally, the two wind stress
equations, \eqref{eq:wpo-3} and \eqref{eq:wpo-4}, respectively relate
off-equatorial wind stress in the western Pacific to the thermocline
depth there (via the relationship between thermocline depth, SST and
sea level pressure), and central Pacific equatorial wind stress to the
thermocline depth in the NINO3 region.

As for the \citet{picaut-adv-refl} model, the western Pacific
oscillator is not dependent on the efficiency of wave reflection at
the western boundary of the Pacific basin.  For suitable choices of
model parameters, which can be justified by non-dimensionalising
\eqref{eq:wpo-1}--\eqref{eq:wpo-4} and imposing physically reasonable
balances between the forcing terms, the western Pacific oscillator
displays self-sustaining oscillations of an amplitude and frequency
consistent with observed ENSO variability.

Apart from these four models and their derivatives, a number of other
approaches to conceptual modelling of ENSO have been pursued.  Studies
covering some of these other approaches include the work of
\citet{vallis-enso-models}, who considered simple ODE models based on
different finite differencing schemes in a simple western
Pacific/eastern Pacific two box configuration, and the work of
\citet{saunders-boolean-delay}, who constructed a Boolean delay
equation model of ENSO variability, an approach that permits the
development of models at a conceptual level that is, in some sense,
even coarser than the level of the four main conceptual models
described above.  Also of interest is the study of \citet{wang-uo},
who constructed a model incorporating most of the mechanisms treated
in each of the four main conceptual models.  Different parameter
regimes of \citeauthor{wang-uo}'s model are able to capture the
behaviour of each of the delayed oscillator, recharge oscillator,
advective-reflective oscillator and western Pacific oscillator.

\section{Previous applications of nonlinear dimensionality reduction to ENSO}

Independent of whether the basic mechanism of variability underlying
ENSO is based on intrinsically nonlinear chaotic dynamics or
stochastically forced linear dynamics, the spatial coherence of \eln
and \lan episodes in the Pacific leads us to expect that there should
be a low-dimensional model that captures at least some of the
variability in the tropical ocean-atmosphere system.  Here, in
Chapters~\ref{ch:nlpca}--\ref{ch:hessian-lle}, I approach the
assessment of ENSO in coupled GCMs by attempting to identify such
low-dimensional structures in the dynamics of the tropical Pacific
atmosphere and ocean.  It should be noted that, in general, the
mechanisms leading to ENSO and ENSO-like variability in current
coupled atmosphere-ocean GCMs show significant differences compared to
the mechanisms contributing to ENSO variability in the real
atmosphere-ocean system.  For instance, \citet{vanoldenborgh-enso}
report that most of the models that they examined show a response of
the zonal wind field to equatorial SST anomalies that is weaker and
more confined to equatorial latitudes than seen in observations.  This
weak wind response is compensated by a stronger direct response of
SSTs to changes in the wind field and a weaker damping of SST
variations than observed.  This different balance of factors in the
models compared to the observations should lead us to view conclusions
drawn from models about ENSO variability in the real atmosphere and
ocean with some caution.  However, it is still of interest to examine
how well we can characterise what low-dimensional dynamics is seen in
the models, and to see if this characterisation can provide any
further insight into the behaviour of the models.  For instance, based
on the Bjerknes-Wyrtki mechanism described in
Section~\ref{sec:enso-mechanisms}, earlier studies have indicated that
ENSO variability can be approximated as a two-dimensional oscillation,
one degree of freedom being associated with the NINO3 SST index, the
mean SST anomaly across the region 150\degree{}W--90\degree{}W,
5\degree{}S--5\degree{}N, and the other with the equatorial Pacific
warm water volume, a proxy for the zonal mean thermocline depth
\citep{burgers-stoch-enso,kessler-events,mcphaden-wwv}.  These two
degrees of freedom vary in approximate quadrature during \eln events.
One would hope that any analysis method aimed at characterising ENSO
variability in observational or simulated data would be able to
identify these two degrees of freedom.

Application of linear dimensionality reduction methods to the
characterisation of ENSO behaviour in equatorial Pacific SST,
thermocline and wind fields is common.  For example, PCA is very
widely used in the climatological community
\citep[e.g.,][]{vanoldenborgh-enso,merryfield-co2-enso,vonstorch-zwiers}
and more sophisticated linear methods such as CCA
\citep{barnston-enso-cca}, principal oscillation patterns
\citep{xu-enso-pops,tang-enso-pops} and SSA \citep{keppenne-enso-ssa}
have all been used for studying ENSO, mostly in the context of
statistical modelling for ENSO prediction.  Linear dynamical modelling
using Markov models or stochastic differential equation models for
ENSO prediction is also common, as reported in
Section~\ref{sec:statistical-models}.

However, of the large number of nonlinear dimensionality reduction
schemes that have been developed (Chapter~\ref{ch:nldr-overview}),
only a small number have previously been applied to ENSO data.  The
studies of which I am aware are restricted to five groups of methods:
\begin{description}
  \item[Nonlinear PCA]{A neural network method based on the use of
    multilayer perceptrons, described in detail and applied to
    simulated ENSO variability in the CMIP3 ensemble in
    Chapter~\ref{ch:nlpca}.  This is the nonlinear dimensionality
    reduction method that has been most extensively applied to
    climatological questions, including ENSO
    \citep{hsieh-review-2004,monahan-enso,an-enso-interdecadal,wu-enso-interdecadal}.}
  \item[Self-organising maps]{Another neural network method, based on
    a different network structure and learning strategy to NLPCA
    (Section~\ref{sec:ann-methods}).  It has been applied to detecting
    interdecadal changes in observed ENSO variability
    \citep{leloup-som} and to inter-model comparison of simulated ENSO
    characteristics \citep{leloup-som-2}.}
  \item[Isomap]{A very widely used geometrical dimensionality
    reduction method based on globally isometric data transformations
    and multi-dimensional scaling.  It is described and applied here
    in Chapter~\ref{ch:isomap}, and has previously been used to
    examine ENSO variability in observational Pacific SST data
    \citep{gamez-enso,gamez-enso-2}.}
  \item[Cumulant functions]{The cumulant function of a data set can
    characterise structures describing large deviations from the mean
    better than the linear decomposition offered by PCA, essentially
    because the cumulant function for a random variable $X$,
    \begin{equation}
      \log \left\langle \exp (s X) \right\rangle = \sum_{n=1}^\infty
      \kappa_n \frac{s^n}{n!},
    \end{equation}
    with $s \in \mathbb{R}$ and $\kappa_n$ called the $n$th cumulant,
    encodes information about all of the moments of the distribution
    of the data.  For a multivariate random variable $\vec{X} \in
    \mathbb{R}^m$, the cumulant function is defined in the obvious way
    as $G(\vec{s}) = \log \langle \exp (\vec{s} \cdot \vec{X})
    \rangle$, with $\vec{s} \in \mathbb{R}^m$.  The data analysis
    method described by \citet{bernacchia-cumulant-1} and applied to
    the analysis of ENSO variability in \citep{bernacchia-cumulant-2}
    maximises $G(\vec{s})$ over all possible directions of $\vec{s}$
    to find data patterns that are, in some sense, most extreme,
    measured not only by the direction of greatest variance, as in
    PCA, but also incorporating some degree of influence from the
    higher moments of the data.}
  \item[Nonlinear regression]{A number of dynamical reduction
    strategies have been applied to ENSO variability based on
    nonlinear regression, most notably those reported in
    \citet{timmermann-ace-model} and \citet{kondrashov-data-based}.}
\end{description}

Comparing this short list to the range of nonlinear dimensionality
reduction methods presented in Chapter~\ref{ch:nldr-overview}, it is
clear that there is some scope for exploring the application of these
methods to climatological questions, and in particular, to ENSO
variability.  As well as being of intrinsic scientific interest, the
problem of characterising ENSO variability in observational and
simulated data provides a good test case for nonlinear dimensionality
reduction methods, primarily because the expected results are
relatively easy to interpret.  ENSO is by far the strongest mode of
climate variability after the annual cycle and has both a clear
signature of temporal variability and easily recognisable spatial
patterns.  This makes it an ideal testbench for the methods described
in Chapter~\ref{ch:nldr-overview}, and I will devote
Chapters~\ref{ch:nlpca}--\ref{ch:hessian-lle} to exploring the
application of some of these ideas to observed and simulated ENSO
variability.


%% file: 05-obs-cmip-enso.tex
\chapter{Tropical Pacific Variability in Observations and the CMIP3 Models}
\label{ch:obs-cmip-enso}

In this chapter, I present results comparing tropical Pacific sea
surface temperature and thermocline depth variability in observed data
and pre-industrial control simulations from the CMIP3 model ensemble.
The goal here is to set the scene for the nonlinear dimensionality
reduction analyses presented in
Chapters~\ref{ch:nlpca}--\ref{ch:hessian-lle} by examining some more
conventional views of ENSO behaviour.  The results here overlap to
some extent with those presented by \citet{capotondi-cmip3-enso}, but
I show results concerning the phasing of equatorial warm water volume
and SST variations, while \citeauthor{capotondi-cmip3-enso}
concentrate on effects on the period of ENSO variability of ocean
advective processes and the spatial structure of surface wind stress
anomalies.

\section{Equatorial Pacific sea surface temperature}

\subsection{Basic equatorial Pacific SST variability}
\label{sec:basic-eq-sst-var}

\begin{figure}
  \begin{center}
    \includegraphics[width=0.48\textwidth]{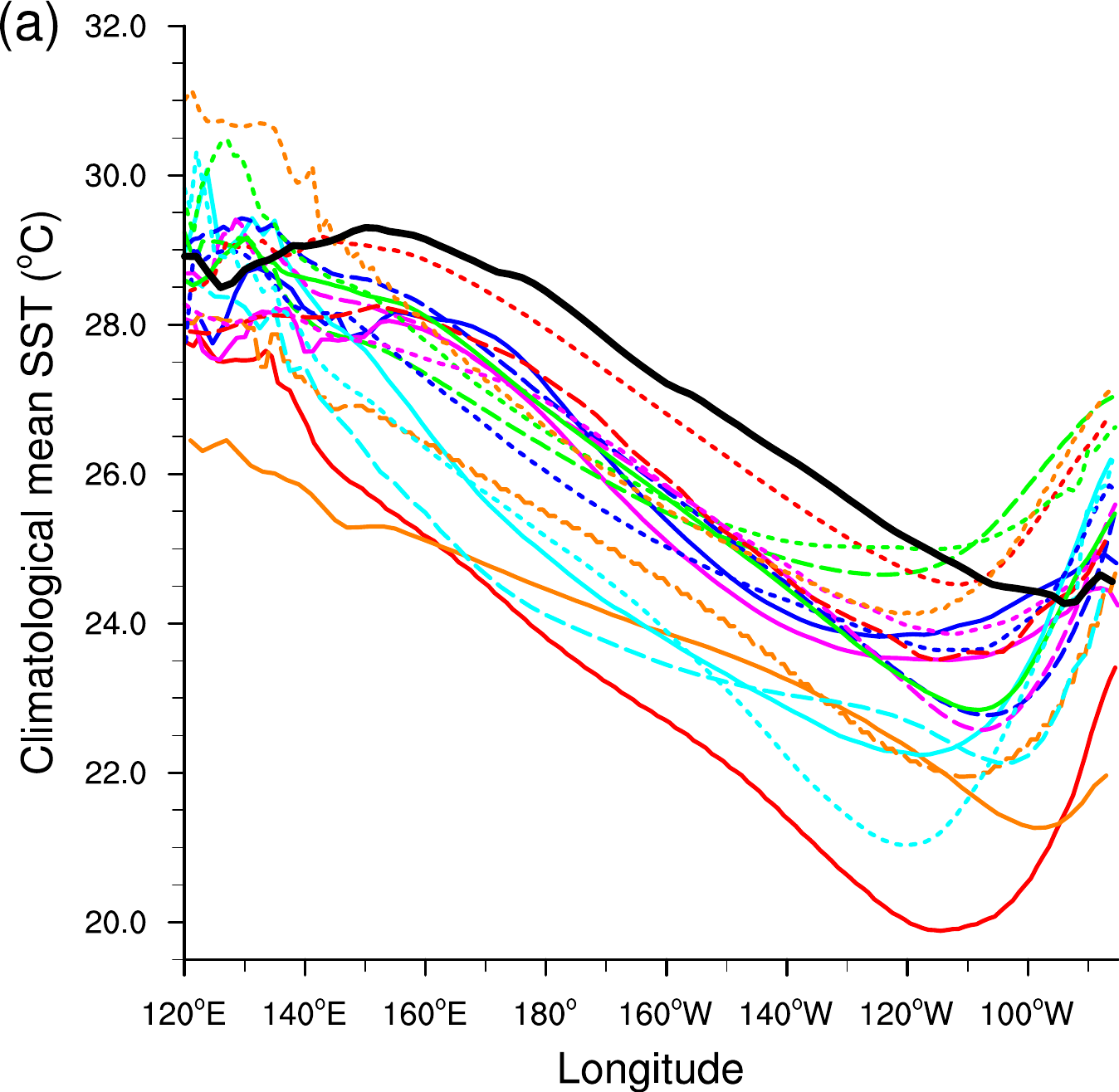}%
    \hspace{0.04\textwidth}%
    \includegraphics[width=0.48\textwidth]{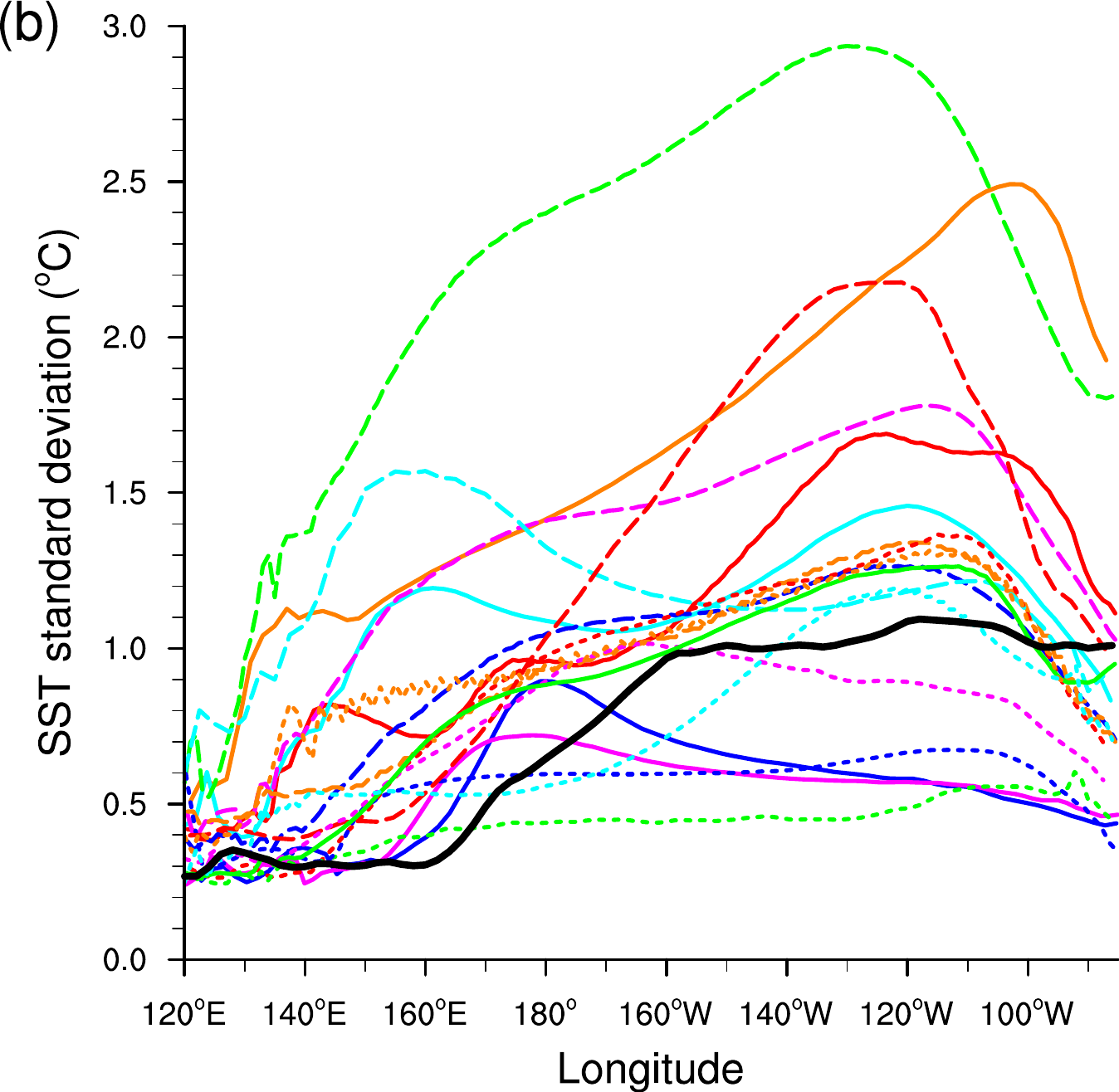}
  \end{center}
  \caption[Equatorial Pacific climatological mean SST and annual SST
    standard deviation]{Climatological mean SST (a) and annual
    standard deviation of SST (b) across the equatorial Pacific from
    observations (thick black line) and models (coloured lines --- see
    Table~\ref{tab:models} for key).  Values shown are averaged
    between 2\degree{}S and 2\degree{}N.}
  \label{fig:basic-sst}
\end{figure}

We first consider the climatology and magnitude of interannual
variability of SSTs in the equatorial Pacific.
Figure~\ref{fig:basic-sst}a shows annual mean SST across the Pacific,
averaged between 2\degree{}S and 2\degree{}N.  Although most models
have a cold bias across the Pacific basin, with SSTs up to 4\degree C
cooler than observed, they do simulate the gradient of mean SST from
the Western Warm Pool around Indonesia (120\degree{}E) to the cooler
waters of the eastern Pacific (90\degree{}W).  However, most of the
models do not show a monotonic eastwards decline in SST across the
basin, instead exhibiting an upturn in mean SST from
100--120\degree{}W until the eastern edge of the basin.  These higher
temperatures near the eastern basin boundary have been observed in
previous inter-model comparisons of tropical Pacific SST variability
\citep{mechoso-enso,latif-ensip,achutarao-enso-cmip} and have been
ascribed to difficulties in modelling marine stratus clouds in this
region, the steep orography near the South American coast and the
narrow coastal upwelling zone in the eastern Pacific.  Relatively
little progress appears to have been made in correcting this
deficiency in current coupled GCMs.

Figure~\ref{fig:basic-sst}b shows the annual standard deviation of SST
across the Pacific in the same latitude band.  Here, observations show
low variability in the western Pacific and higher variability in the
east, where conditions vacillate between the normal cold tongue state
and \eln conditions, characterised by the incursion of warmer water
from the western Pacific.  Some models represent this pattern
reasonably well, although the gradient in variability is represented
less well than the gradient in mean SST.  Again there are problems for
all of the models at the far eastern end of the Pacific basin,
probably for the same reasons as for the mean SST.  The range of
modelled SST variability is wide, with one model (FGOALS-g1.0) showing
variability as much as 2.5 times the observed values.  Some models
(CGCM3.1(T47), CGCM3.1(T63), MIROC3.2(hires) and MIROC3.2(medres))
simulate essentially no variability gradient across the basin.

%
%
%

The SST variability data of Figure~\ref{fig:basic-sst}b can be
summarised using the NINO3 SST index, the mean SST anomaly across the
region 150\degree{}W--90\degree{}W, 5\degree{}S--5\degree{}N.  High
values of this index reflect \eln conditions and low values \lan
conditions.  The fifth column of Table~\ref{tab:models} shows the
standard deviation of NINO3 SST anomalies for each of the models.  For
comparison, the standard deviation of NINO3 SST anomalies for the
ERSST v2 observational data set is 1.26\degree C for the period
1900--1999.  The results in Table~\ref{tab:models} indicate that most
of the models have a reasonable level of NINO3 SST variability, with
CGCM3.1(T47), CGCM3.1(T63) and UKMO-HadGEM1 having too little and
CNRM-CM3 and FGOALS-g1.0 too much.  (As noted in
Section~\ref{sec:cmip3-models}, a few models in the CMIP3 model
ensemble were not used in this study because of unrealistically low
NINO3 SST variability.  Only models with a NINO3 SST anomaly standard
deviation of 0.5\degree C or greater are considered here.)  There is
no obvious link between the degree of cold bias in the mean
climatology (Figure~\ref{fig:basic-sst}a) and the strength of SST
variability, measured either from Figure~\ref{fig:basic-sst}b or the
NINO3 SST index variability.  For instance, one of the models with the
greatest NINO3 SST variability, FGOALS-g1.0, has relatively little
cold bias, while another, CNRM-CM3, is among the models with the
greatest cold bias.  It is perhaps not so surprising that there should
be no clear relationship, since, as we will see below, several
characteristics of ENSO variability differ from model to model,
leading us to suspect that the mechanisms at work in each case may be
somewhat different.

\begin{figure}
  \begin{center}
    \includegraphics[width=0.6\textwidth]{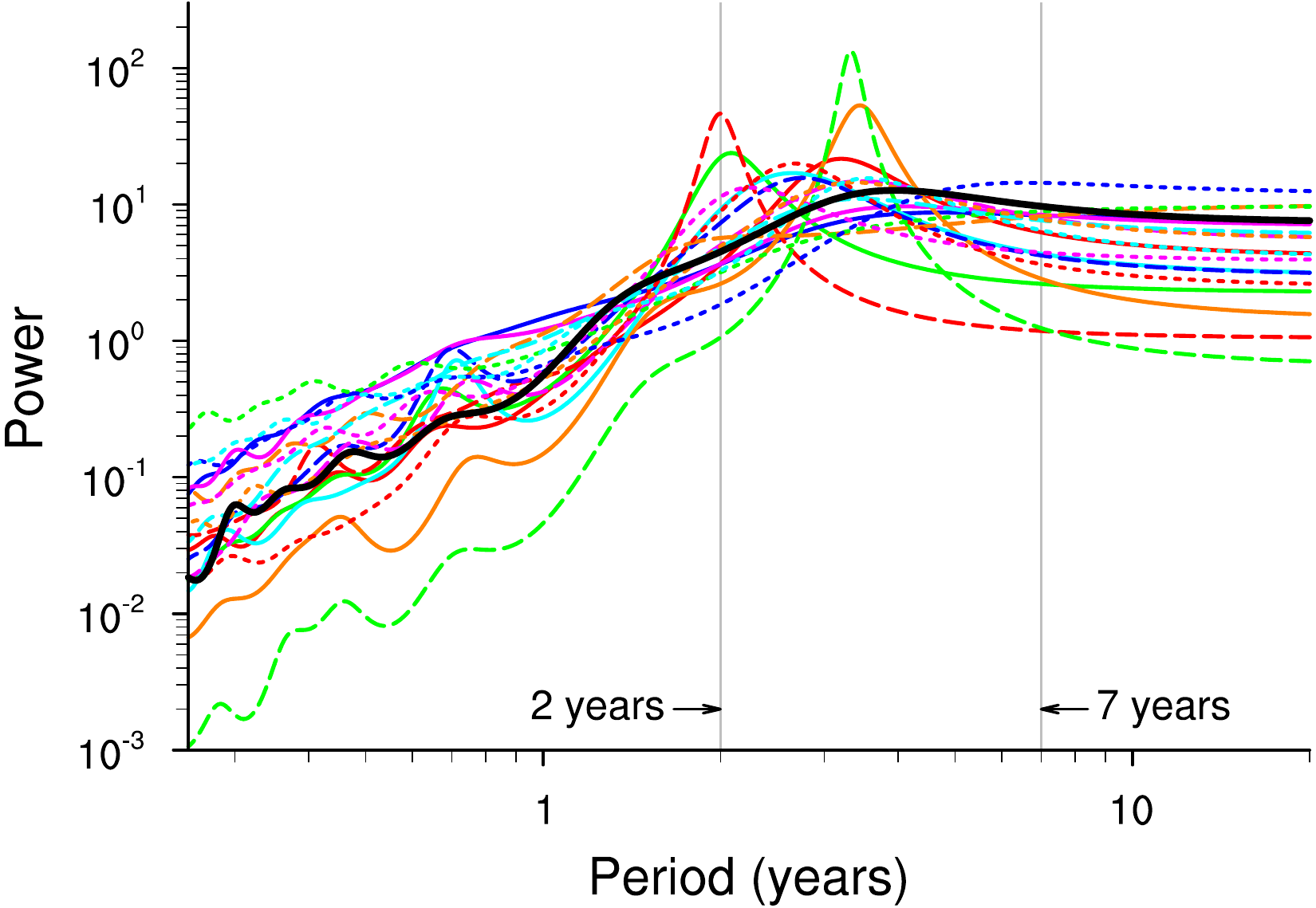}
  \end{center}
  \caption[NINO3 SST index power spectra]{Maximum entropy power
    spectra of NINO3 SST index variability from observations (thick
    black line) and models (coloured lines: see Table~\ref{tab:models}
    for key).  All spectra are calculated using 20 poles.}
  \label{fig:nino3-spectra}
\end{figure}

The temporal aspect of ENSO variability can be examined using power
spectra of the NINO3 SST anomaly time series.
Figure~\ref{fig:nino3-spectra} shows such spectra calculated using a
maximum entropy method \citep[Section~13.7]{press-nr}.  The
observations show a broad and low peak for periods between about 2 and
7 years, indicating the temporal irregularity of ENSO.  Among the
models, this pattern is replicated most closely in the GFDL-CM2.1,
INM-CM3.0 and UKMO-HadCM3 simulations.  Other models show either
weaker variability in the ENSO frequency band, or variability that is
too strongly peaked around a single frequency.  This latter feature is
particularly evident for CCSM3, CNRM-CM3, ECHO-G and FGOALS-g1.0.  For
the more extreme of these models, one might question whether these
narrowband signals can really be identified with ENSO, since they lack
the characteristic broad power spectrum of observed ENSO variability.
Although the basic feedback mechanisms permitting ENSO oscillations
may be represented in these models, the factors leading to the
observed irregular behaviour of ENSO are clearly missing or
under-represented.  For CCSM3 and FGOALS-g1.0, the excessive
regularity of ENSO-like variability may be due to the relatively
narrow meridional extent of both the atmospheric and oceanic response
in the eastern Pacific (e.g., the first SST EOFs for these models in
Figures~\ref{fig:sst-eofs}d~and~m below).  A hypothesis of
\citet{kirtman-1997} suggests that the period of ENSO variability is
modulated by the meridional extent of anomalous zonal wind stress in
the central and eastern Pacific.  This is because the meridional
extent of the anomalous zonal wind stress controls the spectrum of
Rossby waves excited by the anomalous winds.  These Rossby waves
propagate westwards, are reflected from the western boundary and
propagate back into the central and eastern Pacific as Kelvin waves,
where they act to modify the zonal thermocline gradient and turn the
ENSO oscillation over into its next phase.  A more equatorially
confined zonal wind stress excites only faster-moving lower Rossby
wave modes, while a more extensive region of anomalous wind stress
excites a wider range of Rossby wave modes, including off-equatorial
modes that propagate more slowly.  The collective action of these
different, slower moving wave modes acts to produce a slower (and
presumably less regular) turnover of ENSO into its next phase than the
small number of low meridional wavenumber Rossby wave modes excited by
a narrower zonal wind stress.  For CCSM3 at least, some evidence in
this direction is provided by the study of \citet{deser-ccsm3-enso},
which compares the ENSO behaviour of CCSM3 to observations in some
detail and illustrates the small meridional extent of most aspects of
both atmospheric and oceanic variations in CCSM3.

%
%
%

\subsection{Principal component analysis of SST data}
\label{sec:sst-pca}

As described in Chapter~\ref{ch:nldr-overview}, the most commonly used
dimensionality reduction technique of all is principal component
analysis (PCA).  This is widely used in the analysis of climate data,
where it is generally referred to as empirical orthogonal function
(EOF) analysis \citep{vonstorch-zwiers}.  The relationship between
this linear method and the nonlinear NLPCA and Isomap methods will be
explored in
Sections~\ref{sec:pca-vs-nlpca}~and~\ref{sec:isomap-algorithm}, but
here I present PCA results for the SST data sets.  In climate data
analysis applications of PCA, we generally have a time series of $N$
geographical maps of some climatological variable, each with $m$
spatial points.  We use the values from each map to construct data
vectors $\vec{x}_i \in \mathbb{R}^m$, with $i = 1, \dots, N$.  This
setup is very flexible, in that we can easily discard missing values
from our input data when we construct the $\vec{x}_i$ (e.g., for sea
surface temperature maps, we discard all land points).  I calculated
area-weighted EOFs and principal component time series for SST
anomalies from all data sets across the region
125\degree{}W--65\degree{}W, 20\degree{}S--20\degree{}N.  The
latitudinal range is selected here to restrict attention to regions of
the Pacific where seasonal variability is relatively weak, as
determined by the analysis of thermocline depth reported in
Section~\ref{sec:thermocline-calc-methods} below.  Grid box area
weighting is used to account for the variation in grid box size with
latitude in uniform latitude/longitude grids.  This is not a
particularly important effect for the equatorial data considered here,
but is essential for analysis in the higher latitudes.  For the
observed ERSST v2 SSTs, data for the period 1900--1999 was used, while
for the models, all of the available output was used, with simulation
lengths as listed in Table~\ref{tab:models}.  In each case, after
computation, the overall sign ambiguity of each EOF is removed by
requiring each EOF to have its maximum excursion from zero be
positive.  Each EOF is normalised to have unit maximum amplitude for
ease of plotting; the corresponding principal component time series
are rescaled accordingly.  Although the choice of normalisation used
here may appear arbitrary, in the case of equatorial Pacific SST
variability, it turns out to be quite convenient.  All of the models
capture the same leading pattern of ENSO-related SST variability in
the equatorial Pacific as appears in the observational data, so that
this normalisation choice results in EOFs for all of the models with
large-scale patterns that match those of the observations.

\begin{figure}
  \begin{center}
    \includegraphics[width=\textwidth]{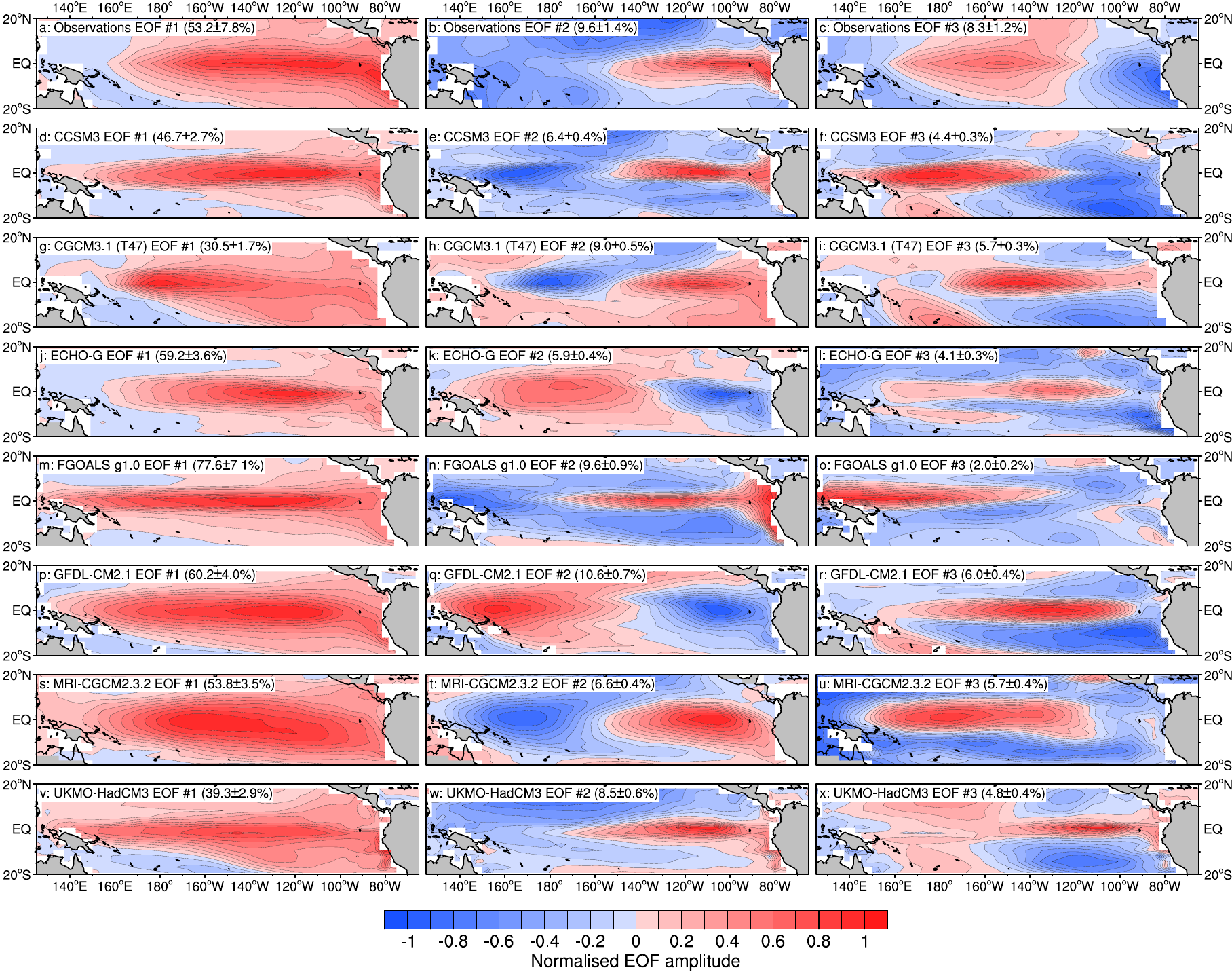}
  \end{center}
  \caption[Equatorial Pacific SST EOFs]{Sea surface temperature EOFs
    for the ERSST v2 observational data set (a--c), CCSM3 (d--f),
    CGCM3.1 (T47) (g--i), ECHO-G (j--l), FGOALS-g1.0 (m--o),
    GFDL-CM2.1 (p--q), MRI-CGCM2.3.2 (r--u) and UKMO-HadCM3 (v--x).
    Each EOF is normalised to have unit maximum amplitude.  Explained
    variance for each EOF is shown in parentheses, with 95\%
    confidence intervals calculated using North's ``rule of thumb''
    \citep{vonstorch-zwiers}.}
  \label{fig:sst-eofs}
\end{figure}

For convenience, we restrict our attention to the first three SST EOFs
which, for most data sets, capture the bulk of the data variance.  The
first three EOFs from the observations are shown in
Figures~\ref{fig:sst-eofs}a--c.  The first EOF
(Figure~\ref{fig:sst-eofs}a) shows an SST pattern similar to that of a
fully developed \eln event, with higher temperatures stretching across
the equatorial Pacific, replacing the normal tongue of cooler water in
the eastern Pacific.  This first EOF explains 53.2\% of the total SST
variance.  The second EOF (Figure~\ref{fig:sst-eofs}b) has a positive
centre of action on the equator near the west coast of South America,
reaching west as far as 150\degree{}W, with some indication of a
balancing negative centre of action near 140\degree{}W, 20\degree{}N,
and explains 9.6\% of the total variance, while the third EOF
(Figure~\ref{fig:sst-eofs}c) explains 8.3\% of the variance and has an
east-west dipole lying along the equator with centres of action around
160\degree{}W and near the coast of South America.

%
%
%

These patterns of observed spatial variability can be compared to
results from the model simulations.  Selected results are shown in
Figures~\ref{fig:sst-eofs}d--x.  The patterns seen represent a
cross-section of behaviour seen in the models.  In each case, the
first EOF is of approximately the right shape, but stretches too far
west across the Pacific.  In the observed data, the region of greatest
weight in the first EOF lies well to the east of the date line, while
in the model results it extends westwards to 150\degree{}E or further.
Also, few of the models display a pattern with a reasonable shape in
the far eastern sector of the Pacific.  GFDL-CM2.1 does a good job,
but other models have a pattern either not properly connected to South
America (CCSM3, ECHO-G and UKMO-HadCM3), or with too much spread of
the EOF pattern near the western coast of South and Central America
(CGCM3.1 (T47), FGOALS-g1.0 and MRI-CGCM2.3.2).  The proportion of the
total SST variance explained by the first EOF differs widely between
the models.  CCSM3 (explained variance 46.7\%), ECHO-G (explained
variance 59.2\%), GFDL-CM2.1 (explained variance 60.2\%) and
MRI-CGCM2.3.2 (explained variance 53.8\%) are closest to the range
seen in the observational data, while the other models lie outside the
observed range, reflecting the unrealistically high (FGOALS-g1.0:
77.7\%) and low (CGCM3.1 (T47): 30.5\%, UKMO-HadCM3: 39.3\%) ENSO
variability seen in the NINO3 SST index in these models
(Table~\ref{tab:models}, column 5).  The second and third EOFs from
the model simulations present a less clear picture.  Their spatial
patterns are variable: CCSM3, FGOALS-g1.0 and UKMO-HadCM3 have a
second EOF bearing some resemblance to that of the observational data,
with a northwest-southeast dipole centred at about 145\degree{}W,
5\degree{}N, while the second EOF pattern seen in CGCM3.1 (T47),
ECHO-G, GFDL-CM2.1 and MRI-CGCM2.3.2 has a distinct equatorial dipole
pattern, more like the third EOF of the observational data than the
second.  There is great variability in the pattern of the second and
third EOFs in the other models (not shown).

How much importance should be attached to differences in EOFs 2 and 3
is not clear.  For the observational data, EOFs 2 and 3 are degenerate
within the established confidence intervals for the explained variance
(EOF 2: $\mathrm{9.6 \pm 1.4\%}$, EOF 3: $\mathrm{8.3 \pm 1.2\%}$), so
that any direction in the linear subspace spanned by the EOFs,
i.e. any linear combination of the spatial patterns of EOFs 2 and 3,
is as good as any other for the purposes of capturing variance in the
data.  For the models, the explained variance confidence intervals are
smaller (because the SST time series are longer), so there is
apparently no degeneracy, but it is still difficult to evaluate the
higher model EOFs in comparison to the observations.  It should be
noted that some care is required in interpreting the confidence
intervals provided for the PCA eigenvalues here.  These are calculated
from a commonly used asymptotic ``rule of thumb''
\citep[Section~13.3.5]{vonstorch-zwiers} based on an equivalent sample
size that aims to take account of serial correlations in the input
data \citep{zwiers-serial-corr}.  Because of residual temporal
correlations and spatial dependencies in the data, this approach is
not necessarily applicable to the data sets here.  Confidence
intervals are provided here purely as a guide, and cannot be trusted
to give completely accurate information about degeneracy of PCA
eigenvectors.

One further comment should be made about comparisons between the
observational EOFs shown in Figures~\ref{fig:sst-eofs}a--c and the
model results in Figures~\ref{fig:sst-eofs}d--x.  The observational
data that I use here spans the period 1900--1999, a time during which
there is a major shift in the behaviour of ENSO: \eln events before
about 1976 are generally weaker and shorter than events after that
time, a change attributed either to variations in the Pacific Decadal
Oscillation or to intrinsic variability in ENSO \citep{mcphaden-enso}.
One result of this shift in behaviour is that some of the modes of
variability extracted by the PCA analysis performed here may be
related to nonstationarity in the input data time series rather than
to statistically stationary ENSO variability (recall that the ERSST v2
NINO3 SST index wavelet spectrum plot in
Figure~\ref{fig:ersst-wavelets} on page~\pageref{fig:ersst-wavelets}
shows some nonstationarity even for the period 1900--1999 selected to
avoid the early observation-sparse part of the data set).  This means
that some care must be exercised in making statements on the basis of
comparison between these observational EOFs and EOFs derived from the
model integrations (taken from control simulations that are nominally
in an equilibrium state).  Whether the CMIP3 model simulations display
regime shifts like the observed mid-1970s change in ENSO variability
is a question that deserves further analysis --- if decadal
variability in ENSO can be seen in a GCM simulation with constant
external forcing, it may lend some weight to the idea that the
observed decadal variability in ENSO is an intrinsic internal mode of
variability, rather than something forced from outside by, for
instance, a decadal climate mode outside the equatorial Pacific, or
anthropogenic global warming.

\begin{figure}
  \begin{center}
    \includegraphics[width=\textwidth]{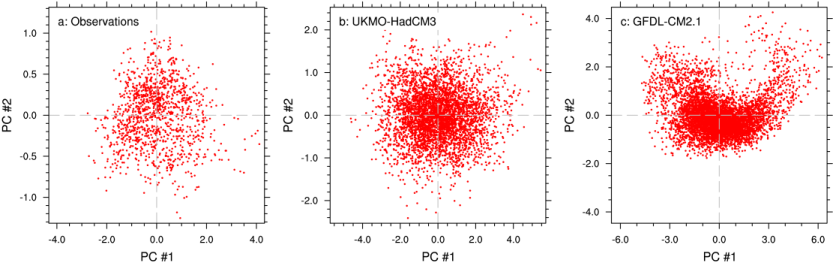}
  \end{center}
  \caption[Equatorial Pacific SST principal component scatter
    plots]{Scatter plots of SST PC \#1 versus PC \#2 for ERSST v2
    observations (a), UKMO-HadCM3 (b) and GFDL-CM2.1 (c).}
  \label{fig:sst-pc-scatter}
\end{figure}

In principal component analysis, the EOFs represent the spatial
patterns of different modes of variability (for real-valued EOFs,
actually standing oscillations), while temporal variability is
captured in the principal component (PC) time series.  Each PC time
series gives the projection of the input data time series onto its
corresponding EOF, and because of the orthogonality of the EOFs, the
PC time series are linearly uncorrelated by construction.  Despite the
lack of linear correlation, there are clear nonlinear relationships
between the PC time series in the Pacific SST data sets examined here.
This can be seen in Figure~\ref{fig:sst-pc-scatter}, which shows
selected scatter plots of PC time series values.
Figure~\ref{fig:sst-pc-scatter}a shows PC \#1 plotted versus PC \#2
for the observational ERSST v2 data set.  Although the two PC time
series are not linearly correlated, the asymmetry in the PC scatter
plot indicates that they may not be truly independent, and that there
may be a nonlinear relationship between the values of PC \#1 and PC
\#2, with large positive and negative values of PC \#1 being
associated with larger negative values of PC \#2.  This nonlinearity
is due to the asymmetry between \eln and \lan events.  On average,
warm anomalies along the equator east of 150\degree{}W during \eln
events are of greater magnitude than cold anomalies during \lan
events.  It is difficult to ascribe this asymmetry to a particular EOF
in this case because of the near degeneracy of EOFs 2 and 3.  This
asymmetric relationship has previously been discussed in the context
of applying nonlinear PCA to Pacific SST data \citep{monahan-enso}.
Similar, and in some cases, even stronger, nonlinear relationships are
seen between the PC time series for model SSTs.
Figure~\ref{fig:sst-pc-scatter}b shows a scatter plot of PC \#1 versus
PC \#2 from the UKMO-HadCM3 model.  Here, there is a similar
asymmetric pattern to that seen in the observations.  Again, it is
difficult to ascribe this to any specific physical mechanism in the
model, but whatever the origin of the relationship, the scatter plot
is not the Gaussian cloud expected for PC time series derived from a
simple linear process.  Similar comments can be made about the more
extreme nonlinearity displayed in Figure~\ref{fig:sst-pc-scatter}c, a
scatter plot of PC \#1 versus PC \#2 for GFDL-CM2.1.  This is
particularly striking because GFDL-CM2.1 is among the CMIP3 models
assessed as having the most realistic ENSO variability by
\citet{vanoldenborgh-enso}.  Here, the greater asymmetry in the PC
scatter plot may be partially due to the wide meridional spread of the
first SST EOF (Figure~\ref{fig:sst-eofs}p) and the very distinct zonal
dipole pattern in the second SST EOF (Figure~\ref{fig:sst-eofs}q).
Similarly nonlinear PC \#1/PC \#2 scatter plots are seen for other
models with similar structures in their first two EOFs (GFDL-CM2.0 and
ECHO-G and, to a lesser extent, MRI-CGCM2.3.2).  Any mechanistic
explanation of this nonlinearity would require a more detailed
analysis of the different ocean-atmosphere feedbacks in the GFDL-CM2.1
model, along the lines of \citet{vanoldenborgh-enso}.  Overall, these
results indicate that PCA may not be the most appropriate tool to use
here, because of these strong nonlinear relationships between the
different PC time series.

%
%
%

\subsection{Asymmetry of SST variability}
\label{sec:sst-asymmetry}

It is of some interest to develop quantitative measures of the spatial
asymmetry between \eln and \lan events
(Figures~\ref{fig:eq-pacific-el-nino}~and~\ref{fig:eq-pacific-la-nina})
in observational and model data.  The nonlinear PCA method described
in Chapter~\ref{ch:nlpca} extracts a measure of asymmetry in a data
set without additional prompting, but there is also some utility in
simpler measures based on the construction of SST composites.  Such a
measure was presented by \citet{monahan-dai}.

Consider a spatial field $\vec{x}(t_i)$ defined at a discrete set of
times $t_i$ with $i = 1, \dots, N$, composited using a time series
$\lambda(t_i)$ in a manner to be defined below.  (For definiteness, in
what follows, $\vec{x}$ will be the Pacific SST anomaly field and
$\lambda$ will be the first PC time series of the SST anomalies,
normalised by its standard deviation.)  Define two subsets of time as
\begin{equation}
  t^{(+)} = \{ t_i \, | \, \lambda(t_i) > c \}, \qquad
  t^{(-)} = \{ t_i \, | \, \lambda(t_i) < -c \},
\end{equation}
for a threshold value of the compositing time series, $c$: we will use
one standard deviation of $\lambda$ as the threshold.  Positive and
negative components of $\vec{x}(t)$, $\vec{x}^{(+)}$ and
$\vec{x}^{(-)}$ are defined as averages over the appropriate time
subsets:
\begin{equation}
  \vec{x}^{(+)} = \langle \vec{x} \rangle_+, \qquad
  \vec{x}^{(-)} = \langle \vec{x} \rangle_-,
\end{equation}
where the positive and negative time averaging operators $\langle
\bullet \rangle_+$ and $\langle \bullet \rangle_-$ are defined as
\begin{equation}
  \label{eq:asymm-means}
  \langle f \rangle_+ = \frac{1}{\# t^{(+)}} \sum_{t \in t^{(+)}} f(t), \qquad
  \langle f \rangle_- = \frac{1}{\# t^{(-)}} \sum_{t \in t^{(-)}} f(t).
\end{equation}
(As usual, we also denote averaging over the whole time series by
$\langle \bullet \rangle$.)

In our case, the positive and negative patterns $\vec{x}^{(+)}$ and
$\vec{x}^{(-)}$ represent \eln and \lan conditions respectively.  In
general, they differ by more than just a sign, since they are means
over different partitions of the time series $\vec{x}(t)$.  Given
these positive and negative patterns, we wish to determine their
components symmetric and asymmetric under a change in sign in
$\lambda$.  Here, a change in sign in $\lambda$ represents the
difference between a typical \eln state and a typical \lan state,
where ``typical'' means one standard deviation of PC \#1.

To do this, we construct an approximation to our time series based on
a nonlinear model sensitive to the type of asymmetries represented by
the difference between the $\vec{x}^{(+)}$ and $\vec{x}^{(-)}$
patterns.  Consider first a linear model, where we expand the time
series, $\vec{x}(t)$, in terms of EOFs, as in
\eqref{eq:pca-expansion}, for simplicity truncating after two EOFs, as
$\vec{x}(t) = \vec{q}_1 \lambda_1(t) + \vec{q}_2 \lambda_2(t) +
\vec{\varepsilon}(t)$.  Here, $\vec{q}_1$ and $\vec{q}_2$ are the
first and second EOFs, $\lambda_1(t)$ and $\lambda_2(t)$ are the first
and second PC time series, and $\vec{\varepsilon}(t)$ is a residual
error term.  This linear approximation, using two expansion time
series, is optimised to explain as much of the data variance as
possible.  Instead of this EOF expansion, consider an alternative,
nonlinear, model to represent variability in $\vec{x}(t)$.  The
simplest non-linear model, based on a second order Taylor series, is
\begin{equation}
\label{eq:asymm-1}
  \vec{x}(t) = \vec{a}_0 + \vec{a}_A \lambda(t) + \vec{a}_S
      [\lambda(t)]^2 + \vec{\varepsilon}(t),
\end{equation}
where, as before, $\vec{\varepsilon}(t)$ is an error residual (with
zero time average), $\lambda(t)$ is an expansion time series, and now
the $\vec{a}_\bullet$ represent spatial patterns in $\vec{x}(t)$:
$\vec{a}_A$ is a pattern that reverses sign under a change in sign in
$\lambda$ (the \emph{anti-symmetric component}) and $\vec{a}_S$ a
pattern that retains the same sign under a change in sign in $\lambda$
(the \emph{symmetric component}).

Assuming, without loss of generality, that the time averages of both
$\vec{x}(t)$ and $\lambda(t)$ vanish, we see that $\vec{a}_0 +
\vec{a}_S \langle \lambda^2 \rangle = 0$, meaning that
\eqref{eq:asymm-1} may be rewritten as
\begin{equation}
  \vec{x}(t) = \vec{a}_A \lambda(t) + \vec{a}_S ([\lambda(t)]^2 -
  \langle \lambda^2 \rangle) + \vec{\varepsilon}(t).
\end{equation}
Now, assuming that $\langle \vec{\varepsilon} \rangle_+ = \langle
\vec{\varepsilon} \rangle_- = 0$, we have from \eqref{eq:asymm-means}
that
\begin{equation}
  \vec{x}^{(+)} = \vec{a}_A \langle \lambda \rangle_+ + \vec{a}_S
  (\langle \lambda^2 \rangle_+ - \langle \lambda^2 \rangle), \qquad
  \vec{x}^{(-)} = \vec{a}_A \langle \lambda \rangle_- + \vec{a}_S
  (\langle \lambda^2 \rangle_- - \langle \lambda^2 \rangle).
\end{equation}
This is a linear system that can be solved to give
\begin{equation}
\begin{aligned}
  \vec{a}_A &= \frac{1}{\Delta} [ (\langle \lambda^2 \rangle_- -
    \langle \lambda^2 \rangle) \vec{x}^{(+)} - (\langle \lambda^2
    \rangle_+ - \langle \lambda^2 \rangle) \vec{x}^{(-)} ], \\
  \vec{a}_S &= \frac{1}{\Delta} [ \langle \lambda \rangle_+
    \vec{x}^{(-)} - \langle\lambda \rangle_- \vec{x}^{(+)} ],
\end{aligned}
\end{equation}
where
\begin{equation}
  \Delta = \langle \lambda \rangle_+ (\langle \lambda^2 \rangle_- -
  \langle \lambda^2 \rangle) - \langle \lambda \rangle_- (\langle
  \lambda^2 \rangle_+ - \langle \lambda^2 \rangle).
\end{equation}
These relations can be used to calculate the $\vec{a}_A$ and
$\vec{a}_S$ patterns from $\vec{x}(t)$ and $\lambda(t)$.  The ratio of
norms of the symmetric and asymmetric patterns, $||\vec{a}_S|| /
||\vec{a}_A||$, provides a measure of asymmetry between positive and
negative excursions of $\vec{x}(t)$.  As well as this norm ratio, one
can also calculate pattern correlations between the asymmetric
component $\vec{a}_A$ and the first SST EOF, and between the symmetric
component $\vec{a}_S$ and the second SST EOF.  The correlation between
$\vec{a}_A$ and EOF \#1 should be unity by construction.  The pattern
correlation between $\vec{a}_S$ and EOF \#2 is not constrained by the
definition of $\vec{a}_S$, and gives a measure of how much of the
asymmetry in the ENSO response is captured by a linear PCA
decomposition.

%
%
\begin{table}
  \begin{center}
    \input{figs/05/asymmetry-table.tex}
  \end{center}
  \caption[\eln/\lan asymmetry measures]{Measures of SST anomaly
    pattern asymmetry for observational and model data: pattern
    correlation coefficients between antisymmetric composite
    $\vec{a}_A$ and first EOF $\vec{q}_1$ and between symmetric
    composite $\vec{a}_S$ and second EOF $\vec{q}_2$, and ratio of
    norms of symmetric and asymmetric SST composites.}
  \label{tab:asymmetry-measures}
\end{table}
%

Table~\ref{tab:asymmetry-measures} shows results of this analysis for
observational SST data and each of the models.  The model asymmetry
results are similar to those of \citet{monahan-dai}: as measured by
this method, relatively few models have as much \eln/\lan asymmetry as
the observational data.  Most models (all but six) have an
$||\vec{a}_S|| / ||\vec{a}_A||$ norm ratio rather smaller than the
observations, while five models have norm ratios comparable to the
observations.  (One model, MIROC3.2(hires), has a large norm ratio
that is almost certainly spurious, as indicated by the small pattern
correlation between the first SST EOF and the antisymmetric composite
$\vec{a}_A$.)  The range of $\Corr(\vec{q}_2, \vec{a}_S)$ is also
similar to that seen by \citet{monahan-dai}.  It appears, at least by
this measure, that relatively few models have a particularly strong
\eln/\lan asymmetry.  This result will be explored further in
Chapter~\ref{ch:nlpca}.

\section{Equatorial Pacific thermocline depth}
\label{sec:thermocline-var-results}

\subsection{Comparison of thermocline calculation methods}
\label{sec:thermocline-calc-methods}

Section~\ref{sec:thermocline-calc-desc} described the two methods used
here for calculating thermocline depth, one, denoted \ztw, based on
the 20\degree C isotherm of ocean temperature, and one, denoted \zgr,
based on the depth of the maximum vertical ocean temperature gradient.
The former, although rather arbitrary and less dynamically motivated
than \zgr, provides a good measure for the quantity of warm water in
the upper ocean.  There are significant differences between the
results of these calculations in some regions of the Pacific, although
most differences lie outside the region used for the calculation of
the equatorial warm water volume (WWV), believed to be the important
factor as far as ENSO variability is concerned.  Regions with
particularly large \ztw/\zgr differences include the Caribbean and
parts of the western Pacific off the coast of China, where there are
deep regions with temperature decreasing more or less monotonically
with depth and no obvious gradient-based thermocline, and regions of
the central Pacific south of about 10\degree{}S and north of about
10\degree{}N, where gradient-based thermocline values are consistently
rather deeper than those derived from the 20\degree C isotherm.  This
is true both for the model results and for observational thermocline
depths derived from NCEP GODAS potential temperature data.  One or two
models have clear problems in their ocean temperature fields that make
the thermocline depth computations difficult, most notably
UKMO-HadCM3, which displays numerical artefacts in its ocean
temperature, with large temperature oscillations apparent from
grid-cell to grid-cell along the equator throughout the IPCC
pre-industrial control simulation.  The same problem appears to affect
UKMO-HadGEM1.  These artefacts largely invalidate the \zgr thermocline
depth calculated for these models.

\begin{figure}
  \begin{center}
    \includegraphics[width=0.8\textwidth]{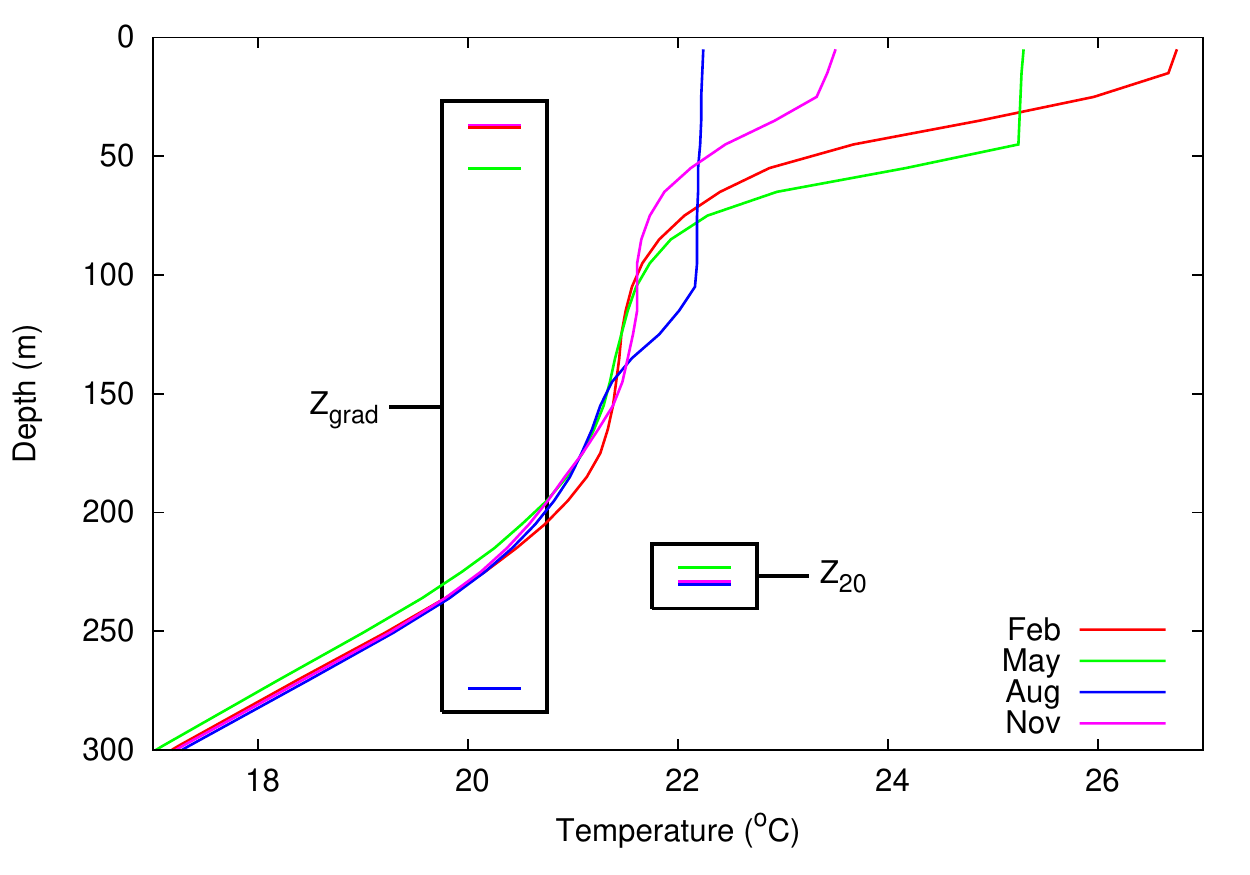}
  \end{center}
  \caption[Seasonal variability of \zgr thermocline
    depths]{Temperature profiles and \ztw and \zgr thermocline depths
    for a single point from the GFDL-CM2.1 model for four months, one
    from each season of a single year.}
  \label{fig:gfdl-profile}
\end{figure}

The \ztw/\zgr differences seen in the southern central Pacific are due
to seasonal variations in temperature structure in the upper mixed
layer of the ocean.  In summer, the ocean surface heats up, producing
a large vertical temperature gradient in the upper 50--100m, which is
held at a temperature rather greater than 20\degree C.  Under these
conditions, the 20\degree C isotherm is relatively deep (around 200m)
while the depth of maximum temperature gradient is closer to the
surface (around 50m).  In winter, the surface layer cools, to give a
more isothermal upper ocean.  The 20\degree C isotherm then lies at
about the same depth as in summer, while the depth of maximum
temperature gradient is much deeper (about 250m).  The cause of the
\ztw/\zgr differences is thus the fast fluctuation of temperatures in
the upper layer of the ocean, in response to heating in the local
spring and summer and cooling in autumn and winter.
Figure~\ref{fig:gfdl-profile} illustrates this effect, showing
temperature profiles and thermocline depths in different seasons for a
single point (165\degree{}E, 20\degree{}S) for the GFDL-CM2.1 model.
This phenomenon also appears to be responsible for the shallower \zgr
thermocline depths seen in many models in the north central Pacific.

%
%
%

Since most \ztw/\zgr differences are confined to regions with
significant seasonal variation in the upper ocean temperature
structure, equatorial waters are less susceptible to this variation,
and, particularly in the narrow equatorial region used for calculation
of warm water volume, there is less difference between the \ztw and
\zgr thermocline depths.

Within the region where WWV is calculated, differences between \ztw
and \zgr are relatively small, as summarised in
Table~\ref{tab:relevant-diffs}.  Compared to the mean and standard
deviation of the differences in this region, the maximum values are
clear outliers, and the overall impact of the differences on the WWV
calculations is expected to be small.  Comparable localisation of
\ztw/\zgr differences is also seen in thermocline depth anomalies,
with differences within the WWV region rather smaller than for the raw
thermocline data, as would be expected if most of the differences
arise from regions of stronger seasonal variability.  These
consistently smaller differences are clear in
Table~\ref{tab:relevant-diffs}.

%
%
\begin{table}
  \begin{center}
    \begin{tabular}{lccc!{\quad\,}ccc}
      \toprule

      \multirow{2}{2cm}{\bf Model} & \multicolumn{3}{c}{\bf Raw
        thermocline} & \multicolumn{3}{c}{\bf Anomalies} \\
      & {\bf Max.} & {$\bm{\langle z \rangle}$} & {$\bm{\sigma}$} &
      {\bf Max.} & {$\bm{\langle z \rangle}$} & {$\bm{\sigma}$} \\

      \midrule

      Observations      & 183.20 & 15.52 & 19.26 & 160.19 & 11.90 & 14.19 \\
      BCCR-BCM2.0       & 260.46 & 16.94 & 22.74 & 231.59 & 12.80 & 17.90 \\
      CCSM3             & 174.87 & 12.39 & 15.87 & 157.47 &  7.14 &  9.15 \\
      CGCM3.1(T47)      & 148.43 & 36.79 & 34.50 & 176.74 &  9.72 & 12.72 \\
      CGCM3.1(T63)      & 132.39 & 33.24 & 33.05 & 167.59 & 10.34 & 14.26 \\
      CNRM-CM3          & 311.13 & 16.30 & 19.18 & 296.81 & 13.10 & 14.14 \\
      CSIRO-Mk3.0       & 190.11 & 15.38 & 19.00 & 174.74 &  9.03 & 11.18 \\
      ECHO-G            & 184.56 & 10.47 & 11.98 & 176.49 &  8.15 &  9.88 \\
      FGOALS-g1.0       & 174.29 & 12.26 & 12.74 & 165.98 &  9.96 &  9.92 \\
      GFDL-CM2.0        & 192.33 & 22.51 & 27.39 & 189.90 &  9.57 & 11.92 \\
      GFDL-CM2.1        & 195.36 & 18.49 & 23.49 & 189.09 & 10.04 & 12.92 \\
      GISS-EH           & 192.12 & 12.85 & 15.57 & 164.55 &  9.84 & 14.40 \\
      INM-CM3.0         & 201.67 & 21.83 & 26.47 & 212.71 & 16.34 & 21.18 \\
      MIROC3.2(hires)   & 247.00 & 38.02 & 38.40 & 241.01 & 12.54 & 15.81 \\
      MIROC3.2(medres)  & 198.45 & 24.84 & 25.89 & 239.42 &  9.44 & 13.08 \\
      MRI-CGCM2.3.2     & 134.77 & 13.57 & 15.22 & 122.21 &  6.60 &  7.93 \\
      UKMO-HadCM3       & 251.15 & 53.13 & 63.35 & 256.61 & 51.04 & 37.25 \\
      UKMO-HadGEM1      & 252.28 & 34.81 & 54.70 & 240.35 & 37.52 & 40.27 \\

      \bottomrule
    \end{tabular}
  \end{center}
  \caption[\ztw/\zgr differences in WWV region]{Maximum, mean and
    standard deviation of differences between \ztw and \zgr
    thermocline depth and thermocline depth anomalies in the region
    used for the calculation of WWV (120\degree{}E--80\degree{}W,
    5\degree{}S--5\degree{}N), all in metres.  (Observations use NCEP
    GODAS potential temperature data.)}
  \label{tab:relevant-diffs}
\end{table}
%

The conclusion of this analysis is that we expect there to be
relatively little difference between results derived using \ztw and
\zgr, at least in the region used for calculating the equatorial
Pacific warm water volume.  This is the quantity (essentially
equivalent to equatorial ocean heat content) of importance in the
dynamics of ENSO variability.  Since there are problems with
calculating \zgr for some models because of numerical issues in the
model temperature fields, I will generally use \ztw data in what
follows.  One can also observe that \emph{warm} water volume is more
intuitively defined in terms of a temperature limit (i.e. water above
20\degree C) than in terms of water above the thermocline as
determined by some other method (i.e., \zgr).

One other aspect of the model \ztw thermocline values deserves some
comment.  As shown in Figure~\ref{fig:basic-sst}a, almost all of the
models have a significant cold bias in their SST climatology in the
equatorial Pacific compared to observations.  These lower temperatures
mean that there is also a consistent bias to shallower thermocline
depth as measured by \ztw.  One result of this bias is that the \ztw
thermocline shoals all the way to the surface more frequently in the
models than in observations.  This surface outcropping of the
thermocline does occur in the real Pacific ocean, for instance during
\lan events following strong \eln{}s (e.g. September 1999), but more
infrequently and with smaller spatial extent than in models with a
large cold bias.  This difference should be kept in mind when
comparing model thermocline and WWV calculations to observations.

\subsection{Principal component analysis of thermocline data}
\label{sec:thermocline-pca}

We can use PCA to explore spatial and temporal variability in
thermocline depth in the same way as was done for SSTs in
Section~\ref{sec:sst-pca}.  I calculated area-weighted EOFs and
principal component time series for \ztw thermocline depth anomalies
from all data sets across the region 125\degree{}W--65\degree{}W,
20\degree{}S--20\degree{}N.  For the observations from the NCEP GODAS
data set, the data analysed runs from January 1980 to March 2008,
while for the model results, all of the available data was used, with
simulations lengths as listed in Table~\ref{tab:models}.

\begin{figure}
  \begin{center}
    \includegraphics[width=\textwidth]{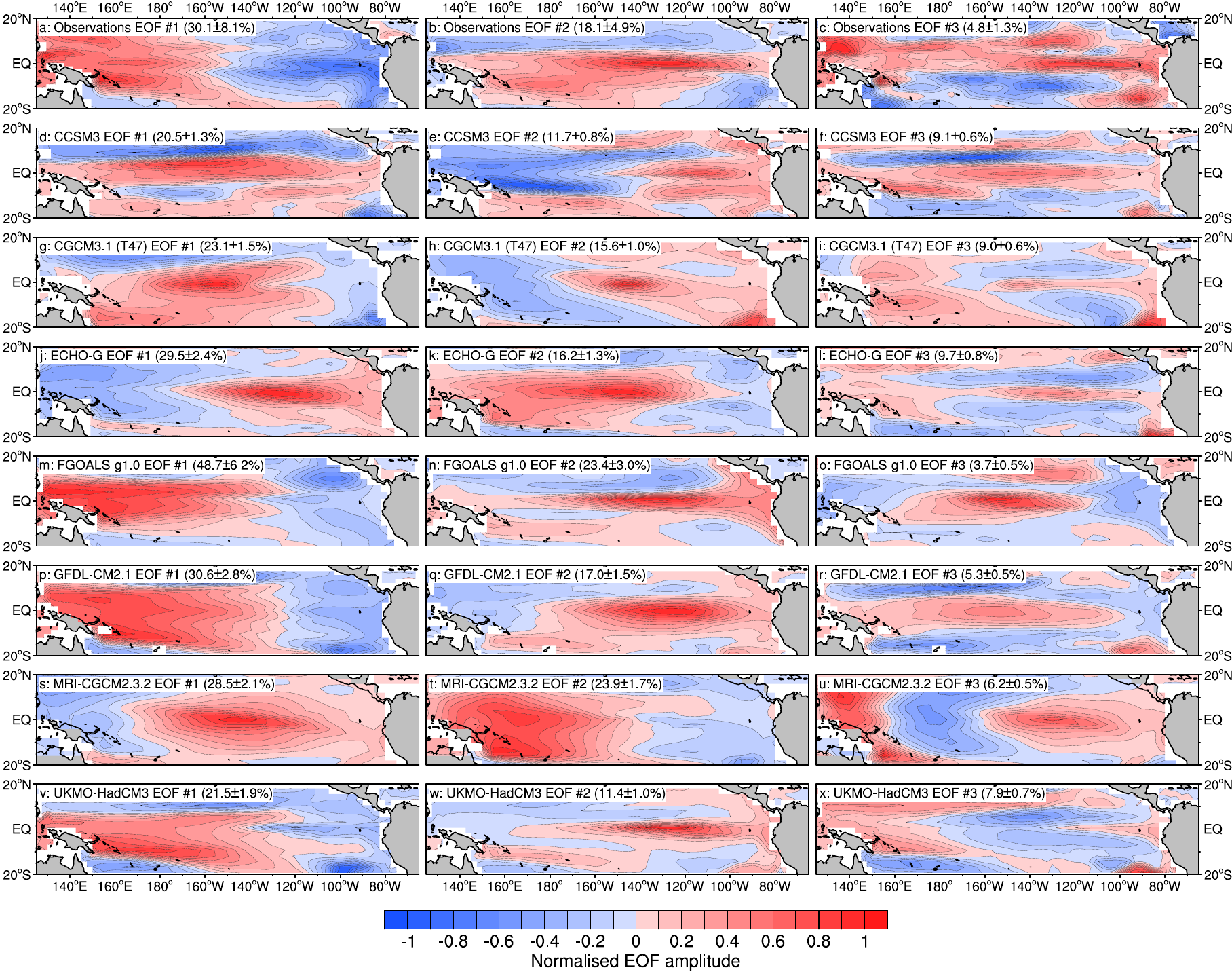}
  \end{center}
  \caption[Equatorial Pacific thermocline depth EOFs]{Thermocline
    depth (\ztw) EOFs for the NCEP GODAS observational data set
    (a--c), CCSM3 (d--f), CGCM3.1 (T47) (g--i), ECHO-G (j--l),
    FGOALS-g1.0 (m--o), GFDL-CM2.1 (p--q), MRI-CGCM2.3.2 (r--u) and
    UKMO-HadCM3 (v--x).  Each EOF is normalised to have unit maximum
    amplitude.  Explained variance for each EOF is shown in
    parentheses, with 95\% confidence intervals calculated using
    North's ``rule of thumb'' \citep{vonstorch-zwiers}.}
  \label{fig:z-eofs}
\end{figure}

The first three thermocline depth EOFs from the observational data are
shown in Figures~\ref{fig:z-eofs}a--c.  The first EOF
(Figure~\ref{fig:z-eofs}a) has a strong zonal dipole pattern related
to ``seesaw'' variations in thermocline depth that occur during \eln
and \lan events, with a deeper eastern thermocline and shallower
western thermocline during \eln, as warm water floods across the whole
of the Pacific basin, and a shallower eastern thermocline and a deeper
western thermocline during \lan, as strong easterly trade winds
reinforce the mean zonal tilt of the equatorial thermocline
(cf. Figure~\ref{fig:enso-cartoons}).  The first EOF explains $30.1
\pm 8.1\%$ of the total thermocline depth variance in the study region
--- the relatively wide confidence interval is due to the short time
series of ocean temperature observations on which the thermocline
depth is based (only 28 years).  The second EOF
(Figure~\ref{fig:z-eofs}b) has a more zonally symmetric pattern, with
a strong contrast between equatorial and off-equatorial regions and a
distinct centre of action on the equator between 160\degree{}W and
100\degree{}W.  This pattern explains $18.1 \pm 4.9 \%$ of the total
data variance and is related to changes in the zonal mean equatorial
thermocline depth, a measure of the equatorial warm water volume (or
equivalently, ocean heat content).  This fluctuates between \eln and
\lan conditions, according to the feedback mechanism proposed by
\citet{wyrtki-1985} (again, cf. Figure~\ref{fig:enso-cartoons}).  The
third EOF (Figure~\ref{fig:z-eofs}c) shows a slightly more confused
pattern, but again displays a degree of zonally symmetric
differentiation between equatorial and off-equatorial regions, so is
again probably related to variations in the zonal mean thermocline
depth and the second, warm water volume, degree of freedom associated
with ENSO variability.

\begin{figure}
  \begin{center}
    \includegraphics[width=\textwidth]{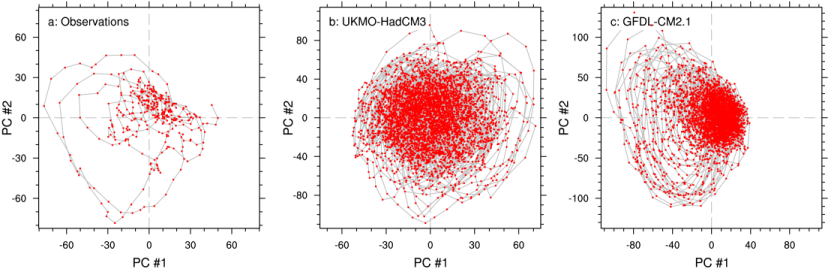}
  \end{center}
  \caption[Equatorial Pacific thermocline depth principal component
    scatter plots]{Scatter plots of \ztw PC \#1 versus PC \#2 for NCEP
    GODAS observations (a), UKMO-HadCM3 (b) and GFDL-CM2.1 (c).
    Points adjacent in time are connected by lines to highlight the
    phasing of PC \#2 variations relative to variations in PC \#1.}
  \label{fig:z-pc-scatter}
\end{figure}

The first and second principal components of observed thermocline
depth evolve in quadrature, reflecting the coherent operation of the
Bjerknes and Wyrtki feedbacks involved in sustaining ENSO
oscillations.  Figure~\ref{fig:z-pc-scatter}a, a scatter plot of \ztw
PC \#1 versus PC \#2 for the observational data, uses lines drawn
between points adjacent in time to clarify this coherent variation.
The large looping excursions towards the bottom left hand corner of
the plot are \eln events, and the quadrature relation between PC \#1
(zonal thermocline tilt) and PC \#2 (mean zonal thermocline depth,
i.e. the total quantity of warm water in the equatorial Pacific) is
shown by the ``circular'' shape of these excursions.  This plot should
be compared with the NINO3 SST index versus warm water volume phasing
plots shown later in Section~\ref{sec:nino3-wwv-phasing}.

%
%
%

%
%
%

PCA results for selected models are shown in
Figures~\ref{fig:z-eofs}d--x, with the first three EOFs for CCSM3
(Figures~\ref{fig:z-eofs}d--f), CGCM3.1 (T47)
(Figures~\ref{fig:z-eofs}g--i), ECHO-G (Figures~\ref{fig:z-eofs}j--l),
FGOALS-g1.0 (Figures~\ref{fig:z-eofs}m--o), GFDL-CM2.1
(Figures~\ref{fig:z-eofs}p--r), MRI-CGCM2.3.2
(Figures~\ref{fig:z-eofs}s--u) and UKMO-HadCM3
(Figures~\ref{fig:z-eofs}v--x).  These are the same models whose SST
EOFs are displayed in Figure~\ref{fig:sst-eofs}.  The patterns seen
are typical of results across the model ensemble.  Of the whole
ensemble, seven models do a reasonable job of replicating the patterns
of thermocline variability seen in observations.  These are GFDL-CM2.1
(best of all the models, Figures~\ref{fig:z-eofs}p--r), FGOALS-g1.0
(also very good, Figures~\ref{fig:z-eofs}m--o), CNRM-CM3, UKMO-HadCM3
(Figures~\ref{fig:z-eofs}v--x), and slightly less good, GFDL-CM2.0,
MIROC3.2 (medres) and ECHO-G (Figures~\ref{fig:z-eofs}j--l).  Six of
the models, CCSM3 (Figures~\ref{fig:z-eofs}d--f), CGCM3.1 (T47)
(Figures~\ref{fig:z-eofs}g--i), CGCM3.1 (T63), CSIRO-Mk3.0, GISS-EH
and MRI-CGCM2.3.2 (Figures~\ref{fig:z-eofs}s--u) show an interesting
phenomenon where the first and second \ztw EOFs show the same patterns
as the observations, but are swapped over.  The first EOF for these
models shows the zonally symmetric ``warm water volume'' mode of
variability while the second mode shows the zonally asymmetric
``thermocline tilt'' pattern.  The explained variance fraction
confidence intervals calculated for these models do not indicate any
degeneracy between the first two EOFs in any of these cases.  It thus
appears that these models do capture the basic mechanisms of
thermocline variability in the equatorial Pacific, but place the wrong
emphasis on zonally symmetric warm water discharge/recharge dynamics
compared to zonally asymmetric equatorial wave dynamics.  The other
models (BCCR-BCM2.0, INM-CM3.0, MIROC3.2 (hires), UKMO-HadGEM1) show
patterns of thermocline depth variability that do not correspond
particularly closely to the observed patterns.  One conclusion we can
draw from these results is that most of the models appear to have a
reasonable representation of at least some of the dynamical processes
affecting thermocline depth in the equatorial Pacific.

Thermocline depth PC scatter plots are shown in
Figure~\ref{fig:z-pc-scatter} for two models, UKMO-HadCM3
(Figure~\ref{fig:z-pc-scatter}b) and GFDL-CM2.1
(Figure~\ref{fig:z-pc-scatter}c).  There are two things to draw from
these images.  First, there is significant nonlinearity in both the
observations and the GFDL-CM2.1 results.  This indicates that,
although the PC time series are linearly uncorrelated, there is a
nonlinear relationship between the different principal components.  It
is not particularly surprising to see nonlinearities in the
thermocline depth PC time series, simply because the SST and
thermocline variations arise from the same dynamical system, so that
one would expect to see nonlinear behaviour in any variables used to
characterise the system.  The second aspect of note in
Figure~\ref{fig:z-pc-scatter} is highlighted by the lines connecting
adjacent points in the PC time series.  In the observations
(Figure~\ref{fig:z-pc-scatter}a), there are clear loops towards the
bottom left hand corner of the plot.  These excursions correspond to
\eln events, when the thermocline tilt (PC \#1, approximately) and
zonal mean thermocline depth (PC \#2, approximately) vary in
quadrature as illustrated in Figure~\ref{fig:enso-cartoons}.  When the
system is not engaged in one of these \eln-related excursions, most of
the time is spent ``loitering'' near the origin in the PC scatter
plot, i.e. anomalies of thermocline tilt and zonal mean thermocline
depth are small.  Of the two models illustrated, GFDL-CM2.1
(Figure~\ref{fig:z-pc-scatter}c) matches this pattern well, showing
clear \eln excursions where PC \#1 and PC \# 2 vary in quadrature, and
loitering near the origin between \eln events.  Reference to the
GFDL-CM2.1 thermocline depth EOFs in Figures~\ref{fig:z-eofs}p--r
confirms that these principal components measure the influence of the
same kind of patterns as the observational PC \#1 and PC \#2 time
series, making direct comparison of
Figures~\ref{fig:z-pc-scatter}a~and~\ref{fig:z-pc-scatter}c
reasonable.  The UKMO-HadCM3 model results in
Figure~\ref{fig:z-pc-scatter}b correspond less well to observations.
There are some \eln-like excursions, but these tend to be much more
symmetrically distributed between positive and negative values of PC
\#1 and PC \#2 than in the observations.  There is also much less
localisation to the region of the plot near the origin between \eln
events than is seen in the observations and in the GFDL-CM2.1 results.
Most of the other models show variations on these two types of
behaviour: either they have reasonably asymmetric PC scatter plots
with clear \eln events and points strongly localised to the origin
between \eln events, or they show a much more symmetric distribution
of points, with some appearance of \eln-like excursions in the
thermocline structure, but much less localisation to a small
thermocline anomaly regime between \eln events.  Of all the models,
the results from GFDL-CM2.1 are the best from this point of view.

\section{Equatorial Pacific warm water volume}

\subsection{Warm water volume calculation methods}

Section~\ref{sec:thermocline-calc-methods} above presented a detailed
comparison of \ztw and \zgr thermocline depth results from
observations and the CMIP3 models.  In terms of ENSO variability, the
most important quantity related to thermocline depth is the equatorial
warm water volume (WWV).  As described in
Section~\ref{sec:thermocline-calc-desc}, this is defined, following
\citet{meinen-wwv}, as the volume of water lying above the thermocline
in the region 120\degree{}E--80\degree{}W, 5\degree{}S--5\degree{}N.
The results of the \ztw/\zgr comparison above would indicate that
there is likely to be little difference between WWV values calculated
using \ztw and those calculated using \zgr.

As one indication that this is indeed the case,
Table~\ref{tab:wwv-correlations} shows, for each model, correlation
coefficients between WWV anomalies calculated from \ztw and from \zgr.
For about half of the models, and the observations, the correlation
coefficients are very high (greater than 0.9 for seven out of the 17
models), while the rest are mixed: six models have correlation
coefficients of 0.5 or greater, and the only two real outliers are
UKMO-HadGEM1 and UKMO-HadCM3, which, as mentioned in
Section~\ref{sec:thermocline-calc-methods}, cause trouble in the \zgr
calculation because of numerical artefacts in the modelled ocean
temperature fields.

%
%
\begin{table}
  \begin{center}
    \begin{tabular}{p{3.6cm}c}
      \toprule
      \textbf{Model} & \textbf{Corr.} \\
      \midrule
      Observations     & 0.908 \\
      BCCR-BCM2.0      & 0.779 \\
      CCSM3            & 0.940 \\
      CGCM3.1(T47)     & 0.515 \\
      CGCM3.1(T63)     & 0.495 \\
      CNRM-CM3         & 0.909 \\
      CSIRO-Mk3.0      & 0.898 \\
      ECHO-G           & 0.952 \\
      FGOALS-g1.0      & 0.921 \\
      \bottomrule
    \end{tabular}
    \hspace{1.5cm}%
    \begin{tabular}{p{3.6cm}c}
      \toprule
      \textbf{Model} & \textbf{Corr.} \\
      \midrule
      GFDL-CM2.0       & 0.916 \\
      GFDL-CM2.1       & 0.903 \\
      GISS-EH          & 0.646 \\
      INM-CM3.0        & 0.492 \\
      MIROC3.2(hires)  & 0.673 \\
      MIROC3.2(medres) & 0.606 \\
      MRI-CGCM2.3.2    & 0.931 \\
      UKMO-HadCM3      & 0.082 \\
      UKMO-HadGEM1     & 0.189 \\
      \bottomrule
    \end{tabular}
  \end{center}
  \caption[\ztw/\zgr WWV anomaly time series correlation
    values]{Correlations between equatorial Pacific warm water volume
    anomaly time series based on \ztw and \zgr.  (Observations use
    NCEP GODAS potential temperature data.)}
  \label{tab:wwv-correlations}
\end{table}
%

On the basis of these results, and the more detailed comparison of
thermocline depth results shown earlier, I will use equatorial Pacific
WWV calculated from \ztw in what follows.

\subsection{NINO3 SST index/warm water volume phasing}
\label{sec:nino3-wwv-phasing}

A commonly adopted method of illustrating the relative phasing between
NINO3 SST index variations (representing the mean SST anomaly in the
eastern equatorial Pacific) and the equatorial Pacific warm water
volume (a proxy for the total equatorial Pacific heat content) is a
phase plot.  Here, one expects to see coherent variations of NINO3 SST
index and WWV in quadrature during \eln events.  These phase plots are
comparable to the thermocline depth PC scatter plots of
Figure~\ref{fig:z-pc-scatter}, with the NINO3 SST index standing in
for the first principal component and the WWV for the second. (The
first principal component represents zonal tilt of the equatorial
thermocline, which adjusts rather quickly to SST anomaly variations in
the eastern Pacific via the Bjerknes feedback mechanism.)

%
%
\begin{figure}
  \begin{center}
    \includegraphics[width=\textwidth]{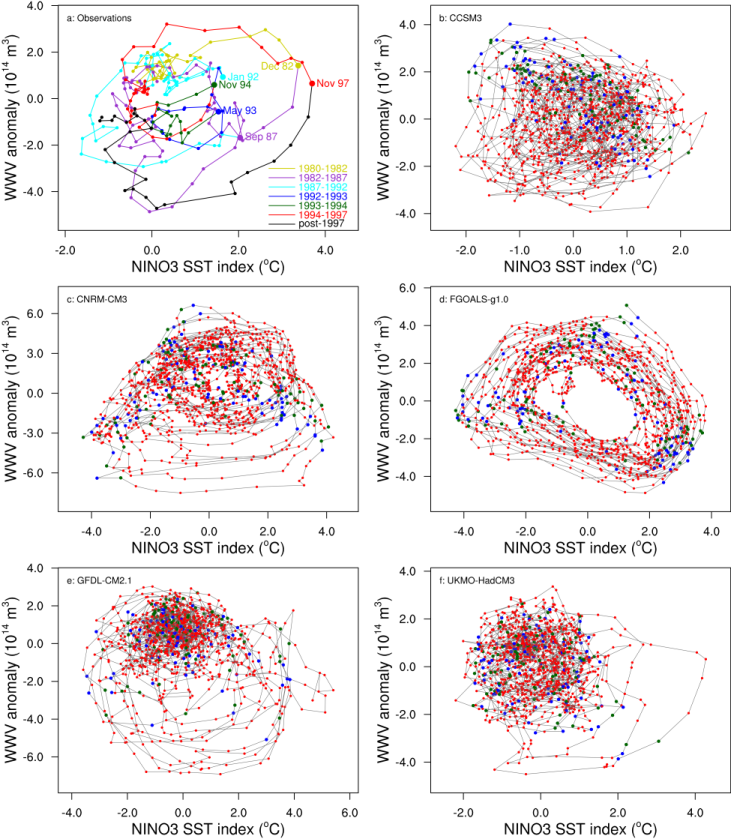}
  \end{center}
  \caption[NINO3 SST index/WWV phase plots]{Phase plots of NINO3 SST
    index versus \ztw equatorial Pacific warm water volume for
    observations (a: NINO3 SST index from ERSST v2 data set and WWV
    derived from NCEP GODAS potential temperature observations) and
    models CCSM3 (b), CNRM-CM3 (c), FGOALS-g1.0 (d), GFDL-CM2.1 (e),
    and UKMO-HadCM3 (f).  For each panel, each point denotes a single
    month, and for the model results, each January is highlighted with
    a larger dark green dot and each February with a larger blue dot.
    (Panel a adapts an idea from Figure~2 of \citep{kessler-events},
    and marks maximum temperature excursions corresponding to \eln
    events, as well as distinguishing the segments of the time series
    between each \eln.)}
  \label{fig:nino3-wwv-phasing}
\end{figure}
%

Figure~\ref{fig:nino3-wwv-phasing} shows NINO3 SST index/WWV phase
plots for observational data (Figure~\ref{fig:nino3-wwv-phasing}a:
NINO3 SST index from the ERSST v2 data set and thermocline depth from
NCEP GODAS data) and several models: CCSM3
(Figure~\ref{fig:nino3-wwv-phasing}b), CNRM-CM3
(Figure~\ref{fig:nino3-wwv-phasing}c), FGOALS-g1.0
(Figure~\ref{fig:nino3-wwv-phasing}d), GFDL-CM2.1
(Figure~\ref{fig:nino3-wwv-phasing}e) and UKMO-HadCM3
(Figure~\ref{fig:nino3-wwv-phasing}f).  Turning to the observations
first, there is a well-defined phase relationship between variations
in NINO3 SST and warm water volume, particularly during \eln events.
In Figure~\ref{fig:nino3-wwv-phasing}a, large \eln events, phase
locked to occur in boreal winter, are clearly identified as loops in
the plot, with large excursions to positive NINO3 SST index being
associated with corresponding coherent variations in WWV.  Also
visible is the ``loitering'' of the system during the recharge of
equatorial warm water volume before the beginning of the next \eln
event, a period during which predictability is generally lower
\citep{kessler-events,mcphaden-wwv}.  In the model results, some
variety of coherent phasing of a form similar to the observations is
seen in some of the models, notably CNRM-CM3
(Figure~\ref{fig:nino3-wwv-phasing}c), GFDL-CM2.1
(Figure~\ref{fig:nino3-wwv-phasing}e) and UKMO-HadCM3
(Figure~\ref{fig:nino3-wwv-phasing}f).  Most of the other models whose
phase plots are illustrated in Figure~\ref{fig:nino3-wwv-phasing} have
a small degree of coherent variation between the NINO3 SST index and
WWV, but there are no clear looping excursions as seen in the
observations during \eln.  An interesting exception to the general
pattern is FGOALS-g1.0 (Figure~\ref{fig:nino3-wwv-phasing}d), which
shows an extremely regular cycle involving the NINO3 SST index and
WWV, corresponding to the sharp spectral peak in the NINO3 SST index
spectrum for this model, at a period of about 3.5 years
(Figure~\ref{fig:nino3-spectra}).

A further characteristic of the behaviour of the observational data is
that \eln events tend to occur in the boreal winter months.  This is
the well-known seasonal phase locking of ENSO \citep{rasmusson-enso}.
By highlighting each January and February in the model phase plots in
Figure~\ref{fig:nino3-wwv-phasing}, we can clearly distinguish between
cases where \eln events occur in boreal winter and cases where \eln
events occur at other times of the year.  For the models with the best
phasing behaviour (CNRM-CM3, GFDL-CM2.1 and UKMO-HadCM3), it does
appear that the large looping \eln excursions occur mostly during
boreal winter, although the two events shown for UKMO-HadCM3 in
Figure~\ref{fig:nino3-wwv-phasing}f occur some months before the peak
observed \eln season.

The final feature to highlight in Figure~\ref{fig:nino3-wwv-phasing}
is the existence of a period of ``loitering'' of the equatorial
Pacific ocean-atmosphere system immediately following an \eln event.
During this time, both NINO3 SST and WWV anomalies remain relatively
small, seen on the phase plots as a concentration of points near the
origin.  This behaviour is most clear in the observational data
(Figure~\ref{fig:nino3-wwv-phasing}a) and the model results for
GFDL-CM2.1 (Figure~\ref{fig:nino3-wwv-phasing}e) and UKMO-HadCM3
(Figure~\ref{fig:nino3-wwv-phasing}f), and is related to the phase of
the ENSO cycle where the store of warm water in the western Pacific is
recharging through the action of westward-propagating off-equatorial
Rossby waves in the tropical Pacific.  During this period, ENSO is
significantly less predictable than during the development and
turnover of an \eln event.  To some degree, this is intuitively
obvious from the phase plots: once an \eln event has started, the
development of the event follows a relatively regular pattern, while
the slower and more diffuse ocean heat recharge that occurs between
\eln events is intrinsically less predictable.  This intuition is
confirmed by studies that show a distinct variation in ENSO
predictability as a function of ENSO phase \citep{kumar-predict-var}.
Some of the models that show coherent excursions of the NINO3 SST
index and WWV do not show this loitering behaviour (for example,
CNRM-CM3 and FGOALS-g1.0), and this is generally associated with ENSO
variability that is too regular.

One thing that seems fairly clear from this analysis is that models
with good surface ENSO behaviour, i.e. models that appear to have a
good ENSO based on examining SST variability, also have reasonable
sub-surface behaviour, showing good phasing between the NINO3 index
and the equatorial Pacific thermocline state, as measured by the
equatorial warm water volume.


%% file: figs/05/asymmetry-table.tex
\begin{center}
\begin{tabular}{lccc}
\toprule
Model & $\Corr(\vec{q}_1, \vec{a}_A)$ & $\Corr(\vec{q}_2, \vec{a}_S)$
& $||\vec{a}_S|| / ||\vec{a}_A||$ \\
\midrule
Observations      & 1.00 & 0.74 & 0.11 \\
BCCR-BCM2.0       & 1.00 & 0.66 & 0.06 \\
CCSM3             & 1.00 & 0.29 & 0.06 \\
CGCM3.1(T47)      & 1.00 & 0.26 & 0.11 \\
CGCM3.1(T63)      & 0.99 & 0.11 & 0.13 \\
CNRM-CM3          & 1.00 & 0.84 & 0.05 \\
CSIRO-Mk3.0       & 1.00 & 0.53 & 0.07 \\
ECHO-G            & 1.00 & 0.83 & 0.05 \\
FGOALS-g1.0       & 1.00 & 0.26 & 0.04 \\
GFDL-CM2.0        & 1.00 & 0.62 & 0.09 \\
GFDL-CM2.1        & 1.00 & 0.80 & 0.10 \\
GISS-EH           & 1.00 & 0.84 & 0.10 \\
INM-CM3.0         & 1.00 & 0.54 & 0.08 \\
IPSL-CM4          & 1.00 & 0.29 & 0.08 \\
MIROC3.2 (hires)  & 0.92 & 0.50 & 0.26 \\
MIROC3.2 (medres) & 1.00 & 0.31 & 0.06 \\
MRI-CGCM2.3.2     & 1.00 & 0.57 & 0.07 \\
UKMO-HadCM3       & 1.00 & 0.02 & 0.06 \\
UKMO-HadGEM1      & 1.00 & 0.41 & 0.04 \\
\bottomrule
\end{tabular}
\end{center}

%% file: 06-nlpca.tex
\chapter{Nonlinear Principal Component Analysis}
\label{ch:nlpca}

\section{Description of method}

Nonlinear principal component analysis (NLPCA) is an extension of the
ideas of principal component analysis to settings where there is a
nonlinear relationship between data variables.  As explained in
Section~\ref{sec:pca-definition}, PCA identifies linear subspaces of
the data space representing the largest proportion of the total data
variance.  The calculations performed in PCA can be expressed as an
error minimisation problem, and this provides the most straightforward
approach for extension to NLPCA.  A number of other methods related to
PCA, such as canonical correlation analysis and singular spectrum
analysis can also be extended to this nonlinear setting
\citep{bretherton-coupled,hsieh-nlcca,ghil-ssa}.  Reviews of these
techniques are available in \citep{hsieh-review-1998},
\citep{hsieh-review-2001} and \citep{hsieh-review-2004}.

\subsection{Extension of ideas from PCA to a nonlinear setting}
\label{sec:pca-vs-nlpca}

In Section~\ref{sec:pca-definition}, the PCA calculation was defined
in terms of finding mutually orthogonal directions of greatest
variance in an input data space, or, equivalently, finding the
eigendecomposition of the data covariance matrix.  It can also be
expressed as a least squares minimisation problem.  Consider data
vectors $\vec{x}_i \in \mathbb{R}^m$, with $i = 1, \dots, N$,
(normally, in climate data analysis, the $\vec{x}_i$ represent
geographical maps of some quantity with $m$ spatial points each) and
seek a linear transformation of the data as
\begin{equation}
  \label{eq:pca-linear-xform-1}
  u_i = \vec{a} \cdot \vec{x}_i
\end{equation}
such that the cost function
\begin{equation}
  \label{eq:pca-linear-xform-2}
  J_{\mathrm{PCA}} = \langle || \, \vec{x}_i - \vec{a} u_i \, ||^2
  \rangle
\end{equation}
is minimised.  The $\vec{a} \in \mathbb{R}^m$ and the $u_i \in
\mathbb{R}$ that satisfy this minimisation problem are called,
respectively, the first empirical orthogonal function (EOF) and the
first principal component (PC), together comprising the first PCA
mode.  Given this first mode, residuals $\tilde{\vec{x}}_i = \vec{x}_i
- \vec{a} u_i$ may be formed, and another minimisation problem solved
to find the second PCA mode.  Modes may be successively projected out
following this pattern to find successive orthogonal subspaces of the
data space.  This formulation is exactly equivalent to the formulation
in terms of the data covariance matrix of
Section~\ref{sec:pca-definition} \citep{burges-review}.  As already
noted in Section~\ref{sec:pca-definition}, approximation of the
$\vec{x}_i$ by linear subspaces does not necessarily work well if the
original data arises from a nonlinear process.

The essence of the NLPCA extension of these ideas is to replace the
linear functions in
\eqref{eq:pca-linear-xform-1}~and~\eqref{eq:pca-linear-xform-2} with
nonlinear functions better able to represent low-dimensional structure
in the original data \citep{kramer-nlpca,monahan-thesis}.  Suppose
that we write $u_i = f(\vec{x}_i)$, where $f: \mathbb{R}^m \to
\mathbb{R}$ is some nonlinear function (the ``reduction function''),
and then consider a cost function
\begin{equation}
  \label{eq:nlpca-minimisation}
  J_{\mathrm{NLPCA}} = \langle || \, \vec{x}_i - (\vec{g} \circ f) \,
  \vec{x}_i \, ||^2 \rangle,
\end{equation}
where $\vec{g}: \mathbb{R} \to \mathbb{R}^m$ is another nonlinear
function (the ``reconstruction function'') which is inverse to $f$ in
a least squares sense.  Through appropriate selection of $f$ and
$\vec{g}$, we may arrive at an optimal nonlinear one-dimensional
representation of the original data, with each $u_i$ giving the
one-dimensional representation of each $\vec{x}_i$.  This approach is
equivalent to PCA when $f$ and $\vec{g}$ are linear functions of the
form $f(\vec{x}) = \vec{a} \cdot \vec{x}$ and $\vec{g}(u) = \vec{a}
u$.  Generalisation to reduced representations of dimensionality
greater than one is done in the natural way; in this case, $\vec{u}_i
\in \mathbb{R}^p$ say, and $\vec{f}$ and $\vec{g}$ are functions
$\vec{f}: \mathbb{R}^m \to \mathbb{R}^p$ and $\vec{g}: \mathbb{R}^p
\to \mathbb{R}^m$.  In most of what follows, we adopt this more
general notation.

Identifying an NLPCA ``mode'' here is a little more tricky than in
PCA, since there is no single spatial pattern that may be identified
as an EOF.  One can instead examine the reduced representations of
each of the original data points, $u_i = f(\vec{x}_i)$, and the
reconstructed data vectors, $\vec{x}'_i = (\vec{g} \circ f) \,
\vec{x}_i$.  These, in some sense, constitute the NLPCA ``mode''.
Since $u_i \in \mathbb{R}$, the $\vec{x}'_i$ lie on a one-dimensional
manifold in $\mathbb{R}^m$.  The smoothness of this manifold depends
on the smoothness of $f$ and $\vec{g}$.

The essential idea is thus clear: identify functions $f$ and $\vec{g}$
such that the reconstructed data vectors $\vec{x}'_i$ lie as close to
the original data vectors $\vec{x}_i$ in a least squares sense.  In
the form of NLPCA pursued here, this is done by representing $f$ and
$\vec{g}$ using neural networks.  The minimisation
\eqref{eq:nlpca-minimisation} then becomes a parametric problem, where
we minimise over possible values of neural network parameters in the
networks that represent $f$ and $\vec{g}$.

\subsection{Auto-associative neural networks}
\label{sec:nlpca-anns}

Neural networks are a means by which arbitrary nonlinear functions may
be approximated in a fashion that allows for relatively
straightforward parametric optimisation (``learning'').  They were
originally developed in the machine learning community, but have since
been applied in many different fields where efficient approximation of
structurally complex functions is required.  \citet{haykin-nns} and
\citet{bishop-nn-book} provide general introductions to neural network
techniques, while \citet{krasnopolsky-neural-nets} reviews
applications of these methods in climate science.  The form of neural
network used in NLPCA is referred to as a multi-layer perceptron
\citep[Chapter~4]{haykin-nns} and consists of three layers of neurons:
an input layer, a hidden layer and an output layer.  Each neuron in
each layer has a number of real-valued inputs and a single real-valued
output.  To represent a function $\vec{f}: \mathbb{R}^m \to
\mathbb{R}^p$ requires a network with $m$ input neurons (one for each
component of the input) and $p$ output neurons (one for each component
of the output).  The number of neurons in the hidden layer determines
the complexity of the function that may be represented by the network.
The transfer functions of the individual neurons are linear or simple
monotonic nonlinear functions, so that a small number of neurons may
approximate only simple functions.  A larger number of neurons
provides the capability of approximating more complicated functions,
but at the cost of requiring more parameters to represent the network,
so resulting in a more difficult optimisation problem to choose those
parameters.  In fact, for arbitrary numbers of hidden layer neurons,
the functions that may be approximated by this type of neural network
are dense in the space of continuous functions with support in the
unit hypercube \citep{cybenko-sigmoidal}; in this sense, networks of
this type are universal approximators, although the practicalities of
estimating the network parameters vitiates this feature in
applications.  (\citeauthor{cybenko-sigmoidal}'s result is in the same
spirit as the Weierstrass approximation theorem \citep{rudin-analysis}
that states that, for any continuous function on a real interval, $f :
[a,b] \to \mathbb{R}$, and any $\varepsilon > 0$, a polynomial $p(x)$
may be found such that $\sup_{x\in[a,b]} | f(x) - p(x) | <
\varepsilon$, i.e. any continuous $f$ can be uniformly approximated by
a polynomial on $[a,b]$.)

\begin{figure}
  \begin{center}
    \subfloat[Network with single bottleneck neuron.]{
      \includegraphics[width=0.45\textwidth]{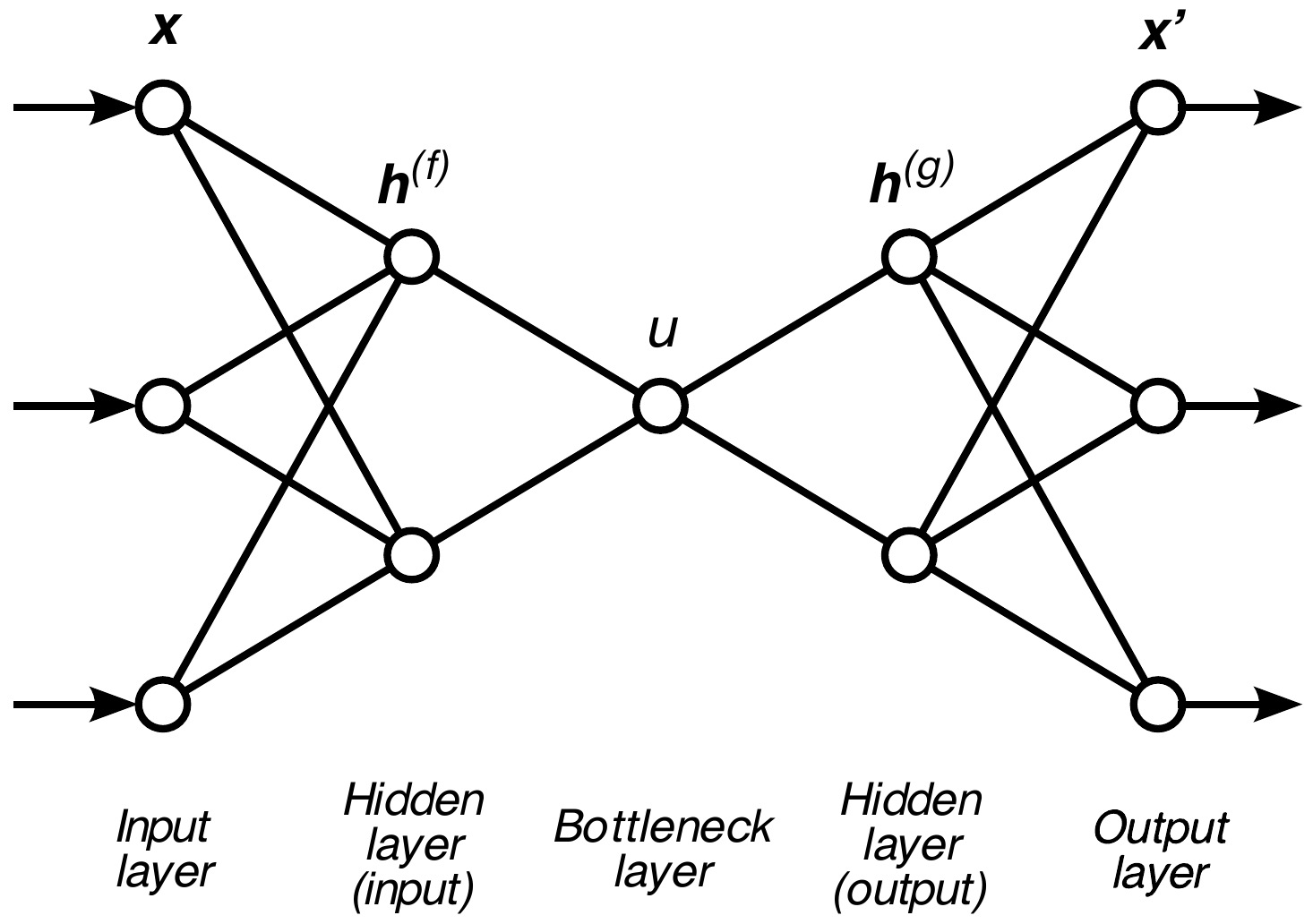}}%
    \qquad%
    \subfloat[Network with two bottleneck neurons, with values
      constrained to lie on the unit circle, allowing representation
      of a one-dimensional periodic variable.]{
      \includegraphics[width=0.45\textwidth]{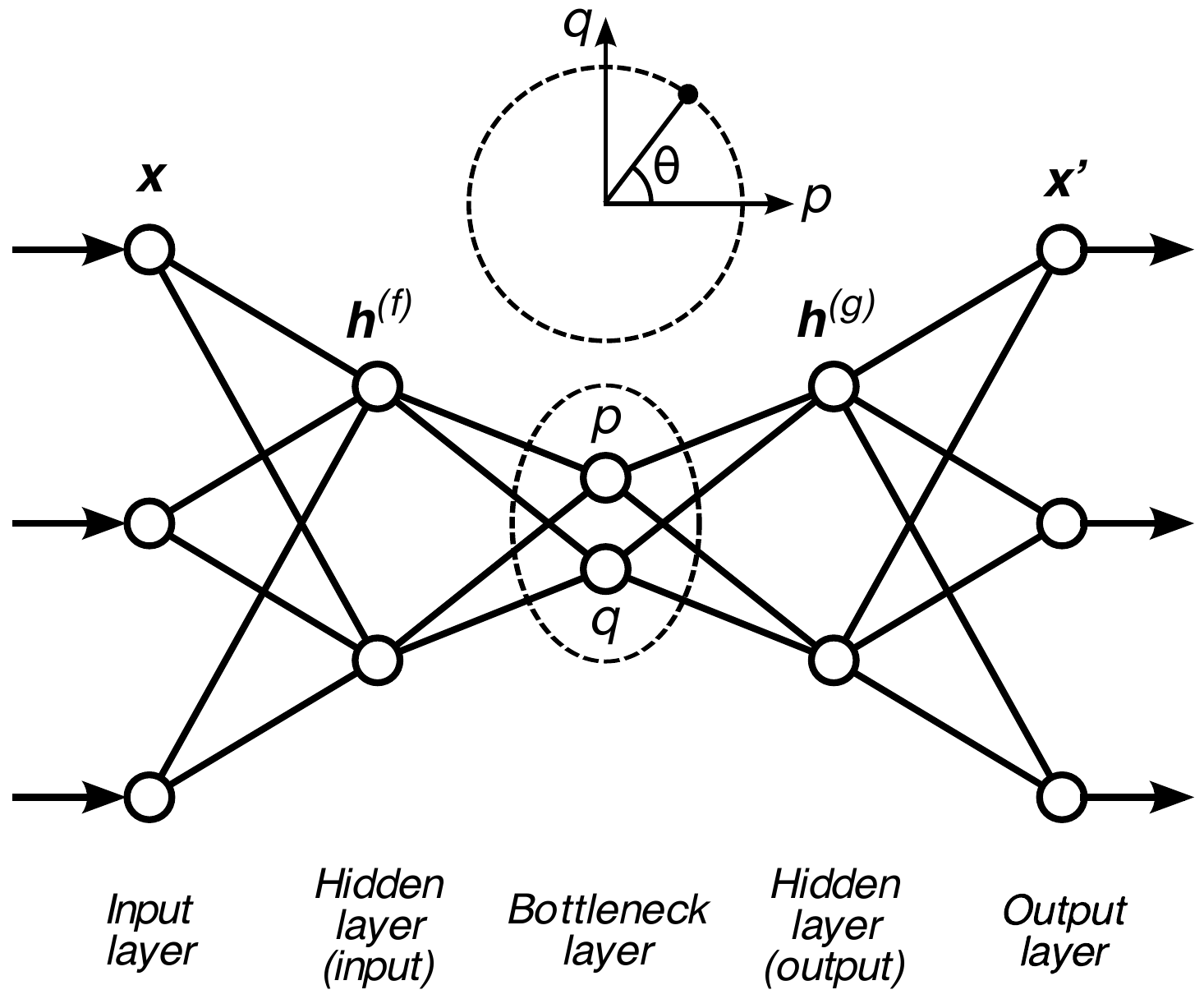}}
  \end{center}
  \caption[NLPCA neural network architecture]{Neural network
    architectures for NLPCA.  (After Figure~2 of
    \citet{hsieh-review-2004}.)}
  \label{fig:nn-architecture}
\end{figure}

In NLPCA, the functions $\vec{f}: \mathbb{R}^m \to \mathbb{R}^p$ and
$\vec{g}: \mathbb{R}^p \to \mathbb{R}^m$ are both approximated by
neural networks.  One can then think of an overall network consisting
of a ``reduction'' network, representing the function $\vec{f}$,
mapping from $m$ inputs to $p$ outputs, connected directly to a
``reconstruction'' network, representing the function $\vec{g}$,
mapping from $p$ inputs to $m$ outputs.  The layer in the combined
network that joins the reduction and reconstruction networks is called
the ``bottleneck layer'' and contains $p$ bottleneck neurons, where $p
< m$, giving a reduced dimensionality representation of the input
data.  Figure~\ref{fig:nn-architecture}a shows such a network with
$p=1$, giving a one-dimensional reduced representation of the
three-dimensional input data.  This architecture, called an
\emph{autoassociative network}, permits the reconstructed outputs to
be compared directly to the inputs to determine how well the network
is able to reproduce the original data from the bottleneck reduced
layer.  A single bottleneck neuron produces a one-dimensional reduced
representation of the input data, two neurons in the bottleneck layer
a two-dimensional reduced representation, and so on.

Beyond choosing $p$, the number of neurons in the bottleneck layer,
and $l$, the number of neurons in the hidden layers, other constraints
may be placed on the architecture of the neural network for particular
purposes.  Figure~\ref{fig:nn-architecture}b illustrates an example,
where a bottleneck layer with two neurons is used, but the values of
those two neurons are constrained to lie on the unit circle
\citep{kirby-circular-neurons}.  The bottleneck layer thus represents
a one-dimensional periodic variable, and the network is suitable for
applications where some form of periodic oscillatory behaviour is
expected.

%
%
%

The type of neural network employed in NLPCA uses a simple combination
of linear and monotonic nonlinear functions to represent the
relationships between the values of different nodes in the network.
Writing $\vec{x}, \vec{x}' \in \mathbb{R}^m$ for the input and output
data of the overall network, $\vec{h}^{(f)}, \vec{h}^{(g)} \in
\mathbb{R}^l$ for the values of the neurons in the input and output
hidden layers and $\vec{u} \in \mathbb{R}^p$ for the values of the
bottleneck neurons, the basic equations for the neural network
transfer function are:
\begin{subequations}
  \label{eq:nlpca-nn-eqs}
  \begin{gather}
    \vec{h}^{(f)} = \tanh \left( \mathbf{W}^{(f)} \vec{x} +
    \vec{b}^{(f)} \right), \label{eq:nlpca-nn-eqs-1} \\
    \vec{u} = \mathbf{W}_u^{(f)} \vec{h}^{(f)} +
    \vec{b}_u^{(f)}, \label{eq:nlpca-nn-eqs-2} \\
    \vec{h}^{(g)} = \tanh \left( \mathbf{W}_u^{(g)} \vec{u} +
    \vec{b}_u^{(g)} \right), \label{eq:nlpca-nn-eqs-3} \\
    \vec{x}' = \mathbf{W}^{(g)} \vec{h}^{(g)} +
    \vec{b}^{(g)}, \label{eq:nlpca-nn-eqs-4}
  \end{gather}
\end{subequations}
where the notation $\vec{a} = \tanh \vec{b}$ means that $a_i = \tanh
b_i$ for each component of the vectors $\vec{a}$ and $\vec{b}$.  Here,
the factors multiplying and translating the neuron values are referred
to as the \emph{weights} of the network, with $\mathbf{W}^{(f)}$,
$\vec{b}^{(f)}$, $\mathbf{W}_u^{(f)}$ and $\vec{b}_u^{(f)}$ being
those for the reduction network representing the function $\vec{f}:
\mathbb{R}^m \to \mathbb{R}^p$ and $\mathbf{W}_u^{(g)}$,
$\vec{b}_u^{(g)}$, $\mathbf{W}^{(g)}$ and $\vec{b}^{(g)}$ being those
for the reconstruction network representing the function $\vec{g}:
\mathbb{R}^p \to \mathbb{R}^m$.  For a fixed network architecture,
these weights completely determine the response of the network to
varying inputs.  We write $\mathbf{W}$ to represent the collection of
all the weight values required to specify a network, i.e., $\mathbf{W}
= \{ \mathbf{W}^{(f)}, \vec{b}^{(f)}, \mathbf{W}_u^{(f)},
  \vec{b}_u^{(f)}, \mathbf{W}_u^{(g)}, \vec{b}_u^{(g)},
  \mathbf{W}^{(g)}, \vec{b}^{(g)} \}$.

The number of parameters required to specify any particular network is
easily calculated.  The weight matrices and vectors have the following
dimensions (for $m$-dimensional inputs, and a network with $p$
bottleneck neurons and $l$ neurons in each of the hidden layers):
\begin{center}
  \begin{tabular}{ll@{\hspace{2cm}}ll}
    $\mathbf{W}^{(f)}$    & $l \times m$ &    $\mathbf{W}_u^{(g)}$  & $l \times p$ \\
    $\vec{b}^{(f)}$       & $l$          &    $\vec{b}_u^{(g)}$     & $l$ \\
    $\mathbf{W}_u^{(f)}$  & $p \times l$ &    $\mathbf{W}^{(g)}$    & $m \times l$ \\
    $\vec{b}_u^{(f)}$     & $p$          &    $\vec{b}^{(g)}$       & $m$
  \end{tabular}
\end{center}
giving a total number of parameters $C = 2l(m + p + 1) + m + p$.
Table~\ref{tab:nlpca-param-counts} shows the number of parameters
required to specify networks of some indicative sizes.  Those with
$m=3$ correspond to configurations used for handling geometrical test
data (Section~\ref{sec:nlpca-test-data}) and those with $m=10$
configurations used for processing Pacific SST data, preprocessed
using PCA (Section~\ref{sec:nlpca-sst-results}).

%
%
\begin{table}
  \begin{center}
    \begin{tabular}{cccc}
      \toprule
      $m$ & $l$ & $p$ & $C$ \\
      \midrule
       3  &  2  &  1  &  24 \\
       3  &  4  &  1  &  44 \\
       3  &  4  &  2  &  53 \\
       \bottomrule
    \end{tabular}%
    \hspace{1cm}%
    \begin{tabular}{cccc}
      \toprule
      $m$ & $l$ & $p$ & $C$ \\
      \midrule
        3  &  6  &  2  &  77 \\
       10  &  4  &  1  & 107 \\
       10  &  6  &  2  & 168 \\
       \bottomrule
    \end{tabular}
  \end{center}
  \caption[Parameter counts for NLPCA network architectures]{Numbers
    of values in network weight vectors for a number of NLPCA network
    configurations ($m$ is the number of input and output neurons, $p$
    the number of bottleneck neurons and $l$ the number of neurons in
    each of the hidden layers).}
  \label{tab:nlpca-param-counts}
\end{table}
%

It was shown by \citet{monahan-thesis} that NLPCA has a variance
partitioning property similar to that of PCA
(Section~\ref{sec:pca-definition}), i.e., if we define residual
vectors $\tilde{\vec{x}}_i = \vec{x}_i - \vec{x}'_i$, then
\begin{equation}
  \label{eq:nlpca-variance-partition}
  \var(\vec{x}) = \var(\vec{x}') + \var(\tilde{\vec{x}}).
\end{equation}
This means that the total data variance is partitioned into a portion
explained by the variance of the first NLPCA mode and the variance of
the residuals.  The same explained variance measure can thus be used
for the nonlinear NLPCA modes as for linear PCA modes.  This result
leads to two different approaches to the practical use of NLPCA,
referred to by \citet{monahan-thesis} as the \emph{modal} and
\emph{nonmodal} approaches.  In modal NLPCA, starting with input data
vectors $\vec{x}_i$, a neural network with a single bottleneck neuron
is used to find the first one-dimensional NLPCA mode, yielding
bottleneck neuron values $u_i$ and reconstructed data vectors
$\vec{x}'_i$.  Residuals $\tilde{\vec{x}}_i = \vec{x}_i - \vec{x}'_i$
are then calculated as above and the NLPCA procedure is applied again,
using the $\tilde{\vec{x}}_i$ as input data, to yield a second
one-dimensional NLPCA mode.  Again, as for PCA, residuals can be
calculated and the NLPCA procedure applied repeatedly to give a series
of NLPCA modes, whose variance partitions the total data variance in
the same way as do PCA modes (see \eqref{eq:pca-variance-partition-2}
in Section~\ref{sec:pca-definition}).  In the second, \emph{nonmodal},
approach, a neural network with more than one bottleneck neuron is
used, yielding a more than one-dimensional reduced representation of
the input data, via the values of the bottleneck neurons.  For
instance, in Section~\ref{sec:nlpca-sst-nonmodal}, I use a
two-dimensional nonmodal analysis of tropical Pacific SST data --- the
two bottleneck neurons used in the NLPCA network give a
two-dimensional reduced representation of the input data.

\subsection{Model fitting considerations}

As shown in Table~\ref{tab:nlpca-param-counts}, the neural network
models used in NLPCA can require a rather large number of parameters.
Fitting so many parameters from observational or model data requires
some care, for two reasons.  The first is the problem of overfitting.
Using a model with many parameters, it is easy to fit any data set to
a high degree of accuracy, but this fit will be specific to that data
set, including whatever idiosyncratic features it may have due to
noise, observational error or other sources.  Some means is required
to avoid this overfitting so that the parameter estimation procedure
forces the resulting network to capture as far as possible only the
``essential'' features of the input data set, ignoring small-scale
features due to noise.  The second problem is that the function to be
minimised, described in detail below, generally has a very large
number of local minima, and may not even have a unique global minimum.
No simple search procedure is likely to converge to the global minimum
if one exists.  The methods adopted here to circumvent these problems
substantially follow \citep{hsieh-review-2004}, with some details
adapted from the Matlab codes available from William Hsieh's
website\footnote{\url{http://www.ocgy.ubc.ca/projects/clim.pred/index.html}}.
Other references reporting applications of NLPCA to climate
applications use similar approaches
\citep{an-enso-interdecadal,monahan-thesis,monahan-enso,wu-enso-interdecadal}.

Before addressing these issues, let us define the cost function to be
minimised to find the network weights.

\subsubsection{Cost function}

We write $\vec{x}_i$ for the input data vectors and $\vec{x}'_i$ for
the reconstructed data vectors produced by the operation of the neural
network, i.e. $\vec{x}'_i = (\vec{g} \circ \vec{f}) \vec{x}_i$, where
the functions $\vec{f}$ and $\vec{g}$ are defined by the reduction and
reconstruction sides of the autoassociative NLPCA neural network.  For
$N$ input vectors $\vec{x}_i$, we define a scaling factor by $S^{-1} =
\langle || \vec{x} || \rangle$.  The scaling factor $S$ is defined so
as to make the basic error term in the cost function of comparable
magnitude to the penalty terms introduced below.  The basic cost
function measuring the mismatch between the data reconstructed by the
neural network and the original input data is then
\begin{equation}
  \label{eq:nlpca-cost-function-basic}
  J_0 = S \langle || \vec{x}' - \vec{x} ||^2 \rangle.
\end{equation}
It is possible to use the 1-norm of the error vector here,
corresponding to measuring the mismatch between the original and
reconstructed data using mean absolute error (MAE) instead of mean
squared error (MSE).  This may provide more robust behaviour in the
presence of very noisy data
\citep{cannon-robust-nlpca,hsieh-noisy-nlpca}, but did not prove
necessary in the examples studied here.  Use of the mean squared error
measure is also more convenient for comparison with earlier work
(particularly \citep{monahan-enso} and \citep{an-enso-interdecadal}).

In addition to this basic expression, we use several extra penalty
terms in the full cost function to attempt to impose constraints on
the values of the bottleneck neurons and on the network weights.  We
cannot guarantee that any constraints imposed by the addition of
penalty terms in the cost function will be satisfied exactly, but this
simple approach is significantly more convenient than the
alternatives, which would involve attempting to minimise a simpler
cost function in a sub-manifold of the full network parameter space, a
sub-manifold defined by a set of highly non-linear equations of the
network weights.  First, we would like for the bottleneck neuron
values to have zero mean, i.e., writing $\vec{u}_i =
\vec{f}(\vec{x}_i)$, we require that $\langle \vec{u} \rangle = 0$.
An approximate form of this constraint can be incorporated into the
cost function by adding a penalty term of the form
\begin{equation}
  J_{C,\mathrm{mean}} = \langle \vec{u} \rangle^2.
\end{equation}
Similarly, we would like the bottleneck neuron values to have unit
variance, i.e., $\langle \vec{u}^2 \rangle = 1$ (assuming zero mean
for the $\vec{u}_i$).  Similarly to the requirement on $\langle
\vec{u} \rangle$, an approximate form of this constraint can be
represented as a penalty term in the cost function of the form
\begin{equation}
  J_{C,\mathrm{var}} = \left( \langle \vec{u}^2 \rangle - 1 \right)^2.
\end{equation}
(This term is not used for networks with circular bottleneck layers,
as in Figure~\ref{fig:nn-architecture}b, since the bottleneck neuron
values in this case are normalised as part of the network transfer
function calculation.  This is an example of the type of treatment of
constraints mentioned above, where a constraint is imposed exactly by
restricting the cost function minimisation to a sub-manifold of the
total network weight space.)

Next, in situations where we use more than one bottleneck neuron,
though again, not for ``circular'' networks, we would like for the
values of different bottleneck neurons to be uncorrelated, in analogy
to the decorrelation of principal component time series in PCA.
Labelling the individual bottleneck neurons values of $\vec{u}_i$ as
$u_i^{(j)}$, with $j = 1, \dots, p$, this decorrelation condition can
be expressed by including the following penalty term in the cost
function:
\begin{equation}
  \label{eq:cost-function-corr}
  J_{C,\mathrm{corr}} = \left\langle \sum_{j=1}^{p-1} \sum_{k=j+1}^p
  (u^{(j)} u^{(k)})^2  \right\rangle.
\end{equation}
This constraint is only of use in multi-dimensional nonmodal
applications of NLPCA.  In fact, in one of the few published accounts
of this approach, it appears that this form of constraint is not
applied \citep{monahan-enso}.

Finally, we incorporate a penalty term to restrict the magnitude of
the network weights produced by the fitting procedure.  This takes the
form
\begin{equation}
  \label{eq:weight-penalty-constraint}
  J_{C,\mathrm{weights}} = P_W || \mathbf{W}^{(f)} ||_F^2
\end{equation}
where $P_W$ is an adjustable penalty coefficient and
$||\mathbf{A}||_F$ denotes the \index[not]{Frobenius norm@$||A||_F$,
  Frobenius norm}Frobenius norm of a matrix $\mathbf{A}$, defined by
\begin{equation}
  || \mathbf{A} ||_F^2 = \sum_{i, j} A_{ij}^2 = \Tr ( \mathbf{A}^T
  \mathbf{A} ),
\end{equation}
with \index[not]{trace operator@$\Tr$, trace operator}$\Tr$ denoting
the trace operator.  The importance of this weight penalty term,
called a weight decay term in the neural network literature
\citep[Section~9.2.1]{bishop-nn-book}, can be understood by
considering the form of the nonlinear functions in the network.  If
the magnitudes of the weights are not restricted, the arguments of the
hyperbolic tangent functions in the expression for $\vec{h}^{(f)}$ in
\eqref{eq:nlpca-nn-eqs} may become arbitrarily large, rendering the
overall network transfer function overly sensitive to small changes in
inputs.  This term provides a degree of regularisation to the transfer
function.  The penalty term serves as an additional foil against
overfitting, and proves to be particularly important for some of the
more nonlinear test examples, although the best fit networks for most
of the Pacific SST analyses have small or zero penalty terms (the
thermocline data analyses using circular networks all use $P_W = 1$
for easy comparison with \citet{an-enso-interdecadal}, who used this
configuration in their analyses of equatorial Pacific thermocline
variability).

The total cost function to be minimised, as a function of the network
weights, treating the input data set as model parameters, is then
\begin{equation}
  \label{eq:nlpca-cost-function}
  J(\mathbf{W}; \vec{x}) = J_0 + J_{C,\mathrm{mean}} +
  J_{C,\mathrm{var}} + J_{C,\mathrm{corr}} + J_{C,\mathrm{weights}}.
\end{equation}

\subsubsection{Basic minimisation issues}

In order to find the network weights that minimise the cost function
\eqref{eq:nlpca-cost-function} for a given set of input data, a simple
quasi-Newton minimisation procedure is used, essentially the function
{\tt dfpmin} from Section~10.7 of \citet{press-nr}.  The gradient
$\nabla_{\mathbf{W}} J$ required for the quasi-Newton algorithm is
approximated by first order finite differences, and, if convergence of
the quasi-Newton optimisation fails, the minimisation tolerance is
expanded several times until convergence is attained.  The
quasi-Newton optimisation is initialised with random values for the
weights $\mathbf{W}$.  An ensemble of initial conditions is used as
part of the strategy to avoid local minima in the cost function (see
below).

Experiments were also conducted using a simulated annealing downhill
simplex minimisation method, following a standard logarithmic
annealing schedule \citep[Section~10.9]{press-nr}.  A final
quasi-Newton minimisation step was applied to get to the bottom of any
local minimum, since simulated annealing methods, while good at
finding deep local minima in a complex landscape, are not good at this
final minimisation to the bottom of a local minimum.  Since the NLPCA
cost function appears to have a large number of local minima (see the
test example results in Section~\ref{sec:nlpca-test-data} below,
particularly Figure~\ref{fig:nlpca-spiral-errors} and associated
discussion), a global stochastic search method such as simulated
annealing would appear to offer some advantages over the local
deterministic quasi-Newton search.  However, in the particular case
examined here, global search does not seem to offer any advantage, a
phenomenon that has been observed before in investigations of the
determination of neural network weights for function approximation
problems \citep{hamm-nn-optimisation}.  Because of the nature of the
cost function here, any simulated annealing approach still needs to
use an ensemble of initial conditions to get good results, and since
simulated annealing methods are generally much slower than
quasi-Newton minimisation, better results can be gained from using
local optimisation combined with a larger ensemble of initial
conditions.  Consequently, all results quoted below are for the
quasi-Newton minimisation algorithm only.

\subsubsection{Avoidance of overfitting}
\label{sec:nlpca-overfitting}

To address the problem of overfitting, we follow the method outlined
by \citet{hsieh-review-2004}, where the input data set is stratified
randomly into \emph{training} and \emph{test} (or \emph{validation})
subsets with $N_{\mathrm{train}}$ and $N_{\mathrm{test}}$ data items
respectively (the fraction of data withheld for testing is fixed at
15\% throughout the results reported here).  The network is trained
using the training data set, so we actually minimise $J(\mathbf{W};
\vec{x}_{\mathrm{train}})$ rather than $J(\mathbf{W}; \vec{x})$, to
find the minimising weight vector $\mathbf{W}_{\mathrm{min}} =
\argmin_{\mathbf{W}} J(\mathbf{W}; \vec{x}_{\mathrm{train}})$.  We now
decide whether to accept or reject the weight vector
$\mathbf{W}_{\mathrm{min}}$ by comparing the mean square error
calculated for the training and test data sets (i.e., the basic cost
function value, $J_0$, without any constraint terms).  If
$J_0(\mathbf{W}_{\mathrm{min}}; \vec{x}_{\mathrm{test}}) > (1 +
\varepsilon_{\mathrm{overfit}}) J_0(\mathbf{W}_{\mathrm{min}};
\vec{x}_{\mathrm{train}})$, we reject the minimisation solution as
overfitted.  Here, $\varepsilon_{\mathrm{overfit}}$ is an overfitting
tolerance factor.  The point of doing this is to have some sort of
out-of-sample test for the fit resulting from the cost function
minimisation.  An overfitted solution will have a small cost function
for the training data set, representing a close fit to the
idiosyncrasies of the training data, while providing a less good fit
to the independent test data set.

Out of the ensemble members that are accepted by the above criterion,
we choose as our ``best fit'' solution the one with the minimum mean
squared error between the original and reconstructed data points.  The
same training/test split of the input data is used for all ensemble
members.  Only the random initial weight vector for the cost function
minimisation differs between the members of the ensemble.

In order to avoid specifying a value for
$\varepsilon_{\mathrm{overfit}}$, it is possible to calculate a post
facto absolute overfitting tolerance as
\begin{equation}
\label{eq:overfit}
  \Delta J_{\mathrm{Acc}} =
  \left( \frac{1}{\#\text{Acc}} \sum_{j \in \text{Acc}}
  \frac{J_0^{(\mathrm{train},j)} - J_0^{(\mathrm{test},j)}}{J_0^{(j)}} \right)
  \Bigg/
  \left( \frac{1}{\#\text{Acc}} \sum_{j \in \text{Acc}}
  \frac{1}{J_0^{(j)}} \right)
\end{equation}
where $J_0^{(\mathrm{train},j)}$ and $J_0^{(\mathrm{test},j)}$ are the
mean squared errors from the $j$th training and validation data sets
respectively, $J_0^{(j)}$ is the overall mean squared error across all
the data in ensemble member $j$, Acc refers to ensemble members
originally accepted with no overfitting tolerance, and the notation
\index[not]{cardinality@$\#S$, set cardinality}$\#S$ denotes the
cardinality of a set $S$.  The numerator in \eqref{eq:overfit} is the
mean relative error difference between training and test data across
the accepted members of the ensemble, which is a measure of, on
average, how ``underfitted'' ensemble members are.  The denominator in
\eqref{eq:overfit} is a normalisation factor.  Once this new
overfitting tolerance has been calculated, acceptance of ensemble
members can be retried by accepting ensemble members for which the
condition $J_0^{(\mathrm{test},j)} - J_0^{(\mathrm{train},j)} < \Delta
J_{\mathrm{Acc}}$ is satisfied, i.e. the $\Delta J_{\mathrm{Acc}}$
calculated in \eqref{eq:overfit} is taken to be the greatest absolute
permitted value of overfitting in the ensemble.  In this way, a larger
number of acceptable solutions is potentially available from the
minimisations performed.

\subsubsection{Avoidance of local minima}

The second problem with minimising the cost function
\eqref{eq:nlpca-cost-function} is the existence of local minima.  To a
great extent, the neural network transfer function, $\vec{g} \circ
\vec{f}$, is a ``black box'': networks with many hidden layer neurons
are able to approximate extremely complicated functions, there is
little opportunity to gain insight into how individual elements in the
weight vector $\mathbf{W}$ affect the form of the overall transfer
function, minor adjustments to network weights can disproportionately
modify the transfer function, and there is no easy way to determine
initial conditions for a minimisation procedure that will result in a
good minimum.

The approach taken here, again following \citet{hsieh-review-2004}, is
to use an ensemble of random initial weight vectors, minimising the
cost function for each, using segregation of the input data into
training and validation subsets to avoid overfitting, and then
selecting the best non-overfitted solution from the ensemble.  Even
following this approach, if the ensemble size is not sufficiently
large, it is quite possible for the fitting procedure to end up in a
poor local minimum.  For example, for some of the test examples shown
below, a linear reduction, essentially equivalent to normal PCA,
appears to have a large basin of attraction.  This can usually be
picked out after the fact by comparing the mean squared error of the
best NLPCA fit to a simple PCA reduction and reconstruction, but, as
is usual in this type of high-dimensional optimisation problem, there
is no a priori means of identifying and avoiding local minima.

Initial network weight vectors $W^{(0)}_j$ are allocated for each
ensemble member as
\begin{equation}
  \label{eq:nlpca-initial-weights}
  W^{(0)}_j = R_W S_j \xi_j,
\end{equation}
where $R_W$ controls the overall spread of weights used in each member
of the ensemble (all of the results quoted here use $R_W = 1$), $S_j$
is a scaling factor for each individual weight value and the $\xi_j$
are random variables distributed as $\xi_j \sim U(-1, 1)$.  The
per-weight scalings are intended to nondimensionalise and
re-dimensionalise inputs and outputs to the network, so, referring to
\eqref{eq:nlpca-nn-eqs}, we use $S_j = S_{\mathrm{train}}$ for
$\mathbf{W}^{(f)}$, $S_j = S_{\mathrm{train}}^{-1}$ for
$\mathbf{W}^{(g)}$ and $\vec{b}^{(g)}$ and $S_j = 1$ for all other
weight vector entries, where $S_{\mathrm{train}}^{-1} = \langle ||
\vec{x} || \rangle_{\mathrm{train}}$ is the mean data norm across the
training data.

\subsubsection{Parameter selection}
\label{sec:nlpca-param-selection}

In order to perform the fitting procedure described above, the
following parameters need to be specified:
\begin{description}
  \item[Network architecture]{The number of bottleneck neurons used is
    determined by the type of reduction required.  In circumstances
    where the dimensionality of the underlying data manifold is known,
    this dimensionality provides a natural choice for the number of
    neurons in the bottleneck layer.  In more realistic problems,
    where the dimensionality of the data manifold is not known, the
    choice is generally between a modal decomposition into a set of
    one-dimensional NLPCA modes, or a low-dimensional nonmodal
    approximation of the data.  For the geometrical test data sets,
    the number of bottleneck neurons is thus chosen to match the
    intrinsic dimensionality of the input data, while for the Pacific
    SST and thermocline data, either one or two bottleneck neurons are
    used as appropriate
    (Sections~\ref{sec:nlpca-sst-results}~and~\ref{sec:nlpca-z-results}).
    The choice of the number of neurons to use in the hidden layers of
    the network is more difficult.  Theoretically, using more neurons
    in the hidden layers should permit the network to approximate more
    complicated functions, and so do a better job of reducing and
    reconstructing the input data exactly.  However, increasing the
    number of hidden layer neurons quickly increases the number of
    weight parameters required to describe the network, and this
    increases the difficulty of minimising the network cost function.
    With larger numbers of neurons in the hidden layers of the
    network, the fitting algorithm can produce networks that fit
    spurious noise ``features'' in the data.  The initial development
    of the ideas of NLPCA is framed in terms of finding an
    \emph{optimal} nonlinear reduction and reconstruction of the input
    data, i.e. finding the global minimum of the cost function
    \eqref{eq:nlpca-minimisation} over all functions $\vec{f}:
    \mathbb{R}^m \to \mathbb{R}^p$, $\vec{g}: \mathbb{R}^p \to
    \mathbb{R}^m$.  However, this is not a particularly well-posed
    problem.  Some restriction on the form of $\vec{f}$ and $\vec{g}$
    has to be imposed to arrive at a practical optimisation problem,
    and the difficulties of fitting large networks mean that it is
    better to abandon the goal of finding an optimal solution even
    within the class of functions that can be approximated by neural
    networks, and to work instead with networks with a relatively
    small number of hidden layer neurons.  This corresponds to
    restricting the functions $\vec{f}$ and $\vec{g}$ to a much
    smaller class of nonlinear functions.  Although the nonlinearity
    of the functions approximated by the network is governed by the
    magnitudes of the network weights, so is essentially unlimited
    even for a network with few hidden layer neurons, the number of
    distinct features that can be fitted by the network increases
    rapidly with the number of hidden layer neurons.  For the analysis
    of Pacific SST and thermocline data here, I follow the choices
    made in \citep{monahan-enso} and \citep{an-enso-interdecadal} for
    network sizes, since these have been shown to produce reasonable
    results for this type of data.}
  \item[Test fraction]{The proportion of the input data withheld for
    overfitting testing is fixed at 15\% for all of the results
    presented here.  This is similar to the values used in
    \citep{hsieh-review-2004} and \citep{monahan-enso}.}
  \item[Initial network weights]{Initial weights for each ensemble
    member network are allocated according to
    \eqref{eq:nlpca-initial-weights}.  This provides weights with
    suitable relative scales for the input data.}
  \item[Ensemble size]{Generally speaking, a larger ensemble size is
    better, giving more chance of avoiding local minima in cost
    function.  However, this has to be traded off against
    computational time.  For all of the results reported here,
    ensemble sizes of at least 25 have been used, with larger
    ensembles used in some test cases where the fitting algorithm had
    trouble.  This approach works well for the geometrical test cases,
    since it is immediately clear whether or not a solution is good,
    but is not so satisfactory for real applications such as the
    Pacific SST and thermocline data examples, where one has only
    simple error measures to decide whether or not a solution is
    reasonable.  In general, because of the complexity of the cost
    function landscape for \eqref{eq:nlpca-cost-function}, there is
    little that can be said a priori about good ensemble size choices
    for these more complex cases.}
  \item[Overfitting tolerance]{The normal approach here is to use the
    post facto calculated overfitting tolerance described in
    Section~\ref{sec:nlpca-overfitting}.  This obviates the need for
    determining a suitable value beforehand, and generally works quite
    well.}
\end{description}

\section{Application to test data sets}
\label{sec:nlpca-test-data}

Before applying NLPCA to tropical Pacific SST and thermocline data, we
can get some feeling for the behaviour of the algorithm using the
simple geometrical data sets described in
Section~\ref{sec:test-data-sets}.  Because NLPCA makes available both
a reduced form of the input data (the values of the bottleneck
neurons) and a reconstruction that should correspond to the original
data (the output of the overall NLPCA neural network), we can examine
the performance of NLPCA on test data sets in two different ways.

\begin{figure}
  \begin{center}
    \begin{tabular}{cc}
      \includegraphics[width=0.49\textwidth]{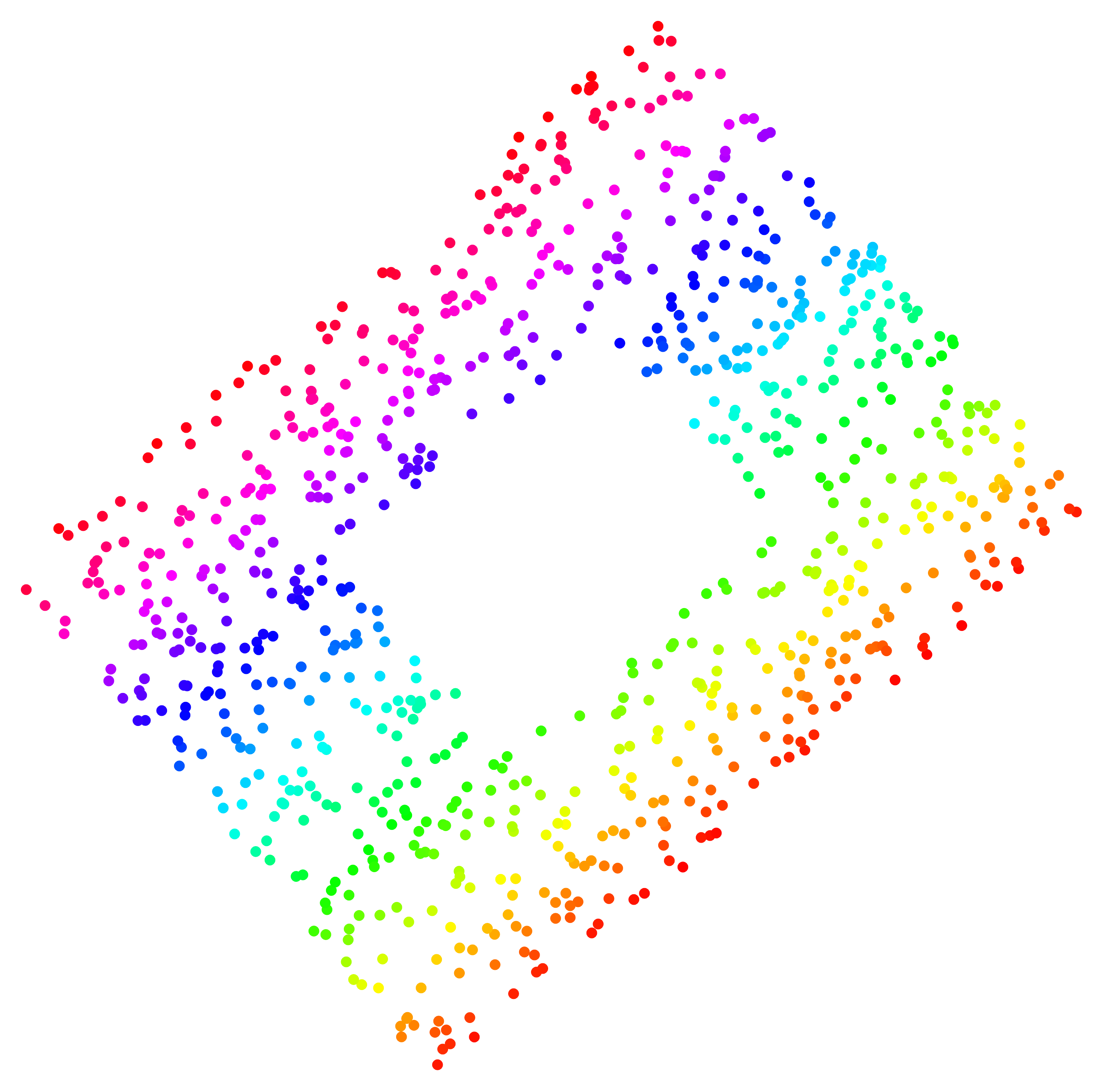} &
      \includegraphics[width=0.49\textwidth]{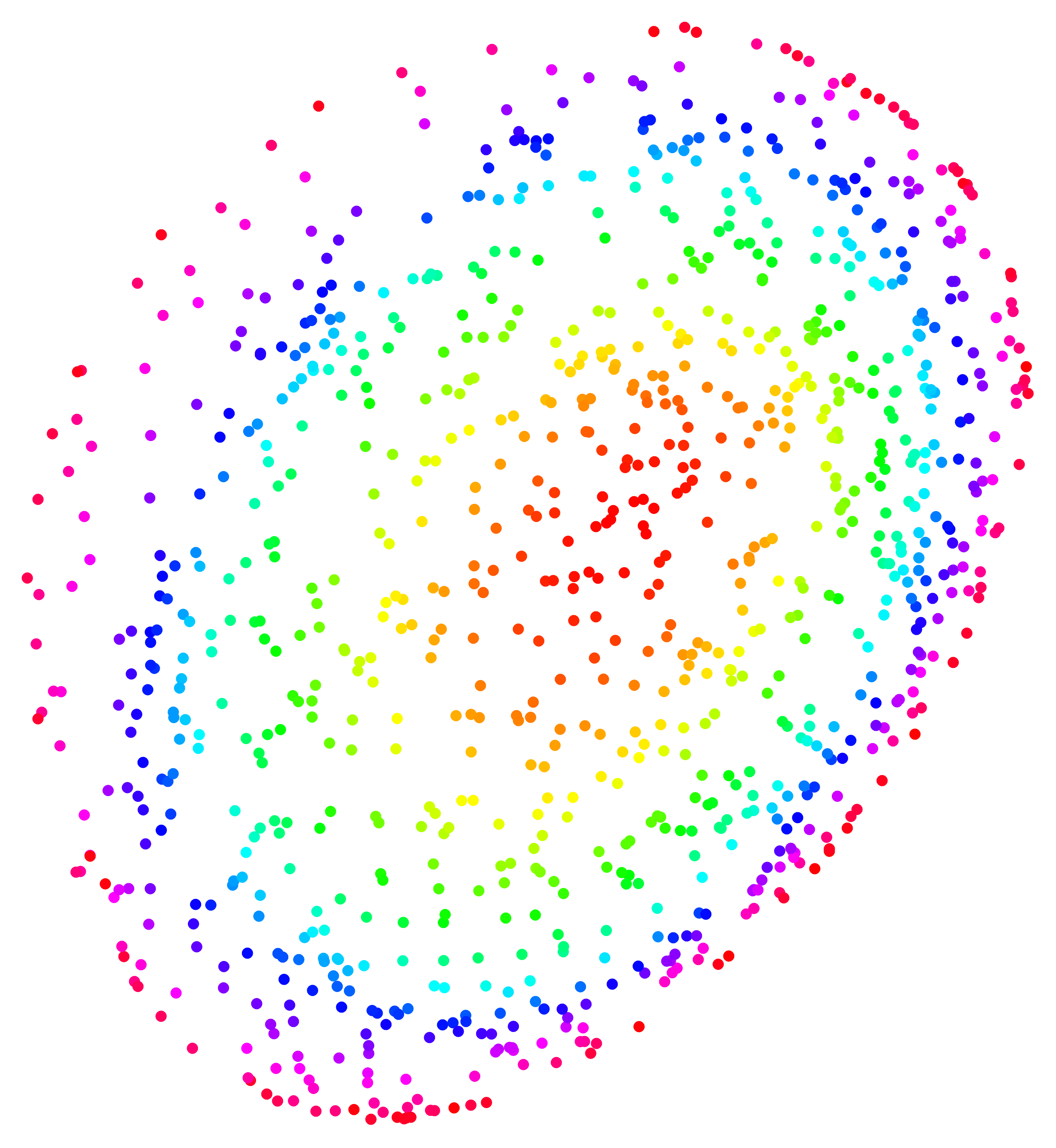} \\
      (a). Plane with hole ($l=5$) & (b). Fishbowl ($l=9$) \\
      \\

      \includegraphics[width=0.49\textwidth]{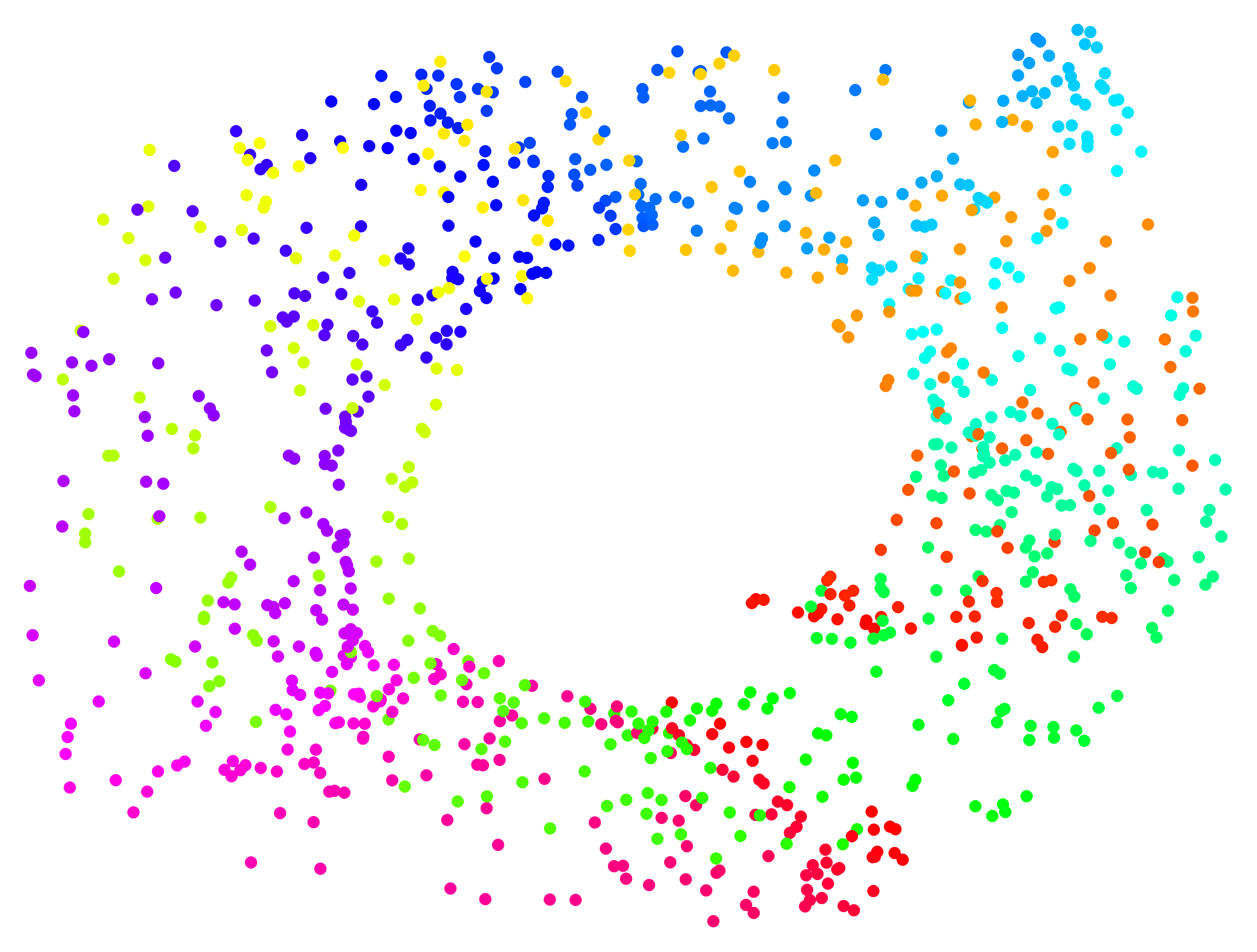} &
      \includegraphics[height=0.49\textwidth]{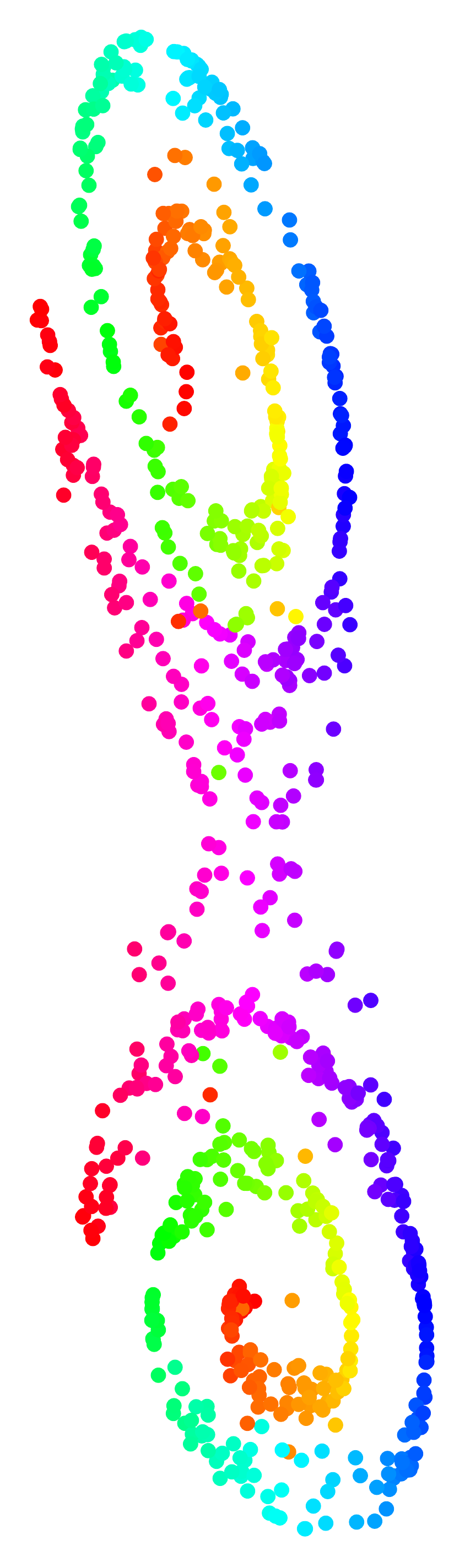} \\
      (c). Swiss roll ($l=12$) & (d). Swiss roll with hole ($l=12$)
    \end{tabular}
  \end{center}
  \caption[NLPCA reductions of geometrical test data sets]{Application
    of NLPCA to geometrical test data sets from
    Section~\ref{sec:test-data-sets} (reduced representations, i.e.,
    bottleneck neuron outputs).  The number of hidden layer neurons
    for the best fit is shown in parentheses in the subcaptions.}
  \label{fig:nlpca-test-data-sets-reduce}
\end{figure}

First, in Figure~\ref{fig:nlpca-test-data-sets-reduce}, we can look at
the effectiveness of the data reduction.  Each plot shows the
bottleneck neuron values of the NLPCA network, displayed by plotting
each point in either $\mathbb{R}$, for one-dimensional input data
manifolds, or $\mathbb{R}^2$, for two-dimensional input data manifolds
(in fact, all the reduced results shown in
Figure~\ref{fig:nlpca-test-data-sets-reduce} are two-dimensional).
Each point is labelled with the same hue as used in the
three-dimensional embedding views in Figure~\ref{fig:test-data-3d}.  A
perfect reduction for a one-dimensional data set would show points
along a line segment with hue varying smoothly from one end to the
other --- this is in fact what is seen for the one-dimensional test
data sets (helix and noisy helix: not shown).  For two-dimensional
data sets, a perfect reduction would show a clearly separated
distribution of hues, with points with the same hue appearing in
geometrical relationships equivalent to those in the original data set
with as little distortion as possible, so as to provide a good mapping
from the original data space, $\mathbb{R}^3$, to the reduced space,
$\mathbb{R}^2$.  For instance, a perfect reduction of the Swiss roll
data set in Figure~\ref{fig:test-data-3d}c would show points
distributed over a rectangular region of the plane, with the hue of
points varying smoothly from one end of the rectangle to the other,
and with points of corresponding hue lying along lines orthogonal to
the direction of variation of hue.

\begin{figure}
  \begin{center}
    \begin{tabular}{cc}
      \includegraphics[width=0.49\textwidth]{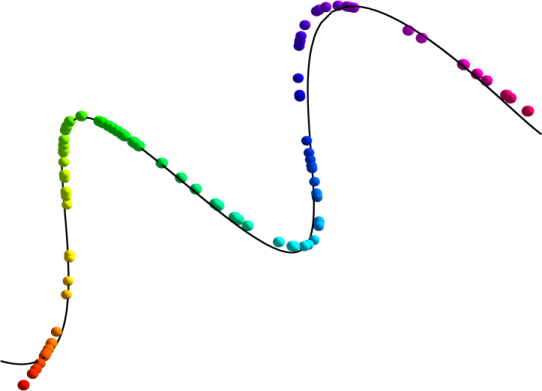} &
      \includegraphics[width=0.49\textwidth]{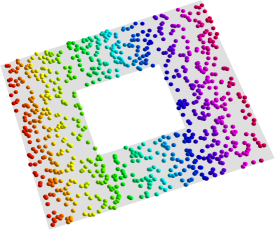} \\
      (a). Helix ($l=7$) & (b). Plane with hole ($l=5$) \\
      \\

      \includegraphics[width=0.49\textwidth]{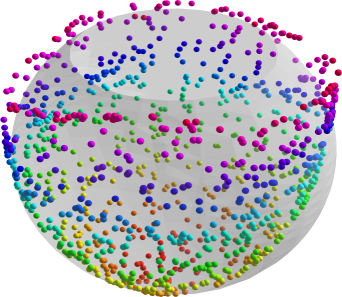} &
      \includegraphics[width=0.49\textwidth]{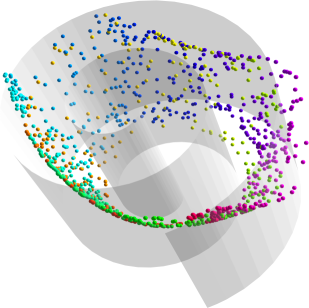} \\
      (c). Fishbowl ($l=9$) & (d). Swiss roll ($l=12$)
    \end{tabular}
  \end{center}
  \caption[NLPCA reconstructions of geometrical test data
    sets]{Application of NLPCA to geometrical test data sets from
    Section~\ref{sec:test-data-sets} (reconstructed representations:
    points show NLPCA reconstructions, original data manifolds shown
    in grey).}
  \label{fig:nlpca-test-data-sets-recon}
\end{figure}

To go with the reduced results of
Figure~\ref{fig:nlpca-test-data-sets-reduce},
Figure~\ref{fig:nlpca-test-data-sets-recon} shows three-dimensional
reconstructions of some of the data sets in the same format as
Figure~\ref{fig:test-data-3d} --- reconstructed data points are shown
as coloured points, with the same hues as used in
Figure~\ref{fig:test-data-3d}, while the underlying manifold from
which the original data points were sampled is shown in grey.

The results shown here are the result of testing with a number of
different neural network architectures (i.e. different numbers of
hidden layer neurons) and different weight penalty terms (factor $P_W$
in \eqref{eq:weight-penalty-constraint}).  The number of bottleneck
neurons in the network is set by the known intrinsic dimensionality of
each data set, i.e. one for the helix and noisy helix data sets, and
two for the two-dimensional surface examples.  The range of hidden
neuron layer counts tested and the size of the ensemble of random
initial conditions varied depending on the perceived difficulty of
fitting each data set, with some experimentation required to get good
results (the ensemble sizes varied between 10, for simple examples
like the plane, up to 50 for the Swiss roll data sets).

%
%
%

Most of the test data sets of Section~\ref{sec:test-data-sets} present
no problem for NLPCA, with good reduction and reconstruction.  See,
for example, the helix example shown in
Figure~\ref{fig:nlpca-test-data-sets-recon}a, or the plane with hole
example shown in Figures~\ref{fig:nlpca-test-data-sets-reduce}b
(reduced representation) and
Figures~\ref{fig:nlpca-test-data-sets-recon}b (reconstruction).  In
these cases, the reduction and reconstruction is either essentially
perfect (for all of the plane and plane with hole examples) or very
nearly so (for the helix), and the reduction and reconstruction are
unaffected by noise in the data.  A slightly stiffer test of the
method is presented by the fishbowl example
(Figures~\ref{fig:nlpca-test-data-sets-reduce}b~and~\ref{fig:nlpca-test-data-sets-recon}c).
Here, a perfect reduction would have a ``bullseye'' pattern with
concentric rings of points of hue varying smoothly from the centre to
the circumference.  This is more or less what is seen here, and the
reconstruction is also reasonable, although there is some distortion.
In this case, NLPCA does a significantly better job than either Isomap
(Figure~\ref{fig:isomap-test-data-sets}b) or the Hessian LLE method
(Figure~\ref{fig:hlle-test-data-sets}b).  Again, the NLPCA reduction
and reconstruction is relatively unaffected by noise added to the
input data.  The final test data sets, all based on a Swiss roll
shape, present a different picture.  Here, NLPCA has great difficulty
finding a good fit to the data manifold.  The reduced data
(Figure~\ref{fig:nlpca-test-data-sets-reduce}c) shows the adjacent
``leaves'' of the Swiss roll piled up one on top of the other,
indicating that the NLPCA reduction has not extracted the intrinsic
structure of the data manifold.  This impression is reinforced by the
reconstruction (Figure~\ref{fig:nlpca-test-data-sets-recon}d), where
the surface onto which the original data points have been mapped bears
little or no resemblance to the original data manifold.  The situation
is even worse for the Swiss roll with hole data sets.
Figure~\ref{fig:nlpca-test-data-sets-reduce}d shows reduced results
for this case.  Here, the NLPCA algorithm seems to have fallen into a
minimum of the cost function representing a mapping from the original
data to bottleneck neuron values that does not at all reflect the
intrinsic structure of the data manifold.

%
%
%

In order to gain some understanding of exactly why the Swiss roll
example proves to be such a problem for the NLPCA algorithm, I
conducted some experiments on a very simple one-dimensional example: a
segment of an Archimedean spiral embedded in the plane.  For hidden
neuron counts ranging from $l=2$ to $l=19$, NLPCA fits were done using
ensembles of 50 initial random weights for penalty values $P_W = 0,
10^{-4}, 10^{-3}, 0.01, 0.1, 1$ and the best fit (in a squared error
sense) was chosen.  Error information was collected for each ensemble
member for further processing.

\begin{figure}
  \begin{center}
    \begin{tabular}{c@{\hspace{1cm}}c}
      \includegraphics[width=0.4\textwidth]{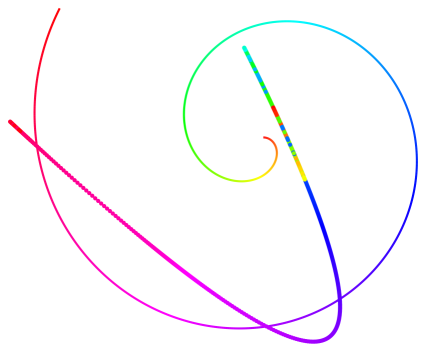} &
      \includegraphics[width=0.4\textwidth]{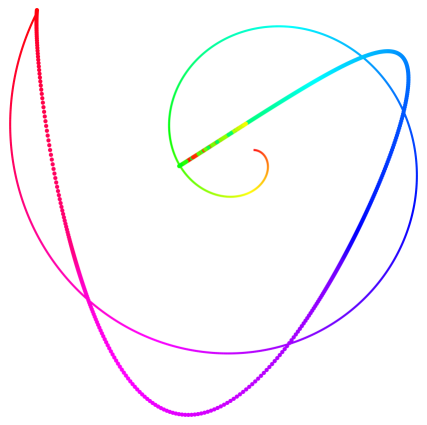} \\
      (a). $l = 2$ & (b). $l = 3$ \\
      \\

      \includegraphics[width=0.4\textwidth]{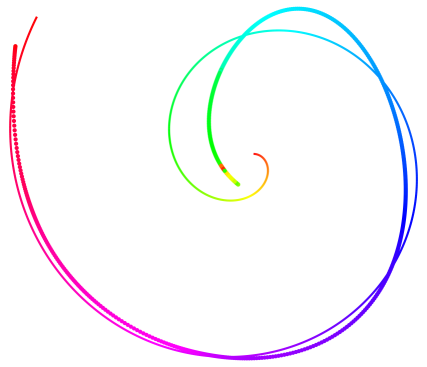} &
      \includegraphics[width=0.4\textwidth]{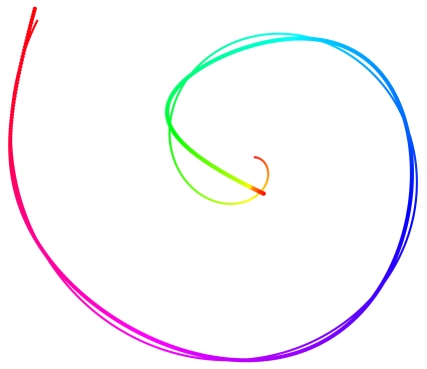} \\
      (c). $l = 4$ & (d). $l = 7$ \\
      \\

      \includegraphics[width=0.4\textwidth]{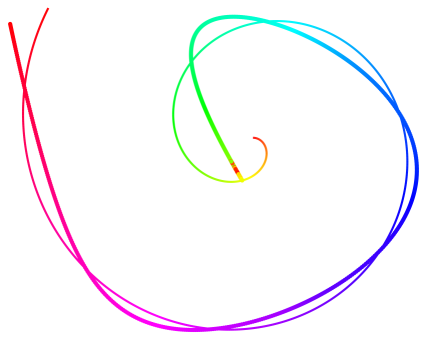} &
      \includegraphics[width=0.4\textwidth]{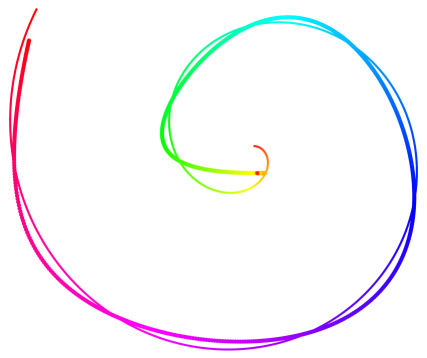} \\
      (e). $l = 8$ & (f). $l = 16$
    \end{tabular}
  \end{center}
  \caption[NLPCA spiral example reconstructions]{NLPCA reconstructions
    of a one-dimensional spiral curve embedded in $\mathbb{R}^2$, for
    different numbers of hidden layer neurons.  Each plot shows the
    best fit attained from an ensemble of 50 random initial weights,
    with the reconstructed data displayed as dots, and the original
    spiral shown as a continuous curve.}
  \label{fig:nlpca-spiral-recon}
\end{figure}

Figure~\ref{fig:nlpca-spiral-recon} shows results of this process for
a number of hidden layer counts, displaying the reconstructed spiral
as dots along with the original spiral as a continuous curve.
Increasing the number of hidden neurons used from 2
(Figure~\ref{fig:nlpca-spiral-recon}a) to 3
(Figure~\ref{fig:nlpca-spiral-recon}b) or 4
(Figure~\ref{fig:nlpca-spiral-recon}c) gives a clear increase in the
complexity of curves that can be represented by the neural network.
The $l=3$ fit has more points of inflection and already comes quite
close to most of the points on the spiral, although geometrically it
looks relatively little like a spiral.  The $l=4$ fit is the first to
look more obviously spiral-like.  This increasing geometrical
complexity of the functions produced by the neural network fitting
arises from the nature of the functions used in the network definition
\eqref{eq:nlpca-nn-eqs}.  The only nonlinearities in the network
transfer function are the hyperbolic tangent functions used in
\eqref{eq:nlpca-nn-eqs-1} and \eqref{eq:nlpca-nn-eqs-3}.  The
hyperbolic tangent function is monotonic, with $\tanh 0 = 0$,
$\lim_{x\to\pm\infty} \tanh x = \pm 1$, so it is necessary to compose
several such functions to produce a result with maxima or minima.  The
more hidden layer neurons used in the NLPCA neural network, the more
complex the functions that the network can approximate.

Once the number of hidden layer neurons used gets up to 7
(Figure~\ref{fig:nlpca-spiral-recon}d) or 8
(Figure~\ref{fig:nlpca-spiral-recon}e), the fit to the spiral data
becomes relatively good.  However, the improvement in fit as the
network complexity increases is not predictable.  The RMS error for
the $l=8$ solution (0.040) is worse than that for the $l=7$ fit
(0.020), and there is, in general, no deterministic pattern to the fit
errors as a function of network complexity, primarily because of the
finite size of the ensembles of random initial weights used in the
fitting procedure.  The plot shown for $l=16$ in
Figure~\ref{fig:nlpca-spiral-recon}f is the best fit that was achieved
over all the experiments done.  Further increase of the number of
hidden layer neurons does not lead to significantly better
reconstructions of the spiral.

%
%
%

The behaviour of the NLPCA fitting of this simple example can be
understood in terms of the structure of the cost function used for the
fitting.  Once the network has enough hidden layer neurons to
represent the basic features of the spiral data set
(Figure~\ref{fig:nlpca-spiral-recon} indicates that this is the case
for $l \gtrsim 4$), the final fit error found depends to a large
extent on the distribution of local minima of the cost function.  It
seems clear that, were it possible to reliably identify the global
minimum of the cost function, networks with more hidden layer neurons
would have an advantage.  One piece of evidence for this is the
persistent failure of all the networks tested to capture the greater
curvature of the inner segment of the spiral: a more complex network
should presumably be able to represent this structure as well as the
lower curvature outer part of the spiral.  However, in the absence of
a method for identifying global cost function minima, local search
methods based on an ensemble of initial conditions must be used.  This
can be a serious problem because, for a neural network using a
sigmoidal nonlinear function in its transfer function (equivalent, for
these purposes, to the hyperbolic tangent function used here), the
number of local minima can increase exponentially with the number of
neurons in the network \citep{sontag-nn-minima,auer-nn-minima}.  Thus,
not only does the parameter estimation problem become more delicate as
the number of hidden layer neurons is increased (involving a search in
a higher dimensional weight space) but the number of local minima in
which the minimisation may become trapped increases exponentially,
requiring exponentially more ensemble members to locate good fits, and
hence exponential time.

\begin{figure}
  \begin{center}
    \includegraphics[width=\textwidth]{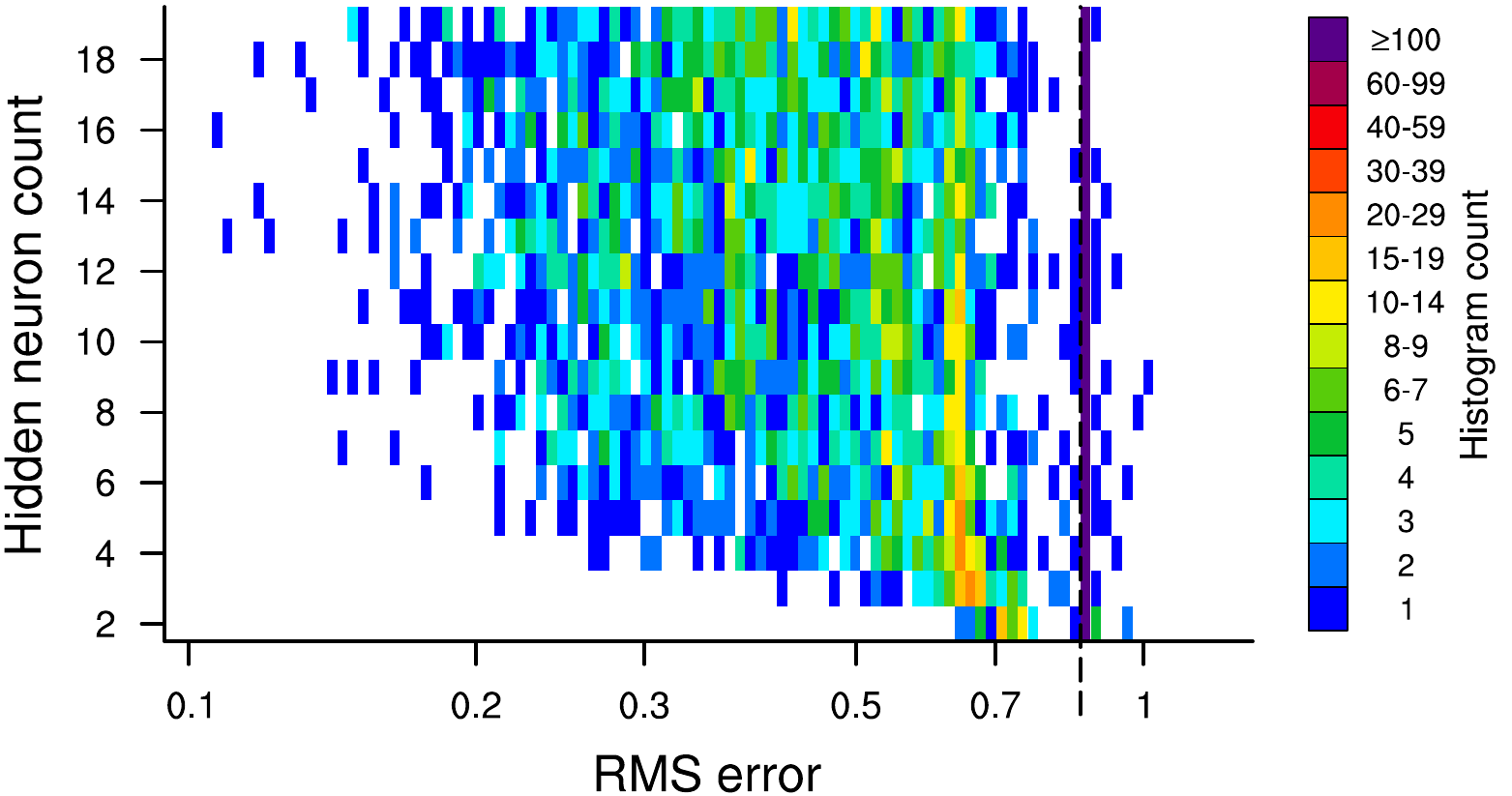}
  \end{center}
  \caption[NLPCA spiral example error histogram]{Raster histogram of
    RMS error in NLPCA reconstructions of a one-dimensional spiral
    curve embedded in $\mathbb{R}^2$, for different numbers of hidden
    layer neurons.  The histogram data is constructed from ensembles
    of a total of 300 random initial weights for each hidden layer
    neuron count (ensembles of 50 initial weights for each of six
    possible weight penalty coefficients).  The dashed black line
    shows the RMS error for a linear PCA reconstruction of the spiral
    data.  Note the logarithmic axis for the RMS error and the
    irregular scale for the histogram values.}
  \label{fig:nlpca-spiral-errors}
\end{figure}

Some further appreciation of the behaviour of the NLPCA optimisation
can be gained by examining the statistics of the error values for the
ensemble members.  Recall that the results in
Figure~\ref{fig:nlpca-spiral-recon} represent, for each hidden layer
neuron count, the best result over ensembles of 50 random initial
weights for each of the possible weight penalty values $P_W = 0,
10^{-4}, 10^{-3}, 0.01, 0.1, 1$.  In
Figure~\ref{fig:nlpca-spiral-errors}, I have plotted the ensemble
members' best fit error values for each hidden layer neuron count as a
raster histogram (note the logarithmic axis for error values and the
irregular scale for the histogram counts).  For comparison, the error
for a PCA reduction of the spiral data is shown by the thick black
dashed line.  The absolute minimum errors achieved for each network
configuration are clear from the blue boxes at the left hand edge of
the plot for each hidden layer neuron count.  The histogram shows a
composite of results for all weight penalty term multipliers, so that,
for each hidden layer neuron count, there are 300 ensemble members
shown.  Ensemble member error values are recorded in the histogram
regardless of whether or not they represent an overfitted solution.
There are two main points to draw from this.  First, a very large
proportion of ensemble members have error values very close to the PCA
reduction.  Second, the best results are really rather rare, even
though overfitted results are included, which should produce a bias
towards lower error values.

%
%
%

These results illustrate a feature of the type of neural network used
in NLPCA that is very difficult to handle in practice.  Theoretically,
under reasonable restrictions on the input data, one ought to be able
to derive an optimal nonlinear reduction and reconstruction of any
function using a neural network with sufficiently many hidden layer
neurons, according to the function approximation theorem of
\citet{cybenko-sigmoidal}.  In practice, this is impossible because of
the number of local minima in the cost function to be minimised.  In
the current instance, there are very many local minima with error
values close to those of PCA (admittedly here there is probably a bias
due to the larger $P_W$ ensembles, which are likely to lead to fits
with small weight values that may give relatively linear behaviour),
and comparatively few minima with lower errors representing good
nonlinear fits.  It would appear that the basins of attraction in the
network fitting procedure for good minima with small errors are very
small indeed, i.e. initial conditions for which the fitting procedure
converges to these good minima are rare.  This problem is exacerbated
if one attempts to use more hidden layer neurons to achieve a better
nonlinear fit.  As noted above, for a neural network using a sigmoidal
transfer function, the number of local minima can increase
exponentially with the number of neurons in the network
\citep{sontag-nn-minima,auer-nn-minima}.  It thus becomes extremely
difficult to locate good minima with larger networks.  Although
Figure~\ref{fig:nlpca-spiral-errors} indicates that more complex
networks do have more local minima with errors significantly smaller
than the PCA error compared to simpler networks, it is exponentially
more costly to locate good minima for these larger networks compared
to simpler networks.

The overall conclusion to draw from this for the purposes of applying
NLPCA to climatological data seems fairly clear, and was stated in a
more general setting earlier, in
Section~\ref{sec:nlpca-param-selection}.  It does not seem reasonable
to hold on to the hope of NLPCA being able to find any sort of optimal
nonlinear reduction of data, since the problems of overfitting and of
locating global (or even good local) minima become intractable for
complex networks.  Instead, it makes more sense to use relatively
simple networks, and to treat NLPCA as a weakly nonlinear extension to
PCA.  This is implicitly the approach that has been followed in
applications of NLPCA to climate data analysis in the literature
\citep[e.g.,][]{an-enso-interdecadal,monahan-enso,wu-enso-interdecadal}
and is the approach I follow in the computations presented here.

\section{Previous applications in climate data analysis}

NLPCA is the nonlinear dimensionality reduction method that has been
used most in climate applications.  Along with its extensions to
correlation and spectral analysis, NLPCA has been applied to questions
as diverse as tropical Pacific climate variability, regimes in
Northern Hemisphere atmospheric dynamics, and the quasi-biennial
oscillation.

The first climate application where NLPCA was used was the question of
tropical variability in the Pacific.  Application of NLPCA with one-
and two-dimensional bottleneck layers to tropical Pacific
observational SST data demonstrated that low-dimensional NLPCA
approximations can characterise variability in this data better than
linear PCA approximations, and that NLPCA approximations are able to
represent the observed \eln/\lan asymmetry
\citep{monahan-thesis,monahan-enso}.  Use of networks with a circular
bottleneck layer successfully captured the oscillatory nature of
observed ENSO-related thermocline depth variations in the equatorial
Pacific, and identified differences in the behaviour of the recharge
and discharge phases of the oscillation \citep{an-enso-interdecadal}.
There have been several applications of NLCCA, the nonlinear analogue
of canonical correlation analysis to tropical climate variability.
NLCCA determines nonlinearly coupled modes of variability in multiple
fields, and has been applied to identify nonlinear correlations
between SST, sea level pressure and wind stress variations in the
equatorial Pacific
\citep{hsieh-tropical-nlcca,wu-tropical-nlcca,wu-enso-interdecadal}.
The extension of singular spectrum analysis (SSA) to a nonlinear
neural network setting \citep{ghil-ssa,hsieh-review-2004} has been
used for the examination of periodic variability in tropical Pacific
SST by \citet{hsieh-tropical-nlssa}.

The quasi-biennial oscillation (QBO) is the most important mode of
variability in the equatorial stratosphere, manifesting itself as
quasi-periodic downwards propagation of easterly and westerly zonal
wind anomalies with a mean period of about 28 months
\citep{baldwin-qbo}.  Different NLPCA-based methods have been used to
examine the QBO in observational tropical wind data.
\citet{hamilton-nlpca-qbo} used ordinary NLPCA with a circular
bottleneck layer to analyse zonal stratospheric winds, while
\citet{hsieh-nlssa-qbo} used the nonlinear SSA method to find
interactions between a dominant 28-month QBO mode, its first harmonic
and an annual cycle mode.  The anharmonic nature of the QBO was not
represented by any individual SSA mode, but was captured by the
nonlinear SSA results, which characterise nonlinear relationships
between different SSA modes.

Low-frequency variability in atmospheric flows and the properties of
atmospheric flow regimes have been of interest for a long time
\citep[e.g.,][]{dole-gordon-regimes,ghil-robertson-regimes}.  A wide
range of statistical methods have been employed to identify and
characterise these regimes, and NLPCA appears to provide some
capability in this area.  The earliest application of NLPCA to
atmospheric flow regimes appears to be the study of
\citet{monahan-nh-regimes-1}, where NLPCA was applied to Northern
Hemisphere wintertime 500\,hPa geopotential height fields from a GCM
simulation, in order to characterise the leading nonlinear mode of
variability.  Here, the interest was in finding some means of
representing the leading patterns of atmospheric variability more
realistically than by conventional linear patterns, of which the
well-known Arctic Oscillation is the leading example for the
wintertime Northern Hemisphere troposphere \citep{thompson-ao}.  The
results indicated the existence of two rather different flow regimes,
a more persistent regime characterised by a standing oscillation
representing modulation of the climatological ridge over Northern
Europe, and a second more episodic regime with split flow south of
Greenland.  In a greenhouse gas forced climate change simulation, the
occupation probabilities of these regimes were observed to change
\citep{palmer-regimes}.  A later study based on observational data,
but using similar analysis methods \citep{monahan-nh-regimes-2}, found
three regimes in Northern Hemisphere wintertime tropospheric flow,
whose occupation frequencies exhibit substantial interdecadal
variability, some of which could be linked to ENSO variability, and
three regimes in the stratospheric flow, associated with vacillations
of the polar vortex and sudden stratospheric warmings.  Other studies
using NLPCA in this context include \citep{teng-nh-regimes} and
\citep{casty-nh-regimes}.  The latter study used NLPCA to examine
Northern Hemisphere wintertime regimes of coupled 500\,hPa
geopotential height, land surface temperatures and precipitation.  The
use of NLPCA in this area has generated some controversy.
\citet{christiansen-nlpca} raised concerns about the use of NLPCA to
decide on the existence of regimes in near-Gaussian data, based on an
earlier analysis by \citet{malthouse-nlpca-problems}.  Under some
conditions, it appears that it is possible for the NLPCA algorithm to
produce reduced manifolds that contain significantly more structure
than is justified by the input data, as shown, in particular, by
Figure~2 of \citep{christiansen-nlpca}.  Some care is certainly
required in the use of NLPCA in this setting, as highlighted by the
further correspondence of \citet{monahan-chr-response} and
\citet{christiansen-reply}, which seems to indicate that the problems
in \citeauthor{christiansen-nlpca}'s study arise from inadequate
overfitting control in the cost function minimisation procedure.

An extension of NLPCA that has proven useful in some applications uses
complex-valued networks \citep{rattan-complex-nlpca}.  These can be
used for the analysis of vector field data such as winds or currents,
where it is possible to exploit spatial correlations between different
vector components.  This approach has been applied in the study of
tropical Pacific winds \citep{rattan-complex-winds}.

Two other interesting, although not strictly climate-related
applications of NLPCA deserve mention.  \citet{herman-complex-tides}
applied NLPCA with a circular bottleneck layer to the analysis of the
temporal and spatial structure of tidal variability in a shallow
coastal sea, while \citet{delfrate-nlpca-rs-inversion} applied NLPCA
to satellite remote sensing, to help solve the inverse problem
required to retrieve temperature and water vapour profiles from
satellite radiance measurements.

\section{Application to analysis of Pacific SSTs}
\label{sec:nlpca-sst-results}

In this section, I report results of the application of NLPCA to the
analysis of Pacific SST variability in both observational and model
data.  This type of analysis has been performed for observational SST
data in the Pacific by \citet{monahan-enso} and has become a standard
problem for exploring modifications to the basic NLPCA algorithm (for
instance, \citet{hsieh-noisy-nlpca} examines the effect of using an
information-based error measure to rank NLPCA fits, using Pacific SSTs
as an example problem).  Here, I will follow the approach developed in
these earlier studies, and will apply the NLPCA method to the problem
of intercomparison of modelled ENSO behaviour in the CMIP3 models.
The main emphasis in these results is on the first NLPCA mode
extracted from the SST data, with only a little about subsequent modes
(Section~\ref{sec:nlpca-sst-mode-2}) and a two-dimensional nonmodal
analysis (Section~\ref{sec:nlpca-sst-nonmodal}).

\subsection{SST NLPCA mode 1}
\label{sec:nlpca-sst-mode-1}

Unlike some other nonlinear dimensionality reduction methods, NLPCA
cannot be used directly with observed or modelled SST data sets.  An
initial dimensionality reduction step is required to reduce the
dimensionality of the inputs so that a neural network of reasonable
size can be used.  This contrasts with methods such as Isomap
(Chapter~\ref{ch:isomap}), where gridded data sets can be handled
without preprocessing.  Here, I follow \citet{monahan-enso} and use
PCA to preprocess the SST data, projecting SST anomalies with respect
to the seasonal cycle in the region 125\degree\,E--65\degree\,W,
20\degree\,S--20\degree\,N onto the first 10 EOFs for each data set.
The proportion of the total data variance explained by these 10 EOFs
is shown, for each data set, in the final column of
Table~\ref{tab:nlpca-sst-mode-1-results}.  This initial PCA step means
that, in all cases, the input and output layers of the NLPCA networks
used here have 10 neurons, one for each EOF.

Since a one-dimensional reduction of the SST variability is required,
a single bottleneck neuron is used.  As noted in
Section~\ref{sec:nlpca-test-data}, the choice of the number of neurons
for the hidden layer is a more difficult problem.  Here again, I
follow \citet{monahan-enso}, using networks with 4 hidden layer
neurons.  This seems to give a reasonable balance between nonlinearity
and excessive overfitting to noise features in the data.  For each
data set, ensembles of 30 random initial conditions were used for each
of the weight penalty values $P_W = 0, 10^{-4}, 10^{-3}, 0.01, 0.1,
1$; 15\% of the input data was reserved as a test sample for checking
for overfitting; and the best non-overfitted solution was selected
according to a minimum squared error criterion comparing the original
input principal component values with those reconstructed by the NLPCA
network.

We first examine the observational data, where we expect to see
results close to those reported in \citet{monahan-enso}.  For both
observations and model results, there are three main things that we
look for in the NLPCA reductions.

\begin{figure}
  \begin{center}
    \includegraphics[width=\textwidth]{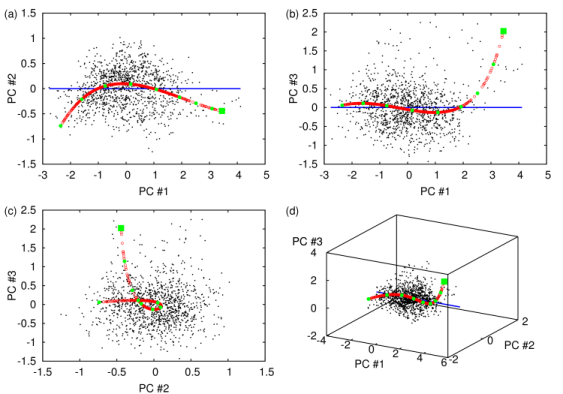}
  \end{center}
  \caption[ERSST NLPCA SST mode 1 reconstruction]{Reconstruction plots
    for NLPCA SST mode 1 for observational ERSST v2 data.  Panels
    (a)--(c) show two-dimensional projections of the reconstruction
    into, respectively, the spaces spanned by EOFs 1 and 2, EOFs 1 and
    3 and EOFs 2 and 3, while panel (d) shows a three-dimensional
    projection of the reconstruction into the space spanned by EOFs 1,
    2 and 3.  Original data points are shown as black dots, projection
    onto the first EOF is shown as a blue line, the NLPCA
    reconstructions are plotted as red circles, and points along the
    reconstruction curve with particular values of $\alpha_1$ are
    highlighted in green (the maximum $\alpha_1$ value is highlighted
    with a square, while the other highlighted values are the minimum
    of $\alpha_1$ and values at 12.5\%, 25\%, 37.5\%, 50\%, 62.5\%,
    75\% and 87.5\% of the way from the minimum to the maximum).}
  \label{fig:nlpca-sst-mode-1-ersst-recon}
\end{figure}

First, is the data variance explained by the first NLPCA mode
significantly greater than that explained by the first PCA mode,
i.e. does a single nonlinear mode do a better job of representing
variance in the data than a single linear mode?  This can be
determined from the first row of
Table~\ref{tab:nlpca-sst-mode-1-results}, which shows RMS SST errors
from PCA and NLPCA reconstructions of the input SST data, along with
explained variance fractions for the first PCA and NLPCA modes.  The
first ``PCA RMS error'' column in
Table~\ref{tab:nlpca-sst-mode-1-results} shows the RMS temperature
difference between the input SST anomaly data and the SST anomaly
field reconstructed by multiplying the first SST EOF by the first SST
principal component time series.  The second ``NLPCA RMS error''
column shows the RMS temperature difference between the input SST
anomaly data and the SST anomaly field reconstructed from the first
NLPCA mode.  This reconstructed SST field is a composite of the first
10 EOFs of the SST anomaly input data, with each EOF scaled by the
corresponding NLPCA output, of which there is one for each EOF (the
reconstruction plots below,
e.g. Figure~\ref{fig:nlpca-sst-mode-1-ersst-recon}, show projections
of these results into the space spanned by just the first three EOFs,
parameterised by PC \#1, PC \#2 and PC \#3).  The third ``RMS error
fraction'' column shows the ratio between the NLPCA and PCA RMS
errors, to give some idea of whether the temperature reconstruction
from the first NLPCA mode is significantly more faithful to the
original input data than the reconstruction using the first PCA mode
alone.  The last three columns of
Table~\ref{tab:nlpca-sst-mode-1-results} show explained variance
fractions (as a fraction of the total SST anomaly variance) for the
first PCA mode and first NLPCA mode and the total variance explained
by the first 10 EOFs used as input to the NLPCA fitting procedure.
Data sets that show a particular improvement in the representation of
SST variability using NLPCA compared to PCA are highlighted in bold
(specifically, data sets for which there is both a 1.5\% or better
reduction in the RMS temperature error, and a 1.5\% or greater
increase in the explained variance).

Comparing the PCA variance and NLPCA variance columns of
Table~\ref{tab:nlpca-sst-mode-1-results} for the observations, we see
that the NLPCA mode does slightly better than the first PCA mode
(53.9\% of total data variance explained by NLPCA SST mode 1 compared
to 52.3\% for the first PCA mode), suggesting some degree of
nonlinearity in the SST behaviour.  This is something that will become
more obvious after comparison with the model results --- as would be
expected, NLPCA does a better job of representing variability in the
input data when there is a strong nonlinear relationship between the
SST principal component time series
(cf. Figure~\ref{fig:sst-pc-scatter} and accompanying discussion in
Chapter~\ref{ch:obs-cmip-enso}).  Related to the better representation
of the SST variance, there is also a slight improvement in the RMS
temperature error of the reconstruction, with a reduction of 1.9\% for
NLPCA compared to PCA.

%
%
\begin{table}
  \begin{center}
    \input{figs/06/nlpca-sst-mode-1-results.tex}
  \end{center}
  \caption[NLPCA SST mode 1 error and variance comparison]{Error and
    variance results for NLPCA SST mode 1.  The first three data
    columns, labelled ``RMS Error'', show RMS SST errors between the
    original data and single mode PCA and NLPCA reconstructions, with
    the ``Frac.'' column showing the NLPCA error as a fraction of the
    PCA error.  The rightmost three columns show explained variance
    fractions for PCA mode 1 and NLPCA mode 1, and the total variance
    explained by the 10 principal components used as input to the
    NLPCA algorithm (explained variance is expressed as a fraction of
    the total data variance).  Entries highlighted in bold show a
    1.5\% or greater reduction in RMS SST error and a 1.5\% or greater
    increase in explained variance for NLPCA as compared to PCA.}
  \label{tab:nlpca-sst-mode-1-results}
\end{table}
%

The second thing to look at is the reconstruction of the data by
NLPCA.  As for the test data shown in
Section~\ref{sec:nlpca-test-data}, we can view the output of the NLPCA
algorithm both in terms of the reduced data (in this case, the value
of the single bottleneck neuron), or in terms of the output of the
NLPCA network, which is compared to the input data to determine the
cost of any particular solution.  The reconstruction output from the
NLPCA network has the same dimension as the input, which, in the case
here, means that the output is 10-dimensional, one value for each of
the input principal component time series.  In order to visualise
these results, Figure~\ref{fig:nlpca-sst-mode-1-ersst-recon} shows
this reconstruction for the observational data, projected into the
space spanned by the first three principal components.  On this and
subsequent reconstruction plots, the individual input data points are
shown as small black points, the first PCA mode is shown as a blue
line, and the NLPCA reconstruction is shown as red circles.  Some
points along the curve defined by the NLPCA reconstruction are
highlighted in green for use below.  The highlighted points are
defined in terms of a standardised time series \citep{monahan-enso},
$\alpha_1(t)$, derived from the bottleneck neuron values for each data
point, $u(t)$, as
\begin{equation}
  \label{eq:nlpca-alpha1-def}
  \alpha_1(t) = s \frac{u(t) - \langle u \rangle}{\sigma_u},
\end{equation}
where $\sigma_u$ is the standard deviation of the bottleneck layer
values.  The factor $s$ is a sign correction factor introduced to
remove an ambiguity inherent in the NLPCA bottleneck neuron values.
There is no intrinsic orientation assigned to the one-dimensional
reduced manifold produced by the NLPCA algorithm, and this proves to
be slightly awkward when it comes to interpreting composite maps of
SST spatial patterns for different $\alpha_1$ values.  We thus
calculate $s$ as
\begin{equation}
  s = \sgn \lambda_1\left(\argmax_t u(t)\right)
\end{equation}
where $\lambda_1(t)$ is the first principal component time series,
used as the first component of the input to the NLPCA network, and
\index[not]{signum@$\sgn x$, signum of $x$}$\sgn x$ denotes the signum
of a value $x$, i.e. $\sgn x = 1$ if $x > 0$, $-1$ if $x < 0$ and zero
otherwise.  The consequence of this choice of orientation of the NLPCA
manifold is that positive $\alpha_1$ values correspond to positive
excursions of the first principal component, and because of the choice
of normalisation used here for PCA (Section~\ref{sec:sst-pca}), this
generally corresponds to \eln conditions.  The $\alpha_1$ values
provide a convenient way to parameterise points along the curve
defined by the NLPCA reconstruction.  The points highlighted in green
in the reconstruction plots indicate the maximum $\alpha_1$ (green
square), the minimum, and a range of values between the minimum and
maximum (12.5\%, 25\%, 37.5\%, 50\%, 62.5\%, 75\% and 87.5\% of the
way from the minimum to the maximum, all shown with green circles).

%
%
%

The main feature to note from
Figure~\ref{fig:nlpca-sst-mode-1-ersst-recon} is the way that the
NLPCA reconstruction diverges quite strongly from the linear PCA
projection.  This reflects the nonlinear relationship between the
principal component time series used as input to the NLPCA algorithm.
In cases where the SST time series arises from a linear Gaussian
process, the NLPCA algorithm should not be able to find a better fit
to the data than the linear PCA fit.  In such a case, the
reconstructed manifold for the first NLPCA SST mode would lie close to
the blue line on the reconstruction plot, and the explained variance
for the first NLPCA mode and first PCA mode shown in
Table~\ref{tab:nlpca-sst-mode-1-results} would be similar.  This
situation does arise for some of the model results examined below, but
for the observations, there is a clear difference between the NLPCA
reconstruction and the linear PCA projection.

\begin{figure}
  \begin{center}
    \includegraphics[width=\textwidth]{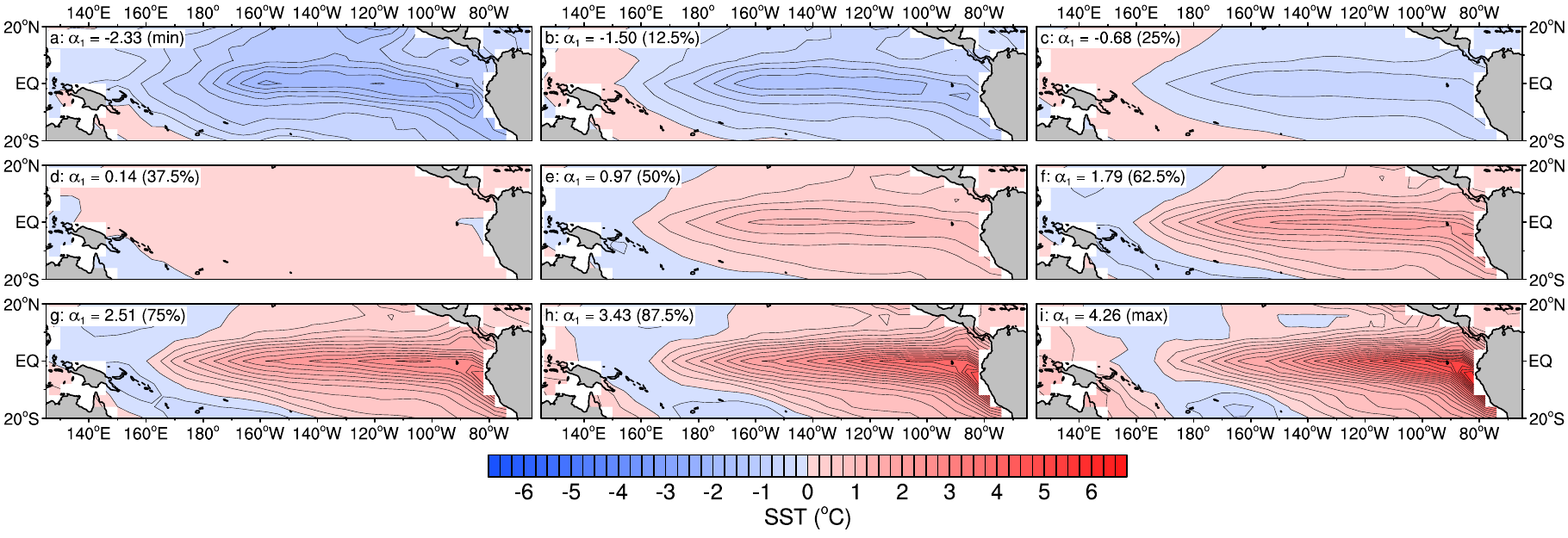}
  \end{center}
  \caption[ERSST NLPCA SST mode 1 spatial patterns]{Spatial pattern
    plots for NLPCA SST mode 1 for observational ERSST v2 data.  Each
    panel shows the SST anomaly composite formed from the point along
    the one-dimensional NLPCA reduced manifold with the corresponding
    $\alpha_1$ value.  These values are highlighted on the
    reconstruction plots in
    Figure~\ref{fig:nlpca-sst-mode-1-ersst-recon}.}
  \label{fig:nlpca-sst-mode-1-ersst-patterns}
\end{figure}

The third way of looking at the NLPCA results is to examine the
patterns of SST variability captured by the one-dimensional NLPCA
reduction.  This is the analogue of looking at map plots of EOFs in
normal linear PCA, although the situation is a little more complicated
here.  As mentioned in Section~\ref{sec:enso-phenomenology}, the
patterns of variability that can be represented by PCA are confined to
simple standing oscillations, so that the pattern of spatial
variability associated with a PCA mode can be represented by a single
EOF map.  In the NLPCA case, the range of variability that can be
expressed by a single NLPCA mode is much wider, limited only by the
range of spatial patterns spanned by all of the EOFs of the principal
component time series used as input.  This means that there is no
single map that can be displayed to express the spatial pattern of
variability of an NLPCA mode.  Instead, we can show spatial patterns
of the NLPCA reconstructions at different points along the
one-dimensional NLPCA manifold \citep{monahan-enso,hsieh-review-2004}.
At each point in time, we can form a composite, $\vec{p}(t)$ of the
first 10 EOFs of the SST data, $\vec{q}_i$, as
\begin{equation}
  \vec{p}(t) = \sum_{i=1}^{10} x'_i(t) \vec{q}_i,
\end{equation}
where $x'_i(t)$ is the $i$th reconstruction output from the NLPCA
network at time $t$.  Instead of parameterising points by time, it is
convenient to use \eqref{eq:nlpca-alpha1-def} to parameterise points
along the NLPCA manifold as $\vec{p}(\alpha_1)$.  Here, and below for
model output, we plot SST maps based on this compositing method for
the $\alpha_1$ values highlighted in each of the three-dimensional
reconstruction plots.  This gives a view of the SST variability
captured by NLPCA SST mode 1 along the length of the one-dimensional
reduced manifold.  Figure~\ref{fig:nlpca-sst-mode-1-ersst-patterns}
shows these SST map plots for the observational data.  Comparison of
the end members for minimum and maximum $\alpha_1$
(Figures~\ref{fig:nlpca-sst-mode-1-ersst-patterns}a~and~i
respectively) clearly shows the difference between a fully developed
\eln (maximum $\alpha_1$) and a fully developed \lan (minimum
$\alpha_1$).  In the \eln state
(Figure~\ref{fig:nlpca-sst-mode-1-ersst-patterns}i), there are
positive SST anomalies across the eastern part of the equatorial
Pacific, stretching southwards along the western coast of central and
South America.  In contrast, the cool SST anomalies associated with
\lan (Figure~\ref{fig:nlpca-sst-mode-1-ersst-patterns}a) are more
confined to the central equatorial Pacific, with only a weak
connection to the American continent.  This asymmetry cannot be
captured by a conventional PCA analysis --- compare the patterns shown
in Figure~\ref{fig:nlpca-sst-mode-1-ersst-patterns} with the pattern
of the first SST EOF presented in Figure~\ref{fig:sst-eofs}a on
page~\pageref{fig:sst-eofs}.  Individual PCA modes can represent only
a single spatial pattern with standing oscillations.  Other methods
have been proposed for measuring this asymmetry using positive and
negative NINO3 SST composites, such as the method of
\citet{monahan-dai} used in Section~\ref{sec:sst-asymmetry}, but NLPCA
has the nice feature of capturing a range of variability between the
asymmetric end states, something that is difficult to do using
index-based SST composites.

%
%
%

The results shown here for observational SST data match quite closely
to those reported by \citet{monahan-enso}, where the relevant figures
for comparison are Figures~3 (cf. my
Figure~\ref{fig:nlpca-sst-mode-1-ersst-recon}) and 5 (cf. my
Figure~\ref{fig:nlpca-sst-mode-1-ersst-patterns}).  The reconstruction
plots presented here are slightly smoother than those of
\citet{monahan-enso}, probably as a result of different approaches to
the regularisation of the NLPCA network transfer function, since no
weight decay term or other regulariser is used in
\citep{monahan-enso}.  These results provide some confidence that the
NLPCA method has been implemented correctly here, thus permitting
application to the CMIP3 model outputs.

%
%
\begin{figure}
  \begin{center}
    \includegraphics[width=\textwidth]{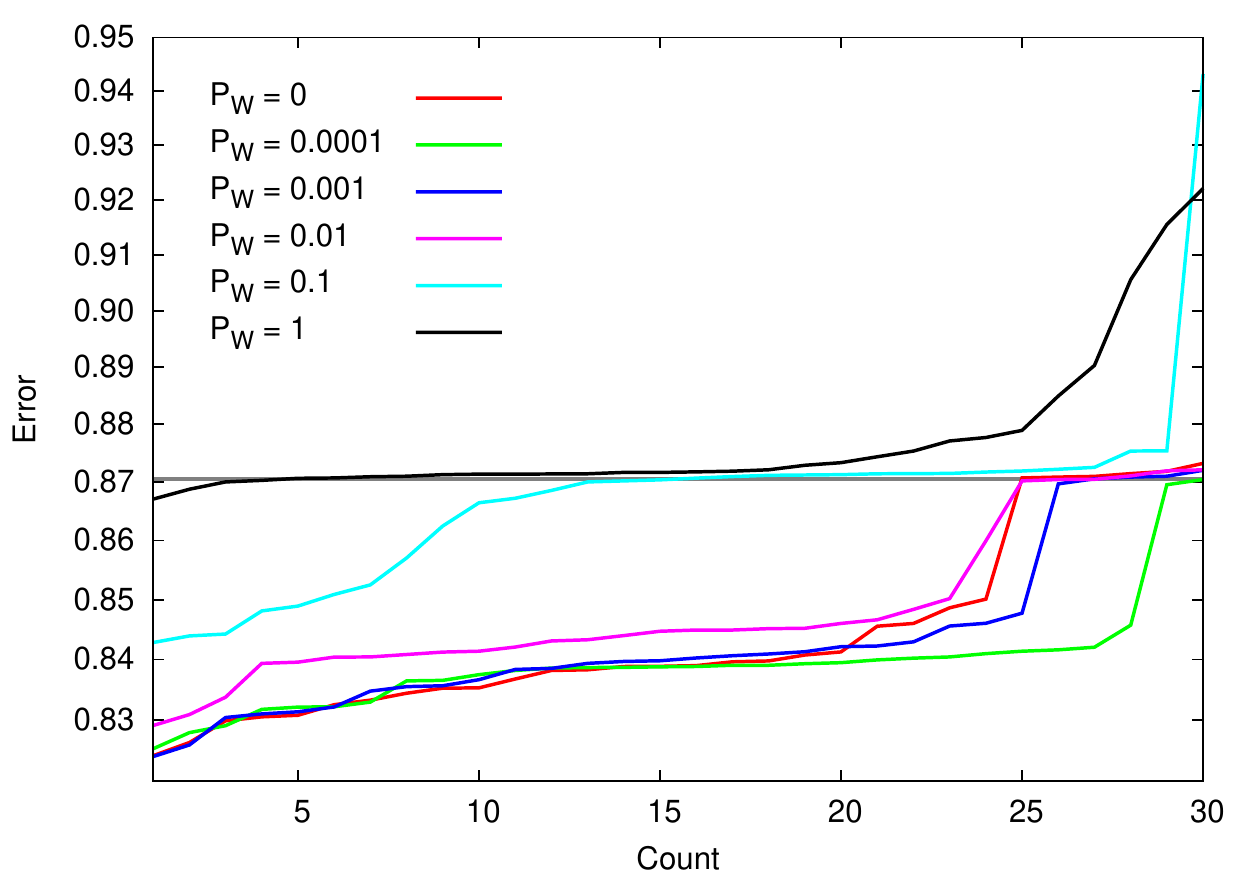}
  \end{center}
  \caption[ERSST NLPCA SST mode 1 error statistics]{Sorted RMS error
    values for NLPCA SST mode 1 fits to observational SST data.  Each
    line shows RMS error values for one weight decay penalty value,
    with error values sorted from smallest to largest.  The horizontal
    grey line shows the RMS error for a linear PCA fit to the data.}
  \label{fig:nlpca-ersst-errors}
\end{figure}
%

As for the spiral example (Figure~\ref{fig:nlpca-spiral-errors}), we
can examine error statistics across the ensembles of random initial
weights used for fitting the NLPCA network parameters.
Figure~\ref{fig:nlpca-ersst-errors} shows one view of this information
for the NLPCA SST mode 1 fit of the observational SST data.  Because
the ensemble sizes used here are smaller than the ensembles that were
possible for the simple spiral example, the error characteristics are
displayed in a different way.  For each weight decay coefficient $P_W$
used in the fitting, we plot the final RMS errors of each fit, sorted
from smallest to largest.  Each ensemble has 30 members, hence the
$x$-axis range.  The RMS error for the linear PCA fit is shown as a
horizontal grey line for comparison.  Note that the error values shown
here are the scaled values used internally to the NLPCA fitting code,
not RMS SST errors.  A few aspects of this plot deserve comment.
First, for larger weight decay coefficients, the fit is constrained to
be closer to a linear fit, giving an error for most random initial
conditions that is close to that of the linear PCA fit.  For smaller
weight decay coefficients, this constraint is lifted somewhat, and
smaller error values are attained.  However, as for the spiral
example, the most striking feature of the error statistics here is the
relative rarity of good solutions.  The overall range in error values
from the fits here is narrower than the range observed for the spiral
example, primarily because the amount of noise in the data here
precludes any very good fits.

Turning now to the analysis of modelled SSTs, consider first the
results summarised in Table~\ref{tab:nlpca-sst-mode-1-results}.  Some
of the models (highlighted in bold in the table) show substantial
improvements in the representation of SST variability using NLPCA mode
1 compared to PCA mode 1, while others (for instance, BCCR-BCM2.0 and
CGCM3.1(T63)) show very little difference in either RMS SST error or
explained variance between the NLPCA and PCA modes.  This seems to
indicate that some models have a more nonlinear response than others:
the models with a linear response have SST variability that is well
represented by the PCA mode, while more nonlinear models do not.  In
two cases, CSIRO-Mk3.0 and MIROC3.2(hires), the NLPCA fit is
significantly worse than the PCA result.  It is not completely clear
why this is the case for CSIRO-Mk3.0, although the NLPCA manifold is
rather close to the PCA linear fit, and the distribution of data
points is rather symmetrical, suggesting a situation close to linear,
in which case NLPCA would at least not be expected to do any better
than PCA.  In the case of MIROC3.2(hires), the distribution of data
points is also relatively symmetrical, but a more important factor may
be the relatively small amount of data available: the MIROC3.2(hires)
simulation is only 100 years long (Table~\ref{tab:models}) while most
of the other models have 4--5 times as much data.  Fitting complex
models like the neural networks used in NLPCA to noisy data is a
delicate process, and requires a large amount of input data to produce
a reasonable result.  It may be that the short time series for
MIROC3.2(hires) is simply not long enough for our purposes here.  In
another case, GISS-EH, the NLPCA mode is quite overfitted, as can be
seen in a reconstruction plot (not shown), where the one-dimensional
manifold produced by the NLPCA fitting has strong ``wiggles'' that,
while providing a good fit to the data, are strongly biased (in the
bias versus variance sense \citep[Section~9.1]{bishop-nn-book}).

We now concentrate on reconstruction and spatial pattern results for
five models, selected based on the results of
Table~\ref{tab:nlpca-sst-mode-1-results} and a subjective
classification of the different types of behaviour seen in the
three-dimensional reconstruction plots for each model.  For each
model, we will examine reconstruction plots and spatial pattern plots,
as was done for the observational data.

\begin{figure}
  \begin{center}
    \includegraphics[width=\textwidth]{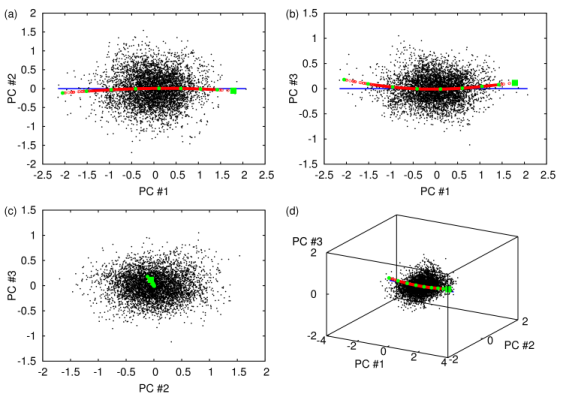}
  \end{center}
  \caption[CGCM3.1(T63) NLPCA SST mode 1
    reconstruction]{Reconstruction plots for NLPCA SST mode 1 for
    CGCM3.1(T63).  All details as for
    Figure~\ref{fig:nlpca-sst-mode-1-ersst-recon}.}
  \label{fig:nlpca-sst-mode-1-cccma_cgcm3_1_t63-recon}
\end{figure}

\begin{figure}
  \begin{center}
    \includegraphics[width=\textwidth]{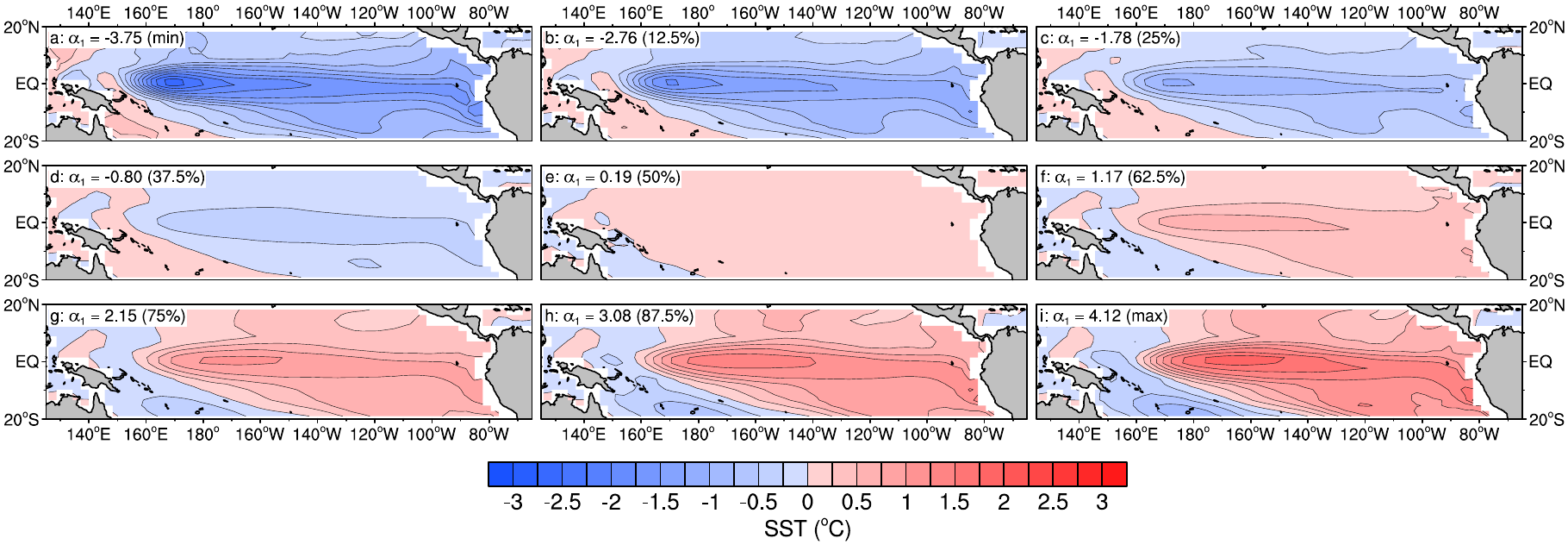}
  \end{center}
  \caption[CGCM3.1(T63) NLPCA SST mode 1 spatial patterns]{Spatial
    pattern plots for NLPCA SST mode 1 for CGCM3.1(T63), corresponding
    to highlighted points in
    Figure~\ref{fig:nlpca-sst-mode-1-cccma_cgcm3_1_t63-recon}.  All
    details are as for
    Figure~\ref{fig:nlpca-sst-mode-1-ersst-patterns}.}
  \label{fig:nlpca-sst-mode-1-cccma_cgcm3_1_t63-patterns}
\end{figure}

The first model we consider is CGCM3.1(T63), for which the
reconstruction plot is shown in
Figure~\ref{fig:nlpca-sst-mode-1-cccma_cgcm3_1_t63-recon} and the
spatial patterns of SST corresponding to the highlighted $\alpha_1$
values in
Figure~\ref{fig:nlpca-sst-mode-1-cccma_cgcm3_1_t63-patterns}.  The
results from Table~\ref{tab:nlpca-sst-mode-1-results} for this model
indicate that there is little difference between NLPCA SST mode 1 and
PCA mode 1, either in terms of the RMS SST error compared to the
original data, or in terms of the proportion of the data variance
explained.  A natural conclusion to draw would be that the SST anomaly
data from CGCM3.1(T63) has a rather symmetric Gaussian distribution
that can be easily represented by principal component analysis.  The
reconstruction plot shown in
Figure~\ref{fig:nlpca-sst-mode-1-cccma_cgcm3_1_t63-recon} confirms
this picture.  The distribution of the original data points in each of
the two-dimensional projections shown in
Figures~\ref{fig:nlpca-sst-mode-1-cccma_cgcm3_1_t63-recon}a--c are
relatively symmetrical Gaussian clouds, with only a small amount of
nonlinearity appearing in the PC \#1 versus PC \#3 plot in
Figures~\ref{fig:nlpca-sst-mode-1-cccma_cgcm3_1_t63-recon}b.  The
result is that the NLPCA reconstruction manifold is almost coincident
with the projection onto PC \#1 indicated by the blue line in the
plots.  One important consequence of this is that spatial patterns of
SST variation captured by the NLPCA mode are very symmetric between
positive and negative temperature excursions
(Figure~\ref{fig:nlpca-sst-mode-1-cccma_cgcm3_1_t63-patterns}), more
like the variation captured by a PCA mode than the nonlinear,
asymmetric variation shown in the spatial patterns for the
observational data.  In
Figure~\ref{fig:nlpca-sst-mode-1-cccma_cgcm3_1_t63-patterns}, apart
from a slight difference in the western equatorial Pacific around
160\degree\,E, the spatial patterns for maximum and minimum $\alpha_1$
are nearly identical up to sign reversal.  This is very different to
the situation in observed data, where there is a distinct spatial
asymmetry between \eln and \lan conditions.  The close correspondence
between the PCA and NLPCA results here, along with the absence of
\eln/\lan asymmetry, indicates that the interannual SST variability in
CGCM3.1(T63) is rather closer to linear than in the observations and
in some of the other models.  In fact, this model has among the
smallest NINO3 SST index variability of all the models
(Table~\ref{tab:models}).  It is no great surprise to find that this
weak variability is close to linear.

%
%
%

%
%
%

\begin{figure}
  \begin{center}
    \includegraphics[width=\textwidth]{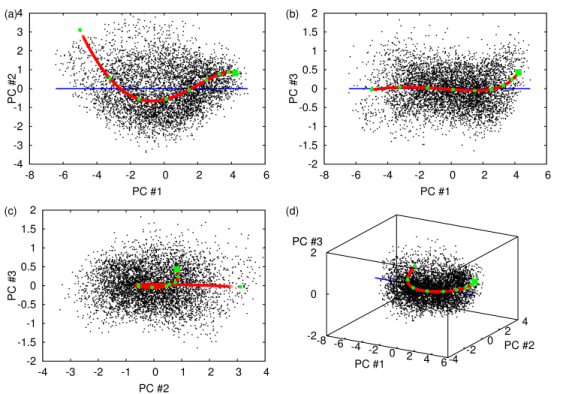}
  \end{center}
  \caption[CNRM-CM3 NLPCA SST mode 1 reconstruction]{Reconstruction
    plots for NLPCA SST mode 1 for CNRM-CM3.  All details as for
    Figure~\ref{fig:nlpca-sst-mode-1-ersst-recon}.}
  \label{fig:nlpca-sst-mode-1-cnrm_cm3-recon}
\end{figure}

\begin{figure}
  \begin{center}
    \includegraphics[width=\textwidth]{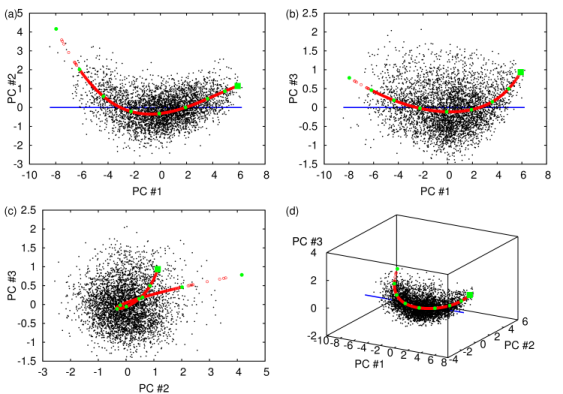}
  \end{center}
  \caption[ECHO-G NLPCA SST mode 1 reconstruction]{Reconstruction
    plots for NLPCA SST mode 1 for ECHO-G.  All details as for
    Figure~\ref{fig:nlpca-sst-mode-1-ersst-recon}.}
  \label{fig:nlpca-sst-mode-1-miub_echo_g-recon}
\end{figure}

\begin{figure}
  \begin{center}
    \includegraphics[width=\textwidth]{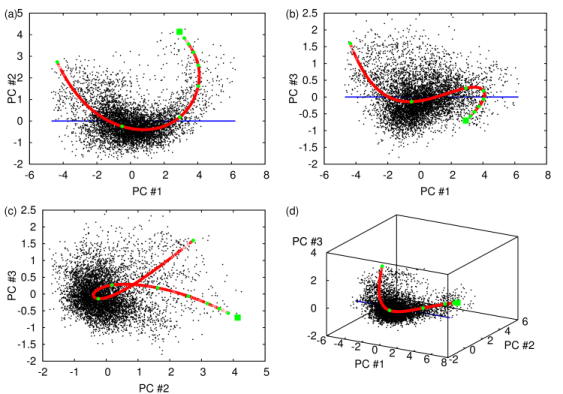}
  \end{center}
  \caption[GFDL-CM2.1 NLPCA SST mode 1 reconstruction]{Reconstruction
    plots for NLPCA SST mode 1 for GFDL-CM2.1.  All details as for
    Figure~\ref{fig:nlpca-sst-mode-1-ersst-recon}.}
  \label{fig:nlpca-sst-mode-1-gfdl_cm2_1-recon}
\end{figure}

The next three models we consider, CNRM-CM3, ECHO-G and GFDL-CM2.1,
all show a substantially better representation of their SST
variability in terms of the first NLPCA mode than in terms of the
first PCA mode.  These models are all highlighted in bold in
Table~\ref{tab:nlpca-sst-mode-1-results} to indicate this improvement,
which is seen both in the RMS SST error and in the explained variance
fractions.  The explanation for the better performance of the NLPCA
reduction here compared to PCA is exactly the converse of that for the
poor performance for CGCM3.1(T63).  For each of the models here, the
distribution of input data is markedly different from a symmetric
Gaussian cloud of points, as can be seen in the reconstruction plots,
Figure~\ref{fig:nlpca-sst-mode-1-cnrm_cm3-recon} (CNRM-CM3),
Figure~\ref{fig:nlpca-sst-mode-1-miub_echo_g-recon} (ECHO-G) and
Figure~\ref{fig:nlpca-sst-mode-1-gfdl_cm2_1-recon} (GFDL-CM2.1).  The
asymmetry is similar to that seen in the observational data in
Figure~\ref{fig:nlpca-sst-mode-1-ersst-recon}, although of greater
magnitude, particularly for GFDL-CM2.1.  In these more nonlinear
cases, a reconstruction using a single PCA mode is a poor
approximation to the input data (compare how poorly the blue lines in
the reconstruction plots match the original input data, shown as black
points), and the greater freedom available to the NLPCA network allows
it to produce a nonlinear fit that does a better job of representing
the variability in the input data.

%
%
%

%
%
%

%
%
%

\begin{figure}
  \begin{center}
    \includegraphics[width=\textwidth]{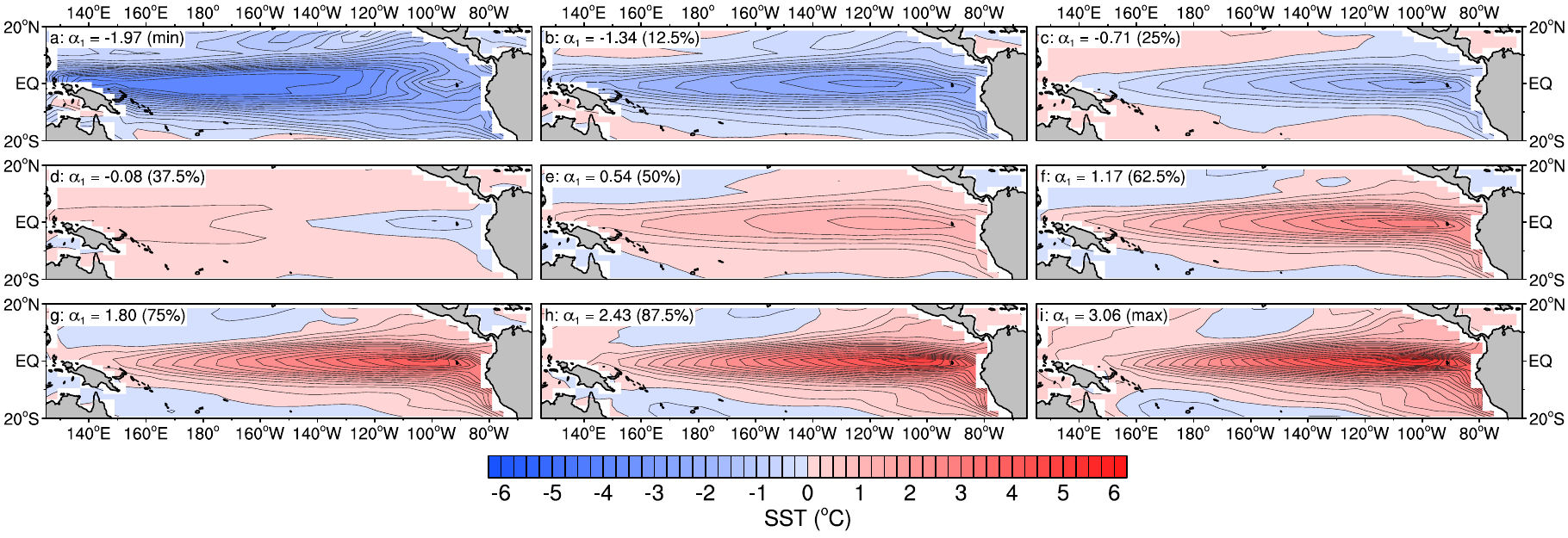}
  \end{center}
  \caption[CNRM-CM3 NLPCA SST mode 1 spatial patterns]{Spatial pattern
    plots for NLPCA SST mode 1 for CNRM-CM3, corresponding to
    highlighted points in
    Figure~\ref{fig:nlpca-sst-mode-1-cnrm_cm3-recon}.  All details are
    as for Figure~\ref{fig:nlpca-sst-mode-1-ersst-patterns}.}
  \label{fig:nlpca-sst-mode-1-cnrm_cm3-patterns}
\end{figure}

\begin{figure}
  \begin{center}
    \includegraphics[width=\textwidth]{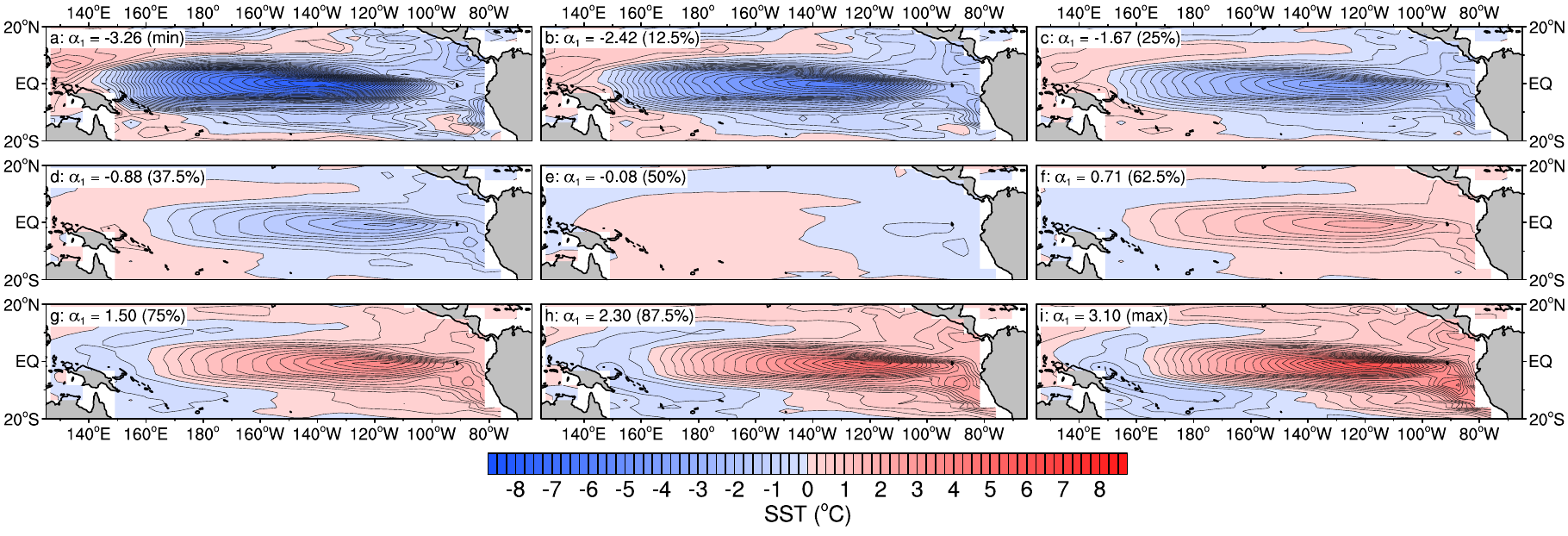}
  \end{center}
  \caption[ECHO-G NLPCA SST mode 1 spatial patterns]{Spatial pattern
    plots for NLPCA SST mode 1 for ECHO-G, corresponding to
    highlighted points in
    Figure~\ref{fig:nlpca-sst-mode-1-miub_echo_g-recon}.  All details
    are as for Figure~\ref{fig:nlpca-sst-mode-1-ersst-patterns}.}
  \label{fig:nlpca-sst-mode-1-miub_echo_g-patterns}
\end{figure}

\begin{figure}
  \begin{center}
    \includegraphics[width=\textwidth]{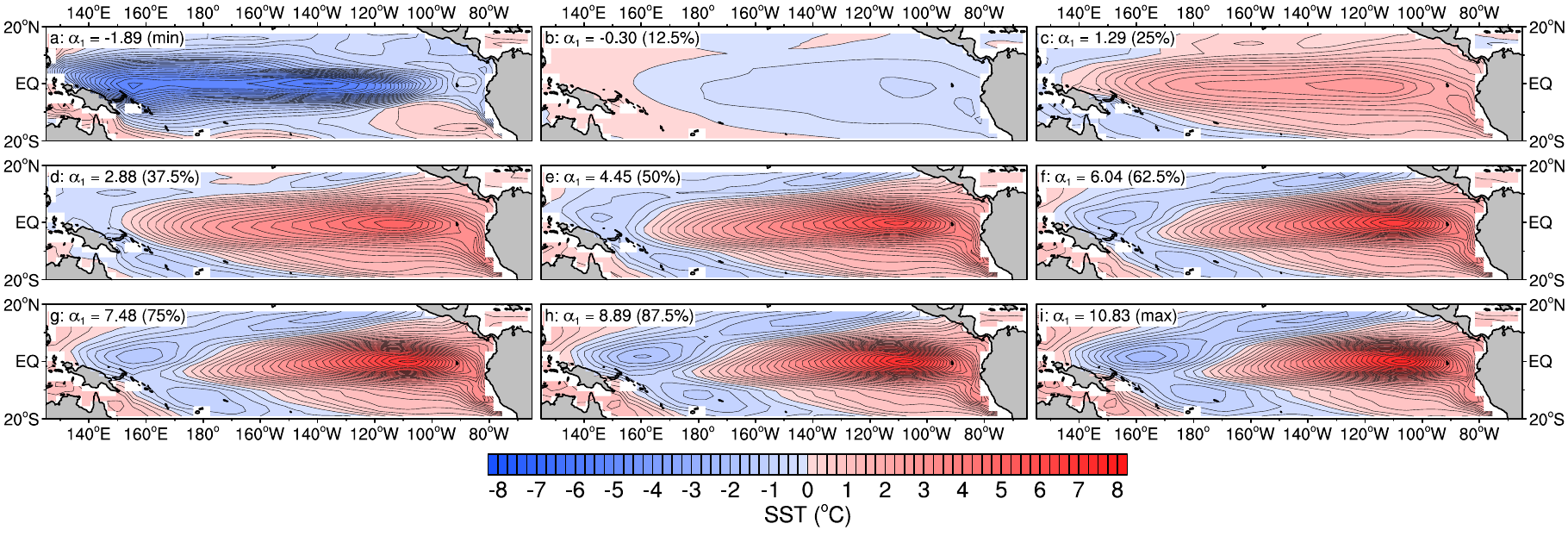}
  \end{center}
  \caption[GFDL-CM2.1 NLPCA SST mode 1 spatial patterns]{Spatial
    pattern plots for NLPCA SST mode 1 for GFDL-CM2.1, corresponding
    to highlighted points in
    Figure~\ref{fig:nlpca-sst-mode-1-gfdl_cm2_1-recon}.  All details
    are as for Figure~\ref{fig:nlpca-sst-mode-1-ersst-patterns}.}
  \label{fig:nlpca-sst-mode-1-gfdl_cm2_1-patterns}
\end{figure}

As before, the nonlinearity in the manifold reconstructed by NLPCA is
manifested clearly in the SST composites constructed for points along
the manifold.  The same spatial SST composites as shown for the
observations and for CGCM3.1(T63) are shown for these three models in
Figure~\ref{fig:nlpca-sst-mode-1-cnrm_cm3-patterns} (CNRM-CM3),
Figure~\ref{fig:nlpca-sst-mode-1-miub_echo_g-patterns} (ECHO-G) and
Figure~\ref{fig:nlpca-sst-mode-1-gfdl_cm2_1-patterns} (GFDL-CM2.1).
For all of these models, there is a distinct asymmetry between the
\eln end-point pattern (warm temperatures, maximum $\alpha_1$, panel i
in the plots) and the \lan end-point pattern (cool temperatures,
minimum $\alpha_1$, panel a in the plots).  Here, the results for
CNRM-CM3 and ECHO-G are quite comparable to those for the
observational data.  The patterns for both models are shifted somewhat
to the west compared to those for the observations, but the meridional
extent of the \eln and \lan patterns is quite good, as is the overall
range of SST variability seen for CNRM-CM3, although this is a little
too large for ECHO-G.  The situation with GFDL-CM2.1 is slightly
different and quite interesting.  The spatial patterns are not bad
compared to the observations, with some of the same slight
deficiencies as seen for CNRM-CM3 and ECHO-G, but the stronger
nonlinearity seen in the reconstruction plot for GFDL-CM2.1
(Figure~\ref{fig:nlpca-sst-mode-1-gfdl_cm2_1-recon}a) skews the
distribution of $\alpha_1$ values along the NLPCA reduced manifold,
with most of them clustered near the maximum value (the green square
in Figure~\ref{fig:nlpca-sst-mode-1-gfdl_cm2_1-recon}).  This skew
leads to a concomitant skew in the spatial patterns for GFDL-CM2.1,
with a strong imbalance towards warm conditions for all but the most
extremely negative $\alpha_1$ values
(Figure~\ref{fig:nlpca-sst-mode-1-gfdl_cm2_1-patterns}).  This effect,
which is purely an artefact of the NLPCA reduction algorithm,
highlights a potential problem with this type of neural network based
method.  The $\alpha_1$ values used to parameterise points along the
reduced NLPCA manifold are derived from the bottleneck neuron layer
values produced by the NLPCA network for each input data point.  The
bottleneck neuron values $u(t)$ are a strongly nonlinear function of
the input principal component time series, and there is no particular
reason why they should provide a uniform parameterisation along the
reduced manifold.  The $u(t)$ do provide a coordinate for the
manifold, but there is no guarantee that this has a simple
(particularly a linear) relationship with the original input data
coordinates.  If a set of reference points uniformly distributed along
the reduced manifold were required, the manifold could be
parameterised by arclength, instead of $\alpha_1$, as done by
\citet{newbigging-nlpca-param}.  Since GFDL-CM2.1 is the only model
for which this proved to be an issue, I have not done this, but it
would be a straightforward modification to the NLPCA data
post-processing.

%
%
%

%
%
%

%
%
%

\begin{figure}
  \begin{center}
    \includegraphics[width=\textwidth]{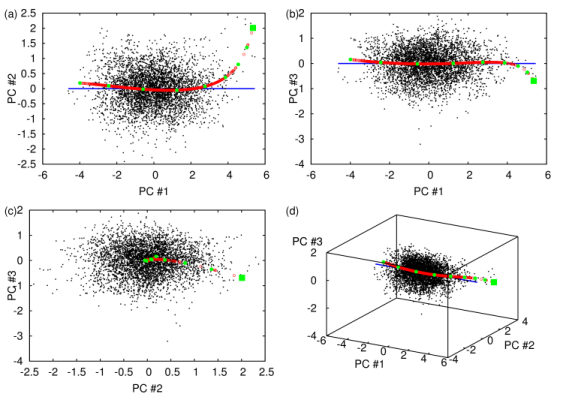}
  \end{center}
  \caption[UKMO-HadCM3 NLPCA SST mode 1 reconstruction]{Reconstruction
    plots for NLPCA SST mode 1 for UKMO-HadCM3.  All details as for
    Figure~\ref{fig:nlpca-sst-mode-1-ersst-recon}.}
  \label{fig:nlpca-sst-mode-1-ukmo_hadcm3-recon}
\end{figure}

\begin{figure}
  \begin{center}
    \includegraphics[width=\textwidth]{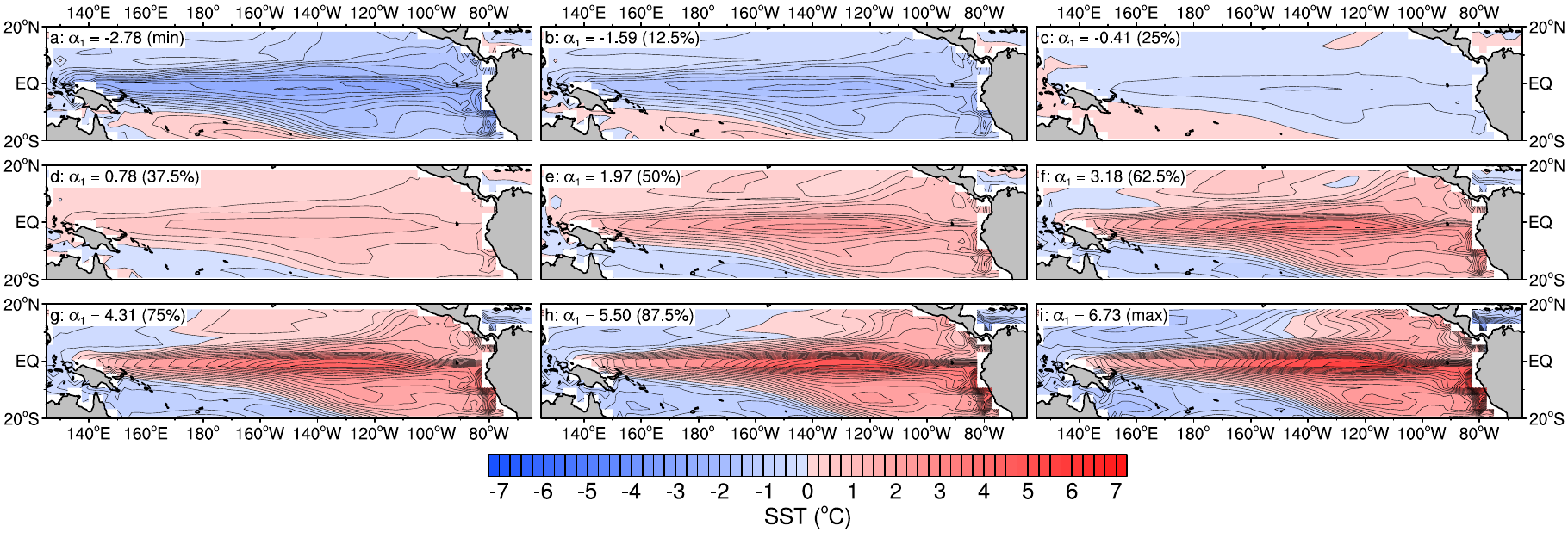}
  \end{center}
  \caption[UKMO-HadCM3 NLPCA SST mode 1 spatial patterns]{Spatial
    pattern plots for NLPCA SST mode 1 for UKMO-HadCM3, corresponding
    to highlighted points in
    Figure~\ref{fig:nlpca-sst-mode-1-ukmo_hadcm3-recon}.  All details
    are as for Figure~\ref{fig:nlpca-sst-mode-1-ersst-patterns}.}
  \label{fig:nlpca-sst-mode-1-ukmo_hadcm3-patterns}
\end{figure}

The final model we will consider here is UKMO-HadCM3.  The main reason
for choosing this model is that its reconstruction plot
(Figure~\ref{fig:nlpca-sst-mode-1-ukmo_hadcm3-recon}) is subjectively
the closest in appearance, in terms of the degree of nonlinearity
seen, to the reconstruction plot for the observations
(Figure~\ref{fig:nlpca-sst-mode-1-ersst-recon}), although the exact
shape of the reduced NLPCA manifold is different.  There is apparently
relatively little difference between the performance of NLPCA and PCA
for this model, based on the results shown in
Table~\ref{tab:nlpca-sst-mode-1-results}, but the spatial patterns in
Figure~\ref{fig:nlpca-sst-mode-1-ukmo_hadcm3-patterns} seem to
indicate at least a degree of asymmetry between \eln and \lan, with
minimum and maximum $\alpha_1$ end-point patterns
(Figures~\ref{fig:nlpca-sst-mode-1-ukmo_hadcm3-patterns}a~and~i)
having rather distinct spatial distributions of SST anomalies.

%
%
%

%
%
%

\subsection{SST NLPCA mode 2}
\label{sec:nlpca-sst-mode-2}

Using the original data vectors, $\vec{x}_i$, used to calculate NLPCA
mode 1, and the corresponding data reconstructions, $\vec{x}'_i$, we
can calculate residual vectors $\tilde{\vec{x}}_i = \vec{x}_i -
\vec{x}'_i$, and can then apply NLPCA to the $\tilde{\vec{x}}_i$ to
find NLPCA mode 2.  Subsequent modes can be calculated in the same
fashion, using the residuals from each NLPCA reconstruction to feed
the NLPCA network for finding the next mode.  The NLPCA reconstruction
is a nonlinear function of the inputs, so the original data cannot be
reconstructed by a linear combination of the NLPCA modes, as is the
case for PCA.  Despite this, as noted in
Section~\ref{sec:pca-vs-nlpca}, the NLPCA modes do satisfy the same
kind of variance partitioning property as linear PCA modes, and it
makes sense to think of a step-wise decomposition of the input data
into a sequence of NLPCA modes.

In \citep{monahan-enso}, the second NLPCA mode of observed equatorial
Pacific SST anomalies is found to represent aspects of ENSO
variability not captured by the first NLPCA mode.  In particular, the
$\alpha_2(t)$ time series, defined in an analogous way to
$\alpha_1(t)$ in \eqref{eq:nlpca-alpha1-def}, is quite non-stationary,
with stronger contributions towards the later portion of the data.
Here, I present NLPCA SST mode 2 results for the observational data
set I use and a small number of models.  The NLPCA network
architectures used for calculating NLPCA SST mode 2 are essentially
the same as those of \citet{monahan-enso}: three hidden layer neurons,
a single bottleneck neuron, and with all training and fitting
parameters identical to the configuration used for calculating NLPCA
SST mode 1.

%
%
\begin{table}[b]
  \begin{center}
    \begin{tabular}{lccccccc}
      \toprule
      {\bf Model} & {\bf PC1} & {\bf PC2} & {\bf 10 PCs} & {\bf NL1} &
      {\bf NL2} & {\bf PC1+2} & {\bf NL1+2} \\
      \midrule
      Observations & 52.3\% &  9.4\% & 88.5\% & 53.9\% & 7.0\% & 61.7\% & 60.9\% \\
      CNRM-CM3     & 62.8\% & 13.1\% & 88.7\% & 64.3\% & 9.7\% & 75.9\% & 75.0\% \\
      ECHO-G       & 58.8\% &  5.9\% & 79.7\% & 60.6\% & 4.7\% & 64.7\% & 65.3\% \\
      GFDL-CM2.1   & 59.8\% & 10.5\% & 87.3\% & 64.8\% & 6.4\% & 70.3\% & 71.2\% \\
      \bottomrule
    \end{tabular}
  \end{center}
  \caption[SST mode explained variance fractions]{Explained variance
    fractions for first ({\bf PC1}) and second ({\bf PC2}) principal
    components, for first ({\bf NL1}) and second ({\bf NL2}) NLPCA SST
    modes, along with the total explained variance for the first ten
    SST principal components ({\bf 10 PCs}) used as input to the NLPCA
    algorithm, and the variance explained by the first two PCs
    together ({\bf PC1+2}) and first two NLPCA modes together ({\bf
      NL1+2}), for comparison purposes.}
  \label{tab:sst-modes-1-2-variance}
\end{table}
%

\begin{figure}
  \begin{center}
    \includegraphics[width=\textwidth]{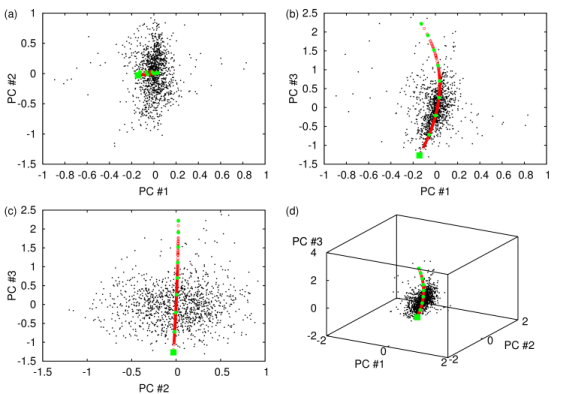}
  \end{center}
  \caption[ERSST NLPCA SST mode 2 reconstruction]{Reconstruction plots
    for NLPCA SST mode 2 for observational ERSST v2 data.  Panels
    (a)--(c) show two-dimensional projections of the reconstruction
    into, respectively, the spaces spanned by EOFs 1 and 2, EOFs 1 and
    3 and EOFs 2 and 3, while panel (d) shows a three-dimensional
    projection of the reconstruction into the space spanned by EOFs 1,
    2 and 3.  Original data points are shown as black dots, the NLPCA
    reconstructions are plotted as red circles, and points along the
    reconstruction curve with particular values of $\alpha_2$ are
    highlighted in green (the maximum $\alpha_2$ value is highlighted
    with a square, while the other highlighted values are the minimum
    of $\alpha_2$ and values at 12.5\%, 25\%, 37.5\%, 50\%, 62.5\%,
    75\% and 87.5\% of the way from the minimum to the maximum).}
  \label{fig:nlpca-sst-mode-2-ersst-recon}
\end{figure}

For the observational SST data, NLPCA SST mode 2 explains 7.0\% of the
total data variance, compared to 53.9\% for NLPCA SST mode 1
(Table~\ref{tab:sst-modes-1-2-variance}).  The first two NLPCA modes
between them explain marginally less of the total data variance than
do the first two PCA modes.  This may indicate that, despite the
observed nonlinearity in the original input data (visible on the
reconstruction plot for the first NLPCA mode in
Figure~\ref{fig:nlpca-sst-mode-1-ersst-recon}), any important
nonlinearity is confined to the space spanned by the first two SST
EOFs.  The reconstruction plot for the NLPCA SST mode 2
(Figure~\ref{fig:nlpca-sst-mode-2-ersst-recon}) shows some
nonlinearity remaining in the residuals from the fitting of the first
NLPCA mode, with a rather asymmetric distribution of outlying points.
The NLPCA fit is quite close to a straight line through the main axis
of variability in the data (note the differing scales on the axes of
Figures~\ref{fig:nlpca-sst-mode-2-ersst-recon}a--c), due in part to
the choice of network structure, which is restricted to three hidden
layer neurons only.

%
%
%

\begin{figure}
  \begin{center}
    \includegraphics[width=\textwidth]{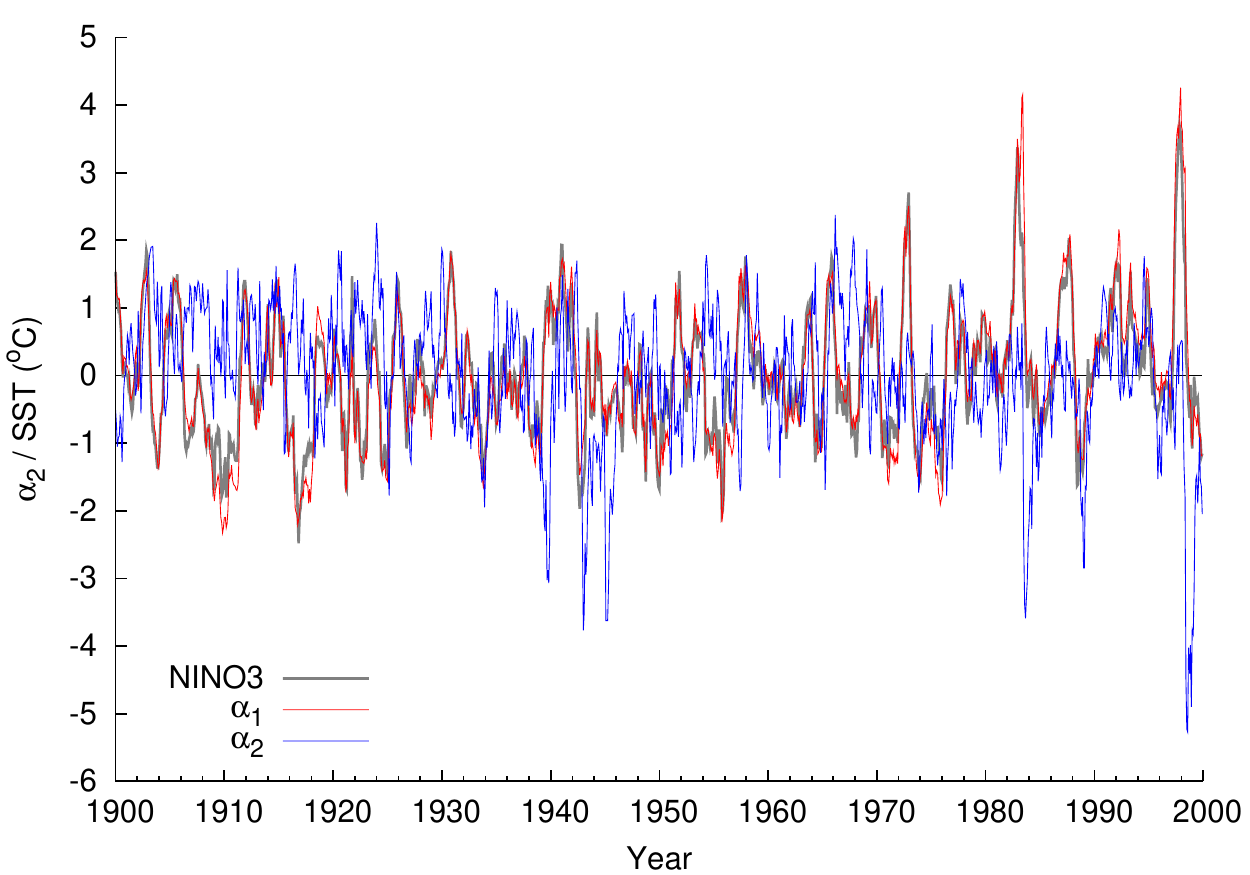}
  \end{center}
  \caption[ERSST NLPCA $\alpha_2$ time series]{Time series plot of
    $\alpha_2$ standardised bottleneck value for NLPCA SST mode 2 for
    observational ERSST v2 data (blue), with corresponding $\alpha_1$
    values for NLPCA SST mode 1 (red) and NINO3 SST index from same
    data set (grey) shown for reference.}
  \label{fig:ersst-nlpca-alpha-2}
\end{figure}

\begin{figure}
  \begin{center}
    \includegraphics[width=\textwidth]{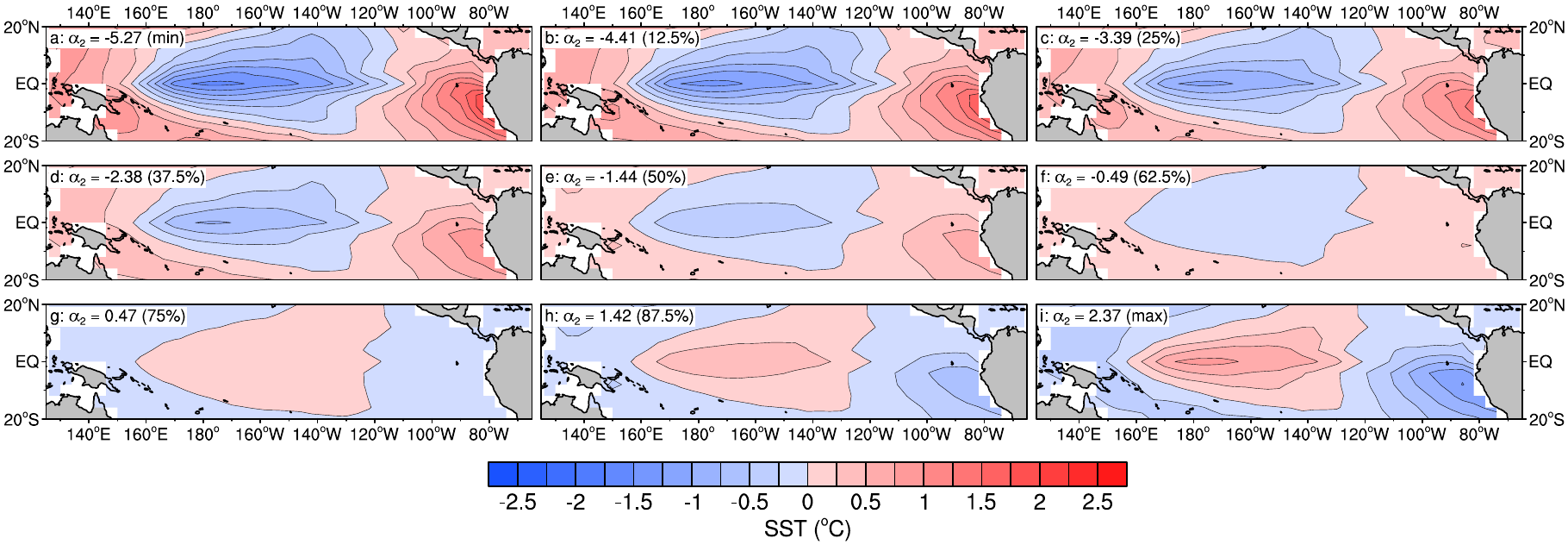}
  \end{center}
  \caption[ERSST NLPCA SST mode 2 spatial patterns]{Spatial pattern
    plots for NLPCA SST mode 2 for observational ERSST v2 data.  Each
    panel shows the SST anomaly composite formed from the point along
    the one-dimensional NLPCA reduced manifold with the corresponding
    $\alpha_2$ value.  These values are highlighted on the
    reconstruction plots in
    Figure~\ref{fig:nlpca-sst-mode-2-ersst-recon}.}
  \label{fig:nlpca-sst-mode-2-ersst-patterns}
\end{figure}

\begin{figure}
  \begin{center}
    \includegraphics[width=\textwidth]{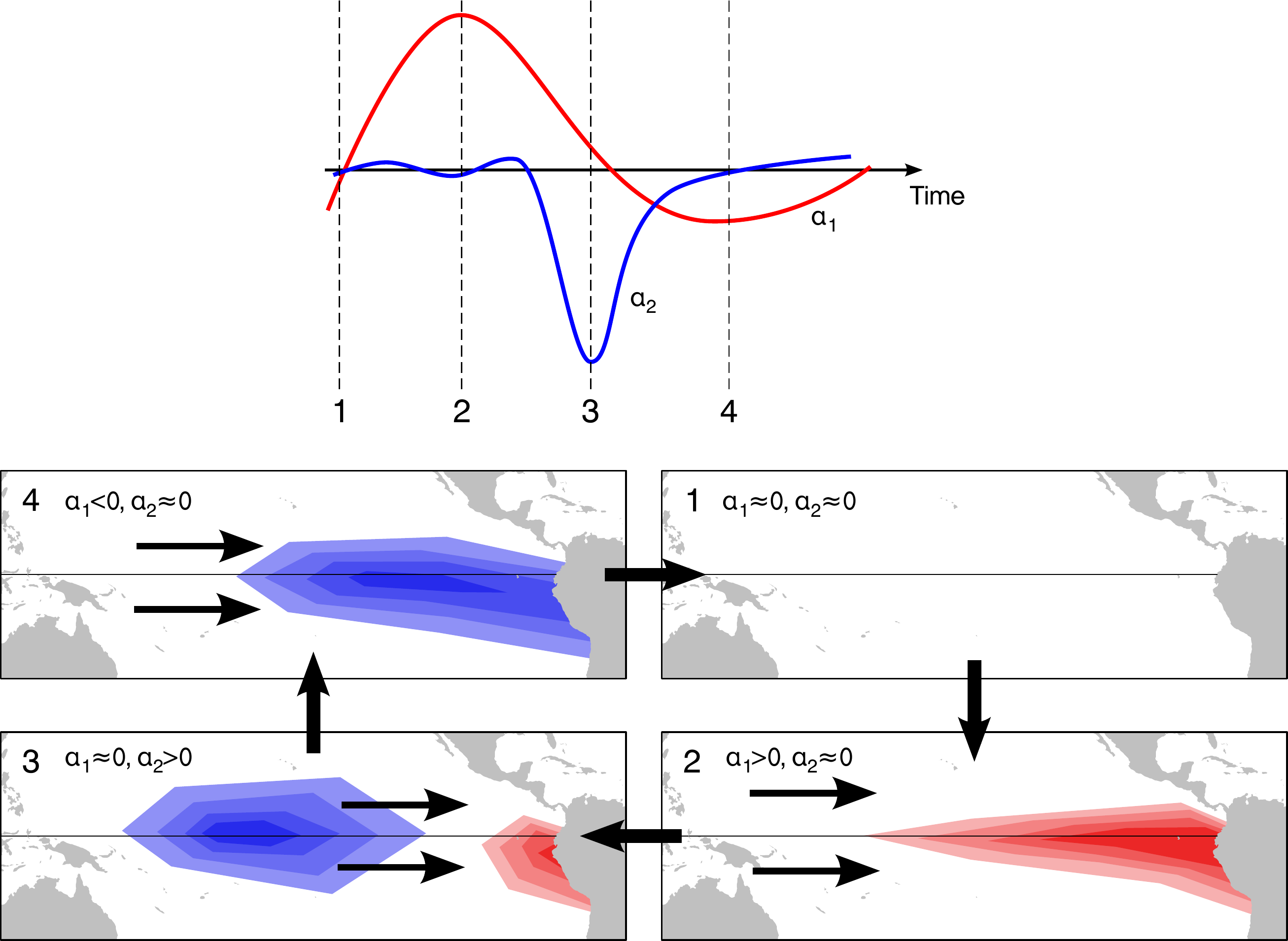}
  \end{center}
  \caption[Strong ENSO events in ERSST data and NLPCA SST mode
    2]{Cartoon view of the evolution of $\alpha_1$ and $\alpha_2$ and
    associated spatial patterns of SST anomalies during strong \eln
    events in the observational ERSST v2 data.  The upper graph shows
    the time evolution of $\alpha_1$ and $\alpha_2$ during a strong
    \eln, with the characteristic negative excursion in $\alpha_2$ at
    the end of the event.  The bottom panels show spatial patterns of
    SST anomalies at four points through the time series plot, with
    warm SST anomalies in red and cool anomalies in blue.  The arrows
    on the maps indicate the evolution of anomalies through time.}
  \label{fig:ersst-strong-events}
\end{figure}

One interesting feature of the fit determined by NLPCA is revealed by
the time series of $\alpha_2$ (Figure~\ref{fig:ersst-nlpca-alpha-2}).
This, as previously observed in \citet{monahan-enso}, displays notable
non-stationarity, with large peaks in the 1940s and the later part of
the time series associated with strong \eln events.  This would lead
us to conclude that, in this instance, NLPCA SST mode 2 captures some
aspect of ENSO variability that is expressed more in stronger \eln
events, particularly in the events that have occurred since the
mid-1970s.  It is not impossible that this is related to the recently
discovered Modoki mode of ENSO variability reported by
\citet{ashok-modoki}.  Careful examination of the time series plot in
Figure~\ref{fig:ersst-nlpca-alpha-2} shows a clear association between
large negative excursions of $\alpha_2$ and the end phase of large
\eln events (for instance, in 1987/88, 1982/83 and 1997/98, as well as
during the long anomalous period between 1939 and about 1944).
Examination of the spatial SST patterns associated with different
values of $\alpha_2$
(Figure~\ref{fig:nlpca-sst-mode-2-ersst-patterns}) goes some way to
explaining this.  The main pattern of variation between large negative
and positive values of $\alpha_2$ is a dipole with centres of action
on the equator at around 170\degree\,W and near the coast of South
America at around 10\degree\,S, 80\degree\,W.  The strength of the
dipole shows greater fluctuations in the negative direction than in
the positive.  Figure~\ref{fig:ersst-strong-events} illustrates, in
cartoon form, the evolution of SST anomalies during a strong \eln
event having a negative excursion of $\alpha_2$ towards the end of the
event.  Although it is not possible to composite individual spatial
patterns found from the NLPCA SST mode 1 and mode 2 analyses, it is
possible to examine the individual patterns in this case as, at
critical points during the evolution of the \eln event, only one of
$\alpha_1$ or $\alpha_2$ is non-zero.  Starting from neutral
conditions ($\alpha_1 \approx 0, \alpha_2 \approx 0$, shown in panel 1
of Figure~\ref{fig:ersst-strong-events}), SST anomalies increase in
the eastern equatorial Pacific to reach the peak pattern of a fully
developed \eln (Figure~\ref{fig:ersst-strong-events}, panel 2).  At
this point, $\alpha_1$ is at its most positive (recall that $\alpha_1$
captures the basic variability between \eln and \lan states; in fact,
the correlation between $\alpha_1$ and the NINO3 SST index time series
here is 0.955), while $\alpha_2$ remains small.  Towards the end of
the \eln, $\alpha_1$ is small, while $\alpha_2$ exhibits a large
negative excursion (Figure~\ref{fig:ersst-strong-events}, panel 3).
The transition between panels 2 and 3, from the mature \eln pattern to
a pattern with cooler waters in the central Pacific and a smaller area
of warm waters near the South American coast, suggests an eastwards
propagation of SST anomalies.  This is reinforced by the transition to
panel 4 in Figure~\ref{fig:ersst-strong-events}, which shows mature
\lan conditions, characterised by negative $\alpha_1$ and small values
of $\alpha_2$.  The transition from panel 2 (\eln conditions) to 3
(cold anomalies propagating eastwards and warm anomalies confined to
the coastal eastern Pacific) is rapid, as seen from the short
timescale of the negative excursions of $\alpha_2$ in
Figure~\ref{fig:ersst-nlpca-alpha-2}.  This is associated with the
rapid discharge of warm water from the equatorial Pacific during an
\eln event, partially explained by the recharge oscillator theory of
\cite{jin-1997a}.  The recharge of equatorial warm water (panels 4
$\to$ 1 $\to$ 2 in Figure~\ref{fig:ersst-strong-events}) is much
slower.

%
%
%

%
%
%

%
%
%

This analysis suggests that more recent stronger \eln events (as well
as the anomalous period in the early 1940s) involve stronger zonal
propagation of surface anomalies than weaker \eln{}s.  The reasons for
this are difficult to discern from the SST dynamics alone, and the
observational thermocline depth data considered in
Section~\ref{sec:nlpca-z-results} extends back only to the beginning
of 1980.  This means that all of the observed \eln events are of the
more recent stronger type.  It is thus not possible to come to any
definite conclusions here.

We now turn to NLPCA SST mode 2 results for some of the models.  These
results turn out to be rather less useful than the mode 1 results, so
we confine our attention to three models only, CNRM-CM3, ECHO-G and
GFDL-CM2.1.  These three models are fairly representative of models
with reasonable ENSO behaviour.  The explained variance results in
Table~\ref{tab:sst-modes-1-2-variance} on
page~\pageref{tab:sst-modes-1-2-variance} show that the first two
NLPCA modes explain either about the same amount of variance as the
first two principal components of the model SST data (CNRM-CM3) or
slightly more (ECHO-G and GFDL-CM2.1).

\begin{figure}
  \begin{center}
    \includegraphics[width=\textwidth]{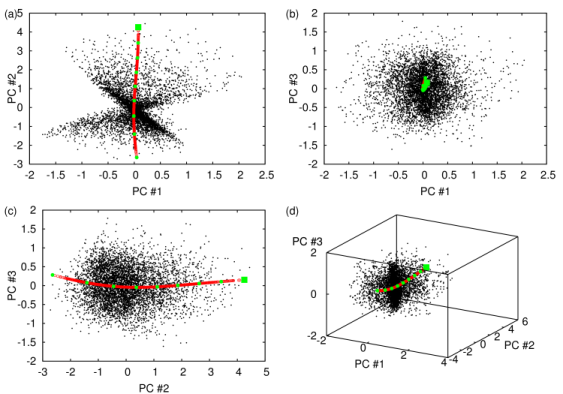}
  \end{center}
  \caption[CNRM-CM3 NLPCA SST mode 2 reconstruction]{Reconstruction
    plots for NLPCA SST mode 2 for CNRM-CM3.  All details as for
    Figure~\ref{fig:nlpca-sst-mode-2-ersst-recon}.}
  \label{fig:nlpca-sst-mode-2-cnrm_cm3-recon}
\end{figure}

\begin{figure}
  \begin{center}
    \includegraphics[width=\textwidth]{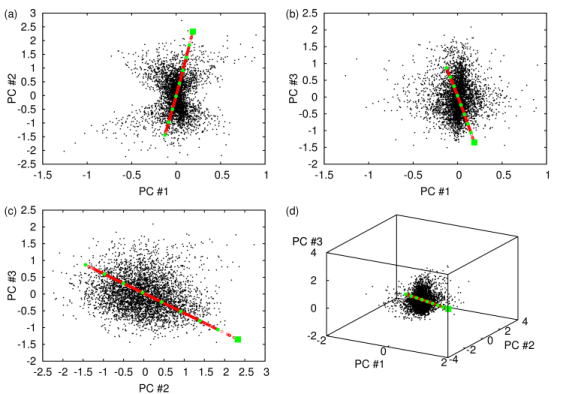}
  \end{center}
  \caption[ECHO-G NLPCA SST mode 2 reconstruction]{Reconstruction
    plots for NLPCA SST mode 2 for ECHO-G.  All details as for
    Figure~\ref{fig:nlpca-sst-mode-2-ersst-recon}.}
  \label{fig:nlpca-sst-mode-2-miub_echo_g-recon}
\end{figure}

\begin{figure}
  \begin{center}
    \includegraphics[width=\textwidth]{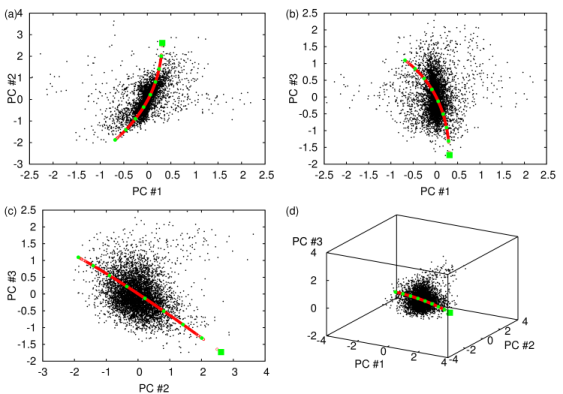}
  \end{center}
  \caption[GFDL-CM2.1 NLPCA SST mode 2 reconstruction]{Reconstruction
    plots for NLPCA SST mode 2 for GFDL-CM2.1.  All details as for
    Figure~\ref{fig:nlpca-sst-mode-2-ersst-recon}.}
  \label{fig:nlpca-sst-mode-2-gfdl_cm2_1-recon}
\end{figure}

Reconstruction plots for NLPCA SST mode 2 for these models are shown
in Figures~\ref{fig:nlpca-sst-mode-2-cnrm_cm3-recon} (CNRM-CM3),
\ref{fig:nlpca-sst-mode-2-miub_echo_g-recon} (ECHO-G) and
\ref{fig:nlpca-sst-mode-2-gfdl_cm2_1-recon} (GFDL-CM2.1).  There is
clear nonlinearity in the distribution of the residuals from NLPCA SST
mode 1, used as inputs for the calculation of SST mode 2, although it
is difficult to see how a one-dimensional manifold could be fit to any
of the data sets in a convincing fashion.  For both CNRM-CM3 and
ECHO-G, there is a strong ``pinch'' in the residual data near the
origin in principal component space, demonstrating the compression of
the data variance in directions tangential to the NLPCA SST mode 1
manifold.  This effect is not so clear in the GFDL-CM2.1 data,
presumably because of the greater nonlinearity of the reconstruction
manifold for NLPCA SST mode 1 for this model
(Figure~\ref{fig:nlpca-sst-mode-1-gfdl_cm2_1-recon}).  For each of the
models, the reconstruction manifold for NLPCA SST mode 2 is close to
linear, since there is no simple nonlinear structure that can easily
be fitted by a one-dimensional manifold, so that the NLPCA fitting
algorithm converges to something close to the first principal
component of the input data.  These somewhat equivocal structures in
the residual data provide some justification for using simple neural
networks to extract the second SST mode --- there is not enough
structure in the data to justify using a more complex network than the
model with three hidden layer neurons used here.  For all of the model
data, the reconstruction manifold for NLPCA SST mode 2 lies mostly in
the direction of the PC \#2 axis.  This contrasts with the results for
the observational data, where the reconstruction manifold, again close
to linear, lies mostly in the direction of PC \#3.  The explanation of
this would appear to lie in the structure of the SST EOFs for the
models compared to the observations.  For the observations, SST EOF 2
(Figure~\ref{fig:sst-eofs}b) has a structure characterised by an
isolated centre of action stretching along the equator from coastal
South America to around 150\degree\,W, while SST EOF 3
(Figure~\ref{fig:sst-eofs}c) has a zonal dipole pattern with centres
of action on the equator near 150\degree\,W and near coastal South
America, at about 10\degree\,S, 80\degree\,W.  In the models
considered here, this zonal dipole pattern appears in EOF 2 rather
than EOF 3 and the pattern seen in EOF 2 of the observations does not
appear in any of the leading EOFs at all.

%
%
%
%
%
%
%
%
%

\begin{figure}
  \begin{center}
    \includegraphics[width=\textwidth]{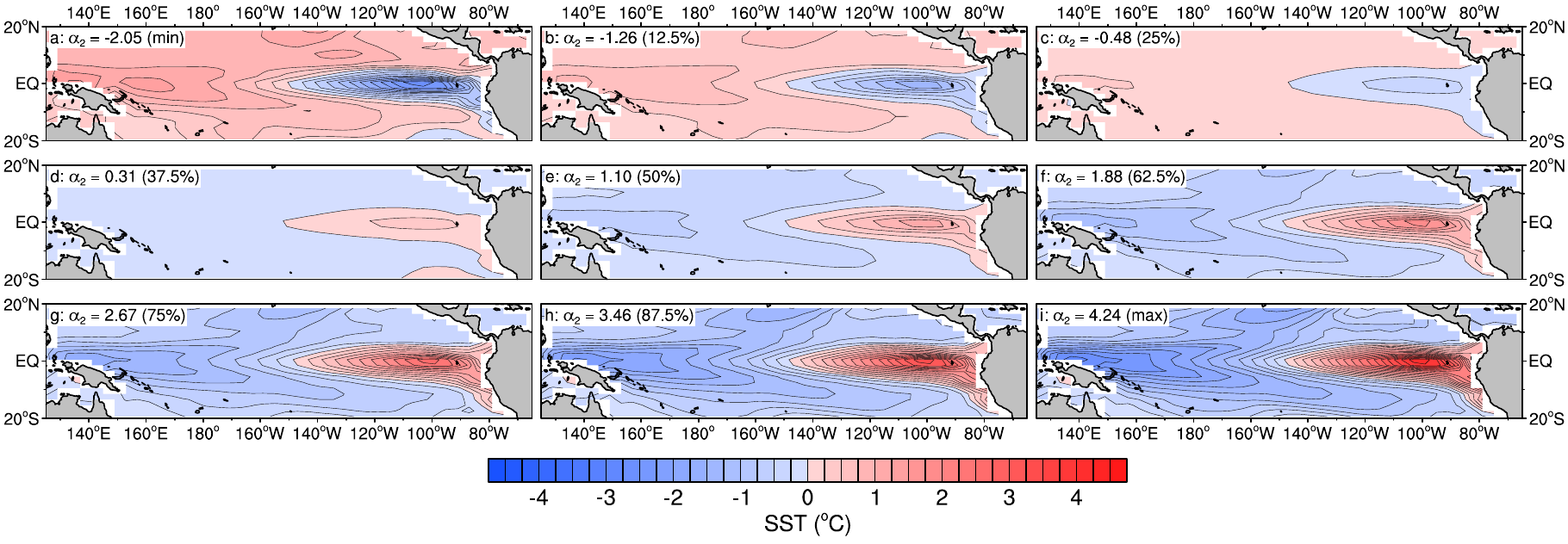}
  \end{center}
  \caption[CNRM-CM3 NLPCA SST mode 2 spatial patterns]{Spatial pattern
    plots for NLPCA SST mode 2 for CNRM-CM3, corresponding to the
    highlighted points in
    Figure~\ref{fig:nlpca-sst-mode-2-cnrm_cm3-recon}.  All details as
    in Figure~\ref{fig:nlpca-sst-mode-2-ersst-patterns}.}
  \label{fig:nlpca-sst-mode-2-cnrm_cm3-patterns}
\end{figure}

\begin{figure}
  \begin{center}
    \includegraphics[width=\textwidth]{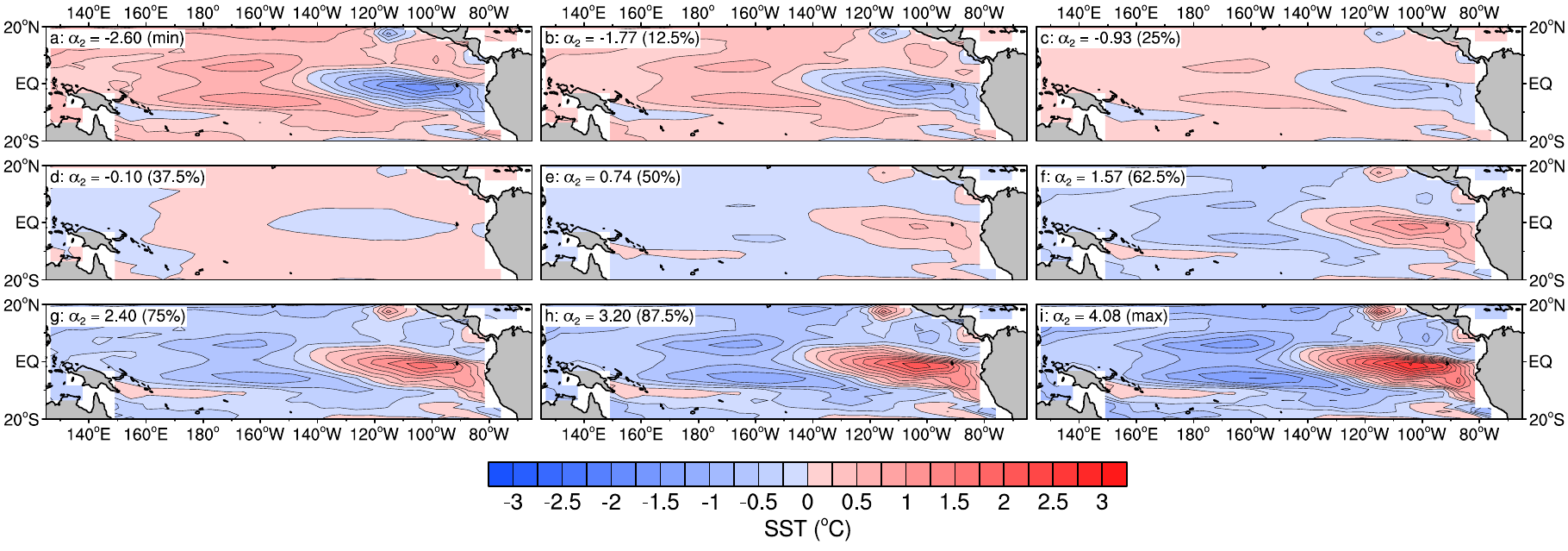}
  \end{center}
  \caption[ECHO-G NLPCA SST mode 2 spatial patterns]{Spatial pattern
    plots for NLPCA SST mode 2 for ECHO-G, corresponding to the
    highlighted points in
    Figure~\ref{fig:nlpca-sst-mode-2-miub_echo_g-recon}.  All details
    as in Figure~\ref{fig:nlpca-sst-mode-2-ersst-patterns}.}
  \label{fig:nlpca-sst-mode-2-miub_echo_g-patterns}
\end{figure}

\begin{figure}
  \begin{center}
    \includegraphics[width=\textwidth]{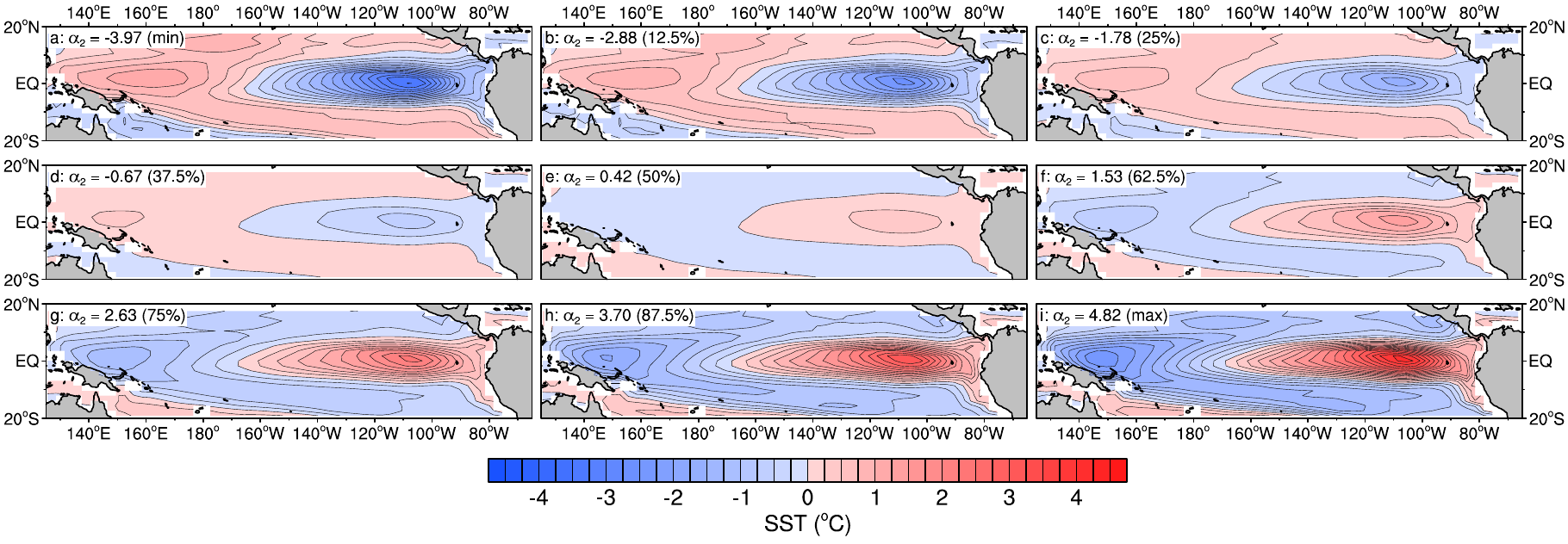}
  \end{center}
  \caption[GFDL-CM2.1 NLPCA SST mode 2 spatial patterns]{Spatial
    pattern plots for NLPCA SST mode 2 for GFDL-CM2.1, corresponding
    to the highlighted points in
    Figure~\ref{fig:nlpca-sst-mode-2-gfdl_cm2_1-recon}.  All details
    as in Figure~\ref{fig:nlpca-sst-mode-2-ersst-patterns}.}
  \label{fig:nlpca-sst-mode-2-gfdl_cm2_1-patterns}
\end{figure}

Given this situation, we might expect the spatial patterns associated
with variations in NLPCA SST mode 2 in the models either to be
reasonably similar to those seen for the observations or to correspond
more closely to the patterns of variability seen in the second SST EOF
for the observations.  The latter turns out to be the case.  Spatial
pattern plots for the three models are shown in
Figures~\ref{fig:nlpca-sst-mode-2-cnrm_cm3-patterns} (CNRM-CM3),
\ref{fig:nlpca-sst-mode-2-miub_echo_g-patterns} (ECHO-G) and
\ref{fig:nlpca-sst-mode-2-gfdl_cm2_1-patterns} (GFDL-CM2.1).  The most
characteristic pattern in all of the models has a centre of
variability in the eastern Pacific, extending from the equatorial
coast of South America west to around 140\degree\,W (CNRM-CM3 and
ECHO-G) or 170\degree\,W (GFDL-CM2.1).  There is little hint of the
dipolar pattern of variability seen in the observational data in
Figure~\ref{fig:nlpca-sst-mode-2-ersst-patterns}.  Some discussion and
comments on the difficulties of interpreting these results are offered
in Section~\ref{sec:nlpca-discussion} below.

%
%
%

%
%
%

%
%
%

\subsection{Nonmodal 2-D NLPCA analysis}
\label{sec:nlpca-sst-nonmodal}

In both of the preceding sections, the neural networks used have had a
single bottleneck neuron only, resulting in one-dimensional
reconstruction manifolds parameterised by the value of this single
bottleneck neuron.  This approach allows for one-dimensional nonlinear
modes to be projected out of a data set one by one, in an approach
referred to as \emph{sequential NLPCA} in \citet{kramer-nlpca}.  The
variance partitioning property of NLPCA represented by
\eqref{eq:nlpca-variance-partition} makes this a meaningful procedure
in terms of interpreting the total variance in the input data.  There
is no particular reason why more neurons may not be used in the
bottleneck layer, and \citet{monahan-enso} presented some results
using a bottleneck layer with two neurons to perform a two-dimensional
nonmodal decomposition of tropical SST data.  Here I briefly present
results of applying this approach, using the same network architecture
as \citet{monahan-enso}, i.e. two neurons in the bottleneck layer and
six neurons in the hidden layers.  Because of the difficulty of
interpreting the results of this type of nonmodal NLPCA, I only show
results for observational data and one model.

\begin{figure}
  \begin{center}
    \includegraphics[width=\textwidth]{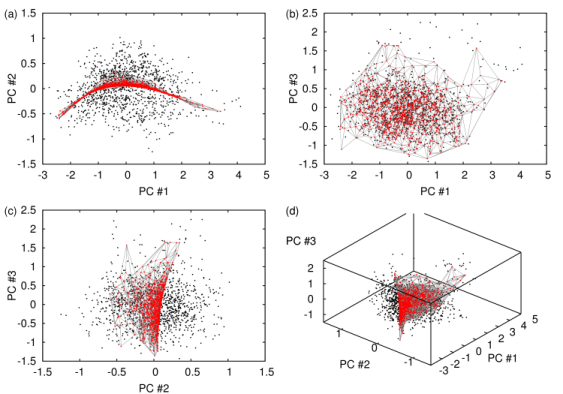}
  \end{center}
  \caption[ERSST NLPCA SST nonmodal reconstruction]{Reconstruction
    plots for NLPCA SST two-dimensional nonmodal analysis for
    observational ERSST v2 data.  Panels (a)--(c) show two-dimensional
    projections of the reconstruction into, respectively, the spaces
    spanned by EOFs 1 and 2, EOFs 1 and 3 and EOFs 2 and 3, while
    panel (d) shows a three-dimensional projection of the
    reconstruction into the space spanned by EOFs 1, 2 and 3.
    Original data points are shown as black dots, the NLPCA
    reconstructions are plotted as red dots, and the two-dimensional
    reconstruction manifold is outlined in grey.}
  \label{fig:nlpca-sst-nonmodal-ersst-recon}
\end{figure}

The essential difficulty in interpreting the two-dimensional nonmodal
NLPCA results can be understood from the reconstruction plot for the
observational SST data shown in
Figure~\ref{fig:nlpca-sst-nonmodal-ersst-recon}.  As for the other
reconstruction plots, this shows the manifold reconstructed by NLPCA
projected into the space spanned by the leading three SST EOFs.  Here,
because there are two neurons in the bottleneck layer, the NLPCA
reconstruction manifold is two-dimensional, one dimension for each
bottleneck neuron value.  This means that, instead of the
one-dimensional $\alpha_1$ and $\alpha_2$ parameterisations of the
reconstruction manifolds used in the modal decomposition, a
two-dimensional parameterisation is required.  It then becomes very
difficult to visualise the spatial patterns of variation captured by
the nonmodal analysis since we would need to display SST patterns from
points sampled from across the two-dimensional reconstruction
manifold.  This is a problem that we will encounter again in
Chapter~\ref{ch:isomap}, when we examine results from the Isomap
dimensionality reduction algorithm.

%
%
%

A second problem with the nonmodal analysis is that the
parameterisation of the two-dimensional reconstruction manifold is not
unique, and so neither is the reduced representation of the data
recovered from the bottleneck neuron values \citep{monahan-enso}.  The
parameterisation of the surface found by NLPCA is ambiguous up to a
homeomorphism.  The bottleneck neuron values will probably not even be
uncorrelated, unless a term is added to the cost function to ensure
that this is the case, as in \eqref{eq:cost-function-corr}.  An
additional analysis step (PCA may be suitable) is required to decouple
the different degrees of freedom in the bottleneck values.

These factors make interpretation of the results of nonmodal NLPCA
analyses much more difficult than the modal decompositions presented
earlier.  This dichotomy between modal decomposition and nonmodal
analysis does not arise with linear methods like PCA.  Since
individual modes can be combined additively, there is no distinction
between performing a modal analysis to project out individual modes of
variability one at a time, and a nonmodal analysis that finds all of
the modes of interest at once (what is usually done in operational
uses of PCA).  The lack of linearity here is a real loss and is a
major handicap for interpreting the results of both NLPCA and other
nonlinear dimensionality reduction methods.

Because of these difficulties, I present only a single example of a
nonmodal analysis of model SST data here.
Figure~\ref{fig:nlpca-sst-nonmodal-ersst-recon} shows reconstruction
plots for such an analysis for CNRM-CM3.  The situation here is
slightly different than for the observational data, since the
two-dimensional manifold discovered by the NLPCA algorithm is more
linear, appearing to indicate that the best two-dimensional
\emph{nonlinear} fit to the data is largely encompassed by the first
two PCA modes.  To some extent, this matches the conclusions drawn
from the explained variance results for the combined first two PCA and
first two NLPCA modes shown in Table~\ref{tab:sst-modes-1-2-variance}.

%
%
\begin{figure}
  \begin{center}
    \includegraphics[width=\textwidth]{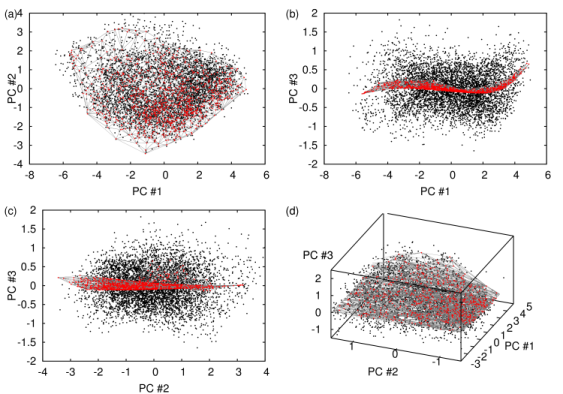}
  \end{center}
  \caption[CNRM-CM3 NLPCA SST nonmodal reconstruction]{Reconstruction
    plots for NLPCA SST two-dimensional nonmodal analysis for
    CNRM-CM3.  All details as for
    Figure~\ref{fig:nlpca-sst-nonmodal-ersst-recon}.}
  \label{fig:nlpca-sst-nonmodal-cnrm_cm3-recon}
\end{figure}
%

\section{Application to analysis of Pacific thermocline variability}
\label{sec:nlpca-z-results}

In this section, I present results of applying NLPCA to observed and
simulated thermocline depth data.  Following the discussion in
Section~\ref{sec:thermocline-calc-methods}, all of the analysis here
is based on the \ztw thermocline depth.  The situation for thermocline
depth is slightly different to that for SST, since ENSO-related
variability in thermocline depth is generally better fitted by a
cyclic variation, i.e. a closed curve in principal component space,
rather than the end-point to end-point variation between \eln and \lan
conditions seen in SST, which is better fitted by an open curve in
principal component space \citep{an-enso-interdecadal}.  This cyclic
variation in the thermocline structure during ENSO events is made
clear in Figure~\ref{fig:z-pc-scatter}a on
page~\pageref{fig:z-pc-scatter}, which shows a scatter plot of PC \#1
versus PC \#2 for observed \ztw.  As noted in
Section~\ref{sec:thermocline-pca}, during \eln{}s, the first two
principal components of \ztw vary in quadrature, representing a cycle
in principal component space.  This observation is the principal
justification for using a circular bottleneck layer
(Figure~\ref{fig:nn-architecture}b) for the NLPCA analysis of
thermocline depth.

To aid comparison with \citet{an-enso-interdecadal}, I use a similar
network configuration for the circular bottleneck layer NLPCA
calculations (for conciseness, referred to as ``NLPCA(cir)'' in what
follows).  As for the SST data, PCA is used as an initial
dimensionality reduction step, and five thermocline depth PCs are used
as input to the NLPCA network.  For all circular bottleneck layer
optimisations, the weight decay coefficient is set to $P_W = 1$ and
ensembles of 30 random initial weight conditions are fitted for
networks with hidden layers with between two and eight neurons.  The
same procedure for overfitting avoidance is used as for the SST
calculations, with a fraction of 15\% of the input data reserved for
overfitting testing.

As well as the circular bottleneck layer fits, a set of single mode
NLPCA calculations were also performed for comparison (referred to as
``NLPCA(1)'' in what follows), to get some idea of how justified we
are in the assumption that a circular network should provide a better
fit to the \ztw data.  These calculations used 5 PCs as input, a
single bottleneck neuron layer, weight decay coefficients of $P_W = 0,
10^{-4}, 10^{-3}, 0.01, 0.1, 1$ and hidden layers with 4, 5 or 6
neurons.  The best non-overfitted solution was selected for comparison
with the circular bottleneck layer results.

Consideration of the results here closely follows the scheme of
Section~\ref{sec:nlpca-sst-mode-1} for SST mode 1 results.  We start
with results for the NCEP GODAS observational data, with the
expectation that these results should be close to those presented by
\citet{an-enso-interdecadal}.  We use the same approach as in the SST
case, looking at error measures, reconstruction plots and spatial
pattern plots at different points in the variation of the thermocline
depth.  Table~\ref{tab:nlpca-z-modes-results} shows RMS error and
explained variance results for the best NLPCA fits, with PCA results
for comparison.  The first ``PCA RMS error'' column shows the RMS \ztw
difference between the input \ztw anomaly data and the \ztw anomaly
field reconstructed by multiplying the first \ztw EOF by the first
\ztw principal component time series.  The next two columns (``NL(1)
RMS error'' and ``NL(c) RMS error'') show, for NLPCA(1) and NLPCA(cir)
respectively, the RMS \ztw difference between the input \ztw anomaly
data and the \ztw anomaly field reconstructed from the best NLPCA
fits.  The reconstructed thermocline depth anomaly fields are
composites of the first 5 \ztw EOFs, with each EOF scaled by the
corresponding NLPCA output, of which there is one for each EOF.  For
each of the NLPCA RMS error columns, the number of neurons $l$ in the
network hidden layer for the best fit is shown in parentheses for each
type of network.  The last four columns in
Table~\ref{tab:nlpca-z-modes-results} show the explained variance
fractions (as a fraction of the total \ztw anomaly variance) for the
first PCA mode, the first NLPCA(1) mode, the first NLPCA(cir) mode,
and the total variance explained by the first 5 EOFs used as input to
the NLPCA fitting procedure.  In each row of the table, the best
result (PCA, NLPCA(1) or NLPCA(cir)) is highlighted in bold for both
RMS error and explained variance.

%
%
\begin{sidewaystable}
  \begin{center}
    \input{figs/06/nlpca-z-modes-results.tex}
  \end{center}
  \caption[NLPCA thermocline error and variance comparison]{Error and
    variance results for NLPCA thermocline depth analyses.  The first
    three data columns, labelled ``RMS Error'', show RMS thermocline
    depth errors between the original data and single mode PCA and
    single mode and circular NLPCA reconstructions.  The rightmost
    four columns show explained variance fractions for PCA mode 1,
    single mode NLPCA and circular bottleneck layer NLPCA, and the
    total variance explained by the 5 principal components used as
    input to the NLPCA algorithm (explained variance is expressed as a
    fraction of the total data variance).  Single mode NLPCA results
    are denoted {\bf NL(1)} and circular bottleneck layer NLPCA
    results as {\bf NL(c)}.  For the NLPCA RMS error columns, the
    number of hidden layer neurons in the best fit network is shown in
    parentheses.  For both the RMS error and explained variance
    measures, the best result of PCA, single mode NLPCA and circular
    bottleneck layer NLPCA is highlighted in bold.}
  \label{tab:nlpca-z-modes-results}
\end{sidewaystable}
%

\begin{figure}
  \begin{center}
    \includegraphics[width=\textwidth]{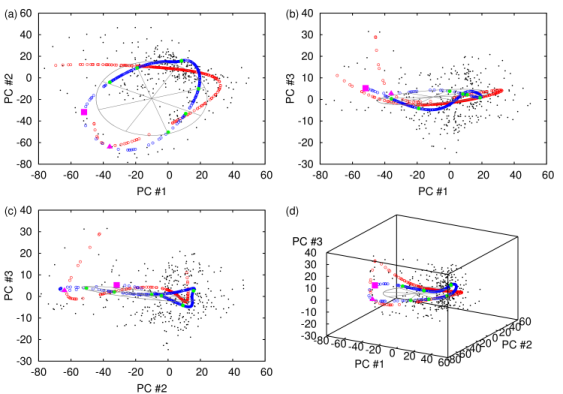}
  \end{center}
  \caption[NCEP GODAS NLPCA thermocline depth mode
    reconstruction]{Reconstruction plots for thermocline depth NLPCA
    modes for observational NCEP GODAS data.  Panels (a)--(c) show
    two-dimensional projections of the reconstruction into,
    respectively, the spaces spanned by EOFs 1 and 2, EOFs 1 and 3 and
    EOFs 2 and 3, while panel (d) shows a three-dimensional projection
    of the reconstruction into the space spanned by EOFs 1, 2 and 3.
    Original data points are shown as black dots, the NLPCA(1)
    reconstruction is plotted as red circles, the NLPCA(cir)
    reconstruction as blue circles, and points along the NLPCA(cir)
    reconstruction curve closest to particular values of
    $\alpha_\circ$ are highlighted in green ($\alpha_\circ = 0$ is
    highlighted with a magenta square, $\alpha_\circ = 45\degree$ with
    a magenta triangle; other highlighted values are $\alpha_\circ =
    90\degree, 135\degree, 180\degree, 225\degree, 270\degree,
    315\degree$).  The segmented grey circle is the best least-squares
    fit to the NLPCA(cir) reconstruction, used in the definition of
    $\alpha_\circ$.}
  \label{fig:nlpca-z-modes-ncep-godas-recon}
\end{figure}

For the observational data, we see from
Table~\ref{tab:nlpca-z-modes-results} that, although both of the NLPCA
analyses do a better job of representing the thermocline depth
variability than the first PCA mode, a better fit is achieved by the
NLPCA(1) mode than the NLPCA(cir) mode.  This is true for both the RMS
\ztw error and the explained variance measures.  We can develop some
understanding of the relationship between the different fits by
examining the same kind of reconstruction plots used for looking at
SST data in Section~\ref{sec:nlpca-sst-mode-1}.  We plot both the
NLPCA(1) and NLPCA(cir) reconstructions, projecting the 5-dimensional
reconstructions in principal component space (one coordinate for each
of the five outputs of the auto-associative NLPCA neural network) into
the space spanned by \ztw EOFs 1--3 for comparison with the original
data.  Figure~\ref{fig:nlpca-z-modes-ncep-godas-recon} shows this
reconstruction for the observed \ztw data.  On this and subsequent
thermocline depth reconstruction plots, the individual input data
points are shown as small black points, the best NLPCA(1)
reconstruction is shown as red circles and the best NLPCA(cir)
reconstruction as blue circles.  The other elements of the plot will
be explained below, as they relate to the calculation of
$\alpha_\circ$, a means of parameterising points around the cycle
fitted by NLPCA(cir), analogous to $\alpha_1$ for the open curve SST
results.

%
%
%

Comparing reconstructions for the NLPCA(1) and NLPCA(cir) methods in
Figure~\ref{fig:nlpca-z-modes-ncep-godas-recon} reveals some aspects
of why the NLPCA(1) reconstruction has a smaller RMS error and greater
explained variance than the NLPCA(cir) fit.  The open curve solution
found by NLPCA(1) appears somewhat overfitted, with large curvature
allowing it to come reasonably close to all of the original data
points.  The NLPCA(cir) reconstruction manifold is rather more
constrained and does not encompass data points lying at larger
negative values of PC \#1 and positive values of PC \#2 (top-left
corner of Figure~\ref{fig:nlpca-z-modes-ncep-godas-recon}a).  To some
extent, this is due to the form of the basic least-squares component
of the NLPCA cost function \eqref{eq:nlpca-cost-function-basic}, which
places equal weight on each data point, thus leading to solutions
biased towards regions of the input data space with a greater density
of input data points.  This is clear from
Figure~\ref{fig:nlpca-z-modes-ncep-godas-recon}, which shows only a
small number of points extending to large negative values of PC \#2
(the looping excursions associated with ENSO events).  The NLPCA cost
function thus assigns less weight to these regions compared to the
region with small PC \#1 and PC \#2 values, associated with
``loitering'' between ENSO events.

There is no clear solution to this problem.  The use of alternative
error norms and minimisation criteria may help to alleviate the
difficulties of selecting between a slightly overfitted NLPCA(1)
solution and a less well-fitting NLPCA(cir) solution
\citep{hsieh-noisy-nlpca,cannon-robust-nlpca}, but these alternative
techniques do little to help with the data density problem.  In fact,
it is difficult to argue for any a priori bias against regions with
high data density.  The only reason we seek a solution that passes
through the relatively rare ``loop'' regions here is that we believe
that there is a periodic variation in the structure of the thermocline
depth and so we would like to enforce conditions that allow us to
extract this mode of variability.

\begin{figure}
  \begin{center}
    \includegraphics[width=\textwidth]{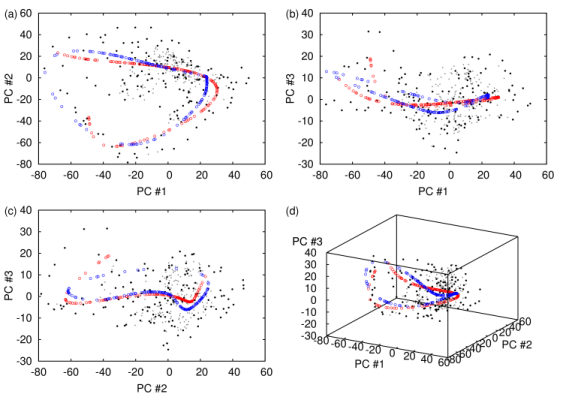}
  \end{center}
  \caption[NLPCA thermocline depth decimation
    reconstruction]{Reconstruction plots for thermocline depth NLPCA
    modes for decimated observational NCEP GODAS data.  Panels
    (a)--(c) show two-dimensional projections of the reconstruction
    into, respectively, the spaces spanned by EOFs 1 and 2, EOFs 1 and
    3 and EOFs 2 and 3, while panel (d) shows a three-dimensional
    projection of the reconstruction into the space spanned by EOFs 1,
    2 and 3.  The data points used for the analysis are shown as black
    dots, data points removed by the decimation procedure are shown in
    grey, the NLPCA(1) reconstruction is plotted as red circles and
    the NLPCA(cir) reconstruction as blue circles.}
  \label{fig:nlpca-decimation-recon}
\end{figure}

Some experiments were performed to examine the effects of decimating
the input data to even out the density of input points.  This
decimation procedure, based on the calculation of nearest-neighbour
sets, using either a neighbour count $k$ or a neighbourhood radius
$\varepsilon$, was as follows:
\begin{enumerate}
  \item{Read 5-dimensional PC time series inputs, and extract the
    first three principal components for each data point as vectors
    $\vec{x}_i \in \mathbb{R}^3$.}
  \item{Calculate, using the Euclidean norm in $\mathbb{R}^3$,
    $k$-neighbourhoods of each point containing $k_{\mathrm{orig}}$
    points, denoted by $\mathcal{N}_k(\vec{x}_i)$.}
  \item{Find the minimum neighbourhood radius, $\rho_{\mathrm{min}} =
    \min_i \rho(\vec{x}_i)$, where the neighbourhood radius is defined
    as $\rho(\vec{x}_i) = \max_{j \in \mathcal{N}_k(\vec{x}_i)} ||
    \vec{x}_j - \vec{x}_i ||$.}
  \item{Calculate $\varepsilon$-neighbourhoods of each point of
    neighbourhood radius $\rho_{\mathrm{min}}$, denoted by
    $\mathcal{N}_\varepsilon(\vec{x}_i)$.}
  \item{Select a point $\vec{x}_l$ randomly from the remaining data
    points, weighting the selection probabilities for each point by
    $\# \mathcal{N}_\varepsilon(\vec{x}_i)$.}
  \item{If $\# \mathcal{N}_\varepsilon(\vec{x}_l) >
    k_{\mathrm{target}}$, remove point $\vec{x}_l$.}
  \item{Repeat steps 4--6 while $\max_i \#
    \mathcal{N}_\varepsilon(\vec{x}_i) > k_{\mathrm{target}}$.}
  \item{Output the original 5-dimensional principal components of the
    points remaining.}
\end{enumerate}
The result of this is that points that started out with
$k_{\mathrm{target}}$ or fewer neighbours are unaffected, while those
with more neighbours are thinned out so that, on average,
neighbourhoods that originally contained $k_{\mathrm{orig}}$ points
contain about $k_{\mathrm{target}}$ points.  The final set of data
points is thus more evenly distributed throughout the input data
space.  For the results shown here, I used $k_{\mathrm{orig}} = 30$
and $k_{\mathrm{target}} = 2$.  The same NLPCA fitting procedure was
applied to the decimated data as for the original data, giving RMS
thermocline depth errors of 11.737\,m (NLPCA(1) with $l=5$) and
11.987\,m (NLPCA(cir) with $l=8$), very slightly greater than for the
undecimated data.  The explained variance fractions were 53.9\% for
NLPCA(1) and 53.5\% for NLPCA(cir), rather larger than for the
original undecimated data.  The reconstruction plots for these
calculations are shown in Figure~\ref{fig:nlpca-decimation-recon}.
Comparison with Figure~\ref{fig:nlpca-z-modes-ncep-godas-recon} shows
that running the NLPCA fits on the decimated data does indeed produce
reconstruction manifolds that better capture the ENSO ``loops''.
However, this improvement is, unsurprisingly, seen in both the
NLPCA(cir) and NLPCA(1) reconstructions, meaning that the decimated
data results do not provide any better guidance for deciding between
the use of a single bottleneck neuron or a circular bottleneck layer.
I use undecimated data when examining the model thermocline depth
variability below, but this data point density-dependent fitting
effect should be borne in mind when examining the reconstruction
plots.

%
%
%

In \citep{an-enso-interdecadal}, NLPCA networks with circular
bottleneck layers were used to examine thermocline depth variability
purely on the basis of the expectation that there should be some sort
of cyclic variability best expressed by a closed curve in principal
component space.  It appears from the results here that there is, in
fact, relatively little in the output from the NLPCA procedure itself
to justify this choice --- the RMS error and explained variance
measures are better for the NLPCA(1) network for both the original and
decimated data.  Choosing to use the NLPCA(cir) network instead is
thus a subjective decision.  More recent work by
\citet{hsieh-noisy-nlpca} applied an information criterion based on a
measure of the consistency of NLPCA reconstructions of points close
together in the input data to select NLPCA solutions, rather than the
simple mean squared error represented by
\eqref{eq:nlpca-cost-function-basic}.  This appears to provide a
slightly better basis for selecting between the use of NLPCA(1) and
NLPCA(cir) solutions in cases of the type seen here.  In order to
maintain a degree of comparability with \citet{an-enso-interdecadal},
I do not use this more sophisticated criterion here.

As for the SST results, it is of interest to extract spatial patterns
of thermocline variability for points around the NLPCA
reconstructions.  To do this for the NLPCA(cir) results, we need an
analogue of the $\alpha_1$ parameter defined in
\eqref{eq:nlpca-alpha1-def} to parameterise points along the
reconstruction manifold.  A natural choice would be to use the angle
defined by the embedding coordinates from the bottleneck layer values
($\theta$ in Figure~\ref{fig:nn-architecture}b).  Unfortunately,
because of the data point density issues described above, the
distribution of $\theta$ about the cycle defined by the NLPCA(cir)
manifold is very non-uniform.  I thus define a parameter
$\alpha_\circ$ based directly on the projection of the NLPCA(cir)
reconstruction into the three-dimensional PC \#1/PC \#2/PC \#3 space.

The goal here is to produce a more or less uniform parameterisation of
points around the NLPCA(cir) reconstruction manifold.  A reasonable
approach seems to be to find the least-squares best fit circle to the
reconstruction points, and then to use angles around this circle as
the $\alpha_\circ$ values.

Consider a circle in $\mathbb{R}^3$, with centre $\vec{c}$ and a unit
axial vector defined by two rotation angles $\alpha$ and $\beta$ as
$\vec{\hat{n}} = R_z(\alpha) R_y(\beta) \vec{\hat{k}} = (\cos \alpha
\sin \beta, \sin \alpha \sin \beta, \cos \beta)$, where
$\vec{\hat{k}}$ is the unit vector in the $z$-coordinate direction and
$R_y(\phi)$ and $R_z(\phi)$ represent rotations by angle $\phi$ about
the $y$- and $z$-axes respectively.  Points on the circle can then be
defined in terms of points on a circle in the $xy$-plane, written as
$\vec{p}' = (\rho \cos \theta, \rho \sin \theta, 0)$, where $\rho$ is
the radius of the circle and $\theta$ an angle parameterising points
around the circle.  Points on the rotated and shifted circle are then
given by $\vec{p} = R_z(\alpha) R_y(\beta) \vec{p}' + \vec{c}$ or
\begin{equation}
  \begin{aligned}
    \vec{p} &= \begin{pmatrix}
      \cos \alpha & \sin \alpha & 0 \\
      -\sin \alpha & \cos \alpha & 0 \\
      0 & 0 & 1
    \end{pmatrix}
    \begin{pmatrix}
      \cos \alpha & 0 & \sin \beta \\
      0 & 1 & 0 \\
      -\sin \beta & 0 & \cos \beta
    \end{pmatrix}
    \begin{pmatrix}
      \rho \cos \theta \\
      \rho \sin \theta \\
      0
    \end{pmatrix} + \vec{c} \\
    &= \begin{pmatrix}
      \cos \alpha \cos \beta & \sin \alpha & \cos \alpha \sin \beta \\
      -\sin \alpha \cos \beta & \cos \alpha & -\sin \alpha \sin \beta \\
      -\sin \beta & 0 \cos \beta
    \end{pmatrix}
    \begin{pmatrix}
      \rho \cos \theta \\
      \rho \sin \theta \\
      0
    \end{pmatrix} + \vec{c} \\
    &= \rho \begin{pmatrix}
      \cos \alpha \cos \beta \cos \theta + \sin \alpha \sin \theta \\
      -\sin \alpha \cos \beta \cos \theta + \cos \alpha \sin \theta \\
      -\sin \beta \cos \theta
    \end{pmatrix} + \vec{c}.
  \end{aligned}
\end{equation}
This expression for $\vec{p}$ can be inverted to transform from points
on the rotated and shifted circle to points on a circle in the
$xy$-plane as
\begin{equation}
  \vec{p}' = R_y(-\beta) R_z(-\alpha) (\vec{p} - \vec{c}).
\end{equation}
Using this, we can construct a simple error function for fitting data
points to the rotated shifted circle.  Treating $(\alpha, \beta,
\vec{c}, \rho)$ as parameters to be varied as part of a minimisation
procedure, given an input set of data points $\vec{x}_i$ with $i = 1,
\dots, N$, we calculate $\vec{x}'_i = (x'_i, y'_i, z'_i)$ as
\begin{equation}
  \vec{x}'_i = R_y(-\beta) R_z(-\alpha) (\vec{x}_i - \vec{c}),
\end{equation}
and then calculate a cost function
\begin{equation}
  J(\alpha, \beta, \vec{c}, \rho; \vec{x}) = \sum_{i=1}^N \left(
  \sqrt{(x'_i)^2 + (y'_i)^2} - \rho \right)^2 + (z'_i)^2.
\end{equation}
This measures both the deviation from a circle of radius $\rho$ of the
shifted, rotated points projected to the $xy$-plane, and the deviation
of the shifted, rotated points from the $xy$-plane.  This is a
suitable cost function for defining a best-fitting circle to the
reconstruction points in a least-squares sense.

Minimising the cost function $J$ gives $\alpha$, $\beta$, $\vec{c}$
and $\rho$ describing the best fit circle to the data points.  The
minimisation is done using a quasi-Newton method
\citep[Section~10.7]{press-nr}, starting from a range of random
initial conditions to provide confidence that a good minimum value of
the cost function has been found, even in the presence of spurious
local minima.  This approach seems to work well, and allows one to
find evenly distributed points around the cycle found by the NLPCA
procedure.  These points are parameterised by the angle $\alpha_\circ$
around the circle, with the zero of $\alpha_\circ$ taken at the
reconstructed data point with the largest magnitude value for PC \#1.
Results are shown for $\alpha_\circ = k \pi / 4$, with $k = 0, \dots,
7$.  Each of the reconstruction plots shown here highlights the points
on the NLPCA(cir) manifold closest to these $\alpha_\circ$ values in
green, and shows the $\alpha_\circ = 0$ and $\alpha_\circ = 45\degree$
points as a magenta square and triangle respectively.  The best fit
circle computed by the procedure described above is displayed in grey,
with ``spokes'' showing $\alpha_\circ$ values of $k\pi/4$ for $k = 0,
\dots, 7$.

\begin{figure}
  \begin{center}
    \includegraphics[width=\textwidth]{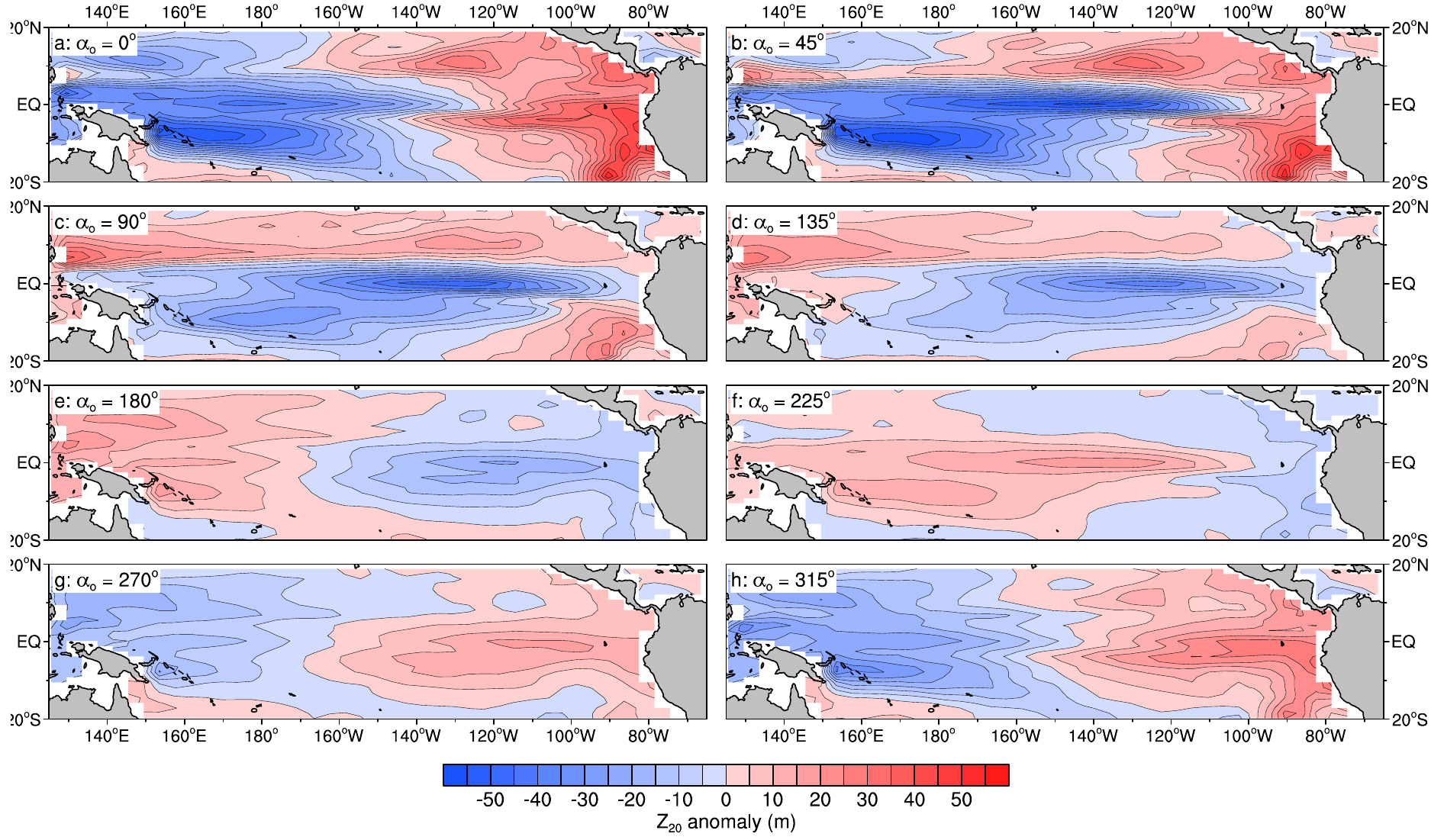}
  \end{center}
  \caption[NCEP GODAS NLPCA thermocline depth mode spatial
    patterns]{Spatial pattern plots for NLPCA(cir) mode for
    observational NCEP GODAS thermocline depth data.  Each panel shows
    the thermocline depth anomaly composite formed from the point
    along the one-dimensional NLPCA(cir) reduced manifold with the
    corresponding $\alpha_\circ$ value.  These values are highlighted
    on the reconstruction plots in
    Figure~\ref{fig:nlpca-z-modes-ncep-godas-recon}.}
  \label{fig:nlpca-z-modes-ncep-godas-patterns}
\end{figure}

Spatial reconstructions derived from this scheme are shown in
Figure~\ref{fig:nlpca-z-modes-ncep-godas-patterns}, and should be
compared with the thermocline depth EOFs plotted in
Figure~\ref{fig:z-eofs}a--c on page~\pageref{fig:z-eofs}.  There is a
clear cycling between conditions with negative thermocline depth
anomalies in the western Pacific and positive anomalies in the east
(Figure~\ref{fig:nlpca-z-modes-ncep-godas-patterns}a, for
$\alpha_\circ = 0$) and conditions with the opposite pattern of
anomalies (Figure~\ref{fig:nlpca-z-modes-ncep-godas-patterns}e, for
$\alpha_\circ = 180\degree$).  There are two observations to make
here.  First, there is a notable asymmetry between the amplitude of
the positive zonal thermocline depth anomaly gradient at $\alpha_\circ
= 0$, corresponding to \eln conditions with an anomalously deep
thermocline in the eastern Pacific, and the corresponding negative
gradient at $\alpha_\circ = 180\degree$, for \lan or normal
conditions, with the former positive \eln gradient being significantly
larger.  In fact, the true opposite state to the $\alpha_\circ = 0$
state appears to be the pattern for $\alpha_\circ = 225\degree$, shown
in Figure~\ref{fig:nlpca-z-modes-ncep-godas-patterns}f --- the
adjacent patterns on either side, at $\alpha_\circ = 180\degree$ and
$\alpha_\circ = 270\degree$, are almost mirror images, indicating that
the $\alpha_\circ = 225\degree$ pattern is an extremum.  The second
observation is that the thermocline depth anomaly patterns for the
transition states between $\alpha_\circ = 0$ and $\alpha_\circ =
180\degree$ are different for the two directions of transition.  The
greater part of this asymmetry is attributable to the different
amplitudes of the $\alpha_\circ = 0$ and $\alpha_\circ = 180\degree$
states, but there does appear to be some difference in the spatial
distributions of anomalies during the phases heading into and out of
\eln conditions.

%
%
%

The results here are quite comparable to those of
\citet{an-enso-interdecadal}.  In particular, the NLPCA reconstruction
plots in Figure~\ref{fig:nlpca-z-modes-ncep-godas-recon} are similar
to \citeauthor{an-enso-interdecadal}'s Figure~3, and the asymmetry of
the spatial patterns around the NLPCA(cir) manifold are also similar,
both in terms of the relative amplitudes of the \eln and \lan
end-member patterns and the asymmetry of the structure of thermocline
depth anomalies during the transitions between \eln and \lan
conditions (my Figure~\ref{fig:nlpca-z-modes-ncep-godas-patterns},
\citeauthor{an-enso-interdecadal}'s Figure~3).

\begin{figure}
  \begin{center}
    \includegraphics[width=\textwidth]{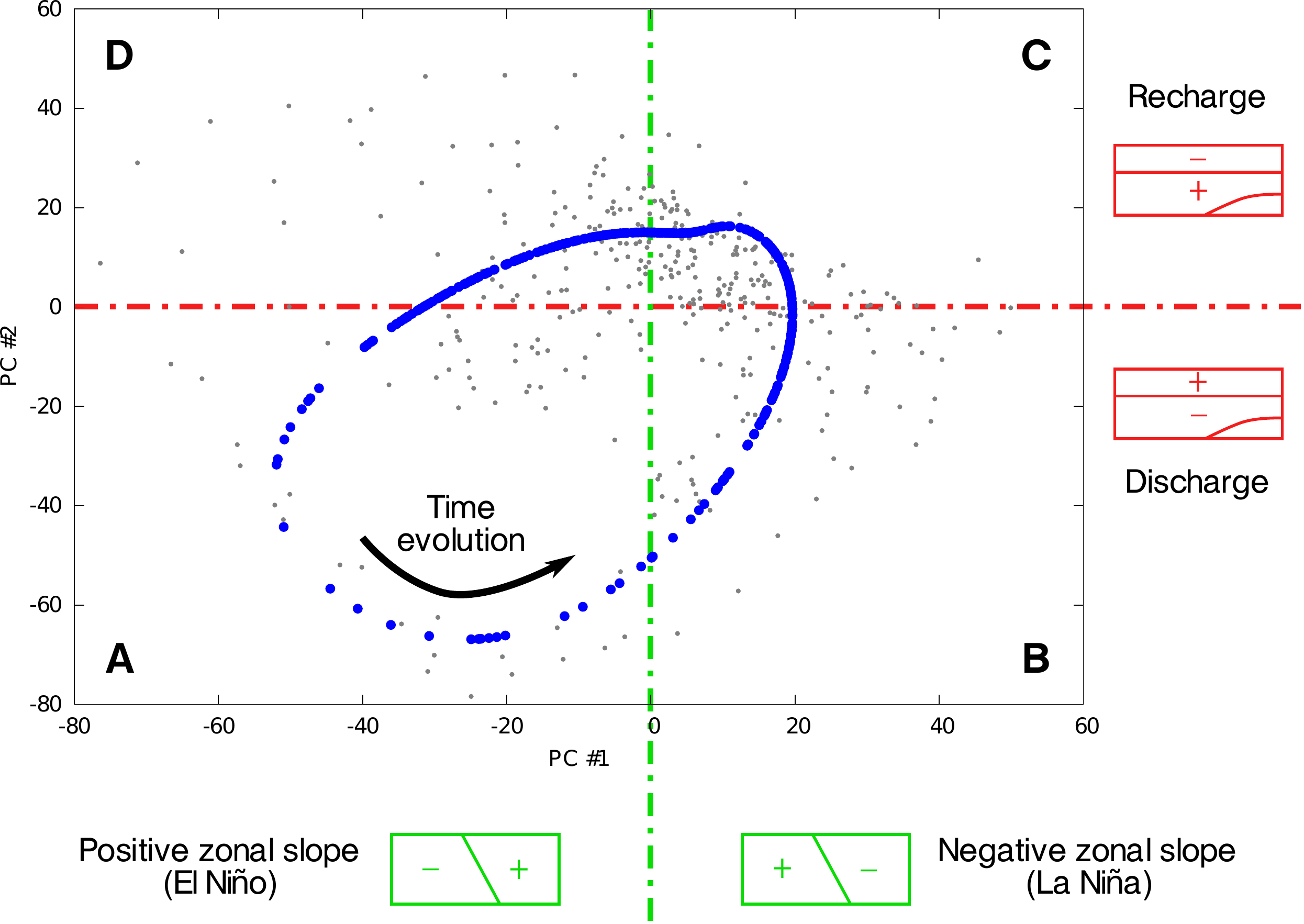}
  \end{center}
  \caption[NCEP GODAS thermocline depth variation schematic]{Schematic
    view, in the PC \#1/PC \#2 plane, of the temporal phasing of
    variations in \ztw anomaly principal components around the
    NLPCA(cir) reduced manifold for NCEP GODAS data.  The black dots
    are the original input data used to construct the NLPCA(cir)
    manifold and the reconstructed points on the NLPCA manifold are
    shown in blue.  In green are shown cartoon representations of the
    positive and negative phases of the first EOF, while in red are
    cartoon representations of the positive and negative phases of the
    second EOF.}
  \label{fig:nlpca-thermocline-progress}
\end{figure}

One can question how useful this analysis really is, since it is plain
from the reconstruction plot
(Figure~\ref{fig:nlpca-z-modes-ncep-godas-recon}) that the bulk of the
variability in the thermocline depth anomalies is captured by the
first two principal components.  Both the NLPCA(1) and NLPCA(cir)
reconstructions mostly lie within the PC \#1/PC \#2 plane, with only
small contributions from PC \#3 (or any other PCs, in fact) in most
cases.  (The axis of the best fit circle to the NLPCA(cir)
reconstruction manifold is tilted from the $z$-axis by only about
6\degree.)  This corresponds quite closely to the idea that the
variability in the thermocline depth anomalies is explained by the
Bjerknes and Wyrtki feedback mechanisms
(Section~\ref{sec:enso-mechanisms}), the Bjerknes feedback affecting
the zonal thermocline gradient (which is captured by the first EOF,
shown in Figure~\ref{fig:z-eofs}a), and the Wyrtki feedback affecting
the equatorial ocean heat content, measured by the zonal mean
equatorial thermocline depth (variations of which are captured by the
second EOF, shown in Figure~\ref{fig:z-eofs}b).  The NLPCA analysis
does clearly identify the asymmetric progress of this interaction
between the zonal thermocline tilt and zonal mean thermocline depth.
Figure~\ref{fig:nlpca-thermocline-progress} shows this, dividing the
influence of the first two EOFs into positive and negative phases
around the NLPCA manifold, which is here projected into the PC \#1/PC
\#2 plane.  For the first EOF, the positive and negative zonal
thermocline depth anomaly gradient phases, shown in schematic form in
green, correspond respectively to \eln and \lan conditions.  For the
second EOF, the positive and negative phases, shown in schematic form
in red, correspond respectively to a recharge and discharge of the
equatorial warm water volume.  The shape of the NLPCA manifold shows
how there is a strong discharge from the equatorial warm water volume
during \eln events and the transition to the following neutral or \lan
conditions (quadrants A and B on
Figure~\ref{fig:nlpca-thermocline-progress}), and a much weaker
recharge during the ``loitering'' \lan phase (quadrants C and D).

%
%
%

Despite some reservations about the ultimate utility of NLPCA in this
situation, the technique does seem to capture the variability that
exists in the data in a relatively clear and unambiguous way, even if,
in this application, we can understand what is happening from the
simpler PCA results.  Examination of results from applying NLPCA to
the model thermocline depth data bears this out.

The results in Table~\ref{tab:nlpca-z-modes-results} for the models
are somewhat mixed.  Apart from two models (CGCM3.1(T47) and
MIROC3.2(medres)) whose thermocline data was fitted better by the
first PCA mode than by either of the NLPCA procedures, the thermocline
data for about half of the models was fitted better by a closed
NLPCA(cir) manifold than an open NLPCA(1) manifold.  There is some
disagreement between the RMS thermocline depth error and explained
variance measures here, with eight models having both better RMS error
and explained variance for NLPCA(cir), four models having both better
RMS error and explained variance for NLPCA(1), and the other three
models having better RMS error for NLPCA(1) and better explained
variance for NLPCA(cir).  For the most part, the error and variance
differences between the NLPCA(1) and NLPCA(cir) are relatively small,
although there are much larger differences between the PCA mode and
the NLPCA modes in some cases (CNRM-CM3, ECHO-G, FGOALS-g1.0,
GFDL-CM2.1 and MRI-CGCM2.3.2).

\begin{figure}
  \begin{center}
    \includegraphics[width=\textwidth]{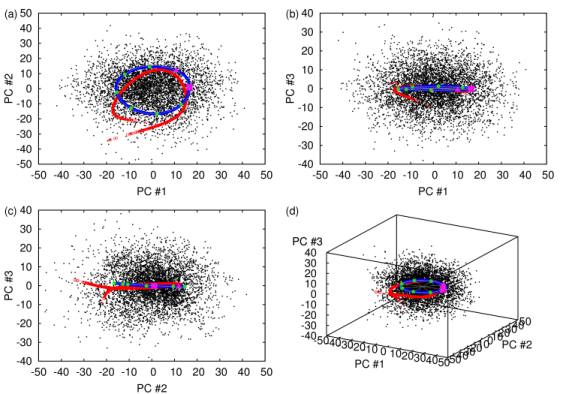}
  \end{center}
  \caption[CGCM3.1(T63) NLPCA thermocline depth mode
    reconstructions]{Reconstruction plots for thermocline depth NLPCA
    modes for CGCM3.1(T63).  All details as for
    Figure~\ref{fig:nlpca-z-modes-ncep-godas-recon}.}
  \label{fig:nlpca-z-modes-cccma_cgcm3_1_t63-recon}
\end{figure}

In the cases where NLPCA(1) does a better job of fitting the input
data, it is usually found that the corresponding NLPCA(cir)
reconstruction manifold is rather degenerate, essentially emulating an
open curve where the cycling between \eln and \lan conditions runs
back and forth along the same track.  In these cases, it is clear that
the assumptions behind the NLPCA(cir) fit are violated: the data are
not well approximated by an open curve in the input data space.  Even
among the cases where NLPCA(cir) does a better job according to both
RMS error and explained variance, there are instances where the
NLPCA(cir) reconstruction is not very useful.  If the input data is
close to a Gaussian cloud (or, more likely, a Gaussian ``blini'' with
larger variance in one or two directions in principal component space
and smaller variance in the other directions), then the NLPCA(cir)
fits tend to result in a near circular path lying in the plane of the
principal components with the greatest variance, i.e. the PC \#1/PC
\#2 plane.  Figure~\ref{fig:nlpca-z-modes-cccma_cgcm3_1_t63-recon}
illustrates this situation for CGCM3.1(T63).  The input data is very
nearly a Gaussian cloud in the space spanned by the first three EOFs
and the NLPCA(cir) procedure produces a near circular manifold whose
axis is tilted at an angle of only 3.4\degree{} to the PC \#3 axis.

%
%
%

\begin{figure}
  \begin{center}
    \includegraphics[width=\textwidth]{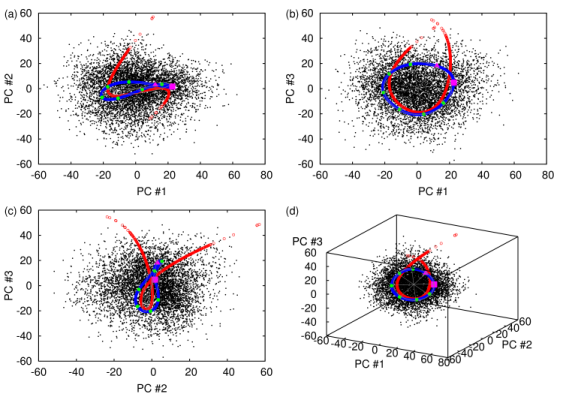}
  \end{center}
  \caption[CCSM3 NLPCA thermocline depth mode
    reconstructions]{Reconstruction plots for thermocline depth NLPCA
    modes for CCSM3.  All details as for
    Figure~\ref{fig:nlpca-z-modes-ncep-godas-recon}.}
  \label{fig:nlpca-z-modes-ncar_ccsm3_0-recon}
\end{figure}

\begin{figure}
  \begin{center}
    \includegraphics[width=\textwidth]{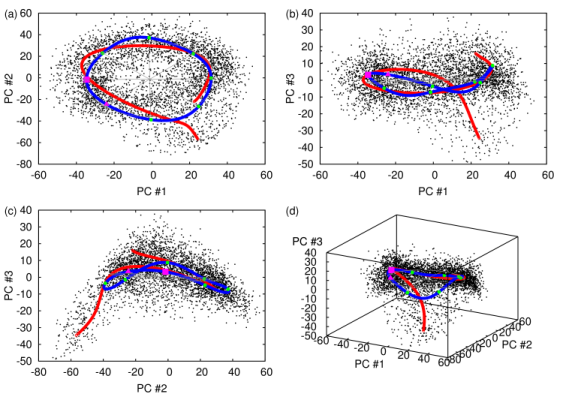}
  \end{center}
  \caption[FGOALS-g1.0 NLPCA thermocline depth mode
    reconstructions]{Reconstruction plots for thermocline depth NLPCA
    modes for FGOALS-g1.0.  All details as for
    Figure~\ref{fig:nlpca-z-modes-ncep-godas-recon}.}
  \label{fig:nlpca-z-modes-iap_fgoals1_0_g-recon}
\end{figure}

\begin{figure}
  \begin{center}
    \includegraphics[width=\textwidth]{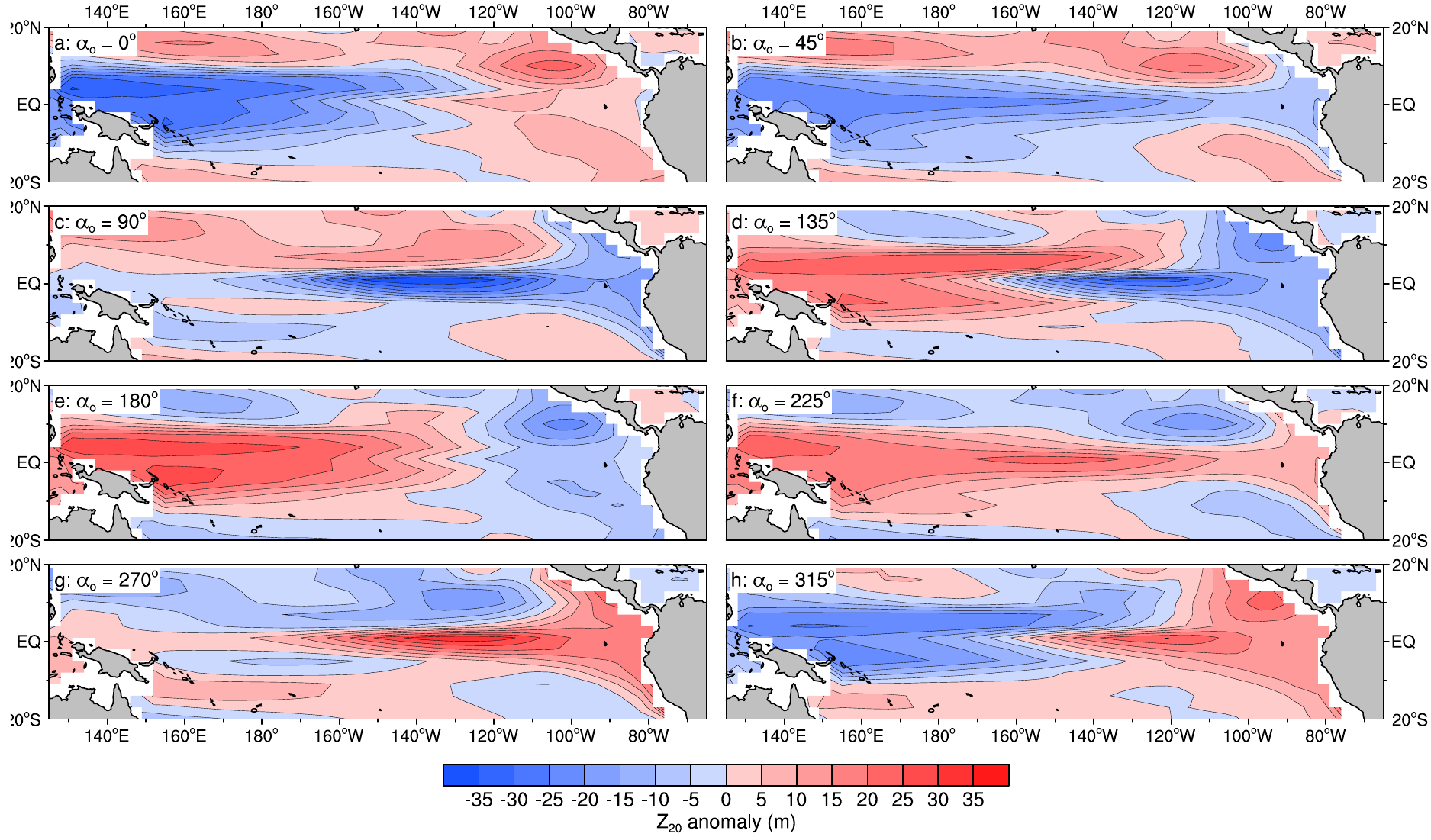}
  \end{center}
  \caption[FGOALS-g1.0 NLPCA thermocline depth mode spatial
    patterns]{Spatial pattern plots for NLPCA(cir) mode for
    FGOALS-g1.0 thermocline depth data.  Each panel shows the
    thermocline depth anomaly composite formed from the point along
    the one-dimensional NLPCA(cir) reduced manifold with the
    corresponding $\alpha_\circ$ value.  These values are highlighted
    on the reconstruction plots in
    Figure~\ref{fig:nlpca-z-modes-iap_fgoals1_0_g-recon}.}
  \label{fig:nlpca-z-modes-iap_fgoals1_0_g-patterns}
\end{figure}

\begin{figure}
  \begin{center}
    \includegraphics[width=\textwidth]{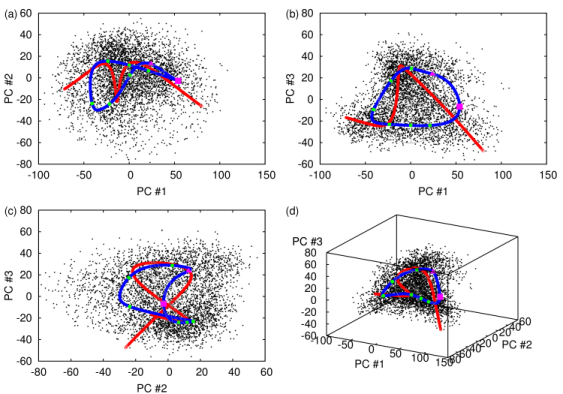}
  \end{center}
  \caption[ECHO-G NLPCA thermocline depth mode
    reconstructions]{Reconstruction plots for thermocline depth NLPCA
    modes for ECHO-G.  All details as for
    Figure~\ref{fig:nlpca-z-modes-ncep-godas-recon}.}
  \label{fig:nlpca-z-modes-miub_echo_g-recon}
\end{figure}

\begin{figure}
  \begin{center}
    \includegraphics[width=\textwidth]{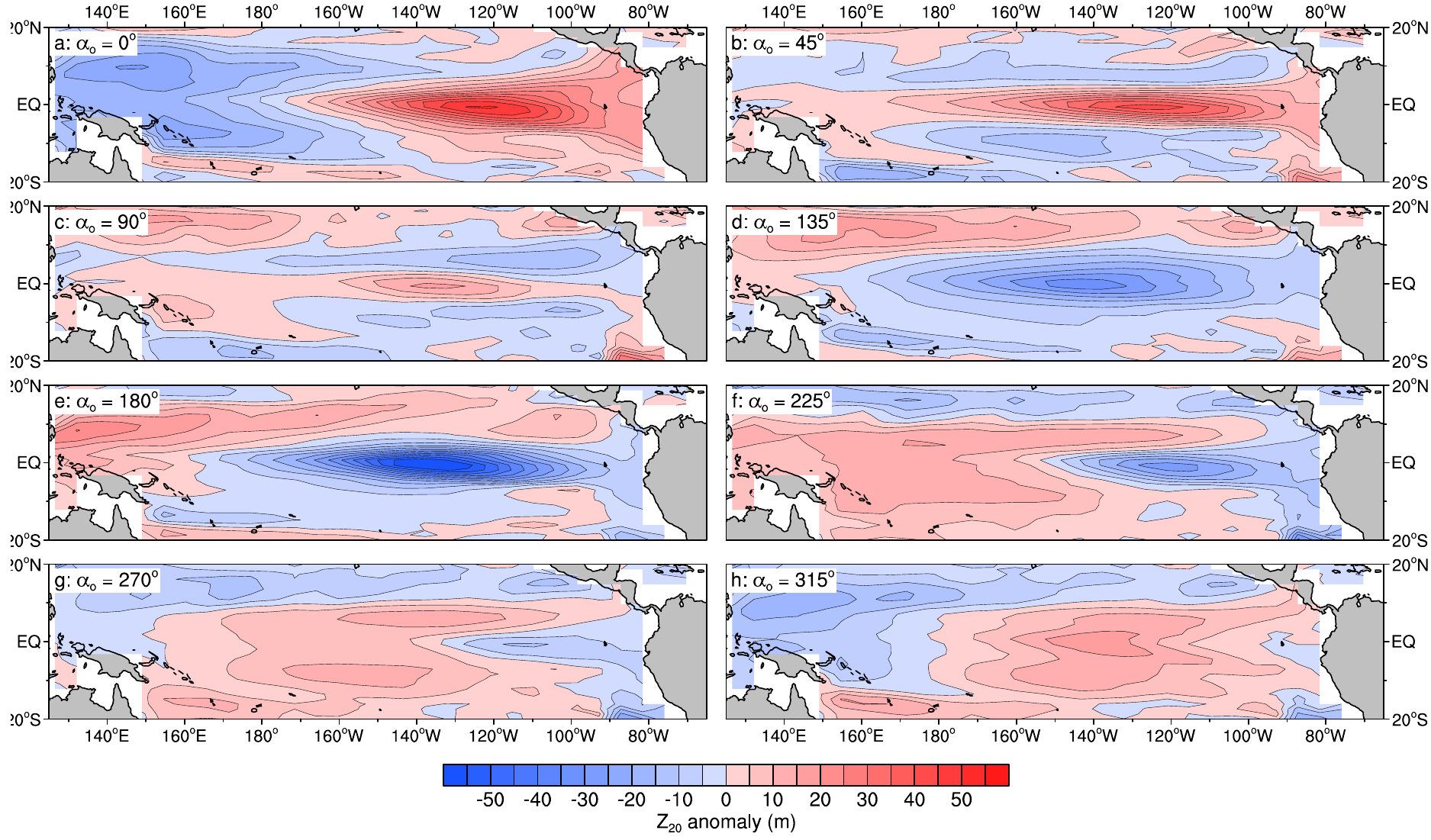}
  \end{center}
  \caption[ECHO-G NLPCA thermocline depth mode spatial
    patterns]{Spatial pattern plots for NLPCA(cir) mode for ECHO-G
    thermocline depth data.  Each panel shows the thermocline depth
    anomaly composite formed from the point along the one-dimensional
    NLPCA(cir) reduced manifold with the corresponding $\alpha_\circ$
    value.  These values are highlighted on the reconstruction plots
    in Figure~\ref{fig:nlpca-z-modes-miub_echo_g-recon}.}
  \label{fig:nlpca-z-modes-miub_echo_g-patterns}
\end{figure}

We will now concentrate on results for three illustrative models, each
displaying important characteristics of the NLPCA(cir) method, and all
of which are fitted better by NLPCA(cir) than NLPCA(1) according to
both measures displayed in Table~\ref{tab:nlpca-z-modes-results}.  For
each model, we will examine reconstruction plots
(Figures~\ref{fig:nlpca-z-modes-ncar_ccsm3_0-recon},~\ref{fig:nlpca-z-modes-iap_fgoals1_0_g-recon}~and~\ref{fig:nlpca-z-modes-miub_echo_g-recon})
and spatial pattern plots
(Figures~\ref{fig:nlpca-z-modes-iap_fgoals1_0_g-patterns}~and~\ref{fig:nlpca-z-modes-miub_echo_g-patterns},
no spatial pattern plot for CCSM3) as was done for the observational
data.  One general feature for all of the results shown is that the
NLPCA(1) manifold is quite overfitted: in most cases, the data we are
dealing with is quite noisy and is close to a Gaussian cloud, making
it difficult for NLPCA using a simple open curve to produce a
reasonable fit to the data without a lot of ``wiggles''
\citep{hsieh-noisy-nlpca}.

The first model we consider is CCSM3, for which the reconstruction
plot is shown in Figure~\ref{fig:nlpca-z-modes-ncar_ccsm3_0-recon}.
The results for CCSM3 in Table~\ref{tab:nlpca-z-modes-results}
indicate that there is only a small difference in the fits produced by
NLPCA(1) and NLPCA(cir).  The reconstruction plot shows a rather
unexpected result, as the NLPCA(cir) manifold lies primarily in the PC
\#1/PC \#3 plane with only small amounts of PC \#2 in any of the
reconstructed points.  The primary reason for this is that the \ztw
EOFs for the CCSM3 model are unlike those for the other models
(Figures~\ref{fig:z-eofs}d--f), with EOF 1 (Figure~\ref{fig:z-eofs}d)
having a structure more reminiscent of EOF 2 in some of the other data
sets, i.e. the zonal mean pattern of variability, once the major
east-west dipole mode of thermocline variability has been removed.
The other CCSM3 \ztw EOFs are rather hard to interpret.  This
highlights a quite severe problem with NLPCA.  Since it requires an
initial dimensionality reduction step in order to reduce the
complexity of the neural networks to be fitted, one has to select a
simple, generally linear, method to do this.  Using PCA is vulnerable
to problems in the variance structure of the original input data, and
some care is required to check that the inputs to the NLPCA algorithm
really do represent the expected form of variability.  To a certain
extent, this kind of result is to be expected for CCSM3.  On the basis
of the NINO3/WWV phasing plot shown in
Figure~\ref{fig:nino3-wwv-phasing}c, CCSM3 does not exhibit coherent
variations of equatorial warm water volume related to thermocline
depth variations.  The same conclusion can be drawn from a scatter
plot of PC \#1 and PC \#2 where points adjacent in time are connected
to help identify phase relationships (not shown).  The moral here is
that, although in this case NLPCA identifies that there is something
anomalous about the behaviour of CCSM3, a certain degree of
interpretation is required to understand exactly what the results
mean.

%
%
%

The second model we examine is FGOALS-g1.0, with reconstruction plot
in Figure~\ref{fig:nlpca-z-modes-iap_fgoals1_0_g-recon} and the
spatial patterns of \ztw anomaly corresponding to the highlighted
$\alpha_\circ$ values in
Figure~\ref{fig:nlpca-z-modes-iap_fgoals1_0_g-patterns}.  This is one
of the models for which NLPCA(cir) does rather better than NLPCA(1),
at least in terms of explained variance (there is little difference
between the RMS thermocline depth error for the two methods).  The
reconstruction plot shows that there is a reasonable level of
nonlinearity in the input data, including the hint of a strange
``ring'' structure in the PC \#1/PC \#2 plane.  This feature is also
visible in PC \#1 versus PC \#2 scatter plots for both thermocline
depth and SST, with the model appearing to avoid a region surrounding
the origin in both plots (not shown).  This ring feature arises as a
result of the excessive regularity of ENSO variability in FGOALS-g1.0
(cf. the NINO3 spectrum in Figure~\ref{fig:nino3-spectra}), with no
``loitering'' between ENSO events as is seen in observational data and
some of the other models.  The NLPCA fits here extend noticeably out
of the PC \#1/PC \#2 plane, even though there is a large spectral gap
in the PCA eigenvalue spectrum, with the first two EOFs together
explaining 72.1\% of the total data variance, while the next EOF
explains only 3.7\%.  By eye, the NLPCA(1) manifold appears to do a
better job of capturing the data variability, including as it does
most of the circuit of the NLPCA(cir) manifold as well as an extension
down into the ``lip'' of data points with more negative PC \#3 values.
This shows how these plots are a little deceptive, since the explained
variance for the first NLPCA(1) mode is 61.4\%, while that for the
first NLPCA(cir) mode is 63.6\%.  This effect arises, once again, from
the data point density-dependent nature of the NLPCA cost function.
Use of the mean square difference as an error measure between the
reconstruction and original data biases the reconstruction away from
more rarely visited regions of data space, regions that may be of
particular dynamical interest.  This problem motivates the use of
different norms for measuring the error between reconstruction and
original data \citep{hsieh-noisy-nlpca,cannon-robust-nlpca}, as well
as completely different statistical methods more effective for
characterising extreme events
\citep{bernacchia-cumulant-1,bernacchia-cumulant-2}.

The spatial patterns for points around the FGOALS-g1.0 reconstruction
manifold (Figure~\ref{fig:nlpca-z-modes-iap_fgoals1_0_g-patterns}) are
qualitatively similar to those for the observational data shown in
Figure~\ref{fig:nlpca-z-modes-ncep-godas-patterns}, although there are
some important differences.  First, the asymmetry seen in the
observational data between the positive and negative zonal thermocline
depth gradient states is absent in FGOALS-g1.0.  The positive zonal
thermocline depth gradient state
(Figure~\ref{fig:nlpca-z-modes-ncep-godas-patterns}a at $\alpha_\circ
= 0$) and the negative zonal thermocline depth gradient state
(Figure~\ref{fig:nlpca-z-modes-ncep-godas-patterns}e at $\alpha_\circ
= 180\degree$) are almost identical apart from their difference in
sign.  In fact this correspondence of spatial patterns between
opposite (i.e. differing by 180\degree{} in $\alpha_\circ$) points
holds all round the cycle of thermocline depth variation for this
model, in contrast to the strong asymmetry between recharge and
discharge phases of the cycle seen in the observations.

%
%
%

%
%
%

The final model we examine is ECHO-G, with reconstruction plot in
Figure~\ref{fig:nlpca-z-modes-miub_echo_g-recon} and spatial patterns
of \ztw anomaly corresponding to the highlighted $\alpha_\circ$ values
in Figure~\ref{fig:nlpca-z-modes-miub_echo_g-patterns}.  This is a
model for which the NLPCA(cir) reconstruction is somewhat better than
the NLPCA(1) reconstruction in terms of explained variance, and there
is little to choose between the two in terms of RMS thermocline depth
error (Table~\ref{tab:nlpca-z-modes-results}).  The reconstruction
plot in Figure~\ref{fig:nlpca-z-modes-miub_echo_g-recon} shows that
the \ztw principal component data used as input to the NLPCA algorithm
is notably nonlinear, which is reflected in the structure of the
NLPCA(cir) reconstruction manifold.  The best fit circle to the
NLPCA(cir) manifold has an axis that is tilted to the PC \#3 axis by
an angle of about 105\degree, and the manifold is strongly distorted
from the near-circular shape seen in some of the other model examples.
On the plots shown here, the fitting of the manifold to the input data
may appear somewhat arbitrary.  Interactive exploration of the data
provides some confidence that the reconstruction manifold does pass
preferentially through regions of high data density and presents a
reasonable fit to the input data, but there remains a sense in which
the fit \emph{is} arbitrary.  Unless the input data is clearly not
well fitted by a closed curve, in which case a degenerate solution is
found, the NLPCA(cir) algorithm is guaranteed to find the open curve
best fitting the data in a least squares sense.  There is an extent to
which reconstruction plots like
Figure~\ref{fig:nlpca-z-modes-miub_echo_g-recon} are susceptible to
pareidoliac interpretation --- we want to see a cycle in the data, so
we see a cycle in the data.  This highlights another disadvantage of
NLPCA and similar neural network based methods.  Extreme care is
required in fitting functions using complex parametric models, and, in
the presence of the large amount of noise found in most climate data
sets, the measures commonly taken to avoid overfitting
(Section~\ref{sec:nlpca-overfitting}) may not be sufficient.  In the
cases presented here, because most of the variability in thermocline
depth is captured by the first few EOFs, one can examine two- or
three-dimensional reconstruction plots to get a feeling for how
reasonable the NLPCA fits to the input data really are, but this
option is not available in problems with a slower fall-off in the PCA
eigenvalue spectrum.  A good example is mid-latitude mid-tropospheric
variability, where estimates of the effective data dimensionality
range from about six for monthly mean data
(Section~\ref{sec:isomap-conclusions} mentions an application of the
Isomap dimensionality reduction algorithm to this type of data) to
several hundred for data measured at a daily timescale
\citep[e.g.,][]{achatz-opsteegh-i,achatz-opsteegh-ii,achatz-branstator-1999}.

Given these caveats, interpretation of the spatial patterns of
thermocline depth variability shown for ECHO-G in
Figure~\ref{fig:nlpca-z-modes-miub_echo_g-patterns} requires some
care.  Reasonably realistic variation is seen between conditions with
a strong zonal \ztw gradient
(Figure~\ref{fig:nlpca-z-modes-miub_echo_g-patterns}a at $\alpha_\circ
= 0$ and Figure~\ref{fig:nlpca-z-modes-miub_echo_g-patterns}f at
$\alpha_\circ = 225\degree$) and more zonally symmetric conditions
(e.g., Figure~\ref{fig:nlpca-z-modes-miub_echo_g-patterns}b at
$\alpha_\circ = 45\degree$), and there is a definite difference in
amplitude between conditions with a positive zonal thermocline depth
gradient (Figure~\ref{fig:nlpca-z-modes-miub_echo_g-patterns}a at
$\alpha_\circ = 0$) and those with a negative zonal gradient
(Figure~\ref{fig:nlpca-z-modes-miub_echo_g-patterns}f at $\alpha_\circ
= 225\degree$).  However, the picture in the transitional states
between the extreme \eln and \lan conditions is rather confused, and
although there appears to be some zonally symmetric warm water
discharge/recharge activity, this does not appear to proceed
consistently around the NLPCA(cir) manifold.  Consider, for example,
the sequence of patterns running from $\alpha_\circ = 0$ to
$\alpha_\circ = 225\degree$,
i.e. Figures~\ref{fig:nlpca-z-modes-miub_echo_g-patterns}a--f.  In
Figures~\ref{fig:nlpca-z-modes-miub_echo_g-patterns}b~and~c
($\alpha_\circ = 45\degree$ and 90\degree), there is an anomalously
positive, zonally symmetric displacement to the equatorial thermocline
depth, while in
Figures~\ref{fig:nlpca-z-modes-miub_echo_g-patterns}d~and~e
($\alpha_\circ = 135\degree$ and 180\degree), this is replaced by a
negative displacement.  It is difficult to tie these variations to any
consistent idea of equatorial warm water volume discharge during this
phase of transition from \eln to \lan conditions.

%
%
%

%
%
%

\section{Discussion and conclusions}
\label{sec:nlpca-discussion}

In this chapter, I have examined results of NLPCA applied to both
simple geometrical test data and rather noisy climate data sets.
NLPCA does a reasonable job of representing nonlinear and asymmetric
variability in both SST and thermocline depth data.  It is possible
that NLPCA may provide a clearer picture of some nonlinear and
asymmetric climate phenomena than is available from linear methods
such as PCA.  Although this conclusion might have been inferred from
the strong non-Gaussianity of some of the principal component scatter
plots of Chapter~\ref{ch:obs-cmip-enso} (e.g.,
Figure~\ref{fig:sst-pc-scatter}c), it is made clearer by the error and
explained variance statistics in
Tables~\ref{tab:nlpca-sst-mode-1-results},
\ref{tab:sst-modes-1-2-variance} and \ref{tab:nlpca-z-modes-results}.
A nonlinear fit really does appear to be better for some of the cases
studied here, and NLPCA is able to identify such nonlinear fits well.

Computation of correlations between the NINO3 SST index and $\alpha_1$
values for the first NLPCA SST mode, and between equatorial warm water
volume anomalies and $\alpha_2$ values for the second NLPCA SST mode
(Table~\ref{tab:alpha-correlations}) shows that the first NLPCA SST
mode does indeed represent the leading mode of variability associated
with ENSO, with, for both observations and most models, correlation
coefficients of greater than 0.9.  There is also a suggestion that the
second NLPCA SST mode is associated with the second main mode of ENSO
variability, here represented as anomalies of equatorial warm water
volume.  There is much more variability between the models in the
second NLPCA mode.

%
%
\begin{table}
  \begin{center}
    \input{figs/06/alpha-correlations.tex}

  \end{center}
  \caption[NLPCA $\alpha_1$ and $\alpha_2$ correlations]{Correlation
    coefficients between the NINO3 SST index and $\alpha_1$ and
    between the equatorial Pacific warm water volume and $\alpha_2$
    for observations and all models.}
  \label{tab:alpha-correlations}
\end{table}
%

There is wide variation in the degree of nonlinearity seen in the
models, some of which seems to be related to the strength of their
ENSO variability.  CGCM3.1 (T63), for instance, has among the smallest
NINO3 SST index variability of all models in the CMIP3 ensemble
($\sigma_{\mathrm{NINO3}} = \mathrm{0.64\degree C}$ from
Table~\ref{tab:models}), and also has a very Gaussian SST anomaly
distribution
(Figure~\ref{fig:nlpca-sst-mode-1-cccma_cgcm3_1_t63-recon}).  In
contrast, GFDL-CM2.1 ($\sigma_{\mathrm{NINO3}} = \mathrm{1.52\degree
  C}$) has a very non-Gaussian SST anomaly distribution.  This is
reflected in the NLPCA reconstruction plots
(Figure~\ref{fig:nlpca-sst-mode-1-gfdl_cm2_1-recon}) and the relative
magnitudes of the variance explained by the first PCA mode and first
NLPCA mode (Table~\ref{tab:nlpca-sst-mode-1-results}).  This clear
difference in behaviour of the NLPCA fits between models makes NLPCA a
plausible tool for inter-model comparison in this context.

The differences in \eln/\lan asymmetry seen in the models can be
compared to the spatial asymmetry calculations from
Section~\ref{sec:sst-asymmetry}, based on \citep{monahan-dai}.  The
simplest approach seems to be to compare the $||\vec{a}_S|| /
||\vec{a}_A||$ norm ratio in Table~\ref{tab:asymmetry-measures}, large
values of which represent greater asymmetry between positive \eln and
negative \lan SST composites, with the improvement in explained
variance going from PCA SST mode 1 to NLPCA SST mode 1 in
Table~\ref{tab:nlpca-sst-mode-1-results}, which is a measure of the
degree of nonlinearity present in the data used as input to the NLPCA
procedure.  Unfortunately, there does not appear to be a clear link
between the results of the two methods.  For the observational data,
which show a 1.6\% increase in explained variance between PCA SST mode
1 and NLPCA SST mode 1, the asymmetry norm ratio is 0.11.  Of the
models that show the greatest improvement in explained variance
between PCA SST mode 1 and NLPCA SST mode 1 (CNRM-CM3, ECHO-G, GISS-EH
and GFDL-CM2.1), two, GFDL-CM2.1 and GISS-EH, have asymmetry norm
ratios in the same range as the observations (both 0.10), while the
other two models have smaller ratios (both 0.05).  Conversely, some of
the models with asymmetry norm ratios closest to the observations (in
particular, CGCM3.1 (T47) and CGCM3.1 (T63)) show little or no
difference in explained variance between PCA SST mode 1 and NLPCA SST
mode 1.  The asymmetry measures presented earlier thus do not appear
to match up with the observed nonlinearity and \eln/\lan asymmetry
seen in the NLPCA results.  One possible explanation for this is that,
because the \citet{monahan-dai} asymmetry analysis is based on
composites of positive and negative excursions of the principal
components of SST, it may weight the spatial patterns of \eln and \lan
differently than the NLPCA analysis, where we are finding extreme end
point patterns along the one-dimensional NLPCA manifold.

From the point of view of assessing which models have ``good'' ENSO
variability, it is not clear how useful the NLPCA results are.  The
observational data indicates that the real atmosphere-ocean system
does exhibit some degree of nonlinearity in its interannual behaviour,
but it is difficult to decide just how to measure this nonlinearity in
the models.  Some clearly have too little, showing very little
interannual variability that can be associated with ENSO (e.g.,
CGCM3.1 (T47) and CGCM3.1 (T63)).  Some others appear to display too
much, in particular FGOALS-g1.0 and GFDL-CM2.1, but it is not clear
exactly what is the relationship between the degree of nonlinearity in
a model and fidelity of its interannual variability --- FGOALS-g1.0
has a very strong and far too regular interannual oscillation in the
Pacific, while GFDL-CM2.1 was assessed by \citet{vanoldenborgh-enso}
as having ENSO variability characteristics among the best of the CMIP3
models.

The circular bottleneck layer networks used in the analysis of
thermocline variability in Section~\ref{sec:nlpca-z-results} are
rather difficult to interpret.  In the first paper where this method
was applied to thermocline depth anomaly data
\citep{an-enso-interdecadal}, no attempt was made to determine whether
the circular bottleneck layer network used provided a better fit to
the data than a network with a single bottleneck neuron.  The
explained variance and error measures shown in
Table~\ref{tab:nlpca-z-modes-results} would seem to indicate that
there is little to choose between the NLPCA(1) and NLPCA(cir) fits in
many cases.  When there is a clear preference for one method over the
other, it is as likely to be in favour of NLPCA(1) as the circular
network method.  To some extent, the use of a circular network rather
begs the question, since the NLPCA fitting algorithm will happily fit
a circular network to any data set, even though we can show that a
Gaussian cloud will often be better fitted in an RMS error sense by a
linear PCA reduction than by a circular manifold.  Suppose that we
have a distribution of points $\rho(\vec{x})$ in $\mathbb{R}^n$, and a
projection to a lower dimensional manifold that we denote by $P$.  For
a normalised distribution $\rho(\vec{x})$, the RMS error between the
projected points and the original points is
\begin{equation}
  \text{RMSE} = \left( \int (\vec{x} - P\vec{x})^2 \rho(\vec{x}) \,
  d^n \vec{x} \right)^{1/2}.
\end{equation}
Consider the two-dimensional case where $\rho(\vec{x})$ is a Gaussian
distribution so that
\begin{equation}
  \rho(\vec{x}) = \frac{1}{2\pi (\det \bm{\Sigma})^{1/2}} \exp \left(
  -\frac{1}{2} \vec{x}^T \bm{\Sigma}^{-1} \vec{x} \right),
\end{equation}
where $\bm{\Sigma}$ is the covariance matrix of the distribution, and
\index[not]{determinant@$\det \mathbf{A}$, matrix determinant}$\det
\mathbf{A}$ denotes the determinant of a matrix $\mathbf{A}$.  Without
loss of generality, we choose the principal axes of the distribution
to be oriented along the coordinate axes, so that the covariance
matrix $\bm{\Sigma}$ is diagonal:
\begin{equation}
  \bm{\Sigma} = \begin{pmatrix}
    \alpha^2 & 0 \\
    0 & \beta^2
  \end{pmatrix}.
\end{equation}
We orient the axes so that $\alpha \geq \beta$.  This means that the
$x$-axis is the leading principal axis of the distribution.  Now
consider two projections, one, $P_1$, that projects all points to the
$x$-axis, which is essentially a linear PCA reduction of the
distribution, and the other, $P_2$, that projects points to a circle
of radius $r$ in the $xy$-plane centred on the origin.  Explicitly,
\begin{equation}
  \begin{gathered}
    P_1 \, (x, y)^T = (x, 0)^T, \\
    P_2 \, (x, y)^T = (r \cos \theta, r \sin \theta)^T,
  \end{gathered}
\end{equation}
where $\theta = \tan^{-1}(y/x)$.  $P_2$ represents the type of
projection produced by a circular NLPCA network.  Calculating the RMS
error for $P_1$,
\begin{equation}
  (\text{RMSE}_1)^2 = \int y^2 \rho(\vec{x}) \, dx \, dy = \beta^2
\end{equation}
and for $P_2$,
\begin{equation}
  (\text{RMSE}_2)^2 = \int [(x - r \cos \theta)^2 + (y - r \sin
    \theta)^2] \rho(\vec{x}) \, dx \, dy.
\end{equation}
This latter integral is rather tricky to evaluate, but its value is
\begin{equation}
  \label{eq:rmse2}
  (\text{RMSE}_2)^2 = \alpha^2 + \beta^2 + r^2 - 2 \alpha
  \sqrt{\frac{2}{\pi}} r E\left[ 1 - \frac{\beta^2}{\alpha^2} \right],
\end{equation}
where $E[m]$ denotes the complete elliptic integral of the second
kind, defined by
\begin{equation}
  E[m] = \int_0^{\pi/2} (1 - m \sin^2 \theta)^{1/2} \, d\theta.
\end{equation}
We can find the minimum value of $\text{RMSE}_2$ by differentiating
\eqref{eq:rmse2} with respect to $r$ and setting the derivative to
zero.  This gives the minimum value of $\text{RMSE}_2$ as
\begin{equation}
  (\text{RMSE}_2^{\mathrm{(min)}})^2 = \beta^2 - \alpha^2 \left(
  \sqrt{\frac{2}{\pi}} E\left[ 1 - \frac{\beta^2}{\alpha^2} \right] -
  1 \right).
\end{equation}
The expression in parentheses is positive so long as $\beta / \alpha
\gtrsim 0.565$, so that when this condition is satisfied
$\text{RMSE}_2 < \text{RMSE}_1$ and the ``circular'' projection does a
better job than the linear PCA projection.  For $\beta / \alpha
\lesssim 0.565$ the linear projection will be better.  (This makes
sense intuitively: for more nearly isotropic distributions, a circular
projection should be possible, while for long thin distributions, then
linear projection should always be better.)  The essential point here
is that, whatever the value of $\beta / \alpha$, the circular network
NLPCA algorithm will still attempt to fit a circular manifold to the
data, even if this is quite inappropriate.  It is thus important to
apply some form of validation to the results, as has been done here,
to decide whether the circular manifold results are better than an
alternative open manifold reduction.

Once a reasonably good NLPCA(cir) fit to a data set has been found,
the natural temptation is to interpret this in terms of an intrinsic
periodic oscillation in the data, but this requires some care.  This
interpretation is almost certainly justified in the thermocline depth
analysis performed by \citet{an-enso-interdecadal} and here, since
there is corroborating evidence that there are regular modes of
thermocline variability that change in quadrature, one associated with
thermocline tilt and one with zonal mean thermocline depth
\citep[e.g.,][]{kessler-events,meinen-wwv,battisti-icm}.  Without this
corroborating evidence, much more caution would be required in
interpreting the reconstructions and spatial pattern variations seen
in the NLPCA(cir) results.  In fact, such caution is probably
justified in interpreting some of the model results, since it is not
at all clear that the thermocline depth variability seen in some of
the models is a good fit to observed thermocline variability.  The
temporal phasing of SST and thermocline depth variability during \eln
events is certainly not represented well in many of the models
(Section~\ref{sec:nino3-wwv-phasing}).

Beyond these specific concerns about the techniques used in NLPCA and
the interpretation of some of the results, there is a larger issue
about the use of neural networks in this type of data analysis.
Training of neural network models requires the fitting of a large
number of parameters, essentially solving a minimisation problem in a
very high-dimensional space (recall the parameter counts for typical
networks shown in Table~\ref{tab:nlpca-param-counts}).  This type of
optimisation problem is notoriously difficult, as evinced by both the
amount of effort that has to be put into ensuring that the solutions
are not overfitted and the relative rarity of good fits among the
random initial weights used in the optimisation process
(Figures~\ref{fig:nlpca-spiral-errors}~and~\ref{fig:nlpca-ersst-errors}).
Several extensions and modifications of NLPCA have been developed to
try to alleviate some of these problems, as well as difficulties
caused by the use of noisy data in climate data analysis.  Most of
these modifications involve the use of alternative norms and error
measures in the overall cost function
\citep{hsieh-noisy-nlpca,cannon-robust-nlpca}.  The assessment of
these modifications to the NLPCA method is difficult.  Although there
exists an extensive mathematical theory for neural networks
\citep{bishop-nn-book,anthony-nn-book}, it is not immediately clear
how to apply this to NLPCA as it stands.  As just one example, one
could argue that NLPCA decompositions should be invariant with respect
to rotation of EOFs, since these correspond to simple orthogonal
transformations in the NLPCA input data space.  However, the simple
weight decay term used in the NLPCA cost function
\eqref{eq:weight-penalty-constraint} is not invariant under linear
transformations of the inputs \citep[Section~9.2.2]{bishop-nn-book},
which means that this simple consistency condition is violated.  What
implications this has for interpretation of NLPCA decompositions is
unclear.

Despite these reservations, NLPCA does provide a relatively simple and
relatively transparent nonlinear extension to standard linear PCA.
The large number of applications NLPCA and related techniques have
found in climate science indicate its worth, at least as a supplement
to more traditional methods of data analysis.


%% file: figs/06/nlpca-sst-mode-1-results.tex
\begin{tabular}{lccc!{\quad\,}ccc}
\toprule
\multirow{2}{3cm}{\bf Model} & \multicolumn{3}{c}{\bf RMS Error (\degree{}C)} & \multicolumn{3}{c}{\bf Variance} \\
 & {\bf PCA} & {\bf NLPCA} & {\bf Frac.} & {\bf PCA} & {\bf NLPCA} & {\bf 10 PCs} \\
\midrule
{\bf Observations} & 0.378 & 0.371 & 98.1\% & 52.3\% & 53.9\% & 88.5\% \\
BCCR-BCM2.0 & 0.539 & 0.537 & 99.7\% & 41.1\% & 41.4\% & 73.0\% \\
CCSM3 & 0.407 & 0.406 & 99.7\% & 46.2\% & 46.6\% & 72.7\% \\
CGCM3.1(T47) & 0.352 & 0.351 & 99.7\% & 29.9\% & 30.3\% & 67.2\% \\
CGCM3.1(T63) & 0.337 & 0.337 & 99.9\% & 31.8\% & 32.0\% & 67.7\% \\
{\bf CNRM-CM3} & 0.593 & 0.582 & 98.3\% & 62.8\% & 64.3\% & 88.7\% \\
CSIRO-Mk3.0 & 0.499 & 0.536 & 107.5\% & 45.9\% & 37.6\% & 76.3\% \\
{\bf ECHO-G} & 0.536 & 0.523 & 97.7\% & 58.8\% & 60.6\% & 79.7\% \\
FGOALS-g1.0 & 0.524 & 0.514 & 98.1\% & 77.2\% & 78.6\% & 94.2\% \\
GFDL-CM2.0 & 0.481 & 0.475 & 98.8\% & 53.1\% & 54.2\% & 80.6\% \\
{\bf GFDL-CM2.1} & 0.579 & 0.542 & 93.6\% & 59.8\% & 64.8\% & 87.3\% \\
{\bf GISS-EH} & 0.492 & 0.482 & 97.8\% & 31.3\% & 33.9\% & 58.8\% \\
INM-CM3.0 & 0.580 & 0.585 & 100.9\% & 40.6\% & 39.9\% & 73.3\% \\
IPSL-CM4 & 0.414 & 0.412 & 99.5\% & 52.8\% & 53.2\% & 78.4\% \\
MIROC3.2(hires) & 0.361 & 0.382 & 106.1\% & 22.4\% & 11.8\% & 57.2\% \\
MIROC3.2(medres) & 0.335 & 0.337 & 100.6\% & 46.6\% & 45.9\% & 78.7\% \\
MRI-CGCM2.3.2 & 0.412 & 0.409 & 99.3\% & 53.3\% & 53.9\% & 81.8\% \\
UKMO-HadCM3 & 0.606 & 0.603 & 99.6\% & 38.7\% & 39.1\% & 68.4\% \\
UKMO-HadGEM1 & 0.538 & 0.540 & 100.4\% & 31.9\% & 31.4\% & 60.0\% \\
\bottomrule
\end{tabular}

%% file: figs/06/nlpca-z-modes-results.tex
\begin{tabular}{lccc!{\quad\,}cccc}
\toprule
\multirow{2}{3cm}{\bf Model} & \multicolumn{3}{c}{\bf RMS Error (m)} & \multicolumn{4}{c}{\bf Variance} \\
 & {\bf PCA} & {\bf NL(1)} & {\bf NL(c)} & {\bf PCA} & {\bf NL(1)} & {\bf NL(c)} & {\bf 5 PCs} \\
\midrule
Observations & 12.426 & {\bf 11.045} (6) & 11.553 (7) & 29.6\% & {\bf 43.4\%} & 38.7\% & 60.7\% \\
BCCR-BCM2.0 & 16.272 & {\bf 16.268} (6) & 16.322 (6) & 17.2\% & {\bf 19.1\%} & 18.2\% & 38.0\% \\
CCSM3 & 11.695 & 11.514 (6) & {\bf 11.458} (3) & 19.5\% & 22.0\% & {\bf 22.6\%} & 51.4\% \\
CGCM3.1(T47) & {\bf 6.929} & 7.271 (6) & 7.246 (7) & {\bf 18.5\%} & 12.8\% & 12.8\% & 59.0\% \\
CGCM3.1(T63) & 6.685 & 6.395 (6) & {\bf 6.292} (6) & 18.8\% & 24.7\% & {\bf 26.9\%} & 58.7\% \\
CNRM-CM3 & 17.087 & {\bf 15.641} (6) & 15.668 (8) & 32.0\% & {\bf 43.9\%} & 42.4\% & 64.6\% \\
CSIRO-Mk3.0 & 11.829 & {\bf 11.746} (5) & 11.798 (7) & 17.7\% & 18.8\% & {\bf 19.2\%} & 47.4\% \\
ECHO-G & 13.523 & {\bf 12.767} (6) & 12.771 (6) & 28.7\% & 36.4\% & {\bf 38.3\%} & 64.1\% \\
FGOALS-g1.0 & 9.912 & 8.317 (5) & {\bf 8.035} (8) & 46.7\% & 61.4\% & {\bf 63.6\%} & 81.6\% \\
GFDL-CM2.0 & 14.518 & 14.129 (4) & {\bf 13.916} (6) & 16.3\% & 20.7\% & {\bf 23.0\%} & 48.1\% \\
GFDL-CM2.1 & 15.233 & {\bf 14.069} (6) & 14.376 (7) & 29.9\% & {\bf 40.4\%} & 35.6\% & 61.3\% \\
GISS-EH & 8.673 & {\bf 8.524} (6) & 8.644 (8) & 34.3\% & {\bf 37.2\%} & 35.0\% & 54.5\% \\
INM-CM3.0 & 9.688 & 9.453 (6) & {\bf 9.330} (5) & 21.8\% & 25.0\% & {\bf 27.0\%} & 51.1\% \\
MIROC3.2(hires) & 12.223 & {\bf 12.170} (6) & 12.217 (7) & 9.1\% & 9.7\% & {\bf 9.8\%} & 21.5\% \\
MIROC3.2(medres) & {\bf 6.755} & 6.834 (6) & 6.768 (7) & {\bf 22.4\%} & 20.7\% & 21.7\% & 53.8\% \\
MRI-CGCM2.3.2 & 7.755 & 7.158 (6) & {\bf 7.023} (8) & 28.0\% & 38.7\% & {\bf 40.6\%} & 66.6\% \\
UKMO-HadCM3 & 13.607 & 14.188 (6) & {\bf 13.474} (5) & 20.3\% & 13.8\% & {\bf 21.2\%} & 50.7\% \\
UKMO-HadGEM1 & 14.592 & 14.539 (5) & {\bf 14.493} (7) & 13.5\% & 14.1\% & {\bf 14.7\%} & 34.1\% \\
\bottomrule
\end{tabular}

%% file: figs/06/alpha-correlations.tex
\begin{tabular}{lcc}
\toprule
{\bf Model} & $\Corr(\text{NINO3}, \alpha_1)$ & $\Corr(\text{WWV}, \alpha_2)$ \\
\midrule
Observations      & 0.955 &  0.497 \\
BCCR-BCM2.0       & 0.978 & -0.161 \\
CCSM3             & 0.979 & -0.358 \\
CGCM3.1 (T47)     & 0.031 & -0.346 \\
CGCM3.1 (T63)     & 0.847 & -0.223 \\
CNRM-CM3          & 0.954 &  0.638 \\
CSIRO-Mk3.0       & 0.797 & -0.200 \\
ECHO-G            & 0.984 &  0.431 \\
FGOALS-g1.0       & 0.992 & -0.828 \\
GFDL-CM2.0        & 0.913 &  0.381 \\
GFDL-CM2.1        & 0.876 &  0.235 \\
GISS-EH           & 0.813 &  0.305 \\
INM-CM3.0         & 0.821 & -0.088 \\
IPSL-CM4          & 0.970 &   n/a  \\
MIROC3.2 (hires)  & 0.496 & -0.196 \\
MIROC3.2 (medres) & 0.842 &  0.181 \\
MRI-CGCM2.3.2     & 0.953 &  0.000 \\
UKMO-HadCM3       & 0.932 & -0.393 \\
UKMO-HadGEM1      & 0.909 & -0.226 \\
\bottomrule
\end{tabular}

%% file: 07-isomap.tex
\chapter{Isomap}
\label{ch:isomap}

\section{Description of method}
\label{sec:isomap-algorithm}

The Isomap algorithm is a two-step process that simultaneously
attempts to find a low-dimensional manifold in which a set of data
points lies, and Euclidean coordinates parameterising this
low-dimensional manifold.  The first step in the algorithm uses a
graph-based approximation to the data manifold to calculate
approximate geodesic distances between data points
(Section~\ref{sec:isomap-geodesics}).  These geodesic distances are
then analysed using multidimensional scaling (MDS) to find an
isometric embedding of the data manifold in a lower-dimensional space
(Section~\ref{sec:isomap-mds}).

\subsection{Geodesic approximation}
\label{sec:isomap-geodesics}

As will be explained below, principal component analysis can be
considered as an application of the same multidimensional scaling
approach used in Isomap.  The key factor that distinguishes Isomap
from PCA is that Isomap uses a distance function that approximates
geodesic distances in the data manifold, while PCA employs Euclidean
distances in the original high-dimensional data space.  The aim of the
Isomap geodesic approximation is to determine the intrinsic structure
of the data manifold, independent of accidental details of the
embedding of this manifold in the data space.

Geodesics in the data manifold are approximated in two stages.  First,
a weighted graph is constructed whose vertices are the data points and
whose edges connect each point to its nearest neighbours, as
determined by Euclidean distances between the data points.  The edge
weights of the graph are the Euclidean distances.  There are two ways
to set this nearest neighbour graph up.  A distance threshold,
$\varepsilon$, can be used, so that edges are included in the graph
from a point to all other points closer than $\varepsilon$.  If the
set of points is denoted by $V \subset \mathbb{R}^m$, the nearest
neighbour graph $\mathcal{G}_\varepsilon$ is then
\begin{equation}
  \mathcal{G}_\varepsilon = (V, E_\varepsilon) = (V, \{ (\vec{x},
    \vec{y}) \, | \, \vec{x}, \vec{y} \in V, \, || \vec{x} - \vec{y}
    || < \varepsilon \}).
\end{equation}
The main benefit of this definition is that it is somewhat insensitive
to inhomogeneities in data point sampling density, and can lead to
more robust MDS results.  Its primary disadvantage is that it is
difficult to establish a reasonable value for $\varepsilon$ without
some experimentation and it may be necessary to select an
inappropriately large value for $\varepsilon$ in order to ensure that
the graph $\mathcal{G}_\varepsilon$ is connected.  The second approach
is to use a nearest neighbour count, $k$, so that the nearest
neighbour graph contains, for each data point, edges to the $k$
nearest neighbours.  The graph $\mathcal{G}_k$ is then defined as
\begin{equation}
  \mathcal{G}_k = (V, E_k) = (V, \{ (\vec{x}, \vec{y}) \, | \,
    \vec{x}, \vec{y} \in V, \, i_{\vec{x}}(\vec{y}) \leq k \}),
\end{equation}
where $i_{\vec{x}}(\vec{y})$ is the index of point $\vec{y}$ in a list
of the points $V \backslash \vec{x}$ sorted in increasing order of
distance from $\vec{x}$.  This method is simple to implement, but does
display a greater degree of sensitivity to variations in data point
sampling density.

Once the distance-weighted nearest neighbour graph has been
constructed, using either the $\varepsilon$-Isomap or $k$-Isomap
method, distances between arbitrary data points, $d_G(\vec{x},
\vec{y})$, are defined by shortest paths in the graph, with the
approximations to the geodesic distances calculated by summing the
edge weights (i.e. inter-point Euclidean distances) along the shortest
paths.  These shortest paths can be determined using standard graph
algorithms; here, I use Floyd's all-sources shortest paths algorithm
\citep{ahu-algorithms}.  Although this algorithm has time complexity
$O(N^3)$, it is good enough for our purposes since the number of data
points is not large ($N \leq 6000$).  More efficient algorithms, for
instance a Fibonacci heap-based implementation of Dijkstra's
algorithm, give better performance for larger data sets.

Asymptotic convergence results exist showing that the difference
between the approximation $d_G(\vec{x}, \vec{y})$ and the true
geodesic distance in the data manifold, $d_M(\vec{x}, \vec{y})$, tends
to zero in a probabilistic sense as the density of data points
increases \citep{bernstein-isomap-proofs}.  From these results, one
can derive a required data point density to achieve any desired
accuracy for $d_G(\vec{x}, \vec{y})$.  Unfortunately, these results
are of limited use in practice.  One usually starts with a set of data
with a given, probably inhomogeneous, sampling density, and one would
like to choose $k$ or $\varepsilon$ so as to produce robust results
from Isomap.  This is difficult, and the best approach seems to be a
brute force sensitivity analysis over reasonable ranges of $k$ and/or
$\varepsilon$ to probe different scales in the data.

\subsection{Multidimensional scaling}
\label{sec:isomap-mds}

Once the approximate geodesic distance function $d_G(\vec{x},
\vec{y})$ has been found, a multidimensional scaling (MDS) procedure
is applied.  This procedure results in an eigenvalue spectrum that can
be examined to determine the dimensionality of the data manifold.  It
also calculates embeddings of the data points into low-dimensional
Euclidean spaces.

MDS \citep{borg-mds} is a statistical technique that takes as input
distance or dissimilarity measures for a set of data points and
attempts to find points in Euclidean space such that the Euclidean
distances between the output points correspond to the distance or
dissimilarity values between the input points.  Both PCA and Isomap
can be considered within this framework.  For PCA, the input distances
are Euclidean distances in the input data, so that MDS leads to an
orthogonal transformation of the data.  For an idealisation of Isomap
where the input distances are exact geodesic distances in the data
manifold, MDS leads to an isometric transformation of the data.

The form of MDS used in Isomap is usually referred to as
\textit{classical scaling} \citep{torgerson-mds,gower-mds,borg-mds}.
As input, we require a distance or dissimilarity measure $d_{ij} =
d(\vec{x}_i, \vec{x}_j)$ calculated between the $N$ data points,
$\vec{x}_i \in \mathbb{R}^m$.  The distance function must satisfy the
usual conditions for distances: $d_{ii} = 0$, $d_{ij} = d_{ji}$,
$d_{ik} \leq d_{ij} + d_{jk}$.

From the distance function, we form a matrix of squared distances
$(\mathbf{D}^{(2)})_{ij} = d_{ij}^2$.  To this matrix we then apply a
double centring transformation, using a centring operator $\mathbf{J}
= \mathbf{I} - n^{-1} \vec{1} \vec{1}^T$, with $\mathbf{I}$ being the
$N \times N$ identity matrix and $\vec{1}$ an $N$ element vector of
ones.  The centring transformation is
\begin{equation}
  \mathbf{Z}^{(2)} = -\frac{1}{2} \mathbf{J} \mathbf{D}^{(2)}
  \mathbf{J}.
\end{equation}

A simple calculation shows that, if $d_{ij}$ is a Euclidean distance
function, then $\mathbf{Z}^{(2)}$ is the matrix of scalar products
between the vectors $\vec{x}_i$, i.e. $(\mathbf{Z}^{(2)})_{ij} =
\vec{x}_i \cdot \vec{x}_j$.  For centred data, i.e. data for which the
mean of the $\vec{x}_i$ is zero, $\mathbf{Z}^{(2)}$ then corresponds
to the covariance matrix normally used for PCA.  For non-Euclidean
distance functions, the matrix $\mathbf{Z}^{(2)}$ encodes comparable
information about the distribution of distances between data points.

Next, the eigendecomposition of the scalar product matrix
$\mathbf{Z}^{(2)}$ is calculated, as $\mathbf{Z}^{(2)} = \mathbf{Q}
\bm{\Lambda} \mathbf{Q}^T$, where $\bm{\Lambda} = \diag(\lambda_1,
\dots, \lambda_N)$ is a diagonal matrix with the eigenvalues of
$\mathbf{Z}^{(2)}$ along its leading diagonal, and $\mathbf{Q}$ is a
matrix with the eigenvectors of $\mathbf{Z}^{(2)}$ as its columns.
The usual hope is that, if the eigenvalues $\lambda_i$ are sorted in
order of decreasing magnitude, $\lambda_p \gg \lambda_{p+1}$ for some
$p < m$ and we can approximate the matrix $\mathbf{Z}^{(2)}$ by
projection onto the subspace spanned by the $p$ leading eigenvectors.
If we denote the matrix of the first $p$ eigenvalues by
$\bm{\Lambda}_+$ and the first $p$ columns of $\mathbf{Q}$ by
$\mathbf{Q}_+$, then the matrix of $p$-dimensional reduced coordinates
for the data points is given by $\mathbf{Y} = \mathbf{Q}_+
\bm{\Lambda}_+^{1/2}$.  Equivalently, denoting the eigenvectors of
$\mathbf{Z}^{(2)}$ by $\vec{q}_k$, the $k$th coordinate of the $i$th
data point in a $p$-dimensional reduced representation is
\begin{equation}
  \label{eq:isomap-mds-coords}
  y_i^k = \sqrt{\lambda_k} q_k^i, \quad k = 1, \dots, p.
\end{equation}

This procedure is essentially that followed in PCA, apart from
possible differences in data normalisation, but there are two
problems, one common to all MDS algorithms and one important only in
the more general setting relevant to Isomap.  First, there is no
guarantee that there is a gap in the eigenvalue spectrum of
$\mathbf{Z}^{(2)}$, making it difficult to decide on a reduced
dimensionality for the data.  Second, the procedure described here is
dependent on the non-negativity of the eigenvalues of the matrix
$\mathbf{Z}^{(2)}$.  In the case of PCA, positive semi-definiteness of
$\mathbf{Z}^{(2)}$ is guaranteed by the use of Euclidean distances
between data points, but in the more general case of Isomap, this is
no longer the case.  For an exact calculation of geodesic distances in
an intrinsically flat manifold, the distance metric is Euclidean and
$\mathbf{Z}^{(2)}$ is positive semi-definite.  In Isomap, geodesics
are calculated only approximately, and errors associated with the
approximation are often enough to render $\mathbf{Z}^{(2)}$
non-positive semi-definite, yielding negative eigenvalues in the MDS
procedure.  Another possible source of negative eigenvalues in Isomap
is the structure of the data manifold.  Isomap can only produce a
faithful reduction of data manifolds that are globally isometric to an
open, connected, convex subset of Euclidean space
\citep{donoho-isometry}.  Data manifolds that are not convex
(i.e. that do not contain all geodesics connecting points lying in the
manifold --- an example is a two-dimensional surface with a hole,
which is then not simply connected) or that possess non-zero intrinsic
curvature do not satisfy these assumptions and have geodesic distance
functions that lead to $\mathbf{Z}^{(2)}$ matrices with negative
eigenvalues.

Eigenvalues in MDS and, in particular, in PCA, are customarily
interpreted as the proportion of the total data variance explained by
a particular mode.  Clearly, negative eigenvalues cannot be
interpreted as variances.  One approach is to ignore any negative
eigenvalues, assuming them to be the result of noise in the data or
errors in the geodesic distance approximation.  A more satisfactory
approach is to observe that negative eigenvalues are always small and,
if the eigenvalues are sorted in order of the absolute value, negative
eigenvalues tend to be paired with positive eigenvalues of similar
magnitude, constituting the tail of the eigenvalue distribution.  This
is because the ``noise'' eigenvalues with smaller absolute value in
the tail of the distribution are as likely to be negative as to be
positive.  The presence of negative eigenvalues can then still be
considered a form of noise, but the position in the eigenvalue
spectrum of the first negative eigenvalue can be used as a cut-off
point for considering the reduced dimensionality of the data.
According to this view, no positive eigenvalue appearing after a
negative eigenvalue can correspond to a real dimension in the reduced
dimensionality data.  The justification for this interpretation is
simply that negative eigenvalues cannot be interpreted as variances,
cannot be used in \eqref{eq:isomap-mds-coords} to calculate reduced
coordinates and so must be neglected.  Some complication is entailed
by this viewpoint, since it is no longer possible to use a simple
measure of explained variance such as $c_p = \sum_{i=1}^p \lambda_i /
\Tr \bm{\Lambda}$ because the trace of the eigenvalue matrix no longer
measures the total variance in the data, due to the presence of the
negative eigenvalues.  It is thus not possible to use an explained
variance threshold to infer the dimensionality of the data and to
choose a set of modes on to which to project.  Here, I use a different
approach, finding a pair of straight lines with a ``knee'' that best
fits the MDS eigenvalue spectrum in a least squares sense and taking
the dimensionality of the data to lie at the knee.  This approach,
which is easy to understand and proves to be reasonably robust, is
explained in detail in Section~\ref{sec:isomap-sens}.

\subsection{Computational complexity}

The two main computational bottlenecks in the Isomap algorithm are
computation of the nearest neighbour graph and the final MDS
eigenvalue problem, which, for $N$ data points, involves finding the
leading eigenvalues and eigenvectors of an $N \times N$ matrix.  An
implementation using a dense eigenvalue solver has computational cost
that scales as $O(N^3)$.  Here, for data sets with $N \leq 6000$, I
use the Anasazi iterative eigenvalue solver from the Trilinos project
\citep{baker-anasazi,heroux-trilinos}.  The block Krylov-Schur scheme
in Anasazi finds the first fifteen eigenvalues and eigenvectors of a
$6000 \times 6000$ matrix in a time entirely negligible compared to
the time required for the all-sources shortest paths calculation used
to approximate geodesic distances in the data manifold.  For still
larger problems, an adaptation of Isomap exists using a smaller number
of \textit{landmark} points \citep{desilva-global-local}, but this
refinement did not prove necessary here.

\section{Application to test data sets}
\label{sec:isomap-test-data}

\begin{figure}
  \begin{center}
    \begin{tabular}{cc}
      \includegraphics[width=0.49\textwidth]{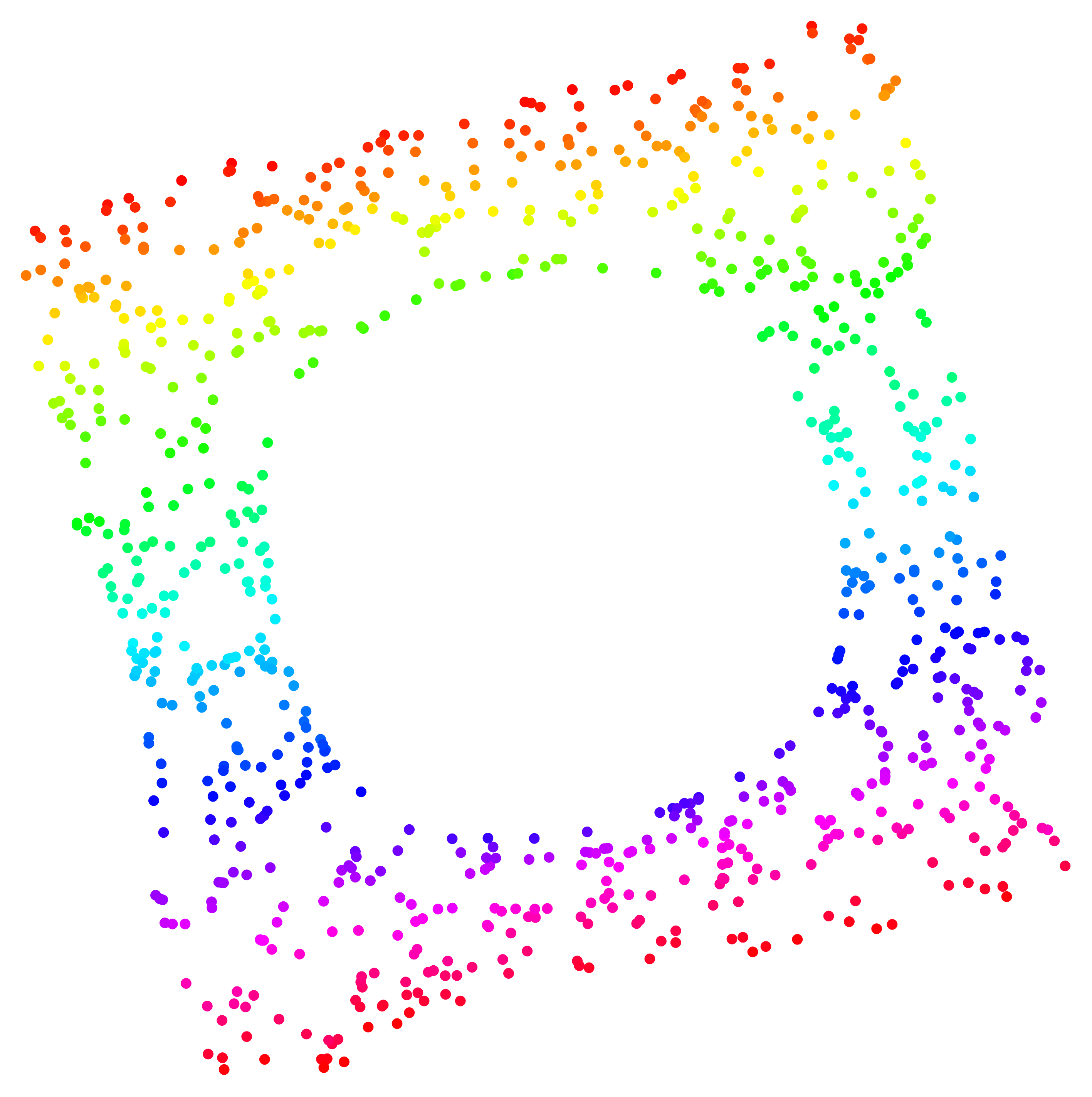} &
      \includegraphics[width=0.49\textwidth]{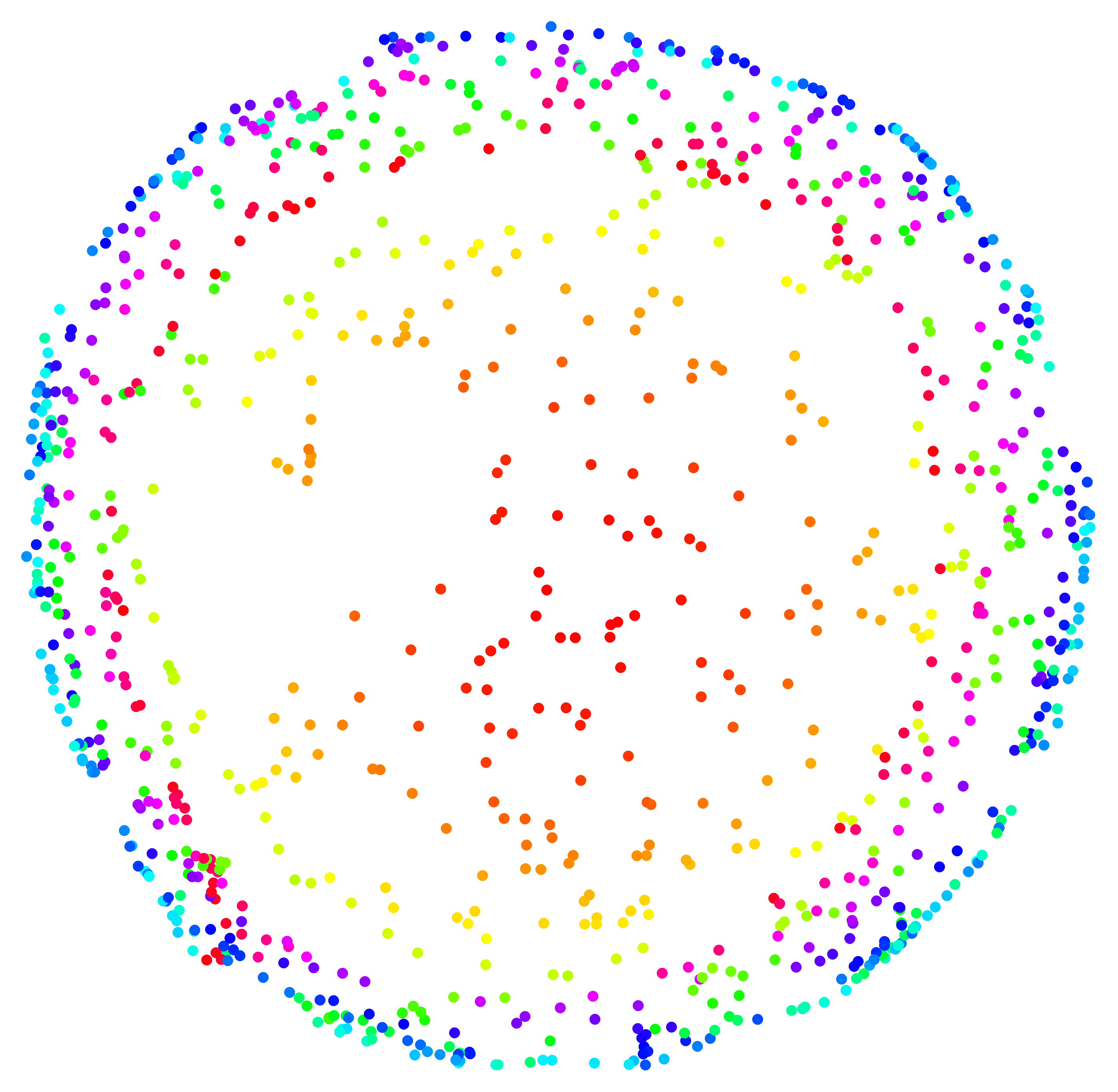} \\
      (a). Plane with hole & (b). Fishbowl \\
      \\

      \includegraphics[width=0.49\textwidth]{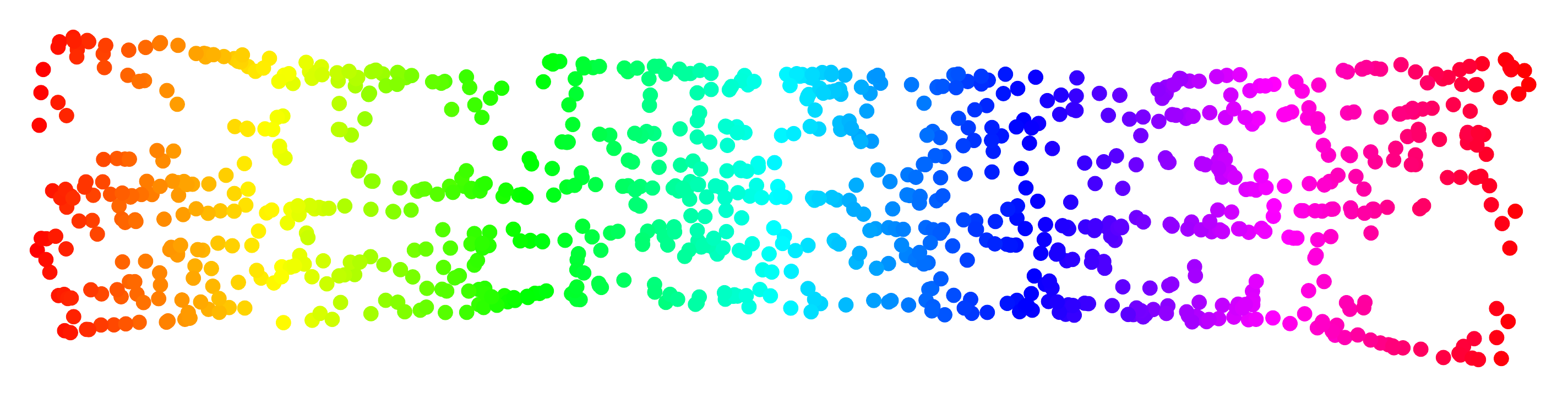} &
      \includegraphics[width=0.49\textwidth]{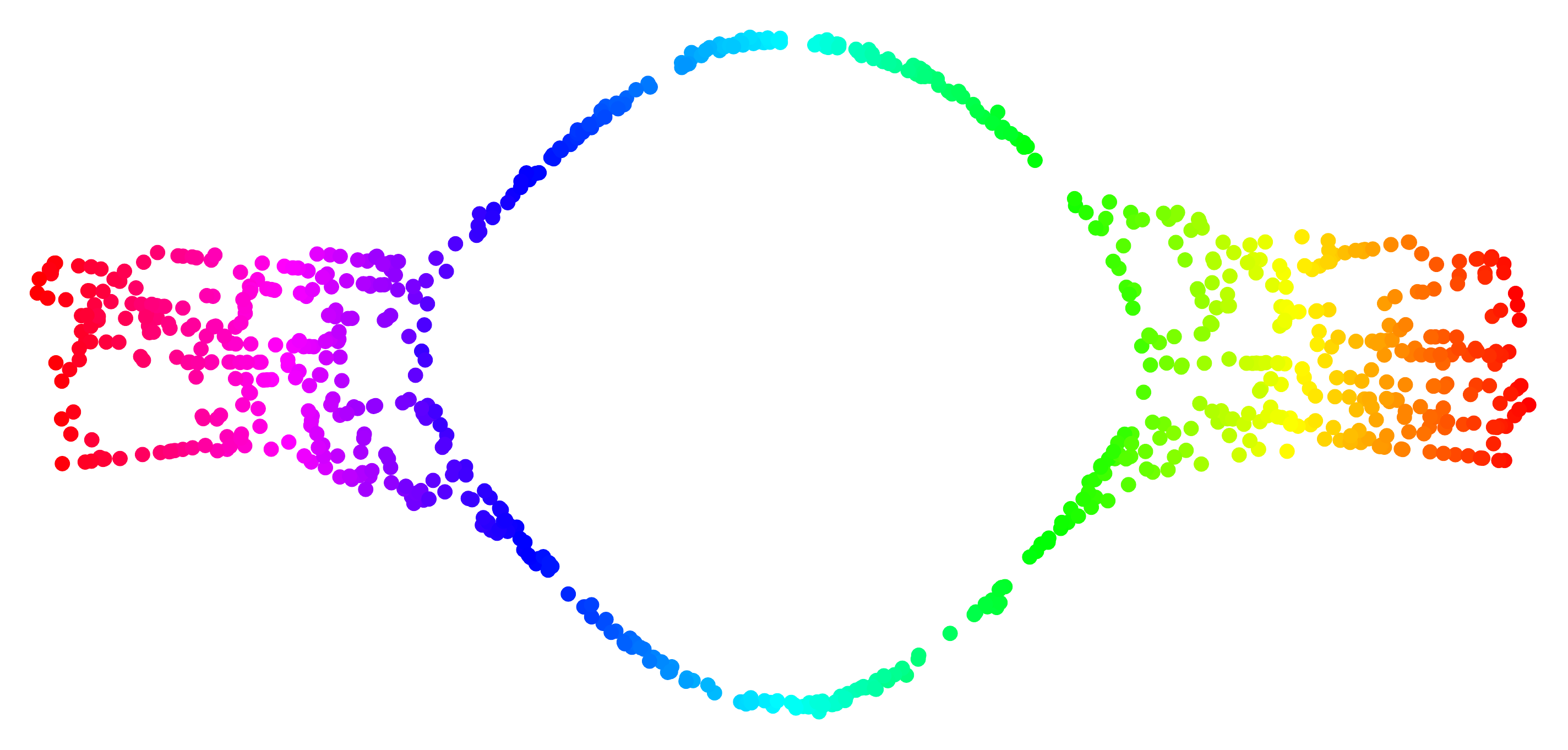} \\
      (c). Swiss roll & (d). Swiss roll with hole
    \end{tabular}
  \end{center}
  \caption[Isomap reductions of geometrical test data
    sets]{Application of Isomap to geometrical test data sets from
    Section~\ref{sec:test-data-sets}.}
  \label{fig:isomap-test-data-sets}
\end{figure}

To give some feeling for the kind of results that Isomap can produce,
it is useful to examine results of applying the algorithm to the
simple geometrical data sets described in
Section~\ref{sec:test-data-sets}.
Figure~\ref{fig:isomap-test-data-sets} presents reduced
representations of some of these data sets.  Most of the data sets
described in Section~\ref{sec:test-data-sets} are reduced by Isomap
without too much trouble, but the examples shown here illustrate some
interesting features of the Isomap algorithm.

%
%
%

The reduced representation of the plane with a hole example shown in
Figure~\ref{fig:isomap-test-data-sets}a gives a reasonable
parameterisation of the surface, but there is noticeable distortion
around the hole that is not seen in the NLPCA reduction
(Figure~\ref{fig:nlpca-test-data-sets-reduce}a).  This distortion is
related to the non-convexity of the original data set: for a
non-simply connected surface, Isomap cannot find a faithful embedding
\citep{donoho-isometry}.  The fishbowl example
(Figure~\ref{fig:isomap-test-data-sets}b) demonstrates another
problem.  There is significant overlap between the different hues used
to identify points on the manifold here, and the Isomap reduced
representation looks very much as though the fishbowl has just been
projected to a plane without taking account of the intrinsic structure
of the manifold.  An extension of Isomap \citep{desilva-global-local},
called conformal Isomap, has been developed to alleviate this problem,
but it is not clear how applicable this method is to more general data
analysis problems where the transformation between the original data
and the reduced manifold is not known to be conformal.  The Swiss roll
example (Figure~\ref{fig:isomap-test-data-sets}c) is handled well by
Isomap, and is used as a basis for exploring the sensitivity of Isomap
results to parameter choices in the following section.  The Swiss roll
with hole example (Figure~\ref{fig:isomap-test-data-sets}d) is handled
much less well and displays very strong distortion associated with the
non-convexity of the original manifold.  The distortion observed for
the examples with holes was one of the motivations for the development
of the Hessian locally linear embedding method
(Chapter~\ref{ch:hessian-lle}), which aims to lift the restriction of
Isomap that the reduction transformation has to be a global isometry
from the original data space to a convex subset of Euclidean space.

Most of the test data sets show relatively little sensitivity to added
noise, but the Swiss roll with a hole is an exception, probably
because of both the distortion seen in
Figure~\ref{fig:isomap-test-data-sets}d and the relative thinness of
the segments of the surface surrounding the hole.  The result is that
the Isomap reduction breaks down completely when noise is added to
this data set (plot not shown).

\section{Isomap sensitivity}
\label{sec:isomap-sens}

The Isomap algorithm has a single tunable parameter, the number of
nearest neighbours used to construct the graph on which the
approximate geodesic calculation is based.  A natural issue to
investigate is how the results inferred from Isomap depend on this
parameter.

To explore some implications of sensitivity to this parameter choice,
I will use the Swiss roll data set, representing a two-dimensional
manifold embedded in $\mathbb{R}^3$ (Figure~\ref{fig:test-data-3d}c).
The important feature of this data set as far as analysis of Isomap
sensitivity is concerned is that the manifold in which the data points
lie is intrinsically flat, but curled up so that points far apart
according to the intrinsic geodesic metric in the manifold are close
together as measured by Euclidean distances in the embedding space.
The implications of this for the construction of the Isomap nearest
neighbour graph are clear: choosing too large a number of nearest
neighbours $k$ or too large a radius $\varepsilon$ will cause points
on adjacent but separate leaves of the manifold to be identified as
nearest neighbours, leading to an incorrect identification of the
topology of the data manifold.

%
%
\begin{figure}
  \begin{center}
    \includegraphics[width=0.48\textwidth]{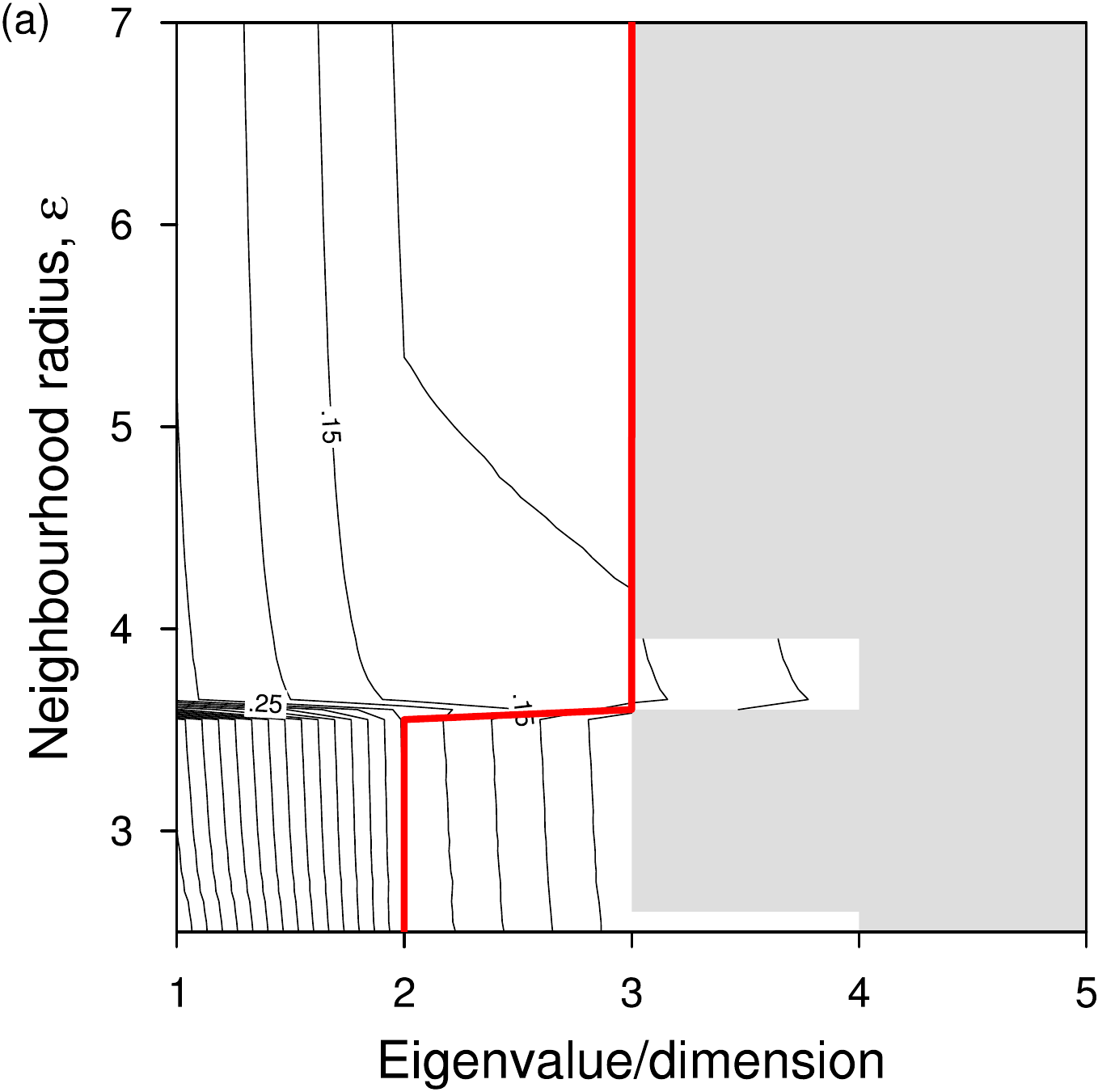}%
    \hspace{0.04\textwidth}%
    \includegraphics[width=0.48\textwidth]{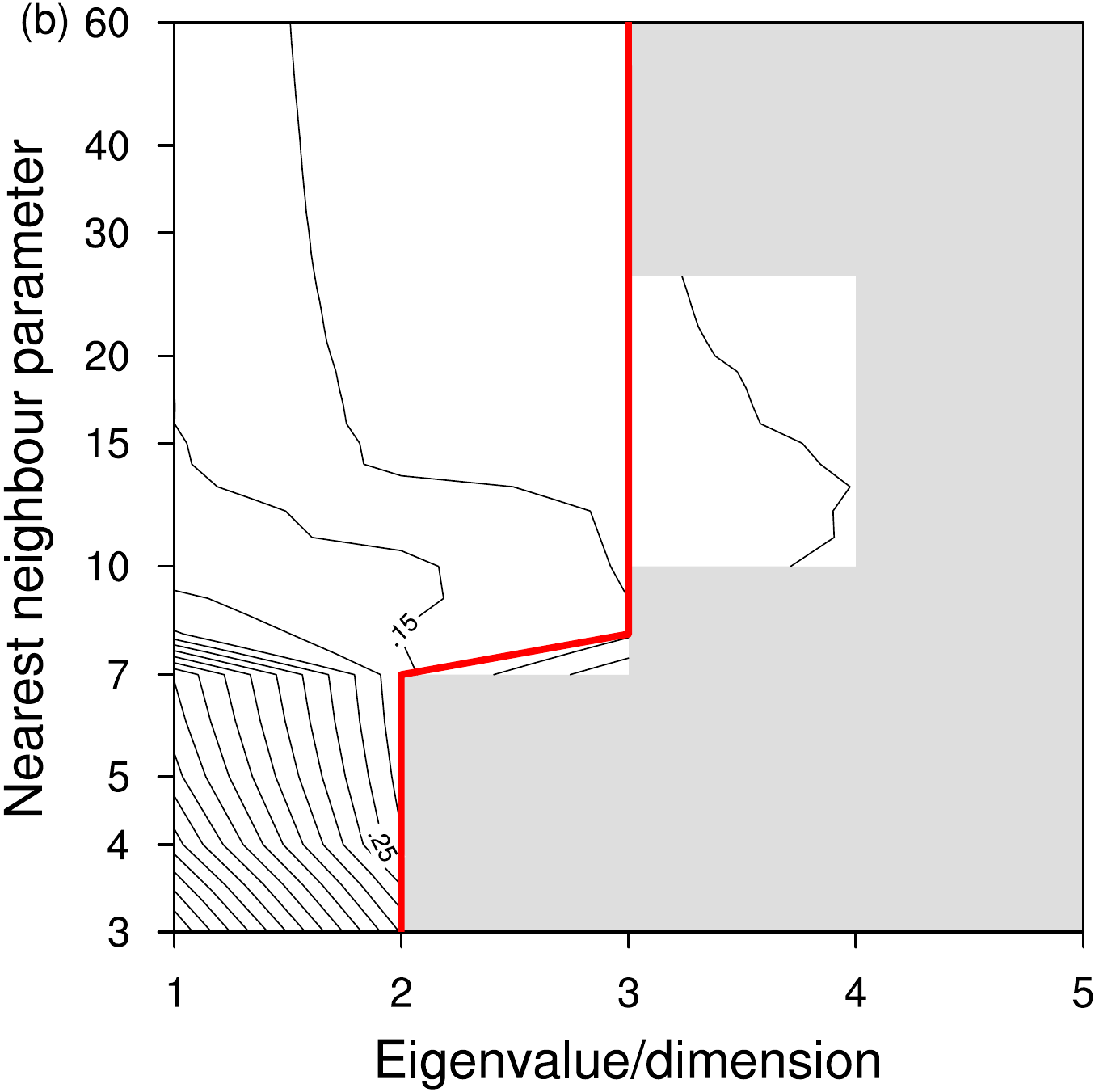}
  \end{center}
  \caption[Isomap eigenvalue convergence: Swiss roll]{Isomap
    eigenvalue convergence and dimension estimates for Swiss roll
    data.  Black contours show MDS eigenvalue spectra normalised by
    the overall largest eigenvalue, as a function of eigenvalue number
    and neighbourhood radius $\varepsilon$ (a) or nearest neighbour
    count $k$ (b) (logarithmic axis).  Grey areas indicate regions of
    the eigenvalue spectra not available for dimensionality reduction
    because of the presence of negative eigenvalues.  The thick red
    line shows the data dimensionality, estimated from the eigenvalue
    spectra as described in the main text.  The true dimensionality of
    the data set is two.}
  \label{fig:swiss-roll-sens}
\end{figure}
%

Figure~\ref{fig:swiss-roll-sens} shows results from Isomap sensitivity
studies using the Swiss roll data, one for $\varepsilon$-Isomap
(Figure~\ref{fig:swiss-roll-sens}a) and one for $k$-Isomap
(Figure~\ref{fig:swiss-roll-sens}b).  Each plot shows MDS eigenvalue
spectra in contour form, as a function of eigenvalue number and the
nearest neighbour parameter ($\varepsilon$ or $k$).  As previously
mentioned, if negative eigenvalues are present in the MDS spectrum,
they must be excluded from any dimensionality reduction, since they
cannot be viewed as measures of explained variance, and cannot be
interpreted in terms of a lower-dimensional real manifold.  The areas
filled in grey in Figure~\ref{fig:swiss-roll-sens} indicate regions of
eigenvalue space that are forbidden by this condition.  No eigenvalues
beyond the first negative eigenvalue can be part of a real
lower-dimensional representation of the data.  Given this constraint,
the dimensionality of the data is estimated by looking for a ``knee''
in the eigenvalue spectrum, and is indicated in
Figure~\ref{fig:swiss-roll-sens} by a thick red line.

In both plots in Figure~\ref{fig:swiss-roll-sens}, there is a change
in behaviour of the eigenvalue spectra as the nearest neighbour
parameter is varied: at $\varepsilon \approx 3.6$ or $k = 7$, there is
a distinct step change in the spectra.  For neighbourhood sizes below
this threshold, convergence of the eigenvalue spectra is quicker than
for values above the threshold.  Consequently, dimensionality
estimates inferred are lower for neighbourhood sizes below the
threshold.  For the $\varepsilon$-Isomap results, this effect reflects
the fact that, in the norm used here, the separation between adjacent
leaves of the Swiss roll manifold is about 3.6.  For neighbourhood
radii smaller than this, the nearest neighbour connections in the
distance-weighted graph used to approximate geodesics are confined to
the surface of the manifold.  For larger neighbourhood radii, the
neighbourhoods spill over between adjacent leaves of the manifold.
Varying the neighbourhood parameter probes different scales in the
data.  Smaller values of $\varepsilon$ pick out smaller scale
structures and detect the separation between the leaves of the
manifold.  Larger values of $\varepsilon$ do not resolve this fine
structure and see the data as an amorphous cloud of points.  Small
values of $\varepsilon$ thus give $p=2$, the true dimensionality of
the embedded manifold, while larger values give $p=3$, the dimension
of the embedding space.

Similar conclusions can be drawn from the $k$-Isomap results
(Figure~\ref{fig:swiss-roll-sens}b), although here the value of $k$ at
which the transition from $p=2$ to $p=3$ occurs is harder to
interpret.  The transitional value $k=7$ is the number of neighbours,
on average, that a data point has within a radius of $\varepsilon
\approx 3.6$, but this number is subject to large sampling
variability, giving a slightly rougher transition for $k$-Isomap than
$\varepsilon$-Isomap.  The data set used here has 1000 points, chosen
to be comparable in size to the equatorial Pacific SST time series
examined below, and this relatively small number of points in
$\mathbb{R}^3$ leads to a wide range of variability in the distance
from a point to its nearest neighbour ($\sim{}0.02$--2.13).  There is
thus a range of values of $k$ for which the $k$ nearest neighbours of
some points all lie on the same leaf of the manifold while the $k$
nearest neighbours of other points span more than one leaf.  Despite
this, the dimensionality estimates are the same as for
$\varepsilon$-Isomap, i.e. $p=2$ for $k \leq 7$ and $p=3$ for $k>7$.

The dimensionality inferred from Isomap depends to a certain extent on
subjective factors.  Although there is no need to choose a total
cumulative explained variance to select the number of leading
eigenvectors to consider, as is sometimes done with PCA, the condition
for locating a ``knee'' in the eigenvalue spectrum is delicate.  Here,
I approximate the spectrum with a pair of lines with a kink at a
selected eigenvalue, then place the knee at that point whose fitted
lines give the smallest RMS error when points on the lines are
compared to the true eigenvalues.  This approach substantially follows
recommendations in \citet{borg-mds}, but there are other methods that
could equally be used.

The main conclusion to draw from this is that, at least in the case of
the simple data set used here, Isomap can probe the dimensionality of
a lower-dimensional data set embedded nonlinearly in a
higher-dimensional space quite well.  In this case, there is
relatively little dependence of the results on the nearest neighbour
parameter $\varepsilon$ or $k$, and what dependence is seen is well
understood in terms of known characteristics of the data set.  The
changes in MDS eigenvalue spectra seen as one varies the neighbourhood
size indicate how the method is probing the data set at different
scales.  This dependence on the parameter $\varepsilon$ or $k$ can be
viewed as a disadvantage (some value of $k$ or $\varepsilon$ needs to
be chosen and there is no clear a priori method to do this) or an
advantage (by varying $k$ or $\varepsilon$, we can probe different
scales to get a better idea of the underlying structure of our data).
The results from $\varepsilon$-Isomap are easier to interpret because
of the propensity for $k$-Isomap results to be influenced by data
sampling variability, although $k$-Isomap is easier to use since there
is no need to determine a suitable range for $\varepsilon$.  The main
impediment to performing the type of sensitivity analysis illustrated
here is computing resource, since Isomap decompositions of the data
for a large number of neighbourhood sizes are needed to form a clear
picture of the structure of the variation in results with
neighbourhood size.

In the sections below showing Isomap results for Pacific SST time
series, sensitivity results are presented in parallel with other
Isomap results to give some feeling for the robustness of the method
and the variability of the results with respect to the neighbourhood
size.  In general, the results are more dependent on neighbourhood
size for the more complex tropical Pacific SST data, and
dimensionality estimates are correspondingly less certain.

\section{Previous applications in climate data analysis}

The only previous application of Isomap to climate data analysis of
which I am aware is the work of \citet{gamez-enso} (also
\citet{gamez-enso-2}), where Isomap was applied to observational SSTs
for the equatorial Pacific to examine ENSO variability.
\citeauthor{gamez-enso}'s results are substantially replicated by my
raw SST analysis of the NOAA ERSST v2 observational data set
(Section~\ref{sec:isomap-raw-sst-results}) and I extend their analysis
to consider simulations using the CMIP3 ensemble of coupled
atmosphere-ocean GCMs.

Although there have to date been no other applications of Isomap to
climate data analysis, there have been some applications in
tangentially related fields.  For instance, Isomap has been used for
the feature identification and dimensionality reduction in the
processing of hyperspectral remote sensing imagery for geophysical
applications
\citep{bachmann-remote-sensing-1,bachmann-remote-sensing-2}, and has
been used for analysing high-dimensional vegetation distribution data
to identify patterns of biodiversity in different ecosystems
\citep{mahecha-biodiversity,mahecha-biogeo-patterns}.  The main
difference between these applications and the use of Isomap in the
type of climate data analysis presented here is the presence of a
dynamical dimension in climate data --- most previous applications of
Isomap have concentrated on analysing static data sets.

\section{Application to analysis of Pacific SSTs}
\label{sec:isomap-sst-results}

All of the results reported here are based on the use of the full
length of the model SST time series available, as listed in
Table~\ref{tab:models}, with 100 years of data from 1900--1999 being
used for the observational SSTs.  Isomap eigenvalue spectra were also
calculated for sub-segments of each data set, consisting of 50, 25 and
10 year segments of the total available data, in order to determine
the sensitivity of Isomap results to time series length.  The results
(data not shown) indicate that there is little variation in the Isomap
eigenvalue spectra, at least for 50 or 25 year sub-segments, leading
us to conclude that the results are reasonably robust with respect to
variations in the amount of data available.

%
%

\subsection{Analysis for raw SSTs}
\label{sec:isomap-raw-sst-results}

In this section, I present Isomap results for tropical Pacific SSTs
from observational and model data sets.  In performing PCA, it is
common to use SST anomalies, so removing the influence of the annual
cycle.  Isomap results for SST anomalies are presented in
Section~\ref{sec:isomap-sst-anom-analysis}, allowing for direct
comparison between PCA and Isomap, but here, one of the things I wish
to explore is the extent to which Isomap is able to determine the
coupling between ENSO and annual variability in the tropical Pacific.
This coupling is one factor lost in the customary anomaly-based PCA
approach.

In this section I use SSTs and in the next, SST anomalies, from the
region 125\degree\,W--65\degree\,W, 20\degree\,S--20\degree\,N,
normalising each data set to zero mean and unit standard deviation at
each spatial point.  This choice of normalisation is used throughout
to permit direct comparison with the earlier work of
\citet{gamez-enso}.

\begin{figure}
  \begin{center}
    \includegraphics[width=0.44\textwidth]{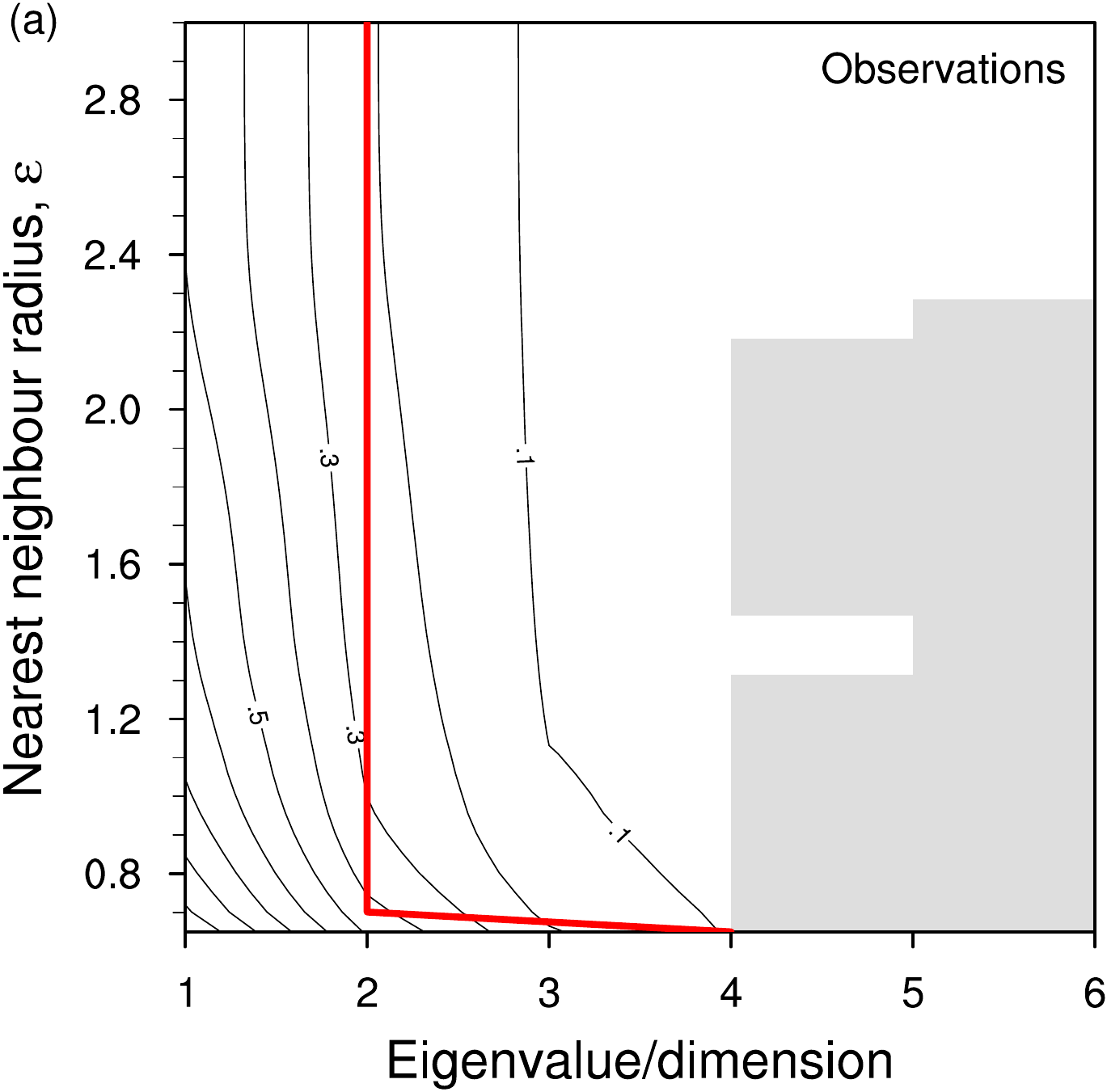}%
    \hspace{0.1\textwidth}%
    \includegraphics[width=0.44\textwidth]{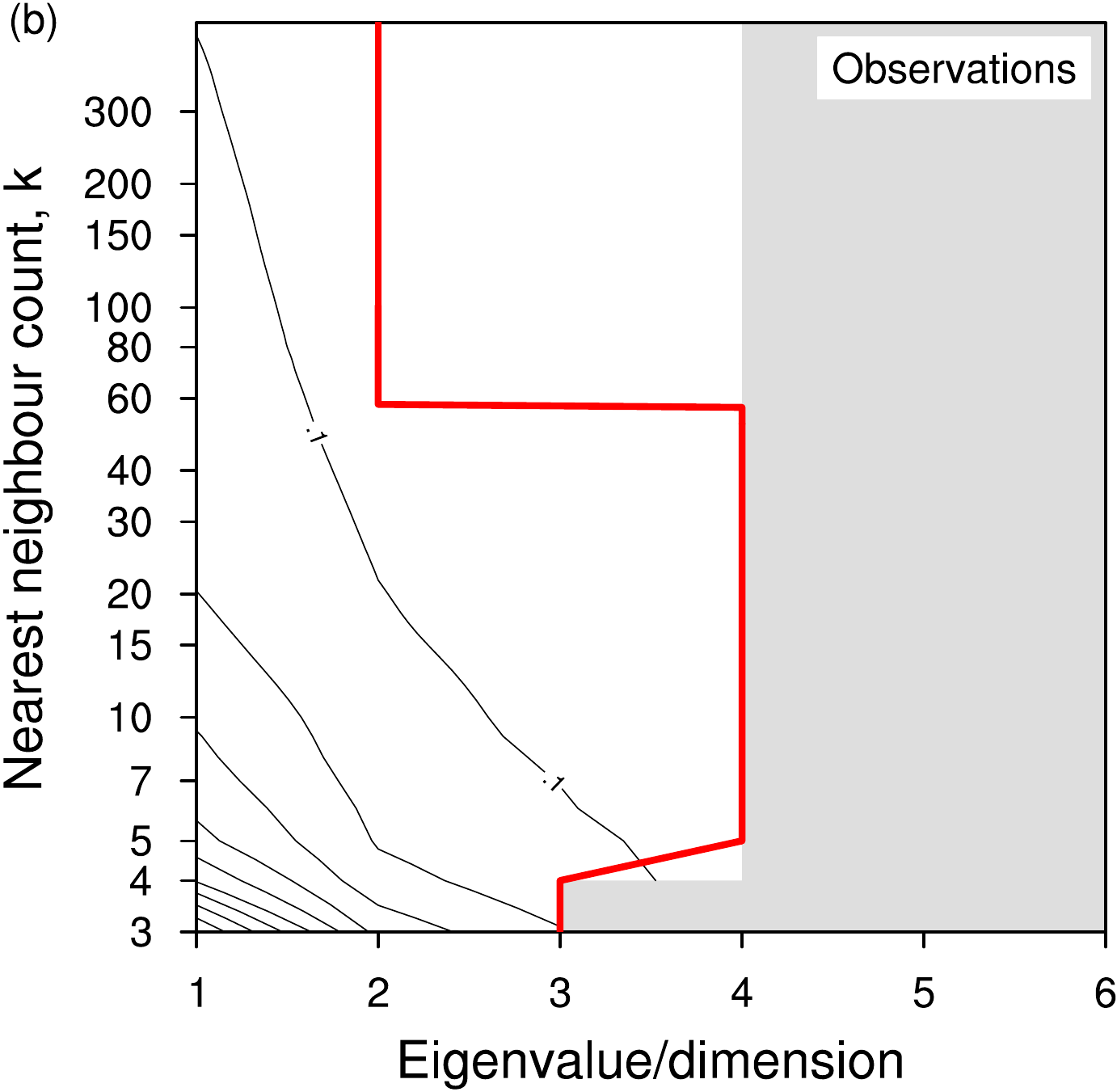}

    \vspace{0.25cm}
    \includegraphics[width=0.44\textwidth]{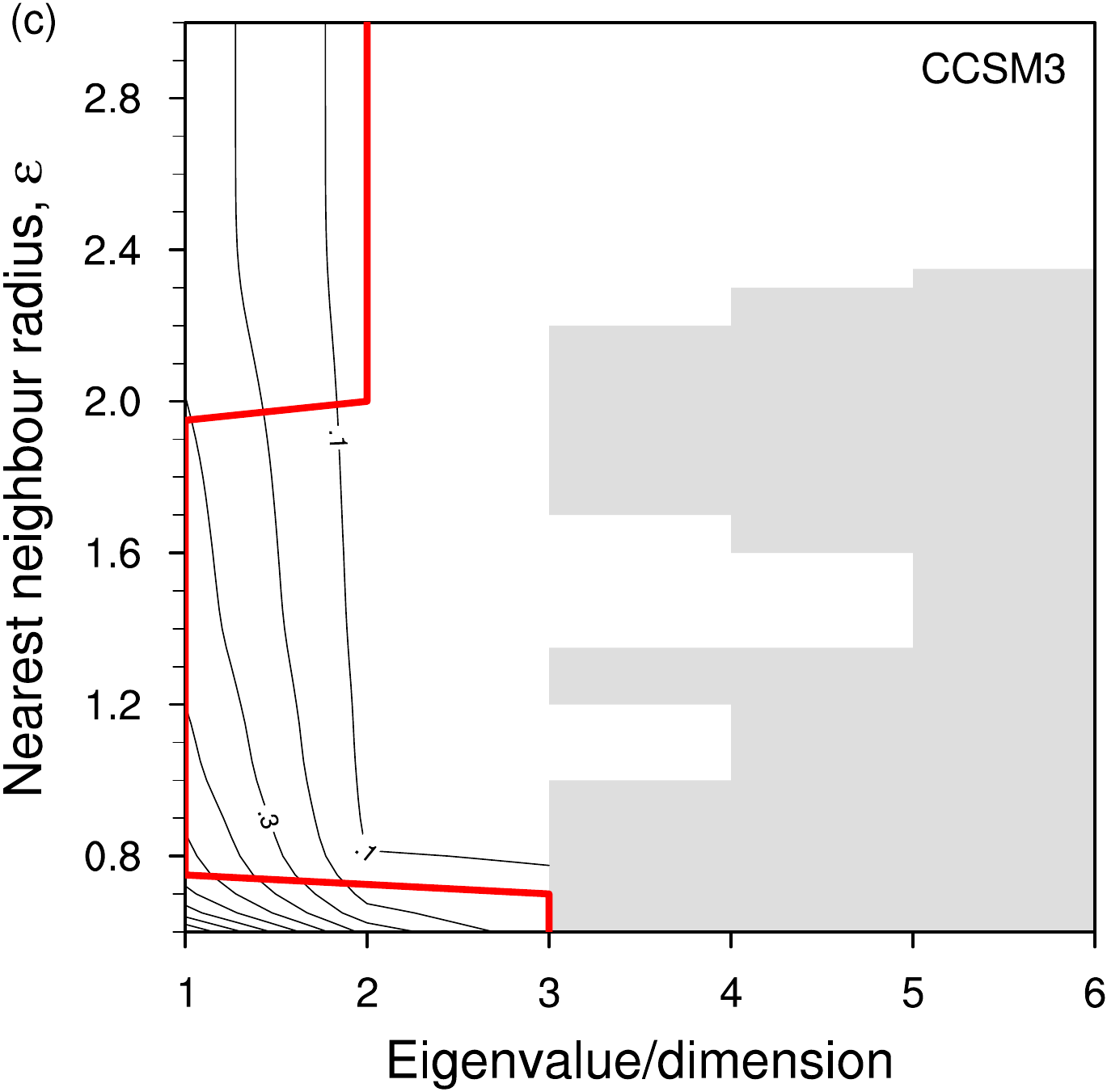}%
    \hspace{0.1\textwidth}%
    \includegraphics[width=0.44\textwidth]{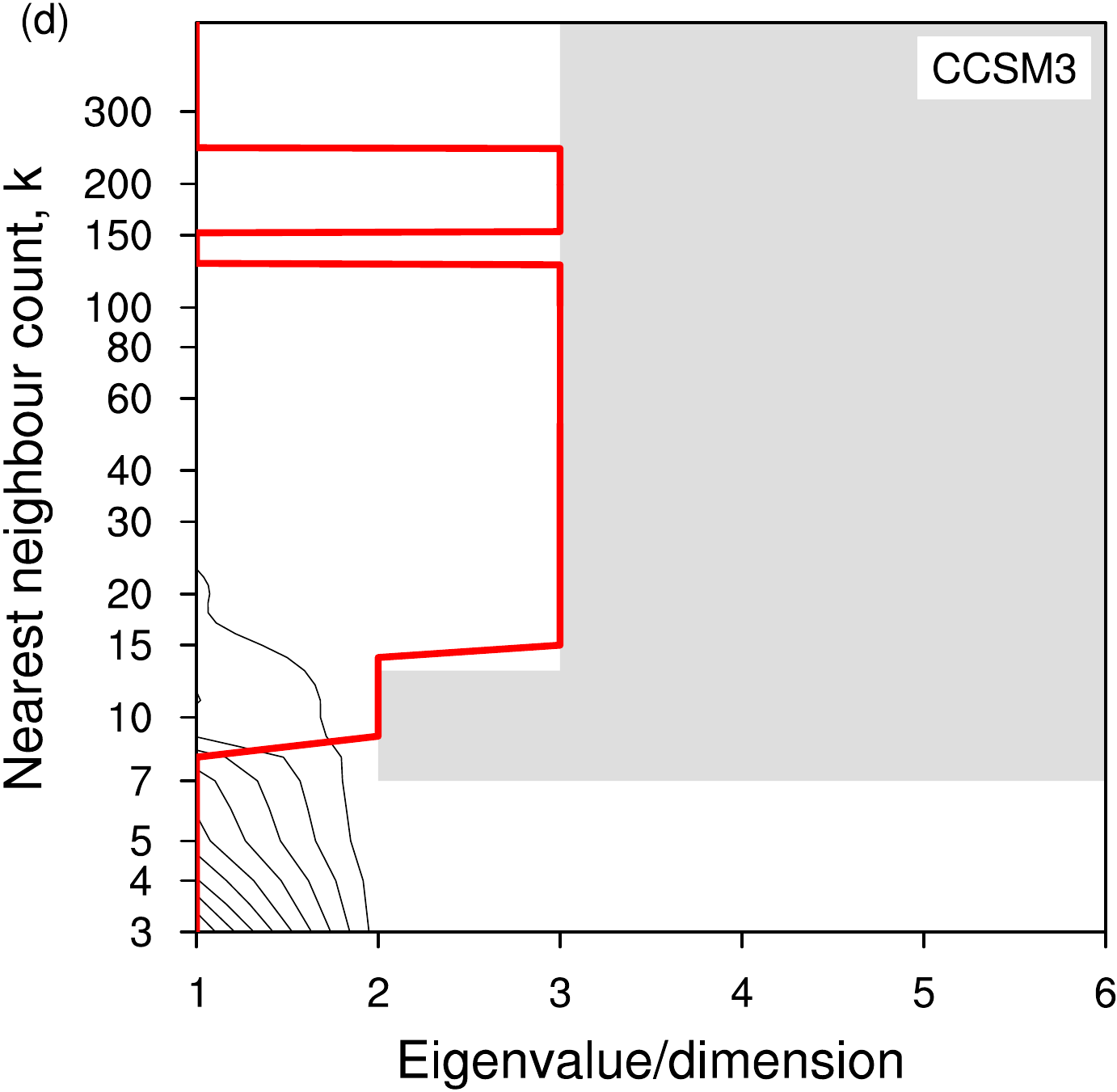}

    \vspace{0.25cm}
    \includegraphics[width=0.44\textwidth]{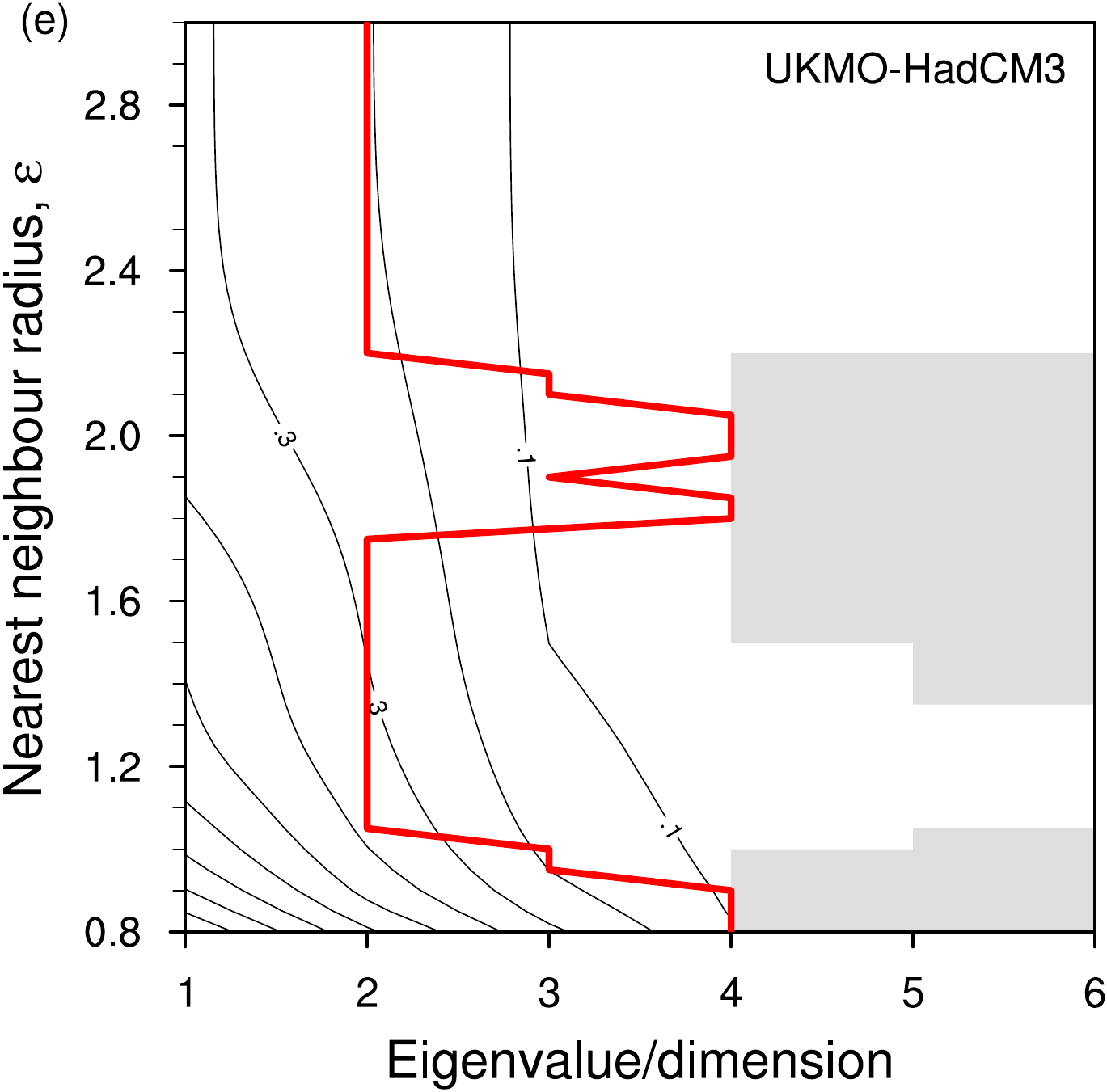}%
    \hspace{0.1\textwidth}%
    \includegraphics[width=0.44\textwidth]{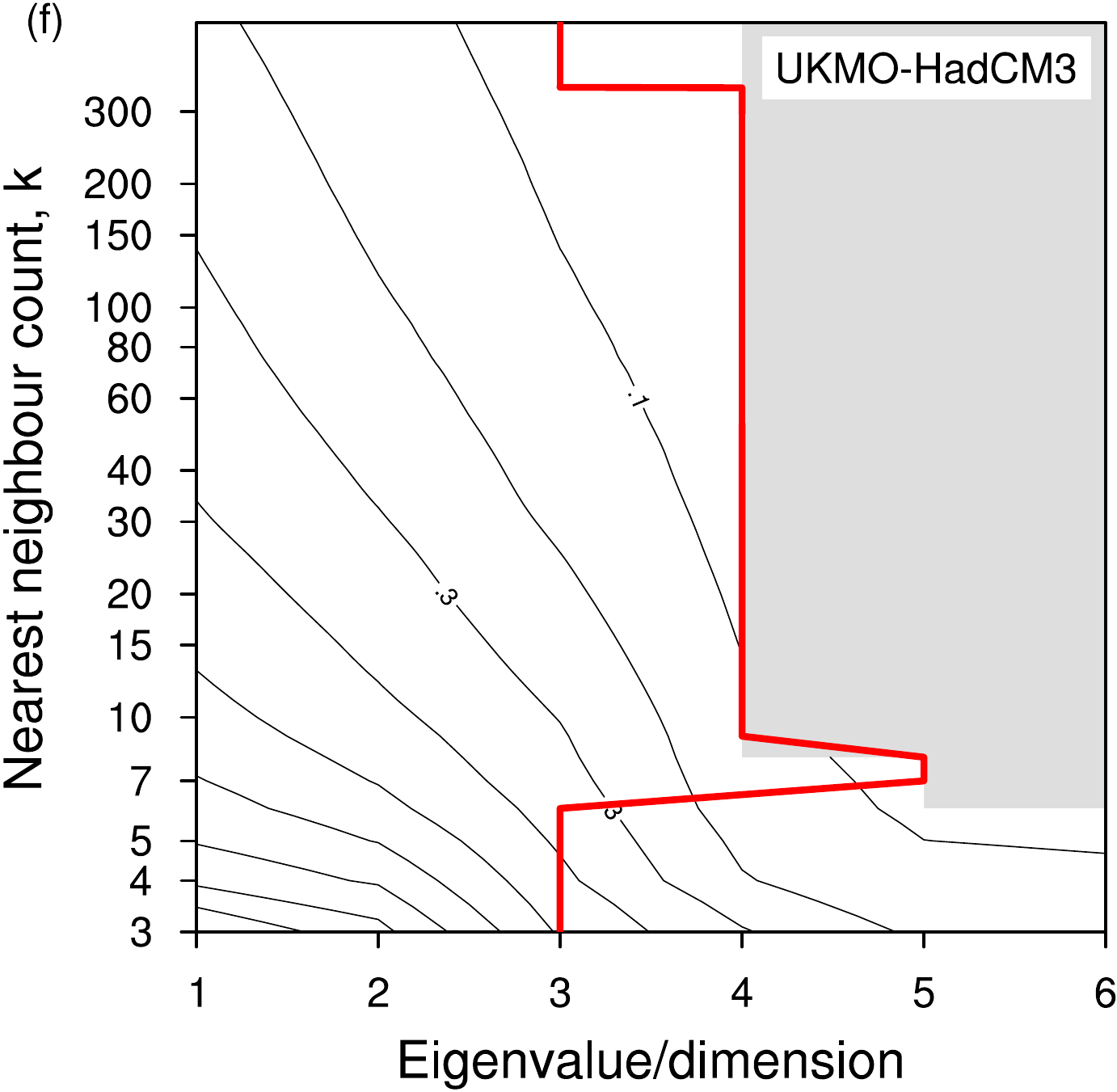}
  \end{center}
  \caption[Isomap eigenvalue convergence: tropical Pacific raw
    SSTs]{Isomap eigenvalue convergence and dimension estimates for
    tropical Pacific raw SSTs, from observations (a and b), CCSM3 (c
    and d) and UKMO-HadCM3 (e and f).  Black contours show MDS
    eigenvalue spectra normalised by the overall largest eigenvalue,
    as a function of eigenvalue number and neighbourhood radius
    $\varepsilon$ (a, c, e) or nearest neighbour count $k$ (b, d, f)
    (logarithmic axis).  Grey areas indicate regions of the eigenvalue
    spectra not available for dimensionality reduction because of the
    presence of negative eigenvalues.  The thick red line shows the
    data dimensionality estimated from the eigenvalue spectra.}
  \label{fig:raw-sst-isomap-sens}
\end{figure}

The leading modes of variability in tropical Pacific SSTs are the
annual cycle and ENSO, and we expect Isomap to pick these out.  As in
the case of the Swiss roll data, it is useful to examine the
sensitivity of Isomap results to variations in the $\varepsilon$ or
$k$ neighbourhood size parameters.
Figure~\ref{fig:raw-sst-isomap-sens} displays Isomap sensitivity plots
for observational SST data (Figures~\ref{fig:raw-sst-isomap-sens}a and
b) and two selected models
(Figures~\ref{fig:raw-sst-isomap-sens}c--f).  Compared to the Swiss
roll results (Figure~\ref{fig:swiss-roll-sens}), the eigenvalue
spectra and corresponding dimensionality estimates for the SST data
show more variation with Isomap neighbourhood size.  The ranges of $k$
and $\varepsilon$ used in Figure~\ref{fig:raw-sst-isomap-sens} are
selected to correspond as far as possible, but it is difficult to
relate results for any particular value of $k$ to those for any
particular value of $\varepsilon$, or vice versa, because of the
variability in distances between data points.  One common feature in
the $\varepsilon$-Isomap plots in Figure~\ref{fig:raw-sst-isomap-sens}
is that the regions of negative eigenvalues in the Isomap spectra
disappear as neighbourhood size increases.  This reflects the
equivalence of Isomap with a large neighbourhood size to PCA under
suitable data normalisation conditions: in the limit of infinite
neighbourhood size, the Isomap geodesic distance approximation
collapses to the use of the original Euclidean distances between data
points, so is equivalent to PCA.  The same effect would also be seen
in the $k$-Isomap results for $k \approx N$.

%
%
%

%
%
\begin{table}
  \caption[Isomap dimensionality estimates for SST data]{Isomap
    dimensionality estimates for tropical Pacific SST data, for raw
    SSTs and SST anomalies.  Values shown are the smallest and largest
    dimensionalities recovered by examining the Isomap eigenvalue
    spectra as the neighbourhood size parameter $k$ or $\varepsilon$
    is varied.}

  \label{tab:isomap-dimensions}

  \begin{center}
    \begin{tabular}{lcccc}
      \toprule

      & \multicolumn{2}{c}{Raw SST} & \multicolumn{2}{c}{SST anomaly} \\
      Data Set & $\varepsilon$ & $k$ & $\varepsilon$ & $k$ \\

      \midrule

      Observations     & 2--4 & 2--4 & 2--3 & 1--2 \\
      BCCR-BCM2.0      & 2--3 & 2--3 & 2--4 & 2--2 \\
      CCSM3            & 1--3 & 1--3 & 2--4 & 2--2 \\
      CGCM3.1(T47)     & 1--2 & 1--3 & 1--4 & 1--1 \\
      CGCM3.1(T63)     & 2--2 & 1--2 & 1--4 & 1--1 \\
      CNRM-CM3         & 2--4 & 2--5 & 2--4 & 2--2 \\
      CSIRO-Mk3.0      & 1--3 & 1--4 & 1--4 & 2--2 \\
      ECHO-G           & 2--4 & 4--5 & 1--4 & 2--2 \\
      FGOALS-g1.0      & 3--4 & 1--4 & 2--3 & 2--5 \\
      GFDL-CM2.0       & 2--3 & 2--3 & 1--1 & 1--2 \\
      GFDL-CM2.1       & 2--3 & 2--4 & 1--2 & 1--2 \\
      GISS-EH          & 2--3 & 1--3 & 1--4 & 1--2 \\
      INM-CM3.0        & 2--3 & 2--3 & 1--4 & 2--2 \\
      IPSL-CM4         & 2--2 & 1--3 & 2--4 & 2--2 \\
      MIROC3.2(hires)  & 2--2 & 1--2 & 1--4 & 2--2 \\
      MIROC3.2(medres) & 2--3 & 1--3 & 1--3 & 1--2 \\
      MRI-CGCM2.3.2    & 2--4 & 3--4 & 1--2 & 1--2 \\
      UKMO-HadCM3      & 2--4 & 3--5 & 2--3 & 2--2 \\
      UKMO-HadGEM1     & 2--3 & 2--3 & 1--4 & 1--2 \\

      \bottomrule
    \end{tabular}
  \end{center}
\end{table}
%

Despite the high embedding dimension of the data (essentially the
number of non-land points in the study region, $m$ in
Table~\ref{tab:models}), the dimensionality estimates inferred from
Isomap in Figure~\ref{fig:raw-sst-isomap-sens} are rather low.  This
is true for the observational data and all models examined.
Table~\ref{tab:isomap-dimensions} shows the range of dimensionality
estimates inferred for each data set.  For raw SSTs, across all data
sets the dimensionality estimates range from 1 to a maximum of about
5.  The eigenvalue spectra here converge rapidly because the leading
modes of variability are overwhelmingly larger in amplitude than the
other modes.  The coherent variation of SST patterns in the tropical
Pacific can easily be represented by a small set of modes.  The
convergence of the Isomap eigenvalue spectra is rather quicker than
the convergence of eigenvalue spectra for PCA performed in a
comparable setting, i.e. using raw SST data rather than SST anomalies,
as shown in \citet{gamez-enso}.  This quicker convergence can be
ascribed to better representation of the nonlinear ENSO variability by
Isomap than by PCA.  The PC scatter plots shown earlier
(Figure~\ref{fig:sst-pc-scatter}) demonstrate that ENSO variability is
probably not a linear Gaussian phenomenon, so this is expected.

The range of dimension estimates shown in
Table~\ref{tab:isomap-dimensions} for SST observations (2--4) is what
we would expect, including two dimensions to describe the periodic
annual cycle and one or two for ENSO variability.  Here, two degrees
of freedom are expected for the annual cycle because of the geometry
of manifolds that can be faithfully represented by Isomap.  The
globally isometric transformations produced by Isomap can represent
only simple Euclidean coordinates and not periodic coordinates,
meaning that any periodic phenomenon requires at least two degrees of
freedom.  There is no equivalent to the ``circular'' bottleneck layer
NLPCA procedure that allows periodic coordinates to be extracted
directly (Section~\ref{sec:nlpca-anns}).  For ENSO variability, as
well as the leading degree of freedom usually represented by the NINO3
SST index, a second degree of freedom associated with zonal mean
thermocline depth variations (Section~\ref{sec:enso-mechanisms}) is
expected to be seen varying in quadrature with the first.  We will
consider this below in terms of comparison to changes in the
equatorial Pacific warm water volume
\citep{burgers-stoch-enso,kessler-events,mcphaden-wwv}.

Some model results show lower dimensional behaviour, including CCSM3
and CGCM3.1 (both T47 and T63).  In the case of CCSM3, the reason for
this behaviour is seen in the NINO3 power spectra in
Figure~\ref{fig:nino3-spectra}.  Here, the observational data show a
broad peak in the ENSO power band (2--7 years).  CCSM3, however, has a
sharper peak at almost exactly 2 years, displaying variability rather
different from observed ENSO variability
(Section~\ref{sec:basic-eq-sst-var} presents a possible explanation
for this narrowband variability in CCSM3).  In the Isomap analysis,
this biannual variability is aliased with the annual cycle, and no
distinct ENSO variability is detected.  The situation with the CGCM3.1
models is different.  Here, the NINO3 power spectrum shows essentially
no peak in the ENSO frequency band.  It is not clear what is happening
here, but it may be relevant that the equatorial SST climatology in
both CGCM3.1 models is poor, showing little or no gradient across the
Pacific (Figure~\ref{fig:basic-sst}).

%
%
\begin{figure}
  \begin{center}
    \includegraphics[width=\textwidth]{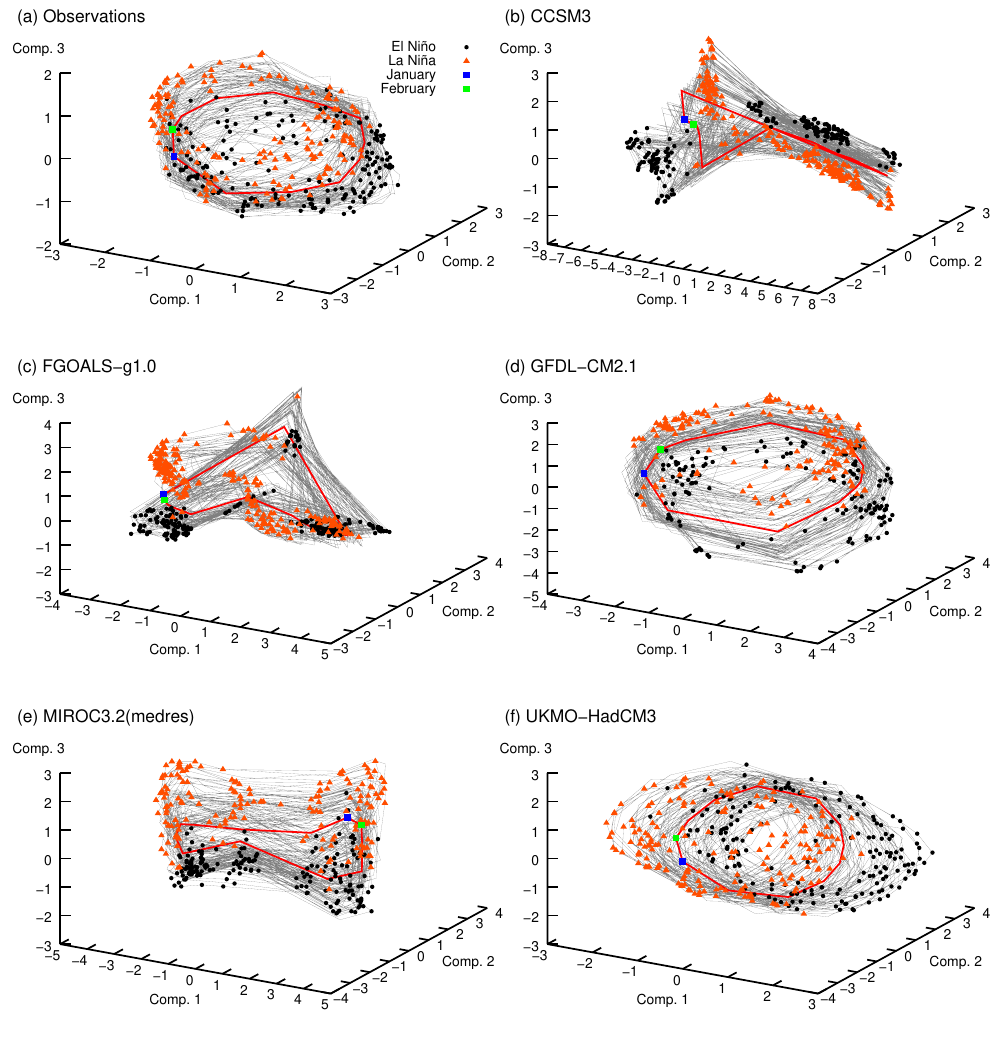}
  \end{center}
  \caption[Three-dimensional Isomap embeddings]{Three-dimensional
    embeddings of Isomap raw SST results for observations (a) and
    selected models (b--f).  Light grey lines join data points
    representing adjacent months in the SST time series.  The mean
    annual cycle is shown as a thicker red line with January and
    February highlighted in blue and green respectively.  Points are
    identified as \eln (black dots) or \lan (red triangles) events
    based on the corresponding NINO3 SST index time series for each
    data set.  For clarity, only 100 years of data is plotted for each
    model.}
  \label{fig:isomap-raw-3d}
\end{figure}
%

%
%
\begin{figure}
  \begin{center}
    \includegraphics[width=0.75\textwidth]{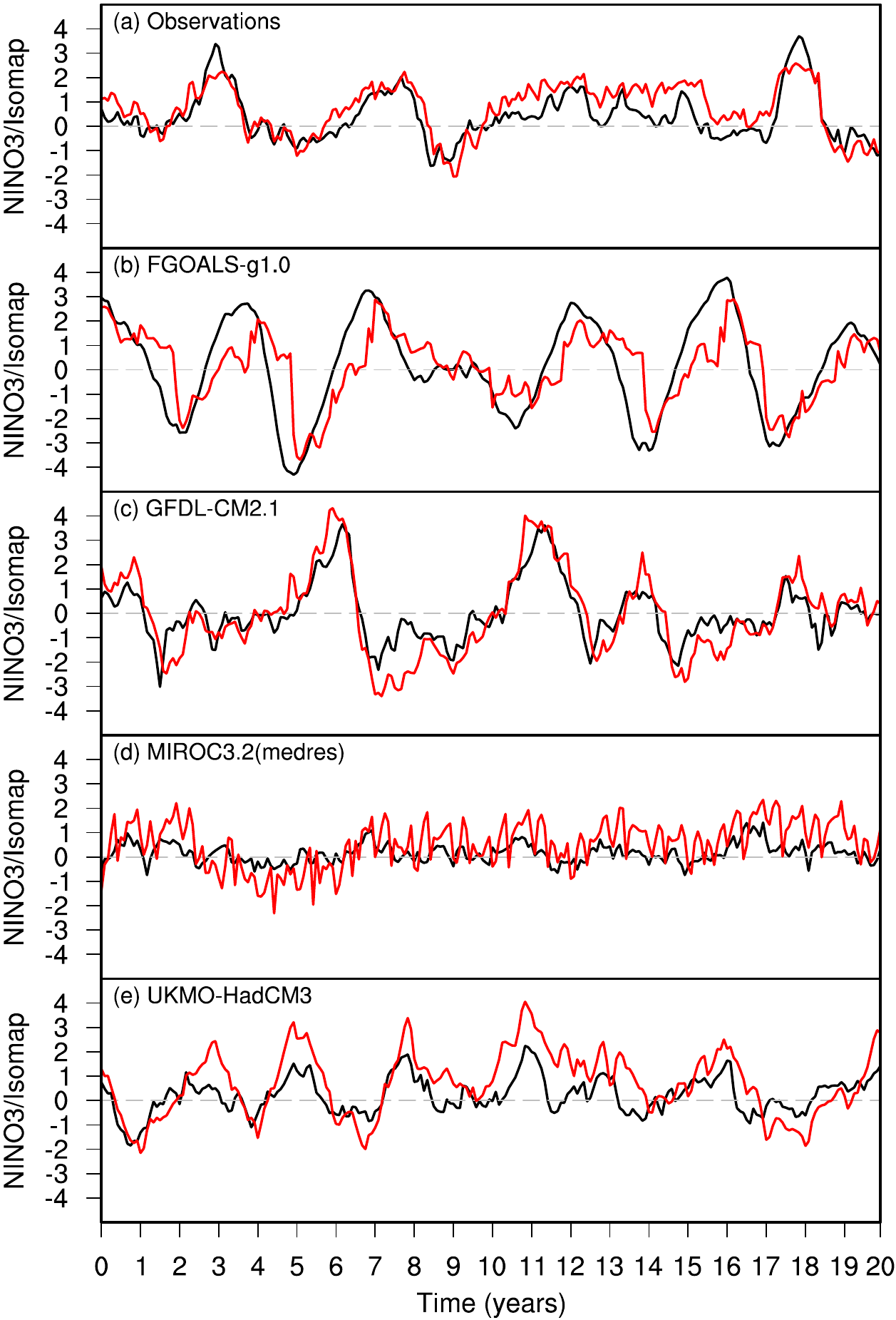}
  \end{center}
  \caption[NINO3 SST index and Isomap rotated component \#3 time
    series]{Time series of NINO3 SST index (black) and rotated Isomap
    component \#3 (red) for observations (a) and selected models
    (b--e).  An arbitrary 20 year slice of data is shown in each
    case.}
  \label{fig:isomap-rot3}
\end{figure}
%

Once we select a dimensionality for embedding of Isomap results, we
can calculate reduced coordinates using \eqref{eq:isomap-mds-coords}.
Here, I select an embedding dimensionality of three, both because this
lies in the range derived from the Isomap eigenvalue spectra and
because it is the highest dimensionality of data we can easily
visualise.  Figure~\ref{fig:isomap-raw-3d} illustrates
three-dimensional embeddings for SST observations and a selection of
models.  The results shown are all for $k$-Isomap with $k=7$.  The
plots show the data as a time series, with points adjacent in time
connected by thin grey lines.  The mean annual cycle is shown as a
thicker red line with January and February highlighted for
orientation.  Points identified as \eln or \lan events on the basis of
the NINO3 SST index are picked out in colour.  For clarity, only 100
years of results are plotted in each case.  Concentrating on the
observations first, it can be seen that Isomap correctly identifies
the annual cycle (represented by motion about the roughly cylindrical
region occupied by the data points) and at least one other form of
variability (represented by motion approximately in the direction of
the axis of the cylindrical region).  The clustering of \eln and \lan
points indicates that this second mode of variability corresponds to
ENSO and generally lies along the direction orthogonal to the annual
cycle in the embedding coordinates.  Following \citet{gamez-enso}, the
role of the ``axial'' mode can be clarified by rotating the Isomap
embedding to bring the mean annual cycle into the $x$-$y$ coordinate
plane.  In this rotated coordinate system, variations in the
$z$-direction record the ``axial'' variability in the original
embedding coordinates (see Section~\ref{sec:isomap-rotation} below for
details of this rotation procedure).  Time series plots of the rotated
third Isomap component for observations and four of the models
selected here are shown in Figure~\ref{fig:isomap-rot3}.  The rotated
Isomap component \#3 time series are plotted in parallel with time
series of the NINO3 SST index, recording ENSO variability.  For the
observations, in Figure~\ref{fig:isomap-rot3}a, it is clear that
rotated Isomap component \#3 quite accurately captures ENSO
variability in the input SST data.  In this case, Isomap has thus
extracted the most important modes of variability in tropical Pacific
SSTs, the annual cycle and ENSO, starting from high-dimensional input
data.  We can also go further and attempt to extract the second degree
of freedom in ENSO variability, usually identified with the equatorial
Pacific ocean heat content or warm water volume
\citep{kessler-events,mcphaden-wwv}, by examining a four-dimensional
embedding of the Isomap results.  The same sort of rotation procedure
can be applied to remove the influence of the annual cycle variability
on both Isomap components \#3 and \#4 (see
Section~\ref{sec:isomap-rotation} for details).  Correlation
coefficients between Isomap rotated component \#3 and the NINO3 SST
index and between Isomap rotated component \#4 and WWV are shown in
Table~\ref{tab:isomap-correlations}.  For the observational data, the
NINO3 correlation is high, as would be expected from
Figure~\ref{fig:isomap-rot3}a, but the correlation between rotated
Isomap component \#4 and WWV is very low.  It thus appears that
rotated Isomap component \#4 here does not capture this second degree
of ENSO variability.

%
%
\begin{table}
  \caption[Correlation between NINO3 SST and WWV and Isomap
    components]{Correlation coefficients between NINO3 SST index and
    warm water volume (WWV) and Isomap components from $k$-Isomap with
    $k=7$: for raw SSTs, the correlation between rotated Isomap
    component \#3 and NINO3 and between rotated Isomap component \#4
    and WWV; for SST anomalies, the correlation between Isomap
    component \#1 and NINO3 and between Isomap component \#2 and WWV.
    Blank entries occur where the Isomap eigenvalue spectrum in a
    particular case does not have enough positive eigenvalues to form
    an embedding of the required dimensionality.}

  \label{tab:isomap-correlations}

  \begin{center}
    \begin{threeparttable}
      \begin{tabular}{lccccc}
        \toprule

        \multirow{3}{2.5cm}{Data Set} & \multicolumn{4}{c}{Correlation} \\
        & \multicolumn{2}{c}{Raw SST} & \multicolumn{2}{c}{SST anomaly} \\
        & NINO3 & WWV & NINO3 & WWV \\

        \midrule

        Observations     & 0.822 & 0.031 & 0.841 & 0.242 \\
        BCCR-BCM2.0      & 0.835 &       & 0.820 & 0.153 \\
        CCSM3            & 0.047 & 0.284 & 0.901 & 0.245 \\
        CGCM3.1(T47)     & 0.223 &       & 0.824 & 0.021 \\
        CGCM3.1(T63)     & 0.225 & 0.148 & 0.814 & 0.284 \\
        CNRM-CM3         & 0.824 & 0.225 & 0.776 & 0.228 \\
        CSIRO-Mk3.0      & 0.746 & 0.227 & 0.717 & 0.407 \\
        ECHO-G           & 0.907 & 0.681 & 0.935 & 0.646 \\
        FGOALS-g1.0      & 0.793 &       & 0.776 & 0.430 \\
        GFDL-CM2.0       &       &       & 0.857 & 0.435 \\
        GFDL-CM2.1       & 0.859 &       & 0.853 & 0.665 \\
        GISS-EH          & 0.652 &       & 0.665 & 0.116 \\
        INM-CM3.0        & 0.730 &       & 0.744 & 0.446 \\
        IPSL-CM4         & 0.844 &       & 0.852 & n/a\tnote{a} \\
        MIROC3.2(hires)  & 0.236 & 0.171 & 0.686 & 0.070 \\
        MIROC3.2(medres) & 0.646 &       & 0.785 & 0.087 \\
        MRI-CGCM2.3.2    & 0.861 &       & 0.909 & 0.511 \\
        UKMO-HadCM3      & 0.804 & 0.093 & 0.809 & 0.012 \\
        UKMO-HadGEM1     & 0.747 &       & 0.752 & 0.274 \\

        \bottomrule
      \end{tabular}
      \begin{tablenotes}
        \item[a]{Ocean temperature data required to calculate
          warm water volume for IPSL-CM4 is not available.}
      \end{tablenotes}
    \end{threeparttable}
  \end{center}
\end{table}
%

Although the fact that Isomap appears to capture the annual cycle
variability and at least some aspects of ENSO variability is
unsurprising, the data-driven nature of Isomap makes it useful for
comparison of model results with observations and for inter-model
comparison.  I apply the same three-dimensional embedding to selected
model results in Figures~\ref{fig:isomap-raw-3d}b--f.  Results for a
number of the models shown (GFDL-CM2.1, MIROC3.2(medres) and
UKMO-HadCM3) are similar to observations, with a clear
three-dimensional structure to the data embedding, cleanly picking out
the annual cycle and ENSO, with distinct clustering of \eln and \lan
events.  For the other two models illustrated, CCSM3 and FGOALS-g1.0,
the three-dimensional Isomap embedding reveals data manifolds of
significantly different form to that of the observations.  As noted
earlier, for CCSM3 this is due to excessively regular interannual
variability in tropical Pacific SSTs that appears to be aliased with
the annual cycle.  For FGOALS-g1.0, the situation appears to be
similar.  The FGOALS-g1.0 NINO3 power spectrum in
Figure~\ref{fig:nino3-spectra} exhibits a narrow peak at a period of
around 3.5 years, rather than a broad peak stretching across the 2-7
year ENSO power band.  This narrowband signal is again likely to
result in lower-dimensional behaviour in the Isomap results.

Time series of rotated Isomap component \#3 alongside the NINO3 SST
index are plotted for a smaller selection of models in
Figures~\ref{fig:isomap-rot3}b--e.  Two of these cases, GFDL-CM2.1
(Figure~\ref{fig:isomap-rot3}c) and UKMO-HadCM3
(Figure~\ref{fig:isomap-rot3}e), are models whose three-dimensional
Isomap embeddings show similar structure to observations.  This is
reflected in the rotated Isomap component \#3 time series, which show
good correlation with the NINO3 SST index.  A good correlation is also
seen for the results for FGOALS-g1.0 (Figure~\ref{fig:isomap-rot3}b),
despite the apparent degeneracy of the 3-D Isomap embedding in
Figure~\ref{fig:isomap-raw-3d}c.  Despite the visual discrepancy
between the FGOALS-g1.0 embedding results and the observations, it
appears that the Isomap algorithm is still able to disentangle the
annual and ENSO variability in the modelled SST data.  The other model
illustrated in Figure~\ref{fig:isomap-rot3} is MIROC3.2(medres)
(Figure~\ref{fig:isomap-rot3}d), which has weaker ENSO variability,
but still shows a reasonable correlation between rotated Isomap
component \#3 and the NINO3 SST index.

For models with strongly degenerate three-dimensional Isomap
embeddings, such as CCSM3 (Figure~\ref{fig:isomap-raw-3d}b),
CGCM3.1(T47), CGCM3.1(T63) and MIROC3.2(hires) (not shown), the
rotated Isomap component \#3 time series show little coherent
variability, and certainly none that correlates with ENSO variability.
Correlation coefficients between rotated Isomap component \#3 and the
NINO3 SST index are shown in Table~\ref{tab:isomap-correlations} for
all models.  The models showing good correlations are those for which
the three-dimensional Isomap embedding displays similar structure to
the observations, i.e. for which Isomap successfully extracts the
annual cycle and an ``orthogonal'' ENSO mode.  As for the
observations, we can also attempt to identify a second degree of
freedom of ENSO variability by examining four-dimensional Isomap
embeddings.  One problem here is that, for some of the models, the
Isomap eigenvalue spectra do not have enough positive leading
eigenvalues to provide a four-dimensional embedding: at least four
positive leading eigenvalues are required.  In cases where a
four-dimensional embedding is possible, I use the same
four-dimensional rotation as for the observations to remove the annual
variability from both rotated Isomap components \#3 and \#4, and
calculate correlation coefficients between the rotated Isomap
components and the NINO3 SST index and simulated WWV time series,
calculated as described in Chapter~\ref{ch:data-and-models}.  As for
the observations, the correlations between rotated Isomap component
\#4 and WWV for the models are generally rather low.

As noted at the beginning of this section, one reason for applying
Isomap to raw SST data, as opposed to SST anomalies, was to determine
the extent to which Isomap is able to identify the coupling between
annual and ENSO variability in the tropical Pacific.  Other, more
direct, analyses of ENSO/annual cycle interactions reveal a strong
influence of the magnitude of the annual cycle in the equatorial
Pacific on ENSO variability \citep{guilyardi-cmip3}.  On the basis of
the results presented here, it appears that an analysis using Isomap
does not provide very much insight into this question.

%
%

\subsection{Analysis for SST anomalies}
\label{sec:isomap-sst-anom-analysis}

%
%
\begin{figure}
  \begin{center}
    \includegraphics[width=0.44\textwidth]{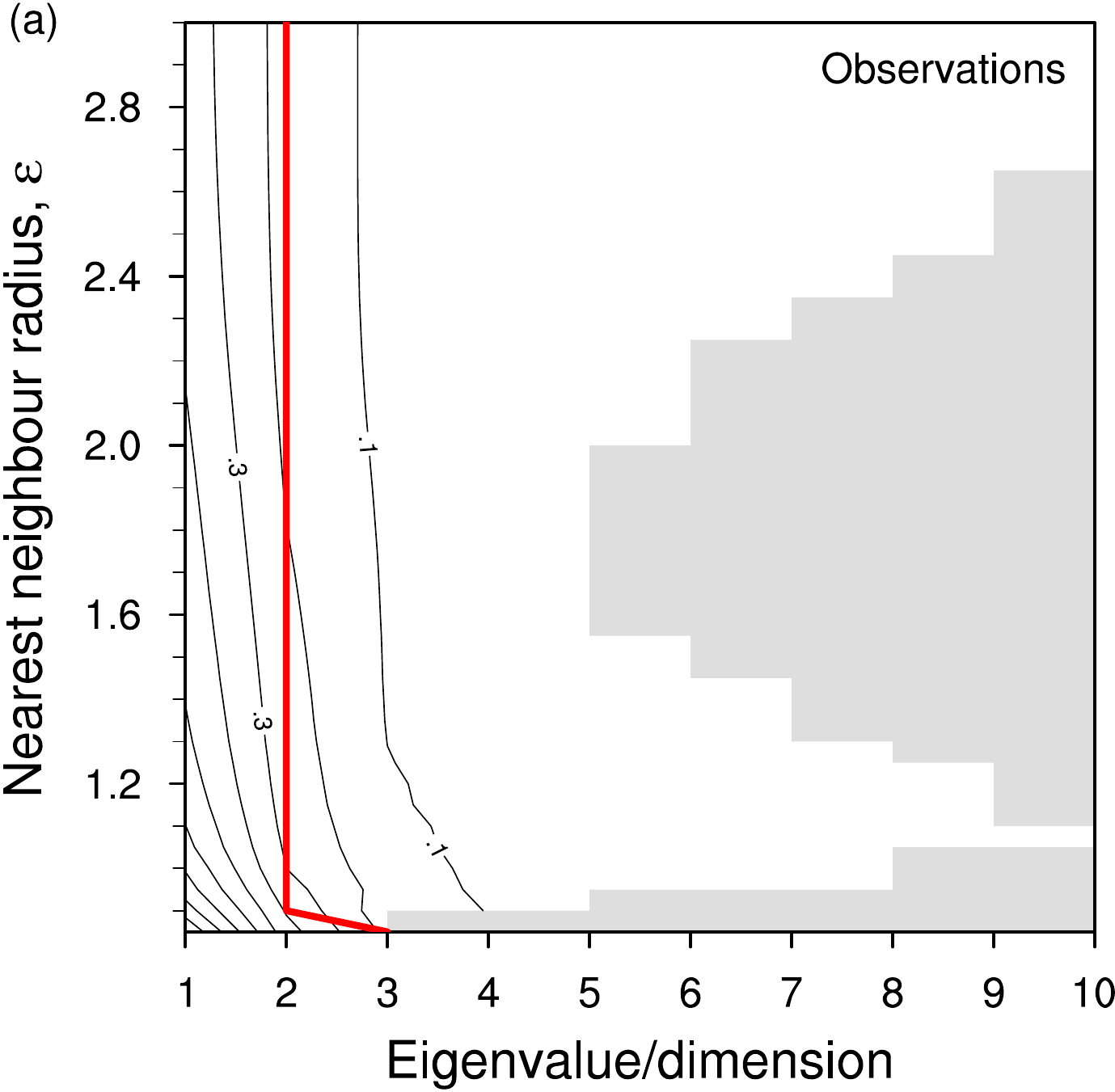}%
    \hspace{0.1\textwidth}%
    \includegraphics[width=0.44\textwidth]{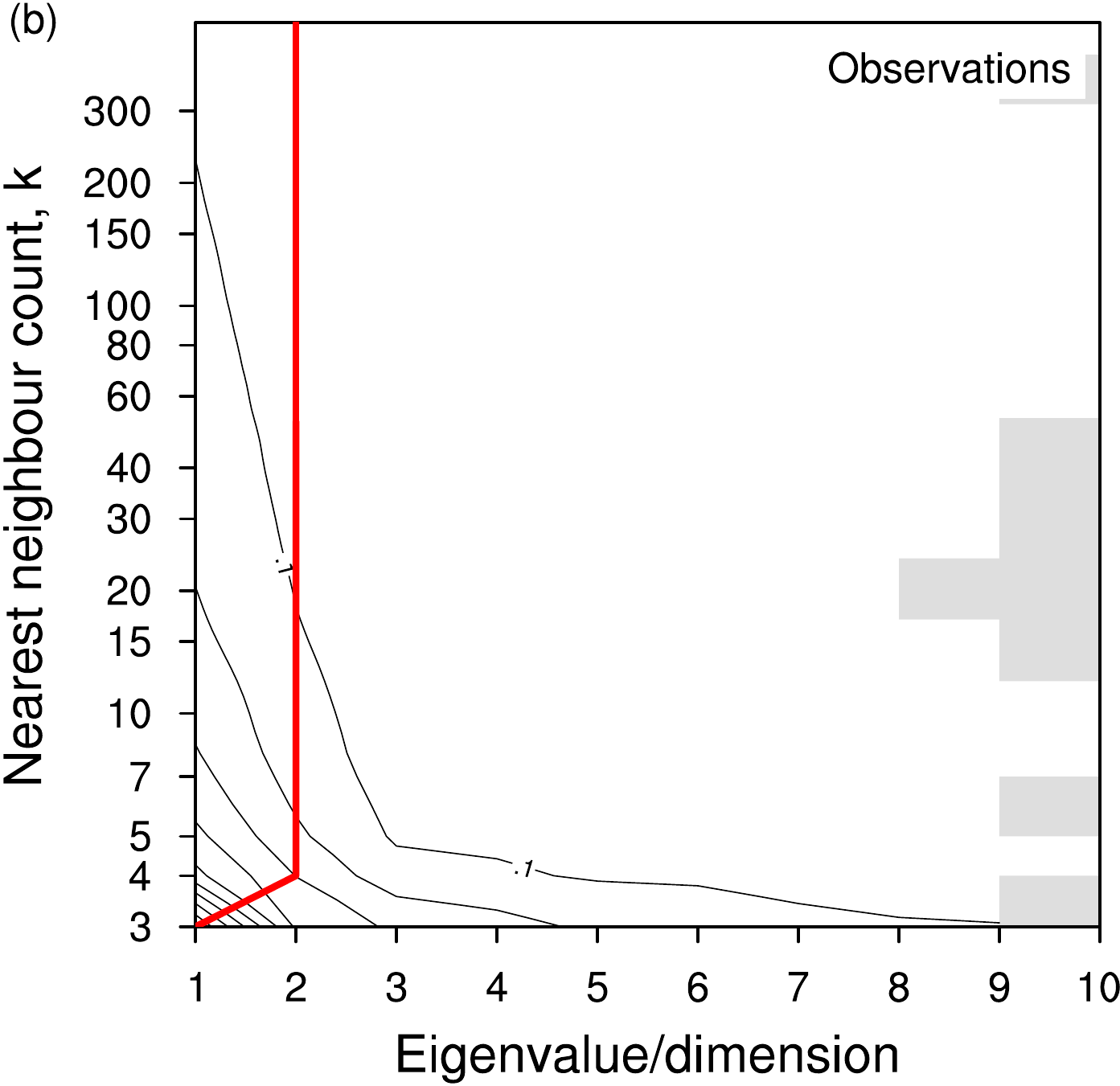}

    \vspace{0.25cm}
    \includegraphics[width=0.44\textwidth]{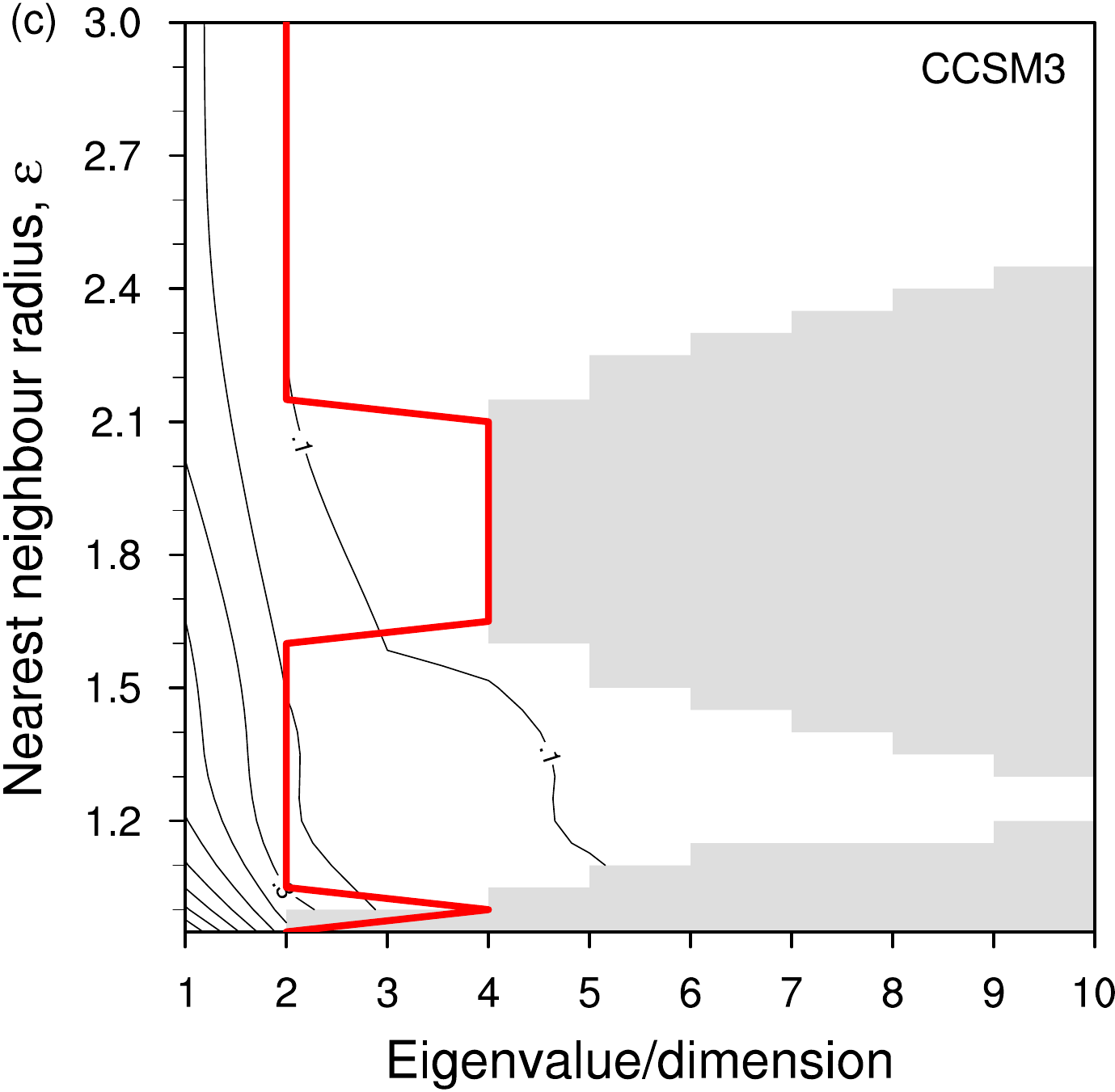}%
    \hspace{0.1\textwidth}%
    \includegraphics[width=0.44\textwidth]{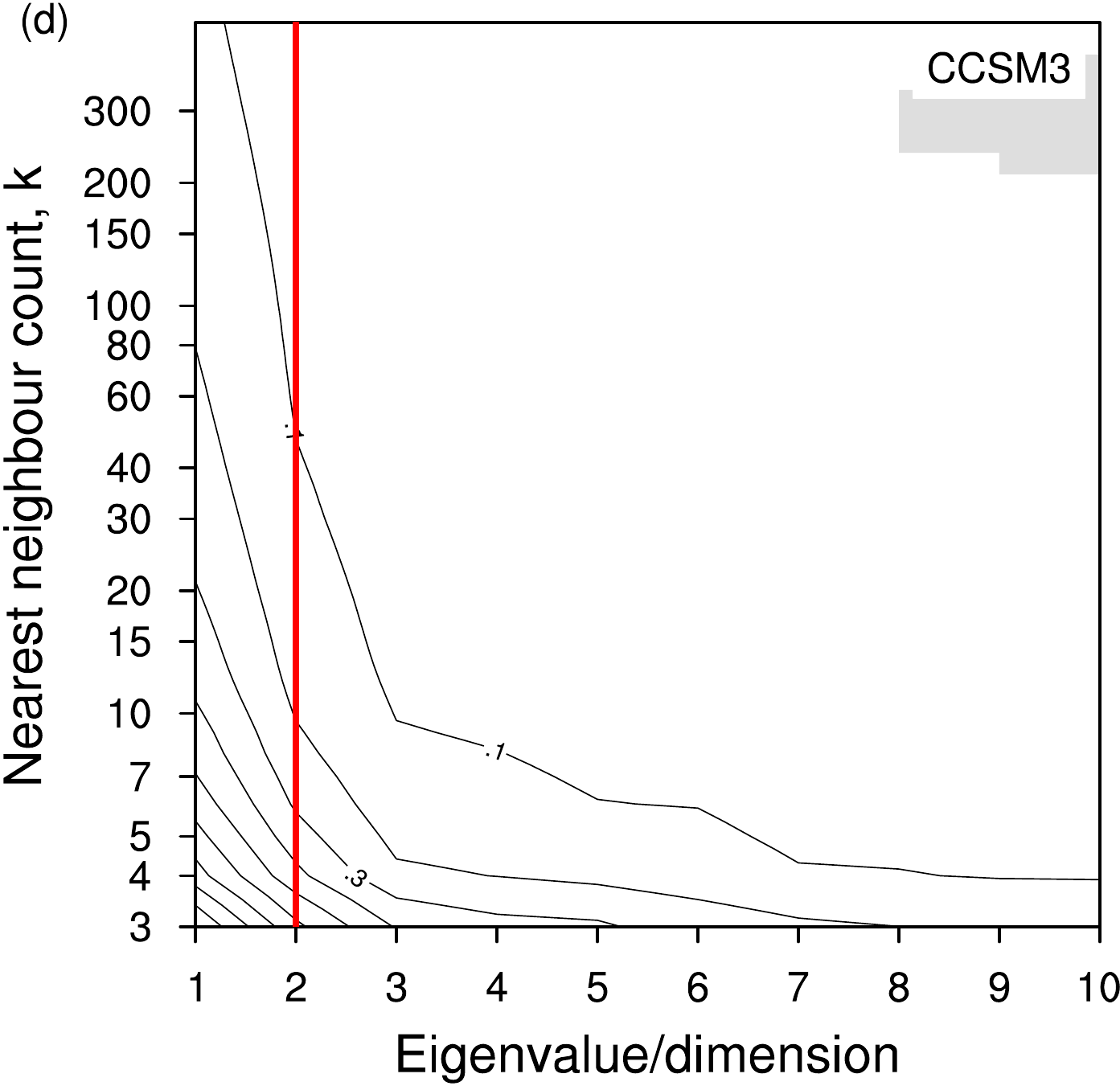}

    \vspace{0.25cm}
    \includegraphics[width=0.44\textwidth]{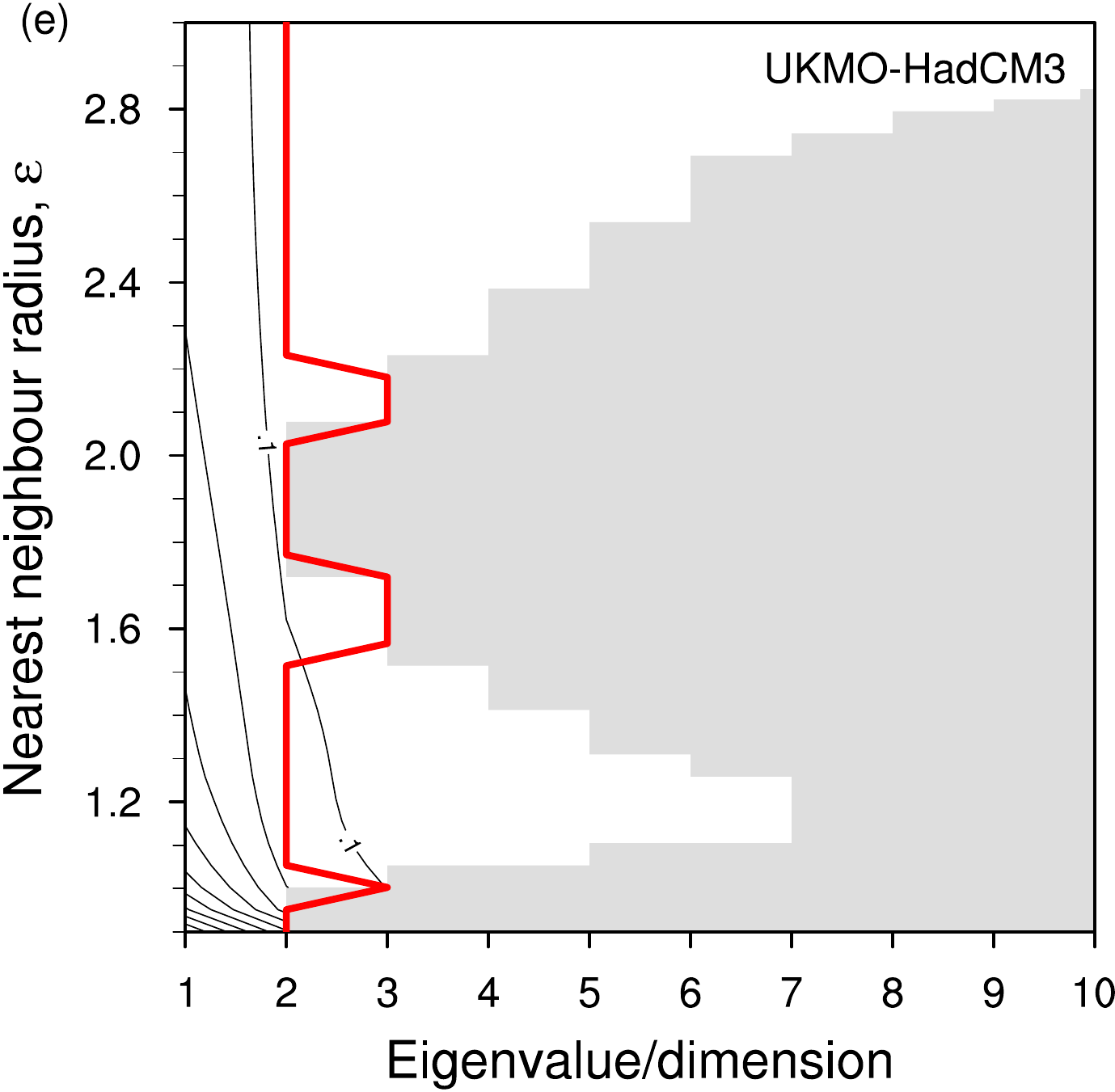}%
    \hspace{0.1\textwidth}%
    \includegraphics[width=0.44\textwidth]{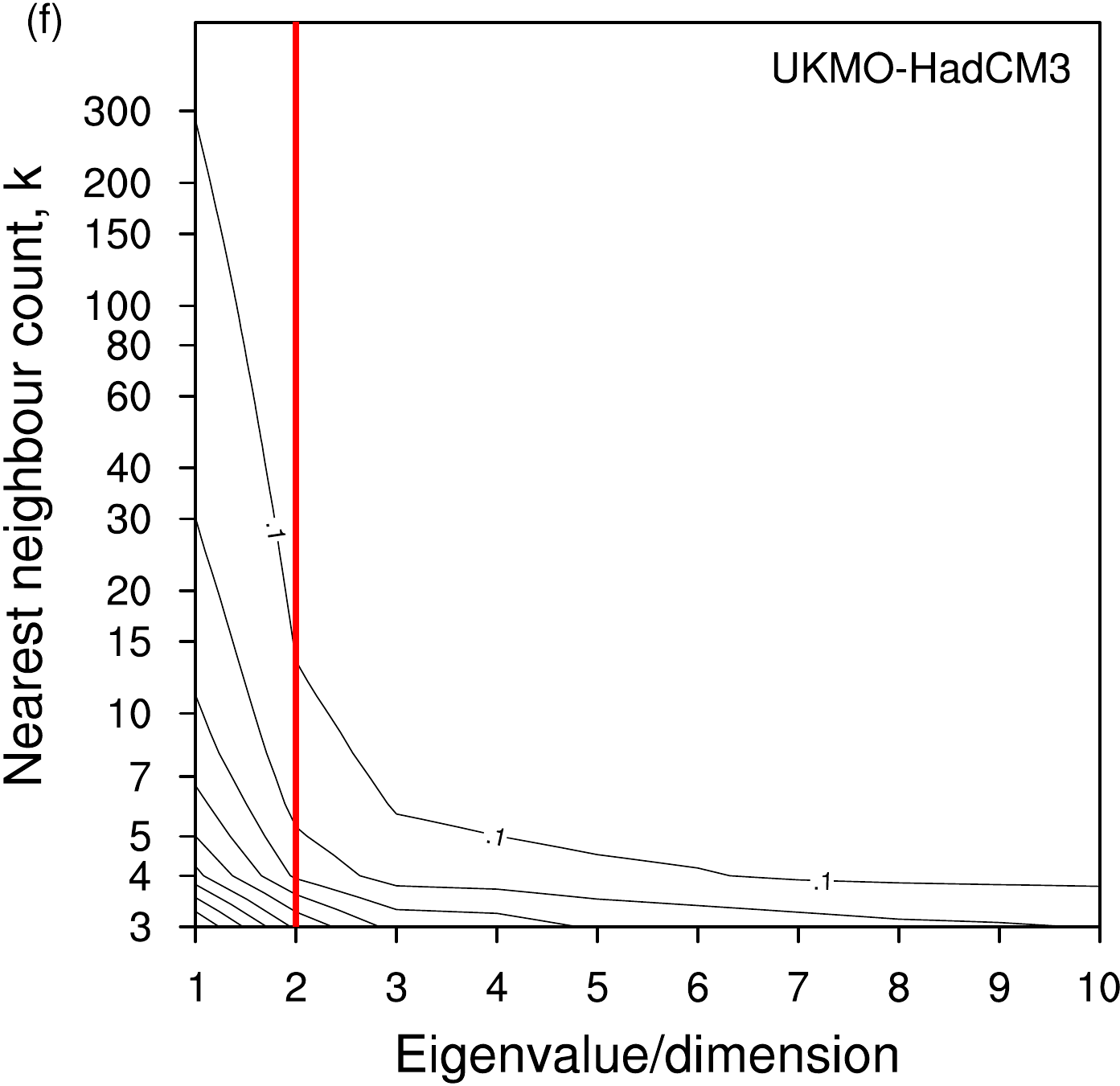}
  \end{center}
  \caption[Isomap eigenvalue convergence: SST anomalies]{Isomap
    eigenvalue convergence and dimension estimates for tropical
    Pacific SST anomalies, from observations (a and b), CCSM3 (c and
    d) and UKMO-HadCM3 (e and f).  Black contours show MDS eigenvalue
    spectra normalised by the overall largest eigenvalue, as a
    function of eigenvalue number and neighbourhood radius
    $\varepsilon$ (a, c, e) or nearest neighbour count $k$ (b, d, f)
    (logarithmic axis).  Grey areas indicate regions of the eigenvalue
    spectra not available for dimensionality reduction because of the
    presence of negative eigenvalues.  The thick red line shows the
    data dimensionality estimated from the eigenvalue spectra.}
  \label{fig:sst-anom-isomap-sens}
\end{figure}
%

In climatological contexts, PCA is normally applied to climate
anomalies, i.e. to data from which the mean annual cycle has been
removed.  This was the case for the equatorial Pacific SST EOFs shown
in Section~\ref{sec:sst-pca}.  We can also apply Isomap to SST
anomalies, thus providing results that are more directly comparable
with the results of PCA than the raw SST Isomap analysis of the
previous section.  These results may also be slightly easier to
interpret because rotation to remove the influence of the annual cycle
is not required.

As for the raw SST Isomap results, the sensitivity of the SST anomaly
Isomap results to variations in the $\varepsilon$ or $k$ parameters
can be examined.  Results for observations and selected models are
plotted in Figure~\ref{fig:sst-anom-isomap-sens} and minimum and
maximum dimensionality estimates derived from these plots are shown in
Table~\ref{tab:isomap-dimensions}.  It can be seen that the
dimensionality estimates for the SST anomaly data are all rather low,
with only one model (FGOALS-g1.0) having a maximum dimensionality
greater than two.  This indicates that only one- or two-dimensional
embeddings of the Isomap results are possible.

I thus examine one-dimensional and where available, two-dimensional,
embeddings of the Isomap results.  The justification for this is that
we expect ENSO to be the major mode of variability in the SST
anomalies, with the first component of any embedding corresponding to
the NINO3 SST index variability, and the second component to the warm
water volume variation --- looking at one- or two-dimensional
embeddings should pick these features out.
Table~\ref{tab:isomap-correlations} shows correlation coefficients
between SST anomaly Isomap components \#1 and \#2 and the NINO3 SST
index and WWV time series respectively.  The strong correlations
between Isomap component \#1 and the NINO3 SST index here indicate
that the one-dimensional Isomap embedding does a good job of
identifying the leading mode of ENSO variability, where it exists.
For most models, the degree of correlation between the SST anomaly
Isomap component \#1 and the NINO3 SST index is similar to the degree
of correlation between the raw SST rotated Isomap component \#3 and
the NINO3 SST index.  For a small number of models (primarily CCSM3,
but also CGCM3.1(T47), CGCM3.1(T63) and to a lesser extent,
MIROC3.2(hires)), the correlation for the SST anomaly Isomap component
\#1 is much higher than for the raw SST results.  A reasonable
explanation for this phenomenon in the case of CCSM3 is that the ENSO
signal in this model is very regular, with a periodicity of almost
exactly two years (Figure~\ref{fig:nino3-spectra}), so is likely to be
strongly aliased with the annual cycle in the raw SST results.
Removing the annual cycle and working with SST anomalies may lift this
degeneracy, allowing the ``true'' ENSO signal to be detected, leading
to a stronger correlation.  This aliasing-based explanation is less
applicable to the other models displaying large differences between
the raw SST and SST anomaly Isomap correlation coefficients, since
they do not have the same sort of very regular ENSO variability as
CCSM3.

As for the raw SST results, we can also seek a second degree of
freedom of ENSO variability by examining correlations between the SST
anomaly Isomap component \#2 and the warm water volume time series
from the models.  Here, some of the correlations between the Isomap
SST anomaly component \#2 and WWV are somewhat better than in the raw
SST case, but there is still great variability in the correlations,
and there is no clear link between a high correlation and a ``good''
ENSO.  For instance, the CMIP3 models identified by
\citet{vanoldenborgh-enso} as having the best ENSO (GFDL-CM2.1,
MIROC3.2(hires), MIROC3.2(medres), UKMO-HadCM3) have correlations
ranging from 0.012 to 0.665.  Again, it seems difficult to draw any
clear conclusions from these results.  This may of course simply be
due to problems in the phasing of variations in eastern equatorial
Pacific SST and zonal mean equatorial Pacific thermocline depth in the
models, described in Section~\ref{sec:nino3-wwv-phasing}.  The NINO3
SST index versus WWV phasing is certainly not particularly clear for
some models, and in the face of this lack of coherent NINO3/WWV
variation, it seems unrealistic to expect Isomap to pick out any
degree of freedom in ENSO variability in most of the models that
displays any coherence with WWV variations.  In this context, it is
perhaps notable that the model with the greatest correlation
coefficient between SST anomaly Isomap component \#2 and WWV (0.665)
is GFDL-CM2.1 (Table~\ref{tab:isomap-correlations}).

Scatter plots of the first two Isomap components display similar
patterns to the principal component scatter plots of
Figure~\ref{fig:sst-pc-scatter}, which seems to indicate that the MDS
eigenvectors produced by Isomap are nonlinearly related, just as are
the EOFs produced by PCA.  I believe that this may be a signal of
intrinsic curvature in the data manifold.  Isomap relies on isometric
transformations of the data points and is therefore only able to
represent embeddings of intrinsically flat manifolds.  Attempting to
project a manifold with non-zero intrinsic curvature to a
lower-dimensional space by an isometric transformation necessarily
leads to distortion of the relationships between points in the
manifold.

\section{Discussion and conclusions}
\label{sec:isomap-conclusions}

I have examined the applicability of Isomap to climate data analysis
in the context of an inter-model comparison of ENSO variability.  This
analysis indicates that Isomap is able to capture some of the
low-dimensional dynamics of ENSO variability in the data sets I have
examined, picking out the gross features in the data.  In some cases,
notably for CCSM3, but also for CGCM3.1(T47), CGCM3.1(T47) and
MIROC3.2(hires), examination of three-dimensional embeddings of the
raw SST Isomap results, both visually
(e.g. Figure~\ref{fig:isomap-raw-3d}b) and via correlations between
rotated Isomap component \#3 and the NINO3 SST index
(Table~\ref{tab:isomap-correlations}) reveals an anomalously low
dimensionality of modelled ENSO variability, apparently caused by too
regular interannual tropical Pacific SST variability, leading to
aliasing of the ENSO signal to the annual cycle.  Although this aspect
of the models can be identified by other means, it is encouraging that
Isomap is able to detect the anomalous behaviour without prompting.
Less encouraging is the fact that Isomap is able to capture only these
gross features of ENSO variability in the models.  The Isomap results
do not show much in the way of variations between models, at least not
in an easily interpreted form.  They also do not capture the sometimes
significant differences between modelled and observed ENSO behaviour
revealed by a simple comparison of model and observational SST EOFs.
Calculation of correlations between Isomap results and WWV time series
for both observations and model simulations do not reveal any strong
relationship between the degrees of freedom found by Isomap and the
second degree of freedom of ENSO variability that is generally
believed to be represented by variations in WWV.  Better results from
this point of view might be found by performing an Isomap analysis
directly on modelled thermocline depths as was done for NLPCA in
Chapter~\ref{ch:nlpca}, rather than simply trying to correlate WWV
with SST Isomap results.  However, unrealistic thermocline depth
variation in some of the models is likely to make this difficult.

A more subtle illustration of differences between PCA and Isomap is
presented by a comparison of the sensitivity of Isomap and
conventional PCA to small changes in the structure of tropical Pacific
SST variability around the mid-1970s shift in ENSO behaviour
\citep{fedorov-change,mcphaden-enso}.  If the observational SST data
set is split into a pre-1976 and a post-1976 component, differences
relating to this change in ENSO behaviour are clearly apparent in SST
EOFs, with a shift to stronger El Ni\~{n}o events.  However, an Isomap
analysis shows no significant differences in eigenvalue spectra
between pre-1976 and post-1976 data (data not shown).  I speculate
that this difference in sensitivity between PCA and Isomap is due to
the fact that the orthogonal transformations associated with PCA,
being more geometrically ``rigid'' than the isometric transformations
of Isomap, are less able to conform to subtle changes in the data
manifold, thus highlighting these relatively small differences.

Although in some senses Isomap is a rather blunt tool, it appears that
it may be useful for exploratory data analysis, particularly if there
is reason to believe that the data in question really is nonlinear and
not too high-dimensional.  In such cases, Isomap may serve a purpose
alongside more conventional techniques.

There are four further issues with the Isomap algorithm that deserve
comment, and that can provide a basis for comparison between Isomap
and the NLPCA method (see Section~\ref{sec:nlpca-discussion} for more
NLPCA-specific discussion).  First is the question of the sensitivity
of the results of nonlinear dimensionality reduction techniques to
parameter choices in the algorithms used.  For Isomap, this means
variations in the $k$ or $\varepsilon$ neighbourhood size parameter.
The possibility of varying this parameter can be viewed as an
advantage, since it provides a mechanism to probe different length
scales in the data in a way that has no analogue in PCA.  How useful
this is depends on the complexity of the data set: for the simple
Swiss roll data, a two-dimensional manifold embedded in
$\mathbb{R}^3$, variation in $k$ or $\varepsilon$ probes the structure
of the data quite successfully.  For the more complex ENSO data sets,
it is not at all clear what sort of structures are being probed as the
neighbourhood size is varied, and there is little consistency between
the results from different models.  In fact, from this point of view,
the sensitivity of Isomap to the neighbourhood size is a clear
disadvantage, since computational requirements generally restrict us
to choosing a particular value of $k$ or $\varepsilon$ for our
analyses, and there is no a priori reason to select one value over
another.  The situation for NLPCA is somewhat more complex than for
Isomap, since there are a larger number of parameters involved: not
only is there a choice of the exact structure of the network to be
used (number of bottleneck nodes, special architectures for the
bottleneck layer, number of nodes in hidden layers), but there are
parameter choices involved in the protocol used to train the network
without overfitting.

Second, results from Isomap are not easy to interpret if the
underlying data manifold has a dimensionality higher than two or
three.  One example is an attempt to apply Isomap to mid-latitude
tropospheric variability.  Here, I performed an Isomap analysis for a
monthly time series of Atlantic sector 500\,hPa geopotential height.
Isomap $k$/$\varepsilon$ sensitivity studies (not shown) indicate a
dimensionality of around 6 for the underlying data manifold.  For
manifolds of such high dimensionality, it is not possible to visualise
the Isomap embeddings as I have done here for ENSO variability.  Two-
or three-dimensional projections are not sufficient to ``unfold'' the
variability in the data, and the data points appear as an amorphous
cloud of points.  This situation also arises with PCA, if the
eigenvalue spectrum converges slowly and many EOFs are required to
explain a sufficient fraction of the data variance, but the linearity
of PCA provides a partial solution.  Linearity permits us to take
single modes, EOFs, and treat them independently.  No such
decomposition is possible for Isomap.  This problem is not an inherent
limitation of all nonlinear dimensionality reduction techniques.  For
instance, NLPCA permits advance selection of the dimensionality to
which the input data is to be reduced by selection of the number of
neurons in the bottleneck layer.  Choosing a one-dimensional reduction
gives the best nonlinear fit of a one-dimensional function to the
input data, independent of the true dimensionality of the underlying
data manifold, which the method makes no direct effort to ascertain.
This approach allows for a modal analysis of the data, where nonlinear
modes are stripped out of the input data one at a time.  This type of
analysis is not possible for Isomap because there is no way to control
the dimensionality of the data reduction.

The third issue is shared with other nonlinear dimensionality
reduction methods and is that it is generally difficult to produce
plots showing spatial patterns of variability for nonlinear
dimensionality reduction in the way that is done for PCA, where map
plots of the leading EOFs are an important analytical tool.  Such maps
can be produced for PCA because real-valued EOFs essentially represent
standing oscillations in the data, so a snapshot at any point in the
oscillation from a positive pattern to a negative pattern records all
the information about the spatial variability in the mode.  For
nonlinear methods, more general temporal variability is possible, and
generally one needs to provide a set of spatial patterns corresponding
to selected points on the reduced data manifold.  This approach was
possible for NLPCA in Chapter~\ref{ch:nlpca} because of the use of
one-dimensional reduced manifolds via a modal analysis, but for
two-dimensional or larger manifolds, the number of spatial patterns
needed becomes prohibitive.

The fourth point to note has been mentioned earlier when discussing
the Isomap Pacific SST results, and this is the question of just what
data manifolds a particular dimensionality reduction technique is
capable of representing.  As noted above, Isomap relies on a global
isometric transformation of the original data space to derive a
reduced Euclidean representation, meaning that only data manifolds
that are globally isometric to Euclidean space can be faithfully
represented by a reduced representation derived from Isomap.  For
NLPCA, the manifolds representable by the reduced representations
depend on the structure of the bottleneck layer in the neural network.
For a single bottleneck neuron, NLPCA can faithfully represent any
open one-dimensional curve, for a ``circular'' bottleneck layer (two
neurons, with values constrained to lie on the unit circle), closed
one-dimensional curves can be represented faithfully, for two
bottleneck neurons, general open two-dimensional surfaces can be
represented, and so on.  The complexity of interpreting the results of
NLPCA increases quickly with the number of neurons in the bottleneck
layer.

The essential problem with nonlinear methods such as Isomap is that
there exist few theoretical results underpinning the numerical
algorithms.  For PCA, there are results identifying EOFs for at least
some systems with normal modes of the system forced by random noise
\citep{north-eofs}.  These findings tie the numerical results of PCA
directly to dynamical characteristics of the system under study.  As
far as I know, there are no corresponding results for Isomap, or
indeed any other nonlinear dimensionality reduction technique.  There
have been applications of Isomap to simple dynamical systems, where
features observed in the Isomap results can be related to the dynamics
of the system \citep{bollt-isomap}, but no such studies exist for
larger systems approaching the complexity of current climate models.
Another approach to gaining analytical understanding is to explicitly
construct data manifolds that can be exactly embedded by Isomap.
\citet{donoho-isometry} did this for an analytic representation of
simple black-and-white images and developed several useful criteria
for recognising classes of images whose data manifolds could be
treated exactly by Isomap.  It is not clear whether a similar approach
to dimensionality reduction of dynamical systems would be fruitful.

\section{Rotation of Isomap components}
\label{sec:isomap-rotation}

As described in Section~\ref{sec:isomap-raw-sst-results},
interpretation of three- and four-dimensional embeddings of raw SST
Isomap results is clarified by rotating the components of the
embeddings to separate the influence of annual variations (represented
by rotated Isomap components \#1 and \#2) from the record of ENSO
variability (as represented by rotated Isomap components \#3 and \#4).
In this section, I explain the details of this rotation procedure,
first for the three-dimensional case, then for the more complex
four-dimensional case.

\subsection{Three-dimensional case}
\label{sec:isomap-rotation-3d}

Consider a three-dimensional Isomap embedding of a monthly time series
of $N$ data items, resulting in a time series of 3-vectors
$\vec{y}_i$, $i = 1, \dots, N$, with components calculated from
\eqref{eq:isomap-mds-coords} of Section~\ref{sec:isomap-mds}.
Assuming that the time series covers a whole number of years, so that
$N$ is a multiple of 12, then the mean annual cycle for the embedding
can be defined as $\bar{\vec{y}}_j$, $j = 1, \dots, 12$, where
\begin{equation}
  \bar{\vec{y}}_j = \frac{1}{N/12} \sum_{i=0}^{N/12-1}
  \vec{y}_{12i+j}.
\end{equation}
In general, the points $\bar{\vec{y}}_j$ of the mean annual cycle will
not lie in a single plane and, in particular, will not lie in the
$x$-$y$ coordinate plane.  This means that each of the three
components of the $\bar{\vec{y}}_j$ will vary over the course of the
annual cycle, i.e. annual variability is ``mixed into'' each of the
three components, even though only two Cartesian coordinates are
strictly needed to represent the periodic annual variation.

%
%
\begin{figure}
  \begin{center}
    \includegraphics[width=0.6\textwidth]{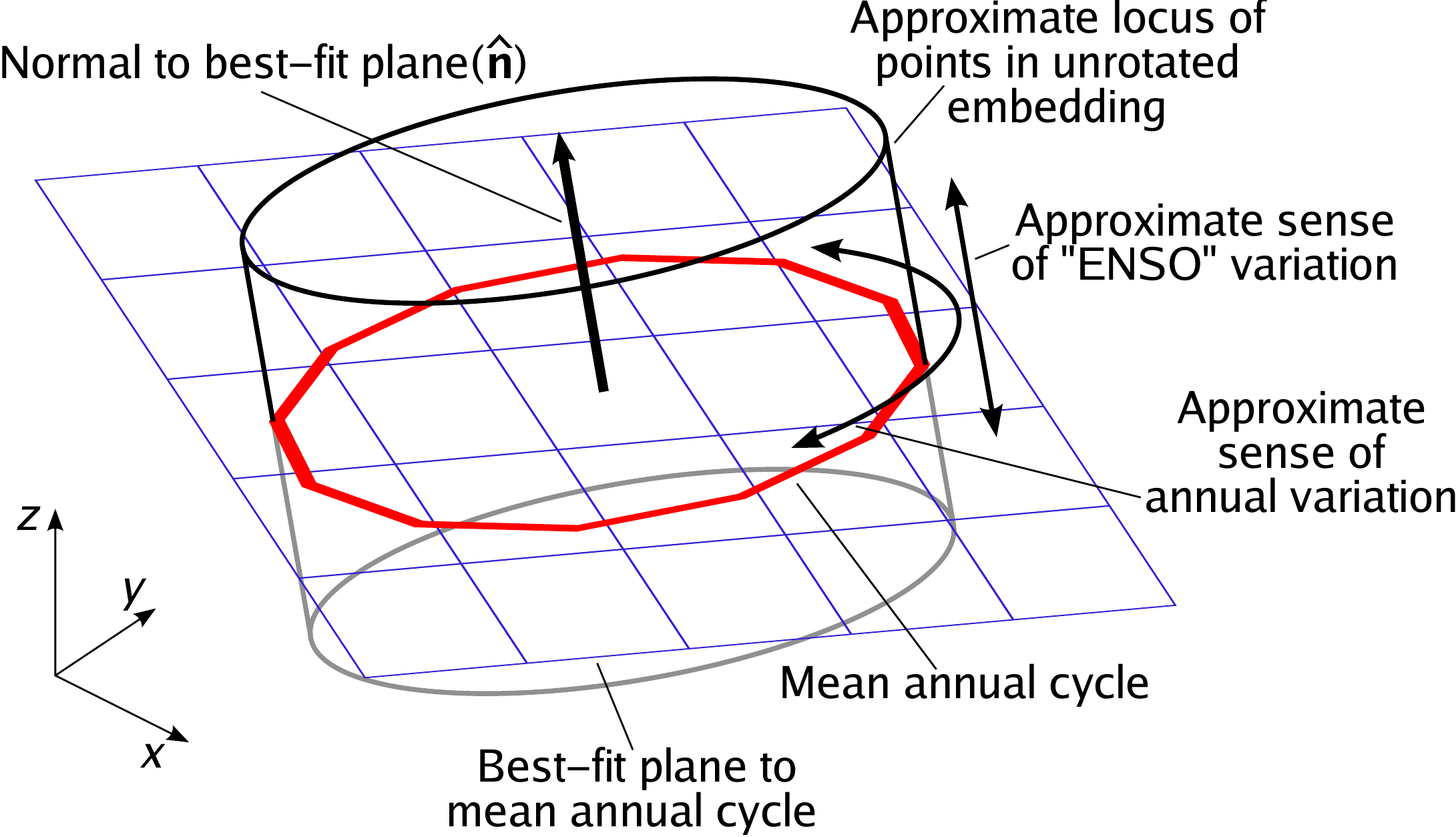}
  \end{center}
  \caption[Isomap component rotation]{Geometry of 3-D Isomap component
    rotation.  The overall view is of an unrotated 3-D Isomap
    embedding.  The thick red curve shows the mean annual cycle, the
    blue grid shows the best-fit plane to the mean annual cycle and
    the black arrow the normal to this plane, which we seek to rotate
    into the direction of the $z$-coordinate axis.  Also illustrated
    is the approximate locus of points in the unrotated Isomap
    embedding (cf. Figure~\ref{fig:isomap-raw-3d}a, for instance) and
    the approximate directions of annual (about the cylindrical locus)
    and ENSO variability (along the axis of the cylinder, orthogonal
    to the annual variation).}
  \label{fig:rotation-diagram}
\end{figure}
%

In order to ``unmix'' the annual cycle from the third Isomap
component, we may rotate the whole of the three-dimensional Isomap
embedding to bring the mean annual cycle into the $x$-$y$ coordinate
plane, the hope then being that variations in the rotated Isomap
component orthogonal to the $x$-$y$ plane, i.e. variations along the
$z$-axis, will represent interannual variability, specifically ENSO
variability.  As noted above, the mean annual cycle points
$\bar{\vec{y}}_j$ do not generally lie in a plane, but we may identify
a best-fit plane in a least-squares sense, and rotate this into the
$x$-$y$ plane.  Although not perfect, this will lead to the most
effective unmixing of annual variability from Isomap component \#3.
Figure~\ref{fig:rotation-diagram} provides a schematic illustration to
complement the description here.

We write the equation of the best-fit plane as $\vec{r} \cdot
\vec{\hat{n}} = d$, with $\vec{r} = x \vec{\hat{i}} + y \vec{\hat{j}}
+ z \vec{\hat{k}}$ being the vector position of a point in the plane,
using the usual notation for the unit vectors in the Cartesian
component directions, $\vec{\hat{n}} = l \vec{\hat{i}} + m
\vec{\hat{j}} + n \vec{\hat{k}}$ being a unit normal to the plane, and
$d$ being the distance of the plane from the origin.  The equation of
the plane then becomes $l x + m y + n z = d$, which can be written as
$z = (d - l x - m y) / n$, or $z = \alpha - \beta x - \gamma y$, with
$\alpha = d/n$, $\beta = l/n$, $\gamma = m/n$.  A least-squares fit of
this model to the mean annual cycle points $\bar{\vec{y}}_j$ allows us
to determine values for $\alpha$, $\beta$ and $\gamma$.  A little
analysis shows that this corresponds to solving the equations
\begin{equation}
  \begin{pmatrix}
    12  & -S_x    & -S_y   \\
    S_x & -S_{xx} & -S_{xy} \\
    S_y & -S_{xy} & -S_{yy}
  \end{pmatrix}
  \begin{pmatrix} \alpha \\ \beta \\ \gamma \end{pmatrix} =
  \begin{pmatrix} S_z \\ S_{xz} \\ S_{yz} \end{pmatrix}
\end{equation}
for $\alpha$, $\beta$ and $\gamma$, where the $S_\bullet$ are sums of
components of the $\bar{\vec{y}}_j$, i.e. $S_x$, $S_y$, $S_z$ are the
sums of the $x$, $y$ and $z$ components, $S_{xx}$, $S_{yy}$ are the
sums of the squared $x$ and $y$ components and $S_{xy}$, $S_{xz}$ and
$S_{yz}$ are the sums of the appropriate component products.

Given the values $\alpha$, $\beta$ and $\gamma$, we can calculate
\begin{equation}
  n = (1 + \beta^2 + \gamma^2)^{-1/2},
\end{equation}
and $l=\beta n$, $m = \gamma n$, $d = \alpha n$, and can then
construct the unit normal to the best-fit plane, $\vec{\hat{n}} = l
\vec{\hat{i}} + m \vec{\hat{j}} + n \vec{\hat{k}}$.  We now wish to
find a rotation taking $\vec{\hat{n}}$ into $\vec{\hat{k}}$ (the unit
vector in the $z$-direction), thus rotating the best-fit plane into
the $x$-$y$ plane.

The required rotation may be determined using Rodrigues' rotation
formula, which states that the result of rotating a vector
$\vec{v}$ through an angle $\theta$ about the axis defined by
another vector $\vec{u}$ is
\begin{equation}
  \vec{v}' = \vec{v} \cos \theta + \vec{u} \times \vec{v} \sin \theta
  + \vec{u} (\vec{u} \cdot \vec{v}) (1 - \cos \theta).
\end{equation}
In the case here, we define a suitable rotation axis as $\vec{u} =
\vec{\hat{n}} \times \vec{\hat{k}} / | \vec{\hat{n}} \times
\vec{\hat{k}} |$, and the angle of rotation is $\theta = \cos^{-1}
(\vec{\hat{n}} \cdot \vec{\hat{k}})$ --- this rotation will take
$\vec{\hat{n}}$ into $\vec{\hat{k}}$ by rotating about a direction
orthogonal to both $\vec{\hat{n}}$ and $\vec{\hat{k}}$.  Some simple
algebra yields expressions for the individual rotated components:
\begin{equation}
  \begin{aligned}
    x' &= nx - lz + \frac{mx - ly}{1 + n} m, \\
    y' &= ny - mz - \frac{mx - ly}{1 + n} l, \\
    z' &= nz + my + lx. 
  \end{aligned}
\end{equation}

Note that the rotation determined by Rodrigues' formula is not unique.
There remains an arbitrary phase to the annual cycle associated with
rotations about the $z$-axis.  For our purposes, this non-uniqueness
is of no consequence --- all we require is some rotation that will, as
far as possible, unmix variations associated with the annual cycle
from Isomap component \#3 to reveal the interannual variability.

\begin{figure}
  \begin{center}
    \includegraphics[width=0.48\textwidth]{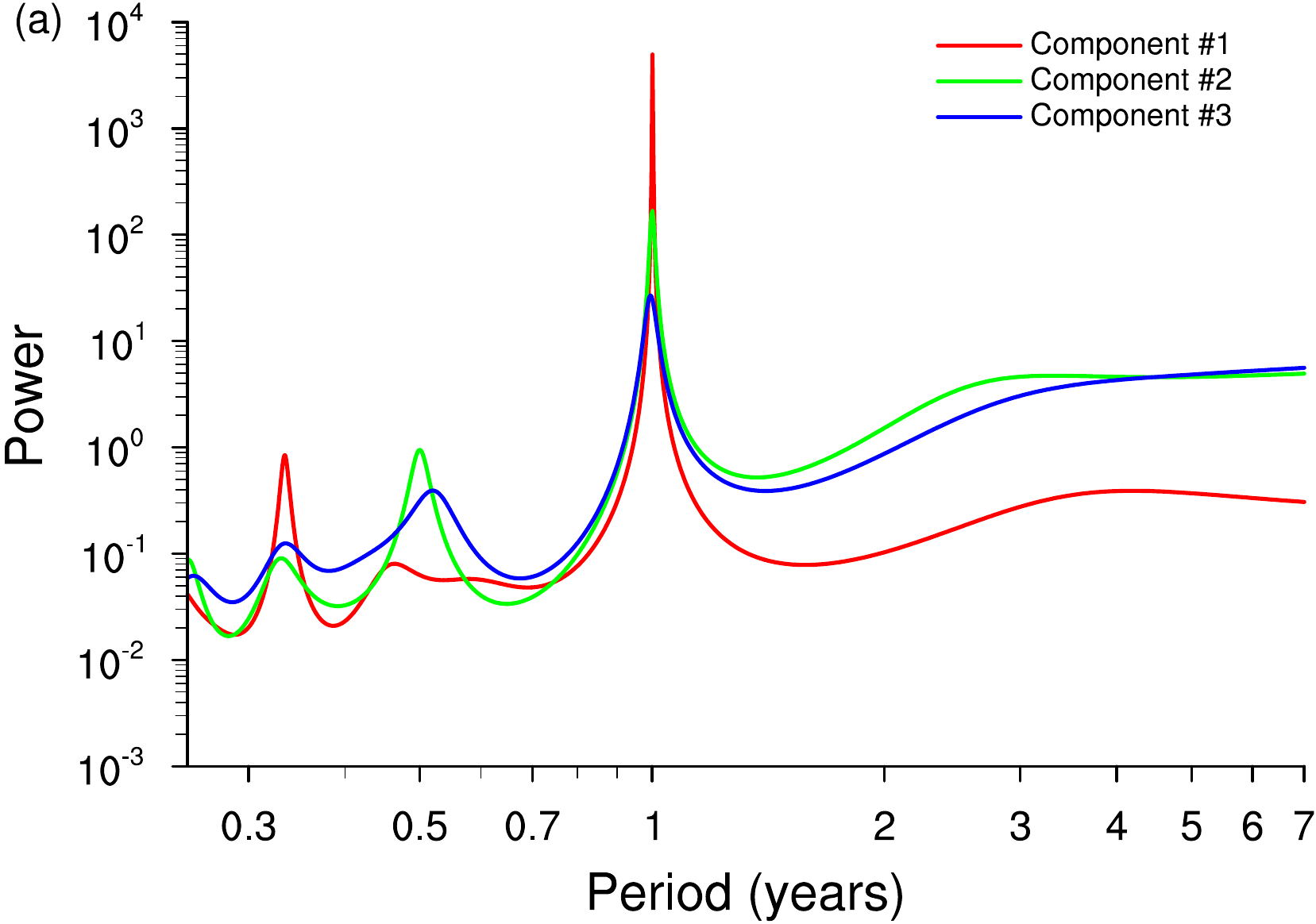}%
    \hspace{0.04\textwidth}%
    \includegraphics[width=0.48\textwidth]{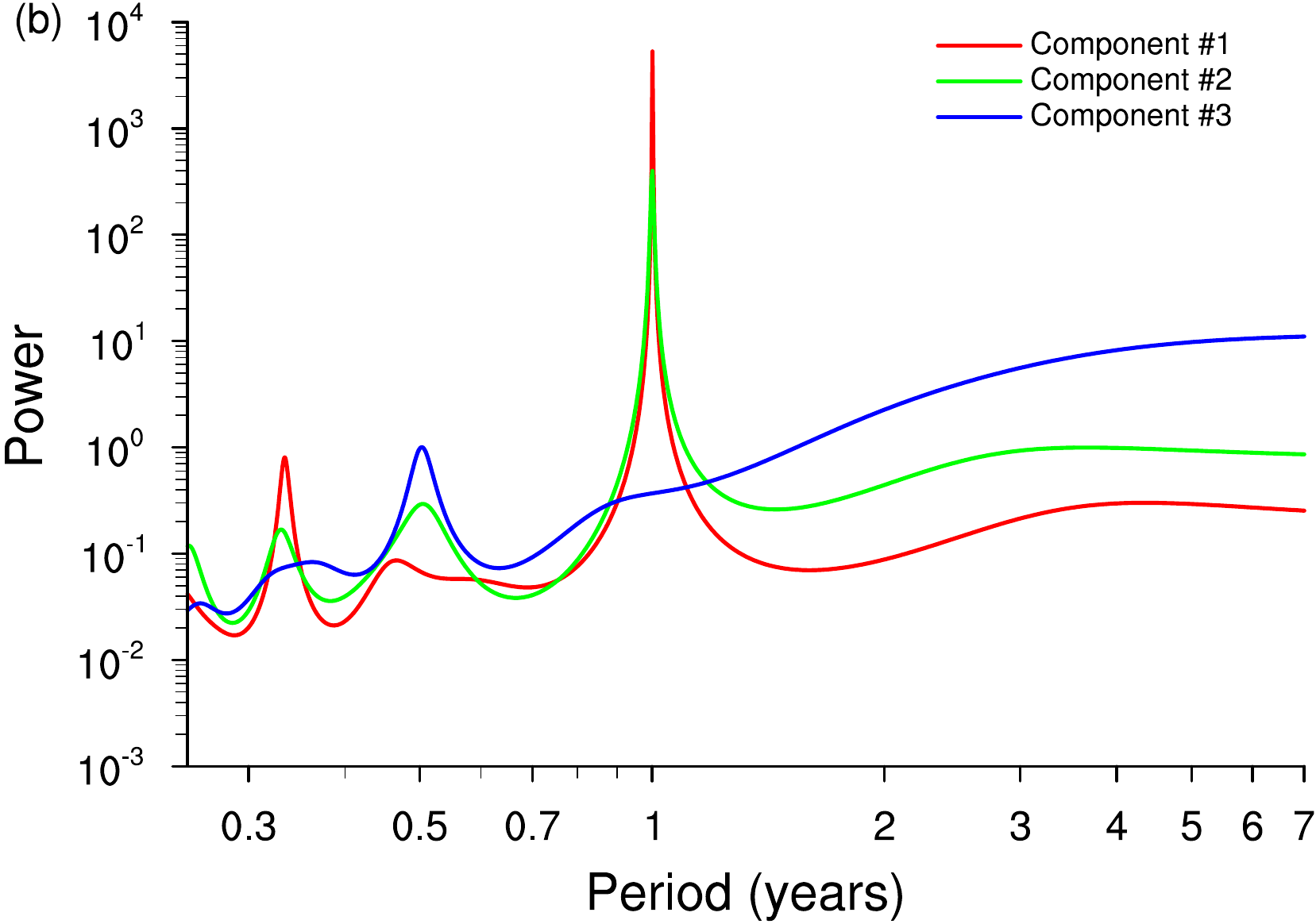}
  \end{center}
  \caption[Isomap component rotation power spectra]{Power spectra for
    Isomap components \#1--3 from a raw SST $k$-Isomap analysis of the
    ERSST v2 observational SST data set, showing the original Isomap
    output (a) and the rotated components (b).}
  \label{fig:rotated-spectra}
\end{figure}

To see that the procedure described does indeed achieve this goal, see
Figure~\ref{fig:rotated-spectra}, where I display power spectra for
Isomap components \#1--3 for $k$-Isomap results ($k=7$) for observed
SSTs.  Figure~\ref{fig:rotated-spectra}a shows spectra for the raw
Isomap components as calculated using \eqref{eq:isomap-mds-coords}.
Here, there is a strong signal at the annual frequency in all three
components.  Figure~\ref{fig:rotated-spectra}b shows spectra for the
rotated Isomap components.  The suppression of the annual signal in
the spectrum of rotated Isomap component \#3 is clear.  Along with the
high correlation between the rotated Isomap component \#3 and the
NINO3 SST index, this indicates that the Isomap algorithm successfully
separates the annual cycle and ENSO variability out of the original
SST field.

%
%
%

\subsection{Four-dimensional case}

The situation for four-dimensional embeddings is significantly more
complicated than the three-dimensional case because of the more
complex structure of the four-dimensional rotation group, SO(4),
compared to SO(3), and the absence of any easy geometrical intuition
in four dimensions.

However, for the purposes of unmixing annual cycle variations from
components \#3 and \#4 of a four-dimensional Isomap embedding, there
are two observations that simplify matters considerably.  (In the
following, I denote the unit vectors in the coordinate directions for
four-dimensional Euclidean space by $( \vec{\hat{e}}_1,
\vec{\hat{e}}_2, \vec{\hat{e}}_3, \vec{\hat{e}}_4 )$.)  The first
observation is that any three-dimensional rotation is also a valid
four-dimensional rotation, i.e. there are proper subgroups of SO(4)
that are isomorphic to SO(3).  If we have a matrix $\mathbf{M}$
representing an element of SO(3), i.e.
\begin{equation}
  \mathbf{M} = \begin{pmatrix}
    m_{11} & m_{12} & m_{13} \\
    m_{21} & m_{22} & m_{23} \\
    m_{31} & m_{32} & m_{33}
  \end{pmatrix}
\end{equation}
with $\mathbf{M}^T \mathbf{M} = \vec{1}$ and $\det \mathbf{M} = 1$,
then we can construct inclusion maps from SO(3) into SO(4) as
\begin{equation}
  \mathbf{M}_3 = \begin{pmatrix}
    m_{11} & m_{12} & m_{13} & 0 \\
    m_{21} & m_{22} & m_{23} & 0 \\
    m_{31} & m_{32} & m_{33} & 0 \\
    0      & 0     & 0      & 1
  \end{pmatrix} \text{ and }
  \mathbf{M}_4 = \begin{pmatrix}
    m_{11} & m_{12} & 0 & m_{13} \\
    m_{21} & m_{22} & 0 & m_{23} \\
    0      & 0     & 1 & 0      \\
    m_{31} & m_{32} & 0 & m_{33} \\
  \end{pmatrix}.
\end{equation}
These matrices represent rotations in the three-dimensional spaces
spanned by $\{ \vec{\hat{e}}_1, \vec{\hat{e}}_2, \vec{\hat{e}}_3 \}$
and $\{ \vec{\hat{e}}_1, \vec{\hat{e}}_2, \vec{\hat{e}}_4 \}$
respectively.

Secondly, since rotations by $\mathbf{M}_3$ do not affect the
$\vec{\hat{e}}_4$ component of any points and rotations by
$\mathbf{M}_4$ do not affect the $\vec{\hat{e}}_3$ components, we can
compose rotations of these two types to unmix the annual variability
from Isomap components \#3 and \#4 independently.  My approach is thus
to use the three-dimensional rotation procedure described in
Section~\ref{sec:isomap-rotation-3d} for each of Isomap components \#3
and \#4 in turn, so as to unmix annual variability from both of these
components.

There is a caveat that should be applied to this procedure.  As in the
three-dimensional case, the rotations I use to unmix the annual
variability from Isomap components \#3 and \#4 are not unique, and
there is still a phase ambiguity present in both of the rotated
components.  Specifically, rotations leaving the
$\vec{\hat{e}}_1$-$\vec{\hat{e}}_2$ plane invariant will not affect
the unmixing of the annual variability from the rotated Isomap
components.  Such rotations, represented by rotation matrices of the
form
\begin{equation}
  \mathbf{M}' = \begin{pmatrix}
    1 & 0 & 0 & 0 \\
    0 & 1 & 0 & 0 \\
    0 & 0 & \cos \phi & -\sin \phi \\
    0 & 0 & \sin \phi & \cos \phi
  \end{pmatrix}
\end{equation}
where $\phi$ is the rotation angle, do not alter the relationship
between the rotated components \#3 and \#4 and the annual cycle
components (\#1 and \#2), but they do alter the relative phasing
between the rotated components \#3 and \#4.  In practice, what this
means is that, if one wishes to identify rotated Isomap components \#3
and \#4 as the ``NINO3'' and ``WWV'' components of ENSO variability,
there is no guarantee that either of the rotated components is purely
one form of ENSO variability or the other.  This makes interpretation
of the correlation results rather difficult.  I have explored a number
of approaches to unmixing the variability of these different degrees
of freedom in this context, but there does not appear to be an easy a
priori way to determine the angle $\phi$ to completely unmix the
components.  One possibility would be to rotate so as to maximise the
correlations between rotated Isomap component \#3 and the NINO3 SST
index and between rotated Isomap component \#4 and the WWV time
series, but this seems to be a rather unsatisfactorily ad hoc
approach.  These difficulties clearly have some bearing on
interpretation of the results on correlations between the rotated
component \#4 and WWV reported in
Section~\ref{sec:isomap-raw-sst-results}.


%% file: 08-hessian-lle.tex
\chapter{Hessian Locally Linear Embedding}
\label{ch:hessian-lle}

\section{Description of method}
\label{sec:hlle-method}

Hessian LLE or Hessian eigenmaps \citep{donoho-hessian} is a
derivative of the locally linear embedding method
(Section~\ref{sec:nonlinear-diff-geom}) that shares much of its
theoretical basis with the method of Laplacian eigenmaps
(Section~\ref{sec:spectral-graph-theory}).  In practical terms, the
computations required by Hessian LLE have more in common with LLE, but
Laplacian eigenmaps provide a clearer framework for understanding how
the method works.  In the same way that Laplacian eigenmaps uses a
graph-based approximation to the Laplace-Beltrami operator on the data
manifold, Hessian LLE relies on a numerical approximation to the
Hessian matrix of a function defined on the data manifold.  The method
leads to a transformation of the original data space coordinates that
is a \emph{local} isometry.  Recall that the Isomap algorithm
(Chapter~\ref{ch:isomap}) produces a \emph{global} isometry of the
original data space, which strongly restricts the manifolds it can
represent faithfully \citep{donoho-isometry}.

Hessian LLE assumes that input points lie on a $p$-dimensional smooth
manifold $M \subset \mathbb{R}^m$.  Tangent spaces $T_x{}M$ can then
be defined at each point $x \in M$ and an orthonormal coordinate basis
can be assigned to each $T_x{}M$ by thinking of the tangent space as
an affine subspace of $\mathbb{R}^m$, tangent to $M$ at $x$ and with
the origin $0 \in T_x{}M$ identified with $x$.  Then, there is a
neighbourhood $\mathcal{N}(x) \subset M$ of $x$ such that each point
$x' \in \mathcal{N}(x)$ has a ``nearest'' point $v' \in T_x{}M$
(thought of as this affine subspace) and the implied mapping $x'
\mapsto v'$ is smooth.  The point $v' \in T_x{}M$ has local
coordinates defined by the choice of orthonormal coordinates for the
tangent space $T_x{}M$, from which we can obtain local coordinates for
the neighbourhood $\mathcal{N}(x)$, which we call tangent coordinates
and denote by $\xi_1^{(\mathrm{tan}, x)}, \dots, \xi_p^{(\mathrm{tan},
  x)}$.

Now consider a function $f : M \to \mathbb{R}$ with $f \in C^2(M)$.
If the point $x' \in \mathcal{N}(x)$ has tangent coordinates
$\xi^{(\mathrm{tan}, x)}$, then we can write $g(\xi^{(\mathrm{tan},
  x)}) = f(x')$ to define a function $g : U \to \mathbb{R}$, defined
on a neighbourhood $U$ of zero in $\mathbb{R}^p$ (which is really the
affine subspace we are identifying with $T_x{}M$).  The map $x'
\mapsto \xi^{(\mathrm{tan}, x)}$ is smooth and so $g \in C^2(U)$.  We
can then define the \emph{tangent Hessian} of $f$ at the point $x \in
M$ to be the normal Hessian matrix of $g$ as
\begin{equation}
  \left( \mathbf{H}_f^{(\mathrm{tan})}(x) \right)_{ij} =
  \left. \frac{\partial^2 g(\xi^{(\mathrm{tan}, x)})}{\partial
    \xi_i^{(\mathrm{tan}, x)} \partial \xi_j^{(\mathrm{tan}, x)}}
  \right|_{\xi^{(\mathrm{tan}, x)}=0}.
\end{equation}
This definition clearly depends on the choice of coordinates for the
tangent space $T_x{}M$.  If we choose a different coordinate basis,
then the Hessian in the new coordinate system, $\mathbf{H}'$ is
related to that in the original coordinates, $\mathbf{H}$, by
$\mathbf{H}' = \mathbf{B} \mathbf{H} \mathbf{B}^T$, where $\mathbf{B}$
is the orthonormal matrix transforming between the two coordinate
systems.  However, we can extract a coordinate invariant quantity by
considering the Frobenius norm of the Hessian, $||\mathbf{H}||_F$,
since
\begin{equation}
  || \mathbf{H}' ||_F^2 = || \mathbf{B} \mathbf{H} \mathbf{B}^T ||_F^2
  = \Tr ( \mathbf{B} \mathbf{H}^T \mathbf{B}^T \mathbf{B} \mathbf{H}
  \mathbf{B}^T ) = \Tr ( \mathbf{H}^T \mathbf{H} ) = || \mathbf{H}
  ||_F^2,
\end{equation}
where we have exploited the permutation property of the trace operator
and the orthogonality of the matrix $\mathbf{B}$.  The Frobenius norm
of the Hessian of $f$ is thus invariant under changes of coordinates
for the tangent spaces of points in the manifold $M$, allowing us to
define a functional
\begin{equation}
  \mathcal{H}(f) = \int_M || \mathbf{H}^{(\mathrm{tan})}_f(p) ||_F^2
  \, d\mu[x]
\end{equation}
based on integrating the Frobenius norm of the Hessian of $f$,
$\mathbf{H}_f$ over the manifold $M$.  Here, $d\mu[x]$ represents a
probability measure over $M$, which must be strictly positive in the
interior of $M$.  Even though the Hessian itself depends on the choice
of basis for $T_x{}M$, because the Frobenius norm of the Hessian does
not, the functional $\mathcal{H}(f)$ is well-defined and independent
of the choice of local coordinates.

The key idea in Hessian LLE, represented by a theorem proved by
\citet{donoho-hessian}, is that the null space of $\mathcal{H}(f)$ is
related to the existence of a locally isometric embedding of the
manifold $M$ in Euclidean space.  Specifically, suppose that $M =
\psi(\Theta)$, where $\Theta$ is an open connected subset of
$\mathbb{R}^p$ and $\psi$ is a locally isometric embedding of $\Theta$
into $\mathbb{R}^m$.  Then $\mathcal{H}(f)$ has a $(p+1)$-dimensional
null space, consisting of the constant function on $M$ and a
$p$-dimensional space of functions spanned by the original isometric
coordinates.  Following from this result, the original isometric
coordinates can be identified, up to a rigid rotation and translation,
by identifying a suitable basis for the null space of
$\mathcal{H}(f)$.

The main theoretical benefit of Hessian LLE over the closely related
Laplacian eigenmaps method is that the Hessian is a better determiner
of linearity.  For a function $f$, $\mathbf{H}_f = 0$ if and only if
$f$ is linear, while the condition $\Delta f = 0$ is much weaker,
being satisfied by any harmonic function on $M$.  To see this,
consider the case in $\mathbb{R}^2$, where the Laplace-Beltrami
operator is just the usual Laplacian $\Delta = \nabla^2 = \partial^2 /
\partial x^2 + \partial^2 / \partial y^2$.  For any analytic function
of a complex variable $f(x + iy) = u(x, y) + i v(x, y)$, the
Cauchy-Riemann conditions ($\partial u / \partial x = \partial v /
\partial y$, $\partial u / \partial y = -\partial v / \partial x$)
imply that both the real and imaginary parts of $f$, $u(x, y)$ and
$v(x, y)$ satisfy Laplace's equation $\nabla^2 u = 0$, $\nabla^2 v =
0$.  We can thus select any analytic complex function, say $f(z) = e^z
= e^x \cos y + i e^x \sin y$, and both the real and imaginary parts of
$f$ are harmonic functions, although they are clearly not linear.  On
the other hand, the Hessian matrix for either of these functions is
clearly non-zero, e.g.
\begin{equation}
  \mathbf{H}(e^x \cos y) = \begin{pmatrix} e^x \cos y & - e^x \sin y
    \\ - e^x \sin y & - e^x \cos y
  \end{pmatrix}.
\end{equation}

Numerically, the Hessian LLE algorithm is implemented as follows.  We
start with input data points $\vec{x}_i \in \mathbb{R}^m$, with $i =
1, \dots, N$, which we assume are sampled from a smooth manifold $M$.
For each point $\vec{x}_i$, we first identify the $k$ nearest
neighbours, denoted by $\mathcal{N}(\vec{x}_i)$.  These nearest
neighbour point sets are then centred with respect to the mean point
for each neighbourhood, $\bar{\vec{x}}_i = \langle \vec{x}_j
\rangle_{j \in \mathcal{N}(\vec{x}_i)}$, and we form a matrix
$\mathbf{X}_i$ whose rows are the centred data points, $\vec{x}_j -
\bar{\vec{x}}_i$ for $j \in \mathcal{N}(\vec{x}_i)$.  An approximate
$p$-dimensional coordinate basis for the tangent space to the manifold
$M$ at point $\vec{x}_i$ is then found via a singular value
decomposition of the matrix $\mathbf{X}_i$: $\mathbf{X}_i =
\mathbf{U}_i \mathbf{D}_i \mathbf{V}_i^T$.  The matrix $\mathbf{U}_i$
is $k \times \min(N, k)$ and the first $d$ columns of $\mathbf{U}_i$
give the tangent coordinates of points in $\mathcal{N}(\vec{x}_i)$ (if
the rank of $\mathbf{U}_i$ is less than $d$, this indicates that a
$d$-dimensional reduction of the input data is not possible).

An estimator for the Hessian $\mathbf{H}^{(\mathrm{tan})}_f(x)$ can
then be developed from these tangent coordinates using a least-squares
estimate in the neighbourhood of each data point.  An approximation to
the functional $\mathcal{H}(f)$ can be constructed from these
estimates for the Hessian, as a symmetric matrix $\mathbf{\tilde{H}}$.
The details of this procedure are covered by \citet{donoho-hessian},
and I closely follow their approach.  An eigendecomposition of the
matrix $\mathbf{\tilde{H}}$ allows an approximation to the null space
of $\mathcal{H}(f)$ to be identified.  To derive a $p$-dimensional
embedding of the input data based on this eigendecomposition, we
identify the subspace spanned by the smallest $p+1$ eigenvalues.
There will be one zero eigenvalue associated with the subspace of
constant functions on $M$, and we construct a basis for our locally
isometric embedding of $M$ from the next $p$ eigenvectors.  We write
the eigendecomposition of $\mathbf{\tilde{H}}$ as
\begin{equation}
  \label{eq:hlle-eigendecomp}
  \mathbf{\tilde{H}} = \mathbf{Q} \bm{\Lambda} \mathbf{Q}^T,
\end{equation}
where $\bm{\Lambda} = \diag(\lambda_1, \dots, \lambda_N)$ is a
diagonal matrix containing the eigenvalues of $\mathbf{\tilde{H}}$,
and $\mathbf{Q}$ is a matrix with the eigenvectors of
$\mathbf{\tilde{H}}$ as its columns.  We then calculate embedding
coordinates for a $p$-dimensional embedding of the original data
points from the $N \times p$ matrix, $\mathbf{Q}_{p\backslash 0}$, the
matrix whose columns are the $p$ nonconstant eigenvectors associated
with the $p+1$ smallest eigenvalues of $\mathbf{\tilde{H}}$, as
\begin{equation}
  \mathbf{Y} = \sqrt{N} \mathbf{Q}_{p\backslash 0}.
\end{equation}
Here $\mathbf{Y}$ is an $N \times p$ matrix of $p$-dimensional
embedding coordinates for the $N$ input points.  This is similar to
the final MDS-based embedding calculation for Isomap (represented by
\eqref{eq:isomap-mds-coords} on page~\pageref{eq:isomap-mds-coords}).
Note that this is somewhat different to the procedure of
\citet{donoho-hessian}.  The method for calculating the final
embedding coordinates described there does not appear to work as well
as this simple approach.  In general, the smaller eigenvalues of the
Hessian are associated with variations in the input data at larger
spatial scales, in just the same way that, for instance, smaller
eigenvalues of the Laplacian are associated with longer wavelength
disturbances.

Hessian LLE has, as far as I am aware, not been used in any
applications to date, and \citet{donoho-hessian} present results only
for simple geometrical data.  The reasons for the lack of uptake of
the method are not clear, but may be related to the greater complexity
of implementation compared to some other methods and to possible
numerical issues relating to the calculation of approximations to the
Hessian, which essentially require the calculation of second
differences between data points.

\section{Application to test data sets}
\label{sec:hlle-test-data}

\begin{figure}
  \begin{center}
    \begin{tabular}{cc}
      \includegraphics[width=0.49\textwidth]{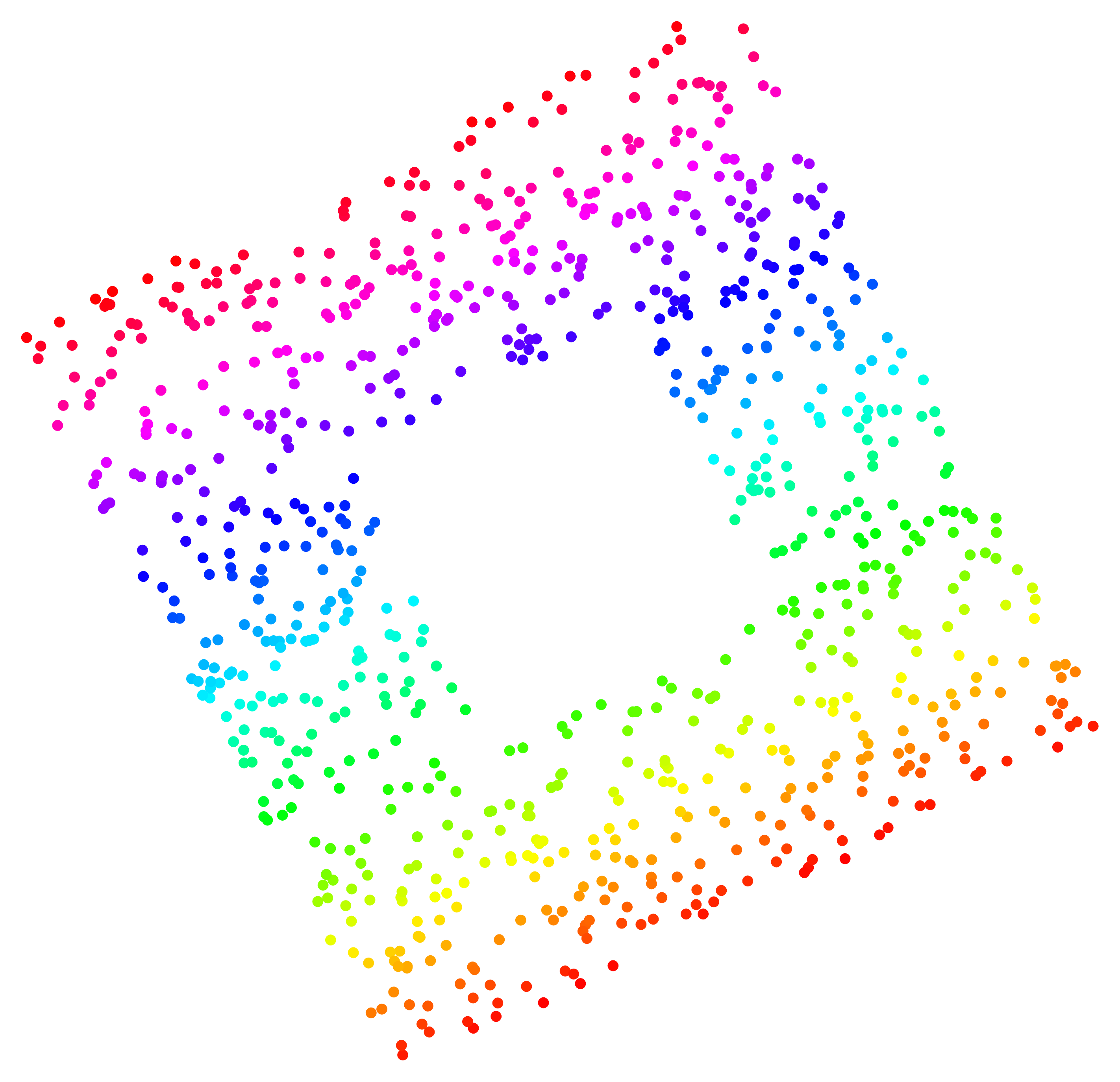} &
      \includegraphics[width=0.49\textwidth]{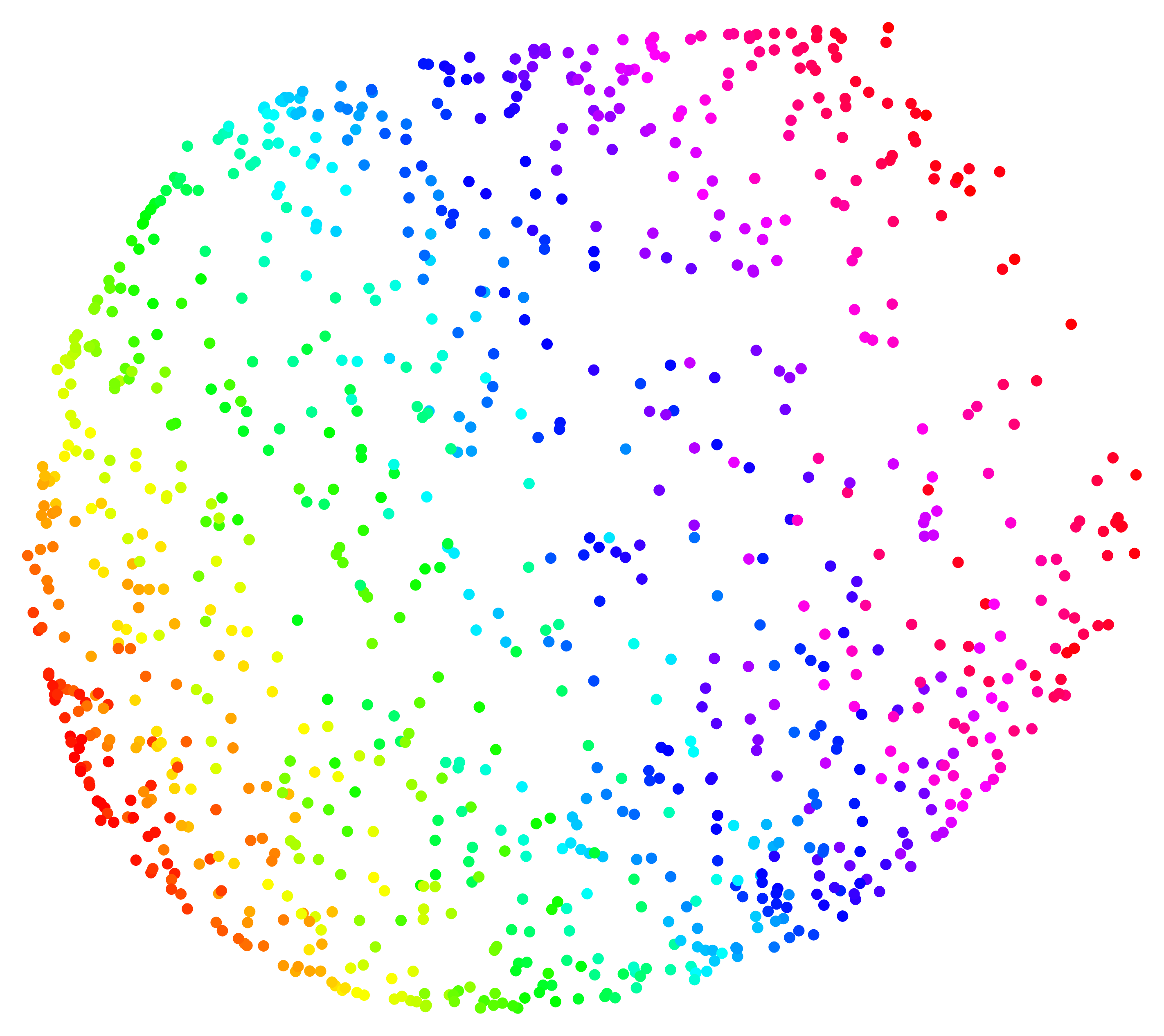} \\
      (a). Plane with hole ($k = 9$ & (b). Fishbowl ($k = 7$) \\
      \\

      \includegraphics[width=0.49\textwidth]{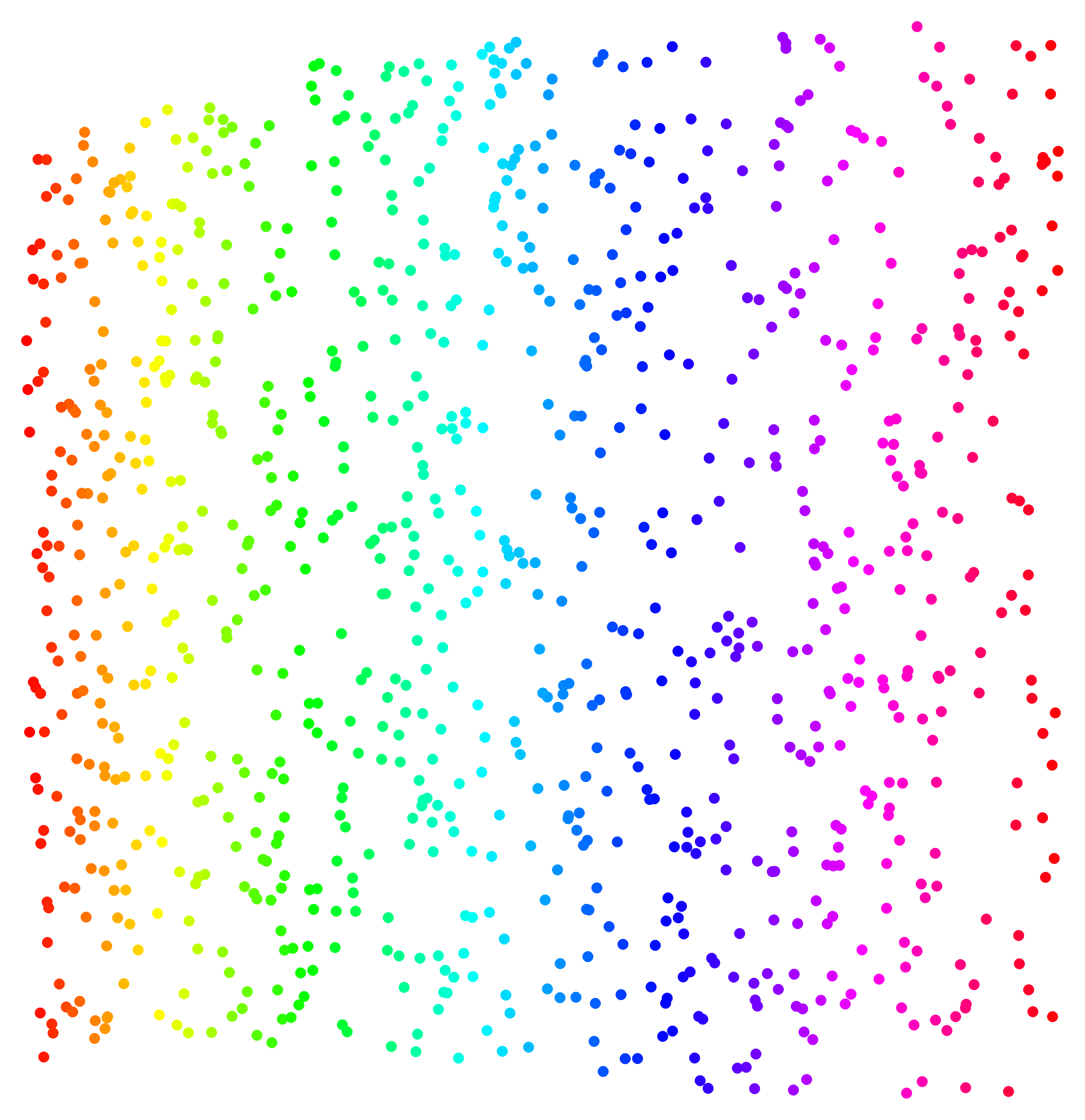} &
      \includegraphics[width=0.49\textwidth]{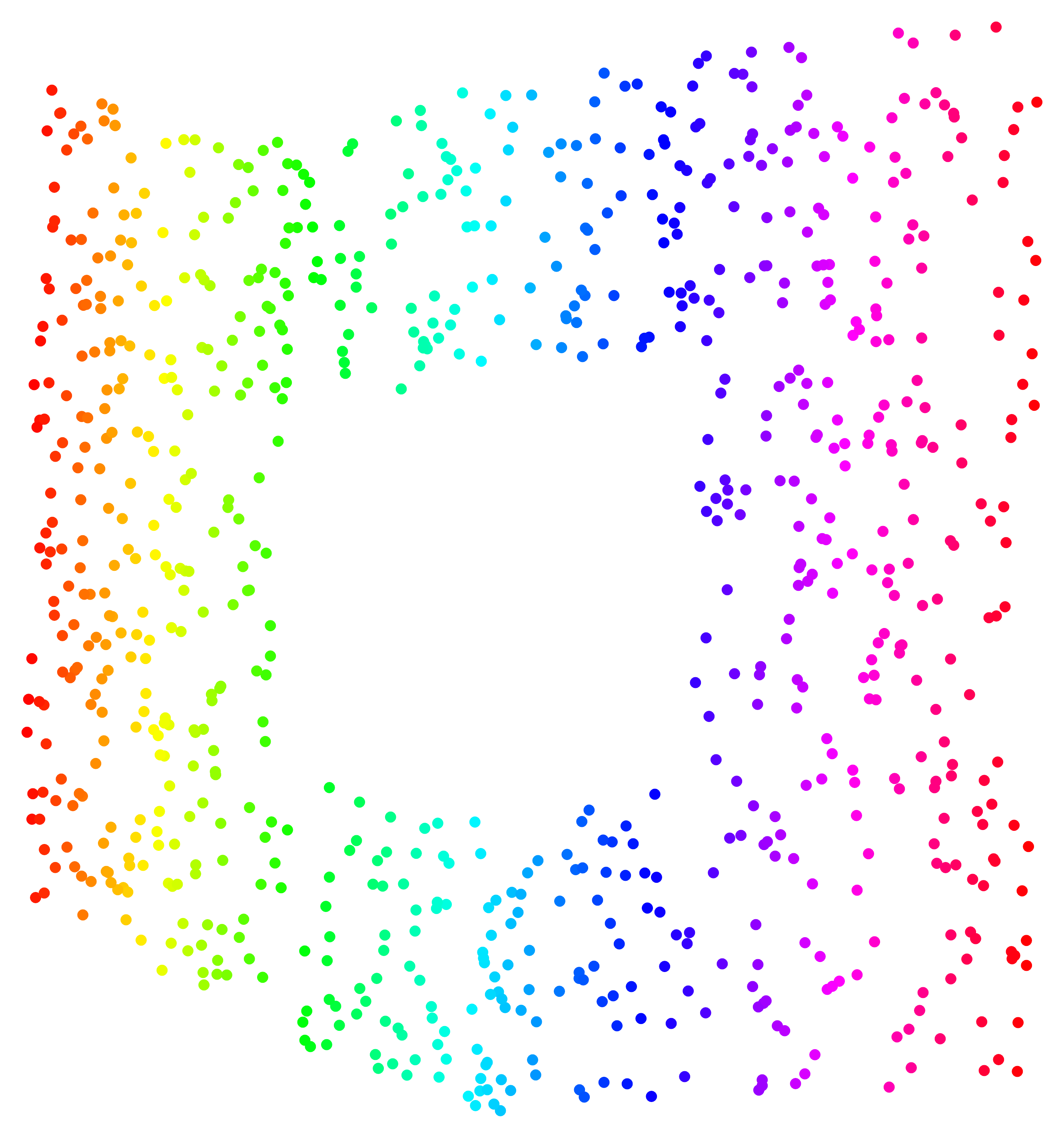} \\
      (c). Swiss roll ($k = 7$) & (d). Swiss roll with hole ($k = 13$)
    \end{tabular}
  \end{center}
  \caption[Hessian LLE reductions of geometrical test data
    sets]{Application of Hessian LLE to geometrical test data sets
    from Section~\ref{sec:test-data-sets}.}
  \label{fig:hlle-test-data-sets}
\end{figure}

As was done for NLPCA (Section~\ref{sec:nlpca-test-data}) and Isomap
(Section~\ref{sec:isomap-test-data}), here we examine the results of
applying Hessian LLE to the simple geometrical test data sets
described in Section~\ref{sec:test-data-sets}.
Figure~\ref{fig:hlle-test-data-sets} shows reduced representations
produced by the Hessian LLE algorithm for four of the test data sets.
Results for the other data sets without noise display similar
characteristics to these examples.  The issue of noise contamination
will be dealt with below.  For each of the data sets shown in
Figure~\ref{fig:hlle-test-data-sets}, the best result is shown for
values of $7 \leq k \leq 50$ ($k$ being the neighbourhood size used
for the computation of tangent space coordinates in the neighbourhood
of each data point).

%
%
%

In general, the Hessian LLE method appears to work well for the
simpler test data sets, but it is quite sensitive to the choice of
$k$.  For the plane with hole (Figure~\ref{fig:hlle-test-data-sets}a),
embeddings produced by Hessian LLE are reasonable for all values of
$k$, in the sense that the reduced representation of the data set is
geometrically identical to the original shape of the data, apart from
a simple shear transformation.  The results shown in
Figure~\ref{fig:hlle-test-data-sets}a, for $k = 9$, have the lowest
such distortion of all the computations examined.  The dependence of
the degree to which the reduced coordinates are distorted by shearing
is very sensitive to the value of $k$, with large changes being
observed between adjacent $k$ values.  By contrast, no good embeddings
are found for the fishbowl example
(Figure~\ref{fig:hlle-test-data-sets}b).  The best result, for $k =
7$, is close to a simple linear projection along the axis of the
fishbowl, and does not succeed in ``unwrapping'' the data manifold at
all.  This result is not too surprising, since the surface of a sphere
is \emph{not} locally isometric to an open subset of the plane,
meaning that Hessian LLE, which is restricted to finding local
isometries of the input data, is unlikely to find a good reduction of
this data set.  In this case, more general transformations of the
input data are required to generate a good embedding, exemplified by
the results of NLPCA (Figure~\ref{fig:nlpca-test-data-sets-reduce}b on
page~\pageref{fig:nlpca-test-data-sets-reduce}), or by the conformal
generalisation of Isomap of \citet{desilva-global-local}.  For the
Swiss roll data sets, good embeddings are found, although again there
is strong dependence on the value of $k$ used in the Hessian LLE
calculations.
Figures~\ref{fig:hlle-test-data-sets}c~and~\ref{fig:hlle-test-data-sets}d
show the best results for the Swiss roll ($k = 7$) and Swiss roll with
hole ($k = 13$).  Good embeddings are found only for relatively few
values of $k$, with much distortion and degeneracy (i.e. embeddings
that map all data points to one or a few points in the reduced space)
for other values.  The good embeddings that are found are notable for
their lack of distortion around non-convex features in the data sets.
Compare Figure~\ref{fig:hlle-test-data-sets}d with
Figure~\ref{fig:isomap-test-data-sets}d on
page~\pageref{fig:isomap-test-data-sets}, which shows the embedding
produced by Isomap for the Swiss roll with hole.  Because the Isomap
algorithm finds only global isometries mapping the original data to an
open \emph{convex} subset of Euclidean space, there is significant
distortion near the hole in the data manifold, which does not arise in
the locally isometric Hessian LLE embedding.

Addition of noise to the test data sets appears to cause problems for
Hessian LLE, almost certainly because of the finite differencing
aspect of the calculation of the approximation to the tangent Hessian.
All finite differencing and numerical differentiation approaches are
somewhat susceptible to problems with noisy data, but the calculation
here is particularly problematic because the Hessian involves second
differences between the data points.

\section{Hessian LLE sensitivity}
\label{sec:hlle-sens}

The Hessian LLE algorithm has a single tunable parameter, the
neighbourhood size $k$ used in the initial assignment of tangent
coordinates in the neighbourhood of each data point.  However, there
is another factor to which Hessian LLE displays more sensitivity than
the other dimensionality reduction methods examined in this thesis.
This is the data point sampling density.  For all of the
two-dimensional geometrical test data sets used here, the results
shown in Sections~\ref{sec:nlpca-test-data} (NLPCA),
\ref{sec:isomap-test-data} (Isomap) and \ref{sec:hlle-test-data}
(Hessian LLE) are based on data sets of 1000 points sampled from the
test manifolds.  For the NLPCA and Isomap methods, this number of data
points appears to provide sufficient resolution for the methods to
produce reasonable embeddings.  For Hessian LLE, however, results
obtained from the algorithm are strongly dependent on the data
sampling density and 1000 points does not appear to be enough data to
guarantee good embeddings.

\begin{sidewaysfigure}
  \begin{center}
    \includegraphics[width=\textwidth]{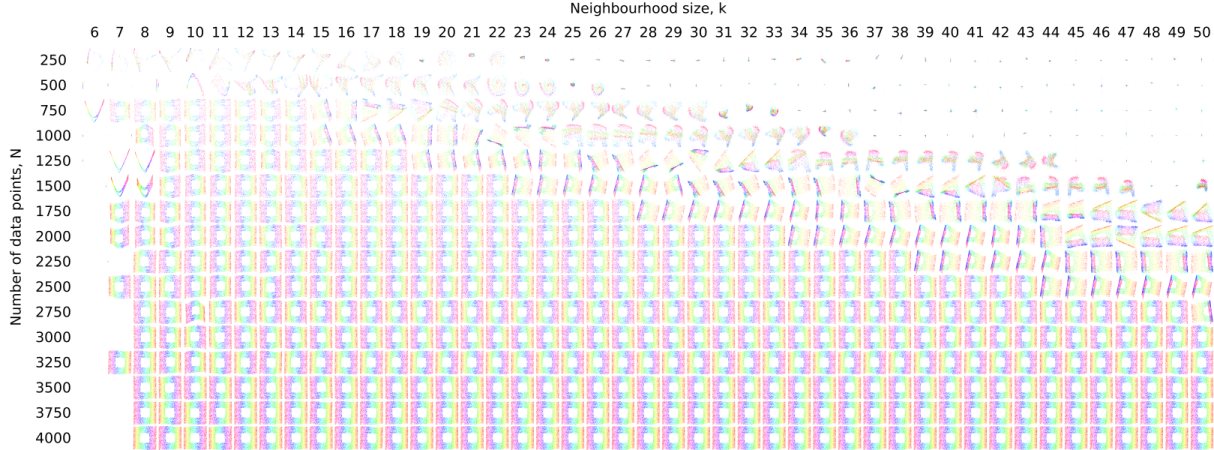}
  \end{center}
  \caption[Hessian LLE sensitivity (no noise)]{Hessian LLE embeddings
    of Swiss roll with hole data sets for varying numbers of data
    points, $N$, and varying neighbourhood size, $k$.  (Blank spaces
    indicate combinations of sampling density and neighbourhood size
    for which no embedding exists because of an insufficient number of
    singular vectors in the computation of tangent coordinates.)}
  \label{fig:hlle-sens-clean}
\end{sidewaysfigure}

\begin{sidewaysfigure}
  \begin{center}
    \includegraphics[width=\textwidth]{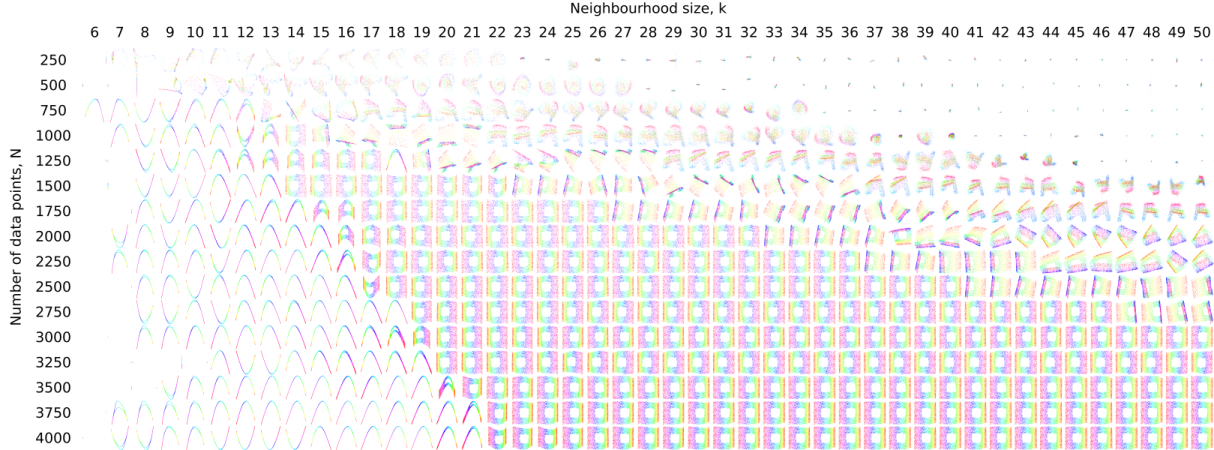}
  \end{center}
  \caption[Hessian LLE sensitivity (with noise)]{Hessian LLE
    embeddings of Swiss roll with hole data sets with added noise, for
    varying numbers of data points, $N$, and varying neighbourhood
    size, $k$.  (Blank spaces indicate combinations of sampling
    density and neighbourhood size for which no embedding exists
    because of an insufficient number of singular vectors in the
    computation of tangent coordinates.)}
  \label{fig:hlle-sens-noise}
\end{sidewaysfigure}

To explore this effect, and also to illustrate the effect of noise on
the Hessian LLE algorithm, I used Swiss roll with hole data sets with
different numbers of data points, ranging from 250 to 4000 points.
For each of these data sets, Hessian LLE embeddings were calculated
for $6 \leq k \leq 50$.  The results are shown in
Figures~\ref{fig:hlle-sens-clean} (no noise) and
\ref{fig:hlle-sens-noise} (added noise).

%
%
%

%
%
%

Consider first the results for data without noise, shown in
Figure~\ref{fig:hlle-sens-clean}.  We see that good embeddings are
obtained for all values of $k$ as long as the data point sampling
density is sufficiently high.  Here, the threshold sampling density is
given by $N \geq 2750$.  For small data point sampling densities, $N
\leq 500$, there are no values of $k$ for which Hessian LLE produces a
good embedding.  In these cases, embeddings for smaller $k$ are highly
distorted while those for larger $k$ are degenerate, i.e. all data
points are mapped to a single point in the reduced dimensionality
space.  For intermediate values of the data point sampling density,
$750 \leq N \leq 2500$, there is a maximum $k$ value below which good
embeddings are seen.  This maximum $k$ value increases more or less
linearly from $k = 14$ when $N = 750$ to $k = 43$ when $N = 2500$.  As
$k$ is increased above the threshold value, embeddings show increasing
distortion and eventually become degenerate.

The Hessian LLE approach thus appears to work for a wide range of
values of the neighbourhood size parameter $k$, but only so long as
the data point sampling density is sufficiently high (and there is no
noise in the data, as we will see below).  The upper limit on the
value of $k$ to get a good embedding is imposed by the requirement
that the neighbourhoods $\mathcal{N}(\vec{x}_i)$ should represent
``locally linear'' subsets of the data manifold.  When $k$ becomes
larger than the threshold seen here, the neighbourhoods start to cover
regions of the data manifold that show significant curvature at the
scale of the mean distance between data points.  Tangent coordinates
derived from the local singular value decomposition procedure become
increasingly distorted as the neighbourhoods increase in size and
encompass more curved regions of the data manifold.  The dependence of
the maximum $k$ value on the data sampling density is a simple result
of the fact that the neighbourhood size grows linearly with $k$ but
inversely with $N$ --- the more data points there are, the closer
together they lie both in the original data space and in the
lower-dimensional data manifold, and a smaller fraction of the total
data manifold is encompassed by any given neighbourhood size, based on
a simple count of neighbours.  At the lowest data point sampling
densities examined here, the Hessian LLE method does not produce good
embeddings for any choice of $k$.  This is unfortunate, particularly
since these data sampling densities are easily handled by other
methods (including the NLPCA and Isomap methods from
Chapters~\ref{ch:nlpca}~and~\ref{ch:isomap}).  This failure would
appear to be attributable to two main effects.  First, the same
dependence of the size of ``locally linear'' neighbourhoods on $k$
extends to the lower data point sampling densities, meaning that it
may become impossible to select a value of $k$ sufficiently small to
represent approximately linear subsets of the data manifold.  Second,
even if a value of $k$ does exist that meets the condition for the
neighbourhoods used to be approximately linear, sampling effects may
come into play when the data point sampling density is low.  Because
data points are distributed randomly across the data manifold, there
will be local fluctuations in the sampling density.  These
fluctuations in local sampling density will be more significant at low
overall data point sampling density because neighbourhoods are defined
in terms of a nearest neighbour count, resulting in larger,
essentially random, variations in the directions of the approximate
tangent spaces identified by the local singular value decomposition
step of the Hessian LLE algorithm.  This makes it impossible to arrive
at a good consistent estimate of the Hessian functional
$\mathcal{H}(f)$ over the data manifold.

In the presence of noise in the input data
(Figure~\ref{fig:hlle-sens-noise}), the same effects in sensitivity to
variations of data sampling density and neighbourhood size are seen as
in the noise-free case.  For instance, there is an upper limit to the
value of $k$ giving a good embedding for each data sampling density,
again because of the failure of local linearity for larger
neighbourhoods.  However, there are some differences from the
noise-free case.  First, the lower limit of the data sampling density
for good embeddings is higher.  No good embeddings are seen for the
noisy data for $N < 1250$, while good embeddings are seen in the
noise-free case for $N = 750$.  Second, there is also a \emph{lower}
limit of $k$ below which no good embeddings are seen.  This limit also
depends on the data sampling density.  For $N = 1500$, the lowest
neighbourhood size for which a good embedding is produced is $k = 14$,
while for $N = 4000$, no good embeddings are seen for $k < 22$.  For
values of $k$ below the lower limit, the Hessian LLE procedure
identifies a one-dimensional manifold, rather than the true
two-dimensional data manifold.  The reason for this lower
neighbourhood size limit is that, for smaller values of $k$, the
neighbourhoods $\mathcal{N}(\vec{x}_i)$ essentially sample only the
noise variability in the input data, and do not capture any of the
structure of the data manifold.  The result is that the local singular
value decomposition step is not able to identify valid local tangent
coordinates.  Interestingly, the hue assignments shown in the
one-dimensional manifolds appearing in these results are consistent,
i.e. hues vary smoothly from one end of the embedded manifold to the
other.  The one-dimensional manifolds recovered from the Hessian LLE
procedure appear to capture at least some aspect of the intrinsic
geometry of the input data, albeit very crudely.  The presence of
noise in the data and the consequent problems with the local singular
value decomposition step used to identify tangent coordinates for the
data manifold conspire to prevent the Hessian LLE algorithm from
identifying any other directions in the data manifold than that
corresponding to the greatest data variance in intrinsic coordinates.
It is not quite clear why this should be, but it appears to be a
consistent feature of the embeddings produced for noisy data.

\section{Application to analysis of Pacific SSTs}
\label{sec:hlle-sst-results}

It is clear that applying Hessian LLE directly to the original
tropical Pacific SST data is unlikely to be successful.  The
dimensionality of the input data in this case is given by the values
of $m$ in Table~\ref{tab:models} on page~\pageref{tab:models}; for all
the CMIP3 models $m \sim O(1000)$, and for the observational SST data,
$m = 1626$.  In order to estimate tangent coordinates via local
singular value decomposition, we need to use a neighbourhood size $k$
that is at least as large as the number of dimensions in the input
data space.  Realistically, $k$ must be significantly larger than $m$
--- for the Swiss roll with hole examples shown in the previous
section, $m = 3$ and good embeddings were found only for $k \gtrsim$
14--22 (depending on the size of the data set).

We thus need to use an initial dimensionality reduction step, much as
was done for the NLPCA method in Chapter~\ref{ch:nlpca}.  As there, we
use PCA to determine the first ten EOFs and principal component time
series, and use the PC time series values as input to the Hessian LLE
method.  We do not follow the normal PCA procedure, based on SST
anomalies, but instead apply PCA to the raw SST data.  This means that
the principal component time series used as input to the Hessian LLE
procedure include both ENSO variability and the annual cycle.  This
will allow us to make some direct comparisons between the Hessian LLE
results and the results obtained from the application of Isomap to raw
SST data reported in Section~\ref{sec:isomap-raw-sst-results}.  (In
fact, we use essentially identical plots to Chapter~\ref{ch:isomap} to
examine eigenvalue sensitivity and three-dimensional embeddings for
Hessian LLE.)  Using this approach, we hope to be able to find
low-dimensional embeddings of the tropical Pacific SST data.  Even
after this linear projection to the space spanned by the first ten
principal components, in all cases, both model results and
observations, the SST data is still much noisier than any of the test
data examples examined in the previous section.  In fact, the noise
reduction resulting from the application of PCA to the SST data is
relatively modest.  For example, for the observational SST data, the
first 10 EOFs already account for 96.5\% of the total variance in the
input data.  For the model data, the explained variance attributable
to the 10 leading EOFs is comparable (the smallest value is 84.5\% for
UKMO-HadCM3).  The relatively high level of noise even in the leading
EOFs can be seen easily from consideration of the principal component
scatter plots shown in Figure~\ref{fig:sst-pc-scatter} on
page~\pageref{fig:sst-pc-scatter} and the related scatter plots used
to display NLPCA results in Chapter~\ref{ch:nlpca}.  This level of
noise may pose some problems for the Hessian LLE procedure.

We can get some idea of the range of values of the neighbourhood size
$k$ that might allow the Hessian LLE method to produce reasonable
embeddings of the Pacific SST data by comparison with the sensitivity
results from Section~\ref{sec:hlle-sens}.  These results also allow us
to make some observations concerning data point sampling density
issues for the SST data.  Clearly, comparison with results for simple
geometrical test data sets will give optimistic bounds on reasonable
values of $k$, since the SST data is noisier and is also of higher
dimension than the Swiss roll-type geometrical test data ($m = 10$ for
the SST data compared to $m = 3$ for the geometrical test data).

First, for the observational data, we have only 1200 data points
(monthly data over the period 1900--1999).  Even in the simple
geometrical test data cases, this appears to be a marginal number of
points to find good embeddings: Figure~\ref{fig:hlle-sens-noise} shows
that, for $N = 1250$, good two-dimensional embeddings of the noisy
Swiss roll with hole data are only found for a few values of the
neighbourhood size $k$.  For the significantly noisier observational
SST data, it seems extremely unlikely that we can find good
embeddings.  For the model results, we have more data points, as many
as 6000 for some simulations.  We can get some idea of a reasonable
(although optimistic) range of $k$ values to use by extrapolating the
upper and lower boundaries of the region of good embeddings from
Figure~\ref{fig:hlle-sens-noise}.  Linearly extrapolating the lower
bound on $k$ gives the condition that, for 6000 data points, good
embeddings might be obtained for $k \gtrsim 28$.  Similarly
extrapolating the upper bound of $k$ for good embeddings leads to the
condition $k \lesssim 110$.  It must be emphasised again that these
are very rough indicative limits and are may be a gross overestimate
of the range of neighbourhood sizes for which good embeddings may be
obtained for the noisier SST data.

\begin{figure}
  \begin{center}
    \includegraphics[width=0.48\textwidth]{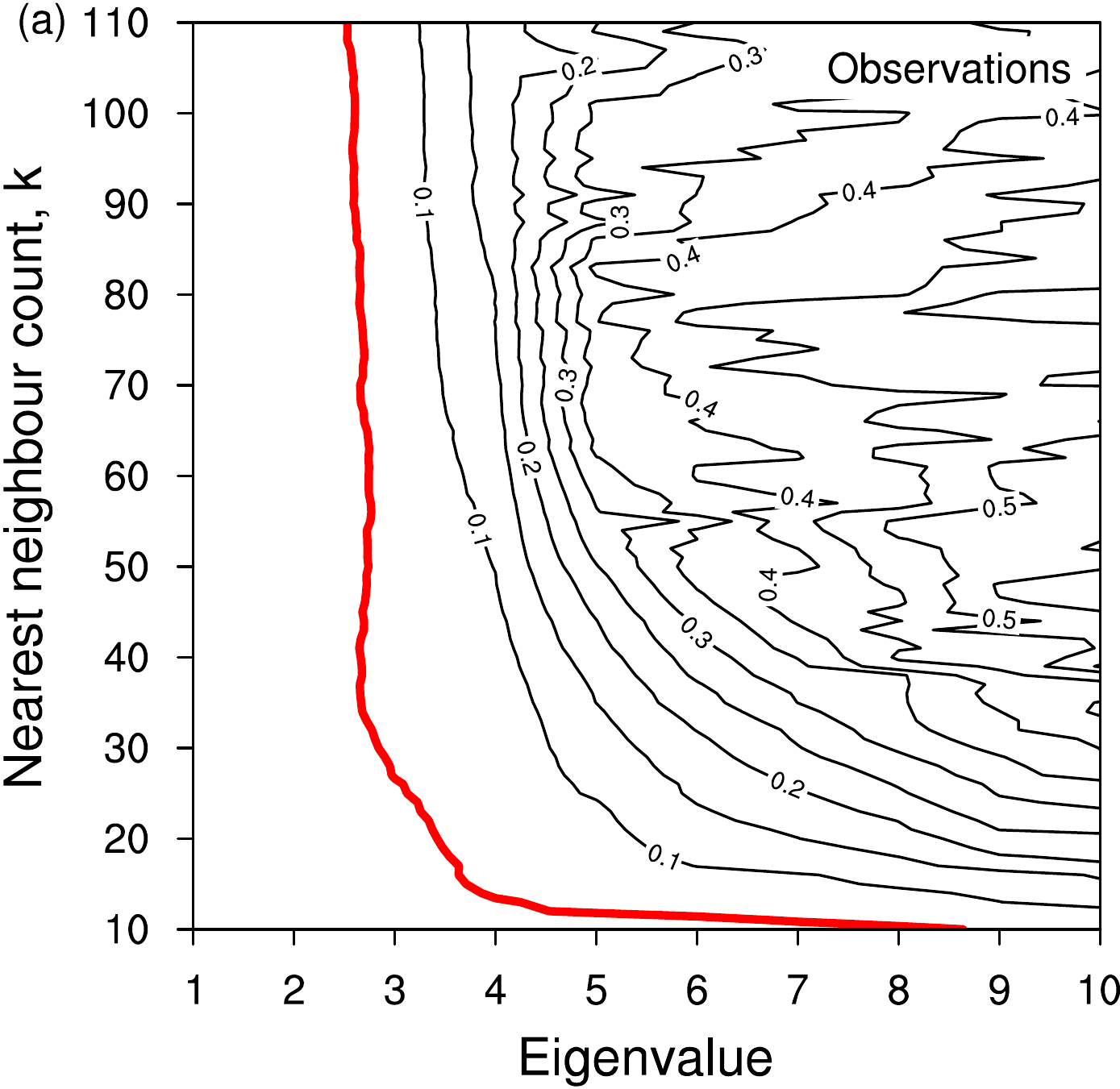}%
    \hspace{0.04\textwidth}%
    \includegraphics[width=0.48\textwidth]{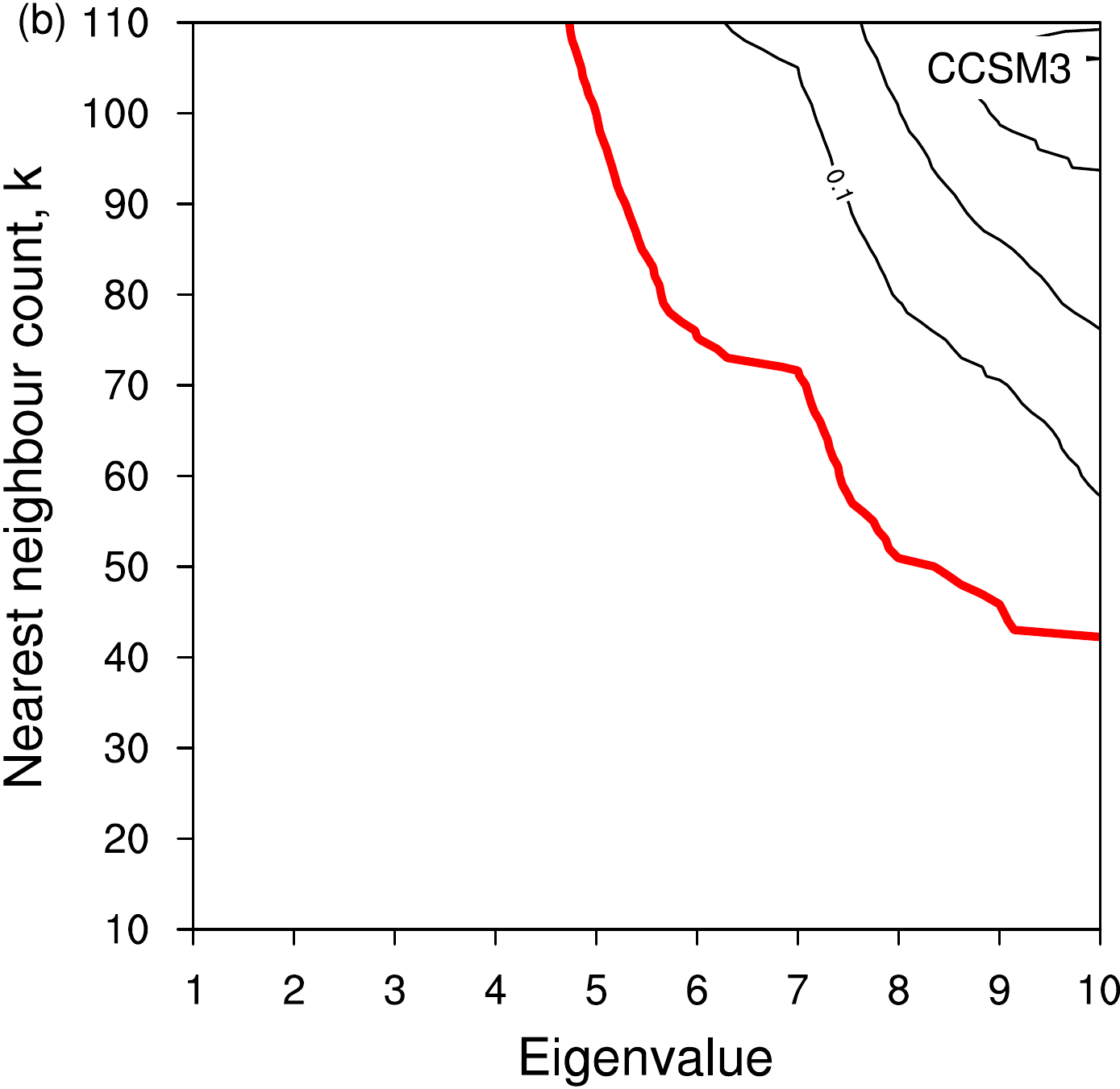}

    \vspace{0.25cm}
    \includegraphics[width=0.48\textwidth]{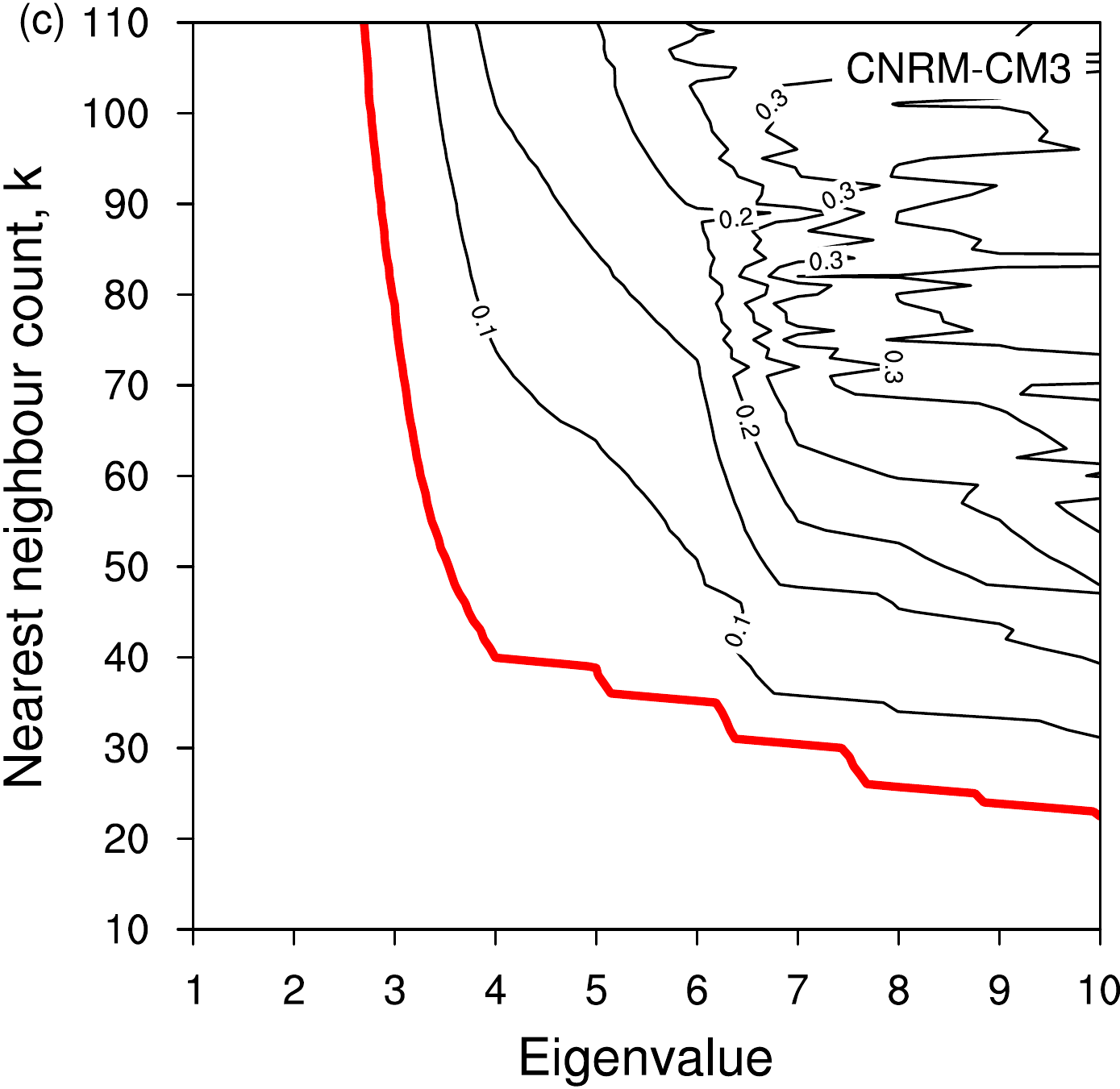}%
    \hspace{0.04\textwidth}%
    \includegraphics[width=0.48\textwidth]{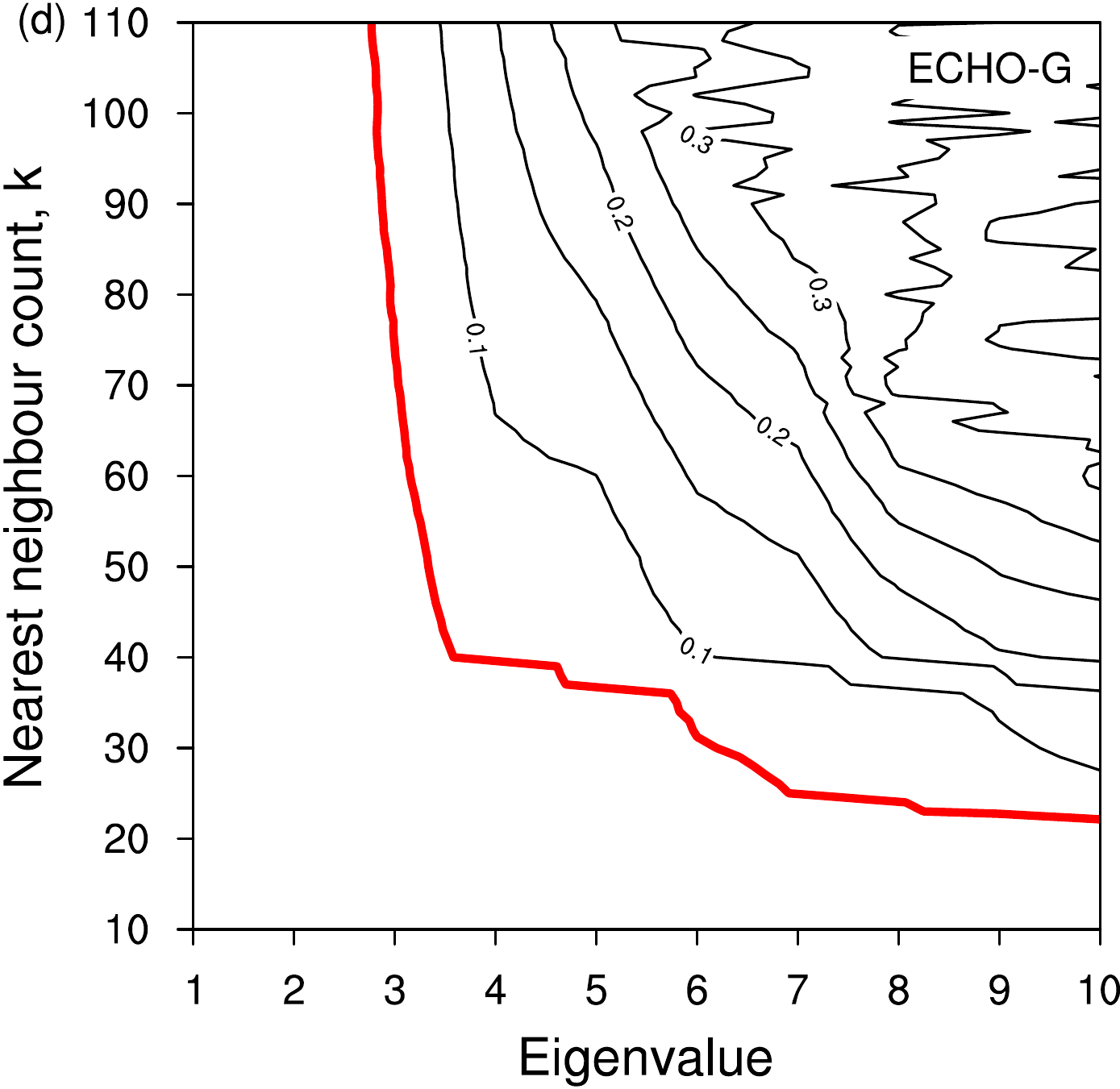}

    \vspace{0.25cm}
    \includegraphics[width=0.48\textwidth]{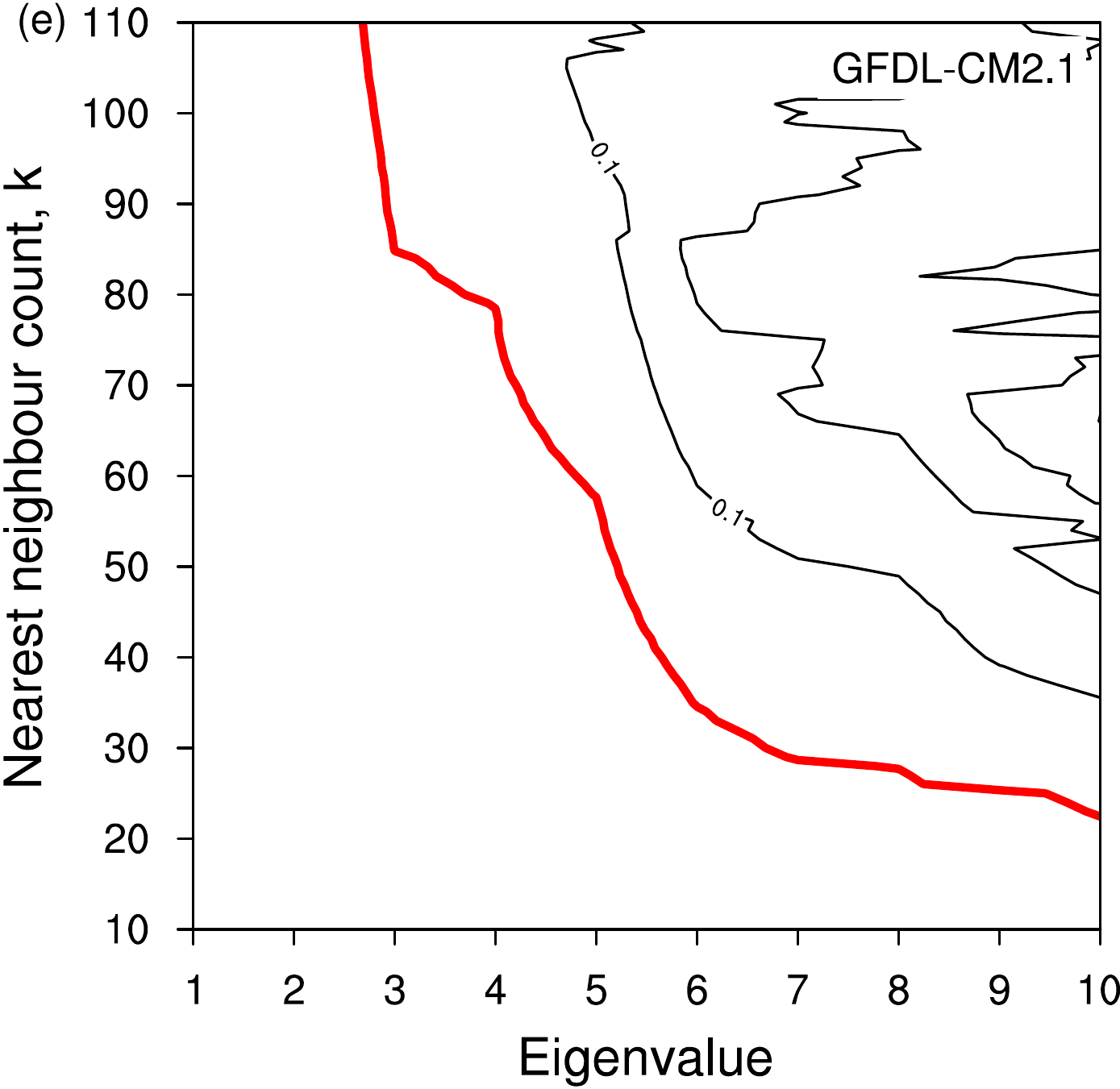}%
    \hspace{0.04\textwidth}%
    \includegraphics[width=0.48\textwidth]{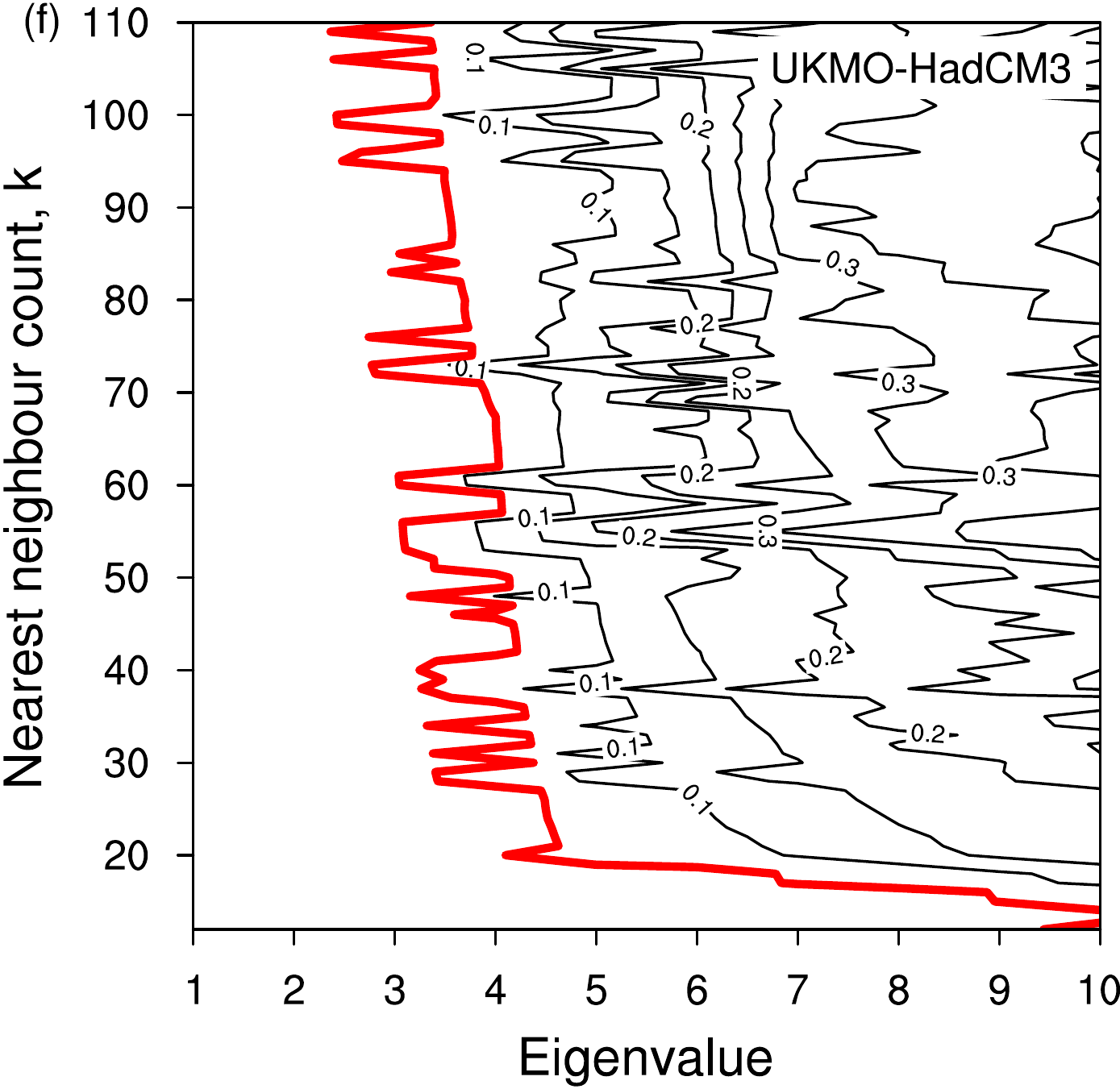}
  \end{center}
  \caption[Hessian LLE eigenvalue convergence: tropical Pacific
    SSTs]{Hessian LLE eigenvalue convergence and dimension estimates
    for tropical Pacific SSTs, from observations (a), CCSM3 (b),
    CNRM-CM3 (c), ECHO-G (d), GFDL-CM2.1 (e) and UKMO-HadCM3 (f).
    Black contours show Hessian LLE eigenvalue spectra as a function
    of eigenvalue number and nearest neighbour count $k$.  The thick
    red line shows the boundary between ``null'' eigenvalues and
    ``non-null'' eigenvalues, set at an (arbitrary) boundary value of
    0.05.}
  \label{fig:raw-sst-hlle-sens}
\end{figure}

Let us consider the observational data first.  As we did for Isomap in
Chapter~\ref{ch:isomap}, we can examine the eigenvalue spectra
obtained in the final embedding calculation, i.e. the diagonal entries
in the matrix $\bm{\Lambda}$ in \eqref{eq:hlle-eigendecomp}.
Theoretically, we expect to find a number of zero eigenvalues,
corresponding to the nullspace of the Hessian on the data manifold,
although numerical issues will clearly mean that we have to content
ourselves with selecting eigenvalues smaller than some finite bound.
Figure~\ref{fig:raw-sst-hlle-sens} shows contour plots of eigenvalue
spectra, as a function of the neighbourhood size $k$ for a number of
data sets, starting with the observational data in
Figure~\ref{fig:raw-sst-hlle-sens}a.  The red line on these plots is a
more or less arbitrary dividing line between ``small'' eigenvalues and
``larger'' eigenvalues, intended to indicate the approximate
dimensionality of the nullspace of the Hessian.  The threshold between
``small'' and ``larger'' here is set at 0.05, although a range of
values could have been chosen.  For most values of the neighbourhood
size $k$, we see from Figure~\ref{fig:raw-sst-hlle-sens}a that the
Hessian for the observational data has an approximately
three-dimensional nullspace.  Since there is always one basis vector
in the nullspace associated with constant functions on the data
manifold, a three-dimensional nullspace should allow for the
computation of two-dimensional embeddings of the observational data.
However, as we will see below, following through with the Hessian LLE
procedure described in Section~\ref{sec:hlle-method} to produce
embedding coordinates for this data produces reasonable
\emph{three}-dimensional embeddings for most values of $k$.  It
appears that the eigenvalue spectra cannot be relied upon to give a
good indication of what embeddings may be obtained, and do not provide
a reliable guide to the underlying data dimensionality.  This is
probably due to the difficulty of accurately estimating small
eigenvalues for the approximate Hessian matrix $\mathbf{\tilde{H}}$
using an iterative eigenvalue solver.  The rather large threshold
value of 0.05 used here to distinguish between ``null'' and
``non-null'' eigenvalues is based on the idea that, numerically, a
reasonable embedding may be obtained from eigenvectors that are not
exactly in the nullspace of the approximate Hessian operator.  From
the embedding results shown below, it seems as though even
eigenvectors associated with eigenvalues failing to meet this
relatively coarse condition can contribute to reasonable embeddings.
In what follows, in order to enable comparison with the Isomap results
of Chapter~\ref{ch:isomap}, we will generally examine only
three-dimensional embeddings of the Hessian LLE results.  This proves
to be a reasonable compromise in terms of the dimensionality estimates
that one might make from the eigenvalue spectra in
Figure~\ref{fig:raw-sst-hlle-sens}.

%
%
%

\begin{figure}
  \begin{center}
    \includegraphics[width=0.48\textwidth]{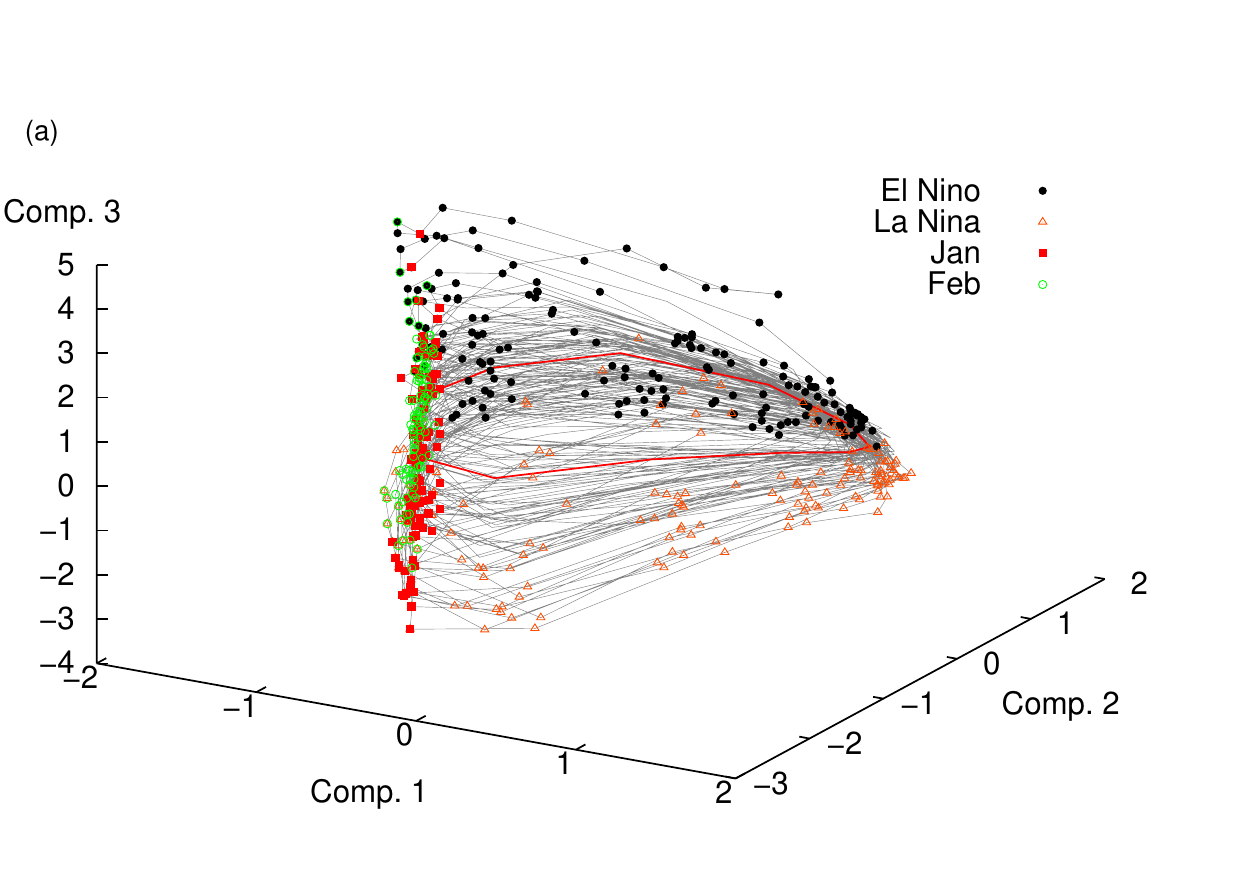}%
    \hspace{0.04\textwidth}%
    \includegraphics[width=0.48\textwidth]{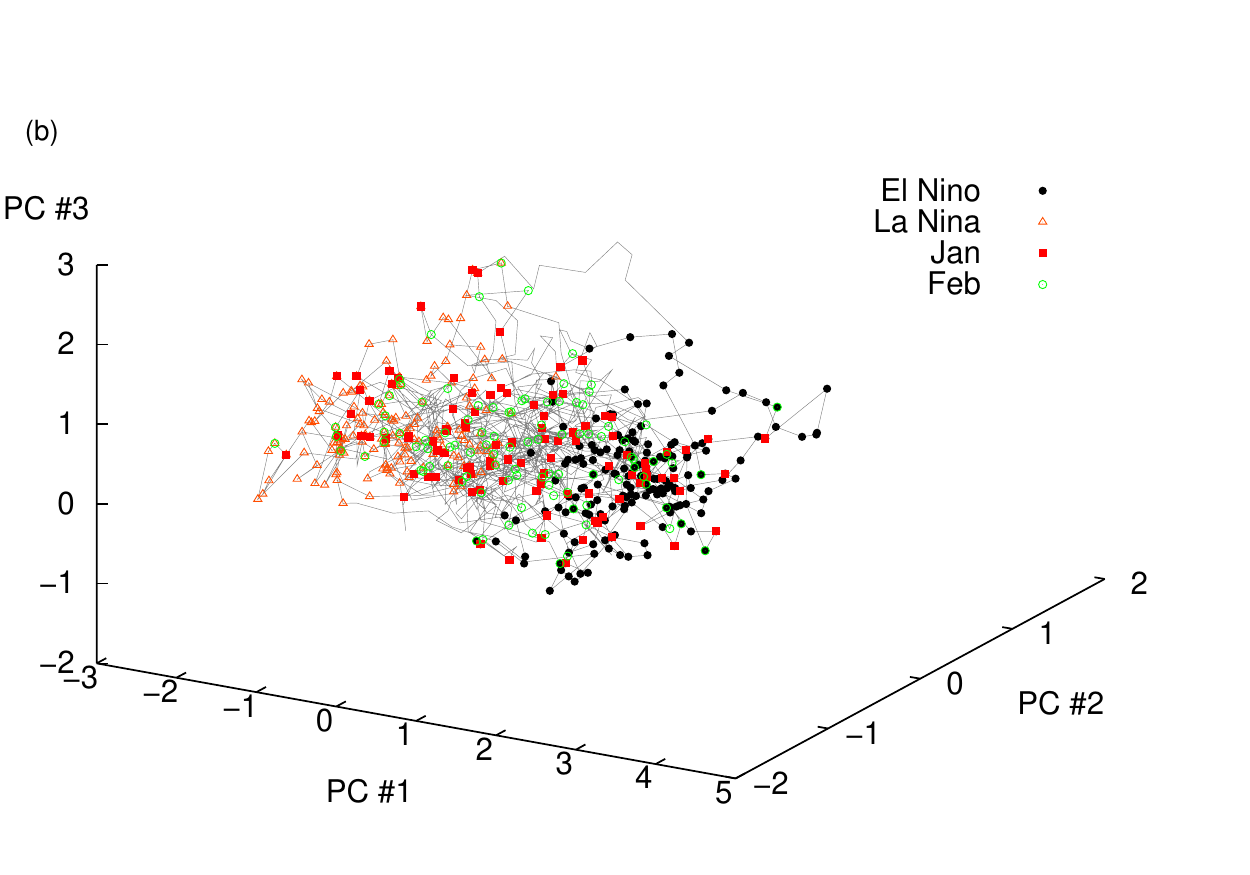}
  \end{center}
  \caption[Three-dimensional Hessian LLE embedding for ERSST
    data]{Three-dimensional Hessian LLE embedding for ERSST
    observational SST data (a, $k = 20$), and equivalent
    three-dimensional embedding plot based on leading three principal
    components of observational SST data (b).  Light grey lines join
    data points representing adjacent months in the SST time series.
    The mean annual cycle is shown as a thicker line with January and
    February highlighted in blue and green respectively.  Points are
    identified as \eln (black dots) or \lan (red triangles) events
    based on the corresponding NINO3 SST index time series for each
    data set.}
  \label{fig:hlle-ersst-embeddings}
\end{figure}

For the observational data, a three-dimensional embedding is shown in
Figure~\ref{fig:hlle-ersst-embeddings}a, in the same format as used in
Figure~\ref{fig:isomap-raw-3d} on page~\pageref{fig:isomap-raw-3d} to
display Isomap embeddings.  In each plot of this type, only 100 years
of data is shown for clarity (some of the model data sets have much
more than this, but the plots become confusing with too many points),
data points adjacent in time are connected by thin grey lines, points
of particular interest (\eln and \lan months, as identified by the
NINO3 SST index and each January and February) are highlighted in
colour, and the mean annual cycle is highlighted as a thick red curve.
Examination of this type of plot for different values of $k$ for the
observation data reveals that for $10 \leq k \leq 11$, the embeddings
obtained by Hessian LLE are degenerate, i.e. all data points are
mapped to only a few points in the lower-dimensional space; for $12
\leq k \leq 15$, the embedding is close to a one-dimensional curve
capturing only variability associated with the annual cycle; for all
$k \geq 16$, the embeddings look similar to
Figure~\ref{fig:hlle-ersst-embeddings}a, with two degrees of freedom
in the embedding showing a clear annual cycle, with motion around the
``cylinder'' formed by the embedding, and a third, more or less
orthogonal, ``axial'' degree of freedom that appears to segregate \eln
from \lan conditions.  This result is very similar to that seen for
Isomap (Figure~\ref{fig:isomap-raw-3d}a for the observational data).
For other values of $k$ in the range $16 \leq k \leq 110$, there are
some changes in orientation of the data manifold in the embedding
space, and some variation in the degree to which the ``axial''
component associated with differences between \eln and \lan is spread
out, but the basic pattern is remarkably insensitive to variations in
$k$.

It could be argued that the achievement of this apparent separation of
the annual cycle and ENSO variability is a much less stringent test of
the Hessian LLE method than it is of Isomap in
Chapter~\ref{ch:isomap}.  There, we started from the original SST
data, without requiring an initial dimensionality reduction step using
PCA.  It would appear to be rather more difficult to go from the
original SST data directly to an embedding such as that of
Figure~\ref{fig:hlle-ersst-embeddings}a than to first reduce the data
to the ten-dimensional space spanned by the first 10 EOFs.  However,
this is perhaps a little deceptive.  The initial step in Isomap is to
construct a matrix of Euclidean distances between the input data
points, which in itself provides a fairly radical reduction in the
data dimensionality.  Furthermore, the structure picked out by the
Hessian LLE method in Figure~\ref{fig:hlle-ersst-embeddings}a is not a
structure that is immediately apparent in the principal component time
series used as input to the Hessian LLE calculations.
Figure~\ref{fig:hlle-ersst-embeddings}b shows the equivalent embedding
plot for the observational data based on the first three principal
components; it is immediately clear that the structure seen in
Figure~\ref{fig:hlle-ersst-embeddings}a is not obviously present here.
This is true for all three-dimensional linear projections of the
10-dimensional PCA results --- the Hessian LLE embedding does not just
project to a linear subspace of the input data, but discovers
intrinsically nonlinear structure.

%
%
%

Moving now to the model data, Hessian LLE eigenvalue spectra for a
selection of models are shown in
Figures~\ref{fig:raw-sst-hlle-sens}b--f.  These are plotted in the
same way as Figure~\ref{fig:raw-sst-hlle-sens}a for the observational
data.  Again, the red lines on the plots show the dividing line
between ``null'' and ``non-null'' eigenvalues, based on a more or less
arbitrary threshold value of 0.05.  Almost all of the model plots, as
do the observations, show a decay to a three-dimensional nullspace for
the Hessian at larger values of $k$, but the model eigenvalue spectra
differ from the observational spectra in the extent to which higher
dimensional nullspaces exist for smaller values of $k$.  For instance,
for two of the models shown, CNRM-CM3
(Figure~\ref{fig:raw-sst-hlle-sens}c) and ECHO-G
(Figure~\ref{fig:raw-sst-hlle-sens}d), the dimensionality of the
nullspace is greater than three for all values of $k$ less than about
40, while for two other models, CCSM3
(Figure~\ref{fig:raw-sst-hlle-sens}b and GFDL-CM2.1
(Figure~\ref{fig:raw-sst-hlle-sens}e), the dimensionality of the
nullspace is significantly larger for most values of $k$.  The other
model, UKMO-HadCM3 (Figure~\ref{fig:raw-sst-hlle-sens}f), has a rather
noisy eigenvalue spectrum.  From the range of results shown here, it
appears that the dimensionality of the nullspace found by the Hessian
LLE algorithm for the model SST data sets is almost always greater
than three, indicating (and it is nothing more than an indication)
that three- and possibly four-dimensional embeddings should be
possible for most models for moderate values of $k$ (a limit of $k
\lesssim 40$ should ensure that the embeddings are reasonable for most
models).  This conclusion is only indicative because of the difficulty
of identifying the dimensionality of the nullspace of the Hessian from
approximate numerical calculations on a finite data set.  Also, as we
will see below, it appears possible to produce reasonably good
embeddings of the data for values of $k$ outside the indicated range,
again dependent on the data set.

%
%
\begin{figure}
  \begin{center}
    \includegraphics[width=0.48\textwidth]{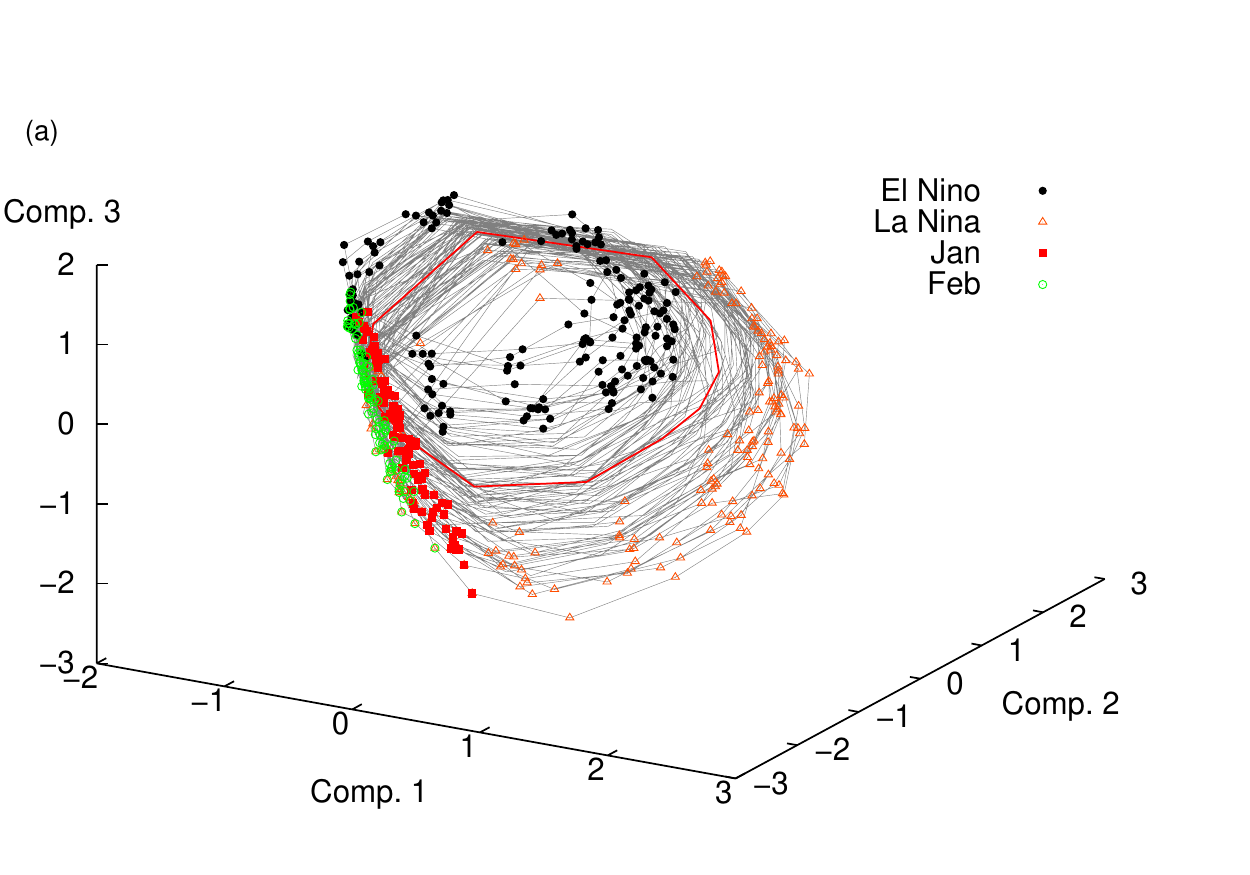}%
    \hspace{0.04\textwidth}%
    \includegraphics[width=0.48\textwidth]{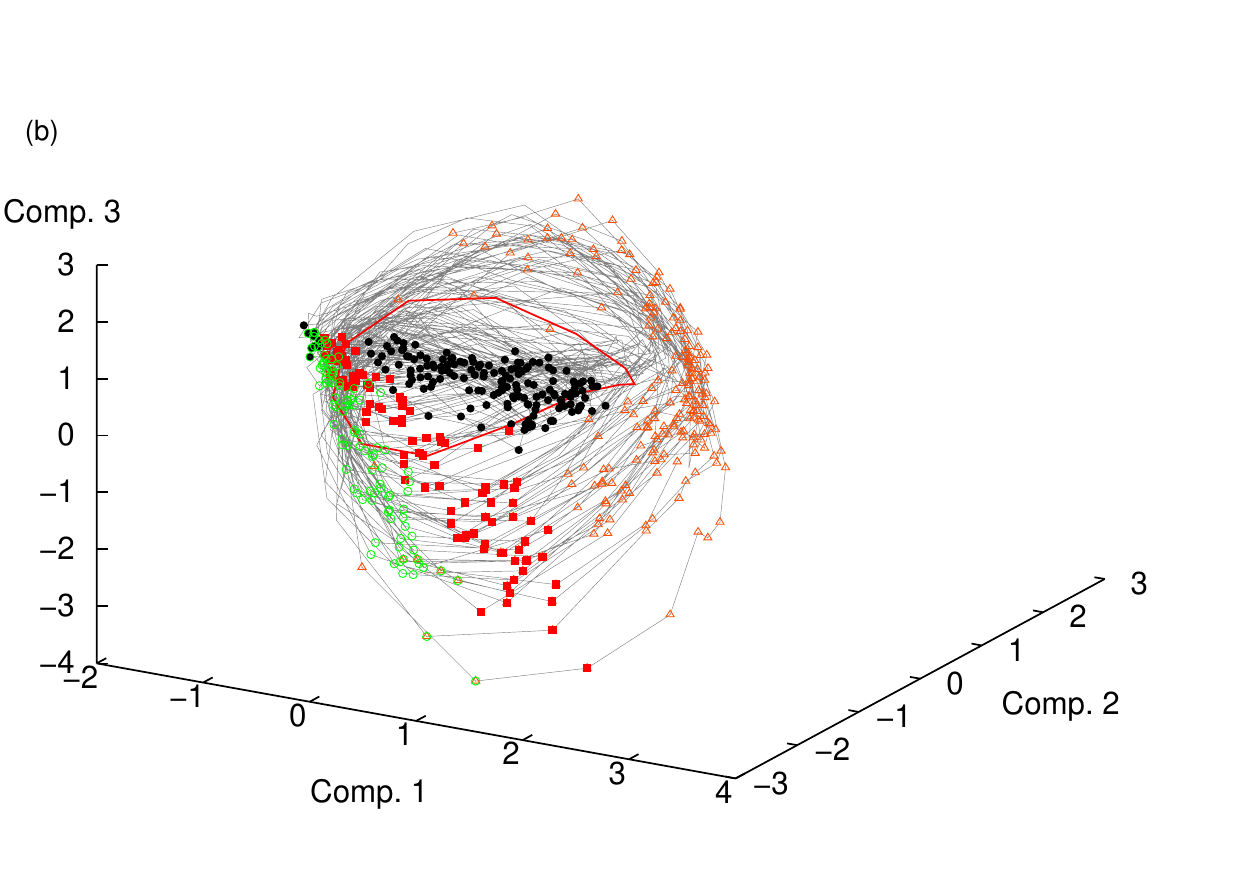}

    \includegraphics[width=0.48\textwidth]{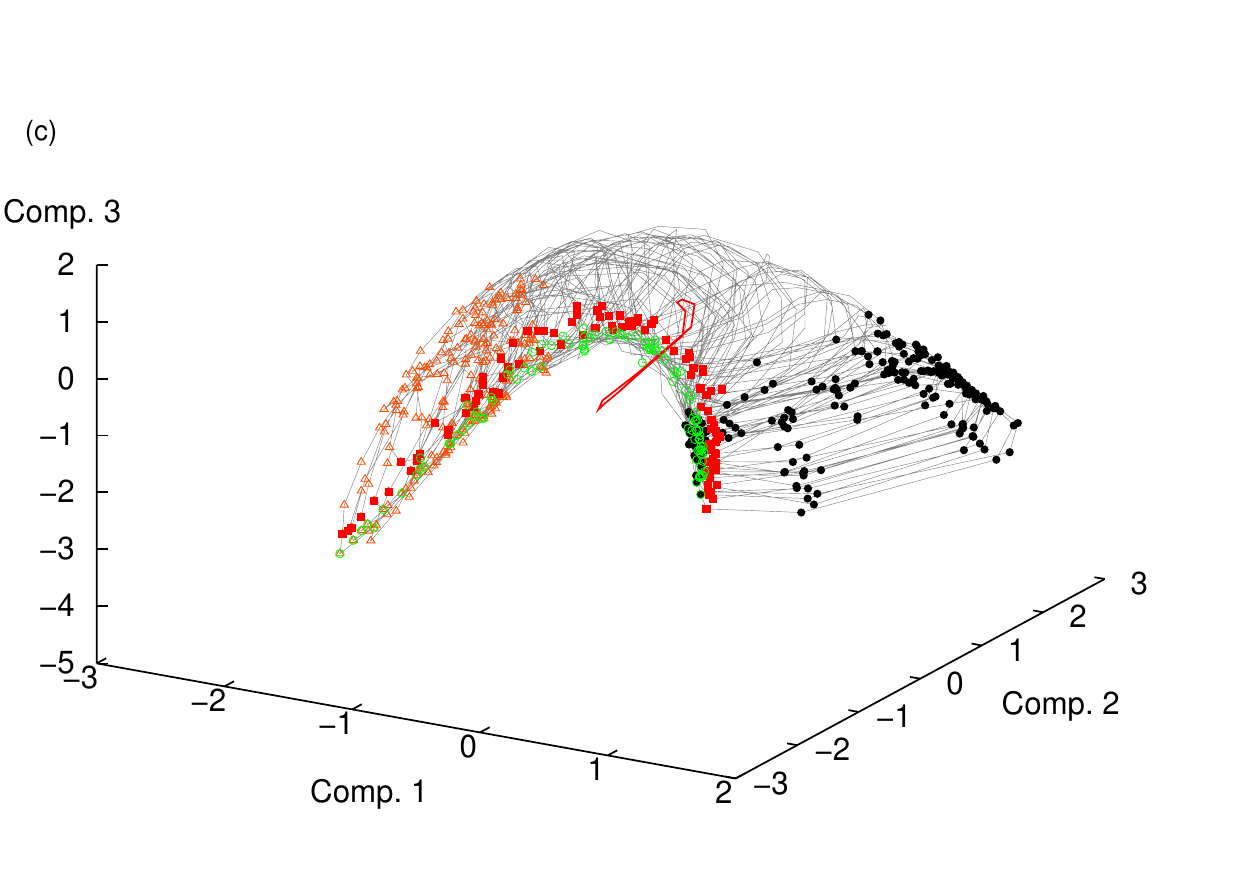}%
    \hspace{0.04\textwidth}%
    \includegraphics[width=0.48\textwidth]{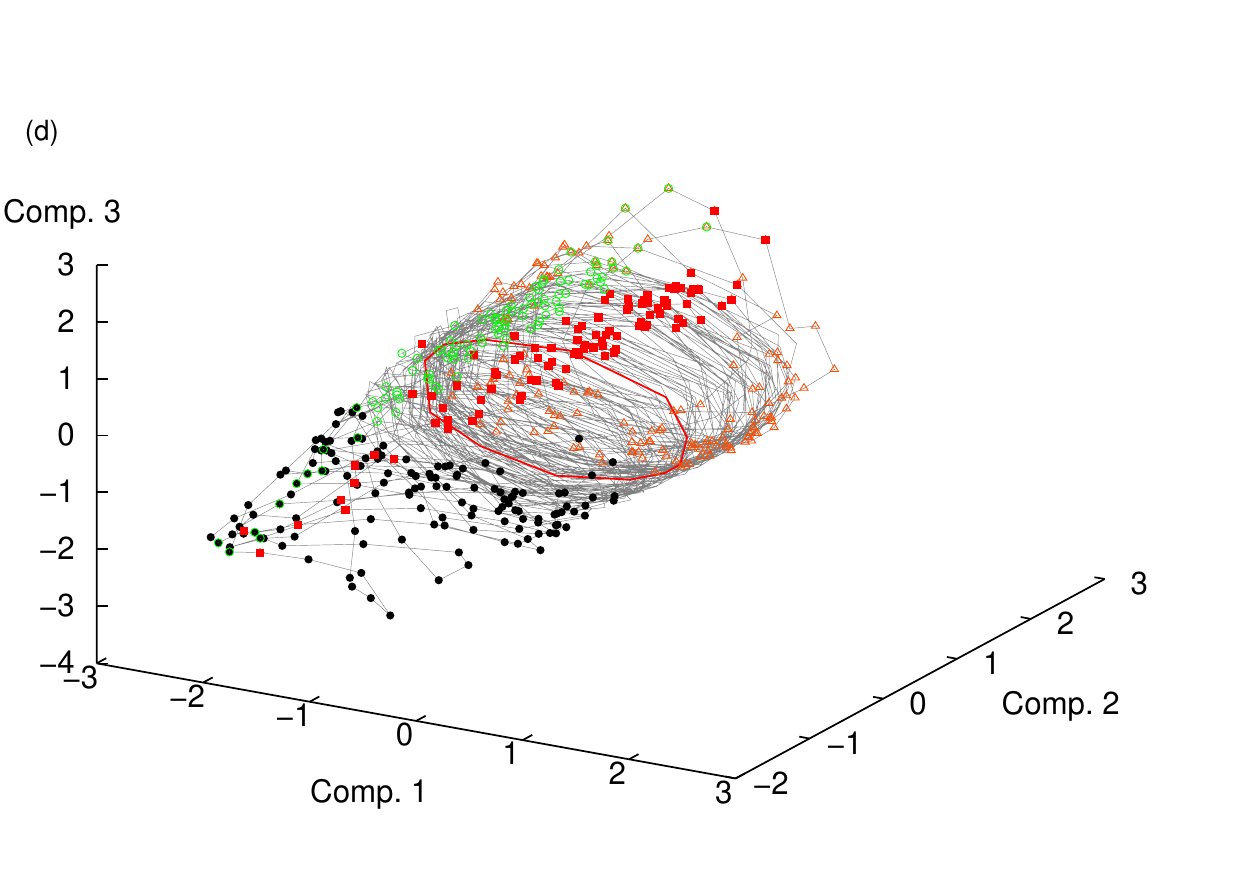}

    \includegraphics[width=0.48\textwidth]{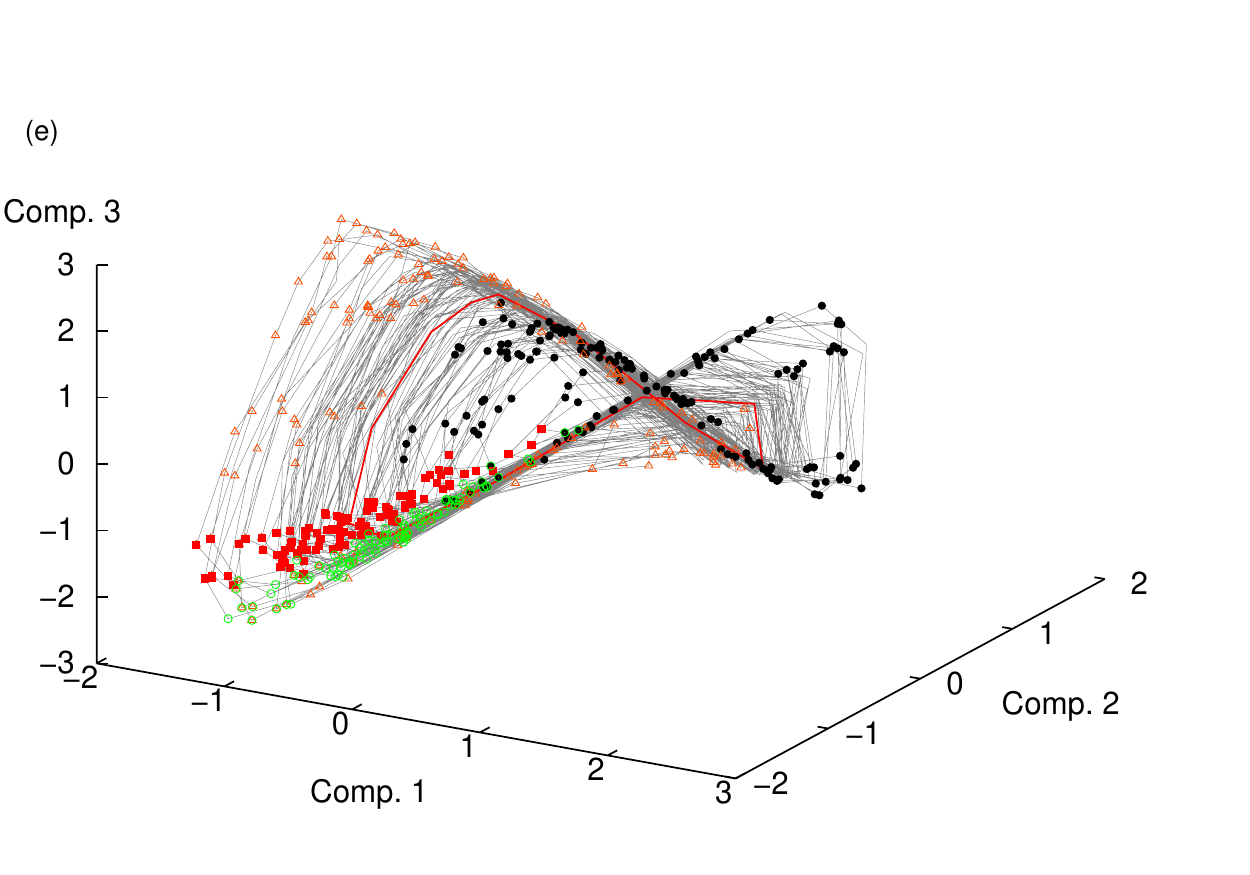}%
    \hspace{0.04\textwidth}%
    \includegraphics[width=0.48\textwidth]{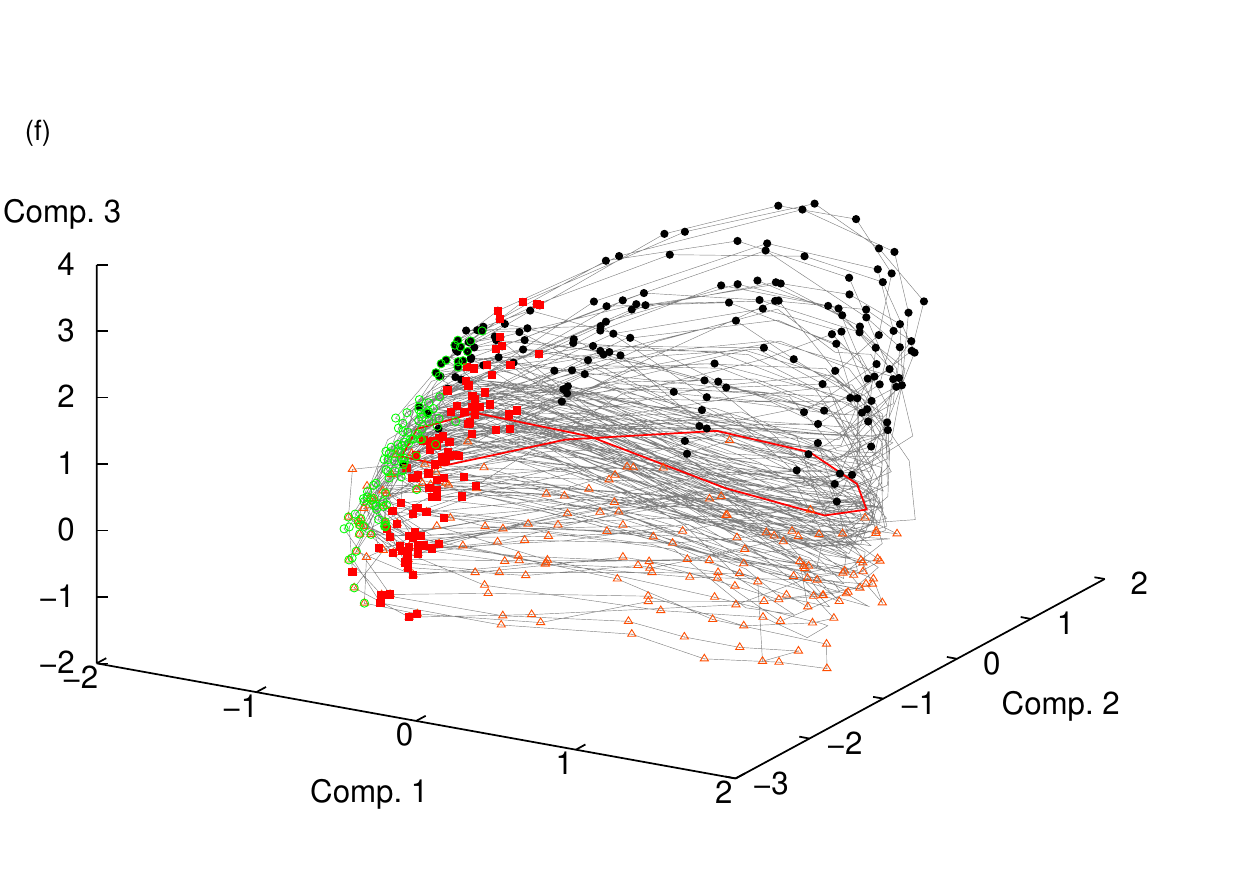}
  \end{center}
  \caption[Three-dimensional Hessian LLE embedding for model SST
    data]{Three-dimensional Hessian LLE embedding for SST data from
    selected models: CCSM3 (a, $k = 90$), ECHO-G (b, $k = 47$),
    FGOALS-g1.0 (c, $k = 35$), GFDL-CM2.1 (d, $k = 60$),
    MIROC3.2(medres) (e, $k = 55$), UKMO-HadCM3 (f, $k = 28$).  Light
    grey lines join data points representing adjacent months in the
    SST time series.  The mean annual cycle is shown as a thicker red
    line with January and February highlighted in blue and green
    respectively.  Points are identified as \eln (black dots) or \lan
    (red triangles) events based on the corresponding NINO3 SST index
    time series for each data set.  For clarity, only 100 years of
    data is plotted for each model.}
  \label{fig:hlle-model-embeddings}
\end{figure}
%

Figure~\ref{fig:hlle-model-embeddings} shows three-dimensional Hessian
LLE embeddings for a range of models.  The dependence of the form of
these embeddings on the neighbourhood size $k$ is rather interesting.
From the eigenvalue spectra shown in
Figure~\ref{fig:raw-sst-hlle-sens}b--f, one might expect that good
embeddings would only be found for $k \lesssim 40$ for most models,
since this is the range of values for which the Hessian LLE algorithm
appears to find a sufficiently high-dimensional nullspace for the
approximate Hessian operator.  However, as mentioned above, these
limits are based on a very rough estimate of which eigenvalues are
``null'' and which are ``non-null''.  In a numerical context, it is
difficult to make an unequivocal judgement about the appropriate
threshold to use for distinguishing between ``null'' and ``non-null'',
and it appears, from the embedding results, that reasonable embeddings
are found for a much wider range of values of $k$ than expected from
the eigenvalue spectra.

For each of the models, the embeddings found for small values of $k$
are degenerate, i.e. all input data points are mapped to a single
point or to a few points in the low-dimensional embedding space.  The
upper limit of $k$ for which this occurs depends on the model:
degenerate embeddings are found for CCSM3 for $k \leq 35$, for
GFDL-CM2.1 for $k \leq 17$, for ECHO-G for $k \leq 36$, and so on.  At
the other end of the scale of values of $k$, good embeddings are found
for most models for larger values of $k$: good embeddings are found
for CCSM3 for $k \geq 38$, for GFDL-CM2.1 for $k \geq 33$, for ECHO-G
for $k \geq 38$.  The exact limits again depend on the model, and for
some models there is variation in the orientation and degree of
distortion of the embedded data manifold as $k$ varies.  All of the
embeddings displayed in Figure~\ref{fig:hlle-model-embeddings} lie in
this ``good'' range of $k$ values, apart from the embedding for
MIROC3.2(medres) in Figure~\ref{fig:hlle-model-embeddings}e, described
below.  The exact values of $k$ for which embeddings are shown were
selected either to give a good orientation for the component rotation
calculations below, or to illustrate interesting features of the
embeddings (Figures~\ref{fig:hlle-model-embeddings}c~and~e).  Between
the lower and upper extremes of $k$ values, the behaviour of the
different model data sets is more varied.  At values of $k$ just above
the range giving degenerate embeddings, some models show
one-dimensional embeddings that capture only the annual cycle in the
input data, without any obvious ENSO variability.  Of the models for
which results are shown here, this is the case for CCSM, GFDL-CM2.1
and MIROC3.2(medres).  For some ranges of $k$, some models show
otherwise good embeddings contaminated by a few outlying points that
are mapped far away from the main body of the embedding.  This effect
is seen for ECHO-G and UKMO-HadCM3.  Other models show an interesting
``funnel'' shaped embedding.  This occurs for GFDL-CM2.1, but is
especially prominent for FGOALS-g1.0, as can be clearly seen in
Figure~\ref{fig:hlle-model-embeddings}c.  Finally, the
MIROC3.2(medres) embeddings (Figure~\ref{fig:hlle-model-embeddings}e)
show a ``butterfly'' shape for most values of $k$, a feature not seen
in any of the other models.

There are a number of things to draw from this.  First, for most of
the model data sets (and the observations), the Hessian LLE procedure
does identify good embeddings for a wide range of values of the
neighbourhood size $k$.  This contrasts with the conclusions that
might be taken from the eigenvalue spectra shown in
Figure~\ref{fig:raw-sst-hlle-sens}, where, for larger values of $k$,
it appears that the nullspace of the approximate Hessian is only
three-dimensional, which would imply that no reasonable
three-dimensional embeddings can be found.  The reason for this
discrepancy is probably that the distinction between a ``null'' and a
``non-null'' eigenvalue is rather coarser than might be suggested from
the eigenvalue plots shown here.  While there is no particularly clear
spectral gap between near zero eigenvalues and other eigenvalues for
most of the models, there is an extent to which eigenvectors which lie
close to the nullspace (i.e. that have small but non-zero eigenvalues)
can be used to form reasonable embeddings, at least to the numerical
accuracy that we can feasibly hope to achieve here.  The second
observation to make about the embeddings shown in
Figure~\ref{fig:hlle-model-embeddings} is that most of them appear to
successfully capture both the annual cycle and ENSO-related
variability in the input data, as did the Isomap embeddings from
Chapter~\ref{ch:isomap}.  We can confirm these conclusions in just the
same way as we did for the Isomap results by attempting to rotate the
embedding plots to ``unmix'' the annual cycle from the ENSO
variability (Section~\ref{sec:isomap-rotation}).

%
%
\begin{figure}
  \begin{center}
    \includegraphics[width=0.75\textwidth]{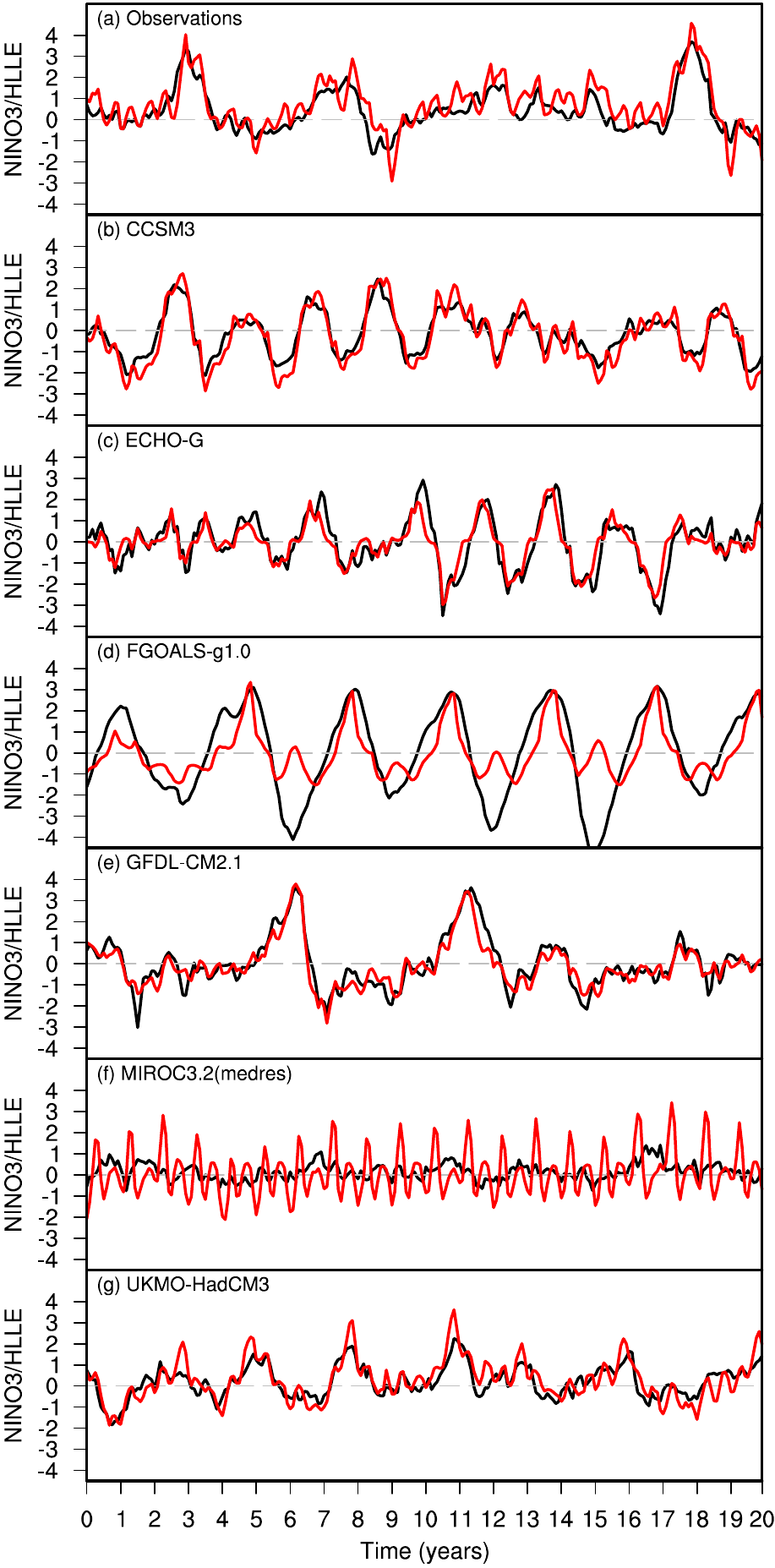}
  \end{center}
  \caption[NINO3 SST index and Hessian LLE rotated component \#3 time
    series]{Time series of NINO3 SST index (black) and rotated Hessian
    LLE component \#3 (red) for observations (a) and selected models
    (b--g).  An arbitrary 20 year slice of data is shown in each
    case.}
  \label{fig:hlle-rot3}
\end{figure}
%

Time series plots of the rotated third Hessian LLE component for
observations and six models are shown in Figure~\ref{fig:hlle-rot3}.
The rotated Hessian LLE component \#3 time series are plotted in
parallel with time series of the NINO3 SST index, recording ENSO
variability.  For the observations, in Figure~\ref{fig:hlle-rot3}a, it
is clear that rotated Hessian LLE component \#3 quite accurately
captures ENSO variability in the input SST data.  In this case, just
as for Isomap in Chapter~\ref{ch:isomap}, Hessian LLE has thus
extracted the most important modes of variability in tropical Pacific
SSTs, the annual cycle and ENSO, starting from higher-dimensional
input data.  (As previously noted, this is slightly less impressive
than the Isomap results because of the need to use an initial linear
dimensionality reduction step.)  The comparable plots in
Figures~\ref{fig:hlle-rot3}b--g for the models show similar results in
most cases, with good correlation between the rotated Hessian LLE
component \#3 and the NINO SST index.  Of the models displayed in
Figure~\ref{fig:hlle-rot3}, the only exception to this picture is
MIROC3.2(medres) (Figure~\ref{fig:hlle-rot3}f), one of the models that
does not show particularly strong ENSO variability according to most
measures.  Correlation coefficients between rotated Hessian LLE
component \#3 and the NINO3 SST index time series shown in
Table~\ref{tab:hlle-correlations} confirm these observations: the
correlation coefficients for all of the models except for
MIROC3.2(medres) are all high.  FGOALS-g1.0 has a slightly smaller
correlation coefficient than the other models, related to an apparent
failure of the rotated Hessian LLE component \#3 to capture the
unusually large negative excursions of the FGOALS-g1.0 NINO3
variability (Figure~\ref{fig:hlle-rot3}d).

%
%
\begin{table}
  \caption[Correlation between NINO3 SST index and Hessian LLE
    component]{Correlation coefficients between NINO3 SST index and
    Hessian LLE rotated component \#3.}

  \label{tab:hlle-correlations}

  \begin{center}
    \begin{tabular}{lc}
      \toprule

      \textbf{Data Set} & \textbf{Correlation} \\

      \midrule

      Observations     & 0.846 \\
      CCSM3            & 0.946 \\
      ECHO-G           & 0.878 \\
      FGOALS-g1.0      & 0.695 \\
      GFDL-CM2.1       & 0.920 \\
      MIROC3.2(medres) & 0.145 \\
      UKMO-HadCM3      & 0.833 \\

      \bottomrule
    \end{tabular}
  \end{center}
\end{table}
%

\section{Discussion and conclusions}
\label{sec:hlle-discussion}

The Hessian LLE method is certainly theoretically very appealing.  It
is based on a more solid mathematical foundation than many other
geometrical/statistical dimensionality reduction methods that have
been proposed, meaning that there is some hope of being able to
analyse the performance of the method rigorously.  From a practical
point of view, Hessian LLE is surprisingly successful in analysing the
relatively noisy SST data sets examined here.  Why this should be the
case is a question that deserves more analysis, but is probably
related to fact that, in the final eigendecomposition of the
approximate Hessian matrix, an approximate nullspace is almost
certainly enough to give a reasonable embedding, and the embedding is
likely to degrade gracefully as the degree of approximation increases.

The results presented here for the geometrical test data sets
(Section~\ref{sec:hlle-test-data}) give some insight into the
dependence of the behaviour of Hessian LLE on data sampling density
and neighbourhood size $k$ although, as already noted, the conclusions
that one might draw from those results turn out to be somewhat too
pessimistic when considering analysis of the tropical Pacific SST
data.  One thing that the geometrical test data sets do help to
elucidate (particularly the sensitivity analyses of
Section~\ref{sec:hlle-sens}) is the effect of data noise on the
method.  This stems from the observation that there appears to be
something of a tradeoff between larger values of the neighbourhood
size $k$, giving better discrimination against noise, and the smaller
values of $k$ required to give neighbourhoods small enough to be
``locally linear''.  If noise in the data is assumed to be
isotropically distributed, both tangential to and normal to the
manifold in which the data points nominally lie, then increasing the
number of points in the neighbourhoods used to calculate local tangent
coordinates will increase the influence of the tangential directions
(i.e. the directions that are of use in determining the intrinsic
structure of the data manifold) at the expense of the normal
directions (which are primarily noise).  Of course, as $k$ increases,
eventually the neighbourhoods become less and less ``local'' and less
and less ``linear'' until they cover a significantly curved portion of
the data manifold and the local singular value decomposition
calculation can no longer find reasonable tangent coordinates.

Overall, the results of applying Hessian LLE to the tropical Pacific
SST data are rather encouraging, producing performance comparable to
the Isomap method explored in Chapter~\ref{ch:isomap}.  The
three-dimensional embeddings derived by the Hessian LLE method are
very similar to those produced by Isomap, and the same component
rotation approach to ``unmixing'' the annual cycle and ENSO signals
from the embedding coordinates works as well here as it does for
Isomap.  For all of the models examined that have reasonable ENSO
variability, the Hessian LLE results look good.  That said, there is
some difficulty in interpreting the eigenvalue spectra produced as
part of the final calculation of the embedding coordinates.  This
difficulty almost certainly arises because of numerical difficulties
in calculating the nullspace of a large ($N \times N$) matrix.  In
contrast to Isomap, where the embedding is computed from the
eigenvectors associated with the \emph{largest} eigenvalues of a
matrix, here we must estimate the eigenvectors associated with the
\emph{smallest} eigenvalues, a calculation more prone to numerical
instability.

One aspect of the tropical Pacific SST results that is slightly
perplexing is that the three-dimensional embeddings obtained are
rather better than one might have expected from comparison with the
geometrical test data set results.  The tropical Pacific SST data is
significantly noisier than any of the geometrical test data (compare
the principal component scatter plots in
Figure~\ref{fig:sst-pc-scatter} and Chapter~\ref{ch:nlpca} with the
noisy Swiss roll with hole shown in Figure~\ref{fig:test-data-3d}d on
page~\pageref{fig:test-data-3d}) and, from studying the range of data
point sampling density and neighbourhood size $k$ for which good
embeddings are obtained in Figure~\ref{fig:hlle-sens-noise}, one might
expect that good embeddings will be obtained for the noisier SST data
only for a very narrow range of neighbourhood sizes.  In fact, for
most of the SST data sets, good embeddings are obtained for most
values of $k$ above a certain threshold.  Why should this be the case?
I believe that this may come down to a question of perception.  The
SST data are certainly noisier but, for the geometrical test data, we
know exactly what a good embedding looks like, since we have sampled
data points from a known manifold.  This means that we can judge the
test data results very carefully and clearly identify cases where the
embedding is not particularly good.  For the ENSO data, on the other
hand, so long as we have an approximate nullspace we can use to build
a projection basis, we get something that looks like a reasonable
embedding.  The annual cycle and ENSO are by far the strongest modes
of variability we expect to see, meaning that it is not difficult for
the Hessian LLE procedure to pick them out.

So, could Hessian LLE be recommended as a dimensionality reduction
method to use for real applications?  Perhaps.  For cases where it is
known that data points are sampled from a manifold and there is little
noise, Hessian LLE definitely outperforms Isomap, since it can deal
with non-convex sets and manifolds that are only locally isometric to
subsets of Euclidean space, rather than requiring that the data
manifold be \emph{globally} isometric to an open, convex subset of
Euclidean space.  For more realistic data sets, where there is noise
and we are not sure that data points are sampled from a reasonable
looking manifold, Hessian LLE would appear to be competitive with
Isomap, despite the worries about numerical issues.  On this problem
presented here, at least, it appears to be as good.


%% file: 09-conclusions.tex
\chapter{Summary and Future Work}
\label{ch:summary}

In this thesis, I have explored the suitability of three
geometrical/statistical nonlinear dimensionality reduction methods for
a relatively simple climate data analysis application.  The methods
presented here, NLPCA, Isomap and Hessian LLE, are just three among a
host of techniques that have been developed, primarily in the machine
learning community, for the purposes of identifying low-dimensional
manifolds in data.  The survey of the dimensionality reduction
literature provided in Chapter~\ref{ch:nldr-overview} was deliberately
restricted to these geometrical/statistical methods, neglecting
dynamical methods, in the belief that geometrical/statistical methods
would be more appropriate for climate data analysis applications,
based on the fact that these methods are intrinsically data-based.
Most dynamical methods require explicit equations for the system under
investigation, which is not practical when dealing with observational
climate data or the results of GCM simulations.

Despite their apparent greater suitability for climate data analysis
problems, there exist a number of difficulties with
geometrical/statistical dimensionality reduction methods.  First,
there has been relatively little theoretical work done to establish
under what conditions particular methods will work well.  This
contrasts with the situation for dynamical methods which have, by and
large, been developed on strong theoretical bases.  This contrast
appears to have arisen both because some geometrical/statistical
methods are intrinsically difficult to analyse, and because of a more
applications oriented viewpoint adopted by workers developing
geometrical/statistical methods.  A related point here is that there
has been relatively little systematic testing of
geometrical/statistical dimensionality reduction methods, with most
presentations of new methods confining themselves to a small set of
simple test examples or a set of standard image recognition problems.
There has certainly not been any systematic attempt to apply these
methods to climate problems or to dynamical systems problems in a more
general sense.

The three methods selected here were all applied to the same
relatively straightforward climate data analysis problem in an
inter-model comparison setting.  From what I have seen, this has not
been done with any of these methods before, and this study provides
the first opportunity to see how well these approaches are able to
pick out distinguishing features of simulations from different models,
simulations that sometimes differ in relatively subtle ways.

The first of the methods used here, nonlinear PCA, has the most
extensive history of applications in climate data analysis of all
nonlinear dimensionality reduction methods.  However, a few open
questions remain.  First, concerning the use of neural networks with
``circular'' nodes for the modelling of periodic variations in data,
it is not clear that there is yet a particularly good rationale for
choosing a ``circular'' network over a normal, single hidden layer
neuron network.  Some of the results of Chapter~\ref{ch:nlpca}
indicate that, in situations where circular networks have previously
been used, simple one-dimensional networks provide a better fit to the
data; attempting to seek a periodic signal in the data is thus a
subjective choice that is not borne out by observed results.  Second,
the usual basis on which NLPCA is presented is as a method of
determining an ``optimal'' nonlinear projection of a data set to a
lower dimensional space.  This appears to be extremely difficult to
achieve in practice, and it seems better to consider NLPCA as
providing a weakly nonlinear extension of PCA, rather than a means of
determining a truly optimal nonlinear reduction.  For most of the
NLPCA results shown in Chapter~\ref{ch:nlpca}, relatively simple
neural network architectures provide the best fit to the input data,
representing this type of weakly nonlinear solution.  Finally, there
is the question of the use of NLPCA on noisy data, as is normally the
situation in climate applications.  Some recent work has been done on
more robust error measures for NLPCA, but there is still an element of
wishful thinking in some of the fits presented as results.  This again
is clear from some of the embedding plots and statistics in
Chapter~\ref{ch:nlpca}, where curves that do not necessarily reflect
any real structure in the data are drawn through the ``middle'' of
point data clouds.

The second method used here, Isomap, has seen one previous application
in climate data analysis, also examining tropical Pacific SST data for
ENSO variability.  Again, as for NLPCA, Isomap has not been used
before in a model intercomparison exercise.  One problem with Isomap
(and indeed many other geometrical/statistical dimensionality
reduction methods) is the selection of parameter values, i.e. the
neighbourhood size used to construct the nearest neighbour graph that
lies at the heart of the method.  Here, I have conducted some
sensitivity studies to attempt to understand how Isomap results vary
as this neighbourhood size changes.  In the case of simple geometrical
test data sets, it seems fairly clear what is happening, with distinct
changes in the Isomap eigenvalue spectra occurring as the
neighbourhoods sample different scales in the input data.  For more
complex data sets, the situation is much less clear, and a certain
degree of experimentation is required to find good parameter values.
There is also an issue surrounding the component rotation method used
to disentangle the different sources of variation seen in the
embeddings produced for the tropical Pacific SST data here.  This
approach is rather specialised for the ENSO problem considered here,
and shares some features with rotated EOF approaches for
``simplifying'' the results of PCA analyses, approaches that are also
ad hoc and difficult to justify on more objective grounds.

The final method examined, Hessian LLE, is relatively new and has not,
as far as I am aware, been applied to anything other than simple
geometrical test data sets before now.  Hessian LLE is theoretically
appealing, but again there are issues with sensitivity to data point
sampling density, algorithm parameters (as for Isomap, a neighbourhood
size parameter) and data noise.  I have explored these issues in
Chapter~\ref{ch:hessian-lle} through sensitivity studies and (to some
extent) brute force.  It would be of interest to develop some clear
theoretical bounds on the required data point sampling density and
noise levels to get reasonable embeddings, but this appears rather
difficult.

In assessing the results obtained in
Chapters~\ref{ch:nlpca}--\ref{ch:hessian-lle}, the first point to
observe is that it is often extremely difficult to do much better than
conventional principal component analysis.  The reduction in
dimensionality provided by PCA is a very natural one, eliminating, as
it does, directions in the data space that are associated with low
variance ``noise''.  Even though the individual PCA modes may not be a
faithful representation of nonlinear degrees of freedom of a data set,
they at least provide a basis for exploring the structure of data
variability in a lower-dimensional and hence much more tractable
setting.  This is reflected in the use of PCA as a preliminary
dimensionality reduction step in many geometrical/statistical
nonlinear dimensionality reduction methods, including NLPCA and, in
the way I apply it here, also Hessian LLE.  (The situation for Isomap
is a little different, since the first step of Isomap, the
construction of a distance matrix from the original data, already
incorporates a large component of dimensionality reduction, obviating
the need for an initial PCA step.)

Once an initial dimensionality reduction step has been performed, the
methods described here do find low-dimensional nonlinear structures in
the data, as shown by the embedding plots in each of
Chapters~\ref{ch:nlpca},~\ref{ch:isomap}~and~\ref{ch:hessian-lle}.  In
each case, these embedding plots capture structure that is not clearly
evident from reduced PCA data.  In this sense, all of the
dimensionality reduction methods examined here provide additional
information that is not directly available from PCA.  On the other
hand, these methods are much more complex to implement than simple
linear methods like PCA, they have much more onerous computational
requirements, and the incremental gain in insight provided by these
methods may not always be enough to justify this extra effort.  For
the problem examined here, there are other data analysis methods that
provide more insight for less cost.

It may in fact be the case that, for the majority of problems in
climate data analysis, the data sets are too noisy, too short (this is
often the case for observational data) and too inhomogeneous in terms
of spatial and temporal sampling for many of these methods to produce
good results.  More generally, this raises a potential problem with
performing a good intercomparison of different dimensionality
reduction methods.  It is not clear that there is a single corpus of
test data from real problems (in any field) to which all available
methods could reasonably be applied.  Fine control over the
characteristics of the test data (sampling density, noise levels,
etc.) would be needed to accurately characterise the behaviour of a
range of different methods, and data from observations or realistic
models may be too difficult to deal with.

Another problem with most nonlinear dimensionality reduction methods
is their dependence on parameters.  Principal component analysis is
simple: construct the data covariance matrix, calculate the
eigendecomposition and you're done.  More complex nonlinear methods
often require the construction of a nearest neighbour graph or
simplicial complex from the input data points, requiring the selection
of a spatial scale for determining which neighbours are ``near''.  For
some data sets, this is not much of a problem, but for situations with
strongly inhomogeneous data sampling, it can be extremely difficult to
select an appropriate single spatial scale.  Similar comments apply to
other tunable parameters in these methods.  Often the only approach to
follow is a brute force parameter sensitivity study followed by
careful ``by eye'' selection of good results.  This is obviously not a
particularly good way to proceed.  As well as parameter sensitivity,
there is also a question around the data sampling requirements for
some of the methods --- as shown in Chapter~\ref{ch:hessian-lle} for
Hessian LLE, there is a strong and systematic dependence of the form
of the reduced data representation found on the sampling density.
Similar investigations in a controlled setting would also be of
interest for NLPCA, Isomap and other dimensionality reduction methods.

Even with these considerations, I believe that there is a place for
some of the methods described here in climate data analysis.  They are
useful for exploratory data analysis, when one is not quite sure
exactly what it is that one is looking for, and there is some benefit
in looking at data in a variety of different ways, in the hope of
discovering some new and pertinent feature.

In terms of further work in this field, an obvious idea is to assess
the suitability for climate data analysis of more nonlinear
dimensionality reduction methods, beyond those examined in
Chapters~\ref{ch:nlpca}--\ref{ch:hessian-lle}.  Whether the type of
multi-model analysis of a particular climate phenomenon (such as ENSO)
as performed here is the appropriate vehicle for this assessment is
not clear.  While there is some benefit to using a realistic problem,
it is easy to get bogged down in the details of the phenomenon under
study, and there is little control over factors in the data that may
affect the performance of the dimensionality reduction methods.

There is however another approach that would make it possible to
examine the performance of both the geometrical/statistical
dimensionality reduction methods described in
Chapter~\ref{ch:nldr-overview} and the dynamical methods I have
neglected here.  This is to pick a well-understood partial
differential equation system, such as the Kuramoto-Sivashinsky
equation or complex Ginzburg-Landau equation, and produce data sets
from numerical integrations of this equation to use as input data for
the geometrical/statistical dimensionality reduction methods.  In this
setting, it is easy to control the spatial and temporal sampling of
the data sets used, permitting side-by-side comparison of different
dimensionality reduction methods under controlled conditions.  By
using a relatively simple system expressed as an explicit equation or
set of equations, it would also be possible to apply some dynamical
reduction methods, and to compare the results of integrating the
resulting reduced dimensionality models with the structures found by
the geometrical/statistical methods from integrations of the original
system.


%% file: glossary.tex
\begin{description}

  \item[autoassociative neural network]{Neural network trained by
    comparing network inputs to outputs, with the goal of finding a
    set of network weights able to reproduce the input signal as
    closely as possible.}

  \item[Cantor set]{Perfect point set, often constructed as a subset
    of an interval, that is uncountable (i.e. has the same cardinality
    as the original interval), of measure zero, and is nowhere dense
    in the interval from which it is constructed.  Cantor sets have
    several strange topological properties: all points in a Cantor set
    are accumulation points, but no point is an interior point; a
    Cantor set is totally disconnected, and so on.  The simplest
    example is the Cantor middle-thirds set, constructed by repeatedly
    deleting the open middle thirds of a set of line segments:
    starting with the unit interval $[0, 1]$, remove the middle third
    to leave the set $[0, 1/3] \cup [2/3, 1]$, then remove the middle
    thirds from each of the remaining sub-intervals, and repeat the
    removal process indefinitely.  The Cantor middle thirds set is
    composed of the points in the interval that are left over after
    this deletion process.}

  \item[cardinality]{Naively, the cardinality of a set $S$, denoted
    $\#S$, is just the number of elements in the set.}

  \item[chaotic dynamics]{Deterministic aperiodic dynamics with
    sensitive dependence on initial conditions.}

  \item[climate variability]{General term used to refer to
    time-dependence at different temporal and spatial scales in the
    climate system.}

  \item[conservative]{A conservative dynamical system is one whose
    evolution preserves phase space volumes.  Equivalently, in common
    cases, evolution of the system preserves energy, meaning that the
    equations of motion of the system can be cast in Hamiltonian
    form.}

  \item[degree of freedom]{For a dynamical system, the number of
    degrees of freedom counts the number of coordinates required to
    uniquely specify a state of the system.}

  \item[dissipative]{The evolution of a dissipative dynamical system
    shrinks phase space volumes.}

  \item[dynamical system]{General term covering most mathematical
    systems for which a concept of \emph{time} is meaningful.
    Continuous examples include differential equation systems of
    various types, while discrete examples include discrete time
    maps.}

  \item[easterly]{For winds, blowing \emph{from} the east.}

  \item[Ekman flow]{Wind-induced motion of ocean near-surface waters.}

  \item[\eln]{Phase of the \eln/Southern Oscillation characterised by
    conditions in the equatorial Pacific showing anomalously warm sea
    surface temperatures in the east, an anomalously weak zonal
    thermocline slope and anomalously weak zonal wind stress.}

  \item[\eln/Southern Oscillation]{Irregular interannual variation in
    ocean and atmosphere conditions, primarily in the tropical
    Pacific, although with impacts in other regions, characterised by
    vacillations in equatorial Pacific sea surface temperature,
    thermocline and wind conditions.}

  \item[flow]{On a manifold $X$, a function $\phi : X \times
    \mathbb{R} \to X$ with the properties that $\phi(x, 0) = x$ and
    $\phi(\phi(x, t), s) = \phi(x, t + s)$ for all $t, s \in
    \mathbb{R}$ and $x \in X$.  These two conditions mean that
    elements of the flow constitute a semigroup parameterised by the
    time parameter $t$.}

  \item[Galerkin projection]{Method of converting a continuous
    mathematical problem (e.g. a partial differential equation) into a
    discrete one (e.g. a system of ordinary differential equations).
    The finite element approach to solving PDE problems is essentially
    a Galerkin method.}

  \item[general circulation model]{Climate or numerical weather
    prediction model based on numerically discretising and integrating
    the primitive equations for fluid motion on a sphere.}

  \item[geopotential height]{Vertical coordinate in the atmosphere
    that simplifies certain calculations, including the representation
    of the primitive equations for flow in the atmosphere.  Define the
    geopotential height at elevation $h$ as $\Phi = \int_0^h g(z) \,
    dz$, where $g(z)$ is the acceleration due to gravity at elevation
    $z$.  Then, the geopotential height $z_g = \Phi / g_0$, where
    $g_0$ is the standard acceleration due to gravity at sea level.}

  \item[global attractor]{Finite-dimensional subspace of the state
    space of an infinite-dimensional dissipative dynamical system to
    which the long term dynamics of the system is confined.  Not all
    infinite-dimensional dynamical systems have global attractors.}

  \item[Hadley cell]{The thermally direct circulation of the
    atmosphere driven by the contrast in solar heating between
    equatorial and off-equatorial regions, characterised by rising
    motion near the equator, meridional motion away from the equator
    at upper levels of the troposphere, downwelling motion in the
    subtropics and a low-level equatorward return flow.}

  \item[inertial manifold]{Smooth finite-dimensional manifold in which
    the global attractor (q.v.) of an infinite-dimensional dynamical
    system may be embedded.}

  \item[Intertropical Convergence Zone]{Low-pressure region near
    equator associated with thermally driven convective upwelling.}

  \item[invariant manifold]{Invariant set (q.v.) which is also a
    manifold.}

  \item[invariant set]{Set of points in the phase space of a dynamical
    system (continuous or discrete) with the property that
    trajectories with initial conditions in the set remain within the
    set for all time.}

  \item[Kelvin wave]{Wave motion in a rotating fluid where the
    Coriolis force is balanced by a boundary.  In the ocean and
    atmosphere, equatorial Kelvin waves are possible, with the equator
    acting as a virtual boundary, and the gradient of the Coriolis
    force (which is zero at the equator and increases in both
    directions away from the equator) providing a restoring force,
    permitting eastwards propagating waves.}

  \item[\lan]{Phase of the \eln/Southern Oscillation characterised by
    conditions in the equatorial Pacific showing anomalously cool sea
    surface temperatures in the east, an anomalously strong zonal
    thermocline slope and anomalously strong zonal wind stress.}

  \item[Lyapunov exponent]{Quantity measuring the rate of separation
    of trajectories for a dynamical system.  Informally, for two
    points with an initial separation $\delta \vec{x}_0$, the
    separation at time $t$ follows $||\delta \vec{x}(t)|| \approx
    e^{\lambda t} ||\delta \vec{x}_0||$, where $\lambda$ is the
    Lyapunov exponent.  Since different rates of expansion (or
    contraction) are possible in different directions in phase space,
    there exists a whole spectrum of Lyapunov exponents.}

  \item[Madden-Julian Oscillation]{Intraseasonal travelling wave
    oscillation in rainfall in the western Pacific, with a timescale
    of 30--60 days, characterised by eastward propagation of regions
    of strong convection and high rainfall, followed by regions of
    weaker convection and suppressed rainfall.}

  \item[manifold]{Topological space, every point of which has a
    neighbourhood homeomorphic to an open set of Euclidean space.
    Informally, a topological space that everywhere locally looks like
    Euclidean space.}

  \item[Markov model]{Any statistical model with the Markov property,
    i.e. for a sequence $x_i$, the conditional probabilities for the
    next value in the sequence satisfy
    \begin{equation*}
      \mathbb{P}(x_{i+1} | x_i, x_{i-1}, \dots, x_0) =
      \mathbb{P}(x_{i+1} | x_i),
    \end{equation*}
    meaning that the system ``has no memory''.}

  \item[multilayer perceptron]{Feedforward neural network with one or
    more internal hidden layers of neurons.}

  \item[ocean heat content]{Mean or integrated ocean temperature in
    the upper 300m of water over a given region.}

  \item[quasi-periodic route to chaos]{Mechanism for producing chaotic
    dynamics in nonlinear systems with periodic forcing, characterised
    by nonlinear resonances and phase locking between intrinsic
    oscillations and the external forcing.}

  \item[Rossby wave]{Wave motion in a rotating fluid where a restoring
    force is provided by the meridional gradient of the Coriolis
    force.  Long wavelength off-equatorial Rossby waves in the
    tropical Pacific propagate westwards.}

  \item[saddle point]{Fixed point of a dynamical system stable in some
    directions and unstable in others.}

  \item[spectral grid]{Basis for the discretisation of fluid dynamical
    equations on the sphere based on spherical harmonics.  A spectral
    grid designation for the resolution of an atmospheric general
    circulation model specifies the highest wavenumbers treated in the
    discretisation and the form of the truncation of the modal
    expansion of the model fields.}

  \item[strange attractor]{Attractor of a dynamical system that is
    also a chaotic invariant set.}

  \item[Walker circulation]{Zonal circulation in the equatorial
    Pacific atmosphere, with rising convective flow over the western
    Pacific, downwelling over the eastern Pacific, westerly high-level
    winds and easterly low level winds, driven by contrast in sea
    surface temperatures from the warm west to the cooler east of the
    Pacific basin.}

  \item[westerly]{For winds, blowing \emph{from} the west.}

  \item[westerly wind burst]{Brief period of intensified westerly
    winds in the western Pacific associated with the Madden-Julian
    Oscillation (q.v.); implicated in the triggering of some ENSO
    events.}

  \item[Western Warm Pool]{The region of warmer waters found in the
    western part of the Pacific Ocean, generated by the action of the
    climatological easterly zonal wind field over the Pacific basin.}

  \item[zonal]{Referring to the longitudinal direction on a sphere,
    e.g. a zonal mean is a mean over longitude values.}

\end{description}